\documentclass[letterpaper,11pt]{article}
\pdfoutput=1

\usepackage{jheppub} 
\usepackage{hyperref} 
\usepackage{epsfig}
\usepackage[normalem]{ulem}
\usepackage{bm}
\usepackage{bbm}
\usepackage{slashed}
\usepackage{xspace}
\usepackage{subfigure}
\usepackage{multirow}

\newcommand{\setcaptionskip}{\setlength\baselineskip{14pt}}
\newcommand{\setmainskip}{\setlength\baselineskip{18pt}}

\usepackage{amsmath,latexsym}
\usepackage{pstricks}
\usepackage{color} 
 
\usepackage{marginnote}

\newcommand{\eq}[1]{Eq.~\eqref{eq:#1}}
\newcommand{\eqs}[2]{Eqs.~\eqref{eq:#1} and \eqref{eq:#2}} 
\renewcommand{\sec}[1]{Sec.~\ref{sec:#1}}
\newcommand{\secs}[2]{Secs.~\ref{sec:#1} and \ref{sec:#2}}

\newcommand{\fig}[1]{Fig.~\ref{fig:#1}}
\newcommand{\figs}[2]{Figs.~\ref{fig:#1} and \ref{fig:#2}}

\newcommand{\app}[1]{App.~\ref{app:#1}}

\newcommand{\tab}[1]{Table~\ref{table:#1}}

\newcommand{\plus}{\!+\!}
\newcommand{\minus}{\!-\!}

\newcommand{\ddslash}{{d\!\!{}^-}}

\newcommand{\id}{\mathbbm{1}}

\newcommand{\leftpartial}{%
  \mathrel{\vbox{\offinterlineskip\ialign{%
    \hfil##\hfil\cr
    $\scriptscriptstyle\leftarrow$\cr
    $\partial$\cr
}}}}
\newcommand{\rightpartial}{%
  \mathrel{\vbox{\offinterlineskip\ialign{%
    \hfil##\hfil\cr
    $\scriptscriptstyle\rightarrow$\cr
    $\partial$\cr
}}}}

\newcommand{\Ecm}{E_\mathrm{cm}}

\newcommand{\nn}{\nonumber}
\newcommand{\mcdot}{\!\cdot\!}
\newcommand{\beq}{\begin{equation}}
\newcommand{\eeq}{\end{equation}}
\newcommand{\bea}{\begin{eqnarray}}
\newcommand{\eea}{\end{eqnarray}}

\newcommand{\lqcd}{\Lambda_\mathrm{QCD}}
\newcommand{\MSbar}{\mbox{${\overline {\rm MS}}$}}

\newcommand{\bn}{{\bar n}}
\newcommand{\bnP}{\bar {\cal P}}

\newcommand{\cP}{{\cal P}}
\def\bnslash{\bar n\!\!\!\slash}
\def\nslash{n\!\!\!\slash}
\newcommand{\SCETa}{\mbox{${\rm SCET}_{\rm I}$}\xspace}
\newcommand{\SCETb}{\mbox{${\rm SCET}_{\rm II}$}\xspace}

\allowdisplaybreaks[1]

\begin{document}



\preprint{\vbox{\hbox{arXiv:1601.04695}\hbox{MIT--CTP 4655}}}

\title{\boldmath An Effective Field Theory for Forward Scattering and Factorization Violation}

\author{Ira Z. Rothstein$^1$ and Iain W. Stewart$^2$}

\affiliation{$^1$Department of Physics, Carnegie Mellon
  University, Pittsburgh, PA 15213, USA}
\affiliation{$^2$Center for Theoretical Physics, Massachusetts Institute of Technology, Cambridge, MA 02139, USA}

\emailAdd{izr@andrew.cmu.edu}
\emailAdd{iains@mit.edu}

\abstract{

  \vspace{0.2cm}
 \setlength\baselineskip{15pt}

Starting with QCD, we derive an effective field theory description for forward scattering and factorization violation as part of the soft-collinear effective field theory (SCET) for high energy scattering.  These phenomena are mediated by long distance Glauber gluon exchanges, which are static in time, localized in the longitudinal distance, and  act as a kernel for  forward scattering where $|t| \ll s$. In hard scattering, Glauber gluons can induce corrections which  invalidate factorization. With SCET, Glauber exchange graphs can be calculated explicitly, and are distinct from graphs involving soft, collinear, or ultrasoft gluons. We derive a complete basis of operators which describe the leading power effects of Glauber exchange. Key ingredients include regulating light-cone rapidity singularities and subtractions which prevent double counting. Our results include a novel all orders gauge invariant pure glue soft operator which appears between two collinear rapidity sectors. The 1-gluon Feynman rule for the soft operator coincides with the Lipatov vertex, but it also contributes to emissions with $\ge 2$ soft gluons. Our Glauber operator basis is derived using tree  level and one-loop matching calculations from full QCD to both \SCETb and \SCETa.  The one-loop amplitude's rapidity renormalization involves mixing of color octet operators and yields gluon Reggeization at the amplitude level.  The rapidity renormalization group equation for the leading soft and collinear functions in the forward scattering cross section are each given by the BFKL equation.  Various properties of Glauber gluon exchange in the context of both forward scattering and hard scattering factorization are described. For example, we derive an explicit rule for when eikonalization is valid, and provide a direct connection to the picture of multiple Wilson lines crossing a shockwave. In hard scattering operators Glauber subtractions for soft and collinear loop diagrams ensure that we are not sensitive to the directions for soft and collinear Wilson lines. Conversely, certain Glauber interactions can be absorbed into these soft and collinear Wilson lines by taking them to be in specific directions. 
We also discuss criteria for factorization violation.   
}



\maketitle
\setmainskip



\section{Introduction}
\label{sec:intro}

Progress in our understanding of interacting four dimensional quantum field theories has come from various directions. The profoundly rich structure of asymptotically free confining theories, such as QCD, has been illuminated through the use of various tools which have been developed over the past 50 years. For certain nonperturbative problems we have lattice calculations at our disposal, while for heavy particles and hard scattering processes perturbative QCD has significant predictive power when used in conjunction with factorization theorems or operator expansions. In  these cases there are effective field theory (EFT) tools for various power expansions, for example, in the lattice spacing, in light and heavy quark masses, and in ratios of kinematic variables, and the field theory formalism to carry out these expansions has been worked out to subleading orders. 

A large number of open questions in QCD arise when considering aspects of near forward scattering,  which dominates the total cross section, and where approximations are often needed to study the leading power term. If we consider two-to-two scattering then we can define near forward scattering with Mandelstam variables as $|t|\ll s$, where  ``near"  is quantified by the small ratio $|t|/s$.  This limit is often referred to as the high-energy limit or the Regge limit.  Beyond two to two scattering there are more possibilities and one or more observables must be chosen to quantify the notion of nearness which take the place of $|t|$.  It could, for instance, be $t$-channel momenta defined through various pairs of particles, or a cut on the rapidity or momentum transverse to the scattering axis.  Near forward scattering can be contrasted with the case of hard scattering where $|t|\sim s$ are both large. 

In forward scattering, if $t$ or a suitable generalization is much larger than the scale of strong coupling, $|t|\gg \Lambda_{\rm QCD}^2$, then there is a well defined  notion of a perturbative scattering vertex and one may attempt to factorize the long distance physics, with wavelengths of order the scale $\Lambda_{\rm QCD}$, from the relatively short distance physics at the scale $t$.  In particular we can aim to find observables for which one can factorize rates into a perturbative scattering kernel and a set of long distance matrix elements like parton distribution functions, in such a way as to preserve a reasonable amount of predictive power.  At leading power in the forward scattering limit, there are no states with virtuality of order $s$ contributing to the cross section, and the scale $s$ arises dynamically through a separation of modes in rapidity space, where the sum of contributions from two sectors $\ln Q_\pm$ can yield $\ln Q_+ + \ln Q_- =\ln s$.  The scale $s$ still plays an important role, due to the existence of parametrically large logs from the hierarchy $|t|/s\ll 1$, which can cause standard perturbation theory to break down. From the relation $x\sim |t|/s$ that appears in DIS and Drell-Yan, the forward scattering limit is also often referred to as the small-$x$ limit. Examples of formalisms designed to sum these logarithms and to treat the associated small-$x$ physics include the classic Balitsky--Fadin--Kuraev--Lipatov (BFKL) equation~\cite{Kuraev:1977fs,Balitsky:1978ic}, the dipole approximation~\cite{Mueller:1994gb}, the Balitsky--Kovchegov (BK) equation~\cite{Balitsky:1995ub,Kovchegov:1999yj}, the Balitsky--Jalilian-Marian--Iancu--McLerran--Weigert--Leonidov--Kovner (BJIMWLK) equation~\cite{JalilianMarian:1997jx,Iancu:2000hn}, as well as general calculational tools like the use of Wilson lines~\cite{Korchemsky:1993hr,Korchemskaya:1994qp,Korchemskaya:1996je}, effective actions~\cite{Lipatov:1995pn,Lipatov:1996ts} and the multi-Wilson line EFT~\cite{Balitsky:1995ub,Balitsky:1998ya,Balitsky:2001gj}. The structure of the BJIMWLK evolution of Wilson lines at higher orders has been analyzed in~\cite{Caron-Huot:2013fea}. These resummations are of phenomenological relevance and are also useful tools for checking all orders ans\"atze which are often made when trying to form theorems regarding the all orders form of amplitudes, see e.g.~\cite{Bret:2011xm,DelDuca:2011ae,Almelid:2015jia}. Finally when $t$ becomes of ${\cal O}(\Lambda_{\rm QCD}^2)$ we are dealing with the non-perturbative scattering of partons that are most often within a bound proton or nucleus, and there is no longer a factorization between $t$ and $\Lambda_{\rm QCD}^2$.

The study of hard scattering processes at hadron colliders is crucial for exploring short distance physics. Here an important role is played by factorization formulae which allow one to define universal functions describing different types of perturbative and nonperturbative physics. The near forward scattering region can play an important role in hard scattering processes as well, since not all of the partons in a hadron are active, i.e. involved in the hard scattering. These additional ``spectator" partons can interact with each other even when they are in hadrons traveling in opposite directions, through processes closely akin to forward scattering. These interactions can spoil factorization for a hard scattering process, since they couple together partons associated to different hadrons, and hence cannot be described solely by single hadron matrix elements, like parton distribution functions. The process through which massless spectators interact is called ``Glauber exchange" and proving that such interactions cancel in various observables has been a subject which, while well appreciated, has often not received the attention it deserves. Derivations of factorization theorems for hadron collisions exist in only a few special cases, namely for Drell-Yan-like process~\cite{Bodwin:1984hc,Collins:1984kg,Collins:1988ig}, for single inclusive hadron production~\cite{Collins:1989gx,Nayak:2005rt}, and recently for double-parton scattering~\cite{Diehl:2015bca} (with arguments for situations with observed jets in~\cite{Aybat:2008ct}), each using the techniques of CSS~\cite{Collins:1988ig}. These results are often used to motivate using factorization to make predictions in other hadron-hadron scattering observables without complete proofs. In general there are many other important ingredients in factorization proofs, including hard-collinear factorization, soft-collinear factorization, ultrasoft-collinear factorization, factorization for the observable, as well as the uniqueness of soft and  collinear Wilson lines.  The Soft-Collinear Effective Theory (SCET)~\cite{Bauer:2000ew,Bauer:2000yr,Bauer:2001ct,Bauer:2001yt,Bauer:2002nz} has allowed this set of questions to be addressed with Lagrangian and operator based methods, facilitating significant advances in the range of processes to which we can consider applying factorization formulae. However, an operator based description of Glauber exchange has not yet been formulated in SCET.

\begin{figure}[t!]
%
%
\raisebox{0.cm}{\hspace{0.3cm} a)\hspace{5.5cm}b)\hspace{7.6cm}}
\raisebox{0.1cm}{\hspace{0.6cm}
\includegraphics[width=0.3\columnwidth]{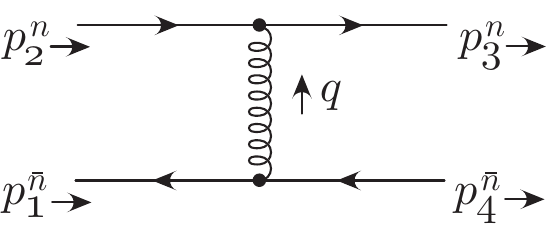}
}
\hspace{0.5cm}
\raisebox{-0.1cm}{\hspace{0.3cm}} 
\raisebox{0.cm}{  
\includegraphics[width=0.23\columnwidth]{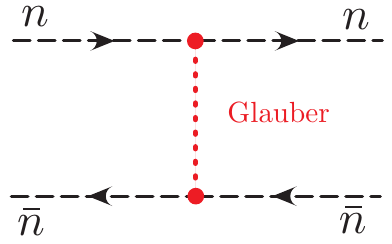}
  }
\raisebox{1.1cm}{\Large $=$}
\raisebox{0.2cm}{  
\includegraphics[width=0.21\columnwidth]{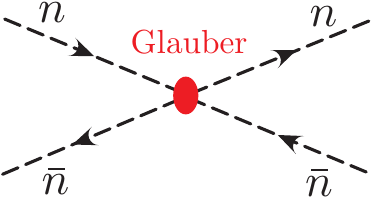}
   }
\vspace{-0.8cm}
\caption{\setcaptionskip
Tree level gluon exchange for $q$-$\bar q$ forward scattering. In a) we show the full QCD graph with a gluon exchange between a quark carrying $n$-collinear momenta $p_{2,3}^n$ and an antiquark carrying $\bn$-collinear momenta $p_{1,4}^\bn$. In b) we show the two notations we will use for this leading power forward scattering in the Effective Theory. 
}
\setmainskip
\label{fig:glaub_tree}
\end{figure}

An example of a Glauber exchange is shown in Fig.~\ref{fig:glaub_tree}. These exchanges are responsible both for leading power forward scattering as well as factorization violation in hard scattering, and are discussed in more detail in \secs{modes}{GlauberEFT}. If Glauber exchange contributions do not spoil factorization, a formalism to treat Glauber exchange can still provide a useful perturbative tool to facilitate the summation of large logs that appear from the forward limit, $\ln(s/t)$ or $\ln(x)$, where $x$ is an appropriate Bjorken-type variable. In situations where factorization is violated a formalism to treat Glauber exchange can be a useful tool for both calculating and characterizing the nature of the violations. 

The purpose of this paper is to set up a systematic effective theory with which to study the near forward scattering region of QCD and factorization violation in hard scattering processes in a single framework. We will work within the framework of SCET. We construct a complete leading power Lagrangian for Glauber exchange and show that it fits seamlessly with the current tools used to study hard, collinear, soft, and ultrasoft factorization in hard scattering processes, without inducing double counting. By working in the framework of an effective field theory, one is able to systematically keep track of terms in the power expansion, exploit symmetries, and derive when certain approximations (like the eikonal approximation) are valid and when they break down. Our EFT will also employ a $\overline{\rm MS}$ style renormalization for rapidity divergences, making it simple to derive rapidity renormalization group equations. Through matching calculations we can also 
directly derive and prove results by calculations in full QCD in the appropriate limit. The formalism presented here gives a starting point for using a field theoretic method to study the physics of the near forward region, even beyond leading power. It also provides a direct method of calculating (possible) factorization violating contributions, and potentially could yield field theoretic methods for handling underlying event contributions in hadronic collisions.

Before proceeding, we briefly comment on the connections of our work to earlier literature. First we note that in the CSS formalism~\cite{Collins:1988ig,Collins:2011zzd} that Glauber contributions are discussed in detail, but are treated as a momentum region and hence are not fully separated from soft and collinear gluon dynamics. This has advantages for certain steps of a  factorization proof, but makes it more difficult to associate unique contributions with Glauber exchange, and also to see how factorization arises for processes that retain nontrivial soft functions. The method of regions~\cite{Beneke:1997zp} has been used to study the Glauber region in Ref.~\cite{Jantzen:2011nz},  but an effective field theory for Glauber exchange has not emerged from this framework. Although Glauber gluons have been formulated as distinct fields in SCET in Refs.~\cite{Idilbi:2008vm,DEramo:2010ak,Ovanesyan:2011xy,Ovanesyan:2012fr,Benzke:2012sz}, this has only been done for cases where they are treated as a background field, such as heavy-ion collisions. This formalism with background fields is referred to as SCET$_{\rm G}$, and we reserve this notation for referring to background Glauber fields. For this case issues with singularities and double counting are easier to control.  For the situation of interest here, Glauber gluon exchange should be treated with operators yielding a scattering potential involving soft and collinear fields rather than explicit Glauber fields, as discussed in~\cite{Iain2010}.   A subtraction formalism that avoids double counting for Glauber gluon exchange in loop diagrams was discussed in Refs.~\cite{Bauer:2010cc,Iain2010}, but  so far only simpler abelian cases have been addressed. (We also note that Glauber interactions cannot in general be completely eikonalized as in Ref.~\cite{Liu:2008cc}.)  In Ref.~\cite{Donoghue:2009cq} the importance of Glauber exchange in SCET for Regge phenomena was emphasized, and in Ref.~\cite{Donoghue:2014mpa} Glauber exchange was analyzed for scalar $\phi^3$ theory, arguing that its absence in the threshold expansion with certain regulators is not indicative of its absence in a unitary effective field theory.
Steps toward deriving an SCET based operator description of Glauber exchange for the full non-abelian case of QCD were taken in Ref.~\cite{Fleming:2014rea}, but a full Lagrangian was not obtained.

Various examples of factorization violation have been studied in the literature. An important type of factorization violation occurs if we cannot disentangle the physics associated to the colliding protons into independent parton distribution functions, as in the CSS analysis~\cite{Collins:1988ig}. Other examples include processes  where the measurement does not factorize in a sufficiently simple manner, such as for the Jade algorithm, see~\cite{Walsh:2011fz}, or where the structure of collinear Wilson lines cannot be uniquely determined for the hard scattering process such as hadron production $H_1+H_2\to H_3+H_4+X$ with a measured small $p_T(H_3H_4)$~\cite{Collins:2007nk,Collins:2007jp,Mulders:2011zt,Aybat:2011vb}.  There are also examples of factorization violation at the amplitude level. This includes factorization violation for splitting functions in space-like collinear limits~\cite{Catani:2011st,Forshaw:2012bi}, which are connected to superleading logarithms~\cite{Forshaw:2006fk}. Another example is Regge Factorization violation from terms that go beyond the Regge amplitude formula~\cite{DelDuca:2001gu,DelDuca:2013ara,DelDuca:2014cya}.  These examples of factorization violation are related to Glauber gluon exchange, and hence can be explored in our formalism.

In the context of small-$x$ physics a multi-Wilson line effective field theory for forward scattering was constructed by Balitsky in Refs.~\cite{Balitsky:1995ub,Balitsky:1998ya,Balitsky:2001gj}. In this framework rapidity factorization separates the amplitude into coefficient functions and matrix elements of multi-Wilson line operators, and the effective Lagrangian has an infinite number of terms. In contrast, in our EFT rapidity factorization separates soft and collinear modes at the level of a Glauber Lagrangian with a fixed number of terms at leading power, and the soft and collinear modes themselves still appear on the same footing as modes in the EFT. This leading Lagrangian can be inserted any number of times when constructing leading power amplitudes. Our Glauber interactions are not a priori eikonal, but become so when it is appropriate.  This makes it possible to use our EFT framework for both forward and hard scattering processes. The soft modes in our Glauber operators have soft Wilson line interactions which are the most relevant at the leading logarithmic order, as well as terms involving dressed soft field strengths that are important at higher orders. Our collinear modes effectively provide an EFT for the Wilson coefficient source terms of Ref.~\cite{Balitsky:2001gj}. We will elaborate on the connection between our EFT and the multi-Wilson line framework in \secs{forwardgraphs}{semiclassical}.

\section{Guide for the Reader}

This paper is written in a fairly modular fashion so that readers can meet their needs without necessarily reading the entire manuscript. In \sec{modes} we introduce the EFT quark and gluon modes for forward and hard scattering, and in \sec{SCET}  we provide a short review of SCET and our notation, which covers the material needed for the remainder of the paper. (Further background information on EFT and SCET can be obtained from the free online 8.EFTx course~\cite{EFTx}.) Readers whose primary interest lies in the construction of the Glauber Exchange EFT should read Secs.~\ref{sec:GlauberEFT}, \ref{sec:match}, \ref{sec:loopmatch}, those interested in forward scattering and resummation should read Secs.~\ref{sec:GlauberEFT}, \ref{sec:loopmatch}, \ref{sec:BFKL}, \ref{sec:phases}, and those who are interested the role of Glaubers in hard scattering and factorization violation  can read Secs.~\ref{sec:GlauberEFT}, \ref{sec:properties}, \ref{sec:spectator}.  

As a further guide, we also summarize individual sections in greater detail.  In \sec{SCETLag} we discuss the Lagrangians involving various soft and collinear fields in \SCETa and \SCETb, and give in \sec{SCETOp}  the standard gauge invariant building blocks for SCET operators.  \sec{GlauberSCET} gives the key results for the gauge invariant Glauber operators in \SCETa and \SCETb. These operators describe offshell Glauber exchanges at leading order in the power counting and to all orders in $\alpha_s$. Many of the results listed in this section are derived systematically in later sections. In \sec{multiglauber} we describe our method for regulating rapidity singularities which uses and builds on Refs.~\cite{Chiu:2011qc,Chiu:2012ir}, and discuss the structure of zero-bin subtractions~\cite{Manohar:2006nz} in the presence of Glauber modes. In \sec{powercount} we give the power counting formula for SCET in the presence of Glaubers that is valid to all orders in $\alpha_s$ and all orders in the power expansion, as well as discuss the completeness of our leading power Glauber Lagrangian. Observables to which the Glauber Lagrangian contributes are discussed in \sec{forward}.  

In~\sec{treematch} we derive the structure of collinear and soft Wilson lines in the Glauber operators by carrying out tree level matching calculations.  Because of time-ordered product contributions from diagrams with onshell propagators in the EFT these matching calculations are more non-trivial than the SCET matching done to yield Wilson lines in Refs.~\cite{Bauer:2000yr,Bauer:2001yt,Bauer:2002nz}. The use of the equations of motion are also crucial to our formulation of gauge invariant Glauber operators. This use of the equations of motion explains why the contributions from these operators appear to be so different from those that one would infer by attempting to use the method of regions to define Glauber interactions in an EFT setting.  In section~\sec{lipatov} we show that the one-gluon Feynman rule of a particular  soft operator (${\cal O}_s^{AB}$) reproduces the Lipatov vertex. In~\secs{basis}{2sgluon} we construct a complete basis of soft operators and carry out tree level matching calculations with up to two soft gluons in order to derive the complete soft operator appearing in the Glauber Lagrangian between $n$ and $\bn$ rapidity sectors from first principles.  

In sections~\sec{loop2match} and~\sec{loop1match}  we carry out one-loop matching calculations for $q\bar q$ forward scattering in \SCETb and \SCETa respectively.  This allows us to demonstrate our use of dimensional regularization and rapidity regulators, showing how the infrared divergences due to an IR mass or offshellness in the full QCD results are exactly reproduced by a sum of contributions in the effective theory.  Each diagram in the effective theory also probes only one scale for the invariant mass $\mu$ and one scale for the rapidity renormalization parameter $\nu$.  We demonstrate that even the full theory one-loop constants are reproduced by the EFT calculation, and explain the connection of this result to the absence of nontrivial dynamics at the hard scale in forward scattering. In \sec{regge} we consider the renormalization of the rapidity divergences that appear in the one-loop  virtual amplitudes in \SCETb, demonstrating that the manner in which they renormalize the octet operators precisely corresponds with gluon Reggeization. This result appears for all operators with octet quantum numbers (both from two quarks and from two gluons), and operator mixing plays an important role in yielding the Reggeization.

In~\sec{BFKL} we square, factorize, and renormalize the first non-trivial amplitude for forward scattering in the presence of Glauber gluon exchange. In particular, we consider processes involving $n$-$\bn$ scattering without phase space restrictions on soft gluons.  For these processes we define novel soft and collinear functions in \sec{factBFKL}, and determine their rapidity renormalization group evolution by extending the earlier soft virtual calculations to include divergences from real soft radiation.  We explicitly compute the one-loop rapidity renormalization group equation for this soft function and show that  it is given by the BFKL equation in \sec{bfkl}. Using renormalization group consistency we also show that the rapidity renormalization group equation (RGE) for each of the collinear functions is given by a BFKL equation in \sec{bfklconsist}. 

In \sec{exponentiation} we calculate graphs appearing from iterations of the Glauber operator. Any crossed diagram carrying Glauber momentum vanishes to all orders in $\alpha_s$ with our rapidity regulator. The sum of all pair-wise iteration diagrams yields the classic eikonal phase, $\exp\big[ {i\phi_G(b_{\perp})}\big]$, where $b_\perp$ is the distance between the particles in transverse impact parameter space (conjugate to the exchanged transverse momenta $q_\perp$). We also give a spacetime picture for the rapidity regulator for Glauber potentials. In \sec{forwardgraphs} we discuss the general structure of iterated Glauber exchange in the presence of   radiation and non-Glauber loops. We show that such instantaneous iterations yield vanishing graphs unless they can be collapsed to a common longitudinal position. We also derive the conditions under which source propagators that accompany Glauber exchange eikonalize.  Then in \sec{semiclassical} we make the connection between Glauber exchange and the semiclassical picture, and derive the connection of our EFT framework to the multi-Wilson line EFT framework and the shock-wave picture.

In~\sec{properties} we consider graphs involving Glauber operators in the presence of a hard scattering vertex.   In \sec{hardmatching} we demonstrate that for interactions between active lines there is an overlap between Glauber exchange and a contribution that is naively present in soft gluon exchange. In particular, the Glauber 0-bin subtraction for soft loop graphs is exactly equal to the Glauber contribution, implying that the same results are obtained for this type of hard scattering diagram whether or not Glauber exchange is included. In \sec{softemission} we extend this analysis to a single soft gluon emission at one-loop in the $ee$, $ep$ and $pp$ channels, showing how Glauber exchange reproduces the $i\pi$ terms in the one-loop soft current of Ref.~\cite{Catani:2000pi}. Then in \sec{higherorder} we extend our discussion of active lines with soft and Glauber diagrams and their overlaps to two-loops.

In \sec{spectator} we consider the additional complications that arise when including interpolating fields for the initial proton states in a collider hard scattering process. This leads to a classification of Glauber exchange diagrams into spectator-spectator (S-S), active-spectator (A-S) and active-active (A-A) cases which are treated in Secs.~\ref{sec:ssfactorization}, \ref{sec:asfactorization}, and \ref{sec:aafactorization}, respectively. For the simplest examples in these categories we show that iterated Glauber exchange  yields either phases or contributions that are related to the direction of soft or collinear Wilson lines in the hard scattering operators (i.e. can be absorbed into Wilson lines), and hence cancel for an inclusive cross section. 

In~\sec{conclusion} we conclude.   Several more technical calculations are included in appendices.  This includes a derivation of a general SCET power counting formula in the presence of Glauber exchange in~\app{powercount}. In~\app{useful} we summarize useful formula for coupling expansions, loop integrals, and a few Feynman rules we use that are not given in~\sec{GlauberEFT}. In~\app{expcalcs} we carry out several iterative Glauber exchange calculations that produce phases in the presence of a hard scattering.

\section{Glauber Exchange and Modes for Forward and Hard Scattering}  
\label{sec:modes}

The mechanism for near forward scattering is often referred to
as ``Glauber exchange''~\footnote{To our knowledge, the use of ``Glauber'' for exchanges with these momenta occurred first in Ref.~\cite{Bodwin:1981fv}.} which  applies to the exchange of an off shell  gluon(s) or photon(s) whose transverse momentum (relative to the incoming beams) is  hierarchically larger than the longitudinal components of the momentum four vector, $k_\perp^2 \gg k^+k^-$. This is distinct from the limit associated to Coulomb exchange, $\vec k^{\,2}\gg (k^0)^2$, where $m+k^0$ is the total energy of a heavy source of mass $m$. (For the Coulomb case with two heavy sources the power counting is done in the relative velocity $v\sim k^0/|\vec k|\sim |\vec k|/m$.)  A tree level example of Glauber exchange between a forward scattered $q\bar q$ pair is shown in Fig.~\ref{fig:glaub_tree}, where the graph gives rise to a potential
\begin{align}
  V_{\rm G}(q_\perp) = - \frac{8\pi\alpha_s(\mu)}{\vec q_\perp^{\,2}  }= 
      \frac{8\pi\alpha_s(\mu)}{t}  \,.
\end{align}
Glauber  and Coulomb exchange share many of the same properties: both are instantaneous in time and lead to poles in scattering amplitudes in the $t/s\to 0$ and $v\to 0$ limits respectively. The Glauber and Coulomb exchanges both generate classical field configurations via summing ladder diagrams (see Sec.~\ref{sec:phases}), and dressing these exchanges with loop graphs gives rise to large logs of the dimensionless parameters $v$ and $t/s$ respectively.  Differences include the fact that Glauber exchange is instantaneous in longitudinal distance and hence more singular, and that Glauber sources can undergo collinear splittings at leading power, unlike  heavy particles.  The structure of modes in SCET that we describe below (with Collinear, Soft, Ultrasoft, and Glauber  modes) also has both similarities and differences to the formulation of NonRelativistic QCD (NRQCD) in Ref.~\cite{Luke:1999kz} (involving potential operators, and simultaneously soft and ultrasoft modes).

The field theory ingredients for our formalism are familiar from hard scattering factorization, namely various soft and collinear fields and their corresponding regions in momentum space.  To introduce some of the key concepts consider as an example the factorization theorem for inclusive Higgs production via gluon fusion, 
\begin{align}  \label{eq:HinclFact}
  \sigma(m_H) &=  \int dY  \sum_{i,j} \int \frac{d\xi_a}{\xi_a} \frac{d\xi_b}{\xi_b} \ 
  H_{ij}^{\rm incl}\Big(\frac{x_a}{\xi_a}, \frac{x_b}{\xi_b}, m_H, \mu\Big)\: f_i(\xi_a,\mu) f_j(\xi_b,\mu) \,,
\end{align}
where $m_H$ is the Higgs mass, $Y$ is the Higgs rapidity, $x_a=m_H e^Y/\Ecm$, and $x_b=m_H e^{-Y}/\Ecm$. Here $f_i(\xi_a,\mu)$ is the parton distribution function (PDF), which is a long-distance matrix element that encodes the probability of finding the parton of type $i$ inside the proton with a light-cone momentum fraction $\xi_a$. The coefficient function $H_{ij}^{\rm incl}$ describes the short-distance hard scattering process which at its core involves gluons fusing with the heavy top-quark loop, and producing the Higgs boson.  The renormalization scale $\mu$ is a gauge and Lorentz invariant cutoff that separates the short distance dynamics at scales $>\mu$ into $H_{ij}^{\rm incl}$, while the long distance dynamics at scales $<\mu$ appears in $f_i$ and $f_j$.  The result in \eq{HinclFact} is valid to all orders in $\alpha_s$, including the dominant nonperturbative corrections through $f_i$ and $f_j$, while corrections to this formula are suppressed by powers of $\lqcd/m_H\ll 1$.

The physics program at a collider like the LHC employs a much richer spectrum of observables than the Drell-Yan-like process in \eq{HinclFact}. For these more general observables there is often belief in the validity of the factorization hypothesis, but no complete proofs.  The differential cross section for observables $a_1$, $a_2$, $a_3$, $\ldots$ is often written as
\begin{align}  \label{eq:abcFact}
   \frac{d\sigma}{da_1\,da_2\, da_3\cdots} & =  \hat\sigma_{ijk\ell\ldots}(a_1,a_2,\ldots) \otimes f_{p/i}\: f_{p/j}  \{ \otimes\, f_{k\to H} \otimes\cdots\otimes  f_{\ell\to H}  \otimes F_{m} \}\,,
\end{align}
where $f_i$ and  $f_j$ are PDFs, we denote convolution integrals by $\otimes$, their is an implicit sum over color channels and flavor indices for the partons, and we may also have fragmentation functions $f_{\ell\to H}$ for a parton $\ell$ converting into hadron $H$.   Here $\hat\sigma_{ijk\ell\ldots}$ encodes corrections that can be calculated perturbatively using quarks and gluons as the degrees of freedom, with or without log resummation, so it has an expansion in $\alpha_s(\mu) \ll 1$, with or without $\alpha_s(\mu) \ln(a_n/a_m) \sim 1$. The function $F_{m}$ denotes final state nonperturbative hadronization corrections for the channel $m$. These corrections are important for some jet observables, and for cases where they arise from soft dynamics can be formulated as vacuum transition matrix elements in quantum field theory~\cite{Korchemsky:1999kt,Korchemsky:2000kp,Ligeti:2008ac,Hoang:2007vb,Abbate:2010xh,Berger:2010xi,Mateu:2012nk}.   For cases with large logarithms between perturbative scales, there is usually a further factorization of the perturbative calculation into components describing different momentum regions.  For example, for an exclusive jet cross section with precisely $N$-jets,
\begin{align}  \label{eq:excljetFact}
  \hat \sigma_{\kappa}  & =  \sum_{\kappa_i}  {\rm tr} \ H^N_{\kappa_H}  {\cal I}_{\kappa_a}  {\cal I}_{\kappa_b}  J_{\kappa_1} \times \cdots \times J_{\kappa_N}  S_{\kappa_S}^N \,,
\end{align}
where the $\kappa$s denote parton channel indices, the hard function $H^N$ denotes short distance dynamics at the collision scale that produces $N$ energetic partons, the ${\cal I}$ encodes initial state energetic perturbative radiation, the $J$s denote final state energetic radiation in the jets, and $S^N$ denotes perturbative soft radiation between the initial and final state partons. For simplicity we have suppressed convolutions between the various functions shown in \eq{excljetFact}, and encoded the color matrix structure of $H^N$ and $S^N$ in the indices $\kappa_H$, $\kappa_S$. Each of these functions encodes physics at different invariant mass scales and rapidities appearing in the process, and renormalization group evolution of these functions can be used to sum large logarithms.  
\begin{table}[t!]
\begin{tabular}{lcrl}
  \hline 
   mode 
   & fields 
   & $p^\mu$ momentum scaling 
   &  \hspace{0.2cm}  physical objects 
 \\
  \hline
  \underline{onshell} \\
   $n_a$-collinear \hspace{0.1cm}
      & $\xi_{n_a}$, $A_{n_a}^\mu$
      &  $(n_a\cdot p, \bar n_a\cdot p, p_{\perp a}) \sim Q (\lambda^2,1,\lambda)$  
      &  \hspace{0.1cm}  collinear initial state jet $a$ 
   \\
   $n_b$-collinear 
     & $\xi_{n_b}$, $A_{n_b}^\mu$
      &  $(n_b\cdot p, \bar n_b\cdot p, p_{\perp b}) \sim Q (\lambda^2,1,\lambda)$ 
      &  \hspace{0.1cm} collinear initial state jet $b$ 
  \\
   $n_j$-collinear 
     & $\xi_{n_j}$, $A_{n_j}^\mu$
      &  $(n_j\cdot p, \bar n_j\cdot p, p_{\perp j}) \sim Q (\lambda^2,1,\lambda)$ 
      &  \hspace{0.1cm} collinear final state jet in $\hat n_j$ 
   \\[-5pt]
    & \ \ \mbox{\footnotesize $\lambda$, $(\lambda^2,1,\lambda)$} \\[-4pt] 
  soft 
     & $\psi_{\rm S}$, $A_{\rm S}^\mu$
      & $p^\mu \sim Q(\lambda,\lambda,\lambda)$
      &  \hspace{0.1cm} soft virtual/real radiation 
   \\[-5pt] 
  &  \phantom{x}\hspace{-0.6cm} \mbox{\footnotesize $\lambda^{3/2}$, $\lambda$} \\[-4pt]      
 ultrasoft 
     & $\psi_{\rm us}$, $A_{\rm us}^\mu$
     & $p^\mu \sim Q(\lambda^2,\lambda^2,\lambda^2)$
      & \hspace{0.1cm}  ultrasoft virtual/real radiation 
   \\[-5pt] 
   &  \phantom{x}\hspace{-0.4cm} \mbox{\footnotesize $\lambda^{3}$, $\lambda^2$} \\[-4pt]  
\underline{offshell}
 \\      
 Glauber
   & -- 
   &  \hspace{0.2cm} $p^\mu \sim  Q(\lambda^a,  \lambda^b, \lambda)$ with $a+b > 2$
    &  \hspace{0.1cm} forward scattering potential
   \\[-5pt]
 & & \mbox{\footnotesize (here $\{a,b\} = \{2,2\}, \{2,1\}, \{1,2\}$)} 
   \\
  hard
    & -- 
     &   $p^2 \gtrsim Q^2$
     & \hspace{0.1cm}  hard scattering
   \\
 \hline
\end{tabular}
  \caption{\setcaptionskip
Infrared momentum regions relevant to hard scattering processes at hadron colliders and the corresponding quark and gluon fields in SCET. Here $Q$ is the scale of the hard interaction, and $\lambda\ll 1$ is a dimensionless power counting parameter. Below the EFT fields we show their power counting (all collinear cases are analogous). The collinear directions $n_i^\mu = (1,\hat n_i)$ are different for the incoming beams $i=a,b$ and for each outgoing identified jet where $\hat n_j$ is the jet direction.  Another standard convention that we will use is the auxiliary vector $\bar n_i = (1, -\hat n_i)$, so that $n_i\cdot \bar n_i = 2$. For Glauber exchange between $n$-collinear, $\bn$-collinear, and soft particles we have the cases $\{a,b\}=\{2,2\}$ ($n$-$\bn$ Glauber), $\{a,b\}=\{2,1\}$ ($s$-$n$ Glauber), and $\{a,b\}=\{1,2\}$ ($s$-$\bn$ Glauber). }
  \label{table:modes}
\setmainskip
\end{table}
(See for example Ref.~\cite{Stewart:2010tn} for a factorization formula for the N-jettiness event shape for $pp$ collisions with invariant mass resummation, or Ref.~\cite{Chiu:2012ir,Neill:2015roa} for a factorization formula for $p_T$ resummation in Higgs production in a $pp$ collision where both invariant mass and rapidity renormalization appear.)   An even less inclusive example of the use of factorization as in \eq{abcFact} is in parton shower Monte Carlos like  $\textsc{Pythia}$~\cite{Sjostrand:2006za} and $\textsc{Herwig}$~\cite{Bahr:2008pv}. Here in principle one can ask for a cross section that is fully differential in all final state hadronic momenta, but it is well appreciated that only some subset of more inclusive observables will satisfy the criteria for factorization. In the Monte Carlo implementation there is also a separation/factorization of the initial state and final state perturbative showers from the PDFs and final state hadronization, making the structure somewhat analogous to \eq{excljetFact}.

The key ingredients underlying the idea of factorization is the separation of momentum fluctuations that occur at different scales, much like the standard idea of scale separation in quantum field theory. For factorization at hadron colliders the key infrared momentum regions can be identified either physically, or with analyses that determine the infrared structure of amplitudes~\cite{Sterman:1978bi}, use the Coleman-Norton theorem~\cite{Coleman:1965xm}, or exploiting the method of regions~\cite{Smirnov:2002pj}. (For a detailed analysis of the method of regions that includes the Glauber region see~\cite{Jantzen:2011nz}.) Some work must still be done to correctly associate these results with degrees of freedom in an effective theory. For most collider observables this results in the momentum regions identified in \tab{modes}, or a subset of these regions for simpler processes.  Indeed, expansions involving these momentum regions are central to utilizing both the factorization methods of CSS where they determine the leading infrared regions, as well as those in  SCET where they determine the relevant low energy fields present in the effective theory. In the 
\begin{figure}[t!]
\begin{center}
\includegraphics[width=0.5\textwidth]{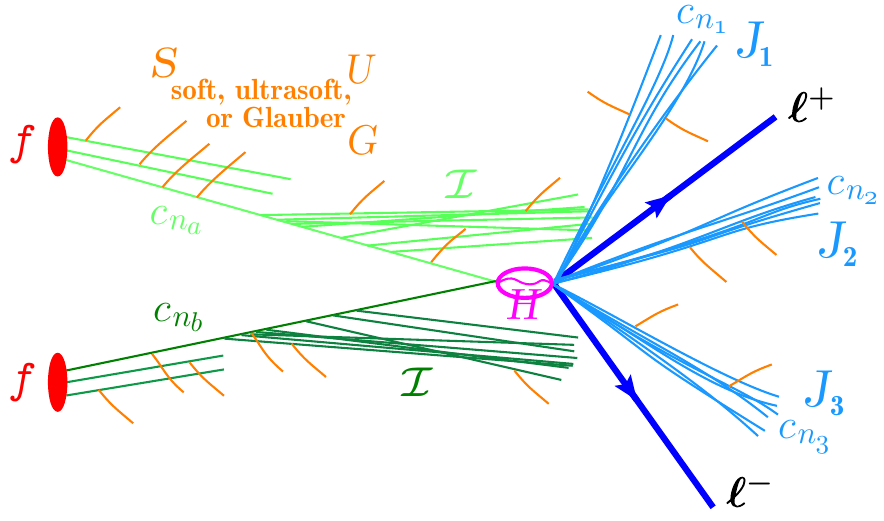}
\end{center}
\vspace{-0.6cm}
\caption{\setcaptionskip
Contributions from different infrared momenta to a hard scattering process at a hardon collider.}
\label{fig:LHCgeneral}
\setmainskip
\end{figure}
SCET literature observables are often divided into two types, those that measure invariant mass type variables which require ultrasoft modes and the use of \SCETa techniques, and those that measure transverse-momenta, which require soft modes and the use of \SCETb ~\cite{Bauer:2002aj}. \SCETb observables sometimes require the use of rapidity renormalization techniques~\cite{Chiu:2011qc}.  For example, measuring the invariant mass of a jet  $p_J^2= p_J^+p_J^- - \vec p_{J}^{\perp\,2}$, induces physical contributions at the ultrasoft scale $p^+\sim \lambda^2$, causing ultrasoft modes to give nontrivial contributions.   Measuring a parametrically small transverse momentum, $p_{T}\sim \lambda$, causes soft modes to give nontrivial contributions.  A pictorial representation of the physics described by these various modes, and their correspondence with functions in \eq{excljetFact}, is given in \fig{LHCgeneral}. In general soft, collinear, and hard modes can talk to each other at leading order in the power counting, but the nature of the hard collision dramatically simplifies these interactions so that they can be put into a factorized form. Of the steps needed to prove hard scattering factorization, the most difficult is the cancellation of contributions from the Glauber region which directly couple modes in different regions together   but which cannot be  captured entirely by Wilson lines.

\begin{figure}[t!]
%
%
\hspace{0.15cm}
\subfigure{
\raisebox{2.5cm}{a)\hspace{-0.4cm}} 
\includegraphics[width=0.16\columnwidth]{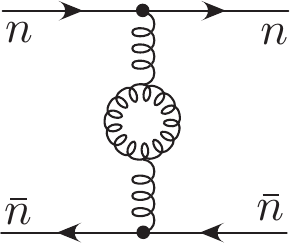}
 \hspace{0.3cm}
\includegraphics[width=0.16\columnwidth]{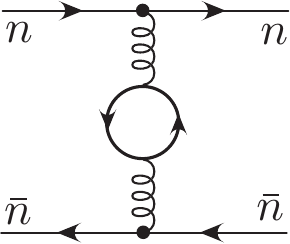}
 \hspace{0.3cm}
\includegraphics[width=0.16\columnwidth]{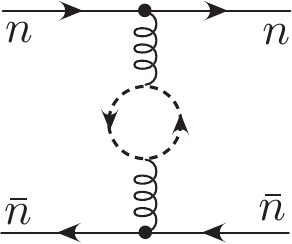}  
}
\hspace{0.3cm}
\subfigure{
\raisebox{2.5cm}{b)\hspace{-0.6cm}} 
\raisebox{-0.05cm}{ 
 \includegraphics[width=0.15\columnwidth]{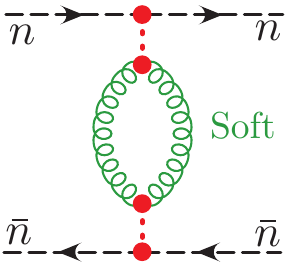} 
 \hspace{0.3cm}
 \includegraphics[width=0.15\columnwidth]{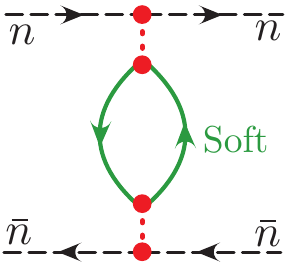}  
}
}
\caption{\setcaptionskip
a) QCD gluon, quark, and ghost vacuum polarization graphs for forward scattering. b)  Soft gluon and quark loop graphs in SCET with Glauber vertices that contain the vacuum polarization (as well as other contributions in the gluon loop graph). Due to the gauge invariance of the soft-collinear Glauber operators inserted at the red ellipse vertices, a soft ghost loop graph does not appear at this order.}
\label{fig:softvac}
\setmainskip
\end{figure}

We will also see that there is a nontrivial interplay between the standard SCET interactions and Glauber operators. Once Glauber operators are present there are also always soft fields even in \SCETa in addition to the ultrasoft fields.\footnote{The presence of both soft and ultrasoft fields with Glauber operators in \SCETa is akin to NRQCD as formulated in Ref.~\cite{Luke:1999kz} with soft and ultrasoft modes. In NRQCD the soft modes are not radiated but play a crucial role in correcting the potentials. In the case of \SCETa forward scattering the softs also provide virtual loop corrections to the Glauber exchange potential.} For example soft fields are responsible for vacuum polarization of the Glauber gluon.  The diagrams that cause the Glauber vacuum polarization are shown in Fig.~\ref{fig:softvac} in both full QCD and in SCET with Glauber operators.  The vacuum polarization is a clear example showing that Glauber gluons cause interactions between the soft and collinear fields in SCET, in addition to giving an interaction between $n$ and $\bn$ collinear fields in different directions as discussed above in Fig.~\ref{fig:glaub_tree}. At leading power Glauber exchange can connect any two modes with different rapidities but the same transverse momentum,  $\{n,\bn,s\}$, as well as simultaneously all three of these.  Interestingly, because we formulate these operators in a gauge invariant fashion there are no ghosts fields in the Glauber operators.  So for example the soft gluon loop in Fig.~\ref{fig:softvac}b produces the full $11 C_A/3$ term in the logarithm related to the QCD $\beta$ function that shows up due to the running of the $\alpha_s(\mu)$ coupling in the tree level Glauber exchange graph of \fig{glaub_tree}.  On the other hand, ultrasoft gluons do not couple to Glauber gluons directly at leading power (nor to soft gluons).  So the only new interactions with ultrasoft gluons appear due to their coupling to collinear fields that show up in the Glauber operators.

In this paper we develop a set of operator based tools within SCET to address these questions.  We construct a Lagrangian that encodes all Glauber effects and show that it fits seamlessly with the current tools used to study hard, collinear, and soft factorization in hard scattering processes, without inducing double counting.  We also explicitly demonstrate the connection of this operator formalism to forward scattering phenomena by showing that the one-loop rapidity renormalization of our operators yields gluon Reggeization at the amplitude level, the BFKL equation at the cross section level, and reproduces the shockwave picture.

\section{Review of SCET and Notation}  \label{sec:SCET}

\subsection{SCET Lagrangians}  
\label{sec:SCETLag}

For each type of onshell momentum mode in \tab{modes} there are both quark and gluon fields in SCET. The small power counting parameter $\lambda \ll 1$ sets the typical size for the momentum fluctuations for each mode.  An $n_j$-collinear mode describes the infrared structure of fluctuations close to a collinear direction $\hat n_j$, where $n_j^\mu=(1,\hat n_j)$, and directions for different collinear fields are distinct, $n_i\cdot n_j \gg \lambda^2$.  

All hard offshell modes are integrated out of the effective theory, leading to operators $O_K$ that describe hard scattering processes.  These operators get inserted only once for each amplitude, but more than one operator may contribute for a given physical process.  The Glauber modes in \tab{modes} are also offshell modes since the scaling of their momenta forbids $p^+p^- = \vec p_\perp^{\,2}$, but they are offshell by an amount of order $p_\perp^2\sim \lambda^2$ rather than $\sim \lambda^0$.  These offshell Glauber modes are still integrated out of the effective theory at the hard scale, much like potential modes in NRQCD~\cite{Luke:1999kz}, since the simultaneous requirements of gauge invariance and homogeneous order-by-order power counting can otherwise not be satisfied.\footnote{In this EFT there is a trade-off between 3 things, 1) having locality at an infrared scale, since the $1/\vec k_\perp^{\,2}$ Glauber potential is non-local, 2) implementing gauge invariance and 3) maintaining a homogeneous power counting in $\lambda$. Since for many calculations and analyses we need to treat our operators non-perturbatively in $\alpha_s$, we choose in favor of maintaining the latter two principals while giving up locality. This is the same choice made for NRQCD in the vNRQCD~\cite{Luke:1999kz,Manohar:1999xd,Manohar:2000hj,Manohar:2000kr,Manohar:2000rz,Hoang:2002yy} or pNRQCD~\cite{Pineda:1997bj,Brambilla:1999xf,Brambilla:2004jw} formalisms.  It is also the same choice made for \SCETb, where the soft Wilson lines are non-local at a scale $p^+\sim p_\perp \sim \lambda$. 
(Without Glauber operators \SCETa maintains locality at infrared scales.)
}   Since the Glauber operators yield a leading order potential there is no power counting restriction on how many times they may appear in the amplitude or cross section for a given process.
 
In a general notation the leading power hard scattering operators $O_K$ for some desired \SCETa process, and the leading power  Lagrangian for any \SCETa process, can be written as 
\begin{align} \label{eq:Lsceta}
   {\cal L}_{\SCETa}^{\rm hardscatter} & = \sum_K    C_K \otimes O_K(\{\xi_{n_i},A_{n_i}\},\psi_{\rm us}, A_{\rm us})  
    \,, \\
  {\cal L}_{\SCETa}^{(0)} &= 
  \Big[ {\cal L}_{us}^{(0)}\big(\psi_{\rm us},A_{\rm us}\big) 
   +  \sum_{n_i}  {\cal L}_{n_i}^{(0)}\big( \xi_{n_i}, A_{n_i}, {n_i}\cdot A_{\rm us}\big )  \Big] \nn\\
  &\quad
   +  \Big\{ {\cal L}_{G}^{{\rm I}(0)}\big(\{ \xi_{n_i}, A_{n_i}\}, \psi_{\rm S}, A_{\rm S}\big)  
       + {\cal L}_S^{(0)}(\Psi_{\rm S}, A_{\rm S}) \Big\}
  \,. \nn
\end{align}
Here $C_K$ are hard Wilson coefficients that depend on large momenta $\bn_i\cdot p$ of collinear gauge invariant products of collinear fields. (Note that ultrasoft gauge fields can appear in the leading order hard scattering operator in \eq{Lsceta} for some \SCETa processes. Although this is usually not the case for collider physics with massless hard scattering producing jets, it is well known in inclusive $B$-meson decays where the Heavy Quark Effective Theory $b$-quark field is ultrasoft or soft.)   The hard scattering operator and two terms in square brackets in \eq{Lsceta} are what we refer to as classic \SCETa, and are the terms usually considered in the SCET literature.  We will discuss $O_K$ further in \sec{SCETOp}.  Glauber operators are contained in ${\cal L}_{G}^{{\rm I}(0)}$ which we discuss in \sec{GlauberSCET}, and must be included when writing down the full \SCETa Lagrangian. A leading power soft Lagrangian ${\cal L}_S^{(0)}$ also appears in \SCETa along with ${\cal L}_{G}^{{\rm I}(0)}$ since it is necessary (for example) to reproduce the vacuum polarization of the Glauber gluon shown in \fig{softvac}. Recall that both ${\cal L}_{\rm us}^{(0)}(\psi_{\rm us},A_{\rm us})$ and ${\cal L}_{\rm S}^{(0)}(\psi_{\rm S},A_{\rm S})$ are each identical to copies of the standard full QCD Lagrangian (prior to subtractions).  Also recall that dropping the coupling to ultrasoft gluons and subtractions, ${\cal L}_n^{(0)}(\xi_n,A_n,0)$ just involves collinear fields in a single sector and is again equivalent to a copy of full QCD (though it is QCD written in the form that is more familiar from light-cone quantization which only involves the ``large'' components of the fermion field $\xi_n$).  The full collinear Lagrangian is a sum of quark and gluon pieces
\begin{align}
  {\cal L}_n^{(0)} &=  {\cal L}_{n\xi}^{(0)}(\xi_n,A_n,n\cdot A_{\rm us})  
    + {\cal L}_{ng}^{(0)}(A_n,n\cdot A_{\rm us}) \,,
\end{align}
where in general ${\cal L}_{ng}^{(0)}$ also contains ghost fields and gauge fixing terms that are gauge covariant with respect to the ultrasoft covariant derivative $in\cdot D_{\rm us}$.  The full expression for ${\cal L}_{ng}^{(0)}$ can be found in Ref.~\cite{Bauer:2001yt}.  The result for the leading power collinear quark Lagrangian is
\begin{align}  \label{eq:Lnxi0}
   {\cal L}^{(0)}_{n\xi} = e^{-i x\cdot {\cal P}}\
 \bar{\xi}_{n}\Big( i n\cdot D+ i \slashed{D}_{n\perp}\frac{1}{i\bn\cdot D_n}
  i \slashed{D}_{n\perp}\Big) \frac{\slashed{\bn}}{2}\,\xi_{n}  \,.
\end{align}
Here the covariant derivatives are given by
\begin{align}
 iD_n^\mu & = i\partial_n^\mu + g A_n^\mu 
  \,, \qquad\qquad
  \partial_n^\mu = {\cal P}^\mu+ \frac{\bn^\mu}{2} in\mcdot\partial
  \,,
  \nn\\
 in\cdot D &= in\cdot\partial + g n\cdot A_n + g n\cdot A_{us}
  \,,
\end{align}
and $i n\cdot D$  contains both the collinear and ultrasoft gauge fields. $\cP$ is the so-called ``label operator'', which picks out the large component of a given momentum. The phase $e^{-ix\cdot {\cal P}}$ is related to the implementation of the multipole expansion that was carried out for ultrasoft $k_\perp$ and $\bn\cdot k$ momenta traveling through collinear propagators. Using momentum labels for the large momenta, and position dependence to encode the small momenta, the full position space field $\hat \xi_n(x)$ has been written as
\begin{align}
  \hat \xi_n(x) = e^{-ix\cdot \cP} \sum_{p_\ell\ne 0} \xi_{n,p_\ell}(x) 
    \equiv  e^{-ix\cdot \cP}  \xi_{n}(x) \,,
\end{align}
where $p_\ell^\mu=\bn\cdot p_\ell\, n^\mu/2 + p_{\ell\perp}^\mu$.  The $\xi_n(x)$ are the fields we use in \eq{Lnxi0}.  In \SCETa we have ${\cal O}(\lambda^2)$ residual momenta in all components, so the argument $(x)$ on the field is a full $x^\mu$. In \SCETb we only have ${\cal O}(\lambda)$ $\perp$-momenta so the fields do not depend on $x_\perp$, and we represent the $\perp$-momenta as a continuous variable in momentum space.

In the form in \eq{Lnxi0} the collinear fermion field satisfies $\nslash \xi_n=0$. If we integrate back in the ``small'' components of the collinear fermion field as an auxiliary field, then the leading power collinear quark Lagrangian in \eq{Lnxi0} becomes
\begin{align} \label{eq:SCET4}
  {\cal L}_{n\xi}^{(0)} = e^{-ix\cdot {\cal P}}\, \bar \psi_n\,  i \slashed{\cal D}_{\!n}\,  \psi_n \,,
  \qquad
   {\cal D}_n^\mu = (\bn^\mu/2)\, n\cdot D + (n^\mu/2)\, \bn\cdot D_n + D_{n\perp}^\mu \,,
\end{align}
where the relation between the  quark fields with four and two active components field is $\psi_n = \big[1+(1/i\bn\mcdot D_n) i\slashed D_{n\perp} (\bnslash/2)  \big]\xi_n$.  In this form it is even more obvious that, for the case where we drop the ultrasoft couplings, each collinear Lagrangian ${\cal L}_n^{(0)}(\xi_n,A_n,0)$ is just a copy of full QCD. Due to the eikonal coupling to ultrasoft gluons in $n\cdot D$, the result in \eq{SCET4} is not the same as full QCD in general. 

With the $\SCETa$ Lagrangian in \eq{Lsceta} the assumption of ignoring rescattering effects between different momentum regions, means dropping ${\cal L}_{G}^{{\rm I}(0)}+{\cal L}_S^{(0)}$.  Since only the fields in ${\cal L}_{G}^{{\rm I}(0)}$ couple to those in the classic $\SCETa$ Lagrangian, it is enough to prove that ${\cal L}_{G}^{{\rm I}(0)}$ interactions do not contribute to a hard scattering observable to prove the decoupling of Glauber gluon effects. If this decoupling occurs then the standard tools of $\SCETa$ can be used to treat hard-collinear factorization, soft-collinear factorization, factorization of the observable, and the uniqueness of various operators with Wilson lines, in order to attempt to derive a factorization formula. 

Next we repeat the above discussion for \SCETb.  Here there are no ultrasoft fields, and soft fields contribute even in the classic SCET framework in both real and virtual diagrams.  In a general notation the leading power hard scattering operators $O_K^{\rm II}$ for some desired \SCETb process, and the leading power  Lagrangian for any \SCETb process, can be written as 
\begin{align} \label{eq:Lscetb}
   {\cal L}_{\SCETb}^{\rm hardscatter} & 
   = \sum_K    C_K^{\rm II} \otimes O_K^{\rm II}( \{\xi_{n_i},A_{n_i}\},\psi_{S}, A_{S})  
    \,, \\
  {\cal L}_{\SCETb}^{(0)} &= 
  \Big[ {\cal L}_{\rm S}^{(0)}\big(\psi_{\rm S},A_{\rm S}\big) 
   +  \sum_{n_i}  {\cal L}_{n_i}^{(0)}\big( \xi_{n_i}, A_{n_i}\big )  \Big] 
   +  {\cal L}_{G}^{{\rm II}(0)}\big(\{ \xi_{n_i}, A_{n_i}\}, \psi_{\rm S}, A_{\rm S}\big)  
  \,. \nn
\end{align}
Here $C_K^{\rm II}$ are hard Wilson coefficients that depend on the large momenta $\bn_i\cdot p$ of collinear gauge invariant products of collinear fields, and on momenta $n_i\cdot p$ of  soft gauge invariant products of soft fields.\footnote{The dependence on $n_i\cdot p$ can be thought of as arising from integrating out hard-collinear propagators in a \SCETa theory, where the hard-collinear fields have an  offshellness of size $\bn_i \cdot p\, n_i\cdot p$~\cite{Bauer:2002aj}. Following this construction we can generate the final \SCETb theory by the matching sequence QCD$\to$ \SCETa $\to$ \SCETb.} We sum over all operators $K$ that are leading order in the power counting and have distinct color and spin indices. The hard scattering operator and the two terms in square brackets are what we refer to as classic \SCETb. We will discuss $O_K^{\rm II}$ further in \sec{SCETOp}.   Glauber operators are contained in ${\cal L}_{G}^{{\rm II}(0)}$ and must be included when writing down the full \SCETb Lagrangian.  In this language the assumption of ignoring Glauber gluons in \SCETb means dropping ${\cal L}_{G}^{{\rm II}(0)}$.  In \eq{Lscetb} the soft Lagrangian ${\cal L}_{\rm S}^{(0)}$ is again just the standard QCD Lagrangian for these fields, and since there are no ultrasoft fields each collinear Lagrangian ${\cal L}_n^{(0)}$ is equivalent to a copy of full QCD, as discussed above for \SCETa. In the leading order classic \SCETb Lagrangian there is no coupling between the soft and collinear sectors. So all the couplings between sectors come either from the hard interaction operators or  $ {\cal L}_{G}^{{\rm II}(0)}$. It is enough to prove that the net effect of ${\cal L}_{G}^{{\rm II}(0)}$ interactions vanishes for a hard scattering observable to prove the decoupling of Glauber gluon effects, and then one can use standard \SCETb to attempt to prove factorization theorems. 

Note that when EFT fields are contracted they can lead to loop diagrams that are referred to as collinear, soft, ultrasoft, or Glauber. The meaning of this language is that the corresponding loop momentum has this type of scaling as in \tab{modes}. 

We will discuss the Glauber operators appearing in ${\cal L}_{G}^{{\rm II}(0)}$ and ${\cal L}_{G}^{{\rm I}(0)}$ in \sec{GlauberSCET}.

\subsection{SCET Operator Building Blocks}
\label{sec:SCETOp}

In this section we discuss gauge invariant operator building blocks for quark and gluon operators in SCET~\cite{Bauer:2000yr,Bauer:2001ct,Bauer:2002nz}.  At any order in the power counting the most general building blocks for $n$-collinear components of SCET operators for QCD (other than the leading power kinetic term) contain three terms~\cite{Marcantonini:2008qn}
\begin{align} \label{eq:nops3}
  \chi_n \,, \qquad  {\cal B}_{n\perp}^\mu \,, \qquad {\cal P}_{n\perp}^\mu \,.
\end{align}
The full expressions for $\chi_n$ and ${\cal B}_{n\perp}^\mu$ are given below in \eqs{opbb}{opbbb} and carry global fundamental and adjoint color indices (also discussed below), but are gauge invariant under local collinear gauge transformations due to the presence of collinear Wilson lines. When expanded these quark and gluon building block fields contain the physical quark and gluon components, 
$\chi_n = \xi_n + \ldots$ and ${\cal B}_{n\perp}^\mu = A_{n\perp}^\mu - (\cP_\perp^\mu/\bnP) \bn\cdot A_n + \ldots$.  To reduce operators down to the three objects in \eq{nops3} we rewrite all $\bn\cdot A_n$'s as $W_n$ Wilson lines, and absorb dependence on $\bn\cdot \cP$ into Wilson coefficients. We also use the equations of motion to remove $in\cdot\partial \,\chi_n$, $in\cdot \partial\, {\cal B}_{n\perp}$, $\bnP\, n\cdot {\cal B}^\mu_n$, $in\cdot\partial\, n\cdot {\cal B}_n$, and use operator identities to remove $[iD_{n\perp}^\mu, iD_{n\perp}^\nu]$ and $[iD_{n\perp}^\mu, in\cdot D_n]$~\cite{Marcantonini:2008qn}.  
Here  $g\,n \cdot  {\cal B}_{n}  = \big[ W_n^\dagger i n \cdot D_{n} W_n \big]$.
Using the scaling of the fields deduced from their kinetic terms, the power counting for these collinear building blocks is $\chi_n\sim \lambda$, ${\cal B}_{n\perp}^\mu\sim \lambda$, and ${\cal P}_{n\perp}^\mu\sim \lambda$. 

We will find it useful to also use the following building blocks for soft fields
\begin{align} \label{eq:sops}
    \psi_s^n\,, \qquad  {\cal B}_{S\perp}^{n\mu}  \,.
\end{align}
Here the $n$ superscript denotes the soft gauge field component $n\cdot A_s$ appearing in the soft Wilson lines in these operators.  For an analysis involving back-to-back $n$-collinear and $\bn$-collinear sectors we will see that $\psi_s^n$, $\psi_s^\bn$, ${\cal B}_{S\perp}^{n\mu}$, and ${\cal B}_{S\perp}^{\bn\mu}$ (defined below in \eqs{opbb}{opbbb}) all appear.  Using the scaling of the fields deduced from their kinetic terms, the power counting for these soft building block fields is $\psi_s^n\sim \psi_s^\bn\sim \lambda^{3/2}$ and ${\cal B}_{S\perp}^{n\mu}\sim {\cal B}_{S\perp}^{\bn\mu}\sim \lambda$.

The collinear and soft building blocks that have a single quark field at lowest order in the coupling are
\begin{align} \label{eq:opbb}
  \chi_n &= W_{n}^\dagger \xi_n \,,
  & W_n & = {\rm FT}\ W_{n}[\bn\cdot A_n]
  =  {\rm FT}\  \: {\rm P} \exp\bigg(  i g\! \int_{-\infty}^0 \!\!\!\!\!\! ds\ \bn\cdot A_n(x+\bn s) \bigg) \,,
   \nn\\
 \chi_\bn &= W_{\bn}^\dagger \xi_\bn \,,
  & W_\bn & = {\rm FT}\ W_{\bn}[n\cdot A_\bn]
  =  {\rm FT}\  \: {\rm P} \exp\bigg(  i g\! \int_{-\infty}^0 \!\!\!\!\!\! ds\ n\cdot A_\bn(x+n s) \bigg) \,,
   \nn\\
  \psi_s^n &= S_n^\dagger \psi_s \,, \quad   
  \psi_s^\bn = S_\bn^\dagger \psi_s \,,
  & S_n & = {\rm FT}\ S_{n}[n\cdot A_S]
  =  {\rm FT}\  \: {\rm P} \exp\bigg(  i g\! \int_{-\infty}^0 \!\!\!\!\!\! ds\ n\cdot A_S(x+n s) \bigg) 
  \,,
\end{align}
where ${\rm FT}$ is for Fourier transform, and P stands for path ordering. The Fourier transform is often written out in momentum space which enables making explicit the notation for the multipole expansion (the lines remain local in the coordinate corresponding to residual momenta, even though they are extended for the larger momentum associated with the $s$ coordinate shown here). Under a collinear gauge transformation $\xi_n\to U_n \xi_n$, $W_n\to U_n W_n$, so $\chi_n$ is invariant, and a similar property holds for the other fields with transformations that have support in their respective momentum sectors.
Although we show only one direction in \eq{opbb} the integrals could instead extend over $[0,\infty]$. Expressions for Wilson lines over $(0,\infty)$ and $(-\infty,0)$ and their Feynman rules are summarized in \app{Wdirection}, and we note that the difference corresponds to the choice of $\bn\cdot k\pm i0$ in eikonal propagators.  In general the direction of the Wilson lines in the fields in \eq{opbb} can be discussed in the context of matching calculations from full QCD. As we will discuss in some detail, in SCET there are soft and Glauber 0-bin subtractions for collinear Wilson lines which cancel this direction dependence (which comes from the region $\bn\cdot k\to 0$). The same thing happens for soft Wilson lines due to Glauber 0-bin subtractions. Ultrasoft Wilson line directions in \SCETa are determined by the physical directions since their diagrams do not have subtractions. At various points below we will discuss the direction of Wilson lines explicitly. 

Note that we follow a convention where the subscript on the collinear field indicates the type of collinear gluon field that the operator contains, rather than the light-like direction of the Wilson line. Thus the $n$ subscript on collinear building blocks means something different than the $n$ superscript on soft building blocks.  

We denote fundamental collinear Wilson lines by $W_n$, where $\bn\cdot A_n = \bn\cdot A_n^A T^A$ in \eq{opbb}, and  adjoint collinear Wilson lines by ${\cal W}_n$, where $\bn\cdot A_n = \bn\cdot A_n^A T^A_{\rm adj}$ with $(T^A_{\rm adj})_{BC} = - i f^{ABC}$. Note that 
\begin{align}
  & W_n^\dagger W_n=\id \,,
  & {\cal W}_n^{AB} {\cal W}_n^{CB} & = \delta^{AC} \,,
\end{align} and
\begin{align}
& (\bn \cdot D) W_n=0 \,, 
  & (\bn \cdot D) {\cal W}_n = 0 \,.
\end{align} 

We also have the following relationship between  Wilson lines in the fundamental and adjoint representations
\begin{align}  \label{eq:Weqtns}
    W_n^\dagger T^A W_n &=  {\cal W}_n^{AB}  T^B \,,
  &   W_n T^A W_n^\dagger & =  {\cal W}_n^{BA} T^B \,.
\end{align}
Their momentum space  expansion with an incoming momentum $k$ for the gluon are
\begin{align}
  W_n &= 1 - \frac{g\, T^A\, \bn\cdot A_{n,k}^A}{\bn\cdot k}  + \ldots \,,
  & W_n^\dagger & = 1 + \frac{g\,T^A\, \bn\cdot A^A_{n,k}}{\bn\cdot k}  + \ldots
  \,,  \nn\\
  {\cal W}_n^{AB} &= \delta^{AB} 
         + \frac{g\, i f^{CAB}\, \bn\cdot A_{n,k}^C}{\bn\cdot k}  + \ldots \,,
 &  ({\cal W}_n^\dagger)^{AB} &= \delta^{AB} 
         - \frac{g\, i f^{CAB}\, \bn\cdot A_{n,k}^C}{\bn\cdot k}  + \ldots \,.
\end{align}
We have analogous results for the fundamental soft Wilson lines $S_n$, $S_n^\dagger$, and adjoint soft Wilson lines ${\cal S}_n$ and ${\cal S}_\bn$.  

The collinear and soft building blocks that involve a single gluon field at lowest order in the coupling are
\begin{align} \label{eq:opbbb}
  {\cal B}_{n\perp}^\mu  &=
      \frac{1}{g}\, \big[ W_n^\dagger i D_{n\perp}^\mu W_n \big] =
      \frac{1}{g}\, \frac{1}{\bn\cdot \cP} \, W_n^\dagger \big[ i\bn\cdot D_n \,, i D_{n\perp}^\mu \big] W_n 
      \,, \nn\\
  {\cal B}_{\bn\perp}^\mu  &=
      \frac{1}{g}\, \big[ W_\bn^\dagger i D_{\bn\perp}^\mu W_\bn \big] =
      \frac{1}{g}\, \frac{1}{n\cdot \cP} \, W_\bn^\dagger \big[ i n\cdot D_\bn \,, i D_{\bn\perp}^\mu \big] W_\bn 
      \,, \nn\\
  {\cal B}_{S\perp}^{n\mu}  &= 
      \frac{1}{g}\, \big[ S_n^\dagger i D_{S\perp}^\mu S_n \big] =
      \frac{1}{g}\, \frac{1}{ n\cdot \cP} \, S_n^\dagger \big[ i  n\cdot D_S\,, i D_{S\perp}^\mu \big] S_n 
      \,, \nn\\
  {\cal B}_{S\perp}^{\bn\mu}  &=
      \frac{1}{g}\, \big[ S_\bn^\dagger i D_{S\perp}^\mu S_\bn \big] =
      \frac{1}{g}\, \frac{1}{\bn \cdot \cP} \, S_\bn^\dagger \big[ i\bn\cdot D_S \,, i D_{S\perp}^\mu \big] S_\bn 
      \,, 
\end{align}
where the Wilson lines here are the same as those in the quark building blocks, again with a direction dependence that as we will discuss is removed by Glauber subtractions, and hence only becomes fixed if we aim to absorb Glauber contributions. The square brackets after the first equalities indicate that the covariant derivatives only act inside the brackets. These gluon operators are  in an adjoint representation so we can write ${\cal B}_{n\perp}^\mu = {\cal B}_{n\perp}^{\mu A} T^A$ etc.  The Wilson lines appearing here can be combined into a single Wilson line in the adjoint representation, for example we have
\begin{align}
    {\cal B}_{n\perp}^{A\mu} &=  
       \frac{1}{\bn\cdot \cP}  \, \bn_\nu i G_{n}^{B\nu\mu_\perp}  {\cal W}_n^{BA}
    \,,
   & {\cal B}_{\bn\perp}^{A\mu} &=  
     \frac{1}{n\cdot \cP}  \, n_\nu i G_{\bn}^{B\nu\mu_\perp}  {\cal W}_\bn^{BA} 
   \,,
\end{align}
with the adjoint collinear Wilson lines ${\cal W}_n^{BA} =  {\cal W}_n^{BA}[\bn\cdot A_n]$ and ${\cal W}_\bn^{BA} =  {\cal W}_\bn^{BA}[n\cdot A_\bn]$, and collinear field strengths $i g G_n^{A \mu\nu} T^A = [ i D_n^\mu , i D_n^\nu ]$.  A useful relation is 
\begin{align}
 W_n^\dagger i D_{n\perp}^\mu W_n = {\cal P}_\perp^\mu + g {\cal B}_{n\perp}^\mu  
 \,.
\end{align} 
To lowest order in the coupling expansion
\begin{align}
  {\cal B}_{n\perp}^\mu = A_{n\perp}^\mu - \frac{k_\perp^\mu}{\bn\cdot k} \bn\cdot A_{n,k} + \ldots  \,.
\end{align}
There are analogous expressions for operators in other sectors, including the soft operators.  The ${\cal B}_{n\perp}^\mu$ operator is gauge invariant under $n$-collinear transformations since $i D_{n\perp}^\mu W_n \to U_n i D_{n\perp}^\mu W_n$ and $W_n^\dagger \to W_n^\dagger U_n^\dagger$.  Again a similar statement holds for the other gluon building block fields with gauge transformations that have support in each of their respective sectors.

We also will make use of fields that are matrices in the color octet space, which we denote with a tilde, such as
\begin{align}  \label{eq:Bmatrix}
  \widetilde {\cal B}_{n\perp}^{AB} &= - i f^{ABC}  {\cal B}_{n\perp}^C 
  \,,
  & \widetilde {\cal B}_{S\perp}^{nAB} &= - i f^{ABC}  {\cal B}_{S\perp}^{nC} \,,
  &\widetilde  {G}_s^{\mu\nu\, AB}&= - i f^{ABC}  {G}_s^{\mu\nu\, C} \,,
\end{align}
where the soft field strength $i g G_s^{A \mu\nu} T^A = [ i D_s^\mu , i D_s^\nu ]$. We also have the adjoint relation 
\begin{align}
{\cal W}_n^T i D_{n\perp}^\mu {\cal W}_n = {\cal P}_\perp^\mu + g \widetilde {\cal B}_{n\perp} \,.
\end{align} 

In the hard scattering operators in both \SCETa and \SCETb we often need to specify the large momenta
for the collinear gauge invariant building blocks, $\chi_n$ and ${\cal B}_{n\perp}$,  for which we use the notation
\begin{align}
  \chi_{n,\omega} &= \delta(\omega-\bn\cdot \cP)  \chi_n \,, 
  & {\cal B}_{n\perp,\omega}^\mu &=  \delta(\omega-\bn\cdot \cP)   {\cal B}_{n\perp}^\mu \,,
   \nn\\
  \chi_{\bn,\omega'} &= \delta(\omega'-n\cdot \cP)  \chi_\bn \,, 
  & {\cal B}_{\bn\perp,\omega'}^\mu &=  \delta(\omega'-n\cdot \cP)   {\cal B}_{\bn\perp}^\mu 
   \,.
\end{align}
In \SCETb we also need to specify one component of the momentum of soft operators, since dependence on this momentum is induced in the Wilson coefficients $C_K^{\rm II}$ by the existence of collinear components in these operators with large momentum in an opposite light-cone component. To encode this dependence we can use the notation
\begin{align}  
\label{label}
   \psi_{S,k}^n  &= \delta(k-n\cdot i\partial_s) \psi_s^n \,, 
 & {\cal B}_{S\perp,k}^{n\mu} & =  \delta(k -n\cdot i\partial_s) {\cal B}_{S\perp}^{n\mu} 
  \,, \nn\\
  \psi_{S,k}^\bn  &= \delta(k-\bn\cdot i\partial_s) \psi_s^\bn \,, 
 & {\cal B}_{S\perp,k}^{\bn\mu} & =  \delta(k -\bn\cdot i\partial_s) {\cal B}_{S\perp}^{\bn\mu} 
  \,.
\end{align}
In \SCETb the general hard scattering operator appearing in \eq{Lscetb} has terms
\begin{align}  \label{eq:HopII}
  & O_K^{\rm II}\big(\{\omega_1^a,\ldots \},\{ k_1^a,\ldots \}\big)  
   \\
  &\ =   
    \big(\chi_{n_1,\omega_1^a}\chi_{n_1,\omega_1^b} \cdots \big)
    \big(\bar \chi_{n_1,\omega_1^c} \bar \chi_{n_1,\omega_1^d}  \cdots\big)
       \big({\cal B}^\perp_{n_1,\omega_1^e}{\cal B}^\perp_{n_1,\omega_1^f}\cdots\big)  
       \big(\chi_{n_2,\omega_2^a} \cdots \big)
       \big(\bar \chi_{n_2,\omega_2^b} \cdots\big)
       \big({\cal B}^\perp_{n_2,\omega_2^c}\cdots\big)  
    \cdots 
    \nn\\
   &\ \ \times    
      \big( \psi_{S,k_1^a}^{n_1} \psi_{S,k_2^b}^{n_1}\cdots\big) 
      \big( \bar\psi_{S,k_1^c}^{n_1} \bar\psi_{S,k_1^d}^{n_1}\cdots\big) 
      \big( {\cal B}_{S,k_1^e}^{\perp n_1}  {\cal B}_{S,k_1^f}^{\perp n_1}  \cdots \big) 
      \big( \psi_{S,k_2^a}^{n_2} \cdots\big) 
      \big( \bar\psi_{S,k_2^b}^{n_2} \cdots\big) 
      \big( {\cal B}_{S,k_2^c}^{n_2\perp}    \cdots \big) \cdots 
   \,, \nn
\end{align} 
where for simplicity we have suppressed color, flavor, and Lorentz indices, and have not displayed factors of $\cP_\perp$ or soft derivatives.

For \SCETa the hard scattering operators $O_K$ are analogous to \eq{HopII} but will contain ultrasoft fields without momentum labels in place of the soft building block fields.  In this EFT the ultrasoft-collinear decoupling is obtained from the BPS field redefinition~\cite{Bauer:2001yt} $\xi_n \to Y_n \xi_n$, $A_n^\mu \to Y_n A_n^{\mu} Y_n^\dagger$, which allows us to factorize the ultrasoft fields into gauge invariant products in the hard scattering operators, such as $(Y_n^\dagger \psi_{us})$, $[Y_n^\dagger i D_{us\perp}^\mu Y_n]$, etc, quite analogous to the combinations that appear inside the soft building block fields in \eq{HopII}.

It is important to understand that when we say that the building blocks are ``collinear gauge invariant'' or ``soft gauge invariant'' we mean up to some transformation at null infinity where the Wilson lines end, which is sometimes synonymous with a global gauge transformation. The quark and gluon building blocks carry color indices which transform covariantly under these remaining transformations. These indices are always fully contracted to give scalar Lagrangians.  The inverse derivatives, $1/\cP_\perp^2$, correspond to a separation in the transverse positions. To ensure the same gauge invariant results obtained here for covariant gauges are also obtained in light-cone gauge, additional transverse Wilson lines at null infinity will presumably be required. In SCET Wilson lines of this type have been considered in ~\cite{Idilbi:2010im,GarciaEchevarria:2011md}. Repeating the tree level matching calculations using $n\cdot A=0$ gauge in QCD it is  straightforward to check that for a single Glauber exchange between $n$ and $\bn$ particles, our same $O_{qq}^{ns\bn}$ operator in \eq{Onsnb} is obtained.

\section{SCET with Glauber Operators}  \label{sec:GlauberEFT}

In this section we present the SCET operators for Glauber Exchange, which determine ${\cal L}_{G}^{{\rm I}(0)}$ and ${\cal L}_{G}^{{\rm II}(0)}$ in \eqs{Lsceta}{Lscetb}, and discuss some of their key properties.

\subsection{Operators for Glauber Exchange in SCET}
\label{sec:GlauberSCET}

Let us now discuss the gauge invariant basis of Glauber operators in SCET which mediate interactions between collinear fields in different sectors and between soft and collinear fields.  The full operators and the lowest order matching are discussed here and the details of the matching calculations needed to derive these operators are left to later sections, including the derivation of the structure of Wilson lines which is given in \sec{treematch}, as well as the derivation of the full structure of the soft mid-rapidity operator which is given in \secs{basis}{2sgluon}.

The Glauber operators can be organized by the number of sectors $\{n,\bn,s\}$ with different rapidities that are involved in each interaction.  Operators describing $n$-$\bn$ interactions also involve soft gluons, and hence three rapidity sectors. These are described first in \sec{ccfwd}, while those for two rapidity sectors, $n$-$s$ or $\bn$-$s$, are described in \sec{SCops}. \sec{matchallpol} carries out the lowest order matching for all gluon polarizations.

\subsubsection{Collinear-Collinear Forward Scattering} \label{sec:ccfwd}

\begin{figure}[t!]
%
%
\subfigure{
\raisebox{-0.2cm}{a)\hspace{0.1cm}} 
\includegraphics[width=0.22\columnwidth]{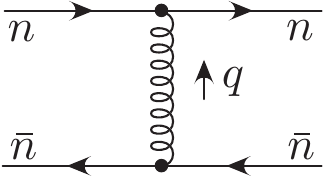}
 \hspace{0.45cm}
\includegraphics[width=0.21\columnwidth]{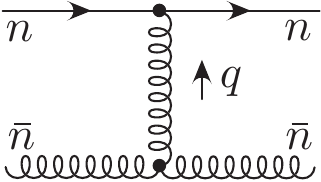}
 \hspace{0.45cm}
\includegraphics[width=0.21\columnwidth]{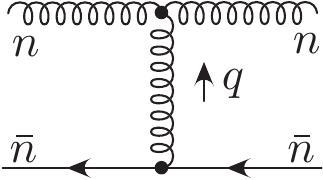}
 \hspace{0.45cm}
\includegraphics[width=0.20\columnwidth]{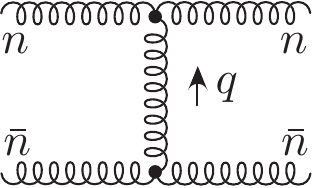}
}
\subfigure{
\raisebox{-0.2cm}{\hspace{0.3cm}  \hspace{0.05cm}} 
\raisebox{0.2cm}{
\includegraphics[width=0.20\columnwidth]{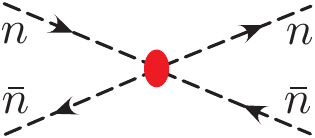}
 }
\hspace{0.23cm}\raisebox{0.8cm}{\Large =}\hspace{0.23cm}  
\includegraphics[width=0.19\columnwidth]{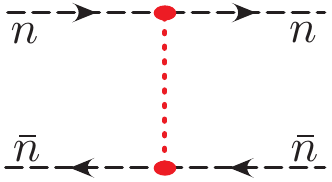}  
\hspace{0.43cm}  
\raisebox{0.2cm}{  
\includegraphics[width=0.20\columnwidth]{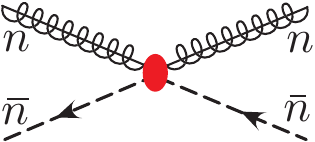}
 }
\hspace{0.15cm}\raisebox{0.8cm}{\Large =}\hspace{0.1cm}  
\raisebox{0.1cm}{  
\includegraphics[width=0.18\columnwidth]{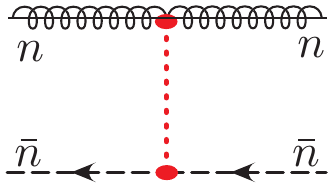}
 }
}
\subfigure{
\raisebox{-0.2cm}{b) \hspace{-0.1cm}} 
\raisebox{0.2cm}{  
\includegraphics[width=0.20\columnwidth]{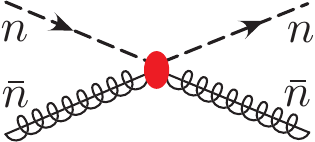}
 }
\hspace{0.23cm}\raisebox{0.8cm}{\Large =}\hspace{0.23cm}  
\includegraphics[width=0.19\columnwidth]{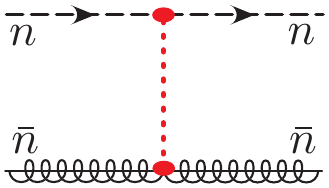}
\hspace{0.45cm} 
\raisebox{0.2cm}{  
\includegraphics[width=0.20\columnwidth]{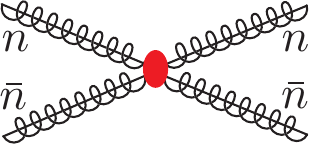}
 }    
\hspace{0.23cm}\raisebox{0.8cm}{\Large =}\hspace{0.23cm}  
\includegraphics[width=0.18\columnwidth]{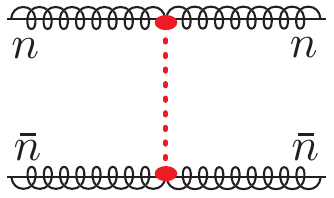}  
}
\caption{\setcaptionskip
Tree level matching for the $nn\bn\bn$ Glauber operators. In a) we show the four full QCD graphs with $t$-channel singularities.  In b) we show the corresponding Glauber operators for the four operators in SCET with two equivalent notations. The notation with the dotted line emphasizes the factorized nature of the $n$ and $\bn$ sectors in the SCET Glauber operators, which have a $1/\cP_\perp^2$ between them. }
\label{fig:treematch_nnb}
\setmainskip
\end{figure}

At tree level we can match onto the Glauber operators in SCET by considering forward scattering diagrams between collinear particles in any two different collinear sectors $n_i\ne n_j$,  or between a soft sector and any collinear sector.  Let us start by consider two collinear sectors $n_i$ and $n_j$.  The relevant forward scattering graphs in QCD with $t$-channel singularities are shown in \fig{treematch_nnb}a.  Each external momentum can be decomposed in light-cone coordinates along the two collinear directions, so
\begin{align} \label{eq:decomposep}
  p^\mu = \frac{n_i^\mu}{2\kappa_{ij}} \, n_j\cdot p + \frac{n_j^\mu}{2\kappa_{ij}}\, n_i\cdot p + p_\perp^\mu 
  \,.
\end{align}
For two generic collinear directions $n_i$ and $n_j$ the vector $p_\perp^\mu$ is defined to be orthogonal to $n_i^\mu$ and $n_j^\mu$, and $\kappa_{ij}= (n_i \cdot n_j)/2 \ne 1$ (and $\kappa_{ij} \gg \lambda^2$ for the two directions to be distinct). With these coordinates a loop measure can be decomposed as
\begin{align}  \label{eq:measure}
  d^4p 
   = \frac{1}{2\kappa_{ij}}\, d(n_i\cdot p)\, d(n_j\cdot p)\, d^2p_\perp
   = \frac{1}{\kappa_{ij}}\, dE_{ij}\, dp^z_{ij}\, d^2p_\perp 
  \,,
\end{align}
where we have defined a generalized ``energy'' and ``longitudinal momentum'' by
\begin{align} \label{eq:genlong}
  E_{ij} &= \frac{n_i\cdot p + n_j \cdot p}{2} \,, 
 & p_{ij}^z & =  \frac{n_i\cdot p - n_j \cdot p}{2}  \,.
\end{align}
For simplicity we will carry out most of our calculations using the back-to-back choice with $n_i=n$, $n_j=\bn$, and $\kappa_{ij} = (n\cdot \bn)/2 = 1$. 
Here we have 
\begin{align} \label{eq:decomposep2}
  p^\mu = 
  \frac{n^\mu}{2} \, \bn\cdot p + \frac{\bn^\mu}{2}\, n\cdot p + p_\perp^\mu  \,,
\end{align}
and the variables in \eq{genlong} reduce to the true energy and longitudinal momentum
\begin{align}  \label{eq:measure2}
  d^4p = \frac{1}{2}\, d(n\cdot p)\, d(\bn\cdot p)\, d^2p_\perp
   =  dp^0\, dp^z\, d^2p_\perp 
  \,.
\end{align}
We will often use the shorthand $p^+=n\cdot p$ and $p^-=\bn\cdot p$. 
All of our calculations, including our final results, will apply equally well to the more general case in \eq{decomposep}. For this more general case factors of $\kappa_{ij}$ must be inserted, but can be inferred by using the invariance to simultaneous rescaling $n_i\to \rho_i n_i$ and $\bn_i\to \bn_i/\rho_i$ for each $i$, which follows from the allowed values for these collinear basis vectors in constructing SCET. This symmetry is called RPI-III invariance~\cite{Manohar:2002fd,Chay:2002vy}.  When we refer to the longitudinal momentum, for this more general case we always mean $p_{ij}^z$. 

We use a common convention for the collinear momenta of the external lines in the $2$--$2$ scattering graphs in \fig{treematch_nnb}a, so $q(p_2^n) + \bar q(p_1^\bn) \to  q(p_3^n) + \bar q(p_4^\bn)$, where the superscripts are included to indicate the type of collinear momentum, and we have the same labeling for the gluon scattering  cases. This is illustrated in \fig{glaub_tree}. When we need to provide further labels to an external particle we will use the same subscript as the momenta, such as for color indices $A_1$, $A_2$, etc, and for vector indices $\mu_1$, $\mu_2$, etc.  Momentum conservation implies $p_1+p_2=p_3+p_4$.  The momentum in the exchange $t$-channel propagator  is $q = p_3-p_2 = p_1-p_4$. The Glauber gluon does not carry large momenta, so the $\sim \lambda^0$ collinear momenta of the particles on the top and bottom of each diagram are conserved:
\begin{align} \label{eq:fwdkinematics}
   \bn\cdot p_2 = \bn\cdot p_3 \,,\qquad\quad   n\cdot p_1 = n\cdot p_4  \,.
\end{align}
These constraints are what ensure the diagrams give forward scattering. To leading power the large Mandelstam invariant is $s= n\cdot p_1\, \bn\cdot p_2  = n\cdot p_4\, \bn\cdot p_3$ and we have the hierarchy  $s\sim \lambda^0 \gg |t|\sim \lambda^2$.  For simplicity we often work in a frame where 
\begin{align} \label{eq:perpkinematics}
  p_1^\perp = -p_4^\perp =q_\perp /2 \,,\quad\qquad
  p_3^\perp = -p_2^\perp =q_\perp/2
  \,.
\end{align} 
Thus for these tree level $2$--$2$ scattering graphs the Mandelstam invariant $t=q_\perp^2 = -\vec q_\perp^{\, 2} < 0$.

For this matching calculation there are four relevant QCD tree graphs, shown in \fig{treematch_nnb}a. They will result in four different Glauber operators, whose Feynman diagrams for this matching are represented by \fig{treematch_nnb}b. For simplicity, here we take $\perp$-polarization for the external gluon fields (leaving the calculation with the full set of polarizations to \sec{matchallpol}).  Expanding in $\lambda$ the results for the top row of diagrams at leading order is
\begin{align}  \label{eq:treennbresult}
 &  i \Big[ \bar u_n \frac{\bnslash}{2} T^B u_n \Big]
       \Big[ \frac{-8\pi\alpha_s(\mu) \delta^{BC}}{\vec q_\perp^{\,2} } \Big]  
         \Big[ \bar v_\bn \frac{\nslash}{2} {\bar T}^C v_\bn \Big] 
  ,\\
 &  i  \Big[ i f^{BA_3A_2} g_\perp^{\mu_2\mu_3} \, \bn\cdot p_2\Big] 
          \Big[ \frac{-8\pi\alpha_s(\mu) \delta^{BC}}{\vec q_\perp^{\,2} } \Big]  
          \Big[ \bar v_\bn \frac{\nslash}{2} {\bar T}^C v_\bn \Big]  
  ,\nn \\
&  i   \Big[ \bar u_n \frac{\bnslash}{2} T^B u_n \Big]
      \Big[ \frac{-8\pi\alpha_s(\mu) \delta^{BC}}{\vec q_\perp^{\,2} } \Big]  
          \Big[ i f^{CA_4A_1} g_\perp^{\mu_1\mu_4} \, n\cdot p_1\Big] 
   ,\nn \\ 
&  i  \Big[ i f^{BA_3A_2} g_\perp^{\mu_2\mu_3} \, \bn\cdot p_2\Big]  
        \Big[ \frac{-8\pi\alpha_s(\mu) \delta^{BC}}{\vec q_\perp^{\,2} } \Big]   
          \Big[ i f^{CA_4A_1} g_\perp^{\mu_1\mu_4} \, n\cdot p_1\Big] 
  . \nn
\end{align}
In writing these results we have written out the collinear quark spinors but left off the collinear gluon polarization vectors $\varepsilon_n^{\mu_2 A_2}(p_2)$ etc, for simplicity.  We use color index $A_i$ for the external gluon of momentum $p_i$. We are also using a $\bar 3$ representation for the antiquark spinors and color generators.\footnote{The relation between our notation and that of Ref.~\cite{Peskin} is $\big[\bar v_\bn(4) \frac{\nslash}{2} \bar T^C v_\bn(1) \big]^{\rm us} =-\big[\bar v_\bn(1) \frac{\nslash}{2} T^C v_\bn(4)\big]^{\rm Peskin}$, where $\bar T^C = - (T^{C})^T$ and $v_\bn(i,\uparrow)^{\rm us} = [ v^*_\bn(i,\downarrow)]^{\rm Peskin}$. Our use of conjugate fields for antifermions puts the quarks and antiquarks on the same footing. Quarks and antiquarks have the same collinear propagators with a momentum flowing forward, we use $\bar T^a$ for the antiquark color generator, and there are no additional signs for external antiquarks.
This setup is commonly used in the spinor-helicity literature~\cite{Dixon:1996wi}. See \app{Antiquarks} for further details.
}

We begin our analysis by discussing the $\SCETb$ operators whose tree level matrix elements reproduce the results in \eq{treennbresult}.   The four \SCETb operators whose matrix elements reproduce \eq{treennbresult} factorize into collinear and soft operators separated by $1/{\cP_\perp^2}$ factors, so we adopt the notation: 
\begin{align}  \label{eq:Onsnb}
     O^{q q}_{ns\bn} &=  {\cal O}_n^{q B}   \frac{1}{\cP_\perp^2} {\cal O}_s^{BC} \frac{1}{\cP_\perp^2} {\cal O}_\bn^{q C} \,,
    &  O^{g q}_{ns\bn} &=  {\cal O}_n^{g B}   \frac{1}{\cP_\perp^2}  {\cal O}_s^{BC} \frac{1}{\cP_\perp^2}{\cal O}_\bn^{q C} \,,
    \nn\\
        O^{q g}_{ns\bn} &=  {\cal O}_n^{q B} \frac{1}{\cP_\perp^2} {\cal O}_s^{BC} \frac{1}{\cP_\perp^2} {\cal O}_\bn^{g C} \,,
    &  O^{g g}_{ns\bn} &=  {\cal O}_n^{g B} \frac{1}{\cP_\perp^2} {\cal O}_s^{BC} \frac{1}{\cP_\perp^2} {\cal O}_\bn^{g C} \,.
\end{align}
On the left-hand side the subscripts indicate that these operators involve three sectors $\{n,s,\bn\}$, while the first and second superscript determine whether we take a quark or gluon operator in the $n$-collinear or $\bn$-collinear sectors. Without soft gluons we have ${\cal O}_s^{BC}=8\pi\alpha_s\delta^{BC} \cP_\perp^2$. 

The $n$-collinear quark and gluon terms, which occur in the first square bracket in each of the four terms in \eq{treennbresult},  are matrix elements of the $n$-collinear operators
\begin{align}  \label{eq:On}
  {\cal O}_n^{q B} &= \overline\chi_{n} T^B \frac{\bnslash}{2} \: \chi_{n} \,,
  & {\cal O}_n^{g B} &= \biggl[ \frac{i}{2} f^{BCD}  {\cal B}_{n\perp\mu}^C \,
   \frac{\bn}{2}\cdot (\cP\!+\!\cP^\dagger)  {\cal B}_{n\perp}^{D\mu} \biggr]  \,.
\end{align}
Each of these operators are bilinears in the quark or gluon building blocks. For the gluon operator, an extra factor of $1/2$ is included to compensate for the symmetry factor from switching the two ${\cal B}_{n\perp}$s when computing the corresponding Feynman rules. The operator ${\cal O}_n^{gB}$ is even under this swap because both the color factor and momentum factor $\bn\cdot (\cP+\cP^\dagger)$ give a change of sign. The $\bn$-collinear quark and gluon terms appear as the contributions in the last square brackets of each of the four terms in \eq{treennbresult}, and are matrix elements of the operators,
\begin{align}  \label{eq:Obn}
  {\cal O}_\bn^{q B} &= \overline\chi_{\bn} T^B \frac{\nslash}{2} \: \chi_{\bn} \,,
  & {\cal O}_\bn^{g B} &= \biggl[ \frac{i}{2} f^{BCD}  {\cal B}_{\bn\perp\mu}^C \,
    \frac{n}{2} \cdot (\cP\!+\!\cP^\dagger) {\cal B}_{\bn\perp}^{D\mu} \biggr]  \,.
\end{align}
Examining \eqs{On}{Obn} we see that the $n$-collinear and $\bn$-collinear results are the same, just with $n\leftrightarrow \bn$.    These collinear operators are bilinears of the fundamental quark and gluon gauge invariant building block operators in SCET. Furthermore, both of these operators are octet combinations of the building blocks.  Due to  momentum conservation, and the fact that there are only two building blocks in each collinear sector, each collinear bilinear has a conserved momentum in its large $\sim\lambda^0$ component. This implements the forward scattering kinematics.  The tree level matching that yields the proper Wilson line structure in the operators in \eqs{On}{Obn} is actually non-trivial due to operator mixing, and is described in detail in Sec.~\ref{sec:treematch}.

The middle terms in square brackets in \eq{treennbresult}, those involving $\alpha_s$, do not have objects like polarization vectors or spinors that correspond to external lines. Nevertheless, they are actually matrix elements of a soft operator which involves soft gluon fields as well as soft Wilson lines.  Accounting for the $1/\cP_\perp^2$ factors in \eq{Onsnb} these operators must reduce to $8\pi\alpha_s \delta^{BC} {\cal P}_\perp^2$ when all soft fields are turned off.  The full soft operator is non-trivial, and is derived in Sec.~\ref{sec:lipatov}, where we obtain
\begin{align}  \label{eq:Os1}
    {\cal O}_s^{BC} 
 & ={8\pi\alpha_s  }
    \bigg\{
    \cP_\perp^\mu {\cal S}_n^T {\cal S}_\bn  \cP_{\perp\mu}
    - \cP^\perp_\mu g \widetilde {\cal B}_{S\perp}^{n\mu}  {\cal S}_n^T  {\cal S}_\bn   
    -  {\cal S}_n^T  {\cal S}_\bn  g \widetilde {\cal B}_{S\perp}^{\bn\mu} \cP^\perp_{\mu}  
    -  g \widetilde {\cal B}_{S\perp}^{n\mu}  {\cal S}_n^T  {\cal S}_\bn g \widetilde {\cal B}_{S\perp\mu}^{\bn}
\nn\\
 &\qquad\qquad
    -\frac{n_\mu \bn_\nu}{2} {\cal S}_n^T   ig \widetilde {G}_s^{\mu\nu} {\cal S}_\bn 
    \bigg\}^{BC} 
    \,.
\end{align}
Here the ${\cal S}_n$ and ${\cal S}_\bn$ Wilson lines are in the adjoint representation as described near Eq.~(\ref{eq:Weqtns})  and the other field objects $\widetilde {\cal B}_{S\perp}^{n}$, $\widetilde {\cal B}_{S\perp}^{\bn}$, and $\widetilde G_s$ are matrices in the adjoint space as in Eq.~(\ref{eq:Bmatrix}).
The adjoint soft Wilson lines ${\cal S}_n$ and ${\cal S}_\bn$ are necessary to maintain soft gauge invariance and  are generated from integrating out off-shell lines, with virtuality $p^2\sim Q^2\lambda \gg Q^2\lambda^2$,  that arise from the emission of a soft line off of a collinear line.
The operator in \eq{Os1} is gauge invariant under soft gauge transformations that vanish at infinity.  The fact
that we have a non-trivial soft operator $O_s^{BC}$ is related to the existence of the soft sector that sits at rapidities between the $n$-collinear and $\bn$-collinear fields.  Here we have been deliberately glib about the multipole expansion for this non-local operator, but will describe this fully in section \sec{LGtransverse} below. The directions for these soft Wilson lines are discussed in \sec{rapidityregulator}.

\begin{figure}[t!]
\begin{center}
\includegraphics[width=0.19\columnwidth]{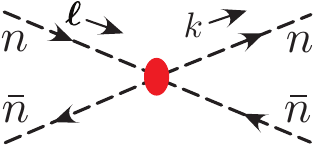}
 \hspace{0.4cm}
 \hspace{-0.05cm}
 \raisebox{0.55cm}{\large $=  
   \mbox{\Large $\frac{-8\pi i \alpha_s }{(\vec \ell_\perp-\vec k_\perp)^2 }$}  
     \Big[ \bar u_n \frac{\bnslash}{2} T^A u_n  \Big]  
     \Big[ \bar v_\bn \frac{\nslash}{2} \bar T^{A} v_\bn \Big] $ \hspace{5cm} }
  \\[10pt]
\includegraphics[width=0.21\columnwidth]{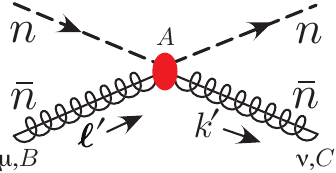}
 \hspace{0.2cm} 
 \hspace{-0.15cm}
 \raisebox{1cm}{\large $=  
    \mbox{\Large $\frac{-8\pi \alpha_s f^{ABC} }{(\vec \ell^\prime_\perp-\vec k^\prime_\perp)^2 }$}  
     \Big[ \bar u_n \frac{\bnslash}{2} T^A u_n  \Big]  
     \Big[ n\mcdot k^\prime\, g_\perp^{\mu\nu} - n^\mu \ell^{\prime\nu}_\perp -n^\nu k^{\prime\mu}_\perp 
      + \mbox{\Large $\frac{\ell^\prime_\perp\cdot k^\prime_\perp n^\mu n^\nu}{n\cdot k^\prime}$} \Big] $  \hspace{-0.3cm}  }
  \\[10pt]
\includegraphics[width=0.21\columnwidth]{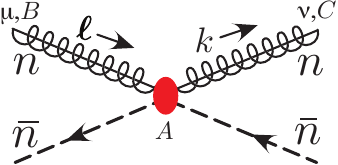}
 \hspace{0.2cm} 
 \hspace{-0.15cm}
 \raisebox{0.6cm}{\large $= 
 \mbox{\Large $\frac{-8\pi \alpha_s f^{ABC} }{(\vec \ell_\perp-\vec k_\perp)^2 }$}  
     \Big[ \bn\mcdot k\, g_\perp^{\mu\nu} - \bn^\mu \ell_\perp^\nu -\bn^\nu k_\perp^\mu 
      + \mbox{\Large $\frac{\ell_\perp\cdot k_\perp \bn^\mu \bn^\nu}{\bn\cdot k}$} \Big] 
    \Big[ \bar v_\bn \frac{\nslash}{2} \bar T^{A} v_\bn \Big] $ }
\\[10pt]
\includegraphics[width=0.21\columnwidth]{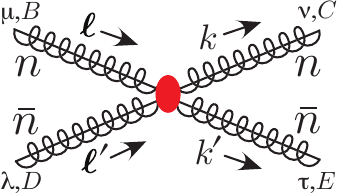}
 \hspace{0.2cm} 
 \hspace{-0.35cm}
 \raisebox{0.9cm}{
  \begin{minipage}{4in}
  \large $=
  \mbox{\Large $\frac{8\pi i \alpha_s f^{ABC}f^{ADE} }{(\vec \ell_\perp-\vec k_\perp)^2 }$}  
     \Big[ \bn\mcdot k\, g_\perp^{\mu\nu} - \bn^\mu \ell_\perp^\nu -\bn^\nu k_\perp^\mu 
      + \mbox{\Large $\frac{\ell_\perp\cdot k_\perp \bn^\mu \bn^\nu}{\bn\cdot k}$} \Big] $
\\
\large $ \times
  \Big[ n\mcdot k^\prime\, g_\perp^{\lambda\tau} 
   - n^\lambda \ell^{\prime\tau}_\perp -n^\tau k^{\prime\lambda}_\perp 
      + \mbox{\Large $\frac{\ell^\prime_\perp\cdot k^\prime_\perp n^\lambda n^\tau}{n\cdot k^\prime}$} \Big] $  
 \end{minipage}
\hspace{1cm}
}
\end{center}
\caption{\setcaptionskip
Lowest order Feynman rules for the Glauber operators $O_{ns\bn}^{ij}$  for $n$-$\bn$ forward scattering.}
\label{fig:LOfeynrule}
\setmainskip
\end{figure}

At lowest order the Feynman diagrams for these operators will be denoted as in \fig{treematch_nnb}b. Two notations are used, one with an extended red dashed line which serves to remind us that the matrix element of ${\cal O}_s^{BC}$ is non-local, giving a potential that scales as $\lambda^{-2}$. The alternate notation collapses this red dashed line to an elliptical blob to indicate that it has no field dependent dynamics.  In general the elliptical red Glauber blob indicates an interaction between either three or two rapidity sectors in this manner,
\begin{align} \label{eq:glaubnotation}
\raisebox{-1cm}{  
\includegraphics[width=0.15\columnwidth]{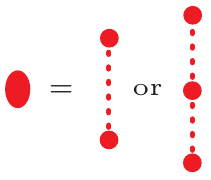} 
} \ \ .
\end{align}
The complete tree level Feynman rule for the quark operator $O_{ns\bn}^{q q}$ is identical to the result used for the matching in \eq{treennbresult}, but this is not the case for the gluon operators since they have terms from other polarizations (derived below in \sec{matchallpol}).  For future use we record the full set of Feynman rules at lowest order in the coupling expansion in Fig.~\ref{fig:LOfeynrule}.

\begin{figure}[t!]
%
%
\includegraphics[width=0.21\columnwidth]{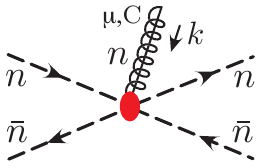} 
\hspace{0.1cm}
 \raisebox{0.7cm}{\large $=$}
 \hspace{-0.05cm}
\raisebox{0.1cm}{
\includegraphics[width=0.15\columnwidth]{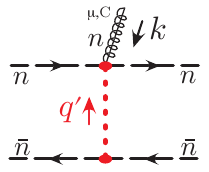}
 }
  \hspace{-0.1cm}
 \raisebox{0.7cm}{\large $=\ i \Big[ \bar u_n \frac{\bnslash}{2} T^A u_n \, 
     \mbox{\normalsize $ig f^{AB'C}$}  \frac{\bn^\mu}{\bn\cdot k} \Big]
       \Big[ \frac{-8\pi\alpha_s\delta^{B'B}}{\vec q_\perp^{\,\prime 2} } \Big]  
         \Big[ \bar v_\bn \frac{\nslash}{2} \bar T^{B} v_\bn \Big]$ }
\\
\hspace{0.cm}
\includegraphics[width=0.21\columnwidth]{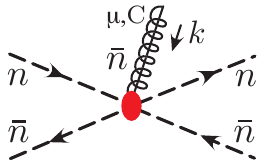} 
 \hspace{0.1cm}
 \raisebox{0.7cm}{\large $=$}
 \hspace{-0.05cm}
\raisebox{-0.5cm}{
\includegraphics[width=0.15\columnwidth]{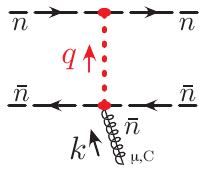}
 }
\hspace{-0.1cm}
 \raisebox{0.7cm}{\large $=\ i \Big[ \bar u_n \frac{\bnslash}{2} T^A u_n  \Big]
       \Big[ \frac{-8\pi\alpha_s\delta^{AA'}}{\vec q_\perp^{\, 2} } \Big]  
         \Big[ \mbox{\normalsize $-ig f^{A'BC}$} \frac{n^\mu}{n\cdot k}\, 
      \bar v_\bn \frac{\nslash}{2} \bar T^{B} v_\bn  \Big] $ }
  \hspace{-0.1cm}
\\
\vspace{-0.1cm} \hspace{-.15cm}
\includegraphics[width=0.21\columnwidth]{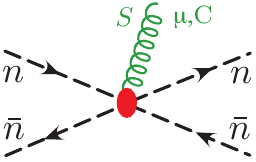}
 \hspace{0.2cm} \raisebox{0.7cm}{\large $=$}  
\raisebox{-0.1cm}{
\includegraphics[width=0.22\columnwidth]{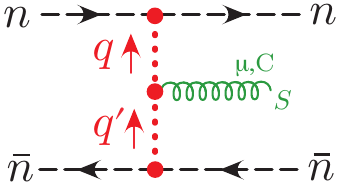} 
}
\hspace{7cm}
\\[8pt]  
\hspace{3.5cm}
 \raisebox{0.3cm}{\large\qquad $=i \Big[ \! \bar u_n \frac{\bnslash}{2} T^A u_n \! \Big] \!
       \bigg[ \frac{8\pi\alpha_s}{\vec q_\perp^{\,2}\, \vec q_\perp^{\,\prime 2} } 
        \:  \mbox{\normalsize $ig f^{ABC}$} 
     \bigg( \mbox{\normalsize $q_\perp^\mu\!+\!q_\perp^{\prime\mu}$} 
            \mbox{\normalsize $\,-\, n\cdot q^\prime$} \frac{\bn^\mu}{2}
          \mbox{\normalsize $\,-\, \bn\cdot q$} \frac{n^\mu}{2}
          \mbox{\normalsize $\,-\,$} \dfrac{n^\mu \vec q_\perp^{\: 2}}{n\cdot q'} 
          \mbox{\normalsize $\,-\,$} \dfrac{\bn^\mu \vec q_\perp^{\: \prime 2}}{\bn\cdot q}  
    \bigg)  \bigg]  
       \! \Big[ \! \bar v_\bn \frac{\nslash}{2} \bar T^{B} v_\bn\!  \Big]
$ } 
\vspace{-0.1cm}
\caption{\setcaptionskip
One gluon with incoming momentum $k$ emitted from the $O_{nS\bn}^{q q}$ Glauber operator.  The first two Feynman rules come from Wilson lines in the $n$-collinear and $\bn$-collinear part of the operator. The last Feynman rule comes from the soft component of the operator, and corresponds with the Lipatov vertex.}
\label{fig:onegluon}
\setmainskip
\end{figure}

There are additional Feynman rules when the operators emit another gluon. For example, consider $O_{ns\bn}^{q q}$ where $q_\perp=p_{1\perp}-p_{4\perp}$ and  $q'_\perp=p_{3\perp}-p_{2\perp}$ are momentum transfers stemming from the  $n$ and  $\bn$-collinear quarks respectively  (following Fig.~\ref{fig:glaub_tree}), and $k$ is the incoming momentum of the gluon.
Then the Feynman rules with one additional $n$-collinear gluon, $\bn$-collinear gluon, or soft gluon emitted are shown in Fig.~\ref{fig:onegluon}.

\begin{figure}[t!]
	%
	%
	%
	\hspace{0.cm}
	\includegraphics[width=0.26\columnwidth]{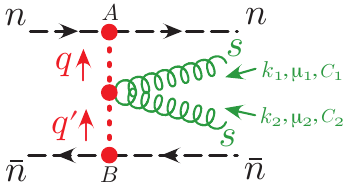} 
 \hspace{0.2cm} \raisebox{1.cm}{\large
  $=i \Big[ \! \bar u_n \frac{\bnslash}{2} T^A u_n \! \Big] \! 
      \Big[ \! \bar v_\bn \frac{\nslash}{2} \bar T^{B} v_\bn\!  \Big] \!
       \Big( \dfrac{8\pi\alpha_s}{\vec q_\perp^{\:2}\, \vec q_\perp^{\:\prime 2} } \Big) 
  $\hspace{6cm} } \\[4pt]
 \phantom{x}\hspace{0.4cm}
 \raisebox{0.7cm}{\large
  $ \times \bigg\{ g^2 f^{C_1 A E} f^{C_2 B E} \bigg[ 
   -g_\perp^{\mu_1\mu_2} - \dfrac{ n^{\mu_1} (2q_\perp^{\prime\mu_2}\!+\!k_{2\perp}^{\mu_2})}{n\cdot k_1}   
    + \dfrac{(2 q_\perp^{\mu_1}\!-\! k_{1\perp}^{\mu_1}) \bn^{\mu_2}}{\bn\cdot k_2}  
    + \dfrac{\bn^{\mu_1} n^{\mu_2}\!-\!n^{\mu_1}\bn^{\mu_2}}{2}
$ } \\[-2pt]
 \phantom{x}\hspace{0.5cm}
\hspace{0.4cm}
 \raisebox{0.7cm}{\large
  $ + \dfrac{n^{\mu_1} \bn^{\mu_2}}{n\cdot k_1\,\bn\cdot k_2} 
   \Big( \vec q_\perp\mcdot \vec q_\perp^{\:\prime} 
  \plus \vec k_{1\perp}\mcdot \vec k_{2\perp} 
  \plus \vec k_{1\perp} \mcdot \vec q_\perp 
  \minus \vec k_{2\perp} \mcdot \vec q_\perp^{\:\prime} 
  \minus \frac12 n\cdot k_2\: \bn\cdot k_2   
  \minus \frac12 n\cdot k_1\: \bn\cdot k_1   \Big)  
$ } \\[-2pt]
 \phantom{x}\hspace{0.5cm}
 \hspace{0.4cm}
 \raisebox{0.7cm}{\large
  $ 
    +\, n^{\mu_1} n^{\mu_2} \Big( \dfrac{\vec q_{\perp}^{\: 2}}{n\cdot q'\: n\cdot k_1}  + \dfrac{\bn\cdot k_2}{2n\cdot k_1}\Big)
    + \bn^{\mu_1} \bn^{\mu_2} \Big( \dfrac{-\vec q_{\perp}^{\:\prime 2}}{\bn\cdot k_2\: \bn\cdot q }  + \dfrac{n\cdot k_1}{2\bn\cdot k_2} \Big)
   \bigg]  
$ }\\[-2pt]
  \phantom{x}\hspace{0.4cm}
 \hspace{0.2cm}
 \raisebox{0.7cm}{\large
  $ + g^2 f^{C_2 A E} f^{C_1 B E}   \bigg[ 
   -g_\perp^{\mu_1\mu_2}  + \dfrac{\bn^{\mu_1}
  (2q_\perp^{\mu_2}\!-\! k_{2\perp}^{\mu_2})}{\bn\cdot k_1}  
  - \dfrac{(2 q_\perp^{\prime\mu_1} \!+\! k_{1\perp}^{\mu_1}) n^{\mu_2}}
  {n\cdot k_2}  
  + \dfrac{n^{\mu_1} \bn^{\mu_2} \!-\! \bn^{\mu_1}n^{\mu_2}}{2}
$ }\\[-2pt]
   \phantom{x}\hspace{0.4cm}
 \hspace{0.5cm}
 \raisebox{0.7cm}{\large
  $ + \dfrac{\bn^{\mu_1} n^{\mu_2}}{n\cdot k_2\: \bn\cdot k_1} 
  \Big( \vec q_\perp\mcdot \vec q_\perp^{\:\prime} 
  \plus \vec k_{1\perp}\mcdot \vec k_{2\perp} 
  \plus \vec k_{2\perp} \mcdot \vec q_\perp 
  \minus \vec k_{1\perp} \mcdot \vec q_\perp^{\:\prime} 
  \minus \frac12 n\cdot k_2\: \bn\cdot k_2   
  \minus \frac12 n\cdot k_1\: \bn\cdot k_1 \Big)  
$ } \\[-2pt]
   \phantom{x}\hspace{0.4cm}
\hspace{0.5cm}
 \raisebox{0.7cm}{\large
  $ 
  + n^{\mu_1} n^{\mu_2} \Big( \dfrac{\vec q_{\perp}^{\: 2}}{n\cdot q'\: n\cdot k_2}  + \dfrac{\bn\cdot k_1}{2n\cdot k_2}\Big)
  + \bn^{\mu_1} \bn^{\mu_2} \Big( \dfrac{-\vec q_{\perp}^{\:\prime 2}}{\bn\cdot k_1\: \bn\cdot q }  + \dfrac{n\cdot k_2}{2\bn\cdot k_1} \Big)
   \bigg]  \bigg\}
$ }
	\vspace{-0.5cm}
	\caption{\setcaptionskip
		Two Soft Gluon Feynman rule for the $O_{ns\bn}^{qq}$ operator. The  terms in $\{\cdots\}$ times $(8\pi\alpha_s)$ are the universal two soft gluon contribution from $O_s^{AB}$. }
	\label{fig:twosoftgluon_feynrule}
	\setmainskip
\end{figure}

The Feynman rule with the soft gluon has contributions from all polarizations and reproduces the Lipatov vertex~\cite{Kuraev:1976ge} used in small-$x$ physics.  Our soft operator has terms beyond the Lipatov vertex from two and more gluon terms which we will discuss and make use of later on. The two soft gluon Feynman rule is shown in \fig{twosoftgluon_feynrule}.   The result in Eq.~(\ref{eq:Os1})  has not previously  appeared in either the QCD or SCET literature, and gives a completely gauge invariant factorized operator that reproduces both forward scattering and the Lipatov vertex.

The scaling for the component operators in \eq{Onsnb} are all identical: $O_n^{i B}\sim \lambda^2$, $O_\bn^{i B}\sim \lambda^2$, and $O_s^{AB}\sim \lambda^2$.  Thus the overall operators in \eq{Onsnb} scale as $O_{ns\bn}^{i j}\sim \lambda^2$.  As we will see below in Sec.~\ref{sec:powercount}, for this type of Glauber operator this scaling yields contributions that are leading order in the power counting for both forward scattering and for hard scattering processes once the scaling of the measure is included. Therefore the operators in \eq{Onsnb} contribute to the leading order Lagrangian in SCET.  Due to our normalization of ${\cal O}_s^{BC}$ the tree level Wilson coefficient for all four of these operators are $1$, and we will later argue that this is true to all orders.  These are the first four terms appearing in our Glauber Lagrangian. We can summarize our matching result, and extend it to other pairs of distinct collinear sectors by writing
\begin{align}  \label{eq:LGsofar}
  {\cal L}_G^{{\rm II}(0)}  & =  e^{-i x\cdot \cP}
  \sum_{n_1,n_2} \sum_{i,j=q,g}  \: 
 {\cal O}_{n_1}^{i B}   \frac{1}{\cP_\perp^2} {\cal O}_s^{BC}    \frac{1}{\cP_\perp^2} {\cal O}_{n_2}^{j C} 
   + \ldots \,,
\end{align}
where the ellipses denote additional leading power terms involving rescattering of soft fields to be discussed below. In this sum $n_1$ and $n_2$ label distinct collinear sectors. (When $n_1\cdot n_2\ne 2$ there are factors of $(2/n_1\cdot n_2)$ in a couple of places which can be inserted using the RPI symmetry.)

\subsubsection{Soft-Collinear Forward Scattering} \label{sec:SCops}

\begin{figure}[t!]
%
%
\raisebox{1.8cm}{a)\hspace{0.1cm}} 
\includegraphics[width=0.22\columnwidth]{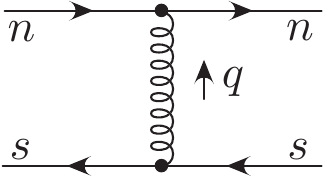}
 \hspace{0.45cm}
\includegraphics[width=0.21\columnwidth]{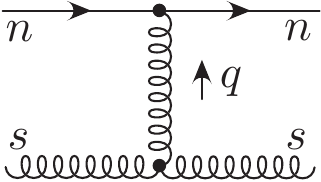}
 \hspace{0.45cm}
\includegraphics[width=0.21\columnwidth]{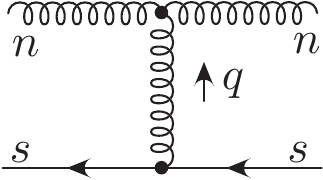}
 \hspace{0.45cm}
\includegraphics[width=0.20\columnwidth]{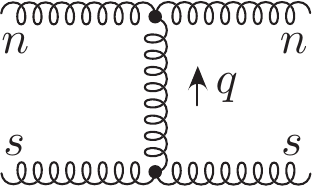}
\\[8pt]
\raisebox{1.4cm}{b)\hspace{0.0cm}} 
\raisebox{0.cm}{  
\includegraphics[width=0.21\columnwidth]{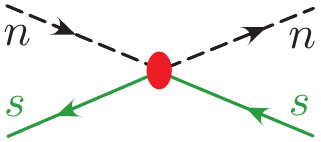}
 \hspace{0.48cm}  
\includegraphics[width=0.21\columnwidth]{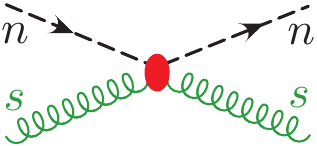}
 \hspace{0.43cm}  
\includegraphics[width=0.21\columnwidth]{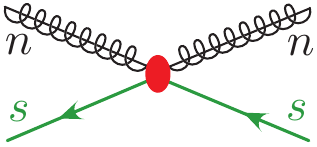}
 \hspace{0.42cm}  
\includegraphics[width=0.21\columnwidth]{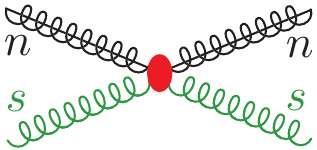}  
}
\caption{\setcaptionskip
Tree level matching for the $nnss$ Glauber operators. In a) we show the four full QCD graphs with $t$-channel singularities.  In b) we show the corresponding Glauber operators in SCET. The matching between the two is given by reading down each column. Results for $\bn\bn ss$ are analogous with $n\leftrightarrow \bn$.}
\label{fig:treematch_ns}
\setmainskip
\end{figure}

An analogous matching calculation can be done for the forward scattering between soft and $n$-collinear fields.   We show the diagrams for this matching calculation in \fig{treematch_ns} and label the momenta for this calculation as $q(p_2^n)+\bar q(p_1^S) \to q(p_3^n) + \bar q(p_4^S)$. We use an analogous labeling for the cases with gluons.  Here the large ${\cal O}(\lambda^0)$ $n$-collinear momentum is conserved as before. Since the soft momenta $n\cdot p_{1,4}\sim \lambda \gg n\cdot p_{2,3} \sim \lambda^2$ they are also conserved by the exchanged Glauber gluon, so we again for these diagrams we have forward scattering with the constraints
\begin{align}  \label{eq:fwdscatterings}
   n\cdot p_1 = n\cdot p_4 \,,\qquad   \bn\cdot p_2 = \bn\cdot p_3  \,.
\end{align}
Or in other words, the $n\cdot p_s$ momentum is conserved on the soft line and the $\bn\cdot p_n$ momentum is conserved on the $n$-collinear line.
 The $\perp$-momenta are the same size here as in the $n$-$\bn$ scattering case, and we follow again the convention that $p_1^\perp = -p_4^\perp = p_3^\perp = -p_2^\perp = q_\perp /2$.  
Again it is convenient to use $\perp$ polarizations to carry out the matching. (In \sec{matchallpol} we will show how all the polarizations can be matched when the on-shell conditions are used.)  Computing the full QCD graphs in \fig{treematch_ns}a and expanding in $\lambda$ gives
\begin{align}  \label{eq:treensresult}
 &  i \Big[ \bar u_n \frac{\bnslash}{2} T^B u_n \Big]
       \Big[ \frac{-1}{\vec q_\perp^{\,2} } \Big]  
         \Big[ 8\pi\alpha_s \bar v_s \frac{\nslash}{2} \bar T^B v_s \Big] 
  , \\
 &   i   \Big[ \bar u_n \frac{\bnslash}{2} T^B u_n \Big] 
      \Big[ \frac{-1}{\vec q_\perp^{\,2} } \Big]  
          \Big[ 8\pi\alpha_s i f^{BA_4A_1} g_\perp^{\mu_1\mu_4} \, n\cdot p_1\Big]  
  ,\qquad 
  i  \Big[ i f^{BA_3A_2} g_\perp^{\mu_2\mu_3} \, \bn\cdot p_2\Big]  
          \Big[ \frac{-1}{\vec q_\perp^{\,2} } \Big]  
          \Big[ 8\pi\alpha_s \bar v_s \frac{\nslash}{2} \bar T^B v_s \Big]  
        , 
\nn\\
&  i  \Big[ i f^{BA_3A_2} g_\perp^{\mu_2\mu_3} \, \bn\cdot p_2\Big] 
        \Big[ \frac{-1}{\vec q_\perp^{\,2} } \Big]   
          \Big[ 8\pi\alpha_s  i f^{BA_4A_1} g_\perp^{\mu_1\mu_4} \, n\cdot p_1\Big] 
      .\nn
\end{align}
Thus despite the differences in the scaling of momenta, the results for the $n$-$s$ scattering are essentially the same as for the $n$-$\bn$ scattering given above in \eq{treennbresult}.  The reason for this is that the comparison of light-cone momenta in these two cases is the same, the $\bn\cdot p$ momenta are largest for the $n$-collinear particles, and the $n\cdot p$ momenta are larger for the $\bn$ or soft particles than they are for the $n$-collinear particles.  For the four SCET operators that are responsible for forward scattering of soft with $n$-collinear particles we write operators with $n$-collinear and soft components separated by a $1/{\cal P}_\perp^2$ factor
\begin{align}  \label{eq:Ons}
    O^{qq}_{ns} &=  {\cal O}_n^{q B} \frac{1}{\cP_\perp^2}  {\cal O}_s^{q_n B} \,,
   & O^{qg}_{ns} &= {\cal O}_n^{q B} \frac{1}{\cP_\perp^2}  {\cal O}_s^{g_n B} \,,
   & O^{gq}_{ns} &=  {\cal O}_n^{g B} \frac{1}{\cP_\perp^2}  {\cal O}_s^{q_n B} \,,
   & O^{gg}_{ns} &= {\cal O}_n^{g B} \frac{1}{\cP_\perp^2}  {\cal O}_s^{g_n B} \,.
\end{align}

\begin{figure}[t!]
\begin{center}
\hspace{0.2cm}
\includegraphics[width=0.2\columnwidth]{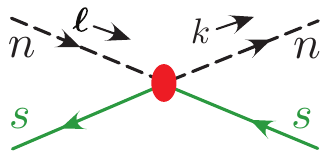} 
\hspace{0.4cm}
 \hspace{-0.25cm}
 \raisebox{0.55cm}{\hspace{-0.2cm}\large $=  
   \mbox{\Large $\frac{-8\pi i \alpha_s }{(\vec \ell_\perp-\vec k_\perp)^2 }$}  
     \Big[ \bar u_n \frac{\bnslash}{2} T^A u_n  \Big]  
     \Big[ \bar v_s \frac{\nslash}{2} \bar T^{A} v_s \Big] $ \hspace{5cm} }
  \\[10pt]
\includegraphics[width=0.21\columnwidth]{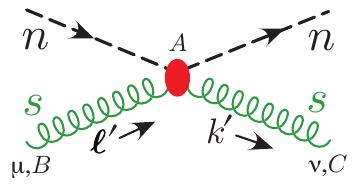}
 \hspace{0.2cm} 
 \hspace{-0.15cm}
 \raisebox{1cm}{\large $=  
    \mbox{\Large $\frac{-8\pi \alpha_s f^{ABC} }{(\vec \ell^\prime_\perp-\vec k^\prime_\perp)^2 }$}  
     \Big[ \bar u_n \frac{\bnslash}{2} T^A u_n  \Big]  
     \Big[ n\mcdot k^\prime\, g_\perp^{\mu\nu} - n^\mu \ell^{\prime\nu}_\perp -n^\nu k^{\prime\mu}_\perp 
      + \mbox{\Large $\frac{\ell^\prime_\perp\cdot k^\prime_\perp n^\mu n^\nu}{n\cdot k^\prime}$} \Big] $  \hspace{-0.3cm}  }
  \\[10pt]
\includegraphics[width=0.21\columnwidth]{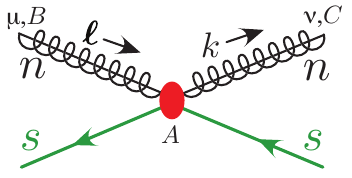}
 \hspace{0.2cm} 
 \hspace{-0.15cm}
 \raisebox{0.6cm}{\large $= 
 \mbox{\Large $\frac{-8\pi \alpha_s f^{ABC} }{(\vec \ell_\perp-\vec k_\perp)^2 }$}  
     \Big[ \bn\mcdot k\, g_\perp^{\mu\nu} - \bn^\mu \ell_\perp^\nu -\bn^\nu k_\perp^\mu 
      + \mbox{\Large $\frac{\ell_\perp\cdot k_\perp \bn^\mu \bn^\nu}{\bn\cdot k}$} \Big] 
    \Big[ \bar v_s \frac{\nslash}{2}\bar  T^{A} v_s \Big] $ }
\\[10pt]
\includegraphics[width=0.21\columnwidth]{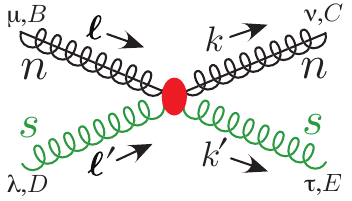}
 \hspace{0.2cm} 
 \hspace{-0.35cm}
 \raisebox{0.9cm}{
  \begin{minipage}{4in}
  \large $=
  \mbox{\Large $\frac{8\pi i \alpha_s f^{ABC}f^{ADE} }{(\vec \ell_\perp-\vec k_\perp)^2 }$}  
     \Big[ \bn\mcdot k\, g_\perp^{\mu\nu} - \bn^\mu \ell_\perp^\nu -\bn^\nu k_\perp^\mu 
      + \mbox{\Large $\frac{\ell_\perp\cdot k_\perp \bn^\mu \bn^\nu}{\bn\cdot k}$} \Big] $
\\
\large $ \times
  \Big[ n\mcdot k^\prime\, g_\perp^{\lambda\tau} 
   - n^\lambda \ell^{\prime\tau}_\perp -n^\tau k^{\prime\lambda}_\perp 
      + \mbox{\Large $\frac{\ell^\prime_\perp\cdot k^\prime_\perp n^\lambda n^\tau}{n\cdot k^\prime}$} \Big] $  
 \end{minipage}
\hspace{1cm}
}
\end{center} 
\vspace{-0.4cm}
\caption{\setcaptionskip
Lowest order Feynman rules for the Glauber operators $O_{ns}^{ij}$  for $n$-$s$ forward scattering. Results for $O_{\bn s}^{ij}$ are analogous with $n\leftrightarrow \bn$.}
\label{fig:LOfeynrulens}
\setmainskip
\end{figure}

The structure of soft Wilson lines in $O_s^{q_n B}$ and $O_s^{g_nB}$ is determined by the direction of the collinear fields, explaining why we add the additional subscript $n$ to the quark and gluon superscripts: $q_n$ and $g_n$.    The SCET operators which reproduce the result in \eq{treensresult} again involve $O_n^{q B}$ or $O_n^{g B}$ from \eq{On} for the $n$-collinear sector terms in the left-most square brackets, just  as was the case for reproducing \eq{treennbresult}. For the soft-collinear scattering there does not exist a set of fields that are between these sectors in rapidity, hence here there is no analog of the soft operator with two adjoint indices in \eq{LGsofar}, and the $1/\cP_\perp^2$ gives the central terms in square brackets in \eq{treensresult}.
The remaining right most terms in square brackets are reproduced by the soft quark and gluon operators:
\begin{align}  \label{eq:Osqgn}
   {\cal O}_s^{q_n B} & =  8\pi\alpha_s\: \Big( \bar\psi^n_{S} \, T^{B} \frac{\nslash}{2} \psi^n_{S} \Big)
   \,, \nn\\
   {\cal O}_s^{g_n B} & = 8\pi\alpha_s\: \Big(
  \frac{i}{2} f^{BCD}  {\cal B}_{S\perp\mu}^{n C}\, 
  \frac{n}{2} \cdot (\cP\!+\!\cP^\dagger)  {\cal B}_{S\perp}^{n D\mu} \Big)  \,.
\end{align}
Here the soft fields with $n$ superscripts carry $S_n$ Wilson lines and were defined in \eqs{opbb}{opbbb} above.   The appearance of these Wilson lines is necessary to preserve soft gauge invariance, and we will see in \sec{treematch} that they arise from integrating out soft attachments to the $n$ collinear lines.  By convention we group the gauge coupling $\alpha_s$ with the soft component of the operator. This is convenient since the running of this $\alpha_s$ occurs from soft loops.\footnote{The apparent symmetry between soft and collinear fields in these forward scattering operators is broken by the fact that the two types of fields have different 0-bin subtractions.} Due to our normalization conventions the total operators in \eq{Ons} have Wilson coefficients that are $1$ at tree level.  To derive the scaling of the operators we note that ${\cal O}_n^{i B}\sim \lambda^2$, and ${\cal O}_s^{i B}\sim \lambda^3$, so with the $1/\cP_\perp^2\sim \lambda^{-2}$ we have the total scaling $O_{ns}^{ij}\sim \lambda^3$. This is the correct scaling for a mixed $n$-$s$ Glauber operator that contributes at leading power in the SCET Lagrangian, once the scaling of the measure is included, as shown below in \sec{powercount}.  The lowest order Feynman rules for $n$-$s$ forward scattering from the operators in \eq{Ons} are shown in \fig{LOfeynrulens}.

If there is another collinear sector, such as our $\bn$, then there will be a set of soft-$\bn$ scattering operators analogous to \eq{Ons}, which we can simply obtain by taking $n\leftrightarrow \bn$ in the above analysis. Here the forward scattering conditions are that the $\bn\cdot p_s$ momentum is conserved on the soft line and the $n\cdot p_\bn$ momentum is conserved on the $\bn$-collinear line. The corresponding operators are
\begin{align}  \label{eq:Obns}
    O^{qq}_{\bn s} &=  {\cal O}_\bn^{q B} \frac{1}{\cP_\perp^2} {\cal O}_s^{q_\bn B} \,,
   & O^{qg}_{\bn s} &= {\cal O}_\bn^{q B} \frac{1}{\cP_\perp^2} {\cal O}_s^{g_\bn B} \,,
   & O^{gq}_{\bn s} &=  {\cal O}_\bn^{g B} \frac{1}{\cP_\perp^2}  {\cal O}_s^{q_\bn B} \,,
   & O^{gg}_{\bn s} &= {\cal O}_\bn^{g B} \frac{1}{\cP_\perp^2}  {\cal O}_s^{g_\bn B} \,,
\end{align}
which now involve the $\bn$-collinear bilinear operators in \eq{Obn}, and the soft operators
\begin{align}  \label{eq:Osqgnb}
   {\cal O}_s^{q_\bn B} & = 8\pi\alpha_s \Big( \bar\psi^\bn_{S} \, T^{B} \frac{\bnslash}{2} \psi^\bn_{S} \Big)
   \,, \nn\\
   {\cal O}_s^{g_\bn B} & = 8\pi\alpha_s \Big(
  \frac{i}{2} f^{BCD}  {\cal B}_{S\perp\mu}^{\bn C}\, 
  \frac{\bn}{2}\cdot (\cP\!+\!\cP^\dagger)  {\cal B}_{S\perp}^{\bn D\mu} \Big)   \,,
\end{align}
where the fields $\psi_s^\bn$ and ${\cal B}_{S\perp}^{\bn D\mu}$ can be found in \eqs{opbb}{opbbb}. Once again with our conventions these operators have tree level Wilson coefficients equal to $1$, and we will later argue that this is true to all orders.  The lowest order Feynman rules for $\bn$-$s$ forward scattering from the operators in \eq{Obns} are given by those in \fig{LOfeynrulens} with $n\leftrightarrow \bn$.

\begin{figure}[t!]
%
%
\hspace{1.5cm}
\includegraphics[width=0.55\columnwidth]{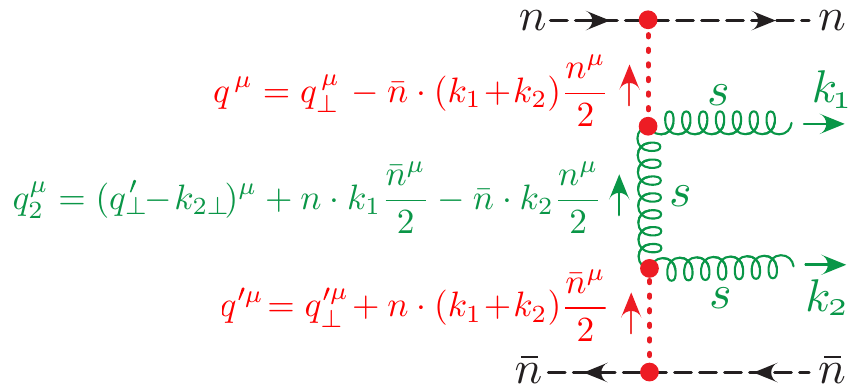} 
\caption{\setcaptionskip
Example of the momentum routing in a T-product of soft-$n$ and soft-$\bn$ Glauber operators which produces two soft gluons. The virtual soft gluon is space-like and the two Glauber operators each still satisfy their forward scattering conditions. 
}
\label{fig:treeSoftTproduct}
\setmainskip
\end{figure}

So far the soft-$n$ and soft-$\bn$ Glauber operators appear to be just like those for $n$-$\bn$ forward scattering, except without an intermediate rapidity sector. However, this is not always the case, due to the fact that the forward scattering constraints in \eqs{fwdkinematics}{fwdscatterings} only restricts one light-cone momentum of each scattering particle, and the light-cone soft momenta are much smaller than either the momentum of the $n$-collinear or $\bn$-collinear particles, $n\cdot k_s \ll n\cdot p_\bn$ and $\bn\cdot k_s \ll \bn\cdot p_n$. In particular we can have diagrams that satisfy the soft forward scattering constraints even though physically they do not appear to be forward scattering soft particles. For example, consider the time-ordered product of an $O_{ns}^{qg}$ and $O_{\bn s}^{qg}$ shown in \fig{treeSoftTproduct}. Here the two soft gluons are produced in the final state and have momenta $n\cdot k_i>0$ and $\bn\cdot k_i>0$ for $i=1,2$. Nevertheless, the two Glauber attachments to the soft gluons still satisfy the forward scattering constraints since $n\cdot q_2 = n\cdot k_1 >0$ and $-\bn\cdot q_2 = \bn\cdot k_2 >0$. This is enabled by the $n$ and $\bn$ collinear particles which can absorb the ${\cal O}(\lambda)$ soft light-cone momenta in one of the two directions. Due to the collinear power counting constraints the momentum $\bn\cdot (k_1+k_2)$ must travel through the  Glauber potential with momentum $q$ into the $n$-collinear particles, and the momentum $n\cdot (k_1+k_2)$ must travel through the Glauber potential with momentum $q^\prime$ from the $\bn$-collinear particles.  This type of time ordered product will play an important role in our calculations later on.

Considering all terms which cause scattering between either colllinear or soft fields we can write the full  Glauber Lagrangian for \SCETb as
\begin{align}  \label{eq:LG}
  {\cal L}_G^{{\rm II}(0)}  
& = 
 e^{-i x\cdot \cP} \sum_{n,\bn} \sum_{i,j=q,g}  \:  O_{ns\bn}^{i j}
   +  e^{-i x\cdot \cP} \sum_n \sum_{i,j=q,g} \:  O_{ns}^{i j} 
  \nn\\
 & \equiv
    e^{-i x\cdot \cP} \sum_{n,\bn} \sum_{i,j=q,g}  \:  
            {\cal O}_n^{i B} \frac{1}{\cP_\perp^2} {\cal O}_s^{BC}  \frac{1}{\cP_\perp^2} {\cal O}_\bn^{j C} 
   + e^{-i x\cdot \cP} \sum_n \sum_{i,j=q,g} \:  {\cal O}_n^{i B} \frac{1}{\cP_\perp^2} {\cal O}_s^{j_n B}   \,.
\end{align}
Thus we see that the Glauber Lagrangian consists of operators connecting 3 rapidity sectors $\{n,s,\bn\}$ and operators connecting 2 rapidity sectors $\{n,s\}$ (and $\{\bn,s\}$). This is the complete result for the Glauber Lagrangian, since as we will explain below in \sec{powercount} there are no loop corrections to this form. For future reference we summarize the operators relevant to forward scattering in \tab{opsummary}.

\begin{table}[t!]
\fontsize{10}{7}\selectfont
\line(1,0){470}
\vspace{-0.2cm}
\begin{align}
  {\cal O}_n^{q B} &= \overline\chi_{n} T^B \frac{\bnslash}{2} \: \chi_{n} 
  \qquad\qquad\qquad\qquad\qquad\qquad
  {\cal O}_n^{g B} = \frac{i}{2} f^{BCD}  {\cal B}_{n\perp\mu}^C \,
  \frac{\bn}{2}\cdot (\cP\!+\!\cP^\dagger)  {\cal B}_{n\perp}^{D\mu}   
   \nn\\
  {\cal O}_\bn^{q B} &= \overline\chi_{\bn} T^B \frac{\nslash}{2} \: \chi_{\bn} 
  \qquad\qquad\qquad\qquad\qquad\qquad
  {\cal O}_\bn^{g B} = \frac{i}{2} f^{BCD}  {\cal B}_{\bn\perp\mu}^C \,
  \frac{n}{2} \cdot (\cP\!+\!\cP^\dagger) {\cal B}_{\bn\perp}^{D\mu}  
\nn\\
    {\cal O}_s^{\! BC} 
    & ={8\pi\alpha_s  }
    \bigg\{
    \cP_\perp^\mu {\cal S}_n^T {\cal S}_\bn  \cP_{\perp\mu}
    - \cP^\perp_\mu g \widetilde {\cal B}_{S\perp}^{n\mu}  {\cal S}_n^T  {\cal S}_\bn   
    -  {\cal S}_n^T  {\cal S}_\bn  g \widetilde {\cal B}_{S\perp}^{\bn\mu} \cP^\perp_{\mu}  
    -  g \widetilde {\cal B}_{S\perp}^{n\mu}  {\cal S}_n^T  {\cal S}_\bn g \widetilde {\cal B}_{S\perp\mu}^{\bn}
    -\frac{n_\mu \bn_\nu}{2} {\cal S}_n^T   ig \widetilde {G}_s^{\mu\nu} {\cal S}_\bn 
    \bigg\}^{BC} 
   \nn\\
  {\cal O}_s^{q_n B} & =  8\pi\alpha_s\: \Big( \bar\psi^n_{S} \, T^{B} \frac{\nslash}{2} \psi^n_{S} \Big)
   \qquad\qquad\qquad\qquad
  {\cal O}_s^{g_n B}  = 8\pi\alpha_s\: \Big(
  \frac{i}{2} f^{BCD}  {\cal B}_{S\perp\mu}^{n C}\, 
  \frac{n}{2} \cdot (\cP\!+\!\cP^\dagger)  {\cal B}_{S\perp}^{n D\mu} \Big)  
  \nn\\
    {\cal O}_s^{q_\bn B} & = 8\pi\alpha_s\: \Big( \bar\psi^\bn_{S} \, T^{B} \frac{\bnslash}{2} \psi^\bn_{S} \Big)
    \qquad\qquad\qquad\qquad
    {\cal O}_s^{g_\bn B}  = 8\pi\alpha_s\: \Big(
    \frac{i}{2} f^{BCD}  {\cal B}_{S\perp\mu}^{\bn C}\, 
    \frac{\bn}{2}\cdot (\cP\!+\!\cP^\dagger)  {\cal B}_{S\perp}^{\bn D\mu} \Big) 
      \nn\\[-20pt]  \nn
\end{align}
\line(1,0){470}
\vspace{-.1cm}
	\caption{\label{table:opsummary}
		Summary of operators appearing in the leading power Glauber exchange Lagrangian in \eq{LG}.}
\end{table}

If we consider the interactions of soft and collinear particles in \SCETa then none of the tree level calculations that we have done in \SCETb change, and hence the Glauber operators are precisely the same as in \SCETb. In this case we are considering \SCETa prior to making the BPS field redefinition, so 
\begin{align} 
  {\cal L}_G^{{\rm I}(0)} ={\cal L}_G^{{\rm II}(0)}
 \,.
\end{align}
However due to the appearance of couplings between the collinear and ultrasoft fields in ${\cal L}_{n_i}^{(0)}$ for \SCETa, and the differences between how momentum sectors are distinguished (via subtraction terms), the precise behavior of these operators in loop diagrams will in general be different.  We will see this explicitly when comparing our  one-loop matching calculations in \secs{loop2match}{loop1match} for \SCETb and \SCETa respectively.

We can also consider the form of the Lagrangian ${\cal L}_G^{I(0)}$ after the BPS field redefinition. This field redefinition only changes the collinear quark and gluon fields, inducing lines $Y_n$ or ${\cal Y}_n$ for $n$-collinear fields, but leaves the soft fields unchanged.  Due to the octet nature of the Glauber operators in ${\cal L}_G^{I(0)}$, only the adjoint lines ${\cal Y}_n$ and ${\cal Y}_\bn$ appear in this Lagrangian. Additional ultrasoft lines can appear from interpolating fields for collinear initial and final states. For a situation where \SCETa is the relevant theory  there are no soft real emissions, since  they are ruled out by restrictions from the observable being measured, and hence the soft gluons appearing in \SCETa due to the presence of Glauber operators can only appear as virtual soft fluctuations.

\subsubsection{Matching for All Polarizations} 
\label{sec:matchallpol}

For completeness, we can also repeat the matching calculations involving external gluons with arbitrary external polarizations. This amounts to not specifying a specific basis for the physical states, and allows us to see how the scattering with non-transverse polarizations are matched by the EFT. To carry out this calculation it is important to use the equations of motion to simplify the gluon matrix elements. For a full theory scattered gluon of momentum $p$ the equations of motion imply $p^2=0$ as well as
\begin{align} \label{eq:polA}
  0 = p^\mu A_\mu(p) 
    =  \frac12 \bn\cdot p\ n\mcdot A(p) + \frac12 n\cdot p\ \bn\mcdot A(p) 
     +  p_\perp \mcdot A_\perp(p) \,.
\end{align}

\begin{figure}[t!]
	%
	%
	\raisebox{-0.2cm}{a)\hspace{-0.3cm}} 
	\includegraphics[width=0.22\columnwidth]{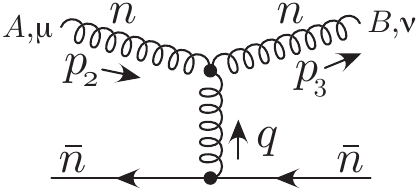}
 \hspace{0.45cm}
	\includegraphics[width=0.20\columnwidth]{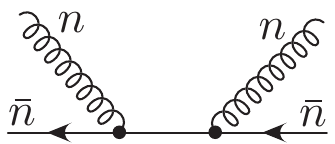}
 \hspace{0.45cm}
	\includegraphics[width=0.20\columnwidth]{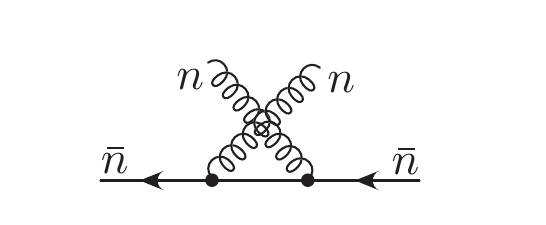}
 \hspace{0.45cm}
	\raisebox{-0.2cm}{b)\hspace{0.1cm}} 
	\includegraphics[width=0.20\columnwidth]{figs/GlaubOp_tree_ggqq}
	\caption{\setcaptionskip
		Tree level matching for the $nn\bn\bn$ Glauber operators, considering all gluon polarizations. In a) we show the three full QCD graphs that contribute, and in b) we show the Glauber operator that they match onto. }
	\label{fig:treematch_nnb_allpol}
	\setmainskip
\end{figure}

As an explicit example we consider the two-gluon two-quark matching calculation given by the diagrams shown in \fig{treematch_nnb_allpol}.  Since the Glauber operator  $O_{\bn s}^{g q}$ obviously only yields $\bn\cdot A$ and $A_\perp$ polarizations, we use \eq{polA} to eliminate the $n\mcdot A$ polarization terms in the  full theory amplitude.  We also will use the forward condition on the amplitude, $\bn\cdot p_2=\bn\cdot p_3$ and the relation $q=p_3-p_2$.   Using \eq{polA} we can set $p_2^\mu \varepsilon_{\mu}(p_2) = 0$, $p_3^\nu  \varepsilon_{\nu}(p_3) = 0$, and write the objects that are dotted into these $\varepsilon$ polarization vectors  as
\begin{align}
  g^{\mu\nu}  
   &\to  g_\perp^{\mu\nu} -  \frac{p_{2\perp}^\mu \bn^\nu}{\bn\cdot p_2}
    -  \frac{\bn^\mu p_{3\perp}^\nu }{\bn\cdot p_2}  
    - \frac{\bn^\mu\bn^\nu n\cdot(p_2+p_3)}{2\bn\cdot p_2} 
    \,, 
 & p_2^\nu &
  \to  (p_{2\perp}^\nu-p_{3\perp}^\nu)  + \frac12 n\cdot (p_2-p_3) \bn^\nu 
    \,, \nn\\
  p_3^\mu 
  & \to  (p_{3\perp}^\mu-p_{2\perp}^\mu)  + \frac12 n\cdot (p_3-p_2) \bn^\mu 
   \,. 
\end{align}
With these manipulations, and canceling various terms, the amplitude from the first of the full theory diagrams in \fig{treematch_nnb_allpol}a is
\begin{align}  \label{eq:Fig8a}
 & \frac{ g^2 f^{ABC} }{q^2}  
    \Big[ \bar v_\bn \frac{\nslash}{2} \bar T^C v_\bn \Big]  
    \Big\{ 2\, \bn\cdot p_2\: g_\perp^{\mu\nu} -2 \bn^\mu p_{2\perp}^\nu -2 p_{3\perp}^{\mu} \bn^\nu - n\mcdot(p_2+p_3) \bn^\mu \bn^\nu
      \Big\}
     \,,
\end{align}
while the leading power contribution from the sum of the two remaining full theory (Compton) diagrams in \fig{treematch_nnb_allpol}a is
\begin{align} \label{eq:Fig8bc}
 & \frac{ g^2 f^{ABC} }{q^2}  
    \Big[ \bar v_\bn \frac{\nslash}{2} \bar T^C v_\bn \Big]  
    \bigg\{ - \frac{q^2}{\bn\cdot p_2}
      \bigg\} \bn^\mu \bn^\nu
     \,.
\end{align}
To obtain \eq{Fig8bc} we have dropped the $+i0$ in the propagators. Keeping the $+i0$ gives rise to an additional term proportional to $\delta(\bn\cdot p_2)$ which we can set to zero, since the large momentum $\bn\cdot p_2>0$ for this matching calculation. The $\bn\cdot p_2=0$ contribution is properly accounted for in Glauber loop graphs, such as those discussed below in \sec{GlauberBox}.  Adding the results for the full theory graphs, and using the equations of motion to carry out the simplification $q^2 + \bn\cdot p_2\, n\cdot (p_2+p_3)= -2 p_{2\perp}\cdot p_{3\perp}$, we find
\begin{align}
  \label{eq:Fig8c}
 & \frac{2 g^2 f^{ABC} }{q^2}  
    \Big[ \bar v_\bn \frac{\nslash}{2} \bar T^C v_\bn \Big]  
    \bigg\{  \bn\cdot p_2\: g_\perp^{\mu\nu} - \bn^\mu p_{2\perp}^\nu - p_{3\perp}^{\mu} \bn^\nu 
   + \frac{p_{2\perp}\mcdot p_{3\perp}}{\bn\cdot p_2} \bn^\mu \bn^\nu
      \bigg\}
     \,.
\end{align}
This result is precisely identical to the complete Feynman rule for the EFT contribution shown in \fig{treematch_nnb_allpol}b, so the complete set of polarizations are reproduced by the EFT operator. (The Feynman rule was given above in \fig{LOfeynrule}.)

This same calculation also demonstrates that the full set of polarizations are present in the operator with two soft gluons (the second graph in \fig{treematch_ns}b). In a similar manner, using the equations of motion the full set of polarizations for gluon-gluon soft-collinear and $n$-$\bn$  scattering are reproduced by the EFT Glauber operators.

\subsection{Formalism for Multi-Glauber Diagrams}
\label{sec:multiglauber}

Here we discuss additional formalism that is needed for diagrams with multiple insertions of Glauber operators. In \sec{GlauberBox} we discuss the regulation of Glauber exchange iterations, then in \sec{LGtransverse} we write the Glauber Lagrangian in transverse momentum space and fully implement its multipole expansion, and finally in \sec{rapidityregulator} we discuss the rapidity regulator for Glauber operators and the implementation of 0-bin subtractions.

\subsubsection{One-Loop Glauber Box and Cross-Box Diagrams} 
\label{sec:GlauberBox}

To illustrate the presence of additional singularities that occur in the presence of Glauber gluons, in this section we will consider the one-loop computation for the iteration of two Glauber operators. We will see that it is necessary to introduce a rapidity regulator into ${\cal L}_G^{\rm II (0)}$ in order to yield well defined results for the various possible contractions of two operators which induce a loop momentum with Glauber scaling. 

\begin{figure}[t!]
	%
	%
	\begin{center}
		\includegraphics[width=0.3\columnwidth]{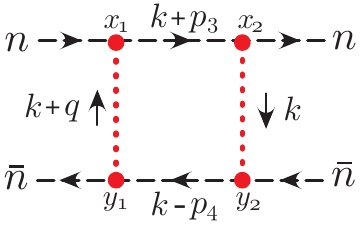} 
		\hspace{2cm}
		\includegraphics[width=0.3\columnwidth]{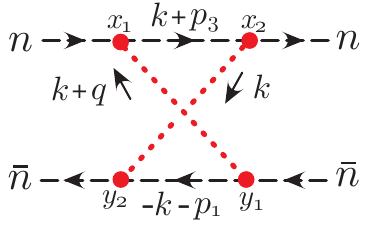}
	\end{center}
	\vspace{-0.6cm}
	\caption{\setcaptionskip
		One loop iterations of the Glauber potential for $n$--$\bn$ forward scattering of a $q\bar q$ pair.}
	\label{fig:glaub_loop}
	\setmainskip
\end{figure}

To see some of the difficulties inherent in having well defined Glauber potentials, we will  start by considering the iteration of two $O_{ns\bn}^{q q}$ potentials to generate a loop graph. We can contract the $n$-collinear quarks and the $\bn$-collinear quarks to  give the ``box'' and ``cross box'' graphs shown in \fig{glaub_loop}.  
To keep the particles onshell in the effective theory the loop momentum $k$ must not spoil the power counting for any of the propagators in the loop.  Therefore we must have $n\cdot k\sim \lambda^2$ and $\bn\cdot k\sim \lambda^2$, but can have $k_\perp\sim \lambda$. We will refer to this as an $n$-$\bn$ Glauber loop momentum.  We decompose 
\begin{align}
  \ddslash\!^dk \equiv \frac{d^dk}{(2\pi)^d} = \frac{1}{2} \, \ddslash\! k^+ \ddslash\! k^-  \ddslash\!^{d-2}k_\perp
\end{align}
where $d=4-2\epsilon$, and recall the forward conditions $p_4^+=p_1^+$ and $p_3^-=p_2^-$. The box and cross-box loop integrals involve two Glauber denominators and two propagators from the collinear quarks. They are
\begin{align}  \label{eq:Gboxes0}
   I_{\rm Gbox} &=\int \!\!    \frac{
     \ddslash\!^{d-2}k_\perp\
     \ddslash\! k^+ \
     \ddslash\! k^- 
     }{ 
    2({\vec k}_\perp^{\,2})({\vec k}_\perp\plus {\vec q}_\perp)^2 
    \Big( k^+ \plus p_3^+ \minus ( {\vec k}_\perp\plus {\vec q}_\perp/2)^{\,2}/p_2^- \plus i0\Big)
    \Big( \minus k^-\plus p_4^-  \minus ( {\vec k}_\perp\plus {\vec q}_\perp/2)^{\,2}/p_1^+ \plus i0\Big)
    }  \,,
   \nn\\
 I_{\rm Gcbox} &=  \int \!\!    \frac{
     \ddslash\!^{d-2}k_\perp\
     \ddslash\! k^+ \
     \ddslash\! k^-
     }{ 
    2({\vec k}_\perp^{\,2})({\vec k}_\perp\plus {\vec q}_\perp)^2 
    \Big( k^+ \plus p_3^+ \minus ( {\vec k}_\perp\plus {\vec q}_\perp/2)^{\,2}/p_2^- \plus i0\Big)
    \Big( \plus  k^- \plus p_1^-  \minus ( {\vec k}_\perp\plus {\vec q}_\perp/2)^{\,2}/p_1^+ \plus i0\Big)
    }  \,.
\end{align}
These graphs involve log divergent integrals of the type $\int dk^+ / (k^+ + \Delta \pm i0)$ and $\int dk^- / (k^- + \Delta \pm i0)$ that are not regulated by dimensional regularization.  These singularities must be dealt with systematically by introducing an additional regulator.

In the case of the potential for two heavy quarks in NRQCD,  the cross-box diagram would be zero because both poles in the energy contour integral would be on the same side, and the box diagram would be convergent since both fermion propagators would carry the loop energy.  Indeed, in NRQCD their are no crossed diagrams for potential iterations at any order in $\alpha_s$, and the iterated box diagrams yield the Coulomb Greens function. In our case the Glauber potential is instead static in both time and longitudinal distance, or equivalently static in the two light-cone times $x^+$ and $x^-$. For the diagrams in \fig{glaub_loop} this implies that we have $x_1^-=y_1^-$, $x_1^+=y_1^+$, and $x_2^-=y_2^-$, $x_2^+=y_2^+$ in position space, where the $x_i$ and $y_i$ coordinates are defined in the figures.  Naively this would seem to imply that only the Glauber box diagram can exist, because in the cross-box diagram the ordering of the $1$ and $2$ Glauber potential vertices is different for the $n$-collinear and $\bn$-collinear lines.  However due to the multipole expansion, which ensures that the  collinear propagators  are homogeneous in the power counting in \eq{Gboxes0}, the $n$-collinear propagator only depends on the $n\cdot k\sim \lambda^2$ Glauber momentum, and not on the $\bn\cdot k \ll \bn\cdot p_n$ component, whereas we have the opposite situation for the $\bn$-collinear propagator. Thus each of the collinear sectors only sees one of these two times $x^+$ or $x^-$, and we must consider both the box with $x_2^+>x_1^+$ and $x_2^->x_1^-$, and the cross-box with $x_2^+>x_1^+$ and $x_1^->x_2^-$ where the Glauber vertices have the opposite ordering on each line.

In the abelian limit we can determine $I_{\rm Gbox}+I_{\rm Gcbox}$ without an additional regulator, by adding the integrands and manipulating them to obtain $\delta(k^+)\delta(k^-)$. We carry out these computations explicitly in \app{abelianexp}, where we also show that this same trick works to all orders in the iterations of Glauber potentials, and leads to the expected eikonal phase result $e^{i\phi} -1$  for the Greens function obtained from the abelian forward scattering potential.  To obtain this result at the integrand level the crossed box type diagrams play a role.   However in QCD the box and crossbox have different color factors, so this type of manipulation does not suffice.

To regulate the integrals in \eq{Gboxes0} for the nonabelian case we will use the rapidity regulator $w^2 |2 q^z|^{-\eta} \nu^\eta$ of Ref.~\cite{Chiu:2012ir}, where $w=w(\nu)$ is a renormalized coupling used to derive RG equations, and in the limit $\eta\to 0$ we set $w(\nu)=1$. 
In terms of light-cone momenta $q^z = (q^- - q^+)/2$, and results and counterterms are identified by taking $\eta\to 0$ prior to expanding for $\epsilon\to 0$. The parameter $\nu$ introduces an extra cutoff parameter that behaves in a similar way to $\mu$ of the $\MSbar$ scheme in dimensional regularization. This regulator acts as 
a factorization scale that separates modes with equal invariant mass but different rapidity.  To regulate multiple iterations of these Glauber potentials we will have one factor of $w|2 q^z|^{-\eta} \nu^\eta$ for each Glauber potential carrying momentum $q$.  We will refer to this as the $\eta$-regulator.\footnote{Including an $\eta$-regulator for each Glauber potential is distinct from the definition used  in Ref.~\cite{Chiu:2012ir}, where it was used for group momenta in soft and collinear Wilson lines.}  In the next section we formulate this regulator for Glauber potentials at the level of the Glauber Lagrangian, and also discuss the regularization of rapidity divergences from soft and collinear loop graphs. In this section the coupling $w(\nu)$ will play no role (since as we will see, the graphs do not have $1/\eta$ poles), so we will from the start set $w(\nu)=1$ below.

For the Glauber loop momentum in \fig{glaub_loop}, $q^z=k^z\sim \lambda^2$, so we have a factor of $|k^z|^{-\eta}(\nu/2)^\eta$ for each of the two potential insertions in these graphs.   The presence of the $|k^z|^{-\eta}$ factor means that the Glauber exchange is no longer static in longitudinal distance. We will recover the static nature of the exchange in this direction only when $\eta\to 0$.  With this regulator the loop integrals become well defined because we are forced to consider the contour integral in the analytic variable $k^0$, rather than using any time slice that involves some amount of $k^z$.  With this regulator the Glauber cross-box integral becomes
\begin{align}  \label{eq:Gcbox}
 I_{\rm Gcbox} &=  \int \!\!    \frac{
     \ddslash\!^{d-2}k_\perp\
     \ddslash\! k^0 \
     \ddslash\! k^z \ |k^z|^{-2\eta}\: (\nu/2)^{2\eta} 
     }{ 
    ({\vec k}_\perp^{\,2})({\vec k}_\perp\plus {\vec q}_\perp)^2 
    \Big( k^0\!-k^z \plus p_3^+ \minus ( {\vec k}_\perp\plus \frac{{\vec q}_\perp}{2})^{2}/p_2^- \plus i0\Big)
    \Big( k^0\!+k^z \plus p_1^-  \minus ( {\vec k}_\perp\plus \frac{{\vec q}_\perp}{2})^{2}/p_1^+ \plus i0\Big)
    } 
   \nn\\
  &= 0 \,,
\end{align}
since the poles are on the same side.  For the Glauber box integral we get
\begin{align}  \label{eq:Gbox}
   I_{\rm Gbox} 
  &=\int \!\!    \frac{ \ddslash\!^{d-2}k_\perp\
     \ddslash\! k^0 \
     \ddslash\! k^z \ |k^z|^{-2\eta}\: (\nu/2)^{2\eta} \qquad\qquad 
     }{ 
    ({\vec k}_\perp^{\,2})({\vec k}_\perp\plus {\vec q}_\perp)^2 
    \Big( k^0\minus k^z \plus p_3^+ \minus ( {\vec k}_\perp\plus \frac{{\vec q}_\perp}{2})^{\,2}/p_2^- \plus i0\Big)
    \Big( \minus k^0\minus k^z\plus p_4^-  \minus ( {\vec k}_\perp\plus \frac{{\vec q}_\perp}{2})^{\,2}/p_1^+ \plus i0\Big)
    }  
  \nn\\
 &= -i \int \!\!    \frac{ \ddslash\!^{d-2}k_\perp \ddslash\! k^z \ |k^z|^{-2\eta}\: (\nu/2)^{2\eta}  }{ 
    ({\vec k}_\perp^{\,2})({\vec k}_\perp\plus {\vec q}_\perp)^2 
    ( \minus 2k^z\minus 2\Delta \plus i0)
    }  
  \nn\\
  &= \frac{-i}{4\pi} \int \!\!    \frac{ \ddslash\!^{d-2}k_\perp }{ 
    ({\vec k}_\perp^{\,2})({\vec k}_\perp\plus {\vec q}_\perp)^2 }
    \Big[ (\nu/2)^{2\eta}\, (-2i\pi)  \csc(2\pi \eta) 
    \sin(\pi\eta) \: ( i \Delta)^{-2\eta} \Big]
\nn\\
  &= \Big(\frac{-i}{4\pi}\Big) \int \!\!    \frac{ \ddslash\!^{d-2}k_\perp }{ 
    ({\vec k}_\perp^{\,2})({\vec k}_\perp\plus {\vec q}_\perp)^2 }
    \ \Big[ -i\pi +{\cal O}(\eta) \Big]   \,,
\end{align}
where the $k^z$ integral is evaluated in \eq{kzint}.
Here 
\beq 2\Delta=\frac{( {\vec k}_\perp\!\!+ {\vec q}_\perp/2)^{\,2}}{p_1^+} \plus \frac{( {\vec k}_\perp\!\!+ {\vec q}_\perp/2)^{\,2}}{p_2^-} \minus p_4^- \minus p_3^+\eeq and the $\eta$ dependent term evaluates to $(-i\pi)$ as $\eta\to 0$ for any value of this $\Delta$.  This extra $(-i)$ is the factor necessary for  the Glauber potential to exponentiate into a phase.   The result in \eq{Gbox} for the $\eta$-regulated box is exactly the same as the result obtained from manipulating the integrands in the sum of the box and cross-box  in the abelian case in \app{abelianexp}.  

Effectively the $\eta$-regulator has decoupled the spacetime constraints so that the box diagram alone  is integrating the two Glauber potentials over $x^\pm$, while the cross box does not contribute. This is the same spacetime picture that is obtained by adding the box and cross-box integrands in the abelian theory to get a $\delta(k^+)\delta(k^-)$ type structure.  In the non-abelian theory it is important as far as the color structure is concerned that it is the box graph alone that contributes. The non-abelian part of the cross-box topology contributes only for another momentum region, namely when we have the loop graph with two soft gluons. In SCET this contribution has the non-abelian color structure and is given by first graph in \fig{softvac}b. (This graph does not correspond solely to vacuum polarization, and encodes the cross box contribution from terms involving soft Wilson lines.)  These soft graphs come from contractions of $O_{ns}^{qg}$ and $O_{\bn s}^{qg}$ with a soft loop momentum. Since the soft gluon terms in the operator involve $f^{ABC}$ they explicitly do not have an abelian contribution, so it is a regulator independent statement that the abelian contribution is entirely carried by the Glauber iterations.  Any consistent regulator for the Glauber singularities must have these properties.   

We will see in~\sec{exponentiation} that the above properties of the $\eta$-regulator extend in a nice way for arbitrary iterations of Glauber potentials with Glauber loop momenta.  Any iteration diagram with crossed Glauber potential lines will give zero in the same manner as the crossed box above, and the $\eta$-regulated iterated boxes alone yield an $e^{i\phi} -1$ Greens function even in the nonabelian theory. We will also give a more physical picture for the action of the $\eta$-regulator in Glauber loops in~\sec{exponentiation}.  In the abelian theory the phase $\phi$ is one-loop exact. In the nonabelian theory there will be one-loop corrections to the forward scattering kernel from graphs involving soft and collinear loop momenta, and the full set of such diagrams will be computed in \sec{loop2match} and \sec{loop1match}.

If we consider quark-quark scattering rather than quark-antiquark scattering, then the same loop integrals in \eqs{Gcbox}{Gbox} appear, and we get the same result other than a modified color structure. We can also extend the above analysis to iterations of Glauber potentials other than $O_{ns\bn}^{q q}$. If we consider $n$-$\bn$ scattering where either or both of the external collinear lines are $\perp$-gluons, then from the form of the Feynman rules in \fig{LOfeynrule} we note that the internal gluon is also $\perp$ and we have the same momentum integrals as those analyzed above, again with a $n$-$\bn$ Glauber loop momentum.  So iterations of these operators also yield the same loops as in $q\bar q$ scattering.  

\begin{figure}[t!]
%
%
\begin{center}
\includegraphics[width=0.3\columnwidth]{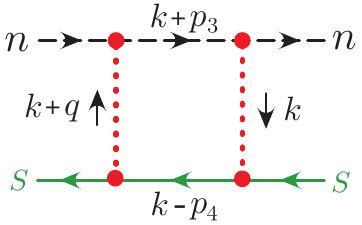} 
\hspace{2cm}
\includegraphics[width=0.3\columnwidth]{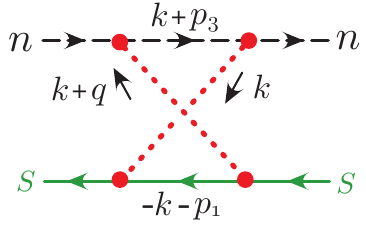}
\end{center}
\vspace{-0.6cm}
\caption{\setcaptionskip
One loop iterations of the Glauber potential for $n$--soft forward scattering of a $q\bar q$ pair.}
\label{fig:glaub_loop_ns}
\setmainskip
\end{figure}
We can also consider Glauber potentials obtained by iterations of the operator $O_{ns}^{ij}$ or by iterations of $O_{\bn s}^{ij}$. In these cases the loop momentum will be Glauber if we have the $s$-$n$ or $s$-$\bn$ Glauber scaling, namely $k^\mu \sim (\lambda^2,\lambda,\lambda)$ for $O_{ns}^{ij}$ iterations, or $k^\mu\sim (\lambda,\lambda^2,\lambda)$ for $O_{\bn s}^{ij}$ iterations. The graphs for two $O_{ns}^{qq}$ iterations are shown in \fig{glaub_loop_ns}. At the level of $4$-point functions the Feynman rules for the scattering involving $n$-$s$ are direct analogs of those for the $n$-$\bn$ scattering. Keeping $p_{2,3}^\mu$ as $n$-collinear, but letting $p_{1.4}^\mu$ be soft, the iteration of two $O_{ns}^{qq}$s gives  the same  denominators as in  the   box and cross-box integrals shown in \eq{Gboxes0}, since the collinear and soft propagator denominators are
\begin{align}
 & \text{$n$-$s$ box:}\qquad\quad\,
  \big[(k^+\plus p_3^+)p_3^- \minus (\vec k_\perp \plus\vec p_{3\perp})^2\plus i0\big] 
  \big[(p_4^-\minus k^-)p_4^+ \minus (\vec k_\perp \minus\vec p_{4\perp})^2\plus i0\big]
   \,, \\
 & \text{$n$-$s$ crossbox:}\quad
  \big[(k^+\plus p_3^+)p_3^- \minus (\vec k_\perp \plus\vec p_{3\perp})^2\plus i0\big] 
  \big[(p_1^-\plus k^-)p_1^+ \minus (\vec k_\perp \plus\vec p_{1\perp})^2\plus i0\big]
   \,. \nn
\end{align}
Here $k^+\sim \lambda^2$, while $k^-\sim \lambda$, but each loop integral scales as $\lambda^5/\lambda^7=\lambda^{-2}$ as before. To regulate these $n$-$s$ Glauber loops we can use $| n\cdot k - \beta \bn\cdot k|^{-\eta}$ instead of $|2k_z|^{-\eta}$ in \eqs{Gcbox}{Gbox}, where $\beta\sim\lambda$ is a boost factor that ensures that the two terms in the regulator have the same $\sim \lambda^2$ scaling.  However, the results for these integrals are the same as if we took $\beta=1$ since they are independent of $\beta$ as long as $\beta> 0$.  To see this simply change variables to $k^{\prime -}=\beta k^-$, noting that this gives back the regulator $|n\cdot k-\bn\cdot k'|^{-\eta}$, and $\ddslash k^-/(k^- -\Delta) = \ddslash k^{\prime -}/(k^{\prime-}-\Delta/\beta)$. The cross-box again vanishes and the box yields the same $\Delta$ independent result that we found in \eq{Gbox} once we expand and drop ${\cal O}(\eta)$ terms. 

\subsubsection{Transverse Momentum Glauber Lagrangian and Multipole Expansion}
\label{sec:LGtransverse}

In the Glauber Lagrangian in \eq{LG} the $\perp$-momenta are all ${\cal O}(\lambda)$ and hence are encoded by continuous label momenta on soft and collinear fields, without residual $\perp$-momenta. The overall $\int d^2x_\perp$ integration then just enforces momentum conservation through $\int d^2x_\perp \exp(-ix_\perp\cdot \cP_\perp) = (2\pi)^2 \delta^2(\cP_\perp)$. It is convenient to make this explicit by writing the Glauber action entirely in transverse momentum space. To do this we insert the identity in the form
\begin{align}
  {\cal O}_n^{iA} = \int d^2q_\perp\: \big[  {\cal O}_n^{iA}\ \delta^2(q_\perp-\cP_\perp^\dagger) \big] = \int d^2q_\perp\:  {\cal O}_n^{iA}(q_\perp) \,,
\end{align}
which allows us to write the Lagrangian in terms of the $n$-collinear bilinear with definite injected transverse momentum $q_\perp$: 
\begin{align}  \label{eq:Onq}
 & {\cal O}_n^{iA}(q_\perp) 
   =\big[ {\cal O}_n^{iA}\ \delta^2(q_\perp-\cP_\perp^\dagger) \big]
   \,.
\end{align}
We make similar definitions for the other bilinear operators with definite transverse momenta
\begin{align}  \label{eq:OnbOsq}
 & {\cal O}_\bn^{jB}(-q_\perp^\prime) 
   =\big[ {\cal O}_\bn^{jB}\ \delta^2(q_\perp^\prime+\cP_\perp^{\dagger}) \big]
   \,, 
 & {\cal O}_s^{j_n A}(-q_\perp) &
   =\big[ {\cal O}_s^{j_n A}\ \delta^2(q_\perp+\cP_\perp^\dagger) \big]
   \,.
\end{align}
Using these definitions for the $n$- and $\bn$-collinear operators, and then moving the overall momentum conserving $\delta^2(\cP_\perp)$ so that it acts only on the soft fields, the  Glauber action with $d=4$ in transverse momentum space is
\begin{align} \label{eq:LGreg}
 \int\!\!  d^4x\, {\cal L}_G^{{\rm II}(0)}  & = 
 \sum_{n,\bn} \sum_{i,j=q,g} \! \int\!\! \frac{dx^+ dx^-}{2}
  e^{-i\tilde x \cdot \cP} (2\pi)^2
  \!\!\int\!\! \frac{d^2q_\perp}{q_\perp^2} \frac{d^2q_\perp^\prime}{q_\perp^{\prime\,2}} \: 
 {\cal O}_{n}^{i A}(q_\perp)\, 
 \, {\cal O}_{s}^{AB}(q_\perp,q_\perp') \, 
  {\cal O}_{\bn}^{jB}(-q_\perp^\prime)
  \nn\\
 &\quad + \sum_n \sum_{i,j=q,g}  \int\!\! \frac{dx^+ dx^-}{2}
  e^{-i\tilde x \cdot \cP} (2\pi)^2
  \!\!\int\!\! \frac{d^2q_\perp}{q_\perp^2} \: 
  {\cal O}_n^{i A}(q_\perp)\, 
  \, {\cal O}_s^{j_n A}(-q_\perp)  
 \nn\\
 &=  \sum_{n,\bn} \sum_{i,j=q,g} \! \int\! [dx^\pm]
  \!\!\int\!\! \frac{d^2q_\perp}{q_\perp^2} \frac{d^2q_\perp^\prime}{q_\perp^{\prime\,2}} \: 
 {\cal O}_{n}^{i A}(q_\perp)\, 
 \, {\cal O}_{s}^{AB}(q_\perp,q_\perp') \, 
  {\cal O}_{\bn}^{jB}(-q_\perp^\prime)
  \nn\\
 &\quad + \sum_n \sum_{i,j=q,g}  \int\! [dx^\pm]
  \!\!\int\!\! \frac{d^2q_\perp}{q_\perp^2} \: 
  {\cal O}_n^{i A}(q_\perp)\, 
  \, {\cal O}_s^{j_n A}(-q_\perp)  
   \,,
\end{align}
where all the operators depend on the positions $x^+$ and $x^-$ and we will make the multipole expansion for these components explicit below. The two color index soft operator in transverse momentum space that appears in \eq{LGreg} is
\begin{align}  \label{eq:Os1q}
    {\cal O}_s^{BC}(q_\perp,q_\perp^\prime)
 & ={8\pi\alpha_s  } \, \delta^2(q_\perp-q_\perp^\prime-\cP_\perp)
    \bigg\{
    q_\perp\cdot  q^\prime_\perp\, {\cal S}_n^T {\cal S}_\bn 
    - q^\perp_\mu g \widetilde {\cal B}_{S\perp}^{n\mu}  {\cal S}_n^T  {\cal S}_\bn   
    -  {\cal S}_n^T  {\cal S}_\bn  g \widetilde {\cal B}_{S\perp}^{\bn\mu} q^\prime_{\perp\mu}  
\nn\\
 &\qquad\qquad
    -  g \widetilde {\cal B}_{S\perp}^{n\mu}  {\cal S}_n^T  {\cal S}_\bn g \widetilde {\cal B}_{S\perp\mu}^{\bn}
    -\frac{n_\mu \bn_\nu}{2} {\cal S}_n^T   
     ig \widetilde {G}_s^{\mu\nu} {\cal S}_\bn 
    \bigg\}^{BC} 
    \,.
\end{align}
For convenience we have defined the short hand notations 
\begin{align}
 &  [dx^\pm] \equiv \frac12\, dx^+dx^- \exp(-i\tilde x\cdot \cP) (2\pi)^2 \,,
  \qquad 
 & \tilde x^\mu & \equiv 
  n\cdot x\, \frac{\bn^\mu}{2} + \bn\cdot x\, \frac{n^\mu}{2} 
  \,.
\end{align}
The factor of $(2\pi)^2$ in $[dx^\pm]$ combine together with the overall $\perp$-momentum conserving $\delta$-function and hence drop out in the Feynman rules when momentum conservation is taken into account explicitly.

Next we consider the flow of $+$-momenta and $-$-momenta in the Glauber Lagrangian.  Unlike the $\perp$-momenta which were always ${\cal O}(\lambda)$, from \tab{modes} we see that here there is a hierarchy between the $n$-collinear, $\bn$-collinear and soft momenta:
\begin{align}
  k_n^+ \ll k_s^+ \ll k_\bn^+ \,, 
  \qquad\qquad 
  k_\bn^- \ll k_s^- \ll k_n^- \,.
\end{align}
These expansions are included in our Glauber Lagrangian due to the presence of a multipole expansion~\cite{Grinstein:1997gv} that is implemented using the mixed momentum-space and position-space label formalism~\cite{Luke:1999kz}, as implemented for SCET in Refs.~\cite{Bauer:2001ct,Bauer:2001yt}. 

Before discussing the Lagrangian in greater detail, we consider a practical application of these expansions for a graph that simultaneously involves soft, $n$, and $\bn$ fields, namely the last diagram in \fig{onegluon}. The momentum space multipole expansion implemented with labels ensures that the light-cone momenta will only be routed in a way which is consistent with the power counting and the fields in SCET remaining nearly onshell.  Here the soft gluon with incoming momentum $k=q-q'$ implies that $k^\pm\sim \lambda$ momenta must flow through the dashed Glauber potentials and into the $n$-collinear and $\bn$-collinear fields.  To be consistent with the power counting the $\bn\cdot k$ soft momentum must flow into the $n$-collinear fields, since $\bn\cdot k \gg \bn\cdot p_\bn$ and would knock the $\bn$-collinear particles offshell. Similarly, the $n\cdot k$ soft momentum must flow into the $\bn$-collinear fields.  Here the Glauber potentials have momenta scaling as $q'\sim (\lambda,\lambda^2,\lambda)$ and $q\sim (\lambda^2,\lambda,\lambda)$ for the $(+,-,\perp)$ components respectively.  The $1/\cP_\perp^2$ potentials still correctly describe these exchanges since we still have $q^+q^-\ll q_\perp^2$ and $q^{\prime +}q^{\prime -}\ll q_\perp^{\prime\,2}$. The $q\sim (\lambda^2,\lambda,\lambda)$ scaling found here also occurred in the potential exchanges in \fig{glaub_loop_ns}. Also note that the $k^-\sim \lambda$ momentum which flows into the $n$-collinear fields is always suppressed relative to the large $p_n^-\sim \lambda^0$ momenta, and hence does not appear in the leading power $n$-collinear propagators or purely $n$-collinear interactions (it will show up in power suppressed terms). Likewise the $k^+\sim \lambda$ momentum flowing into the $\bn$-collinear fields is suppressed relative to $p_\bn^+\sim \lambda^0$ momenta. Thus the presence of these smaller momentum components does not change the homogeneous scaling of collinear propagators. In our formalism the lines carry the smaller momentum components even if they do not show up in the leading power propagators, and we have separate momentum conservation for the ${\cal O}(\lambda^0)$, ${\cal O}(\lambda)$ and ${\cal O}(\lambda^2)$ components of the momenta. 

Next consider how this multipole expansion effects the dependence of various operators in the Glauber Lagrangian.  As already discussed, the large ${\cal O}(\lambda^0)$ momenta carried by collinear fields is conserved within the collinear bilinear operators, corresponding to the near forward scattering constraint in \eq{fwdkinematics}. These momenta are implemented with label momenta, but since they are not exchanged between sectors in the Glauber operators we will not bother to make this explicit in our notation. On the other hand, soft momentum $k^\pm\sim\lambda$ injected by the soft operators will be carried by the collinear fields, and we 
will denote these by momentum labels to distinguish them from the residual collinear momenta that are ${\cal O}(\lambda^2)$. Residual momenta are encoded in the dependence of all operators on the spacetime coordinates $x^\pm$. 
The  Glauber action from \eq{LGreg} with the multipole expansion made explicit is
\begin{align} \label{eq:LGreg_mult}
 \int\!\!  d^4x\, {\cal L}_G^{{\rm II}(0)}  & = 
 \sum_{n,\bn} \sum_{i,j=q,g} \! \int\! [dx^\pm] \! \sum_{k^+,k^-}  
 \int\!\! \frac{d^2q_\perp}{q_\perp^2} \frac{d^2q_\perp^\prime}{q_\perp^{\prime\,2}}\:
 {\cal O}_{n,k^-}^{i A}(q_\perp)\, 
 {\cal O}_{s,-k^\pm}^{AB}(q_\perp,q_\perp') \, 
  {\cal O}_{\bn,k^+}^{jB}(-q_\perp^\prime)
  \nn\\
 & + \sum_n \sum_{i,j=q,g}  \int\! [dx^\pm]  \sum_{k^-}
  \! \int\!\! \frac{d^2q_\perp}{q_\perp^2}\:
  {\cal O}_{n,k^-}^{i A}(q_\perp)\, 
  {\cal O}_{s,-k^-}^{j_n A}(-q_\perp)  
   \,.
\end{align}
In this form derivatives of the position space coordinates $x^+$ and $x^-$ are $\sim \lambda^{2}$.  Here $k^+$ and $k^-$ are ${\cal O}(\lambda)$ soft momenta, that for the collinear operators appear as subleading label momenta underneath the large momenta $p_n^-$ and $p_\bn^+$ in $n$-collinear and $\bn$-collinear operators respectively.  Since they are subleading, they do not appear in the propagators or leading power Feynman for collinear fields, but these labels on the collinear operators are important for conserving momenta.
In terms of transverse momentum space fields for example
\begin{align}
 {\cal O}_{n,k^-}^{qA}(q_\perp) = \int\! d^2p_\perp\: \sum_{k^{\prime -}} \bar\chi_{n,k^{\prime -}+k^-}(p_\perp+q_\perp,x^+,x^-)\frac{\bnslash}{2}T^A 
  \chi_{n,k^{\prime -}}(p_\perp,x^+,x^-) \,.
\end{align}
Here the conserved large ${\cal O}(\lambda^0)$ label momenta $p^-$ for the $\chi_n$ fields are not shown for simplicity.

To understand the form of the original Glauber Lagrangian in \eq{LG} and the equivalent Glauber action given in \eq{LGreg_mult} it is useful to look at mass-dimensions (counted with $Q$s) and power counting dimensions (counted with $\lambda$s) for the various components. For \eq{LG} the collinear operators ${\cal O}_n^{iA} \sim Q^{3}\lambda^2$, ${\cal O}_\bn^{jB} \sim Q^{3}\lambda^2$, the soft operators ${\cal O}_s^{j_n A}\sim Q^3\lambda^3$, ${\cal O}_s^{AB}\sim Q^2\lambda^2$, and $1/\cP_\perp^2\sim Q^{-2}\lambda^{-2}$. Accounting for the $\exp(-ix\cdot \cP)$ the largest momenta determine the scaling of the coordinates in $d^4x$, so for the 3-rapidity operators we have $d^4x \sim Q^{-4}\lambda^{-2}$, whereas for the 2-rapidity operators we have $d^4x \sim Q^{-4} \lambda^{-3}$. Therefore $\int d^4x\, {\cal O}_n^{iA} (1/\cP_\perp^2) {\cal O}_s^{AB} (1/\cP_\perp^2) {\cal O}_\bn^{jB}\sim Q^0\lambda^0$, and $\int d^4x\, {\cal O}_n^{iA} (1/\cP_\perp^2) {\cal O}_s^{j_nA}\sim Q^0 \lambda^0$, as expected. Next consider the Glauber action in \eq{LGreg_mult} where the operators have transverse momentum arguments. Using Eqs.~(\ref{eq:Onq},\ref{eq:OnbOsq},\ref{eq:Os1q}), we have ${\cal O}_n^{iA}(q_\perp) \sim Q\lambda^0$, ${\cal O}_\bn^{jB}(q_\perp) \sim Q\lambda^0$, ${\cal O}_s^{j_n A}(q_\perp)\sim Q\lambda$, and ${\cal O}_s^{AB}(q_\perp,q_\perp')\sim Q^0\lambda^0$. Again the largest momenta determine the measure scaling, so $[dx^\pm]\sim Q^{-2}\lambda^0$ for the 3-rapidity sector operators and $[dx^\pm]\sim Q^{-2}\lambda^{-1}$ for the 2-rapidity sector operators.  Therefore both the 2 and 3-rapidity sector terms in \eq{LGreg_mult} scale as $\sim Q^0\lambda^0$, as before.

\subsubsection{Rapidity Regulator and Zero-Bin Subtractions} 
\label{sec:rapidityregulator}

When there are soft and collinear modes that live at the same invariant mass scale in SCET, we in general need an additional regulator in rapidity space to distinguish these modes and handle divergences~\cite{Manohar:2006nz}. 
This can be achieved using the rapidity regulator of Ref.~\cite{Chiu:2012ir}, which distinguishes modes using a rapidity factorization scale $\nu$. In this subsection we highlight some differences related to the fact that the rapidity regulator must also be introduced to distinguish Glauber contributions. We also discuss zero-bin subtractions~\cite{Manohar:2006nz} from the Glauber region for soft and collinear contributions.  

To regulate rapidity divergences in graphs involving Wilson lines we include  factors of 
\begin{align} \label{eq:rapregulator}
 w \Big| \frac{2 {\cal P}_z}{\nu} \Big|^{-\eta/2}
  ,\qquad\qquad
 w^2 \Big| \frac{ {n \cdot \cal P}}{\nu} \Big|^{-\eta}
 , \qquad\qquad 
 w^2 \Big| \frac{ {\bn \cdot \cal P}}{\nu} \Big|^{-\eta}
 ,
\end{align}
for Wilson lines involving ($n\cdot A_s$ or $\bn\cdot A_s$) soft gluons, $n\cdot A_\bn$ $\bn$-collinear gluons, and $\bn\cdot A_n$ $n$-collinear gluons respectively~\cite{Chiu:2012ir}. At one-loop rapidity divergences will appear as $1/\eta$ poles with a corresponding logarithmic dependence on the cutoff $\nu$. Since $\nu$ is dimensionful, it technically is $\nu/\mu$ that is associated to the rapidity, but we will still follow the common practice of referring to $\nu$ as the rapidity scale.  Here $w$ is a book keeping coupling used to calculate anomalous dimensions through 
\begin{align} \label{eq:wnu}
\nu \frac{\partial}{\partial\nu} w^2(\nu) = -\eta\, w^2(\nu)\,,
  \qquad\qquad 
\lim_{\eta\to 0} w(\nu) = 1 \,.
\end{align}   
The powers of $\eta$ are fixed  to ensure that the rapidity divergences cancel when summing over sectors.
That the correct choice has been made can be seen by regulating the corresponding full theory diagrams
and expanding around the soft and collinear limits.
Counterterms will have both $1/\eta$ and $1/\epsilon$ poles, and are identified by taking $\eta\to 0$ prior to expanding for $\epsilon\to 0$. The regulated expressions for the momentum space Wilson lines are
\begin{align} \label{eq:SWrapreg}
   S_n &= \sum_{\rm perms} \exp\bigg\{ \frac{-g}{n\cdot \cP} \bigg[
   \frac{w |2 \cP^z|^{-\eta/2}}{\nu^{-\eta/2}} n \cdot A_s\bigg] \bigg\} 
   \,,
 & S_\bn &= \sum_{\rm perms} \exp\bigg\{ \frac{-g}{\bn\cdot \cP} \bigg[
   \frac{w |2 \cP^z|^{-\eta/2}}{\nu^{-\eta/2}} \bn \cdot A_s\bigg] \bigg\} 
  \,,  \\
 W_n &= \sum_{\rm perms} \exp\bigg\{ \frac{-g}{\bn\cdot \cP} \bigg[
   \frac{w^2 |\bn\cdot \cP|^{-\eta}}{\nu^{-\eta}} \bn \cdot A_n \bigg]\bigg\}
   \,,
 & W_\bn &= \sum_{\rm perms} \exp\bigg\{ \frac{-g}{n\cdot \cP} \bigg[
   \frac{w^2 |n\cdot\cP|^{-\eta}}{\nu^{-\eta}} n \cdot A_\bn \bigg] \bigg\}
  \,. \nn
\end{align}
Here the regulator momentum operators $\cP$ act only on the gluon field in the square brackets, whereas the inverse momentum operators $-g/\cP$ act on all fields to the right when the exponentials are expanded.  We separately regulate every soft or collinear gluon from the Wilson lines in order to maintain consistency with our use of the rapidity regulator for Glauber loops (rather than introducing the regulator only for the group momentum as in Ref.~\cite{Chiu:2012ir}). We have confirmed that our choice maintains exponentiation for matrix elements that only involve Wilson lines, since the exponentiation can be derived by permutations of momenta under which the regulator is symmetric. An additional complication in the operators we consider is the presence of inverse factors of $\bn\cdot \cP$ and $n\cdot \cP$ that appear outside of the Wilson lines.  Since our operators can be  written in different equivalent forms,  these factors are required for consistency. Examples where this occurs include ${\cal O}_n^{g B}$, ${\cal O}_\bn^{g B}$,  ${\cal O}_s^{g_n B}$, and ${\cal O}_s^{g_\bn B}$, see for example \eq{opbbb}. Here, the inverse power to that in \eq{rapregulator} is used, so for example $\bn\cdot\cP\to \bn\cdot\cP \frac{1}{w^2} \big| \frac{ {\bn \cdot \cal P}}{\nu} \big|^{+\eta}$ in the numerator of the $n$-collinear operator ${\cal O}_n^{g B}$, and $n\cdot\cP\to n\cdot\cP \frac{1}{w} \big| \frac{ 2{\cal P}^z}{\nu} \big|^{+\eta/2}$ in the numerator of the soft operator ${\cal O}_{s}^{g_nB}$. 

We also regulate Glauber loops with the rapidity regulator, by regulating $1/q_\perp^2$ factors in the manner discussed in in \sec{GlauberBox}.  The limit $\eta\to 0$ is always considered first, with the rapidity renormalization carried out at finite $\epsilon$, and then the limit $\epsilon\to 0$ is taken. Graphs without rapidity divergences or sensitivity will give the same answer whether one sets $\eta=0$ before or after the loop integration.
We introduce factors of the $\eta$-regulator for each Glauber potential between the forward scattering components of the operators, so  the  Glauber action with $d=4$ becomes
\begin{align} \label{eq:LGreg_eta}
 \int\!\!  d^4x\, {\cal L}_G^{{\rm II}(0)}  & = 
 \sum_{n,\bn} \sum_{i,j=q,g} \! \int\! [dx^\pm] \!
  \sum_{k_r^+,k_r^-} 
  \int\!\! \frac{d^2q_\perp}{q_\perp^2} \frac{d^2q_\perp^\prime}{q_\perp^{\prime\,2}} \: 
  {\cal O}_{s,-k_r^\pm}^{AB}(q_\perp,q_\perp')
   \\
&\qquad\qquad \times
  \Bigg[ {\cal O}_{n,k_r^-}^{i A}(q_\perp)\,  
  w^2 \Bigg|\frac{in\,\cdot\! \rightpartial
  +i\bn\,\cdot\! \leftpartial
  }{\nu} \Bigg|^{-\eta}
  \, {\cal O}_{\bn,k_r^+}^{jB}(-q_\perp^\prime)
  \bigg] 
  \nn\\
 & + \sum_n \sum_{i,j=q,g}  \int\! [dx^\pm] \,
  \sum_{k_r^-} \!\!   \int\!\! \frac{d^2q_\perp}{q_\perp^2} \: 
  {\cal O}_{n,-k_r^-}^{i A}(q_\perp)\,  w^2 \Bigg|\frac{-\beta_{ns}\,k_r^-
   -i \bn\cdot{\leftpartial} }{\nu}\Bigg|^{-\eta} 
  \, {\cal O}_{s,k_r^-}^{j_n A}(-q_\perp)  
   \,.  \nn
\end{align}
Here $\nu$ is the rapidity renormalization scale and the operators in transverse momentum space are given above in \sec{LGtransverse}.  In the 3-rapidity sector operator, the factor $|in\,\cdot  \rightpartial +\,i\bn\,\cdot \leftpartial |^{-\eta}$ regulates the $n$-$\bn$ Glauber potential, and for graphs where ${\cal O}(\lambda^2)$ momenta do not flow into the soft sector, one can integrate by parts and it becomes $ | 2 i \vec\partial_z |^{-\eta} = |in\cdot\vec\partial-i\bn\cdot\vec\partial|^{-\eta}$. Here these derivatives only pick out ${\cal O}(\lambda^2)$ momenta. In the 2-rapidity sector operator the regulator involves a combination of the $n$-collinear ${\cal O}(\lambda^2)$ momentum and the ${\cal O}(\lambda)$ soft momentum because it is regulating a soft-collinear Glauber potential. The inclusion of the boost parameter $\beta_{ns}>0$ where $\beta_{ns}\sim \lambda$ here ensures that these momenta appear together in a  homogeneous combination in the rapidity regulator.    For the pure Glauber potential in $n$-$\bn$ scattering we have no soft gluons,  so can set 
\begin{align}
 {\cal O}_{s,-k_r^\pm}^{BC}(q_\perp,q_\perp^\prime) = 8\pi\alpha_s\, q_\perp^2\, \delta^2(q_\perp-q_\perp^\prime)\, \delta_{k_r^+,0}\delta_{k_r^-,0} \,,
\end{align}
and \eq{LGreg_eta} gives a factor of $|2i\partial_z|^{-\eta}\nu^\eta\to |2q_z|^{-\eta} \nu^\eta$ for each potential carrying momentum $q$. This then yields the rapidity regulator factors used in the box and cross-box calculations in \sec{GlauberBox}.  For $n$-$s$ scattering the regulator for each potential is made homogeneous by the inclusion of the boost factor $\beta_{ns}\sim\lambda$. As discussed in \sec{GlauberBox}, the result for Glauber loops from $O_{ns}^{ij}$ iterations is independent of $\beta_{ns}$. In  \sec{higherorder} we  encounter two-loop examples where both  the $\eta$ regulator in the Glauber potentials and  in the Wilson lines are needed simultaneously and justifies the choice of the  power of $\eta$ in \eq{LGreg_eta}.

\begin{figure}[t!]
 	%
 	%
\begin{center}
 	\includegraphics[width=0.45\columnwidth]{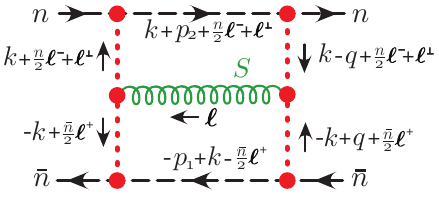} 
    \phantom{x}\hspace{0.5cm}
 	\includegraphics[width=0.35\columnwidth]{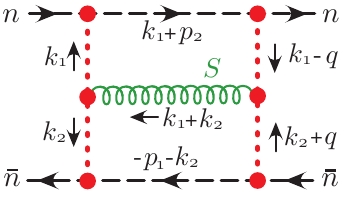}
 \hspace{0.cm}\\
 	\raisebox{0.4cm}{\hspace{-4.5cm} a)\hspace{7.2cm} b) \hspace{2cm} } 
\end{center}
    \vspace{-1.cm}
 	\caption{\setcaptionskip
 	Two different momentum routings for the two loop diagram stemming from the time ordered product of two   ${\cal O}_{ns\bn}^{qq}$  insertions with one soft gluon. The soft momentum $\ell^\mu = k_1^\mu +k_2^\mu$ can only be routed in a way which is consistent with the power counting. In a) the loops are soft and $n$-$\bn$ Glauber with $\ell^\mu \sim (\lambda,\lambda,\lambda)$ and $k^\mu\sim (\lambda^2,\lambda^2,\lambda)$ respectively, while in b) the loops are $s$-$n$ Glauber and $s$-$\bn$ Glauber with $k_1^\mu\sim (\lambda^2,\lambda,\lambda)$ and $k_2^\mu \sim (\lambda,\lambda^2,\lambda)$ respectively.	}
 	\label{fig:hgraph}
 	\setmainskip
 \end{figure}

As a more complicated example of how the rapidity regulators work, we can consider the H-graph involving two Lipatov vertices, which is shown in \fig{hgraph} for two different momentum routings.  In \fig{hgraph}a we have a soft loop momentum $\ell^\mu\sim\lambda$ and a $n$-$\bn$ Glauber loop momentum $k^\mu\sim (\lambda^2,\lambda^2,\lambda)$ for its $(+,-,\perp)$ components. In \fig{hgraph}b the same diagram is shown but now using a $n$-$s$ Glauber loop momentum $k_1^\mu\sim (\lambda^2,\lambda,\lambda)$ and a $\bn$-$s$ Glauber loop momentum $k_2^\mu\sim (\lambda,\lambda^2,\lambda)$. The routing of momentum in the two graphs are related by the changes of variable
\begin{align}  \label{eq:hreroute}
  n\cdot k_1 &= n\cdot k\,,
 & \bn\cdot k_1 &= \bn\cdot (k+\ell) \,,
 & k_{1\perp} &= k_\perp+\ell_\perp \,,
\\
 n\cdot k_2 &= n\cdot (\ell-k) \,,
 & \bn\cdot k_2 &= -\bn\cdot k \,,
 & k_{2\perp} &= -k_\perp 
 \,. \nn
\end{align}
In order that these two momentum routings give the same results, it is important that the rapidity regulators also are transformed into one another under this change of variable, and of course also will regulate the singularities in the diagram. \eq{LGreg_eta} with the $in\cdot\partial$ and $i\bn\cdot\partial$ factors satisfies both these criteria. In particular for the loop integrals in the two routings we have
\begin{align} \label{eq:hintegrands}
 {\rm Fig.}~\ref{fig:hgraph}a:\ 
  & \int\!\! 
   \frac{\ddslash\!^d k\, \ddslash\!^d \ell \, |2k^z|^{-2\eta} |2\ell^z|^{-\eta}\,
   N_a(\ell,k_\perp,q_\perp)\: 
   G_0(k_\perp) G_0(k_\perp\plus\ell_\perp)
   G_0(k_\perp\plus\ell_\perp\minus q_\perp) G_0(k_\perp\minus q_\perp)}
 {\Big[k^+ \plus p_2^+ \minus \frac{(\vec k_{\perp}\plus \vec p_{2\perp}+\vec \ell_\perp)^2}{p_2^-}\plus i0\Big] 
  \Big[\minus k^- \plus p_1^- \minus \frac{(\vec k_{\perp}\minus\vec  p_{1\perp})^2}{p_1^+}\plus i0\Big] \big[ \ell^2\plus i0\big]  }
  ,
  \nn\\[5pt]
 {\rm Fig.}~\ref{fig:hgraph}b:\ 
  & \int\!\! 
   \frac{\ddslash\!^d k_1\, \ddslash\!^d k_2 \, |k_1^+\plus k_2^-|^{-2\eta} |k_1^-\minus k_2^+|^{-\eta}\,
   N_b(k_1^-,k_2^+,k_{1\perp},k_{2\perp},q_\perp)}
 {\Big[k_1^+ \plus p_2^+ \minus \frac{(\vec k_{1\perp}\plus \vec p_{2\perp})^2}{p_2^-}\plus i0\Big] 
  \Big[ k_2^- \plus p_1^- \minus \frac{(\vec k_{2\perp}\plus \vec p_{1\perp})^2}{p_1^+}\plus i0\Big]
  \big[k_2^+k_1^-\minus (\vec k_{1\perp}\plus \vec k_{2\perp})^2\plus i0\big]
  }
  \nn\\[-3pt]
&\qquad\qquad
\times  G_0(k_{1\perp}) G_0(k_{2\perp})
   G_0(k_{1\perp}\minus q_\perp) G_0(k_{2\perp}\plus q_\perp) 
  ,
\end{align}
where for this equation only, $G_0(k_\perp)= (ig^2)/\vec k_\perp^{\:2}$. Here $N_a$ and $N_b$ are functions that are each obtained from the contraction of two Lipatov vertices from \fig{onegluon}. For the two routings the factors of $|2k^z|^{-2\eta}$ and $|k_1^+ + k_2^-|^{-2\eta}$ are each obtained from the $|in\,\cdot  \rightpartial +\,i\bn\,\cdot \leftpartial |^{-\eta}$ regulator in \eq{LGreg_eta}. This regulates the $dk^+dk^-$ integrations in the \fig{hgraph}a routing, and the $dk_1^+dk_2^-$ integrations in the \fig{hgraph}b routing. The other  factors, $|2\ell^z|^{-\eta}$ and $|k_1^--k_2^+|^{-\eta},$ are generated by the regulator in the soft Wilson lines in ${\cal O}_s^{AB}$, and hence only depend on the soft gluons momentum in each case. They regulate eikonal factors that appear inside $N_a$ and $N_b$. Noting that $N_b\to N_a$ under the transformation in \eq{hreroute}, it is easy to see that the two results in \eq{hintegrands} are exactly equivalent under this transformation.

The SCET graphs also have zero-bin subtractions~\cite{Manohar:2006nz} which are necessary to avoid double counting between contributions from the various infrared modes. These subtractions are determined by the SCET propagators appearing in loop diagrams.  For ${\rm SCET}_{\rm II}$ the overlapping modes are collinear, soft, and Glauber.  At leading power collinear gluon propagators have subtractions from the soft and Glauber regions, and soft gluon propagators have subtractions from the Glauber region. At one-loop, if we consider a soft loop diagram $S$ with only soft gluon propagators, or a $n$-collinear loop diagram $C_n$ with only collinear gluon propagators, then the structure of the subtractions is
\begin{align} \label{eq:0bins}
  S &= \tilde S - S^{(G)} \,, 
  & C_n & = \tilde C_n - C_n^{(S)} - C_n^{(G)}  + C_n^{(S)(G)} \,.
\end{align}
Here the superscript indicates the momentum region that the subtraction comes from. The $(G)$ for the soft subtraction can be any one of the three Glauber momentum scalings $(+,-,\perp)$ $\sim (\lambda^2,\lambda^2,\lambda)$ or $(\lambda^2,\lambda,\lambda)$ or $(\lambda,\lambda^2,\lambda)$, while the $(G)$ subtraction for the $n$-collinear case only includes scalings of the form $(+,-,\perp)\sim (\lambda^2,\lambda^2,\lambda)$ or $(\lambda^2,\lambda,\lambda)$. The result for $C_\bn$ is analogously obtained by taking $n\to \bn$ in $C_n$.  If we start with the naive soft loop graph $\tilde S$ with loop momentum $k$, then the Glauber subtraction $S^{(G)}$ is obtained from scaling the $\tilde S$ integrand into the region $k^+k^- \ll \vec k_\perp^{\,2}$ and keeping only terms that are the same order in the $\lambda$ power counting as the original integrand. If we have a naive $n$-collinear loop graph $\tilde C_n$ with loop momentum $\ell$, then there is a soft subtraction $C_n^{(S)}$ from the region $\ell^\mu \sim \lambda$, and a Glauber subtraction $C_n^{(G)}$ from the region $\ell^+\ell^-\ll \vec \ell_\perp^{\,2}$, plus a term $C_n^{(S)(G)}$ that adds back the soft-Glauber overlap region so that it is not over subtracted. This overlap term is constructed from the Glauber limit of the $C_n^{(S)}$ integrand.  Even at one-loop, subtractions other than those in \eq{0bins} are possible, since the subtractions are induced by propagators rather than by the type of loop momentum. For instance, Glauber loops which contain a soft gluon propagator can also have a Glauber subtraction, and we will see examples of this in \sec{softemission}. The zero-bin subtractions are formulated iteratively to all loop orders~\cite{Manohar:2006nz} at the level of the SCET Lagrangian, and a two-loop example with subtractions can be found in \sec{higherorder}. For certain cases at leading power it is known how to formulate subtractions which appear as Wilson line matrix elements together with matrix elements involving full QCD fields, which are equivalent to the zero-bin subtractions, see Refs.~\cite{Lee:2006fn,Idilbi:2007ff}.  It would be interesting to try to extend this to the SCET subtractions that occur in the presence of Glauber loops, but we will not do so here.

Note that when we consider the scaling limits to construct the 0-bin subtractions we do not change the form of the rapidity regulator (the original and subtraction integrals must share the same regulators for the subtraction to properly remove any double counted contributions).   With the rapidity regulator we use here these subtractions often lead to scaleless integrals that just convert whatever divergences occur from the IR or UV, but for some diagrams we will consider they do not give scaleless integrals and play an important role in avoiding double counting. Note that in general, scaleless integrals that are log-divergent in the UV and IR are not treated as vanishing in the EFT. Without the Glauber dependent subtractions the results in \eq{0bins} reduce to the standard soft subtraction on collinear integrands in \SCETb.

For completeness we also discuss here the 0-bin subtractions for the collinear, soft and Glauber loop graphs in \SCETa at one-loop.  Once again the form of these subtractions are determined by propagators. Here collinear gluon propagators can have soft, ultrasoft, and Glauber subtractions, soft gluon propagators can have ultrasoft and Glauber subtractions, and Glauber propagators can have ultrasoft subtractions. If we have loops $C_n$, $S$, and $G$ where the gluons that appear are purely $n$-collinear, soft, or Glauber respectively, then the form of the subtractions are 
\begin{align} \label{eq:scet1subt}
  C_n &= \tilde C_n -  C_n^{(S)} -  C_n^{(G)} -  C_n^{(U)} 
+  C_n^{(S)(G)} +  C_n^{(G)(U)} +  C_n^{(S)(U)} -   C_n^{(S)(G)(U)} 
  \,,  \\
  S &= \tilde S - S^{(G)} - S^{(U)} + S^{(G)(U)}
  \,,\nn \\
  G &= \tilde G - G^{(U)}
  \,, \nn
\end{align}
Subtractions for the analogous $C_\bn$ are the same as those for $C_n$ with $n\leftrightarrow \bn$. Due to the presence of the lower invariant mass ultrasoft modes there are more subtraction terms in \SCETa, and in particular the loops with Glauber exchange propagators also have an ultrasoft subtraction. In the limit that we neglect soft loops in \SCETa, so that there are only ultrasoft gluons (or the soft gluon can be absorbed into the ultrasoft), then the subtractions in \eq{scet1subt} agree with those of Ref.~\cite{Bauer:2010cc}.

In general, the soft and collinear Wilson lines in the operators of the Glauber Lagrangian, \eq{LG}, or in expressions like \eq{SWrapreg}, should have their position space directions $(0,\infty)$ or $(-\infty,0)$ specified. This corresponds with the appearance of $\pm i0$ factors in the momentum space Feynman rules, see \app{Wdirection}. However, the dependence on whether the line extends to $\pm\infty$ will be canceled by the 0-bin subtractions. Soft lines generate propagators such as $(n\cdot k\pm i0)$ with $n\cdot k\sim \lambda$, while it is the Glauber region which properly describes the region of smaller momenta $n\cdot k\sim\lambda^2 $ which includes the pole $n\cdot k=-i0$. The situation is similar for collinear Wilson lines, which have both soft and Glauber 0-bin subtractions.  We will show explicitly the cancellation of Wilson line direction dependence by 0-bins for soft and collinear loop graphs in one-loop and two-loop calculations for forward scattering in \secs{loop2match}{loop1match} and for hard scattering in \secs{hardmatching}{higherorder}. In particular, we explain in \sec{higherorder} that the directions of the soft Wilson lines in the leading power Glauber Lagrangian can be chosen to be either as $(0,\infty)$ or as $(-\infty,0)$ without changing our results. This occurs due to the presence of Glauber region 0-bin subtractions. On the flip side, we will see that Glauber interactions in certain hard scattering diagrams can be absorbed into the direction of soft and collinear Wilson lines in the hard scattering operators. In general, the dependence on these directions may then still cancel out in factorization theorems where infinite Wilson lines are combined into finite lines.

\subsection{Power Counting Theorem and Operator Completeness}
\label{sec:powercount}

In this section we give the all orders power counting formulae for \SCETa and \SCETb that hold in the presence of loops carrying Glauber momenta, and arbitrary power suppressed interactions. We then discuss the complete basis for Glauber exchange at leading power, namely ${\cal O}(\lambda^0)$.  The ingredients needed for this analysis are an SCET power counting theorem valid to any order in $\lambda$ in the presence of Glauber effects, information about the structure of infrared divergences in gauge theory, gauge symmetry, dimensional analysis, and the momentum structure of forward scattering operators in the limit $s\gg t$. 

In \app{powercount} we derive a general power counting formula for an arbitrary diagram with operators at any order in the power counting in both \SCETa and \SCETb.  As shown there, the final formula can be applied to both of these 
theories and says that the graph will scale as $\lambda^\delta$ where
\begin{align} \label{eq:delta}
   \delta 
 & = 6- N^n - N^\bn - N^{nS} - N^{\bn S} +  2 u 
   \,,\\
  & \quad  + 
 \sum_k (k-8) V_k^{us} +  (k\minus 4) \big( V_k^n + V_k^\bn + V_k^S \big) 
       + (k\minus 3)\big(  V_k^{nS} + V_k^{\bn S} \big) 
       + (k\minus 2) V_k^{n\bn} 
     \,. \nn
\end{align}
Note that the scaling includes power counting factors for the external lines in the graph (for example, each collinear fermion or $\perp$ collinear gauge boson gives a factor of $\lambda$). There are various ingredients in this formula. We count the operators whose fields plus derivatives give a scaling of $\lambda^k$, by letting  $V_k^\Omega$ be the number of such operators of type $\Omega$. We use $V_k^{us}$ to count operators that contain only ultrasoft fields, $V_k^n$ for operators with only $n$-collinear and ultrasoft fields, $V_k^\bn$ for only $\bn$-collinear and ultrasoft fields, and $V_k^S$ for only soft and ultrasoft fields. The index $V_k^{n S}$ counts operators with $n$-collinear and soft fields, and possibly ultrasoft fields, but no $\bn$ fields, $V_k^{\bn S}$ for those with both $\bn$ and soft, and possibly ultrasoft fields, but no $n$ fields, and $V_k^{n\bn}$ for $n$ and $\bn$ fields, and possibly soft and ultrasoft fields. Thus operators containing all types of fields are counted by $V_k^{n\bn}$.  The factor $+2u$ in \eq{delta} is relevant for graphs with only ultrasoft fields where one sets $u=1$, and otherwise one sets $u=0$. Since the $V_k^\Omega$ indices count the number of insertions of gauge invariant operators, the power counting formula for $\delta$ is explicitly gauge invariant.

The remaining ingredients in \eq{delta} are topological in nature.
The index $N^n$ counts the number of disconnected $n$-collinear subgraphs if field lines of all other types are erased, $N^\bn$ does likewise for $\bn$-collinear subgraphs. Finally $N^{nS}$ is the number of disconnected subgraphs if just $\bn$-collinear fields are erased, and $N^{\bn S}$ is the number if just $n$-collinear fields are erased.  Note that at leading order in the power counting, that graphs with a loop involving one of the Glauber type loop momenta must involve at least one of the Glauber potential vertices. (This is no longer true at subleading power, for an example see \app{powercount}.) Further details of the derivation of \eq{delta} can be found in \app{powercount}, including how this result reduces to the earlier results given in Refs.~\cite{Bauer:2002uv,Stewart:2003gt} in special cases. In \app{powercount} we also show how \eq{delta} can be used to demonstrate that all time ordered products scale as a power of $\lambda$ that is at least given by the sum of contributions from its constituent operators. The correspondence of our method of power counting with that of CSS~\cite{Sterman:1978bi,Sterman:1978bj,Sterman:1995fz} is also discussed in Ref.~\cite{Bauer:2002uv}.

Consider applying \eq{delta} to  the operators in \eq{LG}.  Counting up the scaling of the building block fields ${\cal O}_n^{jB}\sim {\cal O}_\bn^{jB}\sim\lambda^2$ and $O_s^{AB}\sim \lambda^2$, and counting $1/\cP_\perp^2\sim \lambda^{-2}$ we see that $O_{ns\bn}^{i j} \sim \lambda^2$ and contributes to $V_2^{n\bn}$ for each insertion. Due to the $(k-2)$ prefactor in \eq{delta} the operator $O_{ns\bn}^{i j}$ contributes to the leading power Lagrangian.  Noting that ${\cal O}_s^{j_nB}\sim \lambda^3$ we find $O_{ns}^{ij}\sim \lambda^3$, and this operator contributes to $V_3^{nS}$ which has the prefactor $(k-3)$, so again this is a leading power contribution. Due to the local nature of gauge theories like QCD, we can have at most a quadratic divergence as the difference of the external momentum of two lines goes to zero.  This implies that we have at most a $1/t$ power-law singular structure in our Glauber potentials, and hence at most a $1/\cP_\perp^2$ between operators living in two different rapidity sectors.  (An amplitude can have a $1/(t t')$ where $t$ and $t'$ correspond to two different propagtors, but these are reproduced by a time ordered product of two Glauber operators in the effective theory.)  For a Glauber operator the scattering is forward, and we must therefore have two building blocks for each sector which contributes to the lowest order Feynman rule in order to conserve the large forward momenta and satisfy gauge invariance.  For the $n$ or $\bn$ collinear sectors the lowest order bilinears are therefore $\sim \lambda^2$. The quark bilinear operators will be dimension $3$. Gluon bilinear operators start at dimension $2$ but must include an extra derivative to become dimension $3$ in order to compensate the $-2$ dimension of a $1/t$ insertion in a Glauber operator (and due to the antisymmetry in color).  This derivative is ${\cal O}(\lambda^0)$ for a collinear sector, but adds an additional power of $\lambda$ for a soft bilinear gluon operator. Therefore both the soft quark and gluon octet operators ${\cal O}_s^{j_nB}$ start at $\sim \lambda^3$ in Glauber operators. The lowest order soft operator that appears between two collinear sectors, $O_s^{AB}$, must also be either bilinear in the ${\cal O}(\lambda)$ building blocks $\cP_\perp^\mu$, ${\cal B}_{S\perp\mu}^{nB}$, and ${\cal B}_{S\perp\mu}^{\bn B}$ in order to have an even number of $\perp$ Lorentz indices to contract in the operators, or linear in ${G}_s^{\mu\nu\, AB}\sim \lambda^2$. Therefore $O_s^{AB}$ must start at ${\cal O}(\lambda^2)$ and be purely gluonic, since a fermionic contribution with soft quark fields $\bar \psi_s \psi_s$ starts at ${\cal O}(\lambda^3)$. Mixed operators with $\bar\psi_s \psi_n$ are also at least ${\cal O}(\lambda^{5/2})$. Thus we always have $V_{\le 1}^{n\bn}=0$, $V_{\le 2}^{nS}=0$, and there are no Glauber operators  that are lower order in the power counting than those in \eq{LG}. Obviously there can be no Glauber operators which contribute to $V_k^{us}$, $V_k^{n}$, $V_k^{\bn}$, or $V_k^{S}$ since these indices do not contain fields from two of the $n$, $\bn$, or soft sectors.  

Thus we are left to consider the possibility of additional operators that contribute to the indices $V_2^{n\bn}$, $V_3^{nS}$, and $V_3^{\bn S}$, beyond those given above in \sec{GlauberSCET}.  We assume we have a single power of $1/t$ for any particular $t$, and hence for example with a single $1/t$ that the operator dimensions add to $6$ to give a dimension $4$ Lagrangian.  Since we must use the minimum number of $\lambda$'s, but preserve large momentum conservation for the forward scattering, preserve gauge invariance, and rotational invariance in the transverse plane, only operators bilinear in the building blocks (or with a single $G_s^{\mu\nu}$) are possible for each sector.  We will construct the complete basis of operators for $O_s^{AB}$ below in \sec{basis}. Preserving fermion number  the possible collinear bilinear operators are just $\bar\chi_n^\alpha \chi_n^\beta$ and ${\cal B}_{n\perp}^A \bn\cdot\cP {\cal B}_{n\perp}^B$, and their analogs in the $\bn$ and soft sectors.  Examining the SU(3) quantum numbers we see that $q\bar q$ gives $1\oplus 8$, and $gg$ gives $1_S \oplus 8_A \oplus 8_S \oplus 10_A \oplus \bar{10}_A \oplus 27_S$.  When we combine the operators in different sectors we must produce an overall color singlet, so it is possible to form singlets with other color representations.  However, from tree level matching {\em only} the operators in \eq{LG} (with octet quantum numbers for the collinear bilinears, etc.) are generated by integrating out offshell Glauber exchanges and offshell hard lines. Thus, the key question is whether any other operators governing Glauber exchange can be generated by loop-level matching. The answer to this is no.

Essentially our Lagrangian in \eq{LG} is obtained by simultaneously removing offshell Glauber propagators that have $p_\perp^2\sim \lambda^2$ (representing them as a potential) as well as offshell hard propagators with $p^2\gg \lambda^2$ which contribute to generating Wilson lines.  The loop level matching for forward scattering operators is done to integrate out physics at the scale $s$ and represent offshell non-local physics at the scale $t$ as a potential. But there are no hard-loop diagrams with momenta of order $s$ that have the required overall scaling as $\propto 1/t$.
This can be seem from the fact that a hard loop can be contracted to a point which would generate a higher dimensional operator suppressed by powers of $s$.
 Furthermore, in loop graphs the offshell lines that can have Glauber scaling are instantaneous in time and the $n\cdot x-\bn\cdot x$ longitudinal coordinate, and hence are always sequestered in tree level subcomponents of the graph. These components are tied together by propagators that have onshell scaling for their momenta, which are represented by the onshell fields in the EFT, and hence such contributions are represented by time ordered products of tree level induced Glauber exchange operators in SCET. 
This is true at any loop order, so \eq{LG} is the full Glauber Lagrangian. Thus the form of the Lagrangian is determined at tree-level. In \sec{loopmatch} we explicitly demonstrate this at one-loop for quark-quark scattering, showing that no other operators are generated at one-loop and that the coefficients of the operators in \eq{LG} receives no one-loop corrections.

\subsection{Forward Scattering and Observables}
\label{sec:forward}

In the previous sections we setup the Lagrangian for Glauber exchange between collinear particles traveling in different light-like directions, as well as those involving soft particles, for situations where there are kinematic variables $s\gg |t|$. Here we briefly discuss a few classes of observables where Glauber exchange between these degrees of freedom can play a role.
 
Perhaps the simplest example are situations where we have a hierarchy $|t|/s\ll 1$ from measuring the momentum in an external current, without making direct measurements on the hadronic final state. The canonical example of this is DIS, where we measure the electrons momentum in the final state, and consider Bjorken $x \ll 1$.  Here $x= Q^2/(2 p \cdot q)$ where $p$ is the initial state proton momentum and $q$ is the virtual photon momentum. For DIS the ratio $(-t)/s$ is determined by $x$, and $Q^2=-q^2=-t \gg \Lambda_{\rm QCD}^2$ is a perturbative scale. In this situation the Glauber operators can be used to sum $\ln x$ factors in the cross section and/or parton distribution functions. Indeed the BFKL equation for the rapidity renormalization of the collinear functions in \sec{bfklconsist}, is related to this resummation for both DIS and Drell Yan. A detailed analysis of these small $x$ resummations in SCET will be given elsewhere.

Another application of the Glauber operators is to the study of factorization violation in hard collision processes with initial state hadrons. Examples of how Glauber exchange operators appear in these processes are discussed in \sec{spectator}. Of particular concern is the final state interactions between spectator partons (those not directly participating in the hard scattering). It is known that final state rescattering effects will cancel out at leading power for the inclusive Drell-Yan process~\cite{Collins:1988ig}.  However, observables that make measurements of the final state hadronic radiation, such as transverse-thrust~\cite{Banfi:2004nk} and beam thrust~\cite{Stewart:2009yx}, can be more sensitive to Glauber interactions. Indeed, it has been proposed that the sensitivity of such measurements to the multiple parton interactions (underlying event) in Monte Carlo programs is related to factorization violation from the Glauber momentum regime~\cite{Gaunt:2014ska}. Recently, the presence of factorization violating contributions from the Glauber regime has been demonstrated in~\cite{Zeng:2015iba} for a beam thrust spin asymmetry with scalar quarks.  So far these investigations have not fully accounted for the distinction between perturbative Glauber exchange at the scale of the event shape measurements $|t|\gg \Lambda_{\rm QCD}^2$, that could in principle be treated with forward scattering factorization based methods order-by-order in the strong coupling, and non-perturbative Glauber exchange that couples together hadron-hadron matrix elements at the scale $\Lambda_{\rm QCD}^2$, which can only be handled with di-hadron matrix elements. Factorization violating observables have also been associated to those containing rapidity gaps between jets and super-leading logarithms~\cite{Forshaw:2006fk,Forshaw:2009sf,Forshaw:2012bi}. Our Glauber exchange Lagrangian provides an efficient method for computing
various (potentially) factorization violating contributions, as demonstrated in \secs{properties}{spectator}, and can be utilized to address these observables and questions about factorization more precisely. A related goal would be to build more sophisticated treatments of multi-parton interactions (underlying event) based on calculations utilizing Glauber exchange.

A final category of observables are those typically associated with forward scattering, including things like the total hadronic cross section, and diffractive processes. Since these are dominated by physics from the forward scattering limit they can be described in part with the Glauber exchange operators.  Depending on the precise observable and how inclusive or exclusive the measurements are, the description with SCET will change, in much the same way that the same formalism describes factorization for exclusive and inclusive hard scattering processes in a different manner (such as at the amplitude versus cross section level). In later sections we will discuss the resummation of large rapidity logarithms in the operators for forward scattering, both for amplitudes by Reggeization in \sec{loopmatch} and for scattering cross sections via the BFKL equation in \sec{BFKL}. Note that Reggeization occurs in SCET due to the rapidity factorization of soft and collinear virtual diagrams which are not effected by the final state measurement, whereas the BFKL equation includes also the real emission diagrams. In general this is expected to lead to modified evolution equations which depend on the precise nature of the measurement.

\section{Tree Level Matching Calculations}
\label{sec:match}

In this section we present several tree level calculations which are important for deriving the Glauber EFT presented in \sec{GlauberEFT} above.  In \sec{treematch} we demonstrate how the $W_n$ and $W_\bn$  Wilson lines in the Glauber operators are generated through tree level matching from the full theory, which involves both local and time-ordered product terms in the EFT. In Secs.~\ref{sec:lipatov}, \ref{sec:basis}, and \ref{sec:2sgluon} we construct a complete basis of allowed operators and carry out one-gluon and two-gluon matching calculations to derive the mid-rapidity operator $O_s^{BC}$ given in \eq{Os1}.  Complete one loop calculations for \SCETb and \SCETa are carried out later in \sec{loopmatch}.

\subsection{Wilson Lines $W$ and $S$ from Tree Level Matching}
\label{sec:treematch}

For standard hard scattering operators in SCET the collinear Wilson lines $W_n$ appear in a manner which ensures $n$-collinear gauge invariance in the hard-collinear factorization~\cite{Bauer:2001ct}. These Wilson lines are generated by integrating out hard offshell full QCD quark and gluon propagators, and can be readily derived using the auxiliary Lagrangian formalism presented in the appendices of Refs.~\cite{Bauer:2001yt,Bauer:2002nz}.  

The situation is different for the $W$ Wilson lines appearing in the Glauber operators. In this case we want to integrate out offshell propagators that are either Glauber or hard, but the relevant matching calculation involves diagrams in the full theory with at least one onshell propagator (meaning a propagator whose $p^2$ is such that it is not offshell from the point of view of SCET). In fact, part of the sum of these graphs are nonlocal, whereas another part localizes into a potential. On the EFT side of the calculation there will be both a non-local term involving a T-product that involves an onshell propagator, and the localized potential term involving the Wilson line. The sum of these two terms will reproduce the full theory result.

\begin{figure}[t!]
%
%
\begin{center}
\subfigure{
\raisebox{-0.2cm}{a)\hspace{0.1cm}} 
\includegraphics[width=0.28\columnwidth]{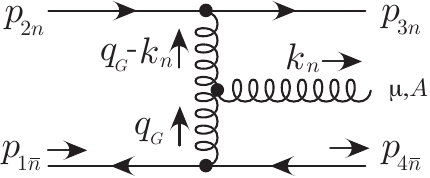}
 \hspace{0.6cm}
\raisebox{-0.2cm}{
\includegraphics[width=0.28\columnwidth]{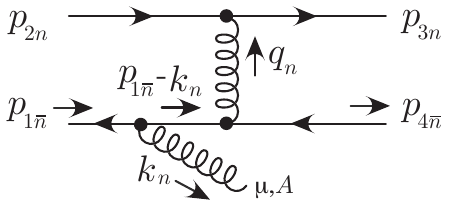}
}
 \hspace{0.6cm}
\raisebox{-0.2cm}{
\includegraphics[width=0.28\columnwidth]{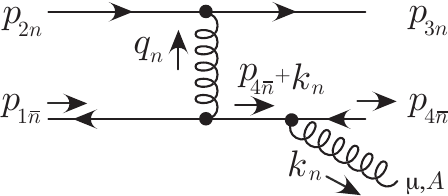}
}
}
\subfigure{
\raisebox{-0.2cm}{b)\hspace{0.1cm}} 
\includegraphics[width=0.25\columnwidth]{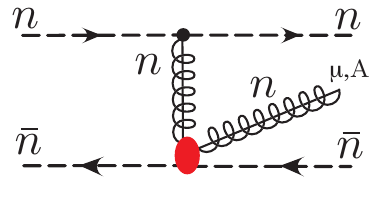}
 \hspace{1cm}
\raisebox{-0.2cm}{
\includegraphics[width=0.25\columnwidth]{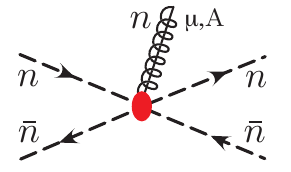} 
} }
\end{center}
\vspace{-0.6cm}
\caption{\setcaptionskip
a) Full theory graphs for the tree level matching of quark-antiquark forward scattering with one extra $n$-collinear gluon. b)  EFT graphs for the tree level matching for the four quark operator with one extra $n$-collinear gluon. Note that the first EFT graph is a time ordered product of a Glauber operator and a collinear Lagrangian interaction.}
\label{fig:match_nnb_nGluon}
\setmainskip
\end{figure}

As a first example, we consider the matching for the $q$-$\bar q$ scattering process with one additional $n$-collinear gluon, $q(p_{2n})+\bar q(p_{1\bn}) \rightarrow q(p_{3n})+\bar q(p_{4\bn}) +g(k_n)$.  The relevant full theory diagrams for this matching calculation are shown in \fig{match_nnb_nGluon}a, while the diagrams in the EFT are shown in \fig{match_nnb_nGluon}b.  There are additional full theory diagram that are not shown, where the $k_n$ gluon attaches, via a Lagrangian insertion,  to the either of the quarks on the top-line, but these on-shell contributions are exactly reproduced by gluon attachments to the $n$-collinear quarks in an $O_{ns\bn}^{q q}$ insertion in the EFT (also not shown). The full theory graphs in \fig{match_nnb_nGluon}a have a gluon with $n$-collinear scaling that either attaches to a triple gluon vertex involving one Glauber propagator and one onshell ($n$-collinear) propagator, or attaches to the $\bn$-collinear quark leading to a hard offshell quark propagator plus an exchange gluon with $n$-collinear scaling. 
To carry out the matching calculation,  we first use the equations of motion relation in \eq{polA} to eliminate $n\cdot A(k_n)$ in terms of $A_\perp(k_n)$ and $\bn\cdot A(k_n)$ in the
first full theory diagram in figure \fig{match_nnb_nGluon}a.
 Then if we consider the $k_n$ external gluon to have $\perp$-polarization for $\mu$,  only the first full theory diagram in \fig{match_nnb_nGluon}a and the first SCET diagram in \fig{match_nnb_nGluon}b are nonzero, and these $\perp$ contributions exactly match. In contrast the  $\bn \cdot A$ polarization for these two diagrams do not match.   This agreement for the $\perp$ polarizations  is very analogous to the agreement we saw earlier for the diagrams in \fig{treematch_nnb_allpol}, just with an extra quark line attached to one of the gluons there, and use of the equations of motion on only one gluon.

When the $k_n$ external gluon has $\bn^\mu$ polarization all the diagrams in \fig{match_nnb_nGluon} contribute. For this case the analogy with simply adding a quark line to one of the gluons in \fig{treematch_nnb_allpol} breaks down, since using the equations of motion on only one gluon line no longer suffices to achieve agreement.  In this case, the result for the sum of the full theory graphs in \fig{match_nnb_nGluon}a is
\begin{align}
{\rm Fig.}~{\ref{fig:match_nnb_nGluon}a}
  =  2 g^3 f^{ABC} \bn^\mu \Big[ \bar u_n \frac{\bnslash}{2} T^B u_n \Big] 
  \Big[ \bar v_\bn \frac{\nslash}{2} \bar T^C v_\bn \Big] 
  \frac{1}{q^2(q-k)^2\, \bn\cdot k} 
  \Big[ q^2 + 2\, n\mcdot k\, \bn\mcdot k \Big] 
  \,.
\end{align}
The result for the first graph in SCET is
\begin{align}
 {\rm Fig.}~{\ref{fig:match_nnb_nGluon}b1}   
  = 2 g^3 f^{ABC} \bn^\mu \Big[ \bar u_n \frac{\bnslash}{2} T^B u_n \Big] 
  \Big[ \bar v_\bn \frac{\nslash}{2} \bar T^C v_\bn \Big] 
  \frac{1}{q^2(q-k)^2\, \bn\cdot k} 
  \Big[ 2 k_\perp\cdot (q_\perp-k_\perp) \Big] 
  \,.
\end{align}
Using $k^2=n\mcdot k\,\bn\mcdot k + k_\perp^2 =0$ and $q=q_\perp$ the difference is
\begin{align}
  {\rm Fig.}~{\ref{fig:match_nnb_nGluon}a}  
   - {\rm Fig.}~{\ref{fig:match_nnb_nGluon}b1} 
  &= 2 g^3 f^{ABC} \bn^\mu \Big[ \bar u_n \frac{\bnslash}{2} T^B u_n \Big] 
  \Big[ \bar v_\bn \frac{\nslash}{2} \bar T^C v_\bn \Big] 
  \frac{1}{q^2(q-k)^2\, \bn\cdot k} 
  \Big[ q_\perp^2  - 2 k_\perp\cdot q_\perp   \Big]
 \nn\\
  &= 2 g^3 f^{ABC} \bn^\mu \Big[ \bar u_n \frac{\bnslash}{2} T^B u_n \Big] 
  \Big[ \bar v_\bn \frac{\nslash}{2} \bar T^C v_\bn \Big] 
  \frac{1}{q_\perp^2\, \bn\cdot k} 
\nn\\
  &= {\rm Fig.}~{\ref{fig:match_nnb_nGluon}b2} \,,
\end{align}
which, as indicated, is precisely the contribution from the $W$ Wilson lines in the second graph in \fig{match_nnb_nGluon}b.  Thus we validate the presence of both the nonlocal T-product and local Wilson line contributions in SCET.

\begin{figure}[t!]
%
%
\begin{center}
\subfigure{
\raisebox{-0.2cm}{a)\hspace{-0.1cm}} 
\includegraphics[width=0.22\columnwidth]{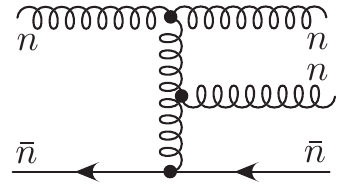}
 \hspace{0.22cm}
\raisebox{-0.1cm}{
\includegraphics[width=0.22\columnwidth]{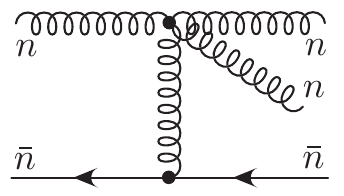}
}
 \hspace{0.22cm}
\raisebox{-0.2cm}{
\includegraphics[width=0.22\columnwidth]{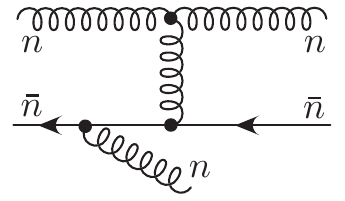}
}
\hspace{0.22cm}
\raisebox{-0.2cm}{
\includegraphics[width=0.22\columnwidth]{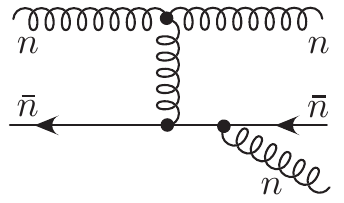}
}
}
%
\subfigure{
\raisebox{-0.2cm}{b)\hspace{0.1cm}} 
\includegraphics[width=0.23\columnwidth]{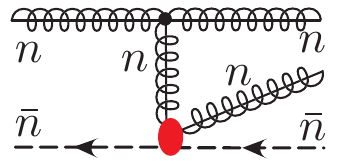}
 \hspace{1cm}
\raisebox{-0.2cm}{
\includegraphics[width=0.23\columnwidth]{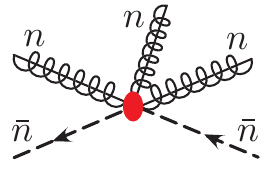}
 } 
}
\end{center}
\vspace{-0.6cm}
\caption{\setcaptionskip
a) Full theory graphs for the tree level matching of gluon-quark forward scattering with one extra $n$-collinear gluon. b)  EFT graphs for the tree level matching for the gluon-quark scattering operator with an extra $n$-collinear gluon. }
\label{fig:match_nnbGluon_nGluon}
\setmainskip
\end{figure}

The same matching calculation can also be carried out for the gluon-quark $n$-$\bn$ scattering, to validate the appearance of additional $\bn\cdot A_n$ fields in the ${\cal B}_{n\perp}$ building blocks of the Glauber operator $O_{ns\bn}^{g q}$. The necessary full theory diagrams are shown in \fig{match_nnbGluon_nGluon}a, while the SCET diagrams are shown in \fig{match_nnbGluon_nGluon}b. As above we work  in Feynman gauge for the internal gluon propagators, and remove $n\cdot A_n$ polarizations using the equations of motion. If all three external $n$-collinear gluons have $\perp$-polarization, then the first and second graphs in \fig{match_nnbGluon_nGluon}a precisely matches with the first graph in \fig{match_nnbGluon_nGluon}b.  To test the Wilson line contribution we can take two gluons to have $\perp$-polarization and one to be a $\bn\cdot A_n$.  In this case  there are contributions from all four full theory graphs, and both SCET diagrams. Once again the sum of contributions in the full and effective theories exactly match up after using the equations of motion to simplify terms. For gluon-gluon $n$-$\bn$ scattering, one can carry out a similar matching calculation to check the structure of $W_n$ Wilson lines in the Glauber forward scattering operator, and once again the full and effective theories agree. 

Due to the symmetry under $n\leftrightarrow \bn$, the above analysis also immediately yields the anticipated result for the 1-gluon part of the $W_\bn$ Wilson lines in the Glauber operators. Carrying out these low order matching calculations for the $W_n$ and $W_\bn$ Wilson lines is important for determining their directions ($i\epsilon$ prescription), and general structure. By power counting we know that only the ${\cal O}(\lambda^0)$ fields $\bn\cdot A_n$ and $n\cdot A_\bn$ fields can appear in these Wilson lines, and that they must appear in a manner that makes the Glauber operators $n$-collinear gauge invariant and $\bn$-collinear gauge invariant.  This suffices to fix the structure of these Wilson lines beyond the one-gluon level in the Glauber operators.

Furthermore, the same matching calculations can also be done for the Glauber operators involving soft-collinear forward scattering, namely $O_{ns}^{ij}$ and $O_{\bn s}^{ij}$. As discussed in \sec{SCops}, the results here are very analogous to $n$-$\bn$ forward scattering, because we still have the same hierarchy of momenta in each component. For example, for $O_{ns}^{ij}$ the only difference is that the overall size of the conserved $n\cdot p_s$ soft momenta is smaller than the $\bn\cdot p_n$ collinear momenta. For this reason the $n$-$\bn$ collinear-collinear scattering calculations discussed above carry over verbatim to the soft-collinear case, and we will not write out the analysis in detail. We have carried out explicit matching calculations to test the soft and collinear Wilson lines, confirming that they are correctly included in these operators.  For $O_{ns}^{ij}$, examples of the necessary diagrams can be obtained by replacing $A_\bn$ gluons in \figs{match_nnb_nGluon}{match_nnbGluon_nGluon} by $A_s$ gluons. In this analysis it is the large momentum direction $n^\mu$ of the remaining $A_n$ fields that determines the components of the soft gluons that show up in the soft $S_n$ Wilson lines.

\subsection{Soft Operator from Tree Level Matching}
\label{sec:lipatov}

\begin{figure}[t!]
	%
	%
	\begin{center}
		\subfigure{
			\raisebox{-0.2cm}{a)\hspace{.8cm}} 
			\hspace{-0.5cm}
  \includegraphics[width=0.20\columnwidth]{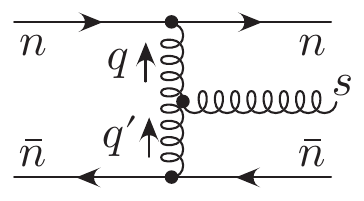}
  \hspace{0.2cm}
			\raisebox{0.4cm}{
  \includegraphics[width=0.20\columnwidth]{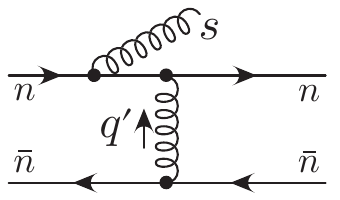}
}
  \hspace{0.2cm}
			\raisebox{0.45cm}{
   \includegraphics[width=0.20\columnwidth]{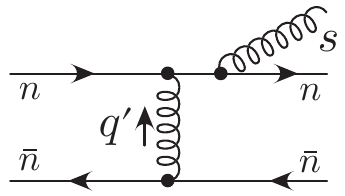}
}
		}
		\\[-20pt]
		\subfigure{
			\hspace{4.5cm}
			\raisebox{1cm}{
	\includegraphics[width=0.20\columnwidth]{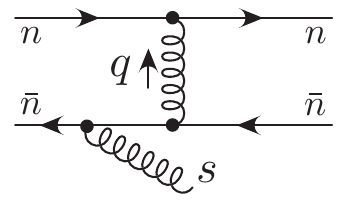}
}
			\hspace{0.2cm}
			\raisebox{1.1cm}{
	\includegraphics[width=0.20\columnwidth]{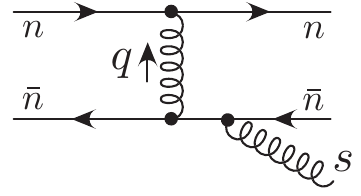}
}
		}
		\\[-30pt]
		%
		\subfigure{
			\hspace{0.9cm}
			\raisebox{-0.1cm}{b)\hspace{0.4cm}} 
	\includegraphics[width=0.2\columnwidth]{figs/GlaubOp_tree_qqqq_sGluon}  
			\hspace{0.2cm} \raisebox{0.6cm}{\Large $=$}  
			\raisebox{-0.1cm}{
	\includegraphics[width=0.22\columnwidth]{figs/Glaub_tree_qqqq_sGluon} 
			}
			\hspace{3.7cm}
		}
	\end{center}
	\vspace{-0.6cm}
	\caption{\setcaptionskip
		One Soft Gluon Matching for the Mid-Rapidity Operator in SCET appearing in quark-antiquark scattering. a) Full theory graphs. b)  EFT Mid-Rapidity Operator graph with one soft gluon, shown by two equivalent diagrams which exploit a localized or factorized notation. }
	\label{fig:match_Lipatov_1gluon}
	\setmainskip
\end{figure}

In the $O_{ns}$ and $O_{\bn s}$ operators there are two different rapidity sectors present, and the full structure of the operators is determined by the analysis of \secs{GlauberSCET}{treematch}, whereas for  the operator  $O_{ns\bn}$  there are three rapidity sectors, and we can have a non-trivial soft operator in addition to the $n$-collinear and $\bn$-collinear components. While the structure of the collinear part of these operators was derived through the analysis of the \sec{treematch}, the matching corrections considered so far have not probed the soft operator.  To do that we must consider soft gluon emission in the presence of $n$-$\bn$ forward scattering, which we will do in this section.

At the one soft gluon level, this emission is governed by the famous Lipatov vertex. This vertex is the combined Feynman rule for the emission of a soft gluon in the presence of the forward scattering of energetic quarks or gluons. For quark-antiquark scattering the corresponding full theory diagrams are shown in \fig{match_Lipatov_1gluon}a, and the contribution from SCET is in \fig{match_Lipatov_1gluon}b. At leading power the full theory diagrams give
\begin{align} \label{eq:fullLipatov1gluon}
  {\rm Fig.}~{\ref{fig:match_Lipatov_1gluon}a}
  &= i 
  \Big[ \bar u_n \frac{\bnslash}{2} T^A u_n \Big] 
  \Big[ \bar v_\bn \frac{\nslash}{2} \bar T^B v_\bn \Big] 
  \\ 
 &\quad
 \times\frac{8\pi\alpha_s}{ \vec q_\perp^{\,2} \vec q_\perp^{\:\prime 2} } \,
   i g f^{ABC} \Big[ q_\perp^\mu \! + q_\perp^{\prime\mu} \!
   - n\cdot q' \frac{\bn^\mu}{2} - \bn\cdot q \frac{n^\mu}{2} 
   - \frac{\bn^\mu \vec q_\perp^{\:\prime 2}}{\bn\cdot q}
   - \frac{n^\mu \vec q_\perp^{\:2}}{n\cdot q^\prime} \Big]
 , \nn
\end{align}
where the 3-gluon vertex graph gives the first four terms, and the soft gluon attachments to the quark lines give the last two. To obtain this result we have used $n\cdot k=-n\cdot q'$ and $\bn\cdot k = \bn\cdot q$. These momenta are ${\cal O}(\lambda)$ whereas $n\cdot q\sim \bn\cdot q'\sim \lambda^2$. Note that we have not used the gluon equations of motion to simplify the result obtained here.
Comparing \eq{fullLipatov1gluon} with the Feynman rule from the $O_{ns\bn}^{qq}$ operator 
of Fig.~{\ref{fig:match_Lipatov_1gluon}b}  (shown above in \fig{onegluon}), we see that the two precisely agree.  Thus, the one-gluon Feynman rule from the soft component of this Glauber operator, which is $O_s^{AB}$ in \eq{Os1}, directly generates the full Lipatov vertex without use of the equations of motion.

\begin{figure}[t!]
	%
	%
	\begin{center}
		\subfigure{
			\hspace{-.4cm}\raisebox{2cm}{a)\hspace{.5cm}} 
			\hspace{-0.7cm}
			\raisebox{0.4cm}{
	\includegraphics[width=0.23\columnwidth]{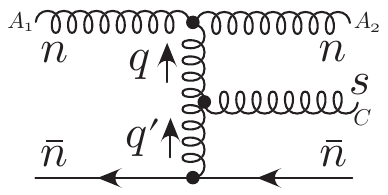}
}
 \hspace{0.cm}
			\raisebox{0.4cm}{
	\includegraphics[width=0.20\columnwidth]{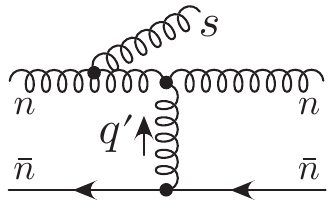}
}
 \hspace{0.2cm}
			\raisebox{0.45cm}{
	\includegraphics[width=0.20\columnwidth]{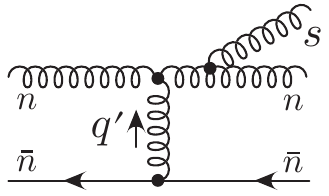}
}
			\hspace{0.4cm}
			\raisebox{2cm}{b)\hspace{0.cm}}  
			\raisebox{0.4cm}{
	\includegraphics[width=0.24\columnwidth]{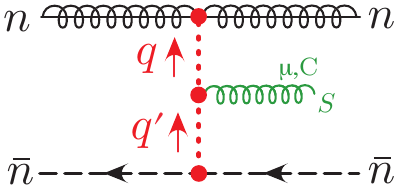}
 }
		}
		\\[-10pt]
		\subfigure{
			\hspace{-4.3cm}
			\raisebox{1.cm}{
	\includegraphics[width=0.20\columnwidth]{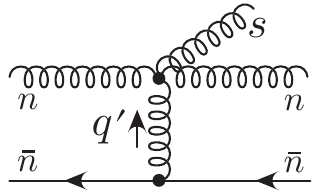}
}
			\hspace{0.2cm}
			\raisebox{1cm}{
	\includegraphics[width=0.20\columnwidth]{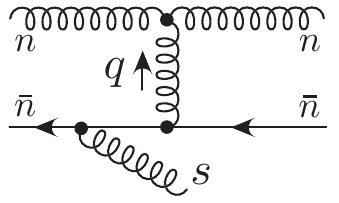}
}
			\hspace{0.2cm}
			\raisebox{1.1cm}{
	\includegraphics[width=0.20\columnwidth]{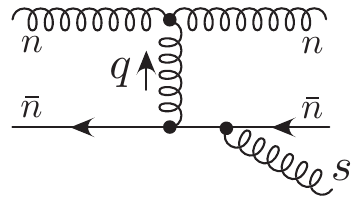}
}
		}
		%
	\end{center}
	\vspace{-1.6cm}
	\caption{\setcaptionskip
		One Soft Gluon Matching for the Mid-Rapidity Operator in SCET appearing in gluon-quark scattering. a) Six full theory graphs. b)  EFT graph from Mid-Rapidity Operator with one soft gluon of momentum $k$. }
	\label{fig:match_LipatovGluon_1gluon}
	\setmainskip
\end{figure}

The same matching calculation can be carried out when one or both of the collinear quark lines in \fig{match_Lipatov_1gluon} are replaced by collinear gluons. The corresponding graphs for the matching calculation with the top line replaced by an $n$-collinear gluon  are shown in \fig{match_LipatovGluon_1gluon}. Taking this $n$-collinear gluon to have $\perp$ polarization, the full theory result is again the same as the SCET Feynman rule \allowdisplaybreaks[0]
\begin{align}
  {\rm Fig.}~{\ref{fig:match_LipatovGluon_1gluon}a}
  &= i 
  \Big[ i f^{A_2 A_1 A} g_\perp^{\alpha\beta}\bn\cdot p_2 \Big] 
  \Big[ \bar v_\bn \frac{\nslash}{2} \bar T^B v_\bn \Big] 
  \\
& \quad
 \times \frac{8\pi\alpha_s}{ \vec q_\perp^{\,2} \vec q_\perp^{\:\prime 2} } \,
   i g f^{ABC} \Big[ q_\perp^\mu + q_\perp^{\prime\mu} 
   - n\cdot q' \frac{\bn^\mu}{2} - \bn\cdot q \frac{n^\mu}{2} 
   - \frac{\bn^\mu \vec q_\perp^{\:\prime 2}}{\bn\cdot q}
   - \frac{n^\mu \vec q_\perp^{\: 2}}{n\cdot q'} \Big]
 \nn\\ 
  & =   {\rm Fig.}~{\ref{fig:match_LipatovGluon_1gluon}b}
  \nn  \,,
\end{align}
where the SCET graph is given by the Feynman rule for $O_{ns\bn}^{gq}$. Here the graph with the 4-gluon vertex does not contribute at this order in the power expansion (it is suppressed by ${\cal O}(\lambda)$) and hence can be neglected. Once again the same universal soft operator $O_s^{AB}$ is responsible for the soft gluon Lipatov vertex in this scattering. A key ingredient in deriving this is the universal nature of the eikonal coupling for soft gluons. The same universal result also holds when a soft gluon is added to quark-gluon scattering with $O_{ns\bn}^{qg}$ and gluon-gluon scattering with $O_{ns\bn}^{gg}$. Essentially, all that changes between these calculations is the color generators for the collinear lines, which still obey the same algebra. \allowdisplaybreaks[1]


\subsection{The Basis of All Possible Soft Components in the $O_{ns\bn}^{ij}$ Glauber Operator}  \label{sec:basis}

In section~\ref{sec:GlauberSCET} we wrote down the  final form of the soft  piece  which sits between the collinear sectors of the Glauber operator, and in the last section we showed that it is consistent with the matching when including an external gluon. In this and the next section we give the complete derivation of the ${\cal O}(\lambda^2)$ mid-rapidity operator in \eq{Os1}. We write the general expansion of the soft piece of the mid-rapidity operator as
\begin{align} \label{eq:OsABbasis}
{\cal O}_s^{AB} = 8\pi\alpha_s \sum_i C_i\, O_i^{AB},
\end{align}
where $O_i^{AB}$ is the full set of operators which are consistent with soft gauge invariance, have mass dimension $2$, and scale as ${\cal O}(\lambda^2)$.  To build an operator in the adjoint matrix space we make use of pairs of adjoint Wilson lines ${\cal S}_n^T {\cal S}_\bn$ or ${\cal S}_\bn^T {\cal S}_n$, the adjoint matrix gluon building blocks $\widetilde {\cal B}_{S\perp}^{nAB}$ and $\widetilde {\cal B}_{S\perp}^{\bn AB}$, the soft gluon field strength made invariant with Wilson lines ${\cal S}_n^T  G_s^{\mu\nu} {\cal S}_\bn$ or ${\cal S}_\bn^T  G_s^{\mu\nu} {\cal S}_n$, plus $\cP_\perp$. Note that soft fermions $\psi_S^n \sim \lambda^{3/2}$ do not contribute to terms in the operator basis, since these quark fields must come in pairs, and hence soft fermion terms are at least ${\cal O}(\lambda^3)$. 

We can also reduce the list of possible operators in the basis using hermiticity, since ${\cal L}_G^{\rm II(0)}$ is hermitian. Examining $({\cal L}_G^{\rm II(0)})^\dagger$ we have
\begin{align}
  & \sum_{n,\bn} \sum_{i,j} \int\!\! 
   \frac{d^2q_\perp d^2q_\perp'}{q_\perp^2\: q_\perp^{\prime 2}}
  \Big[ {\cal O}_n^{iA}(q_\perp) {\cal O}_s^{AB}(q_\perp,q_\perp')    
    {\cal O}_\bn^{j B}(-q_\perp^\prime) \Big]^\dagger 
   \\
  &\quad = \sum_{n,\bn} \sum_{i,j} \int\!\! 
   \frac{d^2q_\perp d^2q_\perp'}{q_\perp^2\: q_\perp^{\prime 2}}\:
  {\cal O}_\bn^{j B}(q_\perp^\prime) 
  \Big[ {\cal O}_s^{AB}(q_\perp,q_\perp^\prime) \Big]^\dagger 
  {\cal O}_n^{i A}(-q_\perp)  
  \nn \\
  &\quad = \sum_{n,\bn} \sum_{i,j} \int\!\! 
   \frac{d^2q_\perp d^2q_\perp'}{q_\perp^2\: q_\perp^{\prime 2}}\:
  {\cal O}_n^{i A}(q_\perp) 
  \Big[ {\cal O}_s(q_\perp^\prime,q_\perp) \Big]^{\dagger AB}_{n\leftrightarrow \bn}
  {\cal O}_\bn^{j B}(-q_\perp^\prime)  
  \nn \,,
\end{align}
where to obtain the last line we swapped $n\leftrightarrow \bn$, $q_\perp\leftrightarrow q_\perp^\prime$, and $A\leftrightarrow B$. If we write factors of $q_\perp$ and $q_\perp^\prime$ using the operator ${\cal P}_\perp$ then swapping of these momenta is automatically accounted for in the hermitian conjugation, so we see that hermiticity requires that the soft operators satisfy 
\begin{align}  \label{eq:hermitian}
   O_i^\dagger \big|_{n\leftrightarrow \bn} = O_i  \,.
\end{align}
For simplicity we left off the adjoint color labels $AB$ for the operators, and will continue to do so below with the understanding that they are matrices in this space. Next, note that each term in the Lagrangian conserves $\perp$-momentum, so the total $\perp$-momentum is zero and we can freely let a $\cP_\perp$ operator act in either direction, $\cP_\perp^\mu =  \cP_\perp^{\dagger\mu}$. We use this freedom to eliminate all $\cP_\perp^\dagger$s.  Finally, whenever possible we will use the operator identities
\begin{align} \label{eq:opidentity}
\big[ {\cal P}_\perp^\mu( {\cal S}_n^T {\cal S}_{\bn}) \big]
  &= -g \widetilde {\cal B}_{S\perp}^{n\mu} ({\cal S}_{n}^T {\cal S}_{\bn})
  +({\cal S}_{n}^T {\cal S}_{\bn})g \widetilde {\cal B}_{S\perp}^{\bn\mu}
  \,, \\
\big[ {\cal P}_\perp^\mu( {\cal S}_\bn^T {\cal S}_n) \big]
  &= -g \widetilde {\cal B}_{S\perp}^{\bn\mu} ({\cal S}_{\bn}^T {\cal S}_{n})
  +({\cal S}_{\bn}^T {\cal S}_{n})g \widetilde {\cal B}_{S\perp}^{n\mu}
  \,, \nn
\end{align}
to eliminate $\cP_\perp$s in terms of $\widetilde {\cal B}_{S\perp}$s. Here the ${\cal P}_\perp^\mu$ acts only inside the square brackets and these relations follow immediately from the definition of ${\cal B}_{S\perp}^{n\mu}$ and ${\cal B}_{S\perp}^{\bn\mu}$ in \eq{opbbb}. 

In addition to the above constraints, we will also impose the restriction that at most one ${\cal S}_n$ Wilson line and one ${\cal S}_\bn$ Wilson line appear in the soft operators $O_i$.  Note that the non-local products $({\cal S}_n^T {\cal S}_\bn)$ and $({\cal S}_\bn^T {\cal S}_n)$ are dimensionless, have power counting $\lambda^0$, and are soft gauge invariant (up to the global transformation at $\infty$). If we did not adopt the restriction of having only one soft line of each type, then it would be possible to insert multiple products of these two-line structures, and the set of potential operators would be substantially larger.  The correct picture is that the ${\cal S}_n$ and ${\cal S}_\bn$ adjoint Wilson lines are generated by integrating out offshell lines attaching to the color octet $n$-collinear and $\bn$-collinear sector operators respectively, at the same time that we remove propagators associated with Glauber exchange.  Therefore the restriction we impose that only one of each type of soft Wilson line appears is very natural. In standard SCET applications to hard scattering, the presence of only one soft line for each collinear operator in a given representation follows immediately from the use of the BPS field redefinition~\cite{Bauer:2001yt} in \SCETa, with subsequent $\SCETa$ to $\SCETb$ matching by lowering the $p^2$ scale for the collinear fields to that of the soft fields.  This method becomes more complicated in the current case, because we are simultaneously removing offshell and Glauber propagators, and when doing the matching we must consider time order product graphs on the SCET side of the calculation rather than just the localized operator whose Wilson lines we want to determine. Based on the simple structure of the collinear operators, we do not expect more than one ${\cal S}_n$ or ${\cal S}_\bn$ to appear in the soft operators at any order in the $\alpha_s$ expansion.

We decompose the basis into operators with zero, one, or two $\widetilde {\cal B}_{S\perp}$ fields, or one $G_s^{\mu\nu}$ field, and consider these classes in turn. Without any $\widetilde {\cal B}_{S\perp}$ fields the minimal basis satisfying the constraints discussed above is
\begin{align} \label{eq:Ops0B}
 O_1 &= {\cal P}_\perp^\mu {\cal S}_{n}^T {\cal S}_{\bn}{\cal P}_{\perp \mu}
 \,,
& O_2 &= {\cal P}_\perp^\mu {\cal S}_{\bn}^T {\cal S}_{n} {\cal P}_{\perp \mu}  
  \,.
\end{align}
Both of these operators satisfy the hermiticity condition in \eq{hermitian} individually. Note that we do not include the operator $\cP_\perp^2$ because it does not contain any soft Wilson lines.\footnote{It turns out that if we did add this $\cP_\perp^2$ operator to our basis, that it would actually be ruled out by the full matching calculation discussed in the next section.} Also note that we have not included operators with ${\cal P}_\perp^2$ acting on soft Wilson lines since they can be eliminated using the identity
\begin{align}
  \cP_\perp^2 ( {\cal S}_n^T {\cal S}_\bn) +  ( {\cal S}_n^T {\cal S}_\bn)  \cP_\perp^2
  =  \big[ \cP_\perp^2  ( {\cal S}_n^T {\cal S}_\bn) \big] 
   + 2 \cP_\perp^\mu ( {\cal S}_n^T {\cal S}_\bn) \cP^\perp_\mu \,,
\end{align}
plus using \eq{opidentity} to eliminate $\big[ \cP_\perp^2  ( {\cal S}_n^T {\cal S}_\bn) \big]$ in terms of operators with at least one $\widetilde {\cal B}_{S\perp}$. In addition, we do not need to include $\big[ \cP_\perp^\mu ( {\cal S}_n^T {\cal S}_\bn) \big] \cP_\perp^\mu$ since $\big[ \cP_\perp^\mu ( {\cal S}_n^T {\cal S}_\bn) \big] \cP_\perp^\mu + \text{h.c.} = \cP_\perp^2 ( {\cal S}_n^T {\cal S}_\bn) +  ( {\cal S}_n^T {\cal S}_\bn) \cP_\perp^2 - 2 \cP_\perp^\mu ( {\cal S}_n^T {\cal S}_\bn) \cP^\perp_\mu$. Direct analogs of these relations are also used to eliminate operators when the Wilson lines are in the other order, $({\cal S}_\bn^T {\cal S}_n)$. 

Accounting for the fact that the fields $\widetilde {\cal B}_{S\perp}^n$ and $\widetilde {\cal B}_{S\perp}^\bn$ are Hermitian, the minimal basis with just a single $\widetilde {\cal B}_{S\perp}$ includes four operators,
\begin{align} \label{eq:Ops1B}
 O_3 & =
 {\cal P}_\perp \mcdot ( g \widetilde {\cal B}_{S\perp}^n) 
 ({\cal S}_n^T {\cal S}_{\bn})
 +({\cal S}_n^T {\cal S}_{\bn}) ( g \widetilde {\cal B}_{S\perp}^\bn)\mcdot {\cal P}_\perp
  \,,
 & O_4 & =
 {\cal P}_\perp \mcdot ( g \widetilde {\cal B}_{S\perp}^\bn) 
 ({\cal S}_\bn^T {\cal S}_{n})
 +({\cal S}_\bn^T {\cal S}_{n}) ( g \widetilde {\cal B}_{S\perp}^n)\mcdot {\cal P}_\perp
  \,,\nn\\
O_5 &={\cal P}^\perp_\mu ({\cal S}_n^T {\cal S}_{\bn}) 
   (g \widetilde {\cal B}_{S\perp}^{\bn\mu})
  + (g \widetilde {\cal B}_{S\perp}^{n\mu}) ({\cal S}_n^T {\cal S}_{\bn}) 
    {\cal P}^\perp_\mu
 \,,
 & O_6 &={\cal P}^\perp_\mu ({\cal S}_\bn^T {\cal S}_{n}) 
   (g \widetilde {\cal B}_{S\perp}^{n\mu})
  + (g \widetilde {\cal B}_{S\perp}^{\bn\mu}) ({\cal S}_\bn^T {\cal S}_{n}) 
    {\cal P}^\perp_\mu
 \,. 
\end{align}
These operators all satisfy the hermiticity requirement in \eq{hermitian} because they each have two terms.  To see that they satisfy the restriction of having only one soft Wilson line in each direction we note that
\begin{align}  \label{eq:cancelSs}
  ({\cal S}_n^T {\cal S}_{\bn}) ( g \widetilde {\cal B}_{S\perp}^{\bn\mu})  
  &=  ({\cal S}_n^T {\cal S}_{\bn}) \big[ {\cal S}_\bn^T i D_{s\perp}^\mu {\cal S}_\bn \big] 
   = \big[ {\cal S}_n^T i D_{s\perp}^\mu {\cal S}_\bn \big] 
  \,, \\
  ( g \widetilde {\cal B}_{S\perp}^n)  ({\cal S}_n^T {\cal S}_{\bn})
  &= \big[ {\cal S}_n^T (-i) \overleftarrow D_{s\perp}^\mu {\cal S}_n \big] 
   ({\cal S}_n^T {\cal S}_{\bn})
   = \big[ {\cal S}_n^T (-i) \overleftarrow D_{s\perp}^\mu {\cal S}_{\bn} \big]
  \,, \nn\\
   ({\cal S}_\bn^T {\cal S}_{n}) ( g \widetilde {\cal B}_{S\perp}^{n\mu})  
  &=  ({\cal S}_\bn^T {\cal S}_{n}) \big[ {\cal S}_n^T i D_{s\perp}^\mu {\cal S}_n \big] 
   = \big[ {\cal S}_\bn^T i D_{s\perp}^\mu {\cal S}_n \big] 
  \,, \nn \\
  ( g \widetilde {\cal B}_{S\perp}^\bn)  ({\cal S}_\bn^T {\cal S}_{n})
  &= \big[ {\cal S}_\bn^T (-i) \overleftarrow D_{s\perp}^\mu {\cal S}_\bn \big] 
   ({\cal S}_\bn^T {\cal S}_{n})
   = \big[ {\cal S}_\bn^T (-i) \overleftarrow D_{s\perp}^\mu {\cal S}_{n} \big]
  \,. \nn
\end{align}
Thus to satisfy the rule of only having a single ${\cal S}_n$ and ${\cal S}_\bn$ in our operators, we must group ${\cal B}_{S\perp}^{n}$ next to an ${\cal S}_n$ and ${\cal B}_{S\perp}^\bn$ next to an ${\cal S}_\bn$. For example, this rules out the operator ${\cal P}_\perp  \mcdot ( g \widetilde {\cal B}_{S\perp}^n) + ( g \widetilde {\cal B}_{S\perp}^\bn)\mcdot {\cal P}_\perp$ (it has two ${\cal S}_n$s in the first term, and two ${\cal S}_\bn$s in the second term). It also eliminates ${\cal P}^\perp_\mu ({\cal S}_n^T {\cal S}_{\bn}) (g \widetilde {\cal B}_{S\perp}^{n\mu}) + (g \widetilde {\cal B}_{S\perp}^{\bn\mu}) ({\cal S}_n^T {\cal S}_{\bn}) {\cal P}^\perp_\mu$ as an operator in the basis (it has four irreducible soft Wilson lines). An additional thing to note about \eq{Ops1B} is that $\cP_\perp$ factors are always on the outside.  We do not include additional operators with $\cP_\perp^\mu$ in the middle since they can always be eliminated in terms of operators in \eq{Ops1B}, plus terms with two $\widetilde {\cal B}_{S\perp}$s. For example,
\begin{align}
 ({\cal S}_n^T {\cal S}_{\bn}) \cP^\perp_\mu  (g \widetilde {\cal B}_{S\perp}^{\bn\mu})
 &= \cP^\perp_\mu ({\cal S}_n^T {\cal S}_{\bn})  (g \widetilde {\cal B}_{S\perp}^{\bn\mu})
  - \big[ \cP^\perp_\mu ({\cal S}_n^T {\cal S}_{\bn}) \big]
  (g \widetilde {\cal B}_{S\perp}^{\bn\mu}) \,,
\end{align}
where the last term can be reduced with \eq{opidentity}. Also combinations with $\big[ \cP_\perp^\mu ({\cal S}_n^T {\cal S}_\bn)\big]$ are directly removed with \eq{opidentity}, and combinations with $\big[ \cP^\perp_\mu (g \widetilde {\cal B}_{S\perp}^{\bn\mu})\big]$ are removed in terms of the other operators by integration by parts.

Next we turn to the operator basis with two $\widetilde {\cal B}_{S\perp}$s. The minimal basis here is given by just two operators
\begin{align} \label{eq:Ops2B}
O_{7}&= ( g \widetilde {\cal B}_{S\perp}^{n \mu}){\cal S}_{n}^T {\cal S}_{\bn} ( g \widetilde {\cal B}_{S\perp \mu}^{\bn}) \,,
 & O_{8} &= ( g \widetilde {\cal B}_{S\perp}^{\bn \mu}){\cal S}_{\bn}^T {\cal S}_{n} ( g \widetilde {\cal B}_{S\perp \mu}^{n})
  \,.
\end{align}
These operators each satisfy \eq{hermitian} alone. Due to the grouping of soft Wilson lines next to appropriate ${\cal B}_{S\perp}$s in \eq{Ops2B}, the operators again have only one soft Wilson line in each direction once we use \eq{cancelSs}. This restriction eliminates operators such as $g {\cal B}_{S\perp}^{n } \cdot  g {\cal B}_{S\perp}^{\bn}$ (4 lines total) and
$g {\cal B}_{S\perp}^{n } \cdot  g {\cal B}_{S\perp}^{n}+g {\cal B}_{S\perp}^{\bn } \cdot  g {\cal B}_{S\perp}^{\bn}$ (two ${\cal S}_n$ lines in the first term, two ${\cal S}_\bn$ lines in the second term). It also eliminates operators like $( g {\cal B}_{S\perp}^{n \mu}) (S_{\bn}^T S_{n}) ( g {\cal B}_{S\perp \mu}^{\bn})$ and $( g {\cal B}_{S\perp}^{n \mu}) (S_{\bn}^T S_{n}) ( g {\cal B}_{S\perp \mu}^{n}) + ( g {\cal B}_{S\perp}^{\bn \mu}) (S_{\bn}^T S_{n}) ( g {\cal B}_{S\perp \mu}^{\bn})$.

Finally we have the operator with a single soft gluon field strength, of which there are two
\begin{align} \label{eq:Ops1G}
O_{9} &= {\cal S}_{n}^T n_\mu \bn_\nu( ig \widetilde G_s^{\mu \nu}) {\cal S}_\bn 
  \,,
 & O_{10} &= {\cal S}_{\bn}^T n_\mu \bn_\nu( ig \widetilde G_s^{\mu \nu}) {\cal S}_n 
  \,,
\end{align}
In principle this operator could be eliminated in terms of ${\cal B}_{S\perp}^n$, ${\cal B}_{S\perp}^{bn}$, $\cP_\perp$, $\psi_S^n$, and $\psi_S^\bn$ fields using the soft gluon equations of motion. However doing so would introduce non-local factors of $1/in\cdot\partial_s$ and $1/i\bn\cdot\partial_s$ which we have not allowed in our construction. Therefore we must keep these two field strength operators.

All together the 10 operators in Eqs.~(\ref{eq:Ops0B},\ref{eq:Ops1B},\ref{eq:Ops2B},\ref{eq:Ops1G}) give a complete basis for the soft operator $O_s^{AB}$. Note that the odd and even operators in the basis are related by $O_{i+1} = O_i \big|_{n\leftrightarrow \bn}$, and that this differs from the hermiticity condition in \eq{hermitian}. In the next section we consider the constraints obtained by matching with up to two soft external gluons in order to fix the corresponding coefficients $C_{1,\ldots,10}$ in \eq{OsABbasis}.

\subsection{All Orders Soft Operator by Matching with up to Two Soft Gluons}
\label{sec:2sgluon}
 
Here we consider the basis of operators $O_{1,\ldots,10}$ determined above in Eqs.(\ref{eq:Ops0B},\ref{eq:Ops1B},\ref{eq:Ops2B},\ref{eq:Ops1G}), 
\begin{align}
 O_1 &= {\cal P}_\perp^\mu {\cal S}_{n}^T {\cal S}_{\bn}{\cal P}_{\perp \mu}
 ,
& O_2 &= {\cal P}_\perp^\mu {\cal S}_{\bn}^T {\cal S}_{n} {\cal P}_{\perp \mu}  
 ,
  \\
 O_3 & =
 {\cal P}_\perp \mcdot ( g \widetilde {\cal B}_{S\perp}^n) 
 ({\cal S}_n^T {\cal S}_{\bn}) \plus
 ({\cal S}_n^T {\cal S}_{\bn}) ( g \widetilde {\cal B}_{S\perp}^\bn)
  \mcdot {\cal P}_\perp 
 ,
& O_4 & =
 {\cal P}_\perp \mcdot ( g \widetilde {\cal B}_{S\perp}^\bn) 
 ({\cal S}_\bn^T {\cal S}_{n}) \plus
 ({\cal S}_\bn^T {\cal S}_{n}) ( g \widetilde {\cal B}_{S\perp}^n)
  \mcdot {\cal P}_\perp
 , \nn\\
 O_5 &={\cal P}^\perp_\mu ({\cal S}_n^T {\cal S}_{\bn}) 
   (g \widetilde {\cal B}_{S\perp}^{\bn\mu})  \plus
  (g \widetilde {\cal B}_{S\perp}^{n\mu}) ({\cal S}_n^T {\cal S}_{\bn}) 
    {\cal P}^\perp_\mu
  ,
& O_6 &={\cal P}^\perp_\mu ({\cal S}_\bn^T {\cal S}_{n}) 
   (g \widetilde {\cal B}_{S\perp}^{n\mu}) \plus
  (g \widetilde {\cal B}_{S\perp}^{\bn\mu}) ({\cal S}_\bn^T {\cal S}_{n}) 
    {\cal P}^\perp_\mu
  , \nn\\
  O_{7} &= ( g \widetilde {\cal B}_{S\perp}^{n \mu}){\cal S}_{n}^T {\cal S}_{\bn} ( g \widetilde {\cal B}_{S\perp \mu}^{\bn}),
 & O_{8} &= ( g \widetilde {\cal B}_{S\perp}^{\bn \mu}) {\cal S}_{\bn}^T {\cal S}_{n} ( g \widetilde {\cal B}_{S\perp \mu}^{n})
 , \nn\\
O_{9} &= {\cal S}_{n}^T n_\mu \bn_\nu( ig \widetilde G_s^{\mu \nu}) {\cal S}_\bn 
   ,
  & O_{10} &= {\cal S}_{\bn}^T n_\mu \bn_\nu( ig \widetilde G_s^{\mu \nu}) {\cal S}_n 
   , \nn
\end{align}
and determine their corresponding Wilson coefficients through matching calculations involving 0, 1, or 2 soft gluons. For this analysis it suffices to consider quarks for the $n$-collinear and $\bn$-collinear external lines. If one or both of the forward collinear external lines are taken to be gluons then the same result will obtained. This equality was discussed for one soft gluon in \sec{lipatov}, and is also true for two soft gluons, essentially resulting from the presence of the eikonal approximation that occurs for soft gluons attached to collinear lines, and the universality of the soft attachments to the exchanged gluon which has Glauber momentum scaling.

With zero soft gluons the resulting amplitude was given in \eq{treennbresult}, and requires that the soft operators $\sum_i C_i O_i$ reduce to $\cP_\perp^2 \delta^{AB}$ when no gluons are present.  Only $O_1$ and $O_2$ have this property, so the constraint from the zero soft gluon emission amplitude is
\begin{align} \label{eq:con0sgluon}
C_1+C_2=1  \,.
\end{align}

For the matching with one external soft gluon of incoming momentum $k$ we consider the five full theory  diagrams in figure \fig{match_Lipatov_1gluon}a, and consider all possible projections of the gluon's polarization with respect to $\{n,\bn,\perp\}$, without exploiting the equations of motion, which gives \eq{fullLipatov1gluon}. While we have already verified in \sec{lipatov} that the combination of operators given in \eq{Os1} reproduces this 1 soft gluon result, we have not yet proven that it is the unique combination which can do so. Using momentum conservation $k=q-q'$,  the one soft gluon matching generates the structures $\{ q_\perp^{\prime\mu}, n^\mu q_\perp \cdot q_\perp^\prime/n\cdot q', n^\mu q_\perp^{2}/n\cdot q', n^\mu q_\perp^{\prime 2}/n\cdot q', n^\mu \bn\cdot q\}$  which give the following five constraints on the operators  $O_{1,\ldots,10}$ in our basis,  
\begin{align} \label{eq:con1sgluon}
C_3+C_4+C_5+C_6&=-1 \,, \\
C_1-C_2+C_3-C_4-C_5+C_6 &=0 \,, \nn \\
-C_3-C_6  &= +1\,, \nn \\
C_4 + C_5 &= 0   \,, \nn \\
C_9+ C_{10} &= -\frac12 \,,\nn
\end{align}
respectively. \eq{con1sgluon} reproduces the full theory amplitude for one soft gluon without using the equation of motion, a fact that will come in handy when we consider the two soft gluon matching below. Other momentum structures with $q_\perp^{\mu}$ or $\bn^\mu$ are related to these by the hermiticity condition in \eq{hermitian}. Simplifying \eq{con1sgluon} and combining it with \eq{con0sgluon} gives
\begin{align} \label{eq:simpcon1sgluon}
  C_1 & = 1 \,,
 & C_2 & = 0 \,,
 & C_3+C_6&=-1 \,, 
 & C_4+C_5 &=0 \,,
 & C_9 + C_{10} &=-\frac12  \,.
\end{align}
Since $C_1=1$ and $C_2=0$, we see that between these two operators, the one with $({\cal S}_n^T {\cal S}_\bn)$ contributes, whereas the one with $({\cal S}_\bn^T {\cal S}_n)$ does not. We will see this pattern continue below for the operators with $\widetilde {\cal B}_{s\perp}$s.

To generate the  remaining constraints we carry out the matching with two external soft gluons of incoming momentum $k_1$ and $k_2$. Matching with two soft gluons goes beyond the level of the Lipatov vertex, and indeed unlike the Lipatov vertex, the soft operator in SCET has Feynman rules with one or two soft $\perp$ gluons, and any number of soft $n\cdot A_s$ and $\bn\cdot A_s$ gluons.   It  is in fact necessary to have at least two soft gluons in order for the operators $O_{7}$ and $O_8$ with two $\widetilde {\cal B}_{s\perp}$ fields to contribute. Since operators with three $\widetilde {\cal B}_{s\perp}$s cannot appear in the basis for $O_s^{AB}$ (due to the dimensionality and power counting constraints), the matching with 3 or more soft gluons is not necessary to determine the coefficients of the operators in the basis. Feynman rules for three or more soft gluons are determined by symmetry once those up to two soft gluons are fixed.

\begin{figure}[t!]
%
%
\subfigure{
\raisebox{1.7cm}{a)\hspace{.0cm}} 
\hspace{-0.55cm}
\includegraphics[width=0.17\columnwidth]{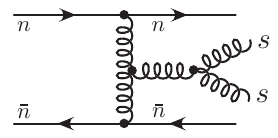}
\hspace{0.1cm}
\raisebox{-0.2cm}{
\includegraphics[width=0.22\columnwidth]{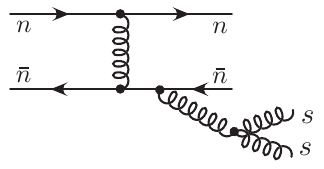}
}
\hspace{-0.35cm}
\raisebox{-0.2cm}{
\includegraphics[width=0.17\columnwidth]{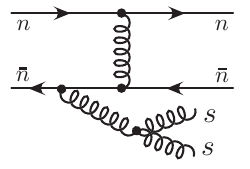}
}
\hspace{0.05cm}
\raisebox{0.cm}{
\includegraphics[width=0.22\columnwidth]{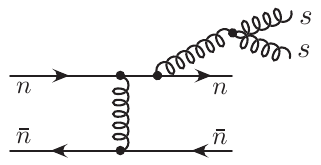}
}
\hspace{-0.35cm}
\raisebox{0.cm}{
\includegraphics[width=0.17\columnwidth]{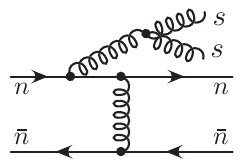}
}
}
\\[-5pt]
\subfigure{
\hspace{-0.15cm}
\raisebox{0.cm}{
\includegraphics[width=0.17\columnwidth]{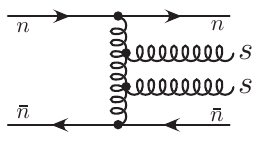}
}
\hspace{0.1cm}
\raisebox{0.cm}{
\includegraphics[width=0.17\columnwidth]{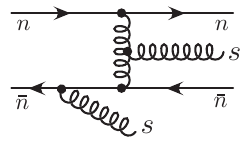}
}
\hspace{0.1cm}
\raisebox{0.cm}{
\includegraphics[width=0.17\columnwidth]{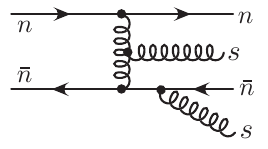}
}
\hspace{0.1cm}
\raisebox{0.cm}{
\includegraphics[width=0.17\columnwidth]{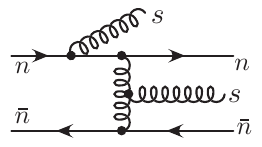}
}
\hspace{0.1cm}
\raisebox{0.cm}{
\includegraphics[width=0.17\columnwidth]{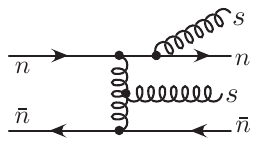}
}
}
\\[-5pt]
\subfigure{
\hspace{-0.15cm}
\raisebox{0.45cm}{
\includegraphics[width=0.17\columnwidth]{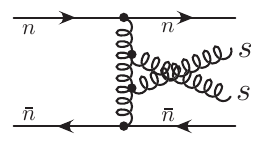}
}
\hspace{0.1cm}
\raisebox{0.45cm}{
\includegraphics[width=0.15\columnwidth]{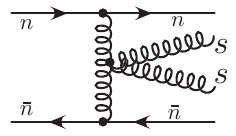}
}
\hspace{0.1cm}
\raisebox{0.02cm}{
\includegraphics[width=0.13\columnwidth]{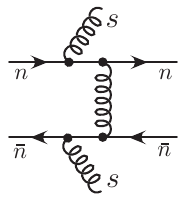}
}
\hspace{0.1cm}
\raisebox{0.cm}{
\includegraphics[width=0.13\columnwidth]{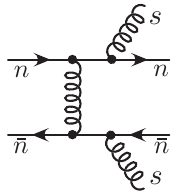}
}
\hspace{0.1cm}
\raisebox{0.15cm}{
\includegraphics[width=0.14\columnwidth]{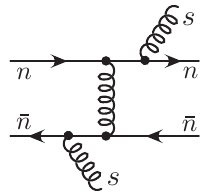}
}
\hspace{0.1cm}
\raisebox{0.15cm}{
\includegraphics[width=0.14\columnwidth]{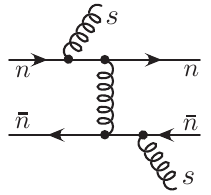}
}
}
\\[-5pt]
\subfigure{
\hspace{-0.1cm}
\raisebox{0.45cm}{
\includegraphics[width=0.13\columnwidth]{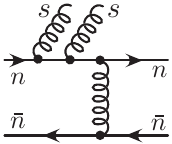}
}
\hspace{0.15cm}
\raisebox{0.45cm}{
\includegraphics[width=0.135\columnwidth]{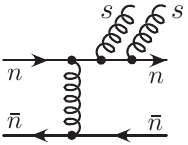}
}
\hspace{0.15cm}
\raisebox{0.45cm}{
\includegraphics[width=0.125\columnwidth]{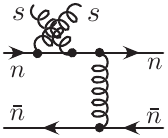}
}
\hspace{0.15cm}
\raisebox{0.45cm}{
\includegraphics[width=0.13\columnwidth]{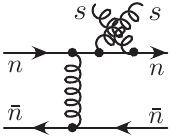}
}
\hspace{0.15cm}
\raisebox{0.45cm}{
\includegraphics[width=0.13\columnwidth]{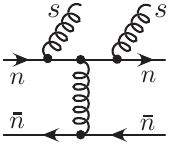}
}
\hspace{0.15cm}
\raisebox{0.45cm}{
\includegraphics[width=0.13\columnwidth]{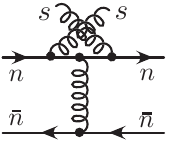}
}
}
\\[-10pt]
\subfigure{
\hspace{-0.1cm}
\raisebox{0.45cm}{
\includegraphics[width=0.13\columnwidth]{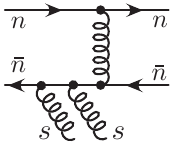}
}
\hspace{0.15cm}
\raisebox{0.45cm}{
\includegraphics[width=0.135\columnwidth]{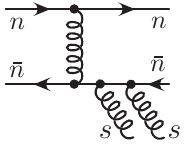}
}
\hspace{0.15cm}
\raisebox{0.45cm}{
\includegraphics[width=0.125\columnwidth]{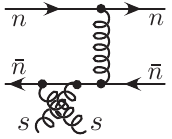}
}
\hspace{0.15cm}
\raisebox{0.45cm}{
\includegraphics[width=0.13\columnwidth]{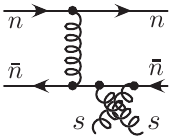}
}
\hspace{0.15cm}
\raisebox{0.45cm}{
\includegraphics[width=0.13\columnwidth]{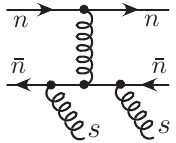}
}
\hspace{0.15cm}
\raisebox{0.45cm}{
\includegraphics[width=0.13\columnwidth]{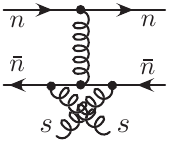}
}
}
\\[-5pt]
%
\subfigure{
\raisebox{1.8cm}{b)\hspace{0.2cm}}  
\raisebox{-0.1cm}{
\includegraphics[width=0.22\columnwidth]{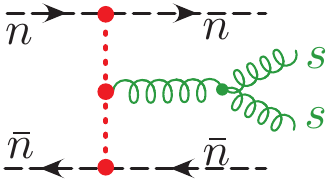}
 }
\hspace{0.15cm}
\raisebox{-0.3cm}{
\includegraphics[width=0.17\columnwidth]{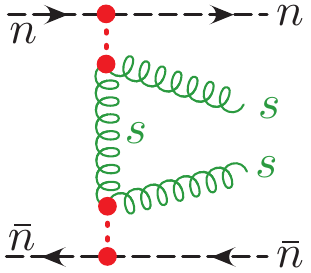}
 }
\hspace{0.15cm}
\raisebox{-0.3cm}{
\includegraphics[width=0.17\columnwidth]{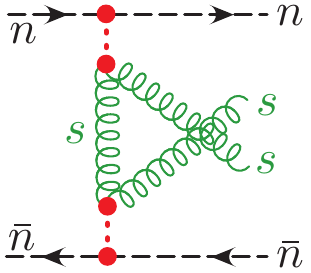}
 }
\hspace{0.15cm}
\raisebox{-0.1cm}{
\includegraphics[width=0.2\columnwidth]{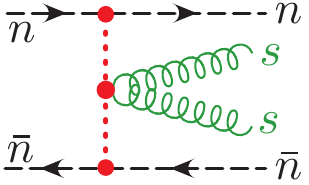}
 }
\hspace{3.7cm}
}
\caption{\setcaptionskip
Two Soft Gluon Matching for the Mid-Rapidity Operator. a) Full theory graphs with scaling for external particles labeled. b)  SCET graphs involving the mid-rapidity Operator and two soft gluons. The first three graphs are T-products while the last is the direct Mid-Rapidity Operator two gluon term.}
\label{fig:match_Lipatov_2gluon}
\setmainskip
\end{figure}

For $n$-$\bn$ quark-antiquark forward scattering with two soft gluons, there are 28 diagrams in the full theory, shown in \fig{match_Lipatov_2gluon}a. 
Since we were able to match with one soft gluon without recourse to the equations of motion, we know that we will automatically reproduce all the graphs in the first row in \fig{match_Lipatov_2gluon}a via the first graph shown in \fig{match_Lipatov_2gluon}b, which is the time ordered product of the one soft gluon $O_s^{AB}$ operator and a triple gluon vertex. The remaining 23 full theory diagrams, from rows other than the first one in \fig{match_Lipatov_2gluon}a, all contribute to the matching in a non-trivial fashion. On the SCET side  there are contributions from the second and third time ordered product graphs in  \fig{match_Lipatov_2gluon}b, which must be included along with the direct $O_s^{AB}$ contribution shown by the fourth graph in \fig{match_Lipatov_2gluon}b, in order to reproduce the full theory.
One would expect that the SCET time ordered product graphs will reproduce more non-local terms in the full theory, in particular those involving full soft propagator denominators $1/k^2$, as opposed to the expected non-locality of the Glauber potential, $1/k_\perp^2$. In fact, after subtracting these SCET T-product graphs from the 23 full theory diagrams, the result remains non-local. However after using the equations of motion in the form $k_1^2=0$, $k_2^2=0$, and
\begin{align}  \label{eq:eom2}
 &\vec k_{1\perp} \mcdot \vec \epsilon_\perp(k_1)
  = \frac{\bn \mcdot k_1\, n \mcdot \epsilon(k_1)+ n \mcdot k_1\,\bn \mcdot \epsilon(k_1)}{2}\,,
 & \vec k_{2\perp} \mcdot \vec \epsilon_\perp(k_2) &
  = \frac{\bn \mcdot k_2\, n \mcdot \epsilon(k_2)+ n \mcdot k_2\,\bn \mcdot \epsilon(k_2)}{2}\,,
\end{align}
and momentum conservation, $k_1+k_2=q-q'$, the results all localize into the form of the Glauber potentials, and can be reproduced by the terms in our basis for $O_s^{AB}$. Notice that using \eq{eom2} has the effect of moving some contributions from $\perp$-$\perp$ to the $\{n$-$\perp$, $\perp$-$n$, $\bn$-$\perp$, $\perp$-$\bn$, $n$-$\bn$, $\bn$-$n$, $n$-$n$, $\bn$-$\bn\}$ final state polarizations, and some from $n$-$\perp$, $\perp$-$n$, $\bn$-$\perp$, and $\perp$-$\bn$ into $\{n$-$\bn$, $\bn$-$n$, $n$-$n$, $\bn$-$\bn\}$.  The constraints on the operator coefficients follow from matching polarizations, independent kinematic factors, and color structures, of which there are two $f^{C_1AE}f^{C_2BE}$ and $f^{C_2AE}f^{C_1BE}$
necessitated by Bose symmetry (after using the Jacobi identity).

The matching of the $\perp$-$\perp$ final states is only sensitive to $O_{7,9}$ and $O_{8,10}$ and receives contributions from the first diagram in the second row and the first two diagrams
in the third row of \fig{match_Lipatov_2gluon}a. The last three SCET graphs all contribute. For this polarization choice the T-product graphs with a propagating soft gluon reproduce the graphs with the 3-gluon vertices in the full theory (after using the equations of motion), and the matching result comes from the four gluon vertex graph, giving
\begin{align}
C_7+C_ 8=-1.
\end{align}

The $n$-$\perp$ 2-gluon state receives contributions from four of the full theory graphs, as well as from equation of motion terms from $\perp$-$\perp$. This gives four constraints, from the coefficients of the four structures $\frac{n^{\mu_1}}{n\cdot k_1} \big\{ q_\perp^{\mu_2} f^{C_1 AE}f^{C_2BE}$, $q_\perp^{\prime\mu_2} f^{C_1 AE}f^{C_2BE}$, $q_\perp^{\mu_2} f^{C_2 AE}f^{C_1BE}$, $q_\perp^{\prime\mu_2} f^{C_2 AE}f^{C_1BE}\big\}$,
\begin{align}  \label{eq:npresults}
C_3+C_5- C_7 &=0 \,,  \\
C_3+C_5+C_7 &=-2 \,,\nn \\
-C_4 - C_6 - C_8 &=0 \,, \nn \\
-C_4 - C_6 +C_8 &=0  \,. \nn
\end{align}
These same constraints also cause the $\bn$-$\perp$ polarization choice to agree between the full theory and SCET, and by symmetry the $\perp$-$n$ and $\perp$-$\bn$ polarizations as well. Simplifying these results we conclude that 
\begin{align} \label{np}
C_3+ C_5 &=-1 \,,
 & C_4+C_6&= 0 \,,
 & C_7 &= -1 \,,
 & C_8 & = 0 \,,
\end{align}
and if we combine these results with those from \eq{simpcon1sgluon} we get
\begin{align} \label{eq:simpcon2sgluon1}
  C_1 &= 1 \,,
 & C_2 & =0 \,,
 & C_3 + C_5 & =-1\,,
 & C_4 & = -C_5 = -C_6 \,,
  \\
  C_7 & =-1 \,,
 & C_8 & =- 0\,,
 & C_9 + C_{10} &= -\frac12 
  \,. \nn
\end{align}
Since not all coefficients are fixed we must proceed to compare additional polarization projections.

The constraints for the $n$-$\bn$ polarization choice are little more tricky because there are 11 full theory diagrams that contribute, and we get contributions from using the equations of motion in the results for $\perp$-$\perp$, $n$-$\perp$, and $\perp$-$\bn$. Also, there are many more kinematic variables involved and thus many more constraints, and one must pick a minimal basis of momentum structures after using the momentum conservation and the equations of motion. We find 14 constraints that need to be satisfied, but 10 of them provide only redundant information. For our choice of independent structures the four that provide new information come from the structures $k_{1\perp}\cdot k_{2\perp} f^{C_1 AE} f^{C_2 BE}$, $k_{1\perp}\cdot k_{2\perp} f^{C_2 AE} f^{C_1 BE}$, $q_\perp^{\prime 2} f^{C_1 AE} f^{C_2 BE}$, and $q_\perp^{\prime 2} f^{C_2 AE} f^{C_1 BE}$, giving respectively
\begin{align}
 C_9 &=- \frac12  \\
 C_{10} &= 0 \,,\nn\\
 C_3 + \frac12 C_7 - C_9 &= -1 \,,\nn\\
 -C_6 + \frac12 C_8 +C_{10} & = 0  \,. \nn
\end{align}
Combining these results with \eq{simpcon2sgluon1} yields a unique solution for all the coefficients, giving our final answer
\begin{align} \label{eq:Ciresult}
 & \boxed{ C_2=C_4=C_5=C_6=C_8=C_{10}=0} 
  \,,\\
& \boxed{ C_1 =-C_3 = -C_7= +1 \,, \qquad C_9 = -\frac12 }
  \,. \nn
\end{align}
Thus we see that all operators in the basis involving $(S_\bn^T S_n)$ have zero coefficients, while all operators with $(S_n^T S_\bn)$ except $O_5$ have nonzero coefficients. As a consistency check, we have verified that the full theory results for the remaining polarizations $\{n$-$\bn$, $n$-$n$, $\bn$-$\bn\}$ are also correctly reproduced. 

Putting the results in \eq{Ciresult} back into \eq{OsABbasis} the final result is
\begin{align} 
{\cal O}_s^{BC} 
& ={8\pi\alpha_s  }
\bigg\{
\cP_\perp^\mu {\cal S}_n^T {\cal S}_\bn  \cP_{\perp\mu}
- \cP^\perp_\mu g \widetilde {\cal B}_{S\perp}^{n\mu}  {\cal S}_n^T  {\cal S}_\bn   
-  {\cal S}_n^T  {\cal S}_\bn  g \widetilde {\cal B}_{S\perp}^{\bn\mu} \cP^\perp_{\mu}  
-  g \widetilde {\cal B}_{S\perp}^{n\mu}  {\cal S}_n^T  {\cal S}_\bn g \widetilde {\cal B}_{S\perp\mu}^{\bn}
\nn\\
&\qquad\qquad
-\frac{n_\mu \bn_\nu}{2} {\cal S}_n^T   ig \widetilde {G}_s^{\mu\nu} {\cal S}_\bn 
\bigg\}^{BC} 
\,.
\end{align}
This is precisely the result for $O_s^{AB}$ that we quoted earlier in \eq{Os1}.

\section{One Loop Matching Calculations}
\label{sec:loopmatch}

In \sec{loop2match} we do a complete one-loop forward scattering matching calculation between the full theory and \SCETb theory with Glauber operators.
The structure of virtual rapidity divergences, and their relation to gluon Reggeization is derived in \sec{regge}.  The one-loop matching calculation is also carried out for \SCETa, and is presented in \sec{loop1match}.

\subsection{One Loop Matching in \SCETb}
\label{sec:loop2match}
 
In this section we carry out the one-loop matching for forward scattering, comparing graphs in the full theory and in SCET.  The goals of this analysis are to check the completeness of our EFT description by checking that all infrared (IR) divergences in the full theory are correctly reproduced by SCET, to understand the structure of ultraviolet and rapidity divergences that appear in the SCET diagrams, and to characterize the type of corrections that can be generated at the hard scale by matching.   

\begin{figure}[t!]
	%
	%
%
\begin{center}
  \raisebox{2cm}{
  \hspace{-0.6cm}
  a)\hspace{3.5cm} 
  b)\hspace{2.9cm} 
  c)\hspace{2.8cm} 
  d)\hspace{2.9cm} 
  e)\hspace{3cm} 
   } \\[-50pt]
\hspace{-.83cm}
\includegraphics[width=0.23\columnwidth]{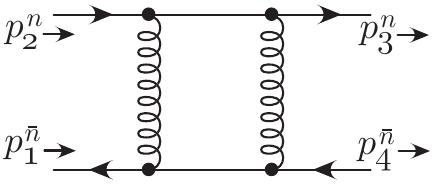}
\hspace{0.2cm}
\includegraphics[width=0.18\columnwidth]{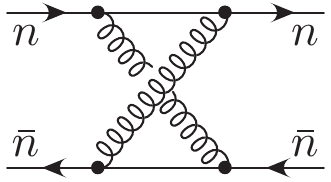}
\hspace{0.2cm}
\includegraphics[width=0.18\columnwidth]{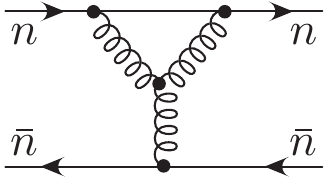}
\hspace{0.2cm}
\includegraphics[width=0.18\columnwidth]{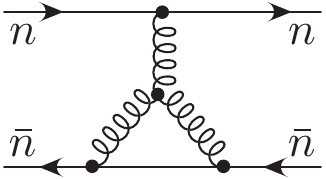}
\hspace{0.cm}
\raisebox{0.5cm}{      
\includegraphics[width=0.16\columnwidth]{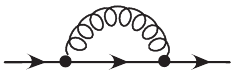}
}
  \\[15pt]
  \raisebox{2cm}{
  \hspace{-0.6cm}
  f)\hspace{3.1cm} 
  g)\hspace{2.95cm} 
  h)\hspace{3cm} 
  i)\hspace{2.9cm} 
  j)\hspace{3cm} 
   } \\[-53pt]
\hspace{-0.55cm}
\includegraphics[width=0.18\columnwidth]{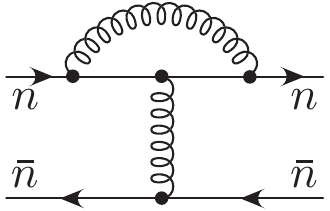}
\hspace{0.2cm}
\raisebox{-0.1cm}{
\includegraphics[width=0.18\columnwidth]{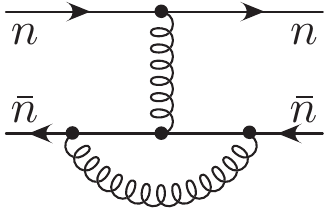}
}
\hspace{0.2cm}
\includegraphics[width=0.18\columnwidth]{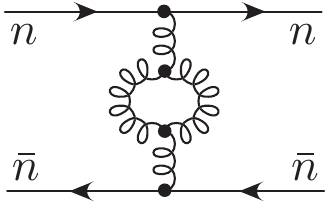}
\hspace{0.2cm}
\includegraphics[width=0.18\columnwidth]{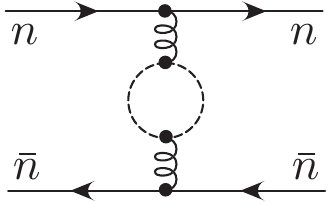}
\hspace{0.2cm}
\includegraphics[width=0.18\columnwidth]{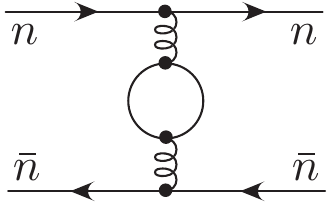}
\hspace{0.2cm}
\end{center}
   \vspace{-0.9cm}
	\caption{\setcaptionskip
		Full theory graphs for the matching calculation of quark-antiquark forward scattering at one-loop. The results for the graphs are expanded with their external momenta as labeled.}
	\label{fig:full_oneloop_matching}
	\setmainskip
\end{figure}

To be definite, we will consider quark-antiquark forward scattering. (This is directly related to quark-quark forward scattering since the extra exchange diagrams for the quark-quark case go as $1/u$ rather than $1/t$, and hence are power suppressed.) The external momentum routing we use is the same as shown labeled on \fig{glaub_tree}, which we repeat for convenience on the first graph of \fig{full_oneloop_matching}. The large forward momenta are conserved, $\bn\cdot p_2=\bn\cdot p_3$ and $n\cdot p_1=n\cdot p_4$, and the large Mandelstam invariant $s=n\cdot p_1\: \bn\cdot p_2=n\cdot p_4\: \bn\cdot p_1$ to leading power. The exchanged momentum is given by the much smaller Mandelstam invariant $t = q_\perp^2 = -\vec q^{\, 2}_\perp$ where $q=p_3-p_2=p_1-p_4$, and we take $p_1^\perp = -p_2^\perp=p_3^\perp = -p_4^\perp = q_\perp/2$. 

To regulate IR divergences in the full theory in a manner that can also be implemented in SCET$_{\rm II}$, we include a small gluon mass $m$. For SCET$_{\rm II}$ the mass $m$ is included for both soft and collinear gluons in loops, as well as for the Glauber potential from $1/\cP_\perp^2$ terms via $1/\vec k_\perp^{\,2} \to 1/(\vec k_\perp^{\,2}+m^2)$.   Since we take $m\to 0$ whenever possible, this does not cause any problems with gauge invariance in this one-loop calculation (for example we set $m= 0$ from the start for the vacuum polarization graphs). The full theory is UV finite after coupling renormalization, and we make use of dimensional regularization with $d=4-2\epsilon$ to regulate divergences in individual diagrams.   For SCET$_{\rm II}$ dimensional regularization with $d=4-2\epsilon$ will be used with factorization scale $\mu$ in $\MSbar$ to regulate invariant mass divergences. 

We also use a rapidity regulator~\cite{Chiu:2012ir} to regulate additional divergences that are associated with distinguishing soft and collinear modes~\cite{Manohar:2006nz}. These divergence arise as a consequence of the fact that the soft and collinear fields have the same virtuality and to distinguish them we must choose a rapidity factorization scale $\nu$.  This regulator is implemented in the manner discussed in \sec{rapidityregulator}, and shows up in both Glauber, soft, and collinear loops.  The limit $\eta\to 0$ is always considered first, with the rapidity renormalization carried out at finite $\epsilon$, and then the limit $\epsilon\to 0$ is taken. Graphs without rapidity divergences or sensitivity will give the same answer whether one sets $\eta=0$ before or after the loop integration. The graphs in SCET have subtractions which ensure there is no double counting, and for the calculations here this corresponds to using \eq{0bins}. At one-loop we will see that graphs with rapidity divergences only have scaleless 0-bin subtractions. However, there are graphs without rapidity divergences for which the 0-bin subtractions are not scaleless integrals and are crucial for avoiding double counting.  In the presence of Glauber gluons, the appropriate 0-bin subtractions for soft and collinear one-loop graphs are given in \eq{0bins}. In the \SCETb calculations we are considering in this section, there are no 0-bin subtractions for the Glauber loop graphs.

\subsubsection{Full Theory Graphs}

Consider first the full QCD graphs shown in \fig{full_oneloop_matching} which we number from a) to j). These graphs are computed exactly, and then the results are expanded in the EFT limit with $|t|\ll s$. There are two additional box-type graphs obtained by rotating \fig{full_oneloop_matching}a,b by $90^\circ$, but neither of these graphs contributes at leading power in this limit. The proper cut structure is obtained with $s=s+i0$ and $t=t+i0$, where we note that for our kinematics $s>0$ and $t<0$. The $i0$ will be implied in the rest of the paper. The group theory and spinor factors come in one of four combinations which we denote
\begin{align} \label{eq:spinors}
  {\cal S}_1^{n\bn} &= -\Big[ \bar u_n T^A T^B \frac{\bnslash}{2} u_n\Big]
    \Big[ \bar v_{\bn} \bar T^A \bar T^B \frac{\nslash}{2} v_{\bn} \Big] 
    \,,
 & {\cal S}_2^{n\bn} &= C_F \Big[ \bar u_n T^A  \frac{\bnslash}{2} u_n\Big]
    \Big[ \bar v_{\bn} \bar T^A\frac{\nslash}{2} v_{\bn} \Big] 
    \,,\nn\\
  {\cal S}_3^{n\bn} &= C_A \Big[ \bar u_n T^A  \frac{\bnslash}{2} u_n\Big]
    \Big[ \bar v_{\bn} \bar T^A\frac{\nslash}{2} v_{\bn} \Big] 
    \,,
 & {\cal S}_4^{n\bn} &= T_F n_f \Big[ \bar u_n T^A  \frac{\bnslash}{2} u_n\Big]
    \Big[ \bar v_{\bn} \bar T^A\frac{\nslash}{2} v_{\bn} \Big]
    \,.
\end{align}
Since the techniques for carrying out one-loop integrals are standard, we will only quote the result for the QCD graphs at the integrand level, and then the final expanded result for each graph. For the full theory box graph we have
\begin{align}
{\rm Fig.}~\ref{fig:full_oneloop_matching}a 
 &= -g^4 \int\!\! \ddslash\!^{d} k\:
 \frac{\big[\bar u(p_3) T^A T^B \gamma^\mu (\slashed{k} + \slashed{p}_3) \gamma^\nu u(p_2)\big]
 \big[\bar v(p_4) \bar T^A \bar T^B \gamma_\mu (\slashed{k} - \slashed{p}_4) \gamma_\nu v(p_1)\big]}
  {[k^2-m^2] (k+p_3)^2 (k-p_4)^2 [(k+q)^2-m^2] }
 \nn\\
 &= \frac{-4i\alpha_s^2}{t} \: {\cal S}_1^{n\bn}
  \bigg[ 2 \ln\Big(\frac{s}{t}\Big) \ln\Big(\frac{-t}{m^2}\Big)+\ln^2\Big(\frac{-t}{m^2}\Big) - \frac{\pi^2}{3}\bigg] +\ldots
 \,,
\end{align}
where the ellipses indicate terms that are higher order in $t/s$. Similarly for the cross-box we have
\begin{align}
{\rm Fig.}~\ref{fig:full_oneloop_matching}b
 &= -g^4 \int\!\! \ddslash\!^{d} k\:
 \frac{\big[\bar u(p_3) T^A T^B  \gamma^\nu (\slashed{p}_2-\slashed{k} ) \gamma^\mu u(p_2)\big]
 \big[\bar v(p_4) \bar T^B \bar T^A \gamma_\mu (\slashed{k} - \slashed{p}_4) \gamma_\nu v(p_1)\big]}
  {[k^2-m^2] (k-p_2)^2 (k-p_4)^2 [(k+q)^2-m^2] }
 \nn\\
 &= \frac{4i\alpha_s^2}{t} \Big( {\cal S}_1^{n\bn} - \frac12 {\cal S}_3^{n\bn}\Big) 
  \bigg[ 2 \ln\Big(\frac{-s}{t}\Big) \ln\Big(\frac{-t}{m^2}\Big)+\ln^2\Big(\frac{-t}{m^2}\Big) - \frac{\pi^2}{3}\bigg] +\ldots
 \,.
\end{align}
For the two Y-graphs with a single three-gluon vertex the graphs give equal contributions and we have
\begin{align} \label{eq:oneloop_full_Y}
{\rm Figs.}~\ref{fig:full_oneloop_matching}c+\ref{fig:full_oneloop_matching}d
 &= \frac{g^4\,C_A}{q^2} \iota^\epsilon \mu^{2\epsilon}\!\!
 \int\!\! \ddslash\!^{d} k\:
 \frac{\big[\bar u(p_3) T^A  \gamma^\nu \slashed{k} \gamma^\lambda u(p_2)\big]
 \big[\bar v(p_4) \bar T^A \gamma^\mu  v(p_1)\big]\: T_{\mu\nu\lambda}(q,-k-p_3,k+p_2)}
  {k^2 [(k+p_2)^2-m^2] [(k+p_3)^2-m^2] } 
 \nn\\
 &
=  \frac{i\alpha_s^2}{t} \: {\cal S}_3^{n\bn}
  \bigg[ \frac{6}{\epsilon} + 6 \ln\Big(\frac{\mu^2}{-t}\Big) 
  + 8 \ln\Big(\frac{-t}{m^2}\Big)+4 \bigg] +\ldots
 \,.
\end{align}
Here we have included the factor that implements the $\overline{\rm MS}$ scheme,
\begin{align}
\iota^\epsilon=(4\pi)^{-\epsilon}e^{\epsilon\gamma_E}
\end{align}
as well as a $\mu^{2\epsilon}$. The triple gluon vertex momentum factor is $T_{\mu\nu\lambda}(k_1,k_2,k_3)=g_{\mu\nu}(k_1-k_2)_\lambda +g_{\nu\lambda}(k_2-k_3)_\mu + g_{\lambda\mu}(k_3-k_1)_\nu$. Since there are four external fermions, the wavefunction renormalization graph shown in \fig{full_oneloop_matching}e contributes through $2(Z_\psi-1)$ multiplying the tree level t-channel exchange diagram, and we will refer to this contribution as the result for \fig{full_oneloop_matching}e,
\begin{align} \label{eq:oneloop_full_wfn}
{\rm Fig.}~\ref{fig:full_oneloop_matching}e 
  &= \frac{i g^2}{q^2} \Big[ \bar u(p_3) T^A \gamma^\nu u(p_2)\: \bar v(p_4) \bar T^A \gamma_\nu v(p_1) \Big] 2 i\frac{d}{d\slashed{p}} 
   \: \iota^\epsilon \mu^{2\epsilon}\!\!
   \int\!\! \ddslash\!^{d} k\:
 \frac{ (-g^2 C_F) \gamma^\mu (\slashed{k}+\slashed{p})\gamma_\mu }
  {[k^2-m^2] (k+p)^2 } 
 \nn\\
 &= \frac{i\alpha_s^2}{t}\:  {\cal S}_2^{n\bn}
  \bigg[ -\frac{4}{\epsilon} - 4 \ln\Big(\frac{\mu^2}{m^2}\Big) +2 \bigg] 
  \,.
\end{align}
The two vertex renormalization graphs give the same contribution, and we find
\begin{align} \label{eq:oneloop_full_vert}
{\rm Figs.}~\ref{fig:full_oneloop_matching}f\!+\!\ref{fig:full_oneloop_matching}g
  &= \frac{g^4\,(2C_F\!-\!C_A)}{q^2}  \iota^\epsilon \mu^{2\epsilon}\!\! \int\!\! \ddslash\!^{d} k\:
 \frac{\big[\bar u(p_3) T^A  \gamma^\nu (\slashed{k}\!+\!\slashed{p}_3)\gamma_\mu (\slashed{k}\!+\!\slashed{p}_2)\gamma_\nu u(p_2)\big]
 \big[\bar v(p_4) \bar T^A \gamma^\mu  v(p_1)\big]}
  {[k^2-m^2] (k+p_2)^2 (k+p_3)^2 } 
 \nn\\
 &=\frac{i\alpha_s^2}{t}\Big(  {\cal S}_2^{n\bn}-\frac12  {\cal S}_3^{n\bn} \Big) 
 \bigg[ \frac{4}{\epsilon} + 4 \ln\Big(\frac{\mu^2}{-t}\Big) 
   -4\ln^2\Big(\frac{m^2}{-t}\Big) -16\ln\Big(\frac{m^2}{-t}\Big) -16 
  \bigg] 
  \,.
\end{align}
Finally for the full theory vacuum polarization graphs we have the standard Feynman gauge results (here we can set $m^2=0$ from the start),
\begin{align}  \label{eq:oneloop_full_vac}
{\rm Figs.}~\ref{fig:full_oneloop_matching}h
  \!+\!\ref{fig:full_oneloop_matching}i
  \!+\!\ref{fig:full_oneloop_matching}j
 &=\frac{i\alpha_s^2}{t}  {\cal S}_3^{n\bn} 
  \bigg[ \frac{10}{3\epsilon} + \frac{10}{3} \ln\Big(\frac{\mu^2}{-t}\Big) 
   +\frac{62}{9} \bigg] 
 +\frac{i\alpha_s^2}{t}  {\cal S}_4^{n\bn}
  \bigg[ -\frac{8}{3\epsilon} - \frac{8}{3} \ln\Big(\frac{\mu^2}{-t}\Big) 
   -\frac{40}{9} \bigg] 
  .
\end{align}
The sum of UV divergences from Eqs.~(\ref{eq:oneloop_full_Y}-\ref{eq:oneloop_full_vac}) only involves ${\cal S}_{3}^{n\bn}$ and ${\cal S}_{4}^{n\bn}$, and is canceled by the $\MSbar$ coupling counterterm $Z_g-1=-(\alpha_s/8\pi)(11C_A/3 -4 T_F n_f/3)$, which adds a contribution
\begin{align}
  \text{$Z_g$ counterterm graph} 
 &= \frac{i\alpha_s^2}{t} \bigg( -{\cal S}_3^{n\bn}\:  \frac{22}{3\epsilon}
  + {\cal S}_4^{n\bn}\: \frac{8}{3\epsilon} \bigg)
  \,.
\end{align}
Adding up all the full theory one-loop graphs plus the coupling counterterm graph we find
\begin{align}\label{eq:full_oneloop_result}
  \text{Full Theory} &= {\rm Figs.}~\ref{fig:full_oneloop_matching} +\text{$Z_g$ c.t.} \nn\\
  &= \frac{i\alpha_s^2}{t}  {\cal S}_1^{n\bn}
   \bigg[ 8 i \pi \ln\Big( \frac{-t}{m^2}\Big) \bigg]
   +  \frac{i\alpha_s^2}{t}  {\cal S}_2^{n\bn}
   \bigg[ -4\ln^2\Big( \frac{m^2}{-t}\Big)-12\ln\Big( \frac{m^2}{-t}\Big)
   -14 \bigg] 
  \nn\\[5pt] 
  &\quad 
  + \frac{i\alpha_s^2}{t}  {\cal S}_3^{n\bn}
   \bigg[ -4\ln\Big(\frac{s}{-t}\Big)\ln\Big(\frac{-t}{m^2}\Big) 
   + \frac{22}{3} \ln\Big(\frac{\mu^2}{-t}\Big) +\frac{170}{9} + \frac{2\pi^2}{3} \bigg]
  \nn \\[5pt]
  &\quad   + \frac{i\alpha_s^2}{t}  {\cal S}_4^{n\bn}
   \bigg[ -\frac{8}{3} \ln\Big(\frac{\mu^2}{-t}\Big) -\frac{40}{9}  
  \bigg] .
\end{align}

\begin{figure}[t!]
	%
	%
	%
%
\begin{center}
  \raisebox{2cm}{
  \hspace{-0.4cm}
  a)\hspace{3.cm} 
  b)\hspace{2.7cm} 
  c)\hspace{3cm} 
  d)\hspace{2.9cm} 
  e)\hspace{3cm} 
   } \\[-53pt]
\hspace{-0.5cm}
\raisebox{0.3cm}{
\includegraphics[width=0.17\columnwidth]{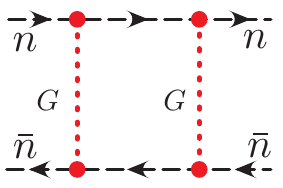}
}\hspace{0.2cm}
\raisebox{0.3cm}{
\includegraphics[width=0.17\columnwidth]{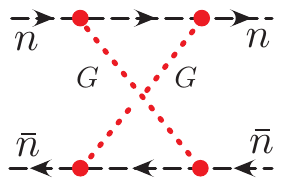} 
}\hspace{0.2cm}
\includegraphics[width=0.16\columnwidth]{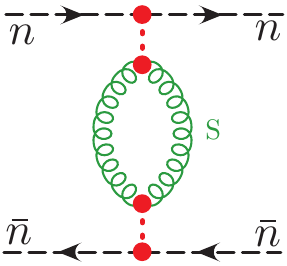}
\hspace{0.2cm}
\includegraphics[width=0.16\columnwidth]{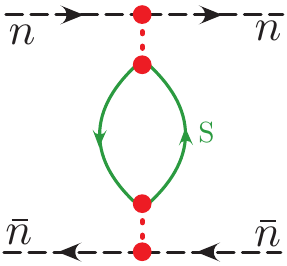}
\hspace{0.2cm}
\raisebox{0.3cm}{
\includegraphics[width=0.17\columnwidth]{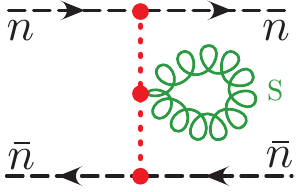}
} \hspace{0.2cm}
	\\[8pt]
  \raisebox{2cm}{
  \hspace{-0.4cm}
  f)\hspace{2.9cm} 
  g)\hspace{3.0cm} 
  h)\hspace{3.2cm} 
  i)\hspace{2.7cm} 
  j)\hspace{3cm} 
   } \\[-55pt]
\hspace{-0.55cm}
\includegraphics[width=0.19\columnwidth]{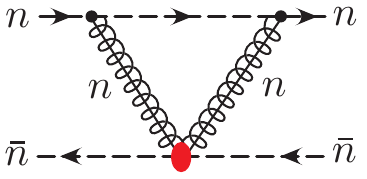}
\hspace{0.2cm}
\includegraphics[width=0.19\columnwidth]{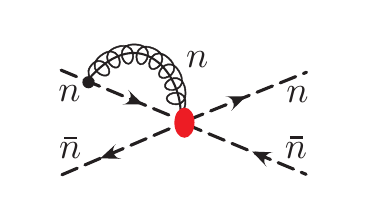}
\hspace{0.2cm}
\includegraphics[width=0.19\columnwidth]{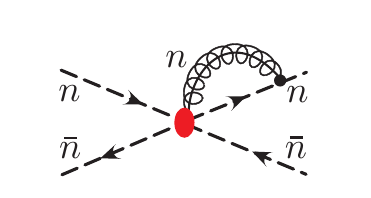}
\hspace{0.2cm}
\includegraphics[width=0.16\columnwidth]{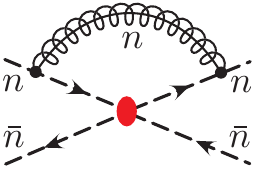}
\hspace{0.2cm}
\raisebox{0.5cm}{
\includegraphics[width=0.17\columnwidth]{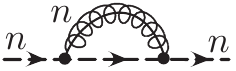}
}\hspace{0.2cm}
\\[10pt]
  \raisebox{2cm}{
  \hspace{-0.4cm}
  k)\hspace{2.9cm} 
  l)\hspace{3.cm} 
  m)\hspace{3.1cm} 
  n)\hspace{2.6cm} 
  o)\hspace{3cm} 
   } \\[-53pt]
\hspace{-0.55cm}
\includegraphics[width=0.19\columnwidth]{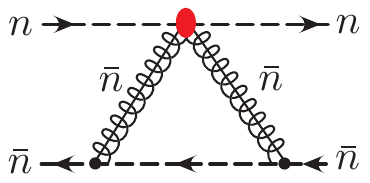}
\hspace{0.19cm}
\raisebox{-0.2cm}{
\includegraphics[width=0.18\columnwidth]{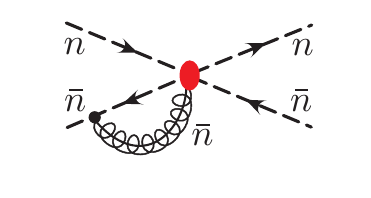}
}\hspace{0.19cm}
\raisebox{-0.2cm}{
\includegraphics[width=0.18\columnwidth]{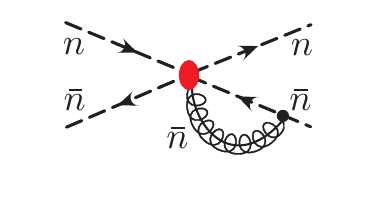}
}\hspace{0.19cm}
\raisebox{-0.2cm}{
\includegraphics[width=0.16\columnwidth]{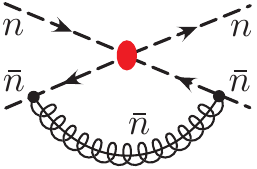}
}\hspace{0.19cm}
\raisebox{0.5cm}{
\includegraphics[width=0.17\columnwidth]{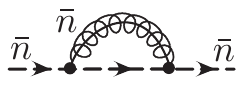}
}\hspace{0.2cm}
	\\[5pt]
\end{center}
	\vspace{-0.4cm}
	\caption{\setcaptionskip
		\SCETb graphs for the matching calculation of quark-antiquark forward scattering at one-loop. The first two graphs involve the Glauber potential. The next three graphs involve soft gluon or soft quark loops.   The second and third rows involve collinear loops with either the quark-gluon Glauber scattering operators or the quark-quark Glauber scattering operator, plus wavefunction renormalization.}
	\label{fig:SCET2_oneloop_matching}
	\setmainskip
\end{figure}

\begin{figure}[t!]
	%
	%
%
\begin{center}
\raisebox{2cm}{ \hspace{-2.8cm}  a) \hspace{5.2cm} b) \hspace{7.5cm} } 
  \\[-57pt]
\includegraphics[width=0.128\columnwidth]{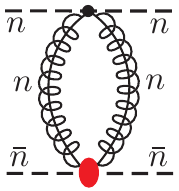}
\hspace{0.4cm}
\raisebox{0.1cm}{	   
\includegraphics[width=0.128\columnwidth]{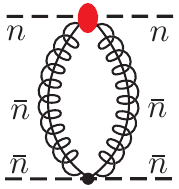}
}
\hspace{0.4cm}
\hspace{0.5cm}
\includegraphics[width=0.128\columnwidth]{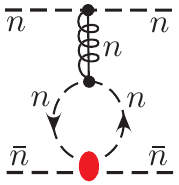}
\hspace{0.4cm}
\includegraphics[width=0.12\columnwidth]{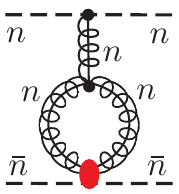}
\hspace{0.4cm}
\includegraphics[width=0.128\columnwidth]{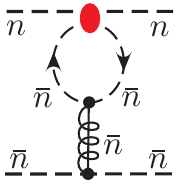}
\hspace{0.4cm}
\includegraphics[width=0.12\columnwidth]{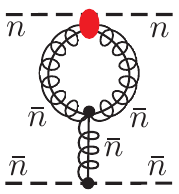}
	%
\end{center}
	\vspace{-0.3cm}
	\caption{\setcaptionskip
		a) Additional collinear graphs with the fermion two-gluon vertex from ${\cal L}_{n,\bn}^{(0)}$ which vanish.  b) Additional tadpole collinear loops graphs for forward scattering.  These graphs do not contribute to the matching calculation since 
		they vanish due to their soft zero-bin subtractions.}
	\label{fig:SCET2_oneloop_matching2}
	\setmainskip
\end{figure}

\subsubsection{\SCETb Loop Graphs and Matching}

Next consider the SCET forward scattering graphs shown in \fig{SCET2_oneloop_matching}. The first two graphs involve loops with Glauber loop momenta, the next three with soft loop momenta, and the remaining ten with $n$- or $\bn$-collinear loop momenta. We number these graphs from a) to o). Both notations for the Glauber operators are used (with and without the dashed red lines, see \eq{glaubnotation}), depending on what is most convenient. Note that we do not draw wavefunction or vertex renormalization graphs involving a soft gluon attached to a collinear quark, since these graphs vanish in Feynman gauge where they are proportional to $n^2=0$. 

The two graphs with an iteration of the Glauber operator, \fig{SCET2_oneloop_matching}a,b, were discussed above in \sec{GlauberBox}.  These graphs require regulation by the rapidity regulator to yield well defined answers, but their results are independent of $\eta$ as $\eta\to 0$. In particular \fig{SCET2_oneloop_matching}b vanishes (with or without the mass IR regulator), and 
\begin{align} \label{eq:glauber_loop}
 \text{Glauber Loops} 
&= \raisebox{-0.6cm}{
\includegraphics[width=0.13\columnwidth]{figs/EFT_loop1_qqqq} 
 }
 \nn\\
 &= (-4 g^4) \:  {\cal S}_1^{n\bn}\:  I_{\rm Gbox}
  = (-4 g^4) \:  {\cal S}_1^{n\bn} \Big(\frac{-i}{4\pi}\Big)  
    \int \!\!   \frac{ \ddslash\!^{d-2}k_\perp\ (-i\pi)  }{ 
    [{\vec k}_\perp^{\,2}+m^2][({\vec k}_\perp\plus {\vec q}_\perp)^2+m^2] }
  \nn\\
 &= \frac{i\alpha_s^2}{t} \:  {\cal S}_1^{n\bn}
   \bigg[ 8 i \pi \ln\Big( \frac{-t}{m^2}\Big) \bigg] .
\end{align}    
Thus we already see that the iterated Glauber exchange reproduces the full ${\cal S}_1^{n\bn}$ piece of \eq{full_oneloop_result}. 

Next we consider the SCET graphs contributing to the $C_F T^A\otimes \bar T^A$ color structure, ie. that have terms involving ${\cal S}_2^{n\bn}$. This occurs only in the collinear loop graphs in Figs.~\ref{fig:SCET2_oneloop_matching}i,j,n,o.  The loops in these graphs involve only Lagrangian insertions and a single collinear sector, therefore it is easy to check that they contribute the same result as in full QCD, which is the sum of \eqs{oneloop_full_wfn}{oneloop_full_vert} (this follows immediately from \eq{SCET4} above). Therefore,
\begin{align} \label{eq:oneloop_scet2_vertwfn}
& 
 \raisebox{-0.4cm}{
\includegraphics[width=0.13\columnwidth]{figs/EFT_loop6_qqqq}
}
\ \ +\ \
\raisebox{-0.6cm}{
\includegraphics[width=0.13\columnwidth]{figs/EFT_loop7_qqqq}
}\ \ +\ \  \bigg(
\raisebox{-0.2cm}{
\includegraphics[width=0.13\columnwidth]{figs/EFT_loop14_qqqq}
} \ +\ 
\raisebox{-0.2cm}{
\includegraphics[width=0.13\columnwidth]{figs/EFT_loop14b_qqqq}
 } \bigg)\
\raisebox{-0.2cm}{
\includegraphics[width=0.11\columnwidth]{figs/GlaubOp_tree_qqqq}
} 
 \nn\\ 
 &= \frac{i\alpha_s^2}{t} \: {\cal S}_2^{n\bn}
   \bigg[ -4\ln^2\Big( \frac{m^2}{-t}\Big)-12\ln\Big( \frac{m^2}{-t}\Big)
   -14 \bigg] 
 \nn \\
  &\quad + \frac{i\alpha_s^2}{t}\: {\cal S}_3^{n\bn} 
 \bigg[ -\frac{2}{\epsilon} - 2 \ln\Big(\frac{\mu^2}{-t}\Big) 
   +2\ln^2\Big(\frac{m^2}{-t}\Big) +8\ln\Big(\frac{m^2}{-t}\Big) +8
  \bigg] 
  . 
\end{align}
Looking at only the  ${\cal S}_2^{n\bn}$ term, we see that the SCET$_{\rm II}$ graphs reproduce the full ${\cal S}_2^{n\bn}$ piece of \eq{full_oneloop_result}. The situation is similar for the $n_f T_F T^A\otimes \bar T^A$ term, ie. ${\cal S}_4^{n\bn}$. The only SCET graph that is proportional to $n_f$ is the soft loop graph in \fig{SCET2_oneloop_matching}d which gives the same results as the quark vacuum polarization in the full theory,
\begin{align} \label{eq:soft_eye_fermion} 
\raisebox{-0.5cm}{
\includegraphics[width=0.13\columnwidth]{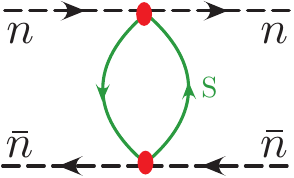}
}
  \ 
  & = \frac{i\alpha_s^2}{t}  {\cal S}_4^{n\bn}
   \bigg[ -\frac{8}{3\epsilon} -\frac{8}{3} \ln\Big(\frac{\mu^2}{-t}\Big) -\frac{40}{9}  
  \bigg] .
\end{align}
So the full theory ${\cal S}_4^{n\bn}$ term in \eq{full_oneloop_result} is also exactly reproduced. 

This leaves the final color structure $C_A T^A\otimes \bar T^A$, ie. ${\cal S}_3^{n\bn}$. Here things are more complicated, many graphs contribute, and there is no one-to-one correspondence between graphs in the full theory and effective theories.  For our SCET$_{\rm II}$ calculation we have contributions from  Figs.~\ref{fig:SCET2_oneloop_matching}i,n given above in \eq{oneloop_scet2_vertwfn}, as well as from Figs.~\ref{fig:SCET2_oneloop_matching}c,e,f,g,h,k,l,m which we will consider in turn. We will encounter rapidity divergences in these diagrams. There are also additional collinear graphs given in Figs.~\ref{fig:SCET2_oneloop_matching2} which we will discuss, but which do not in the end contribute (those in Fig.~\ref{fig:SCET2_oneloop_matching2}a because the integral vanishes, while those in Fig.~\ref{fig:SCET2_oneloop_matching2}b vanish only after accounting for their soft 0-bin subtraction). 

First consider the contribution from the T-product of two Glauber operators, ${\cal O}_{ns}^{qg}$ with ${\cal O}_{\bn s}^{qg}$, which is shown in \fig{SCET2_oneloop_matching}c.  
The Feynman rules for these soft-collinear scattering operators are given in \fig{LOfeynrulens}. Due to the presence of $1/n\cdot k$ and $1/\bn\cdot k$ propagators this soft loop graph will have rapidity divergences, and we must include the rapidity regulator. Note that if we collapse our dashed Glauber propagators to blobs that this graph can also be drawn as
\begin{align}
  \includegraphics[width=0.16\columnwidth]{figs/EFT_loop3_qqqq_ext}
  \hspace{0.3cm}\raisebox{1.05cm}{\Large =}\hspace{0.3cm}
  \raisebox{0.3cm}{
  \includegraphics[width=0.18\columnwidth]{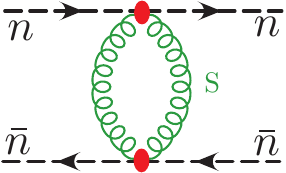}
   }
  \nn \\[-35pt]
\end{align}
and for this reason, and to remind the reader that this graph contains more than just vacuum polarization, we will refer to it as a ``soft eye'' graph. For this ``soft eye'' diagram we find
\begin{align} \label{eq:soft_eye}
\raisebox{-0.9cm}{
\includegraphics[width=0.13\columnwidth]{figs/EFT_loop3_qqqq_ext}
}
  \ 
  &=  \frac{4 g^4}{t^2}\: {\cal S}_3^{n\bn} 
    \iota^\epsilon \mu^{2\epsilon}\!\! \int\!\! 
   \frac{\ddslash\!^{d} k\ |2k_z|^{-\eta}\,\nu^\eta }{ [k^2-m^2][(k+q)^2-m^2]} 
  \: \bigg\{ \frac{4w^2 [k_\perp \cdot (k_\perp+q_\perp)]^2}{\bn\cdot k\: n\cdot k}
    + (d-2) n\cdot k\: \bn\cdot k 
   \nn\\[-12pt]
  &\qquad\qquad\qquad \qquad  + 2 (k_\perp+q_\perp)^2 + 2 k_\perp^2
    \bigg\} \frac{1}{2} 
 \nn\\
  &= - \frac{i\alpha_s^2}{t}\: {\cal S}_3^{n\bn} \bigg\{
    \frac{8}{\eta}\, g(\epsilon,\mu^2/t) + \frac{4}{\epsilon^2} +\frac{4}{\epsilon} \ln\Big(\frac{\mu^2}{\nu^2}\Big) 
    +4 \ln\Big(\frac{\mu^2}{\nu^2}\Big) \ln\Big(\frac{\mu^2}{-t}\Big)
    -2 \ln^2\Big(\frac{\mu^2}{-t}\Big) + \frac{\pi^2}{3}
   \nn\\
  &\qquad\qquad\qquad
   + 2 \Big( -\frac{11}{3\epsilon} -\frac{11}{3}\ln\frac{\mu^2}{-t}-\frac{67}{9} \Big) \bigg\} .
\end{align}
Here inside the integral the denominators in square brackets have a $+i0$, the factor of $(d-2)=g_\perp^{\mu\nu} g^\perp_{\mu\nu}$, and the rapidity divergence comes only from the first term in curly brackets. The factor of $1/2$ in the first line is a symmetry factor. The function multiplying the $1/\eta$ rapidity divergence for the result in \eq{soft_eye} is
\begin{align} \label{eq:g}
  g(\epsilon,\mu^2/t)
    &= e^{\epsilon\gamma_E} \Big(\frac{\mu^2}{-t}\Big)^\epsilon  \cos(\pi\epsilon) \Gamma(-\epsilon) \Gamma(1+2\epsilon) \,.
\end{align}
We have included the bookkeeping parameter $w$ in the first line of \eq{soft_eye}. In this section, where we are concerned with the matching, we will keep these factors only in the integrand, then drop them when quoting results. For the result in \eq{soft_eye}, it is interesting to note that the full $11 C_A/3\epsilon$ factor needed for the 1-loop $\beta$-function for the strong coupling has been generated from a graph only involving gluons, without a ghost contribution. This arises due to the form of the soft gauge invariance of the gluon operator in the EFT. Only the rapidity divergent integral in \eq{soft_eye} is non-standard, and we carry it out in \app{useful}. 

The choice of $\pm i0$ factors in the $(\bn\cdot k\pm i0)$ and $(n\cdot k\pm i0)$ denominators of \eq{soft_eye} does not change the result for this integral, due to the Glauber 0-bin subtraction that must be carried out for this soft graph. The easiest way to see this is to carry out the $k^0$ integration by contours. If the eikonal propagators are $(\bn\cdot k+i0)(n\cdot k + i0)$ or $(\bn\cdot k-i0)(n\cdot k - i0)$ then we can close the $k^0$ contour to only pick the poles from the propagators $[k^2-m^2+i0][(k+q)^2-m^2+i0]$, and doing the integral gives the result quoted in \eq{soft_eye}. In this case the naive soft integral is the full result, $S=\tilde S$, and the Glauber 0-bin subtraction is zero, because these propagators become $[k_\perp^2-m^2+i0][(k_\perp+q_\perp)^2-m^2+i0]$ in the Glauber limit, and the $k^0$ poles in the eikonal propagators are on the same side. The vanishing of this Glauber 0-bin subtraction occurs for the same reason as the vanishing of the Glauber cross-box. On the other hand if the eikonal propagators are taken to have opposite sign $i0$s, $(\bn\cdot k+i0)(n\cdot k - i0)$ or $(\bn\cdot k-i0)(n\cdot k + i0)$, then when we calculate the naive soft loop $\tilde S$ by closing the $k^0$ contour,  relative to the above we have to include an additional additive contribution from an eikonal pole. When we pick this pole, we either set $n\cdot k=0$ or $\bn\cdot k=0$ in the other propagators, so the relativistic propagators are exactly reduced to their form in the Glauber limit. Therefore, in this case this extra contribution in $\tilde S$ is exactly canceled by the fact that the graph now has a nonzero Glauber 0-bin subtraction, $S^{(G)}$,
\begin{align}
 S^{(G)}({\rm Fig.}~\ref{fig:SCET2_oneloop_matching}c)
  &=  \frac{4 g^4}{t^2}\:w^2 {\cal S}_3^{n\bn} 
    \iota^\epsilon \mu^{2\epsilon}\!\! \int\!\! 
   \frac{\ddslash\!^{d} k\ |2k_z|^{-\eta}\,\nu^\eta }{ [k_\perp^2-m^2][(k_\perp+q_\perp)^2-m^2]} 
  \:  \frac{4 [k_\perp \cdot (k_\perp+q_\perp)]^2}{(\bn\cdot k\pm i0)(n\cdot k\mp i0)}
  \,.
\end{align}
Note that the numerator remained unchanged under the Glauber limit, and that the terms without eikonal propagators are power suppressed in this limit and hence dropped. Thus, when we calculate the full soft loop result as $S=\tilde S-S^{(G)}$ the extra term from picking the eikonal pole cancels out, and the result is again the one given in \eq{soft_eye}. This cancellation is the reason that the choice for the direction of the soft Wilson lines is not important in the Glauber Lagrangian.

The remaining soft loop graph is \fig{SCET2_oneloop_matching}e, which is generated by the self contraction of the two gluon Feynman rule from the soft operator $O_s^{AB}$ shown in \fig{twosoftgluon_feynrule}, and which we refer to as the ``soft flower'' graph.  The integrand for this loop graph involves a gluon propagator with the IR regulator, $1/(k^2-m^2)$, times the part of the $O_s^{AB}$ Feynman rule in $\{\cdots \}$  that is obtained by setting $k_1=-k_2=k$, $q'=q$, and contracting with $\delta^{C_1C_2} g_{\mu_1\mu_2}$:
\begin{align}  \label{eq:softeyecontract}
 \frac{ \big\{ \cdots \big\} \delta^{C_1C_2}g_{\mu_1\mu_2} }{(k^2-m^2)}
  &=  \frac{g^2 C_A \delta^{AB}}{(k^2-m^2)} \bigg[ -2(d-2) 
  + \frac{4 \big( k_\perp^2 - q_\perp^2 \big)}{-n\cdot k\: \bn\cdot k} + 4 \bigg]
 \nn\\ 
  &= \frac{4 g^2 C_A \delta^{AB}}{(k^2-m^2)} \bigg[ 1+ \epsilon 
    - \frac{  k^2 }{n\cdot k\: \bn\cdot k}
  - \frac{  \vec q_\perp^{\:2} }{n\cdot k\: \bn\cdot k}
   \bigg] \,.
\end{align}
Looking at the terms in the square brackets,
the $1+\epsilon$ term has its scale set by $m^2$, and hence vanishes as the IR regulator $m\to 0$. The $k^2/(n\cdot k\, \bn\cdot k)$ term has a numerator that cancels the $(k^2-m^2)$ denominator, and hence is scaleless in the $k_\perp$ integral and vanishes. 
 This leaves only the last term which gives a rapidity divergent contribution:
\begin{align} \label{eq:soft_flower} 
& 
\raisebox{-0.6cm}{
  \includegraphics[width=0.13\columnwidth]{figs/EFT_loop5_qqqq}
  }
  \ 
   = \frac{4 g^4}{t}\: {\cal S}_3^{n\bn}w^2 
    \iota^\epsilon \mu^{2\epsilon}\!\! \int\!\! \ddslash\!^{d} k
   \: \frac{|2k_z|^{-\eta}\,\nu^\eta }{ [k^2-m^2](n\cdot k)(\bn\cdot k)} 
   \\[5pt]
 &\qquad
  = -\frac{i \alpha_s^2}{t}\, {\cal S}_3^{n\bn}
    \bigg\{ \frac{8}{\eta} h(\epsilon,\mu^2/m^2)
   - \frac{4}{\epsilon^2}- \frac{4}{\epsilon} \ln\Big(\frac{\mu^2}{\nu^2}\Big) 
   - 4 \ln\Big(\frac{\mu^2}{\nu^2}\Big) \ln\Big(\frac{\mu^2}{m^2}\Big)
   \!+\! 2 \ln^2\Big(\frac{\mu^2}{m^2}\Big) \!+\! \frac{\pi^2}{3} \bigg\}
   , \nn 
\end{align}
where the coefficient of the rapidity divergence involves the function
\begin{align} \label{eq:h}
 h(\epsilon,\mu^2/m^2) &= e^{\epsilon\gamma_E} \Big(\frac{\mu^2}{m^2}\Big)^\epsilon\: \Gamma(\epsilon) \,.
\end{align}
The result in \eq{soft_eye} could have been obtained by just keeping the $\cP_\perp^\mu ({\cal S}_n^T {\cal S}_\bn) \cP^\perp_\mu$ term in $O_s^{AB}$, since the other operators only contributed to the terms identified in \eq{softeyecontract} which vanished. Once again the choice of $\pm i0$ in the eikonal propagators does not change the result for this loop diagram due to its Glauber 0-bin subtraction. The mathematics of this cancellation are exactly the same as discussed above for the soft eye graph.  

The sum of the three one-loop graphs with soft loops from Eqs.~(\ref{eq:soft_eye_fermion},\ref{eq:soft_eye},\ref{eq:soft_flower}) is
\begin{align}  \label{eq:soft_1loop_total}
{\rm Figs.}~\ref{fig:SCET2_oneloop_matching}c,d,e 
 &=  
   -\frac{i \alpha_s^2}{t}\: {\cal S}_3^{n\bn} 
   \bigg[ \frac{8}{\eta} h(\epsilon,\mu^2/m^2)
  +  \frac{8}{\eta}\, g(\epsilon,\mu^2/t) 
    +4 \ln\Big(\frac{\mu^2}{\nu^2}\Big) \ln\Big(\frac{m^2}{-t}\Big)
   + 2 \ln^2\Big(\frac{\mu^2}{m^2}\Big) 
  \nn\\
  &\qquad\qquad\qquad
   -2 \ln^2\Big(\frac{\mu^2}{-t}\Big)  + \frac{2\pi^2}{3}
   + 2 \Big( -\frac{11}{3\epsilon} -\frac{11}{3}\ln\frac{\mu^2}{-t}-\frac{67}{9} \Big) \bigg]
    \nn\\
  & \quad\ 
  + \frac{i\alpha_s^2}{t} \: {\cal S}_4^{n\bn}
   \bigg[ -\frac{8}{3\epsilon} -\frac{8}{3} \ln\Big(\frac{\mu^2}{-t}\Big) -\frac{40}{9}  
  \bigg] .
\end{align}
Note that the $1/\epsilon^2$ and $\ln(\mu^2/\nu^2)/\epsilon$ terms have canceled in this result, leaving only the $1/\eta$ rapidity divergences and $1/\epsilon$ UV divergences. 
Since the bare soft operator $O_s^{AB}$ has a factor of $\alpha_s^{\rm bare}= Z_\alpha \mu^{2\epsilon}\alpha_s(\mu)$ multiplying the fields, there is a $Z_{\alpha}$ coupling counterterm contribution in the operator Feynman rule.  It gives the contribution
\begin{align}
 \text{soft $\alpha_s$ counterterm} 
 &= \frac{i\alpha_s^2}{t} \bigg( -{\cal S}_3^{n\bn}\:  \frac{22}{3\epsilon}
  + {\cal S}_4^{n\bn}\: \frac{8}{3\epsilon} \bigg) .
\end{align}
This result exactly cancels the $1/\epsilon$ terms in \eq{soft_1loop_total}, so with the counterterm the total sum of all soft loop graphs is given by
\begin{align}  \label{eq:soft_total_loop}
 \text{Soft Loops}
 &= {\rm Figs.}~\ref{fig:SCET2_oneloop_matching}c,d,e+Z_\alpha\:\text{c.t.} 
\nn\\
  &=  
   \frac{i \alpha_s^2}{t}\: {\cal S}_3^{n\bn} 
   \bigg\{- \frac{8}{\eta} h(\epsilon,\mu^2/m^2)
   -  \frac{8}{\eta}\, g(\epsilon,\mu^2/t) 
    - 4 \ln\Big(\frac{-t}{\nu^2}\Big) \ln\Big(\frac{m^2}{-t}\Big)
  \nn\\
  &\qquad\qquad\qquad
   - 2 \ln^2\Big(\frac{-t}{m^2}\Big) 
    +\frac{22}{3}\ln\Big(\frac{\mu^2}{-t}\Big)+\frac{134}{9} - \frac{2\pi^2}{3} \bigg\} 
    \nn\\
  & \quad\ 
  + \frac{i\alpha_s^2}{t} \: {\cal S}_4^{n\bn}
   \bigg[ -\frac{8}{3} \ln\Big(\frac{\mu^2}{-t}\Big) -\frac{40}{9}  
  \bigg] .
\end{align}
Thus in \SCETb the sum of graphs in the soft sector only has rapidity divergences.  The logarithms from these soft loops are minimized for $\mu\sim \nu\sim \sqrt{-t}$  which is consistent with our power counting. It is interesting to note that the full two-loop cusp anomalous dimension, which is determined by $K=(67/18-\pi^2/6)C_A - 10 n_f T_F/9$, appears as the constant term for our one-loop soft exchange result\footnote{We thank Hua-Xing Zhu for discussions about this.} in \eq{soft_total_loop},
\begin{align} \label{eq:twoloopcusp}
   \frac{i \alpha_s^2}{t}   \bigg\{ \Big(\frac{134}{9}-\frac{2\pi^2}{3}\Big)C_A -\frac{40}{9} T_F n_f \bigg\} \Big[ \bar u_n T^A  \frac{\bnslash}{2} u_n\Big]
   \Big[ \bar v_{\bn} \bar T^A\frac{\nslash}{2} v_{\bn} \Big] 
    &=  \frac{4 i \alpha_s^2}{t} K \Big[ \bar u_n T^A  \frac{\bnslash}{2} u_n\Big]
    \Big[ \bar v_{\bn} \bar T^A\frac{\nslash}{2} v_{\bn} \Big] \,.
\end{align}
It would be interesting to investigate in detail the reason for this correspondence. 

Finally we consider the remaining collinear diagrams, in \fig{SCET2_oneloop_matching}f,g,h,k,l,m.  The two V-graphs in \fig{SCET2_oneloop_matching}f,k give related contributions, and are induced by the Glauber operator involving $n$-collinear gluons, mixing back into $n$-collinear quarks (and likewise for the $\bn$-collinear loop). The $O_{ns\bn}^{gq}$ Glauber operator only produces $A_{n\perp}$ and $\bn\cdot A_n$ gluons, so for the $n$-collinear V-graph we have
\begin{align} \label{eq:ncollinearV}
& 
\raisebox{-0.4cm}{
  \includegraphics[width=0.14\columnwidth]{figs/EFT_loop8_qqqq}
  }
  \   
  = \frac{-ig^4}{t} f^{ABC}  \Big[ \bar v_{\bn} \bar T^C\frac{\nslash}{2} v_{\bn} \Big]  \int\!\! \ddslash\!^{d} k
  \frac{(\iota^\epsilon \mu^{2\epsilon} |\bn\cdot k|^{-\eta} \nu^\eta) \ \bn\cdot (k+p_3)}{[k^2-m^2][(k+q)^2-m^2](k+p_3)^2}
   \\
  &\hspace{2.8cm}
  \times  \bigg[ 2w^2 \bn\cdot k\, g_\perp^{\mu\nu} 
   + \frac{2 k_\perp\cdot (k_\perp\plus q_\perp)\bn^\mu\bn^\nu}{\bn\cdot k} 
   - 2 \bn^\mu (k_\perp^\nu\plus q_\perp^\nu) - 2 k_\perp^\mu \bn^\nu \bigg]
  \nn\\
  & \hspace{2.8cm}
  \times   \bar u_n T^B T^A  \frac{\bnslash}{2} 
  \bigg( n_\nu + \frac{\gamma^\perp_\nu (\slashed{k}_\perp\plus \slashed{p}_{3\perp})}{\bn\cdot(k\plus p_3)} 
  + \frac{\slashed{p}_{3\perp}\gamma^\perp_\nu}{\bn\cdot p_3}\bigg)
  \bigg(n_\mu + \frac{\gamma^\perp_\mu \slashed{p}_{2\perp}}{\bn\cdot p_3}
  + \frac{(\slashed{k}_\perp\plus \slashed{p}_{3\perp})\gamma^\perp_\mu}{\bn\cdot (k\plus p_3)}\bigg)
  u_n
  \nn\\[10pt]
 & 
  = \frac{-g^4 C_A}{2t} \Big[ \bar v_{\bn} \bar T^A\frac{\nslash}{2} v_{\bn} \Big] \!\!
   \int\!\! 
  \frac{\ddslash\!^{d} k\ (\iota^\epsilon \mu^{2\epsilon} |\bn\cdot k|^{-\eta} \nu^\eta) }{[k^2-m^2][(k+q)^2-m^2](k+p_3)^2}
 \ \bar u_n T^A  \frac{\bnslash}{2} \Bigg\{
  \frac{8 w^2 k_\perp\!\cdot\! (k_\perp\plus q_\perp)\bn\!\cdot\!(k\plus p_3)}{\bn\cdot k}
   \nn\\
 &\ \ + 2 \bn\!\cdot\! k \, \bn\!\cdot (k\plus p_3) \bigg( 
  \frac{\gamma_\perp^\nu (\slashed{k}_\perp\plus \slashed{p}_{3\perp}) }{\bn\cdot(k\plus p_3)} 
  + \frac{\slashed{p}_{3\perp}\gamma_\perp^\nu}{\bn\cdot p_3}\bigg)\bigg(
  \frac{\gamma^\perp_\nu \slashed{p}_{2\perp}}{\bn\cdot p_3}
  + \frac{(\slashed{k}_\perp\plus \slashed{p}_{3\perp})\gamma^\perp_\nu}{\bn\cdot (k\plus p_3)} \bigg)
  \nn\\
 &\ \  -4 \bn\!\cdot (k\plus p_3)  \bigg[ \frac{(\slashed{k}_\perp\! \plus \slashed{q}_{\perp}\!)(\slashed{k}_\perp\!\plus \slashed{p}_{3\perp}\!)}{\bn\cdot (k\plus p_3)} + \frac{\slashed{p}_{3\perp}\!(\slashed{k}_\perp\!\plus \slashed{q}_{\perp})}{\bn\cdot p_3} 
  + \frac{\slashed{k}_\perp\! \slashed{p}_{2\perp}}{\bn\cdot p_3}
  + \frac{(\slashed{k}_\perp\!\plus \slashed{p}_{3\perp}\!) \slashed{k}_\perp}{\bn\cdot (k\plus p_3)} \bigg]
  \Bigg\} u_n 
  \,.
  \nn
\end{align}
Only the first term in curly braces has a rapidity divergence, and we give the result for this integral in \app{useful}. All the other loop integrals are standard. The result for the $\bn$-collinear V-graph in \fig{SCET2_oneloop_matching}k is the same as the final answer with $p_3\to p_4$. Combining the results for these two graphs after doing the integrals we find
\begin{align}   \label{eq:oneloop_collinear_V}
& \raisebox{-0.4cm}{
  \includegraphics[width=0.14\columnwidth]{figs/EFT_loop8_qqqq}
   }
\ +\  
\raisebox{-0.4cm}{
   \includegraphics[width=0.14\columnwidth]{figs/EFT_loop9_qqqq}
  }
 \\
   &= \frac{i\alpha_s^2}{t}\: {\cal S}_3^{n\bn} \bigg[ \bigg\{
    \frac{4}{\eta}\, g(\epsilon,\mu^2/t)  
    -\frac{4}{\epsilon} \ln\Big(\frac{\nu}{\bn\cdot p_3}\Big) 
    -4 \ln\Big(\frac{\nu}{\bn\cdot p_3}\Big) \ln\Big(\frac{\mu^2}{-t}\Big)
    - \frac{3}{\epsilon} - 3 \ln\Big(\frac{\mu^2}{-t}\Big) 
  - 6 + \frac{4\pi^2}{3} \bigg\} 
  \nn\\
  &\qquad\qquad
    + \bigg\{
    \frac{4}{\eta}\, g(\epsilon,\mu^2/t)  
    -\frac{4}{\epsilon} \ln\Big(\frac{\nu}{n\cdot p_4}\Big) 
    -4 \ln\Big(\frac{\nu}{n\cdot p_4}\Big) \ln\Big(\frac{\mu^2}{-t}\Big)
    - \frac{3}{\epsilon} - 3 \ln\Big(\frac{\mu^2}{-t}\Big)    
    - 6 + \frac{4\pi^2}{3} \bigg\}\bigg]
  \nn\\[5pt]
  &= \frac{i\alpha_s^2}{t}\: {\cal S}_3^{n\bn} \bigg\{
    \frac{8}{\eta}\, g(\epsilon,\mu^2/t)  
    -\frac{4}{\epsilon} \ln\Big(\frac{\nu^2}{s}\Big) 
    -4 \ln\Big(\frac{\nu^2}{s}\Big) \ln\Big(\frac{\mu^2}{-t}\Big)
    - \frac{6}{\epsilon} - 6 \ln\Big(\frac{\mu^2}{-t}\Big) 
    - 12 + \frac{8\pi^2}{3} \bigg\}
   \,. \nn
\end{align}
Here the factors of $\ln(s)$ appear from adding the two diagrams and using $\ln(\bn\cdot p_3)+\ln(n\cdot p_4)=\ln s$. 

For the collinear loop integral in \eq{ncollinearV} we must consider the soft and Glauber 0-bin subtractions, $C=\tilde C - C^{(S)} -C^{(G)} + C^{(S)(G)}$, but here we will see that the subtractions give vanishing contributions. In the soft limit $k^\mu \sim \lambda$, so in \eq{ncollinearV} the denominator $(k+p_3)^2\to (n\cdot k\,\bn\cdot p_3)$. Only the rapidity divergent term gives an integral scaling as $\lambda^0$, whereas all the remaining terms in the curly brackets give integrals scaling as ${\cal O}(\lambda)$ that are dropped. The contribution for the soft subtraction is therefore
\begin{align}
  C^{(S)}({\rm Fig.}~\ref{fig:SCET2_oneloop_matching}f)
 &
  = -\frac{g^4}{2t}\: {\cal S}_3^{n\bn}\!\!
  \int\!\! \ddslash\!^{d} k
  \frac{(\iota^\epsilon \mu^{2\epsilon} |\bn\cdot k|^{-\eta} \nu^\eta) }{[k^2-m^2][(k+q)^2-m^2](n\cdot k+i0)}
 \:
  \frac{8w^2 k_\perp\!\cdot\! (k_\perp\plus q_\perp)}{\bn\cdot k}
  \,.
\end{align}
This integral can be performed by contours in $k^+=n\cdot k$. Since $q$ is purely transverse the poles in the two relativistic propagators are on the same side for either $k^->0$ or $k^-<0$, so the full result is obtained from the $k^-<0$ region by closing about the $n\cdot k=-i0$ pole. This leaves a vanishing scaleless integral in $k^-$,
\begin{align}
  \int_{-\infty}^0 dk^-  \frac{(-k^-)^{-\eta}}{k^-} = 
   - \int_{0}^\infty dk^-  \frac{(k^-)^{-\eta}}{k^-} = \frac{1}{\eta} -\frac{1}{\eta} = 0 \,,
\end{align}
so the soft subtraction $C^{(S)}({\rm Fig.}~\ref{fig:SCET2_oneloop_matching}f) =0$. The remaining subtractions come from the Glauber limit, and soft+Glauber limit, and are considered together. Again power counting implies that only the rapidity divergent term must be considered and we find
\begin{align} \label{eq:CGCSG}
    C^{(G)}({\rm Fig.}~\ref{fig:SCET2_oneloop_matching}f)
  - C^{(S)(G)}({\rm Fig.}~\ref{fig:SCET2_oneloop_matching}f)
 & = -\frac{g^4}{2t}\: {\cal S}_3^{n\bn}\!\!
  \int\!\! \ddslash\!^{d} k
  \frac{(\iota^\epsilon \mu^{2\epsilon} |\bn\cdot k|^{-\eta} \nu^\eta) \,\bn\mcdot p_3}{[k_\perp^2-m^2][(k_\perp+q_\perp)^2-m^2]}
 \:
  \frac{8w^2 k_\perp\!\cdot\! (k_\perp\plus q_\perp)}{\bn\cdot k}
 \nn\\
 & \times \bigg[ \frac{1}{\bn\mcdot p_3\,n\mcdot(k\plus p_3)\plus (k_\perp \plus p_{3\perp})^2+i0} - 
  \frac{1}{\bn\mcdot p_3\,n\mcdot k+i0}\bigg]
  \,.
\end{align} 
In the difference we have two poles on the same side in the $n\cdot k$ contour integral, so the contributions from the subtractions in \eq{CGCSG} vanish. Thus, with our regulators all the 0-bin subtractions vanish for the collinear graph and result for the $n$-collinear V-graph loop in \eq{ncollinearV} is simply obtained from the naive integral, $C=\tilde C$. The situation is identical for the 0-bin subtractions for the $\bn$-collinear V-graph.

Next we consider the collinear Wilson line graphs in \fig{SCET2_oneloop_matching}h,i,m,n. Using the Feynman rules from \fig{onegluon}, we see that the contractions with the incoming or outgoing collinear quark give the same contribution. In particular, a sign from the color structure cancels against a sign from the eikonal propagator from flipping the direction of the gluons momentum, and only the large momenta $\bn\cdot p_3=\bn\cdot p_2$ and $n\cdot p_1=n\cdot p_4$ appear in the answer. Due to the presence of the $f^{ABC}$ the color structure simplifies, $-i f^{ABC} T^C T^A = (C_A/2) T^B$, so only the structure ${\cal S}_3^{n\bn}$ appears.  For the collinear Wilson line graphs we find
\begin{align}  \label{eq:oneloop_collinear_W}
& 
\raisebox{-0.4cm}{
  \includegraphics[width=0.13\columnwidth]{figs/EFT_loop10_qqqq}
  }
\ +\ 
\raisebox{-0.4cm}{
  \includegraphics[width=0.13\columnwidth]{figs/EFT_loop11_qqqq}
  }
\ +\ 
\raisebox{-0.4cm}{
  \includegraphics[width=0.13\columnwidth]{figs/EFT_loop12_qqqq}
  }
\ +\ 
\raisebox{-0.4cm}{
  \includegraphics[width=0.13\columnwidth]{figs/EFT_loop13_qqqq}
  }
 \\[5pt]
  &\ \ 
  =  {\cal S}_3^{n\bn}\: \frac{2w^2 g^4 }{t} \bigg[
  \int\!\! \ddslash\!^{d} k
  \frac{(\iota^\epsilon \mu^{2\epsilon} |\bn\cdot k|^{-\eta} \nu^\eta) \ \bn\cdot (k+p_3)}{[k^2-m^2](k+p_3)^2 (\bn\cdot k)}
  +  \int\!\! \ddslash\!^{d} k
  \frac{(\iota^\epsilon \mu^{2\epsilon} |n\cdot k|^{-\eta} \nu^\eta) \ n\cdot (k+p_4)}{[k^2-m^2](k+p_4)^2 (n\cdot k)} \bigg]
  \nn \\
  &\ \
  =  \frac{i\alpha_s^2}{t}\, {\cal S}_3^{n\bn} 
  \bigg\{ \frac{4}{\eta} h(\epsilon,\mu^2/m^2)
   +\frac{4}{\epsilon} \ln\Big(\frac{\nu}{\bn\mcdot p_3}\Big) 
   + 4 \ln\Big(\frac{\nu}{\bn\mcdot p_3}\Big) \ln\Big(\frac{\mu^2}{m^2}\Big)
   + \frac{4}{\epsilon} +4 \ln\Big(\frac{\mu^2}{m^2}\Big) 
   + 4 - \frac{2\pi^2}{3} \bigg\}
  \nn\\
  &\quad
  +  \frac{i\alpha_s^2}{t}\, {\cal S}_3^{n\bn} 
  \bigg\{ \frac{4}{\eta} h(\epsilon,\mu^2/m^2)
   +\frac{4}{\epsilon} \ln\Big(\frac{\nu}{n\mcdot p_4}\Big) 
   + 4 \ln\Big(\frac{\nu}{n\mcdot p_4}\Big) \ln\Big(\frac{\mu^2}{m^2}\Big)
   + \frac{4}{\epsilon} +4 \ln\Big(\frac{\mu^2}{m^2}\Big) 
   + 4 - \frac{2\pi^2}{3} \bigg\}
 \nn\\
  &\ \
  =  \frac{i\alpha_s^2}{t}\, {\cal S}_3^{n\bn} 
  \bigg\{ \frac{8}{\eta} h(\epsilon,\mu^2/m^2)
   +\frac{4}{\epsilon} \ln\Big(\frac{\nu^2}{s}\Big) 
   + 4 \ln\Big(\frac{\nu^2}{s}\Big) \ln\Big(\frac{\mu^2}{m^2}\Big)
   + \frac{8}{\epsilon} +8 \ln\Big(\frac{\mu^2}{m^2}\Big) 
   + 8 - \frac{4\pi^2}{3} \bigg\}
  \,. \nn
\end{align}
The result for this loop integral is described in \app{useful}.
Again the $\ln(s)$ factors here appear from adding the $n$-collinear and $\bn$-collinear graphs, $\ln(\frac{\bn\cdot p_3}{\nu})+\ln(\frac{n\cdot p_4}{\nu})=\ln(\frac{s}{\nu^2})$.  In precisely the same manner as for the collinear V-graphs, the soft 0-bin and Glauber 0-bin subtractions all vanish for these collinear Wilson line graphs.  

Next we consider the graphs in \fig{SCET2_oneloop_matching2}. The diagrams in \fig{SCET2_oneloop_matching2}a arise because there is a two-quark two-gluon Feynman rule in the ${\cal L}_n^{(0)}$ and ${\cal L}_\bn^{(0)}$ collinear Lagrangians. This $n$-collinear loop graph is proportional to a vanishing loop integral
\begin{align}
\int\!\! \ddslash\!^{d} k
  \frac{(\iota^\epsilon \mu^{2\epsilon} |\bn\cdot k|^{-\eta} \nu^\eta) \ \bn\cdot k}{[k^2-m^2][(k+q)^2-m^2] \: \bn\cdot (k+p_3)} = 0
  \,,
\end{align}
and the same is true for the $\bn$-collinear loop graph. On the other hand, the tadpole diagrams in \fig{SCET2_oneloop_matching2}b do not have vanishing collinear integrals. In all these tadpole graphs the three propagators are $q^2 k^2 (k+q)^2$ (with an additional IR regulator $-m^2$ when appropriate), so the large collinear momenta $\bn\cdot p_3$ and $n\cdot p_4$ do not appear. Although the vertex in the gluon loop graphs could introduce an eikonal denominator, for these graphs it is always canceled since the same eikonal factor appears in the numerator. In all diagrams the collinear gluon propagator that is outside the loop gives the $q^2 = q_\perp^2$ which looks like a Glauber potential. Indeed, these tadpole graphs are double counting a contribution that has already been included from the soft diagrams in \fig{SCET2_oneloop_matching}c,d. Therefore it is not surprising that when we consider the soft zero-bin subtractions for each of these diagrams, that we obtain precisely the same loop integrals and
\begin{align}
  & \tilde C(\text{Fig.}\ref{fig:SCET2_oneloop_matching2}b) - C^{(S)}(\text{Fig.}\ref{fig:SCET2_oneloop_matching2}b) = 0 \,,
  & C^{(G)}(\text{Fig.}\ref{fig:SCET2_oneloop_matching2}b) & - C^{(S)(G)}(\text{Fig.}\ref{fig:SCET2_oneloop_matching2}b) = 0
\end{align}
for each diagram in \fig{SCET2_oneloop_matching2}b. Thus the collinear loop diagrams in \fig{SCET2_oneloop_matching2} do not contribute.

The sum of all the collinear graphs from Eqs.~(\ref{eq:oneloop_scet2_vertwfn},\ref{eq:oneloop_collinear_V},\ref{eq:oneloop_collinear_W}) gives
\begin{align}  \label{eq:collinear_total_loop}
 & \text{Collinear Loops} =
 {\rm Figs.}~\ref{fig:SCET2_oneloop_matching}f\text{-}o
  \nn\\
  &\qquad =  
   \frac{i \alpha_s^2}{t}\: {\cal S}_3^{n\bn} 
   \bigg\{\frac{8}{\eta} h\Big(\epsilon,\frac{\mu^2}{m^2}\Big)
    + \frac{8}{\eta}\, g\Big(\epsilon,\frac{\mu^2}{-t}\Big) 
   + 4 \ln\Big(\frac{\nu^2}{s}\Big) \ln\Big(\frac{-t}{m^2}\Big)
   +2\ln^2\Big(\frac{m^2}{-t}\Big) 
   \!+\! 4 \!+\!  \frac{4\pi^2}{3} \bigg\}
\nn\\
   &\qquad\quad
  + \frac{i\alpha_s^2}{t} \: {\cal S}_2^{n\bn}
   \bigg[ -4\ln^2\Big( \frac{m^2}{-t}\Big)-12\ln\Big( \frac{m^2}{-t}\Big)
   -14 \bigg]  
  .
\end{align}
Again there are cancellations that have occurred for the sum of graphs, the $\ln(\nu^2/s)/\epsilon$ terms have canceled, as have all the $1/\epsilon$ terms. (This is also true separately for the $n$-collinear graphs and $\bn$-collinear graphs.) Thus the collinear graphs also only have rapidity divergences.  The logarithms from these collinear loops are minimized with $\mu\sim \sqrt{t}$ and $\nu\sim \bn\cdot p_3 \sim n\cdot p_4 \sim \sqrt{s}$. Once again this is as expected, and consistent with the power counting. 

Finally, we can add up the Glauber, soft, and collinear SCET loop graphs from Eqs.~(\ref{eq:glauber_loop}, \ref{eq:soft_total_loop}, \ref{eq:collinear_total_loop}). In the sum of soft and collinear loops the $1/\eta$ rapidity divergences cancel, as expected since they arose from defining EFT modes that were sensitive to a single rapidity scale to avoid having large logs which are ratios of rapidity scales. Note that the rapidity divergences $h(\epsilon,\mu^2/m^2)/\eta$ from the soft and collinear Wilson line graphs cancel, independent from the rapidity divergences $g(\epsilon,\mu^2/(-t))/\eta$ appearing in the soft eye-graph and collinear V-graphs, which also cancel.  We find
\begin{align}  \label{eq:total_SCET2_result}
 \text{Total SCET} &= {\rm Figs.}~\ref{fig:SCET2_oneloop_matching}a\text{-}o
    +Z_\alpha\:\text{c.t.} 
  \nn\\
  &=  \frac{i\alpha_s^2}{t} \:  {\cal S}_1^{n\bn}
   \bigg[ 8 i \pi \ln\Big( \frac{-t}{m^2}\Big) \bigg] 
    + \frac{i\alpha_s^2}{t} \: {\cal S}_2^{n\bn}
   \bigg[ -4\ln^2\Big( \frac{m^2}{-t}\Big)-12\ln\Big( \frac{m^2}{-t}\Big)
   -14 \bigg]  
\nn\\
  &+ 
   \frac{i \alpha_s^2}{t}\: {\cal S}_3^{n\bn} 
   \bigg\{
    - 4  \ln\Big(\frac{s}{-t}\Big)\ln\Big(\frac{-t}{m^2}\Big)
    +\frac{22}{3}\ln\frac{\mu^2}{-t}
      +\frac{170}{9} + \frac{2\pi^2}{3} \bigg\} 
    \nn\\
  & \quad\ 
  + \frac{i\alpha_s^2}{t} \: {\cal S}_4^{n\bn}
   \bigg[ -\frac{8}{3} \ln\Big(\frac{\mu^2}{-t}\Big) -\frac{40}{9}  
  \bigg] \,.
\end{align} 
This total SCET result agrees exactly with the full theory one-loop result in \eq{full_oneloop_result} for all color structures, all IR divergences, all logs, and all constant terms. Since all IR divergences are correctly reproduced this provides a non-trivial test of our EFT framework. The $\ln\frac{\mu^2}{-t}$ dependence is proportional to the one-loop beta function, and hence exactly corresponds with the $\mu$ dependence in the $\alpha_s(\mu)$ of the tree level Glauber exchange diagram.  This logarithm shows that the scale $\mu^2\simeq -t>0$ is the preferred value for this potential. The various $\ln\frac{m^2}{-t}$  are infrared in origin. Finally, since $s\gg -t$ there is one large logarithm, $\ln\frac{s}{-t}$, which is generated by the separation of rapidity singularities in the soft and collinear diagrams (as opposed to invariant mass singularities).  The resummation of these logarithms leads to gluon Reggeization in the EFT operators, which we discuss in more detail in the next section.

Notice that although we have used Feynman gauge in this calculation the result does not depend upon the choice of which generalized covariant gauge we use. The reason for this is straightforward as the gauge dependent terms are contracted with the light-cone vector associated with the Wilson line which will cancel an eikonal propagator leading to a rapidity finite result. Thus for the gauge dependent pieces the rapidity regulator can be omitted, and these pieces will then cancel in the standard fashion.

It is natural to consider what the difference would be if we had considered quark-quark scattering, rather than quark-antiquark scattering. This corresponds to the crossing of two external lines, $p_1 \leftrightarrow -p_4$, which takes $s\to u=-s$ at leading order in the power expansion $-t\ll s$. In the full theory the only non-trivial change is to the box and cross-box diagrams which are interchanged under the crossing. The full result for quark-quark scattering is obtained from \eq{full_oneloop_result} by the simple replacement $S_i^{n\bn}\to S_{iqq}^{n\bn}$, where 
\begin{align}
S_{1qq}^{n\bn} &=-\Big[\bar u_n T^A T^B \frac{\bnslash}{2} u_n\Big]
  \Big[\bar u_\bn T^A T^B \frac{\nslash}{2} u_\bn \Big] \,,
  & {\cal S}_{2qq}^{n\bn} &= C_F \Big[ \bar u_n T^A  \frac{\bnslash}{2} u_n\Big]
    \Big[ \bar u_{\bn} T^A\frac{\nslash}{2} u_{\bn} \Big] 
    \,,\nn\\
  {\cal S}_{3qq}^{n\bn} &= C_A \Big[ \bar u_n T^A  \frac{\bnslash}{2} u_n\Big]
    \Big[ \bar u_{\bn}  T^A\frac{\nslash}{2} u_{\bn} \Big] 
    \,,
 & {\cal S}_{4qq}^{n\bn} &= T_F n_f \Big[ \bar u_n T^A  \frac{\bnslash}{2} u_n\Big]
    \Big[ \bar u_{\bn} T^A\frac{\nslash}{2} u_{\bn} \Big]
    \,.
\end{align}  
In the SCET calculation the only changes are to the color structures, and the final result is again obtained from \eq{total_SCET2_result} by taking $S_i^{n\bn}\to S_{iqq}^{n\bn}$.  Thus, once again the total SCET and full theory results agree.

The fact that the SCET result in \eq{total_SCET2_result} agrees exactly with the full theory result in \eq{full_oneloop_result} implies that there are no hard matching corrections to the Glauber operator at the scale $\mu^2\sim s$. (The analogous statement in the threshold expansion is that there are no contributions to the forward scattering at leading power from hard loop momenta.)    It is easy to see that the pattern observed here at one loop continues to higher orders in $\alpha_s$ for all leading power terms: there are no individual SCET diagrams that can possibly depend on $\ln(s)$ since the relevant momenta that form this combination cannot occur in any loop integral by the power counting. To get a $\ln(s)$ from a single loop requires a loop that knows about both scattering particles, but a hard loop of this type will give a $1/s$ rather than the leading power $1/t$. 
At leading power the $\ln(s)$ dependence only arises from the rapidity divergences which are sensitive to the large $p_n^-$ and $p_\bn^+$ collinear momenta.  This is a general feature of the EFT for leading power forward scattering, there are no hard matching corrections. Thus, as we argued in \sec{powercount}, the tree level results for all the Glauber exchange operators actually include the complete all-order Wilson coefficients.

One can repeat this one loop forward matching calculation in other kinematic scenarios.  In \sec{loop1match} we repeat this matching for \SCETa kinematics, where there are now simultaneously soft and ultrasoft modes, and the results for the various diagrams above change. Again we find that the full theory result, which now includes a $\ln^2(s)$, is exactly reproduced by the \SCETa calculation, the IR divergences are properly reproduced, and there is no hard matching corrections for the Glauber Lagrangians.  One can also repeat the matching calculation for $n$-soft forward scattering at one-loop in \SCETb, and although the precise form of the diagrams change, drawing parallels with the results obtained in this section, it is again clear that the full theory result will be exactly reproduced by the sum of EFT diagrams.

\subsection{Reggeization from Rapidity Renormalization}
\label{sec:regge}

Given that the large logarithm in the one-loop forward scattering amplitude of \eq{total_SCET2_result} is generated by adding up modes that are separated in rapidity, 
\begin{align}
 \ln\frac{s}{-t}
  =\ln\frac{p_3^+}{\nu} +\ln\frac{p_4^-}{\nu} + \ln\frac{\nu^2}{-t} 
  \,,
\end{align}
one can resum these logarithms by carrying out a separate rapidity renormalization and resummation of the soft and collinear components of the amplitudes. The anomalous dimension in rapidity space is determined by the coefficient of the $1/\eta$ poles, and is connected to the coefficient of the $\ln(\nu^2)$ terms in the soft and collinear amplitudes. From \eqs{soft_total_loop}{collinear_total_loop} we see that the coefficient of the $\ln(\nu^2)$ terms involves the logarithm $\ln(-t/m^2)$, and hence is IR divergent. Obviously an IR divergent anomalous dimension does not make sense.  This infrared divergence is a reflection of the fact that the separate renormalization of soft and collinear objects should be done at the level of the squared amplitude including phase space integrals, where the corresponding soft and collinear functions include both virtual and real radiation diagrams, and are IR finite. Nevertheless, the contribution to this renormalization from virtual diagrams can be examined at the amplitude level, and we will see that it corresponds to the classic result for gluon Reggeization. Therefore for the purpose of this section we put aside the presence of the $\ln(m^2)$ IR divergence, and demonstrate how the classic IR divergent  result emerges in SCET.  In the next section we will carry out the renormalization for the soft function, where the IR divergence is properly resolved.

\subsubsection{Notation for Virtual Counterterms and Anomalous Dimensions}
    \label{sec:adnotations}

The rapidity divergent $n$-collinear loops in \fig{SCET2_oneloop_matching} consist of the V-graphs (\fig{SCET2_oneloop_matching}f,k) and the $W$-Wilson line graphs (\fig{SCET2_oneloop_matching}g,h,l,m). In addition the sum of vertex and wavefunction renormalization graphs (\fig{SCET2_oneloop_matching}i,j,n,o) contribute a $C_A/\epsilon$ pole that cancels that of the V-graphs. From the point of view of the $n$-collinear sector, the V-graphs involve a mixing of $O_n^{gA}$ into $O_n^{qA}$. On the other hand the Wilson line, vertex, and wavefunction graphs take $O_n^{qA}$ back to $O_n^{qA}$.  Thus we see that the $n$-collinear virtual renormalization can be viewed as involving mixing with a $2\times 2$ matrix structure
\begin{align} \label{eq:Cnrenorm}
  \vec {\cal O}_n^{A{\rm bare}} = \hat V_{{\cal O}_n} \cdot \vec {\cal O}_n^A(\nu,\mu) \,,
  \qquad
 \hat V_{{\cal O}_n} 
   = \left( \begin{matrix} 1 + \delta V_n^{qq} & \delta V_n^{qg} \\
   \delta V_n^{gq} &  1 + \delta V_n^{gg}
 \end{matrix} \right) \,,
  \qquad
  \vec {\cal O}_n^A = \left( \begin{matrix} {\cal O}_n^{q A} \\ {\cal O}_n^{g A} \end{matrix}\right)
  \,.
\end{align}
Here we use the notation ``$V$'' rather than a traditional ``$Z$'' for the renormalization factors to remind the reader that these are just the divergent $1/\epsilon$ and $1/\eta$ contributions from virtual graphs and may still involve the IR regulator $m$. They are not the complete renormalization results. The component notation for terms in $\vec {\cal O}_n^A$ in \eq{Cnrenorm} applies to both the bare and renormalized operators.
The terms $\delta V_n^{qq}$ and $\delta V_n^{gq}$ are determined by graphs with external quark fields, whereas $\delta V_n^{gg}$ and $\delta V_n^{qg}$ are analogs with external gluon fields (which therefore cannot be obtained simply from subcomponents of the results in \fig{SCET2_oneloop_matching}).  We have built in the fact that the renormalization is diagonal in color space by using the same index $A$ for the bare and renormalized operators. The same decomposition applies for the $\bn$-collinear sector with $n\to \bn$ for all terms, which we write out just to be definite
\begin{align} \label{eq:Cbnrenorm}
  \vec {\cal O}_\bn^{B{\rm bare}} = \hat V_{{\cal O}_\bn} \cdot \vec {\cal O}_\bn^B(\nu,\mu) \,,
  \qquad
 \hat V_{{\cal O}_\bn} 
   = \left( \begin{matrix} 1 + \delta V_\bn^{qq} & \delta V_\bn^{qg} \\
   \delta V_\bn^{gq} &  1 + \delta V_\bn^{gg}
 \end{matrix} \right) \,,
  \qquad
  \vec {\cal O}_\bn^B = \left( \begin{matrix} {\cal O}_\bn^{q B} \\ {\cal O}_\bn^{g B} \end{matrix}\right)
  \,.
\end{align}

The structure for the rapidity divergent soft sector is more complicated since we have operators ${\cal O}_s^{q_n A}$, ${\cal O}_s^{g_n A}$, ${\cal O}_s^{q_\bn A}$, ${\cal O}_s^{g_\bn A}$, as well as ${\cal O}_s^{AB}$. Phrased in the language of mixing, the single color index operators with $S_n$ Wilson lines, ${\cal O}_s^{q_nA}$ and ${\cal O}_s^{g_nA}$, will mix with themselves, but not with ${\cal O}_s^{q_\bn A}$ and ${\cal O}_s^{g_\bn A}$ which have $S_\bn$ Wilson lines. This occurs because soft loops and emissions from a soft operator alone do not generate Wilson lines. Thus for these single index operators we have
\begin{align} \label{eq:SArenorm}
  \vec {\cal O}_{s_n}^{A{\rm bare}} = \hat V_{{\cal O}_{s_n}} \cdot \vec {\cal O}_{s_n}^A(\nu,\mu) \,,
  \qquad
 \hat V_{{\cal O}_{s_n}} 
   = \left( \begin{matrix} 1 + \delta V_{s_n}^{qq} & \delta V_{s_n}^{qg} \\
   \delta V_{s_n}^{gq} &  1 + \delta V_{s_n}^{gg}
 \end{matrix} \right) \,,
  \qquad
  \vec {\cal O}_{s_n}^A = \left( \begin{matrix} {\cal O}_s^{q_n A} \\ {\cal O}_s^{g_n A} \end{matrix}\right)
  \,,
\end{align}
plus a direct analog for ${\cal O}_s^{q_\bn A}$ and ${\cal O}_s^{g_\bn A}$ obtained with $n\to\bn$. For the double index operator ${\cal O}_s^{AB}$ we have self renormalization as well as mixing from time-ordered products (T-products) with the same color structure, such as $i\int d^4x\, T {\cal O}_s^{g_n A}(x) {\cal O}_s^{g_\bn B}(0)$. Since in ${\cal O}_s^{AB}$ the index $A$ couples to an $n$-collinear sector and the index $B$ couples to a $\bn$-collinear sector, we must maintain this same structure on the T-product terms. Since  ${\cal O}_s^{AB}$ has a no-gluon Feynman rule, to mix into it one must have the fields in the T-product annihilate each other. At ${\cal O}(\alpha_s)$ this can only occur through either a soft quark or gluon loop, which allows only certain products of operators to appear, but at higher orders other terms can also contribute. The full form of the mixing equation is
\begin{align} \label{eq:SABrenorm}
 & \vec {\cal O}_{s}^{AB{\rm bare}} = \hat V_{{\cal O}_{s}} \cdot \vec {\cal O}_{s}^{AB}(\nu,\mu) \,,
  \\
 & \hat V_{{\cal O}_{s}} 
   = \left( \begin{matrix} 1 + \delta V_s &\ \,  0 &\ \, 0 &\ \, 0 & 0 \\
   \delta V_s^{Tqq} &  
   \multicolumn{4}{c}{\multirow{4}{*}{$\hat V_{{\cal O}_{s_n}}\otimes \hat V_{{\cal O}_{s_\bn}}$ }} \\ 
   \delta V_s^{Tgq} & \\
   \delta V_s^{Tgg} & \\
   \delta V_s^{Tqg} & 
 \end{matrix} \right) \,,
\qquad
  \vec {\cal O}_{s}^{AB} = \left( \begin{matrix} {\cal O}_s^{AB} \\ 
  \mbox{$i\int d^4x$}\: T\, {\cal O}_s^{q_n A}(x) {\cal O}_s^{q_\bn B}(0) \\
  \mbox{$i\int d^4x$}\: T\, {\cal O}_s^{g_n A}(x) {\cal O}_s^{q_\bn B}(0) \\
  \mbox{$i\int d^4x$}\: T\, {\cal O}_s^{g_n A}(x) {\cal O}_s^{g_\bn B}(0) \\  
  \mbox{$i\int d^4x$}\: T\, {\cal O}_s^{q_n A}(x) {\cal O}_s^{g_\bn B}(0)   
  \end{matrix}\right)
  \,. \nn
\end{align}
Here the lower-right $4\times 4$ block in $ \hat V_{{\cal O}_{s}}$ is determined by the renormalization factor $\hat V_{{\cal O}_{s_n}}$ in \eq{SArenorm} and its analog with $n\to \bn$, since these terms just involve the renormalization of individual operators appearing in the T-products. For example, the $(2,4)$ entry of $\hat V_{{\cal O}_s}$ is $\delta V_{s_n}^{qg} \delta V_{s_\bn}^{qg}$.  The $1\times 1$ entry of $\hat V_{{\cal O}_s}$ is the self renormalization of ${\cal O}_s^{AB}$, and the four entries below it are due to mixing of the T-products into ${\cal O}_s^{AB}$. The entries with $0$s indicate that the operator ${\cal O}_s^{AB}$ does not mix into the T-products. At one-loop the nonzero entries are $\delta V_s$ (from the soft subgraph in \fig{SCET2_oneloop_matching}e), $\delta V_s^{Tqq}$ (from the soft part of \fig{SCET2_oneloop_matching}d), $\delta V_s^{Tgg}$ (from the soft part of \fig{SCET2_oneloop_matching}c), and the terms in the $4\times 4$ submatrix $\hat V_{{\cal O}_{s_n}}\otimes \hat V_{{\cal O}_{s_\bn}}$. 

The independence of the bare collinear and bare soft operators to the rapidity renormalization scale $\nu$ and to the invariant mass renormalization scale $\mu$, $(\nu d/d\nu) {\cal O}^{\rm bare} = 0$ and $(\mu d/d\mu) {\cal O}^{\rm bare} = 0$, leads to renormalization group equations in the standard fashion. Thus we have
\begin{align} \label{eq:adimOn}
 \nu \frac{\partial}{\partial \nu}\:  \vec {\cal O}_n^A(\nu,\mu)
  &= \hat \gamma_{{\cal O}_n}^{\nu} \cdot  \vec {\cal O}_n^A(\nu,\mu) \,,
 \qquad
 \hat \gamma_{{\cal O}_n}^{\nu} 
  = - \hat V_{{\cal O}_n}^{-1} \cdot\:  \nu \frac{\partial}{\partial \nu}\: 
   \hat V_{{\cal O}_n}
  =  \left( \begin{matrix} \gamma_{n\nu}^{qq} & \gamma_{n\nu}^{qg} \\
   \gamma_{n\nu}^{gq} &  \gamma_{n\nu}^{gg}
 \end{matrix} \right) 
  \,, \\
 \mu \frac{\partial}{\partial \mu}\:  \vec {\cal O}_n^A(\nu,\mu)
  &= \hat \gamma_{{\cal O}_n}^{\mu} \cdot  \vec {\cal O}_n^A(\nu,\mu) \,,
 \qquad
 \hat \gamma_{{\cal O}_n}^{\mu} 
  = - \hat V_{{\cal O}_n}^{-1} \cdot\:  \mu \frac{\partial}{\partial \mu}\: 
   \hat V_{{\cal O}_n}
  =  \left( \begin{matrix} \gamma_{n\mu}^{qq} & \gamma_{n\mu}^{qg} \\
   \gamma_{n\mu}^{gq} &  \gamma_{n\mu}^{gg}
 \end{matrix} \right) 
\,,\nn  
\end{align}
with analogous results for $\hat \gamma_{{\cal O}_\bn}^{\nu}$ and $\hat \gamma_{{\cal O}_\bn}^{\mu}$ by taking $n\to \bn$. In particular we have the relation
\begin{align}
 \gamma_\bn^{ij} = \gamma_n^{ij}\: \Big|_{n\rightarrow \bn} 
\end{align}
between $n$-collinear and $\bn$-collinear anomalous dimensions.  The $\nu$-anomalous dimension is entirely determined by the $1/\eta$ pole in $\hat V_{{\cal O}_n}$, and the $\mu$-anomalous dimension by the $1/\epsilon$ pole in $\hat V_{{\cal O}_n}$. 

For the soft operators with one color index we have similar results
\begin{align} \label{eq:adimOsn}
 \nu \frac{\partial}{\partial \nu}\:  \vec {\cal O}_{s_n}^A(\nu,\mu)
  &= \hat \gamma_{{\cal O}_{s_n}}^{\nu} \cdot  \vec {\cal O}_{s_n}^A(\nu,\mu) \,,
 \qquad
 \hat \gamma_{{\cal O}_{s_n}}^{\nu} 
  = - \hat V_{{\cal O}_{s_n}}^{-1} \cdot\:  \nu \frac{\partial}{\partial \nu}\: 
   \hat V_{{\cal O}_{s_n}}
  =  \left( \begin{matrix} \gamma_{s_n\nu}^{qq} & \gamma_{s_n\nu}^{qg} \\
   \gamma_{s_n\nu}^{gq} &  \gamma_{s_n\nu}^{gg}
 \end{matrix} \right) 
  \,, \\
 \mu \frac{\partial}{\partial \mu}\:  \vec {\cal O}_{s_n}^A(\nu,\mu)
  &= \hat \gamma_{{\cal O}_{s_n}}^{\mu} \cdot  \vec {\cal O}_{s_n}^A(\nu,\mu) \,,
 \qquad
 \hat \gamma_{{\cal O}_{s_n}}^{\mu} 
  = - \hat V_{{\cal O}_{s_n}}^{-1} \cdot\:  \mu \frac{\partial}{\partial \mu}\: 
   \hat V_{{\cal O}_{s_n}}
  =  \left( \begin{matrix} \gamma_{s_n\mu}^{qq} & \gamma_{s_n\mu}^{qg} \\
   \gamma_{s_n\mu}^{gq} &  \gamma_{s_n\mu}^{gg}
 \end{matrix} \right) 
\,,\nn  
\end{align}
and again obtain results for $\hat \gamma_{{\cal O}_{s_\bn}}^{\nu}$ and $\hat \gamma_{{\cal O}_{s_\bn}}^{\mu}$ by taking $n\to \bn$. 
Finally for the soft operators with two-color indices we have
\begin{align} \label{eq:adimOs}
 & \nu \frac{\partial}{\partial \nu}\:  \vec {\cal O}_{s}^{AB}(\nu,\mu)
  = \hat \gamma_{{\cal O}_{s}}^{\nu} \cdot  \vec {\cal O}_{s_n}^{AB}(\nu,\mu) \,,
 \qquad
 \mu \frac{\partial}{\partial \mu}\:  \vec {\cal O}_{s}^{AB}(\nu,\mu)
  = \hat \gamma_{{\cal O}_{s}}^{\mu} \cdot  \vec {\cal O}_{s}^{AB}(\nu,\mu) \,,
 \\
& \hat \gamma_{{\cal O}_{s}}^{\nu} 
  = - \hat V_{{\cal O}_{s}}^{-1} \cdot\:  \nu \frac{\partial}{\partial \nu}\: 
   \hat V_{{\cal O}_{s}}
  \,,
 \qquad 
\hat \gamma_{{\cal O}_{s}}^{\mu} 
  = - \hat V_{{\cal O}_{s}}^{-1} \cdot\:  \mu \frac{\partial}{\partial \mu}\: 
   \hat V_{{\cal O}_{s}}
 \nn\\
 & \hat \gamma_{{\cal O}_{s}}^{\nu} 
  = \left( \begin{matrix} \gamma_{s\nu}^{\rm dir} &\ \,  0 &\ \, 0 &\ \, 0 & 0 \\
   \gamma_{s\nu}^{Tqq} &  
   \multicolumn{4}{c}{\multirow{4}{*}{$\hat \gamma_{{\cal O}_{s_n}^\nu}\otimes \hat \gamma_{{\cal O}_{s_\bn}^\nu}$ }} \\ 
   \gamma_{s\nu}^{Tgq} & \\
   \gamma_{s\nu}^{Tgg} & \\
   \gamma_{s\nu}^{Tqg} & 
 \end{matrix} \right)
  \,, 
 \qquad
 \hat \gamma_{{\cal O}_{s}}^{\mu} 
  =  \left( \begin{matrix} \gamma_{s\mu}^{\rm dir} &\ \,  0 &\ \, 0 &\ \, 0 & 0 \\
   \gamma_{s\mu}^{Tqq} &  
   \multicolumn{4}{c}{\multirow{4}{*}{$\hat \gamma_{{\cal O}_{s_n}^\mu}\otimes \hat \gamma_{{\cal O}_{s_\bn}^\mu}$ }} \\ 
   \gamma_{s\mu}^{Tgq} & \\
   \gamma_{s\mu}^{Tgg} & \\
   \gamma_{s\mu}^{Tqg} & 
 \end{matrix} \right)
\,.\nn  
\end{align}

\subsubsection{Relations between Virtual Anomalous Dimensions in \SCETb}
   \label{sec:adrelations}

Having established notations for the anomalous dimensions, we next consider the constraints imposed by the fact that there is no overall $\nu$ or $\mu$ dependence for the scattering of soft and collinear particles or the scattering of $n$-collinear and $\bn$-collinear particle, since there are no $\nu$ or $\mu$ dependent Wilson coefficients in these Lagrangians.   For simplicity we will work out these constraints at one-loop order, which is the level needed for our analysis.  First consider the scattering between two neighboring rapidity sectors, $n$-soft scattering mediated by one or more $O_{ns}^{ij}$ operators. Here there is no mixing of multiple insertions of ${\cal L}_G^{\rm II(0)}$ back into a single insertion. The only such diagram involves the iteration $O_{ns}^{ik}O_{ns}^{kj}$ with a Glauber loop that has one soft and one collinear propagator, and this loop diagram is identical to the box calculation in \sec{GlauberBox}, and hence is finite. Therefore, at this order we can look at the $O_{ns}^{ij}$ (defined in  Eq.(\ref{eq:Ons}))
alone.

The fact that the tree level matching is exact and that quark and gluon operators have identical coefficients, places
strong constraints on the anomalous dimensions.  Imposing the condition that
\begin{align} \label{eq:rel0}
  \nu \frac{d}{d\nu}\: \sum_{ij=q,g} O_{ns}^{ij} =
  \nu \frac{d}{d\nu}\: ({\cal O}_n^{qA} + {\cal O}_n^{gA}) \frac{1}{\cP_\perp^2} ({\cal O}_s^{q_nA} + {\cal O}_s^{g_nA})    
  = 0 \,
\end{align}
implies that
\begin{align} \label{eq:rel1}
	& \nu \frac{d}{d\nu} ({\cal O}_n^{qA} + {\cal O}_n^{gA}) = \gamma_{n\nu}
	({\cal O}_n^{qA} + {\cal O}_n^{gA}) 
	\,,
	& \nu \frac{d}{d\nu} ({\cal O}_s^{q_nA} + {\cal O}_s^{g_nA})= \gamma_{s_n\nu}  
	({\cal O}_s^{q_nA} + {\cal O}_s^{g_nA}) &
	\,,
\end{align}
i.e. the sum of the two operators must mix into itself. Furthermore, \eqs{adimOn}{adimOs} imply that these constants of proportionality are given by
\begin{align} \label{eq:adrel1}
  \gamma_{n\nu} \equiv \gamma_{n\nu}^{qq} + \gamma_{n\nu}^{gq} 
    &= \gamma_{n\nu}^{gg} + \gamma_{n\nu}^{qg} 
   \,,
   & \gamma_{s_n\nu}\equiv \gamma_{s_n\nu}^{qq} + \gamma_{s_n\nu}^{gq} 
    &= \gamma_{s_n\nu}^{gg} + \gamma_{s_n\nu}^{qg} 
  \,.
\end{align}

These results can also be derived starting only with \eq{rel0}, differentiating all terms, and setting to zero the linear combinations of anomalous dimensions multiplying ${\cal O}_n^{iA} (1/\cP_\perp^2) {\cal O}_s^{j_n A}$ for each choice of $i$ and $j$. 
The result in \eq{adrel1} constrains the sum of entries in the columns of $\hat \gamma_{{\cal O}_n}^\nu$ to be equal. The fact that only the combination $({\cal O}_n^{qA} + {\cal O}_n^{gA})$ appears also implies that $\gamma_{n\nu}$ is the only combination of entries from $\hat \gamma_{{\cal O}_n}^\nu$ that we need, with analogous results for the soft $\hat \gamma_{{\cal O}_{s_n}}^\nu$.
The root of these results is that the rapidity renormalization only depends on the presence of the octet color index $A$, and not on the choice of fermion or gluon building blocks. Nevertheless we will see that mixing between fermions and gluons operators still plays a crucial role in this universality.

Due to the connection between the rapidity cutoffs in the neighboring soft and $n$-collinear sectors for $O_{ns}^{ij}$ as expressed by \eq{rel0}, we also have the additional relation
\begin{align}  \label{eq:adrel2}
 \gamma_{n\nu}^{qq} + \gamma_{n\nu}^{gq}  
    & = - \gamma_{s_n\nu}^{qq} - \gamma_{s_n\nu}^{gq} \,, 
   \qquad \text{or} \ \ 
   \gamma_{n\nu} = - \gamma_{s_n\nu} 
   \,.
\end{align}
Thus the relevant rapidity anomalous dimensions in the $n$-collinear and soft sector are equal with opposite signs. For the anomalous dimensions for operators appearing in $O_{\bn s}^{ij}$ analogous results to \eqs{adrel1}{adrel2} also hold, simply replacing $n\to \bn$. Therefore we also define 
\begin{align}
\gamma_{\bn\nu} \equiv \gamma_{\bn\nu}^{qq} + \gamma_{\bn\nu}^{gq} = \gamma_{\bn\nu}^{gg} + \gamma_{\bn\nu}^{qg} \,,
\qquad\qquad
\gamma_{s_\bn\nu}\equiv \gamma_{s_\bn\nu}^{qq} + \gamma_{s_\bn\nu}^{gq} = \gamma_{s_\bn\nu}^{gg} + \gamma_{s_\bn\nu}^{qg} 
\,.
\end{align}

There is also no overall $\mu$ dependence for the $n$-soft scattering operator
\begin{align} \label{eq:rel1a}
 \mu \frac{d}{d\mu}   \sum_{ij=q,g} O_{ns}^{ij} = 
   \mu \frac{d}{d\mu} \: ({\cal O}_n^{qA} + {\cal O}_n^{gA}) \frac{1}{\cP_\perp^2} ({\cal O}_s^{q_nA} + {\cal O}_s^{g_nA})    
  = 0 \,. 
\end{align}
This relation is ensured by the fact that there is no $\mu$ dependence for the individual soft and collinear sectors at this order, 
\begin{align} \label{eq:rel2}
  & \mu \frac{d}{d\mu} ({\cal O}_n^{qA} + {\cal O}_n^{gA}) =0
  \,,
 & \mu \frac{d}{d\mu} ({\cal O}_s^{q_nA} + {\cal O}_s^{g_nA}) =0 &
  \,,
\end{align}
which implies even simpler relations for the $\mu$ anomalous dimensions,
\begin{align} \label{eq:adrel3}
  \gamma_{n\mu}\equiv \gamma_{n\mu}^{qq} + \gamma_{n\mu}^{gq} 
    &= \gamma_{n\mu}^{gg} + \gamma_{n\mu}^{qg} = 0
   \,,
   & \gamma_{s_n\mu}\equiv \gamma_{s_n\mu}^{qq} + \gamma_{s_n\mu}^{gq} 
    &= \gamma_{s_n\mu}^{gg} + \gamma_{s_n\mu}^{qg} = 0 
  \,.
\end{align}
Again we have analogous results with $n\to \bn$.

Next we consider the consistency equations for the scattering of two rapidity sectors when there is another rapidity sector in between, namely $n$-$\bn$ scattering. In this case multiple insertions of ${\cal L}_G^{\rm II(0)}$ can mix back into a single insertion through the intermediate rapidity sector. At one-loop the graph built from iterations of Glauber potentials, $O_{ns\bn}^{ik}O_{ns\bn}^{kj}$ with one $n$-collinear and one $\bn$-collinear propagator, are again finite. However we also have graphs with soft loops of gluons or quarks from the T-product of two operators, $\sum_{k=q,g} O_{ns}^{ik} O_{\bn s}^{jk}$, that mix back into a single $O_{ns\bn}^{ij}$. These T-products are precisely the graphs shown in \fig{SCET2_oneloop_matching}c,d.  Due to this mixing the consistency equation for $n$-$\bn$ scattering takes place at the level of demanding that there is no $\nu$ dependence for the time evolution operator induced by the ${\cal L}_G^{\rm II(0)}$ Lagrangian, rather than for the Lagrangian itself. At one-loop we therefore have
\begin{align} \label{eq:rel3}
  \nu \frac{d}{d\nu}\: \sum_{ij=q,g} \bigg( O_{ns\bn}^{ij}(0) 
   + \sum_{kk'=q,g} i\, {\rm T} \int\!\! d^4x\:O_{ns}^{ik}(x) O_{\bn s}^{jk'}(0) \bigg) =0
  \,.
\end{align}
Due to the previous result in \eq{rel0} there is no contribution from the individual operators in the T-product, but when $k=k'$ the T-product itself can still mix into $O_{ns\bn}^{ij}$ through the anomalous dimensions $\gamma_{s\nu}^{Tij}$. The operator $O_{ns\bn}^{ij}$ can also mix back into itself.  Since $O_{ns\bn}^{ij}= {\cal O}_n^{iA} (1/\cP_\perp^2) {\cal O}_s^{AB} (1/\cP_\perp^2) {\cal O}_\bn^{jA}$, and the anomalous dimensions for the collinear operators are already constrained by \eq{adrel1}, this relation will provide a constraint on the remaining coefficients in the soft anomalous dimension $\hat \gamma_{{\cal O}_s}^{\nu}$. At one-loop only $\gamma_{s\nu}^{\rm dir}$, $\gamma_{s\nu}^{Tqq}$, and $\gamma_{s\nu}^{Tgg}$ can possibly be nonzero, so \eq{rel3} gives the relation
\begin{align} \label{eq:adrel4}
  \gamma_{s\nu} \equiv \gamma_{s\nu}^{\rm dir} + \gamma_{s\nu}^{Tqq} + \gamma_{s\nu}^{Tgg} = - \gamma_{n\nu} - \gamma_{\bn \nu}
  \,.
\end{align}
This result encodes a cancellation of rapidity cutoff dependence between the $n$-collinear, $\bn$-collinear, and soft sectors.  

For the $\mu$ anomalous dimension we also have
\begin{align} \label{eq:rel4}
  \mu \frac{d}{d\mu}\: \sum_{ij=q,g} \bigg( O_{ns\bn}^{ij}(0) 
   + \sum_{kk'=q,g} i\, {\rm T} \int\!\! d^4x\:O_{ns}^{ik}(x) O_{\bn s}^{jk'}(0) \bigg) =0
  \,.
\end{align}
Given \eqs{rel1a}{adrel3} this implies that at one-loop 
\begin{align}
  \gamma_{s\mu}\equiv  \gamma_{s\mu}^{\rm dir} + \gamma_{s\mu}^{Tqq} + \gamma_{s\mu}^{Tgg} =0 
  \,.
\end{align}

We will see in the next section that if we consider the virtual $\nu$-anomalous dimensions alone, then they depend on a logarithm of the IR regulator, $\ln(m^2)$. This dependence is canceled when we consider the full anomalous dimensions obtained by the sum of divergences in virtual and real radiation graphs.  Because of this cancellation, the relations derived here in \eqs{adrel2}{adrel4} also apply for the corresponding real radiation graphs.

\subsubsection{One-Loop Virtual Anomalous Dimension Results}
\label{sec:oneloopregge}

Here we consider the one-loop calculation of the various virtual contributions to anomalous dimensions discussed in \sec{adnotations} with the goal of identifying the non-trivial terms and cross-checking the various relations discussed in \sec{adrelations}. With external quarks the results can be determined from the SCET$_{\rm II}$ diagrams given in \sec{loop2match}, while our results below with external gluons required additional calculations.

First consider the $n$-collinear sector with the bilinear quark and gluon operators ${\cal O}_n^{qA}$ and ${\cal O}_n^{gA}$.  For ${\cal O}_n^{qA}$ mixing back into ${\cal O}_n^{qA}$ there are $W$ Wilson line graphs and the vertex graph (plus wavefunction renormalization), all of which can be read off  from the results in \sec{loop2match} by stripping off the appropriate prefactor that is related to the other sectors.  We have
\begin{align} \label{eq:OqtoOq}
& \includegraphics[width=0.17\columnwidth]{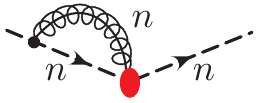}
  \hspace{0.1cm}
\raisebox{0.3cm}{+}  \hspace{0.1cm}
\includegraphics[width=0.17\columnwidth]{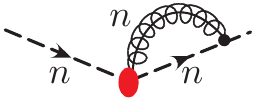}
 \hspace{0.1cm}
\raisebox{0.3cm}{+}  \hspace{0.2cm}
\includegraphics[width=0.16\columnwidth]{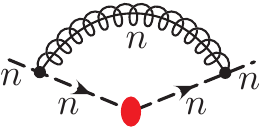}
 \hspace{0.1cm}
\raisebox{0.3cm}{+\ 	\text{w.fn. renorm} \ } 
\nn\\
&\qquad  = {\cal S}^{nq}\, \frac{\alpha_s C_A}{4\pi}  \bigg[
w^2\frac{2\,h\big(\epsilon,\frac{\mu^2}{m^2}\big)}{\eta}  + w^2\frac{2}{\epsilon} \ln\Big(\frac{\nu}{\bn\cdot p}\Big) + \frac{3}{2\epsilon}\bigg]
= {\cal S}^{nq}\, \delta V_n^{qq} 
\,,
\end{align}
where the spinors are contained in the tree level matrix element ${\cal S}^{nq} = \bar u_n T^A \frac{\bnslash}{2} u_n$, and $h(\epsilon,\mu^2/m^2)$ was defined in \eq{h}.  Here and below we will drop the factors of $w$ which multiple the pure $1/\epsilon$ poles. To obtain \eq{OqtoOq} we have taken \fig{SCET2_oneloop_matching}g,h,i, stripped off a prefactor of $i (\bar v_\bn \bar T^A \frac{\nslash}{2} v_\bn) (8\pi\alpha_s)/t$, which includes the factors associated with the tree level matrix element of the non $n$-collinear parts of the operator, namely $(1/\cP_\perp^2){\cal O}_s^{AB}(1/\cP_\perp^2)O_\bn^{B}$, as well as an overall $i$. Since we are interested in determining anomalous dimensions, only the divergent terms that are needed to determine the $\delta V_n^{qq}$ counterterm are shown in \eq{OqtoOq} and the results below, and we now include factors of the rapidity coupling $w^2=w^2(\nu)$ (which is needed to determine the rapidity anomalous dimension, and then can be set to $1$). Just as discussed in the matching calculation, the collinear tadpole loop graphs vanish due to their soft zero-bin subtractions
\begin{align}
\raisebox{-0.8cm}{
  \includegraphics[width=0.13\columnwidth]{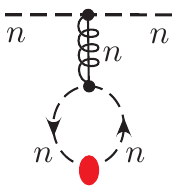}
  } 
\hspace{0.1cm} 
& =0
\,, 
& \raisebox{-0.8cm}{
  \includegraphics[width=0.12\columnwidth]{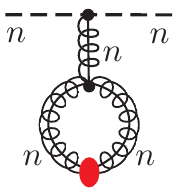}
  } 
\hspace{0.1cm} 
&= 0 \,.
\end{align}
There is only one nonzero graph where the operator ${\cal O}_n^{gA}$ mixes into ${\cal O}_n^{qA}$, namely the V-graph. This result can be again found from the results in \sec{loop2match}, and determines the $\delta V_n^{gq}$ counterterm, 
\begin{align}
\raisebox{-0.6cm}{
  \includegraphics[width=0.17\columnwidth]{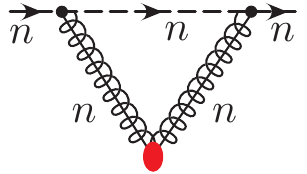}
  }  \hspace{0.1cm} 
= {\cal S}^{nq}\, \frac{\alpha_s C_A}{4\pi}  \bigg[
w^2\frac{2\,g\big(\epsilon,\frac{\mu^2}{-t}\big)}{\eta}  -w^2 \frac{2}{\epsilon} \ln\Big(\frac{\nu}{\bn\cdot p}\Big) - \frac{3}{2\epsilon}\bigg]
= {\cal S}^{nq}\, \delta V_n^{gq}  \,,
\end{align}
where the function $g(\epsilon,\mu^2/(-t))$ is given in \eq{g}. 

To determine the remaining $n$-collinear counterterms we need to consider graphs involving external gluons, which require new calculations.  Rather than giving a detailed discussion of these diagrams we simply relegate non-trivial ingredients like the 3-gluon vertex from ${\cal O}_n^{gg}$ to \app{useful}, and quote here the final results for the divergent terms (using Feynman gauge):
\begin{align} \label{eq:Vggresults}
\raisebox{-0.4cm}{
  \includegraphics[width=0.17\columnwidth]{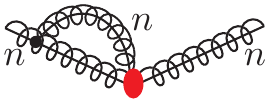}
   }  \hspace{0.1cm} + \hspace{0.1cm}
\raisebox{-0.4cm}{
  \includegraphics[width=0.17\columnwidth]{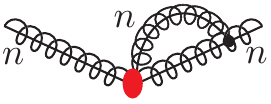}
   }  \hspace{0.1cm} 
&= {\cal S}^{ng}\, \frac{\alpha_s C_A}{4\pi}  \bigg[ w^2\frac{2\,h\big(\epsilon,\frac{\mu^2}{m^2}\big)}{\eta}  + w^2\frac{2}{\epsilon} \ln\Big(\frac{\nu}{\bn\cdot p}\Big) - \frac{1}{2\epsilon}\bigg] 
\,, \nn\\[5pt]
\raisebox{-0.4cm}{
   \includegraphics[width=0.16\columnwidth]{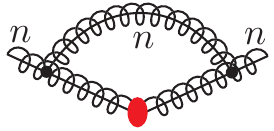}
  } 
\hspace{0.1cm} 
&= {\cal S}^{ng}\, \frac{\alpha_s C_A}{4\pi}  \bigg[  w^2 \frac{2\,g\big(\epsilon,\frac{\mu^2}{-t}\big)}{\eta}  -w^2 \frac{2}{\epsilon} \ln\Big(\frac{\nu}{\bn\cdot p}\Big) - \frac{7}{6\epsilon}\bigg] 
\,, \nn\\[10pt]
\text{gluon w.fn.} &= {\cal S}^{ng}\,  \frac{\alpha_s}{4\pi} \bigg[
\frac{5}{3\epsilon}\, C_A -  \frac{4}{3\epsilon}\, n_f T_F  \bigg] \,,
\end{align} 
where ${\cal S}^{ng} = i f^{ABC} g_\perp^{\beta\gamma} \bn\cdot p\, \varepsilon^{B}_\beta \varepsilon^{C}_\gamma$ is the tree level matrix element of ${\cal O}_n^{gA}$. Just like in the quark calculation, the collinear gluon tadpole graphs give zero due to their soft zero-bin subtraction. There is also a graph with the four-gluon vertex which has a vanishing integral even before the zero-bin subtraction. Thus we have
\begin{align}
  \raisebox{-0.8cm}{
  \includegraphics[width=0.12\columnwidth]{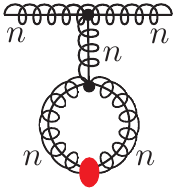}
  } \hspace{0.1cm} 
  &=  {\cal S}^{ng}\, \frac{\alpha_s C_A}{4\pi}  \bigg[  \ \frac{5}{3\epsilon}
  - \frac{5}{3\epsilon} \ \bigg] =0
  \,, 
 & \raisebox{-0.8cm}{
  \includegraphics[width=0.12\columnwidth]{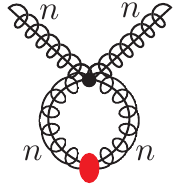}
  }  \hspace{0.1cm} 
   &= 0 \,.
\end{align}
The remaining contributions determine the $\delta V_n^{gg}$ counterterm to be
\begin{align}
\delta V_n^{gg} &= \frac{\alpha_s}{4\pi}  \bigg[ \frac{2w^2h\big(\epsilon,\frac{\mu^2}{m^2}\big)+2w^2g\big(\epsilon,\frac{\mu^2}{-t}\big)}{\eta}\, C_A - \frac{4}{3\epsilon}\, n_f T_F \bigg] 
\,.
\end{align}
It is interesting to note that  that the $C_A/\epsilon$ terms cancel.

Finally, we consider the mixing of ${\cal O}_n^{qA}$ into ${\cal O}_n^{gA}$. The relevant diagrams are
\begin{align}  \label{eq:Vqgresults}
\raisebox{-0.5cm}{
  \includegraphics[width=0.19\columnwidth]{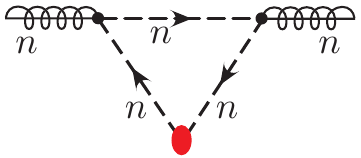}
   }  \hspace{0.1cm} + \hspace{0.1cm}
\raisebox{-0.5cm}{
  \includegraphics[width=0.19\columnwidth]{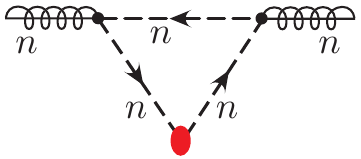}
  }  \hspace{0.1cm} 
&= {\cal S}^{ng}\, \frac{\alpha_s n_f T_F}{4\pi}  \bigg[\ \frac{4}{3\epsilon} \ \bigg] 
\,,
\end{align}
where we have summed over all possible $n_f$ light flavors that can appear in the ${\cal O}_n^{q A}$ operator. Again the collinear quark loop tadpole graph is exactly canceled by the soft zero-bin subtraction,
\begin{align}
 \raisebox{-0.8cm}{
  \includegraphics[width=0.12\columnwidth]{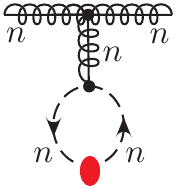}
  } 
 \hspace{0.1cm} 
 &= {\cal S}^{ng}\, \frac{\alpha_s n_f T_F}{4\pi}  \bigg[\ -\frac{4}{3\epsilon}  +\frac{4}{3\epsilon}\ \bigg] =0
 \,.
\end{align}
Thus, the result in \eq{Vqgresults} yields the counterterm for ${\cal O}_n^{qA}$ mixing into ${\cal O}_n^{gA}$,
\begin{align}
\delta V_n^{qg} &= \frac{\alpha_s}{4\pi} \bigg[ \: \frac{4}{3\epsilon}\,  n_f T_F\: \bigg] 
\,.
\end{align}

At one-loop order the $n$-collinear rapidity anomalous dimension contributions are given by $\gamma_{n\nu}^{ij} = - (\nu d/d\nu) \delta V_n^{ij}$. Differentiating both the explicit $\ln \nu$ dependence and the $\nu$ dependence in the coupling $w$ by using $(\nu d/d\nu) w^2 = -\eta w^2$ (then setting the renormalized $w=1$), and expanding to ${\cal O}(\epsilon^0)$, we have
\begin{align}
\gamma_{n\nu}^{qq} &= -\frac{\alpha_s C_A}{4\pi} \bigg[ -2h(\epsilon,\mu^2/m^2) +\frac{2}{\epsilon}\bigg]
= \frac{\alpha_s(\mu) C_A}{2\pi}\: \ln\Big(\frac{\mu^2}{m^2}\Big)
\,, \\
\gamma_{n\nu}^{gq} &= -\frac{\alpha_s C_A}{4\pi} \bigg[
-2g(\epsilon,\mu^2/(-t)) -\frac{2}{\epsilon}\bigg]
= \frac{\alpha_s(\mu) C_A}{2\pi}\: \ln\Big(\frac{-t}{\mu^2}\Big)
\,,\nn\\
\gamma_{n\nu}^{gg} &= -\frac{\alpha_s C_A}{4\pi} \bigg[
-2g(\epsilon,\mu^2/(-t)) -2h(\epsilon,\mu^2/m^2)\bigg]
= \frac{\alpha_s(\mu) C_A}{2\pi}\: \ln\Big(\frac{-t}{m^2}\Big)
\,,\nn\\
\gamma_{n\nu}^{qg} &= 0
\,. \nn
\end{align}
For the $\mu$ anomalous dimensions at one-loop we have $\gamma_{n\mu}^{ij}=-(\mu d/d\mu) \delta V_n^{ij}$. Noting that  $\alpha_s(\mu) g(\epsilon,\mu^2/(-t))$ and $\alpha_s(\mu) h(\epsilon,\mu^2/m^2)$ are both  $\mu$-independent, and recalling that $(\mu d/d\mu) \alpha_s(\mu) = -2\epsilon\alpha_s(\mu) + {\cal O}(\alpha_s^2)$ we find
\begin{align}
\gamma_{n\mu}^{qq} &
= \frac{\alpha_s(\mu) C_A}{2\pi}\: \bigg[ 2\ln\Big(\frac{\nu}{\bn\cdot p}\Big) + \frac{3}{2} \bigg]
\,, \\
\gamma_{n\mu}^{gq} &
= -\frac{\alpha_s(\mu) C_A}{2\pi}\: \bigg[ 2\ln\Big(\frac{\nu}{\bn\cdot p}\Big) + \frac{3}{2} \bigg]
\,,\nn\\
\gamma_{n\mu}^{gg} &
= -\frac{2\alpha_s(\mu) n_F T_F }{3\pi}\:
\,,\nn\\
\gamma_{n\mu}^{qg} &
=  \frac{2\alpha_s(\mu) n_F T_F }{3\pi}\:
\,. \nn
\end{align}
Note that these results satisfy the necessary condition for the paths in $\nu$ and $\mu$ space to be independent~\cite{Chiu:2012ir},  $(\nu d/d\nu) \gamma_{n\mu}^{ij} = (\mu d/d\mu) \gamma_{n\nu}^{ij}$. 
From these results we can immediately check that we reproduce the first relation in each of \eq{adrel1} and \eq{adrel3}, $\gamma_{n\nu}^{qq}+\gamma_{n\nu}^{gq}=\gamma_{n\nu}^{gg}+\gamma_{n\nu}^{qg}$ and $\gamma_{n\mu}^{qq}+\gamma_{n\mu}^{gq}=\gamma_{n\mu}^{gg}+\gamma_{n\mu}^{qg}=0$. Thus there is no overall $\mu$ anomalous dimension for the relevant combination of operators, $({\cal O}_n^{qA} + {\cal O}_n^{gA})$, as anticipated. It is interesting to note that this occurs due to a cancellation of terms between the anomalous dimensions generated by the two individual operators.  We also obtain the relevant rapidity anomalous dimension for $({\cal O}_n^{qA} + {\cal O}_n^{gA})$ which is
\begin{align}  \label{eq:gamnnu}
\gamma_{n\nu} = \frac{\alpha_s C_A}{2\pi} \: \ln\Big(\frac{-t}{m^2}\Big)
\,.
\end{align}
Again mixing plays a key role in obtaining this result. In particular, the graph that contributes the $\ln(-t)$ in the anomalous dimension for ${\cal O}_n^{qA}$ is initiated by gluons, and enters through $\gamma_{n\nu}^{gq}$. 

Next we turn to the soft anomalous dimensions.  For the operators ${\cal O}_s^{q_n A}$ and ${\cal O}_s^{g_n A}$ the contributing diagrams are very similar to our analysis of the $n$-collinear operators above, but there is not a one-to-one correspondence to the diagrams, and the soft calculation also has contributions related to the running coupling that appears explicitly in ${\cal O}_s^{i_n A}$. Due to the similarities we do not bother to give a detailed discussion of the various diagrams. The key difference is that for the soft graphs the rapidity regulator appears as $|\bn\cdot k - n\cdot k|^{-\eta}$ rather than $|\bn\cdot k|^{-\eta}$, which reverses the sign of the $1/\eta$ poles. For this reason, the final rapidity anomalous dimension for the relevant combination of single color index soft operators, (${\cal O}_s^{q_n A}+{\cal O}_s^{g_n A})$ has the opposite sign to the $n$-collinear case,
\begin{align} \label{eq:gamsnu}
\gamma_{s_n\nu} =\gamma^{qq}_{s_n\nu}+\gamma^{gq}_{s_n\nu} =\gamma^{gg}_{s_n\nu}+\gamma^{qg}_{s_n\nu}= - \frac{\alpha_s C_A}{2\pi} \: \ln\Big(\frac{-t}{m^2}\Big)
\,.
\end{align}
Together \eq{gamnnu} and \eq{gamsnu} satisfy the expected relation in \eq{adrel2}, that $\gamma_{s_n\nu}= -\gamma_{n\nu}$. Just like for the collinear operators there is no overall $\mu$ anomalous dimension for (${\cal O}_s^{q_n A}+{\cal O}_s^{g_n A})$. 

The final operator to consider is the two index soft operator ${\cal O}_s^{AB}$. The results needed for the renormalization of this operator at one-loop can all be read off of those from \sec{loop2match}. There are two types of contribution, the mixing of ${\cal O}_s^{AB}$ back into ${\cal O}_s^{AB}$, and contributions from products of the single index soft operators, ${\cal O}_s^{k_n A} {\cal O}_s^{k_\bn B}$ mixing into ${\cal O}_s^{AB}$.  Considering ${\cal O}_s^{AB}$ we have the flower graph and the counterterm from the $\alpha_s$ prefactor
\begin{align}  \label{eq:OabtoOab}
\raisebox{-0.7cm}{
  \includegraphics[width=0.17\columnwidth]{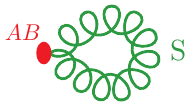}
  } +Z_{\alpha_s}\text{-counterterm}  
 \hspace{0.1cm} 
&= -  4\delta^{AB} C_A\,  \alpha_s^2 w^2\,t \:
\bigg\{ \frac{2}{\eta} h(\epsilon,\mu^2/m^2)
- \frac{1}{\epsilon^2}- \frac{1}{\epsilon} \ln\Big(\frac{\mu^2}{\nu^2}\Big) 
\bigg\} \nn\\
 &\quad - 2\delta^{AB} \alpha_s^2\, t\: 
 \bigg( \frac{11 C_A}{3\epsilon} - \frac{4 n_f T_F}{3\epsilon} \bigg) 
  \nn\\
 &= \delta^{AB}\, (8\pi\alpha_s)\, t\:  \delta V_s
\,, 
\end{align}
where $h(\epsilon,\mu^2/m^2)$ was defined in \eq{h}. To obtain \eq{OabtoOab} we have taken the result for \fig{SCET2_oneloop_matching}c, and stripped off a prefactor of $i (\bar u_n T^A \frac{\bnslash}{2} u_n)(\bar v_\bn \bar T^B\frac{\nslash}{2} v_\bn)/t^2$. This prefactor includes terms associated with the tree level matrix element of the non soft parts of the operator, namely $O_n^{A}(1/\cP_\perp^2)$ on one side and $(1/\cP_\perp^2)O_\bn^{B}$ on the other, as well as an overall $i$. From the remaining terms we again show only the divergences in \eq{OabtoOab}, since only the renormalization is being considered here. This result determines
\begin{align}
  \delta V_s = -\frac{\alpha_s(\mu)}{4\pi} C_A w^2  \bigg\{ \frac{4}{\eta} h(\epsilon,\mu^2/m^2)
- \frac{2}{\epsilon^2}- \frac{2}{\epsilon} \ln\Big(\frac{\mu^2}{\nu^2}\Big) 
\bigg\} 
 - \frac{\alpha_s(\mu)}{4\pi} \bigg( \frac{11 C_A}{3\epsilon} - \frac{4n_f T_F}{3\epsilon} \bigg) \,.
\end{align}
The final contributions come from T-products of two soft operators ${\cal O}_s^{k_n A} {\cal O}_s^{k_\bn B}$ for $k=g,q$, which occurred in our one-loop matching calculation in \fig{SCET2_oneloop_matching}c,d.

\begin{align}  \label{eq:OsOstoOab}
\raisebox{-1.1cm}{
  \includegraphics[width=0.1\columnwidth]{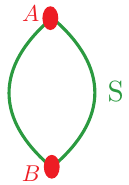}
  }  \hspace{0.15cm} 
&= + \delta^{AB} n_f T_F\,  \alpha_s^2\,t \:
\bigg( -\frac{8}{3\epsilon} \bigg)
 = \delta^{AB}\, (8\pi\alpha_s)\, t\:  \delta V_s^{Tqq}
\,, \\
 \raisebox{-1.1cm}{
   \includegraphics[width=0.1\columnwidth]{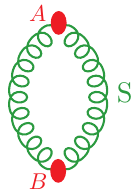}
  }  \hspace{0.15cm} 
 &= -  4\delta^{AB} C_A\, w^2 \alpha_s^2\,t \:
 \bigg\{  \frac{2}{\eta} g(\epsilon,\mu^2/(-t))
+ \frac{1}{\epsilon^2} + \frac{1}{\epsilon} \ln\Big(\frac{\mu^2}{\nu^2}\Big) 
  - \frac{11}{6\epsilon}
 \bigg\} = \delta^{AB}\, (8\pi\alpha_s)\, t\:  \delta V_s^{Tgg}
 \,, \nn
\end{align}
giving the results
\begin{align}
 \delta V_s^{Tqq} &= \frac{\alpha_s(\mu)}{4\pi} \bigg( - \frac{4 n_f T_F}{3\epsilon}
  \bigg)
  \,,\\
 \delta V_s^{Tgg} &= -\frac{\alpha_s(\mu)}{4\pi} C_A w^2 \bigg\{ \frac{4}{\eta} 
    g(\epsilon,\mu^2/(-t)) + \frac{2}{\epsilon^2} + \frac{2}{\epsilon} \ln\Big(\frac{\mu^2}{\nu^2}\Big) - \frac{11}{3\epsilon} \bigg\}
  \,. \nn
\end{align}

At one-loop order the soft rapidity anomalous dimension contributions are given by $\gamma_{s\nu}^X = - (\nu d/d\nu) \delta V_s^{X}$. Differentiating both the explicit $\ln \nu$ dependence and the $\nu$ dependence in the coupling $w$ by using $(\nu d/d\nu) w^2 = -\eta w^2$ (then setting the renormalized $w=1$), and expanding to ${\cal O}(\epsilon^0)$,  we have
\begin{align}
\gamma_{s\nu}^{\rm dir} &= \frac{\alpha_s C_A}{4\pi} \bigg[ -4h(\epsilon,\mu^2/m^2) +\frac{4}{\epsilon}\bigg]
= - \frac{\alpha_s(\mu) C_A}{\pi}\: \ln\Big(\frac{\mu^2}{m^2}\Big)
\,, \\
\gamma_{s\nu}^{Tqq} &= 0
\,,\nn\\
\gamma_{s\nu}^{Tgg} &= \frac{\alpha_s C_A}{4\pi} \bigg[
-4g(\epsilon,\mu^2/(-t)) - \frac{4}{\epsilon} \bigg]
= - \frac{\alpha_s(\mu) C_A}{\pi}\: \ln\Big(\frac{-t}{\mu^2}\Big)
\,. \nn
\end{align}
Thus for the relevant soft virtual contribution to the rapidity anomalous dimension, $ \gamma_{s\nu} = \gamma_{s\nu}^{\rm dir}+\gamma_{s\nu}^{Tqq}+\gamma_{s\nu}^{Tgg}$, we obtain
\begin{align}  \label{eq:gamsnu2}
 \gamma_{s\nu} &= -\frac{\alpha_s(\mu) C_A}{\pi}\: \ln\Big(\frac{-t}{m^2}\Big) 
  \,.
\end{align}
Utilizing \eq{gamnnu}, and noting that $\gamma_{\bn\nu}=\gamma_{n\mu}$, we see that $\gamma_{s\nu} = -\gamma_{n\nu} -\gamma_{\bn\nu}$ as expected.

For the $\mu$ anomalous dimensions at one-loop we have $\gamma_{s\mu}^{X}=-(\mu d/d\mu) \delta V_s^{X}$. Noting that the combinations $\alpha_s(\mu) g(\epsilon,\mu^2/(-t))$ and $\alpha_s(\mu) h(\epsilon,\mu^2/m^2)$ are $\mu$-independent, and recalling that $(\mu d/d\mu) \alpha_s(\mu) = -2\epsilon\alpha_s(\mu) + {\cal O}(\alpha_s^2)$ we find
\begin{align}
\gamma_{s\mu}^{\rm dir} &
= \frac{\alpha_s(\mu)}{2\pi}\: \bigg[ 2 C_A \ln\Big(\frac{\mu^2}{\nu^2}\Big) - \frac{11 C_A}{3} + \frac{4 n_f T_F}{3} \bigg]
\,, \\
\gamma_{s\mu}^{Tqq} &
= \frac{\alpha_s(\mu)}{2\pi}\:  \bigg[ - \frac{4 n_f T_F}{3}  \bigg]
\,,\nn\\
\gamma_{s\mu}^{Tgg} &
= \frac{\alpha_s(\mu)}{2\pi}\:  \bigg[ -2 C_A \ln\Big(\frac{\mu^2}{\nu^2}\Big)
  + \frac{11 C_A}{3}  \bigg]
\,. \nn
\end{align}
Thus for the only relevant combination of $\mu$-anomalous dimensions we find $\gamma_{s\mu} = \gamma_{s\mu}^{\rm dir}+\gamma_{s\mu}^{Tqq}+\gamma_{s\mu}^{Tgg}=0$, so there is no $\mu$-evolution for the \SCETb soft operator as expected.

\subsubsection{Solving the Virtual Rapidity RGE: Reggeization}
\label{sec:RRGEsoln}

As we have seen in the matching calculation, the result for the one loop scattering amplitude for collinear particles, involves a logarithm of  the ratio $s/(-t)$.  To resum these logarithms to higher orders in perturbation theory we can utilize the rapidity renormalization group to encode the large logarithms in an evolution factor and ensure that all of the individual factorized pieces of the amplitude are evaluated at the appropriate rapidity scales $\nu$ where they do not have large logarithms.  From \eq{soft_total_loop} we see that the  soft piece of the factorized amplitude will not have large logarithms if we choose $\nu=\mu=\sqrt{-t}$, while from \eq{collinear_total_loop} we see that the collinear parts of the factorized amplitude will not have large logarithms if we choose $\nu=\sqrt{s}$. In fixed order perturbation theory the large logarithms arise because only one of these two choices for $\nu$ is possible.

Choosing $\nu=\mu=\sqrt{-t}$, the soft piece of the amplitude does not have large logarithms, and the large logs reside in the two collinear amplitudes. Therefore we must use the rapidity renormalization group to connect these collinear amplitudes to the rapidity scale where their logarithms are minimized. Since we are interested in the leading-logarithmic resummation we only need as boundary conditions the matrix elements at tree level, and hence it suffices to sum the logarithms in $O_{ns\bn}^{ij}$ (given in \eq{Onsnb}) with $i,j$ summed over both quarks and gluons. From the first result in \eq{rel1} we have the rapidity evolution equation
\begin{align} \label{eq:rrgeqtn}
\frac{d}{d\log \nu}({\cal O}_n^{qA} + {\cal O}_n^{gA}) = \gamma_{n \nu} ({\cal O}_n^{qA} + {\cal O}_n^{gA}) \,,
\end{align}
where the leading order anomalous dimension $\gamma_{n\nu}=\frac{\alpha_s(\mu) C_A}{2\pi} \ln(-t/m^2)$ was computed in \eq{gamnnu}. Since this anomalous dimension is independent of $\nu$ it is trivial to integrate \eq{rrgeqtn}, and thus obtain the relation between the renormalized collinear operators evaluated at two different rapidity scales $\nu$:
\begin{align}
  & 
  ({\cal O}_n^{qA} + {\cal O}_n^{gA})(\nu_1) = \Big(\frac{\nu_0}{\nu_1}\Big)^{-\gamma_{n\nu}}
    ({\cal O}_n^{qA} + {\cal O}_n^{gA})(\nu_0) \,.
\end{align}
Taking $\nu_1=\sqrt{-t}$ and $\nu_0=\sqrt{s}$ we can now connect the collinear operator to the scale $\nu=\sqrt{s}$ where logarithms in its amplitude are minimized,
\begin{align} 
  & 
  ({\cal O}_n^{qA} + {\cal O}_n^{gA})(\nu=\sqrt{-t}) = \Big(\frac{s}{-t}\Big)^{-\gamma_{n\nu}/2}
    ({\cal O}_n^{qA} + {\cal O}_n^{gA})(\nu=\sqrt{s}) \,.
\end{align}
For ${\cal O}_\bn^{jA}$ we have the same rapidity anomalous dimension equation with $\gamma_{\bn\nu}=\gamma_{n\nu}$, and hence the same resummed result, namely
\begin{align} 
  & 
  ({\cal O}_\bn^{qA} + {\cal O}_\bn^{gA})(\nu=\sqrt{-t}) = \Big(\frac{s}{-t}\Big)^{-\gamma_{n\nu}/2}
    ({\cal O}_\bn^{qA} + {\cal O}_\bn^{gA})(\nu=\sqrt{s}) \,.
\end{align}

Putting these results together the leading logs are summed in the operator $O_{ns\bn}^{ij}$ by, 
\begin{align} \label{eq:Reggeresult}
  &\sum_{i,j=q,g} O_{ns\bn}^{ij} (\nu=\sqrt{-t})
   \\ 
   &\ \ = 
  ({\cal O}_n^{qA} + {\cal O}_n^{gA})(\nu=\sqrt{-t}) 
  \frac{1}{\cP_\perp^2} {\cal O}_s^{AB}(\nu=\sqrt{-t}) \frac{1}{\cP_\perp^2}
  ({\cal O}_\bn^{qB} + {\cal O}_\bn^{gB})(\nu=\sqrt{-t}) 
   \nn\\
  &\ \ = \Big(\frac{s}{-t}\Big)^{-\gamma_{n\nu}}
    ({\cal O}_n^{qA} + {\cal O}_n^{gA})(\nu=\sqrt{s})
   \frac{1}{\cP_\perp^2} {\cal O}_s^{AB}(\nu=\sqrt{-t}) \frac{1}{\cP_\perp^2}
  ({\cal O}_\bn^{qB} + {\cal O}_\bn^{gB})(\nu=\sqrt{s}) 
   \,. \nn
\end{align}
For the renormalized operators on the right-hand side there are no large logarithms in their matrix elements, since they are evaluated at the scales $\nu$ which minimize their respective rapidity logarithms.  (If we had instead started the evolution at $\nu=\sqrt{s}$ then there would be no evolution for the collinear operators, and the soft operator's evolution would generate this same result.)

The factor of $(\frac{s}{-t})^{-\gamma_{n\nu}}$ in \eq{Reggeresult} is the standard Reggeized gluon result, where $\alpha_g = -\gamma_{n\nu}$ is the gluon Regge exponent. At the leading logarithmic resummed order we have this same factor for quarks and gluon channels. At higher order there are distinctions between the channels, see for example~\cite{DelDuca:2014cya}, in particular factors of $(\frac{s}{t})^{-\gamma_{n\nu}}$ also appear.  Since $(\frac{s}{t})^{-\gamma_{n\nu}} =  (\frac{s}{-t})^{-\gamma_{n\nu}} e^{i\pi\gamma_{n\mu}}$ the two factors differ only at next-to-leading logarithmic order.


\subsection{One Loop Matching in \SCETa}
\label{sec:loop1match}

In this section we repeat the matching calculation carried out in \sec{loop2match}, but in the theory \SCETa. Although our focus in the majority of this paper is on \SCETb, we mentioned in \sec{GlauberSCET} that, prior to the BPS field redefinition, the Glauber Lagrangian for \SCETa is identical 
in form to that for \SCETb, and only differs in the form of its 0-bin subtractions. This section will serve to check at one-loop that we have the proper form of the Glauber Lagrangian for \SCETa, and highlight some differences between the results in various sectors between \SCETb and \SCETa.  The main distinction for \SCETa is the presence of ultrasoft modes, which live at a scale parametrically smaller than the soft, collinear, and Glauber modes. Here Glauber exchange graphs also have 0-bin subtractions due to the ultrasoft region, and there are additional subtractions for soft and collinear loop diagrams. Because we are studying Glauber dependent processes in \SCETa, we must simultaneously consider soft and ultrasoft diagrams.

Just as in \sec{loop2match} we consider quark-antiquark forward scattering of energetic particles with the same external momenta.  To regulate IR divergences in the full theory in a manner that is suitable for \SCETa, we take the external particles to be offshell \footnote{Utilizing  an off-shell regulator disallows the use of the BPS field redefinition.} with
\begin{align}
  p_2^2 = p_3^2 \equiv -p^2 < 0 \,, \qquad p_1^2 = p_4^2 \equiv -\bar p^2 < 0 \,.
\end{align}
For this \SCETa matching calculation we no longer have a gluon mass in any diagrams. The hierarchy of invariant mass scales here is $s \gg -t\sim p^2\sim \bar p^2 \gg p^2 \bar p^2 /s$, where the soft, Glauber, and collinear modes live at the intermediate scale, and only the ultrasoft modes live at the small $p^2 \bar p^2/s$ ``see-saw scale''.  When necessary we will also include the $\eta$-regulator to handle rapidity divergences that are not regulated by the combination of the offshellness and dimensional regularization.  In \SCETa the 0-bin subtractions for the collinear, soft and Glauber one loop graphs were given above in \eq{scet1subt}. We will again make use of the Dirac and color decomposition ${\cal S}_{1,2,3,4}^{n\bn}$ given in \eq{spinors}. 

The full theory diagrams for this matching calculation are the same ones shown in \fig{full_oneloop_matching}, but now calculated with the offshellness IR regulator. Here we simply quote the sum of all the diagrams
\begin{align}\label{eq:full_oneloop_result1}
  \text{Full } & \text{Theory} = {\rm Figs.}~\ref{fig:full_oneloop_matching} +\text{$Z_g$ c.t.} \\
  &= \frac{i\alpha_s^2}{t}  {\cal S}_1^{n\bn}
   \bigg[ 8 i \pi \ln\Big( \frac{-s t}{p^2 \bar p^2}\Big) +4\pi^2 \bigg] 
  \nn\\[5pt] 
  &\quad 
  + \frac{i\alpha_s^2}{t}  {\cal S}_2^{n\bn}
   \bigg[ -4 \ln^2\Big(\frac{-t}{p^2}\Big)-4 \ln^2\Big(\frac{-t}{\bar p^2}\Big)
          + 6 \ln\Big(\frac{-t}{p^2}\Big)  + 6 \ln\Big(\frac{-t}{\bar p^2}\Big)
        -4 - \frac{8\pi^2}{3} \bigg]
  \nn \\[5pt]
  &\quad 
  + \frac{i\alpha_s^2}{t}  {\cal S}_3^{n\bn}
   \bigg[ -2 \ln^2\Big(\frac{-st}{p^2\bar p^2}\Big)
    +4 \ln^2\Big(\frac{-t}{\bar p^2}\Big) +4 \ln^2\Big(\frac{-t}{\bar p^2}\Big)
    + \frac{22}{3} \ln\Big(\frac{\mu^2}{-t}\Big)     
    +\frac{170}{9}  + 2\pi^2 \bigg]
  \nn \\[5pt]
  &\quad   + \frac{i\alpha_s^2}{t}  {\cal S}_4^{n\bn}
   \bigg[ -\frac{8}{3} \ln\Big(\frac{\mu^2}{-t}\Big) -\frac{40}{9}  
  \bigg] 
  . \nn
\end{align}
Interesting differences from the case of the mass IR regulator in \eq{full_oneloop_result} include the presence of $\ln^2(s)$ in the $C_A T^A \otimes T^A$ color structure, as well as a $i\pi \ln(s)$ in the phase term. 

\begin{figure}[t!]
	%
	%
	%
%
\begin{center}
  \raisebox{2cm}{
  \hspace{-0.4cm}
  a)\hspace{3.cm} 
  b)\hspace{2.7cm} 
  c)\hspace{3cm} 
  d)\hspace{2.9cm} 
  e)\hspace{3cm} 
   } \\[-53pt]
\hspace{-0.4cm}
\raisebox{0.3cm}{
\includegraphics[width=0.17\columnwidth]{figs/EFT_loop1_qqqq}
}\hspace{0.2cm}
\raisebox{0.3cm}{
\includegraphics[width=0.17\columnwidth]{figs/EFT_loop2_qqqq}
}\hspace{0.2cm}
\includegraphics[width=0.16\columnwidth]{figs/EFT_loop3_qqqq_ext}
\hspace{0.2cm}
\includegraphics[width=0.16\columnwidth]{figs/EFT_loop4_qqqq_ext}
\hspace{0.2cm}
\raisebox{0.3cm}{
\includegraphics[width=0.17\columnwidth]{figs/EFT_loop5_qqqq}
}\hspace{0.2cm}
\\[8pt]
\raisebox{2cm}{
  \hspace{-0.4cm}
  f)\hspace{2.9cm} 
  g)\hspace{3.0cm} 
  h)\hspace{3.2cm} 
  i)\hspace{2.7cm} 
  j)\hspace{3cm} 
   } \\[-55pt]
\hspace{-0.55cm}
\includegraphics[width=0.19\columnwidth]{figs/EFT_loop8_qqqq}
\hspace{0.2cm}
\includegraphics[width=0.19\columnwidth]{figs/EFT_loop10_qqqq}
\hspace{0.2cm}
\includegraphics[width=0.19\columnwidth]{figs/EFT_loop11_qqqq}
\hspace{0.2cm}
\includegraphics[width=0.16\columnwidth]{figs/EFT_loop6_qqqq}
\hspace{0.2cm}
\raisebox{0.5cm}{
\includegraphics[width=0.17\columnwidth]{figs/EFT_loop14_qqqq}
}\hspace{0.2cm}
\\[10pt]
  \raisebox{2cm}{
  \hspace{-0.4cm}
  k)\hspace{2.9cm} 
  l)\hspace{3.cm} 
  m)\hspace{3.1cm} 
  n)\hspace{2.6cm} 
  o)\hspace{3cm} 
   } \\[-53pt]
\hspace{-0.6cm}
\includegraphics[width=0.19\columnwidth]{figs/EFT_loop9_qqqq}
\hspace{0.2cm}
\raisebox{-0.2cm}{
\includegraphics[width=0.18\columnwidth]{figs/EFT_loop12_qqqq}
}\hspace{0.2cm}
\raisebox{-0.2cm}{
\includegraphics[width=0.18\columnwidth]{figs/EFT_loop13_qqqq}
}\hspace{0.2cm}
\raisebox{-0.2cm}{
\includegraphics[width=0.16\columnwidth]{figs/EFT_loop7_qqqq}
}\hspace{0.2cm}
\raisebox{0.5cm}{
\includegraphics[width=0.17\columnwidth]{figs/EFT_loop14b_qqqq}
} 
\\[10pt]
  \raisebox{2cm}{
  \hspace{-0.8cm}
  p)\hspace{3.1cm} 
  q)\hspace{3.3cm} 
  r)\hspace{3.3cm} 
  s)\hspace{2.9cm}  
   } \\[-53pt]
\includegraphics[width=0.19\columnwidth]{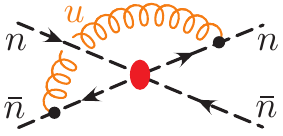}
\hspace{0.3cm}
\includegraphics[width=0.19\columnwidth]{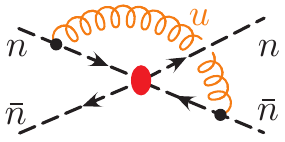}
\hspace{0.3cm}
\includegraphics[width=0.19\columnwidth]{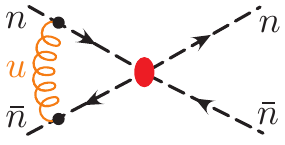}
\hspace{0.3cm}
\includegraphics[width=0.19\columnwidth]{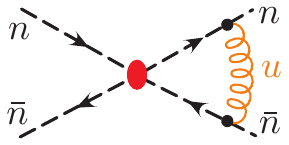}
\hspace{0.3cm}
\end{center}
	\vspace{-0.5cm}
	\caption{\setcaptionskip
		\SCETa graphs for the matching calculation of quark-antiquark forward scattering at one-loop. The first two graphs involve the Glauber potential. The next three graphs involve soft gluon or soft quark loops.   The second and third rows involve collinear loops with either the quark-gluon Glauber scattering operators or the quark-quark Glauber scattering operator, plus wavefunction renormalization. The last row shows the ultrasoft diagrams which contribute. Additional collinear graphs (which vanish) are shown in \fig{SCET2_oneloop_matching2}.
        }
	\label{fig:SCET1_oneloop_matching}
	\setmainskip
\end{figure}

For the \SCETa calculation the EFT diagrams are shown in \fig{SCET1_oneloop_matching}. These are the same diagrams as we had for \SCETb, except for the addition of graphs with an ultrasoft gluon in \fig{SCET1_oneloop_matching}p,q,r,s. Some of the diagrams from the \SCETb calculation will take on different values here, since they are now evaluated with the offshellness regulator and with different 0-bin subtractions. 

We start with the ultrasoft diagrams, shown in the last row of \fig{SCET1_oneloop_matching}. Since we are working in Feynman gauge, graphs where the ultrasoft gluon connects two $n$-collinear lines, or two $\bn$-collinear lines are zero, and hence are not shown. For the sum of the four displayed ultrasoft diagrams we have
\begin{align} \label{eq:usoft_loop}
& 
\raisebox{-0.4cm}{
\includegraphics[width=0.13\columnwidth]{figs/EFT_loop18_qqqq}
}
\ +\ 
\raisebox{-0.4cm}{
\includegraphics[width=0.13\columnwidth]{figs/EFT_loop19_qqqq}
}
\ +\ 
\raisebox{-0.4cm}{
\includegraphics[width=0.13\columnwidth]{figs/EFT_loop20_qqqq}
}
\ +\ 
\raisebox{-0.4cm}{
\includegraphics[width=0.13\columnwidth]{figs/EFT_loop21_qqqq}
}
  \\[5pt] 
  &\quad
  =  \frac{4 g^4}{t} \int  \!\!  \frac{\ddslash\!^{d}k}{k^2\plus i0}
  \bigg[ \frac{2{\cal S}_1^{n\bn}}{(n\cdot k \plus \tilde p_3 \plus i0)
   (\bn\cdot k\minus \tilde p_4 \minus i0)} - 
   \frac{2{\cal S}_{1'}^{n\bn}}{(n\cdot k \minus \tilde p_3 \minus i0)
   (\bn\cdot k\minus \tilde p_4 \minus i0)} \bigg]
  \,, \nn
\end{align} 
where 
\begin{align}
 \tilde p_3 &= p_3^2/\bn\cdot p_3 = p_2^2/\bn\cdot p_2 = -p^2/\bn\cdot p_3
 \,, 
 &\tilde p_4 &= p_4^2/n\cdot p_4=p_1^2/n\cdot p_1=-\bar p^2/n\cdot p_4
 \,,
\end{align} 
and the additional color structure ${\cal S}_{1'}^{n\bn}= - \big(\bar u_n \frac{\bnslash}{2} T^A T^B u_n\big)\big(\bar v_\bn \frac{\nslash}{2} \bar T^B \bar T^A v_\bn\big)$, and satisfies ${\cal S}_{1'}^{n\bn}={\cal S}_1^{n\bn}-{\cal S}_3^{n\bn}/2$. Performing the integrals in \eq{usoft_loop} we find
\begin{align} \label{eq:usoft_loop1}
 \text{Ultrasoft Loops} &=
   {\rm Figs.}~\ref{fig:SCET1_oneloop_matching}p,q,r,s  
  \\
  &=  \frac{i \alpha_s^2}{t}\: {\cal S}_1^{n\bn}
 \bigg[ \frac{8i\pi}{\epsilon} + 8 i \pi \ln\Big(\frac{\mu^2 s}{p^2\bar p^2}\Big) + 4 \pi^2 \bigg] 
  \nn \\
  &\ \ 
  + \frac{i \alpha_s^2}{t}\: {\cal S}_3^{n\bn}
  \bigg[ -\frac{4}{\epsilon^2} - \frac{4}{\epsilon} \ln\Big(\frac{\mu^2 s}{p^2\bar p^2}\Big) - 2 \ln^2\Big(\frac{\mu^2 s}{p^2\bar p^2}\Big) -\pi^2
  \bigg]
  \,. \nn
\end{align}
Since the ultrasoft gluon is the mode with the lowest invariant mass, it has no 0-bin subtractions. Note that the logs are minimized at the smallest see-saw scale, $\mu^2 \sim p^2\bar p^2/s$, as expected.

Next we consider the Glauber loop graphs in \fig{SCET1_oneloop_matching}a,b. As usual these graphs require the rapidity regulator $|2 k^z|^{-\eta}\nu^\eta$ to make them well defined.  Just like for our analysis in \SCETb the Glauber cross-box in \SCETa (\fig{SCET1_oneloop_matching}b) has two poles in $k^0$ on the same side of the contour, and hence vanishes. Thus, once again only the Glauber box graph is nonzero. It is given by
\begin{align} \label{eq:glauber_loop1}
\text{Glauber Loops} 
&= \raisebox{-0.6cm}{
\includegraphics[width=0.13\columnwidth]{figs/EFT_loop1_qqqq} 
 }
 \nn\\
  &= (-4 g^4) \:  {\cal S}_1^{n\bn} \ I_{\rm Gbox}
  = (-4 g^4) \:  {\cal S}_1^{n\bn} \Big(\frac{-i}{4\pi}\Big)  
    \int \!\!   \frac{ \ddslash\!^{d-2}k_\perp\ (-i\pi)  }{ 
    {\vec k}_\perp^{\,2}\:({\vec k}_\perp\plus {\vec q}_\perp)^2 }
  \nn\\
 &= \frac{i\alpha_s^2}{t} \:  {\cal S}_1^{n\bn}
   \bigg[ -\frac{8i\pi}{\epsilon} -8 i \pi \ln\Big( \frac{\mu^2}{-t}\Big) \bigg] \,,
\end{align} 
where the Glauber box integral $I_{\rm Gbox}$ is the same as in \eq{Gbox}, and hence the only difference is the absence of $m$ in the $\vec k_\perp$ integral. Examining this integrand, we see that the $1/\epsilon$ in \eq{glauber_loop1} is an IR divergence from $\vec k_\perp\to 0$ and $\vec k_\perp \to -\vec q_\perp$.  However, in \SCETa we must also subtract the ultrasoft 0-bin subtractions for this Glauber loop to get the full result, $G =\tilde G - G^{(U)}$. The subtractions occur from the regions $k^\mu\sim \lambda^2$ and $k^\mu+q^\mu\sim \lambda^2$, giving two equal contributions, whose sum is
\begin{align}  \label{eq:glauber_loop1subt}
  G^{(U)}({\rm Fig.}~\ref{fig:SCET1_oneloop_matching}a)
  &=  \frac{8 g^4}{t} \:  {\cal S}_1^{n\bn}
    \int \!\!    \frac{ \ddslash\!^{d-2}k_\perp\
     \ddslash\! k^0 \
     \ddslash\! k^z \ |k^z|^{-2\eta}\: (\nu/2)^{2\eta} \qquad\qquad 
     }{ 
    ({\vec k}_\perp^{\,2})
    \Big( k^0\minus k^z \plus p_3^+ \minus (\frac{{\vec q}_\perp}{2})^{\,2}/p_2^- \plus i0\Big)
    \Big( \minus k^0\minus k^z\plus p_4^-  \minus (\frac{{\vec q}_\perp}{2})^{\,2}/p_1^+ \plus i0\Big)
    }  
  \nn\\
  &=  \frac{8 g^4}{t} \:  {\cal S}_1^{n\bn}
    \int \!\!    \frac{ \ddslash\!^{d-2}k_\perp\
     \ddslash\! k^0 \
     \ddslash\! k^z \ |k^z|^{-2\eta}\: (\nu/2)^{2\eta}
     }{ 
    \big({\vec k}_\perp^{\,2}\big) \big(k^0-k^z + \tilde p_3 + i0 \big)
     \big( - k^0-k^z+ \tilde p_4 + i0 \big)
    }  
  \nn\\
  &= \frac{i\alpha_s^2}{t}\: {\cal S}_1^{n\bn}\: \bigg[ 
  -\frac{8i\pi}{\epsilon} + \frac{8i\pi}{\epsilon_{\rm uv}} \bigg] 
  \,.
\end{align}
This result involves the difference of a UV and IR $1/\epsilon$ pole. 
Thus when we subtract this 0-bin contribution from the original integrand in \eq{glauber_loop1}, $\tilde G-G^{(U)}$, we find that the proper interpretation of the $1/\epsilon$ in the result for the \SCETa Glauber box graph is as a UV divergence. 

Examining all the diagrams in \fig{SCET1_oneloop_matching} we see that only the ultrasoft and Glauber loops have contributions with the ${\cal S}_1^{n\bn}$ color structure. Adding the results for these diagrams from \eqs{usoft_loop1}{glauber_loop1} we have
\begin{align}  \label{eq:oneloop_scet1_USG}
 \text{Ultrasoft + Glauber Loops} &=
   {\rm Figs.}~\ref{fig:SCET1_oneloop_matching}a,p,q,r,s  
  \\
  &=  \frac{i \alpha_s^2}{t}\: {\cal S}_1^{n\bn}
 \bigg[ + 8 i \pi \ln\Big(\frac{-st }{p^2\bar p^2}\Big) + 4 \pi^2 \bigg] 
  \nn \\
  &\ \ 
  + \frac{i \alpha_s^2}{t}\: {\cal S}_3^{n\bn}
  \bigg[ -\frac{4}{\epsilon^2} - \frac{4}{\epsilon} \ln\Big(\frac{\mu^2 s}{p^2\bar p^2}\Big) - 2 \ln^2\Big(\frac{\mu^2 s}{p^2\bar p^2}\Big) -\pi^2
  \bigg]
  \,. \nn
\end{align}
Note that the UV divergences have canceled out in the sum of the ultrasoft and Glauber diagrams.  $i\alpha_s^2{\cal S}_1^{n\bn}/t \big[ 8i\pi/\epsilon-8i\pi/\epsilon\big]=0$. Comparing the full ${\cal S}_1^{n\bn}$ term here with that of the full theory in \eq{full_oneloop_result1}, we see that the full theory result is exactly reproduced in the sum of the ultrasoft and Glauber graphs of \SCETa. The presence of the $\ln(s)$ in this term arises because of the presence of the ultrasoft see-saw scale $p^2\bar p^2/s$. 

Next we consider the \SCETa graphs contributing to the $C_F T^A\otimes \bar T^A$ color structure, ie. terms involving ${\cal S}_2^{n\bn}$. This occurs only in the collinear loop graphs in Figs.~\ref{fig:SCET1_oneloop_matching}i,j,n,o.  Just as in \SCETb, the loops in these graphs involve only Lagrangian insertions and a single collinear sector, and are the same result as in full QCD. With the offshellness regulator we have
\begin{align} \label{eq:oneloop_scet1_vertwfn}
&\raisebox{-0.4cm}{
\includegraphics[width=0.13\columnwidth]{figs/EFT_loop6_qqqq}
}
\  + \
\raisebox{-0.6cm}{
\includegraphics[width=0.13\columnwidth]{figs/EFT_loop7_qqqq}
}\ +\  \bigg(
\raisebox{-0.2cm}{
\includegraphics[width=0.13\columnwidth]{figs/EFT_loop14_qqqq}
} +
\raisebox{-0.2cm}{
\includegraphics[width=0.13\columnwidth]{figs/EFT_loop14b_qqqq}
} \bigg)\
\raisebox{-0.2cm}{
\includegraphics[width=0.11\columnwidth]{figs/GlaubOp_tree_qqqq}
} 
 \\
  &\quad
  = \frac{i\alpha_s^2}{t} \: {\cal S}_2^{n\bn}
   \bigg[ -4\ln^2\Big( \frac{p^2}{-t}\Big) 
   -4\ln^2\Big( \frac{\bar p^2}{-t}\Big)
   +6\ln\Big( \frac{p^2}{-t}\Big)+6\ln\Big( \frac{\bar p^2}{-t}\Big)
   -4-\frac{8\pi^2}{3}\: \bigg] 
 \nn  \\
  &\quad\quad + \frac{i\alpha_s^2}{t}\: {\cal S}_3^{n\bn} 
 \bigg[ 2\ln^2\Big( \frac{p^2}{-t}\Big) +2\ln^2\Big( \frac{\bar p^2}{-t}\Big)
  + 4 \ln\Big( \frac{p^2}{-t}\Big)+4\ln\Big( \frac{\bar p^2}{-t}\Big)
  -\frac{2}{\epsilon} - 2\ln\Big(\frac{\mu^2}{-t}\Big)
  +\frac{4\pi^2}{3} \bigg] 
  \,, \nn
\end{align}
where the $1/\epsilon$ is UV. Looking at only the  ${\cal S}_2^{n\bn}$ term, we see that the SCET$_{\rm I}$ graphs reproduce the full ${\cal S}_2^{n\bn}$ piece of \eq{full_oneloop_result1}. The situation is similar for the $n_f T_F T^A\otimes \bar T^A$ term, ie. ${\cal S}_4^{n\bn}$. The only \SCETa graph that is proportional to $n_f$ is the soft graph in \fig{SCET1_oneloop_matching}d which gives the same result as in \SCETb, and as the quark vacuum polarization in the full theory,
\begin{align} \label{eq:soft_eye_fermion1}  
\raisebox{-0.5cm}{
\includegraphics[width=0.13\columnwidth]{figs/EFT_loop4_qqqq}
}
  \ 
  & = \frac{i\alpha_s^2}{t}  {\cal S}_4^{n\bn}
   \bigg[ -\frac{8}{3\epsilon} -\frac{8}{3} \ln\Big(\frac{\mu^2}{-t}\Big) -\frac{40}{9}  
  \bigg] .
\end{align}
So the full theory ${\cal S}_4^{n\bn}$ term in \eq{full_oneloop_result1} is also exactly reproduced. 

This leaves the final color structure $C_A T^A\otimes \bar T^A$, ie. ${\cal S}_3^{n\bn}$. Just as for \SCETb, this color structure involves the most complicated calculations, and there is no one-to-one correspondence between graphs in the full theory and in \SCETa.  

For this SCET$_{\rm I}$ calculation we have contributions from  Figs.~\ref{fig:SCET1_oneloop_matching}i,n,p,q given above in \eqs{oneloop_scet1_USG}{oneloop_scet1_vertwfn}, as well as from Figs.~\ref{fig:SCET1_oneloop_matching}c,e,f,g,h,k,l,m which we will consider in turn. We will encounter rapidity divergences in these diagrams. Again we should consider the additional collinear graphs shown in Figs.~\ref{fig:SCET2_oneloop_matching2}, but they vanish for the same reasons discussed above in our \SCETb calculation, since the results discussed there were independent of the choice of IR regulator.

First consider the contribution from the T-product of two Glauber operators with a soft gluon loop, ${\cal O}_{ns}^{qg}$ with ${\cal O}_{\bn s}^{qg}$, which is shown in \fig{SCET1_oneloop_matching}c.  For this soft eye diagram in \SCETa we find the same result as in \SCETb,
\begin{align} \label{eq:soft_eye1}
\raisebox{-0.9cm}{
\includegraphics[width=0.13\columnwidth]{figs/EFT_loop3_qqqq_ext}
}
  \ 
  &= - \frac{i\alpha_s^2}{t}\: {\cal S}_3^{n\bn} \bigg\{
    \frac{8}{\eta}\, g(\epsilon,\mu^2/(-t)) + \frac{4}{\epsilon^2} +\frac{4}{\epsilon} \ln\Big(\frac{\mu^2}{\nu^2}\Big) 
    +4 \ln\Big(\frac{\mu^2}{\nu^2}\Big) \ln\Big(\frac{\mu^2}{-t}\Big)
   \nn\\[-10pt]
  &\qquad\qquad\qquad
    -2 \ln^2\Big(\frac{\mu^2}{-t}\Big) + \frac{\pi^2}{3}
    + 2 \Big( -\frac{11}{3\epsilon} -\frac{11}{3}\ln\frac{\mu^2}{-t}-\frac{67}{9} \Big) \bigg\} 
  \,,
\end{align}
where $g(\epsilon,\mu^2/(-t))$ was given above in \eq{g}. 
To see why the result for this graph in \SCETa and \SCETb are the same, we can look back at \eq{soft_eye} and notice that the result was independent of $m^2$. For the \SCETa soft eye graph we drop $m^2$ from the start, but this does not change the result. Just as described in detail for \SCETb, the Glauber 0-bin for this \SCETa soft loop again ensures that the sign $\pm i0$ in the eikonal propagators does not effect the result. Finally, we consider the two ultrasoft 0-bins for the integrand in \eq{soft_eye} (with $m^2=0$), first by considering $k^\mu\sim \lambda^2$, and then by switching variables to $k^{\prime\mu}=k^\mu+q^\mu$ and considering $k^{\prime\mu}\sim \lambda^2$. In both cases one of the relativistic propagators becomes a $q^2=t$, so the ultrasoft 0-bin integral scales as a power suppressed term $\lambda^8\lambda^4/[t^3\lambda^4]\sim \lambda^2$ and hence does not contribute. 

The remaining soft loop graph is the flower graph in \fig{SCET1_oneloop_matching}e. The naive integrand for this soft loop graph is the same as that for \SCETb in \eq{soft_flower} just setting $m^2=0$. However, now the soft loop integral also has an ultrasoft 0-bin subtraction, and since there is no scale in the soft integral, this subtraction integral is identical to the original one. Once again the choice of $\pm i0$ in the eikonal propagators does not change the result for this loop diagram or for the ultrasoft 0-bin subtraction, due to the Glauber 0-bin subtractions $S^{(G)}$ and $S^{(G)(U)}$. Thus in \SCETa the soft flower graph vanishes
\begin{align} \label{eq:soft_flower1}
  S({\rm Fig.}~\ref{fig:SCET1_oneloop_matching}e) 
   &=\tilde S({\rm Fig.}~\ref{fig:SCET1_oneloop_matching}e) - 
     S^{(U)}({\rm Fig.}~\ref{fig:SCET1_oneloop_matching}e) 
   \\
   &= \frac{4 g^4}{t}\: {\cal S}_3^{n\bn} 
    \iota^\epsilon \mu^{2\epsilon}\: \bigg[ \int\!\! \ddslash\!^{d} k
   \: \frac{|k_z|^{-\eta}\,\nu^\eta }{ [k^2](n\cdot k)(\bn\cdot k)} -
   \int\!\! \ddslash\!^{d} k
   \: \frac{|k_z|^{-\eta}\,\nu^\eta }{ [k^2](n\cdot k)(\bn\cdot k)} \bigg]
   \nn \\
   &=0 \,.  \nn
\end{align}
This occurs because the contribution of the soft flower graph is already contained in the ultrasoft graphs in \SCETa. The $1/\eta$ rapidity divergence that occurred in the soft flower graph for \SCETb is now regulated by the combination of the offshellness and dimensional regularization in the \SCETa ultrasoft graphs. We will see below that the collinear Wilson line graphs in \SCETa also do not depend on the rapidity regulator.

Since the bare soft operator $O_s^{AB}$ has a factor of $\alpha_s^{\rm bare}$ multiplying the fields, there is also the $Z_{\alpha}$ coupling counterterm contribution
\begin{align}
 \text{soft $\alpha_s$ counterterm} 
 &= \frac{i\alpha_s^2}{t} \bigg( -{\cal S}_3^{n\bn}\:  \frac{22}{3\epsilon}
  + {\cal S}_4^{n\bn}\: \frac{8}{3\epsilon} \bigg) .
\end{align}
The sum of the two nonzero soft loop graphs from \eqs{soft_eye_fermion1}{soft_eye1}, plus this $\alpha_s$ counterterm gives the total soft loop contribution
\begin{align}  \label{eq:soft_total_loop1}
\text{Soft Loops} &=
{\rm Figs.}~\ref{fig:SCET1_oneloop_matching}c,d +Z_\alpha\:\text{c.t.}
  \nn\\
  &= \frac{i\alpha_s^2}{t}\: {\cal S}_3^{n\bn} \bigg\{
    - \frac{8}{\eta}\, g(\epsilon,\mu^2/(-t)) - \frac{4}{\epsilon^2} 
    - \frac{4}{\epsilon} \ln\Big(\frac{\mu^2}{\nu^2}\Big) 
    - 4 \ln\Big(\frac{\mu^2}{\nu^2}\Big) \ln\Big(\frac{\mu^2}{-t}\Big)
   \nn\\
  &\qquad\qquad\qquad
    +2 \ln^2\Big(\frac{\mu^2}{-t}\Big) - \frac{\pi^2}{3}
    +\frac{22}{3}\ln\Big(\frac{\mu^2}{-t}\Big)+\frac{134}{9}  \bigg\} 
 \nn\\
  & \quad\ 
  + \frac{i\alpha_s^2}{t} \: {\cal S}_4^{n\bn}
   \bigg[  -\frac{8}{3} \ln\Big(\frac{\mu^2}{-t}\Big) -\frac{40}{9}  
  \bigg] .
\end{align}
The logarithms from the soft loops are minimized for $\mu\sim \nu\sim \sqrt{t}$  which is consistent with the power counting. 
Note that unlike the situation in \SCETb, here the $1/\epsilon^2$ and $\ln(\mu^2/\nu^2)/\epsilon$ terms do not cancel, so there are both $1/\eta$ rapidity divergences and $1/\epsilon$ UV divergences in the soft contribution. Once again, the one-loop constants that appear in the soft graph are identical to the the two-loop cusp anomalous dimension, just as in \SCETb in \eq{twoloopcusp}.

Finally we consider the remaining collinear diagrams, in \fig{SCET1_oneloop_matching}f,g,h,k,l,m.  The two V-graphs in \fig{SCET1_oneloop_matching}f,k give the same result for \SCETa as they did in \SCETb. Again this occurs because the answer in \eq{oneloop_collinear_V} is independent of the IR regulator $m^2$. So setting $m^2=0$ from the start, the results for these two graphs is once again
\begin{align}   \label{eq:oneloop_collinear_V1}
& \raisebox{-0.4cm}{
\includegraphics[width=0.14\columnwidth]{figs/EFT_loop8_qqqq}
}
\ +\  
\raisebox{-0.4cm}{
\includegraphics[width=0.14\columnwidth]{figs/EFT_loop9_qqqq}
}
 \\
  &\quad 
  = \frac{i\alpha_s^2}{t}\: {\cal S}_3^{n\bn} \bigg\{
    \frac{8}{\eta}\, g(\epsilon,\mu^2/t)  
    -\frac{4}{\epsilon} \ln\Big(\frac{\nu^2}{s}\Big) 
    -4 \ln\Big(\frac{\nu^2}{s}\Big) \ln\Big(\frac{\mu^2}{-t}\Big)
    - \frac{6}{\epsilon} - 6 \ln\Big(\frac{\mu^2}{-t}\Big) 
    - 12 + \frac{8\pi^2}{3} \bigg\}
   \,. \nn
\end{align}
Recall that the factors of $\ln(s)$ appear from adding the two diagrams and using $\ln(\bn\cdot p_3)+\ln(n\cdot p_4)=\ln s$. The 0-bin subtraction contributions for the result in \eq{oneloop_collinear_V1} all vanish. The soft and Glauber 0-bin subtractions vanish for the same reason as in \SCETb. And the ultrasoft 0-bin subtractions from the scalings $k^\mu \sim \lambda^2$, $k^\mu+q^\mu\sim \lambda^2$, $k^\mu+p_3^\mu\sim \lambda^2$ all lead to power suppressed integrals relative to the leading power $\sim \lambda^{-2}$ contribution.


Next we consider the \SCETa collinear Wilson line graphs in \fig{SCET1_oneloop_matching}h,i,m,n. The result is similar to \eq{oneloop_collinear_W} with modifications due to the change of IR regulator. The resulting \SCETa loop integral is the standard one-loop Wilson line integral, which is well defined without the rapidity regulator, so
\begin{align}  \label{eq:oneloop_collinear_W1}
& 
\raisebox{-0.4cm}{
\includegraphics[width=0.13\columnwidth]{figs/EFT_loop10_qqqq}
}
\ +\ 
\raisebox{-0.4cm}{
\includegraphics[width=0.13\columnwidth]{figs/EFT_loop11_qqqq}
}
\ +\ 
\raisebox{-0.4cm}{
\includegraphics[width=0.13\columnwidth]{figs/EFT_loop12_qqqq}
}
\ +\ 
\raisebox{-0.4cm}{
\includegraphics[width=0.13\columnwidth]{figs/EFT_loop13_qqqq}
}
 \nn \\[5pt]
  &\quad 
  = i {\cal S}_3^{n\bn}\: \frac{2 g^4 }{t} \bigg[
  \int\!\! \ddslash\!^{d} k
  \frac{(\iota^\epsilon \mu^{2\epsilon}) \ \bn\cdot (k+p_3)}{[k^2](k+p_3)^2 (\bn\cdot k)}
  +  \int\!\! \ddslash\!^{d} k
  \frac{(\iota^\epsilon \mu^{2\epsilon}) \ n\cdot (k+p_4)}{[k^2](k+p_4)^2 (n\cdot k)} \bigg]
  \nn \\
  &\quad
  =  \frac{i\alpha_s^2}{t}\, {\cal S}_3^{n\bn} 
  \bigg\{ \frac{8}{\epsilon^2}
   + \frac{4}{\epsilon} \ln\Big(\frac{\mu^2}{p^2}\Big) 
   +\frac{4}{\epsilon} \ln\Big(\frac{\mu^2}{\bar p^2}\Big) 
   + 2\ln^2\Big(\frac{\mu^2}{p^2}\Big) 
   + 2\ln^2\Big(\frac{\mu^2}{\bar p^2}\Big) 
\nn\\
 &\quad \qquad\qquad\qquad
   + \frac{8}{\epsilon} + 4 \ln\Big(\frac{\mu^2}{p^2}\Big) 
   + 4 \ln\Big(\frac{\mu^2}{\bar p^2}\Big) 
   + 16 - \frac{2\pi^2}{3} \bigg\}
  \,. 
\end{align}
In precisely the same manner as for the V-graphs the soft and Glauber 0-bin subtractions all vanish for these collinear Wilson line graphs.  

The sum of all the collinear graphs from Eqs.~(\ref{eq:oneloop_scet1_vertwfn},\ref{eq:oneloop_collinear_V1},\ref{eq:oneloop_collinear_W1}) gives
\begin{align}  \label{eq:collinear_total_loop1}
 & \text{Collinear Loops} =
 {\rm Figs.}~\ref{fig:SCET2_oneloop_matching}f\text{-}n
  \\ 
 &\qquad
  = \frac{i\alpha_s^2}{t} \: {\cal S}_2^{n\bn}
   \bigg[ -4\ln^2\Big( \frac{p^2}{-t}\Big) 
   -4\ln^2\Big( \frac{\bar p^2}{-t}\Big)
   +6\ln\Big( \frac{p^2}{-t}\Big)+6\ln\Big( \frac{\bar p^2}{-t}\Big)
   -4-\frac{8\pi^2}{3}\: \bigg] 
  \nn \\
  &\qquad\quad + \frac{i\alpha_s^2}{t}\: {\cal S}_3^{n\bn} 
 \bigg[ 
    \frac{8}{\eta}\, g(\epsilon,\mu^2/(-t))  
    -\frac{4}{\epsilon} \ln\Big(\frac{\nu^2}{s}\Big) 
    -4 \ln\Big(\frac{\nu^2}{s}\Big) \ln\Big(\frac{\mu^2}{-t}\Big)
   + \frac{8}{\epsilon^2}
   + \frac{4}{\epsilon} \ln\Big(\frac{\mu^2}{p^2}\Big) 
   +\frac{4}{\epsilon} \ln\Big(\frac{\mu^2}{\bar p^2}\Big) 
 \nn\\
 &\qquad\qquad\qquad\qquad 
  + 2\ln^2\Big( \frac{p^2}{-t}\Big) +2\ln^2\Big( \frac{\bar p^2}{-t}\Big)
   + 2\ln^2\Big(\frac{\mu^2}{p^2}\Big) 
   + 2\ln^2\Big(\frac{\mu^2}{\bar p^2}\Big) 
    +\frac{10\pi^2}{3} 
   + 4 
\bigg] 
  \,. \nn
\end{align}
Again there are cancellations that have occurred for the sum of graphs, including all the $1/\epsilon$ and single log terms. (This is also true separately for the $n$-collinear graphs and $\bn$-collinear graphs.) Unlike in \SCETb, the collinear graphs alone do have $1/\epsilon^2$ and $\ln(\cdots)/\epsilon$ divergences.  The logarithms from these collinear loops are minimized with  $\nu\sim \bn\cdot p_3 \sim n\cdot p_4 \sim \sqrt{s}$ and $\mu\sim \sqrt{t}$ (taking the offshellness $p^2\sim t \sim \bar p^2$). Once again this is as expected, and consistent with the power counting. 

Finally, we can add up the ultrasoft, Glauber, soft, and collinear SCET loop graphs from Eqs.~(\ref{eq:oneloop_scet1_USG}, \ref{eq:soft_total_loop1}, \ref{eq:collinear_total_loop1}). In the sum of soft and collinear loops the $g(\epsilon,\mu^2/(-t))/\eta$ rapidity divergences cancel, as expected. Furthermore, the $1/\epsilon^2$ terms cancel in the sum of ultrasoft plus soft plus collinear terms.  Adding the terms and simplifying we find a large amount of simplifications to the logarithmic terms, yielding
\begin{align}  \label{eq:total_SCET1_result}
 \text{Total } & \text{SCET}_{\rm I} = {\rm Figs.}~\ref{fig:SCET1_oneloop_matching}a\text{-}s
    +Z_\alpha\:\text{c.t.} 
  \nn\\
&=  \frac{i \alpha_s^2}{t}\: {\cal S}_1^{n\bn}
 \bigg[ + 8 i \pi \ln\Big(\frac{-st }{p^2\bar p^2}\Big) + 4 \pi^2 \bigg] 
  \\
 &\ \ 
  + \frac{i\alpha_s^2}{t} \: {\cal S}_2^{n\bn}
   \bigg[ -4\ln^2\Big( \frac{p^2}{-t}\Big) 
   -4\ln^2\Big( \frac{\bar p^2}{-t}\Big)
   +6\ln\Big( \frac{p^2}{-t}\Big)+6\ln\Big( \frac{\bar p^2}{-t}\Big)
   -4-\frac{8\pi^2}{3}\: \bigg] 
  \nn \\
  &\ \ 
  + \frac{i \alpha_s^2}{t}\: {\cal S}_3^{n\bn}
  \bigg[  - 2 \ln^2\Big(\frac{-st}{p^2\bar p^2}\Big)
  + 4\ln^2\Big( \frac{-t}{p^2}\Big) +4\ln^2\Big( \frac{-t}{\bar p^2}\Big)
    +\frac{22}{3}\ln\Big(\frac{\mu^2}{-t}\Big)
    + \frac{170}{9} + 2\pi^2 
  \bigg] 
 \nn\\
  & \ \
  + \frac{i\alpha_s^2}{t} \: {\cal S}_4^{n\bn}
   \bigg[  -\frac{8}{3} \ln\Big(\frac{\mu^2}{-t}\Big) -\frac{40}{9}  
  \bigg] 
 \,. \nn
\end{align} 
This total \SCETa result agrees exactly with the full theory one-loop result in \eq{full_oneloop_result1} for all color structures, all IR divergences, all logs, and all constant terms. Since all IR divergences are correctly reproduced this provides a non-trivial test of this \SCETa EFT framework with Glaubers. Again, the $\ln\frac{\mu^2}{-t}$ dependence is proportional to the one-loop beta function, and hence shows that the scale $\mu^2\simeq -t>0$ is the preferred value for the $\alpha_s(\mu)$ in the tree level potential.  Since $s\gg -t\sim p^2\sim \bar p^2$ there is one large double logarithm in the \SCETa result, $\ln^2(\frac{-st}{p^2\bar p^2})$, which is generated by combining the hierarchy in rapidity between the soft and collinear diagrams, and the hierarchy in invariant masses between the collinear and ultrasoft diagrams in the following manner:
\begin{align}
 & -2 \ln^2\Big(\frac{-st}{p^2\bar p^2}\Big) 
  + 4 \ln^2\Big(\frac{-t}{p^2}\Big)
  + 4 \ln^2\Big(\frac{-t}{\bar p^2}\Big) 
  = \bigg[ -4 \ln\Big(\frac{s}{-t}\Big)\ln\Big(\frac{-t}{\mu^2}\Big)
     -2 \ln^2\Big(\frac{\mu^2}{-t}\Big) 
     \bigg] 
    \\
  &\qquad\qquad\quad
    +\bigg\{  -2 \ln^2\Big(\frac{\mu^2 s}{p^2\bar p^2}\Big)
     + 2 \ln^2\Big(\frac{\mu^2}{p^2}\Big) 
     + 2 \ln^2\Big(\frac{\mu^2}{\bar p^2}\Big)
     + 2 \ln^2\Big(\frac{-t}{p^2}\Big)
     + 2 \ln^2\Big(\frac{-t}{\bar p^2}\Big)
     \bigg\}
  \,. \nn
\end{align}
Here the terms in square brackets come from the sum of the rapidity divergent collinear V-graphs and soft eye graph, whereas the terms in curly brackets  come from the ultrasoft graphs and collinear Wilson line and vertex graphs. In the sum there is no dependence on the renormalization scale $\mu$. 

Once again the fact that the \SCETa result in \eq{total_SCET1_result} agrees exactly with the full theory result in \eq{full_oneloop_result1} implies that there are no hard matching corrections to the Glauber operator at the scale $\mu^2\sim s$.  In the \SCETa calculation the $\ln(s)$ dependence arises from combining collinear rapidity divergences that involve the large $p_n^-$ and $p_\bn^+$ collinear momenta, just like in \SCETb, as well as from the see-saw scale $p^2\bar p^2/s$ that appears in the ultrasoft diagram result in \eq{usoft_loop1}. Once again, this pattern continues at higher orders in $\alpha_s$ and there are no hard matching corrections for the Glauber Lagrangian at the scale $\mu^2\simeq s$. Therefore the tree level matching results given in \sec{GlauberEFT} yield the complete Glauber Lagrangian also in \SCETa.

\section{BFKL and the Rapidity Renormalization Group}
\label{sec:BFKL}

\subsection{Factorization with the Glauber Lagrangian}
\label{sec:factBFKL}

In this section we consider how to include the Glauber Lagrangian into a factorized analysis for situations where the Glauber exchange is important and does not cancel out, such as in, forward scattering, or to sum logs of $x$ in the small $x$ region.  We will use the example of forward scattering in order to have an explicit context for our calculations, though since many of the results are valid for operators that appear in other processes, the results presented here apply equally well there as well.

Since the Glauber Lagrangian couples together soft and collinear modes we can only factorize the cross section  if we expand the Glauber Lagrangian insertions in a Taylor series.  To organize this factorization we expand the time evolution  operator generated by the Glauber Lagrangian. Written as a path integral the full time evolution operator in SCET is
\begin{align}
  U(a,b;T) & = \int \big[{\cal D}\phi\big] \exp\bigg[ i \int_{-T}^{T}\!\! d^4x\: \big( {\cal L}_{n\bn s}^{(0)}(x) + {\cal L}_G^{\rm II(0)}(x)\big) \bigg] \,,
\end{align}
where ${\cal L}_{n\bn s}^{(0)}={\cal L}_n^{(0)}+{\cal L}_\bn^{(0)}+{\cal L}_s^{(0)}$ is the non-Glauber parts of the SCET Lagrangian,  $a,b$ indicate the field boundary conditions at time $t=-T,+T$, and $[{\cal D}\phi]$ is a short hand to indicate the functional integral over all relevant SCET soft and collinear fields. We will only be interested in the large $T$ limit, $T\to \infty (1-i0)$. All these Lagrangian terms are leading order in the power counting. Using \eq{LGreg} we can expand the Glauber part of the time evolution operator as
\begin{align} \label{eq:Uglabexpn}
  T \exp \, i\! \int\!\! d^4x\, {\cal L}_G^{\rm II(0)}(x)  
   &=\bigg[ 1 + i \int\!\! d^4y_1\: {\cal L}_G^{\rm II(0)}(y_1) + \frac{i^2}{2!} T\!\! \int\!\! d^4y_1\, d^4y_2\: {\cal L}_G^{\rm II(0)}(y_1) {\cal L}_G^{\rm II(0)}(y_2) + \ldots \bigg]
  \nn\\
   &= 1 + 
   T \sum_{k=1}^\infty \sum_{k'=1}^\infty   
     \bigg[  \prod_{i=1}^{k} \int\!\! [dx_i^\pm]\!\! \int\!\! \frac{d^2q_{\perp i}}{q_{\perp i}^2}
      \big[{\cal O}_n^{q A_i}(q_{\perp i}) + {\cal O}_n^{g A_i}(q_{\perp i})\big](x_i) \bigg]
  \nn\\
  &\qquad\qquad\qquad \times 
  \bigg[ \prod_{i'=1}^{k'} \int\!\! [dx_{i'}^\pm]\!\!\int\!\! \frac{d^2q_{\perp i'}}{q_{\perp i'}^2} 
  \big[{\cal O}_\bn^{q B_{i'}}(q_{\perp i'}) +  {\cal O}_\bn^{g B_{i'}}(q_{\perp i'})\big](x_{i'}) \bigg]
  \nn\\
  &\qquad\qquad\qquad \times 
     O_{s(k,k')}^{A_1\cdot A_k,B_1\cdots B_{k'}}(q_{\perp 1},\ldots,q_{\perp k'})(x_1,\ldots,x_{k'})
  \nn\\
&\equiv 1 + \sum_{k=1}^\infty \sum_{k'=1}^\infty \ U_{(k,k')}
  \,, 
\end{align}
where here $T$ is the time-ordering operation.
For simplicity we have suppressed the presence of the rapidity regulator for the Glauber exchanges. 
In the last equality of \eq{Uglabexpn} we have organized the expansion according to the number of $n$-collinear operators $k$, and number  of $\bn$-collinear operators $k'$, rather than according to the number of insertions of the Glauber Lagrangian. Any symmetry factors like $1/k!$ are included in the definition of $O_{s(k,k')}^{A_1\cdot A_k,B_1\cdots B_{k'}}$.

For example, the first nontrivial term with $k=k'=1$ is
\begin{align} \label{eq:U11}
  U_{(1,1)} &=
 i \int [dx^\pm][dx^{\prime\pm}] \sum_{k^\pm} \!\int\!\! \frac{d^2q_\perp}{q_\perp^2} \frac{d^2q_\perp'}{q_\perp^{\prime 2}} 
   \big[{\cal O}_{n,k^-}^{q A}(q_{\perp})
    + {\cal O}_{n,k^-}^{g A}(q_{\perp})\big](\tilde x) 
   \big[{\cal O}_{\bn,k^+}^{q B}(q_{\perp}') 
    + {\cal O}_{\bn,k^+}^{g B}(q_{\perp}')\big](\tilde x') 
  \nn\\
  &\qquad \times
    O_{s(1,1),-k^\pm}^{AB}(q_\perp,q_\perp')(\tilde x,\tilde x') 
  \,.
\end{align}
Here the soft operator includes both a direct contribution from the two index soft operator ${\cal O}_s^{AB}$ from a single insertion of ${\cal L}_G^{\rm II(0)}$, as well as a T-product term from the product ${\cal O}_s^{i_nA} {\cal O}_s^{j_\bn B} $ that comes from two insertions of ${\cal L}_G^{\rm II(0)}$:
\begin{align} \label{eq:Os11}
 & O_{s(1,1),-k^\pm}^{AB}(q_\perp,q_\perp^\prime)(\tilde x,\tilde x')
 \\
 &\ 
 = \frac{1}{(2\pi)^2} \delta^2(\tilde x\minus \tilde x')\: {\cal O}_{s,-k^\pm}^{AB}(q_\perp,-q_\perp')(\tilde x) 
   +i\:T\!\! \sum_{i,j=q,g} \!\!
     {\cal O}_{s,-k^-}^{i_n A}(q_\perp)(\tilde x)\
     {\cal O}_{s,-k^+}^{j_\bn B}(-q_\perp')(\tilde x')
  \,. \nn\\
 &\
  = \frac{1}{(2\pi)^2} \delta^2(\tilde x\minus \tilde x')\: {\cal O}_{s,-k^\pm}^{AB}(q_\perp,-q_\perp')(\tilde x) 
  +i\:T\, e^{i\tilde x'\cdot \hat P}\!\! \sum_{i,j=q,g} \!\!
     {\cal O}_{s,-k^-}^{i_n A}(q_\perp)(\tilde x\minus \tilde x')\
     {\cal O}_{s,-k^+}^{j_\bn B}(-q_\perp')(0)
     e^{-i\tilde x'\cdot \hat P} 
   . \nn
\end{align}
Here $\delta^2(\tilde x-\tilde x') = 2 \delta(x^+-x^{\prime +})\delta(x^--x^{\prime -})$.
Note that we have flipped the $q_\perp'$ sign in ${\cal O}_{s,-k^\pm}^{AB}(q_\perp,-q_\perp')$ when defining $O_{s(1,1)}^{AB}(q_\perp,q_\perp^\prime)(\tilde x,\tilde x')$ so that both $q_\perp$ and $q_\perp'$ are outgoing from the soft operator. For the collinear operators in \eq{U11} the ${\cal O}(\lambda)$ soft momenta $k^\pm$ are residual to the respective large collinear momenta, but show us how these soft momenta are routed in the collinear operators.  

Consider the forward scattering of energetic collinear particles that is mediated by having a single $U_{(1,1)}$ on each side of the cut. Since here the amplitude is linear in the number of Glauber exchanges we can refer to this as the linear approximation. We will see that it is valid to obtain the leading logarithmic resummation from the BFKL equation.  We take color singlet initial states $\langle p p' |$, such as proton-proton or quarkonia-quarkonia scattering, where one hadron is $n$-collinear and $\bn$-collinear. The corresponding non-trivial transition matrix is
\begin{align}
T_{(1,1)} &=  \frac{1}{V_4} \sum_X  \big\langle p p' \big| U_{(1,1)}^\dagger  \big| X \big\rangle
   \big\langle X \big| U_{(1,1)}\big| p p' \big\rangle 
 \,,
\end{align} 
where the volume factor $V_4 =(2\pi)^4 \delta^4(0)$ must be removed since each of these matrix elements gives a momentum conserving $\delta$-function.  Since we are working order by order in the Glauber Lagrangian these squared matrix elements can be factorized into soft and collinear components. Below we will analyze these soft and collinear components to pull out various $\delta$-functions and make explicit the flow of momenta. Although this takes some algebra the answer below in \eq{T11} is quite intuitive.

To carry out the factorization we must consider the implications of the multipole expansions of the various momentum scales appearing in $\langle X | U_{(1,1)} | p p'\rangle$. This is enforced by the presence of label and residual momenta. First consider the $p^-$-momenta and $x^+$ and $x^{\prime +}$ dependence. For the $n$-collinear matrix element we have
\begin{align}
  & \big\langle X_n \big| 
  {\cal O}_{n,k^{-}}^{iA}(q_\perp)(\tilde x) 
  \big| p \big\rangle
  = \delta_{p_\ell^-,P_{X_n\ell}^-} \delta_{p_{\ell_s}^-,P_{X_n\ell_s}^- - k^-} 
   e^{-\frac{i}{2} x^+ (p_r^- - P_{X_nr}^-)} \delta^2(q_\perp-p_{X_n}^\perp) 
   M_n (q_\perp,p_\ell^-,x^-) ,  
  \nonumber \\
  & \big\langle X_s \big| 
  {\cal O}^{AB}_{s(1,1),-k^\pm }(q_\perp,q'_\perp)(\tilde x,\tilde x') 
  \big| p' \big\rangle
  = \frac{\delta^2(\tilde x-\tilde x')}{(2\pi)^2} 
  \delta_{-k^-,P_{X_s\ell_s}^-} e^{\frac{i}{2} x^+ P_{X_s r}^-}
  M_s(q_\perp, q'_\perp,k^+,x^-,\ldots) + \ldots 
  , \nonumber \\
  & \big\langle X_{\bar n} \big| 
  {\cal O}_{{\bar n},k^{+}}^{jB}(q'_\perp)(\tilde x') 
  \big| p' \big\rangle
  =   e^{-\frac{i}{2} x^{\prime +} (p^{\prime -} - P_{X_{\bar n}}^-)} 
  \, \delta^2(q'_\perp-p_{X_{\bar n}}^\perp) 
   M_{\bar n}(q'_\perp,p_\ell^{\prime +},x^{\prime -},\ldots)
  \,, 
\end{align}
where the $+\ldots$ for the soft matrix element indicate the T-product term, which has the same scaling. Here the subscript $\ell$ refers to ${\cal O}(\lambda^0)$ label momenta, the subscript $\ell_s$ refers to ${\cal O}(\lambda)$ sublabel momenta, and the subscript $r$ refers to ${\cal O}(\lambda^2)$ residual momenta. After integrating over $x^+$ and summing on $k^-$ we can combine the label and residual momenta back into a full continuous delta function for the ${\cal O}(\lambda^0)$ momenta of the $n$-collinear states, $\delta_{p_\ell^-,P_{X_n\ell}^-} \delta_{p_{\ell_s}^-,P_{X_n\ell_s}^-+P_{X_s\ell_s}^-} \delta(p_r^- - P_{X_n r}^- -P_{X_s r}^-) = \delta(p^- - P_{X_n}^-) + {\cal O}(\lambda)$, where the smaller minus momenta drop out. For the soft matrix element we have only ${\cal O}(\lambda)$ and ${\cal O}(\lambda^2)$ momenta, and hence we can set $x^+=0$ in the soft matrix element to neglect these smaller components in the recombination and can drop the $k^-\sim \lambda$ sublabel on the $n$-collinear operator. Similarly, for the $x^{\prime +}$ coordinate that connects the soft and ${\bar n}$-collinear matrix elements, it is the ${\cal O}(\lambda)$ momentum of the soft that dominates over the ${\cal O}(\lambda^2)$ momentum of the ${\bar n}$-collinear, so we set $x^{\prime +}=0$ in the ${\bar n}$-collinear matrix element.  If we repeat these considerations for the $p^+$-momenta, and $x^-$, $x^{\prime -}$ dependencies, then we similarly find that we can set $x^{\prime -}=0$ in the soft matrix element, $x^-=0$ in the $n$-collinear matrix element, and can drop the $k^+$ sublabel on the ${\bar n}$-collinear operator. For the $n$-collinear and ${\bar n}$-collinear matrix elements this leaves
\begin{align}
  \int\! \frac{dx^+}{2} \big\langle X_n \big| 
  {\cal O}_{n}^{iA}(q_\perp)\Big(x^+ \frac{{\bar n}}{2}\Big) 
  \big| p \big\rangle
  &=  \delta(p^- - p_{X_n}^-) \delta^2(q_\perp-p_{X_n}^\perp) 
  M_n(p^-,q_\perp) \,,
 \nonumber\\
  \int\! \frac{dx^{\prime -}}{2} \big\langle X_{\bar n} \big| 
  {\cal O}_{{\bar n}}^{jB}(q'_\perp)\Big(x^{\prime -} \frac{n}{2}\Big) 
  \big| p' \big\rangle
  &=  \delta(p^+ - p_{X_{\bar n}}^+) \delta^2(q'_\perp-p_{X_{\bar n}}^\perp) 
  M_{\bar n}(p^+,q'_\perp) \,.
\end{align}
When we square these collinear matrix elements, two copies of the $\delta$-functions with ${\cal O}(\lambda^0)$ momenta will appear, for example $\delta(p^- - p_{X_n}^-) \delta(p^- - p_{X_n}^-) = \delta(0) \delta(p^- - p_{X_n}^-)$. Therefore one part of the volume factor, $V_1=2\pi \delta(0)$, will be canceled in the squared of each collinear matrix element, and we define
\begin{align} \label{eq:CnCnb}
& \frac{1}{V_1} \sum_{X_n}  \Big\langle p \Big| \sum_{j=q,g} 
  \int\!\!dx^{\prime\prime +}
  {\cal O}_{n,k^{\prime -}}^{jA'}(q_\perp'')\Big(x^{\prime\prime +}\frac{{\bar n}}{2}\Big)
  \Big| X_n \Big\rangle
  \Big\langle X_n \Big| 
  \sum_{i=q,g} \int\!\!dx^{+}
  {\cal O}_{n,k^{-}}^{iA}(q_\perp)\Big(x^{+}\frac{{\bar n}}{2}\Big)
  \Big| p \Big\rangle 
  \nonumber \\
  &\qquad 
  = \delta^{AA'}\,2\, 
    \delta^2(q_\perp-q_\perp'') \, \vec q_\perp^{\: 2}\,   
    C_n(q_\perp,p^-)
 \,, \nonumber\\
& \frac{1}{V_1}  \sum_{X_{\bar n}}  \Big\langle p' \Big| \sum_{j=q,g} 
  \int\!\!dx^{\prime\prime\prime -}
  {\cal O}_{{\bar n},k^{\prime +}}^{jB'}(q_\perp''')\Big(x^{\prime\prime\prime -}\frac{n}{2}\Big)
  \Big| X_{\bar n} \Big\rangle \Big\langle X_{\bar n} \Big| 
  \sum_{i=q,g}\int\!\!dx^{\prime -}
  {\cal O}_{{\bar n},k^{+}}^{iB}(q_\perp')\Big(x^{\prime -}\frac{n}{2}\Big) \Big| p' \Big\rangle
  \nonumber \\
  &\qquad 
  =  \delta^{BB'}\, 2\, 
   \delta^2(q_\perp'-q_\perp''')\, \vec q_\perp^{\:\prime 2}\,   C_{\bar n}(q_\perp',p^{\prime +})
  \,,
\end{align}
where we've introduced the functions $C_n$ and $C_{\bar n}$ to encode the nontrivial dependencies.  By including the factors of $\vec q_\perp^{\: 2}$ and $\vec q_\perp^{\:\prime 2}$ on the right-hand-side we are adopting a normalization where the collinear functions $C_n$ and $C_{\bar n}$ include $1/\vec q_\perp^{\: 2}$ and $1/\vec q_\perp^{\:\prime 2}$ Glauber exchange potentials. We see that the matrix elements of the collinear operators gives only one combination of the color indices. 

For the soft matrix element $\langle X_s | \cdots | 0\rangle$  we are left with
the operator 
\begin{align} \label{eq:Os11red}
  O_{s(1,1)}^{AB}(q_\perp,q_\perp^\prime) 
  & \equiv
  \frac{(2\pi)^2}{2} \sum_{k^\pm} \int dx^{\prime +} dx^{-}
  O_{s(1,1),-k^\pm}^{AB}(q_\perp,q_\perp')
  \Big(x^-\frac{n}{2}, x^{\prime+}\frac{{\bar n}}{2}\Big)
  \\
  & = \sum_{k^\pm}
  {\cal O}_{s,-k^\pm}^{AB}(q_\perp,-q_\perp')(\tilde x=0)
  \nonumber\\
  &\
  + \frac{i}{2} (2\pi)^2 \sum_{k^\pm} \int \!\! dx^{\prime +} dx^- 
     \:T  \sum_{i,j=q,g} \!\!
     {\cal O}_{s,-k^-}^{i_n A}(q_\perp)\Big(\frac{n}{2} x^-\Big)\
     {\cal O}_{s,-k^+}^{j_{\bar n} B}(-q_\perp')\Big(\frac{{\bar n}}{2} x^{\prime +}\Big)
   \,. \nonumber
\end{align}
Note that the soft operators here are unrestricted in their $k^\pm$ momenta labels, which can be absorbed back into the $x^-$ and $x^{\prime +}$ coordinate dependence, for example $\sum_{k^-} {\cal O}_{s,-k^-}^{i_n A}(q_\perp)(x^- n/2) = {\cal O}_{s}^{i_n A}(q_\perp)(x^- n/2)$. The squared soft matrix element is then given by
\begin{align} \label{eq:S4q}
 & \frac{1}{V_2} \frac{(2\pi)^4}{q_\perp^2 q_\perp^{\prime 2}  
   q_\perp^{\prime\prime 2}  q_\perp^{\prime\prime\prime 2} }
  \sum_{X_s}  \big\langle 0 \big| 
   O_{s(1,1)}^{\dagger A'B'}(q_\perp'', q_\perp''')
  \big| X_s \big\rangle \big   \langle X_s \big| 
   O_{s(1,1)}^{AB}(q_\perp,q_\perp')\big| 0\big\rangle
 \equiv \frac{1}{q_\perp^2 q_\perp^{\prime 2}}\,
    S_G^{AA'BB'}(q_\perp,q_\perp',q_\perp'',q_\perp''') ,
\end{align}
where $V_2=(2\pi)^2\delta^2(0)$ includes the remaining part of the volume factor, and the prefactor $1/(q_\perp^2 q_\perp^{\prime 2})$ is pulled out in the definition of $S_G$ for later convenience.  The contraction of color indices and the $\perp$ $\delta$-functions from the collinear sectors in \eq{CnCnb} allows us to reduce the form of the required soft function further to
\begin{align} \label{eq:Gdefn}
 S_G(q_\perp,q_\perp')
 &= \int\!\! d^2\!q_\perp'' d^2\!q_\perp'''\, \delta^{AA'} \delta^{BB'} 
 \delta^2(q_\perp\!\minus q_\perp'')\delta^2(q_\perp'\!\minus q_\perp''')
  S_G^{AA'BB'}(q_\perp,q_\perp',q_\perp'',q_\perp''') 
 \nn \\
 &=  \frac{(2\pi)^4}{V_2} \frac{\delta^{AA'}\delta^{BB'}}{(\vec q_\perp^{\:2}\, \vec q_\perp^{\:\prime\,2})} 
   \sum_X  \big\langle 0 \big| O_{s(1,1)}^{AB}(q_\perp,q_\perp^\prime)
       \big| X \big\rangle \big   \langle X \big|  O_{s(1,1)}^{\dagger A'B'}(q_\perp,q_\perp^\prime) \big| 0\big\rangle
    \,.
\end{align}
The $\delta^{AA'}\delta^{BB'}$ contraction in \eq{Gdefn} implies that the combined Glauber exchanges on either side of the cut are in a color singlet state. This linear approximation with one (Glauber) gluon exchange on each side of the cut is sometimes referred to as the Low-Nussinov pomeron. In some applications one may be required to consider a color-octet configuration and/or a $\perp$-momentum configuration with $q_\perp \ne q_\perp^{\prime\prime}$ and $q_\perp'\ne q_\perp^{\prime\prime\prime}$, but we will not examine a case like this here.

Combining all these results, the squared forward transition matrix at lowest order in the Glauber exchange is given by 
\begin{align}  \label{eq:T11}
 T_{(1,1)} 
 &= \int\! d^2q_\perp  d^2q_\perp'\:
 C_n(q_\perp,p^-)  S_G(q_\perp,q_\perp')  C_\bn(q_\perp',p^{\prime +})
  \,,
\end{align}
Here $C_n(q_\perp,p^-)$ and $C_\bn(q_\perp',p^{\prime +})$ are given by the matrix elements in \eq{CnCnb}.
Finally, we note that conjugation relation in \eq{hermitian} implies 
\begin{align}
 O_{s(1,1)}^{AB}(q_\perp,q_\perp')=O_{s(1,1)}^{BA}(q_\perp',q_\perp)\Big|_{n\leftrightarrow \bn} \,.
\end{align}
Since we integrate over soft $\pm$-momenta to define $S_G(q_\perp,q_\perp')$ it only has the trivial $n\cdot \bn=2$ dependence on the collinear directions that show up in the soft operator Wilson lines, and hence its definition implies that it is a symmetric function
\begin{align} \label{eq:Gsymm}
  S_G(q_\perp,q_\perp') &= S_G(q_\perp',q_\perp) \,.
\end{align}
Note that here we have not factorized in the scales $t$ and $\Lambda_{\rm QCD}^2$, so the collinear and soft functions contain both of these scales, with the dependence on $t$ appearing through $q_\perp$ or $q_\perp^{\prime}$. The factorization in \eq{T11} for $T_{(1,1)}$ separates the modes in rapidity, allowing for a resummation of $\ln(s/t)$'s, but does not include a factorization from expanding in $\Lambda_{\rm QCD}^2/t\ll 1$. 

The result in \eq{T11} gives a factorized form for the forward scattering process at lowest order in the Glauber exchange operators, but to all orders in the soft and collinear Lagrangians, ${\cal L}_S^{(0)}$ and ${\cal L}_{n,\bn}^{(0)}$. Therefore the functions $C_n(q_\perp)$, $C_{\bn}(q_\perp')$, and $S_G(q_\perp,q_\perp')$ each have non-trivial series in $\alpha_s$. In the next two sections, \secs{bfkl}{bfklconsist} we will consider the renormalization of the lowest order transition amplitude $T_{(1,1)}$, which at leading logarithmic order simply involves the rapidity renormalization of these soft and collinear functions, and only requires ${\cal O}(\alpha_s)$ real and virtual calculations. For the full scattering correction at this same order in $\alpha_s$, there is also a term with more insertions of the Glauber operators:
\begin{align}
  T_{(2,1)}+T_{(1,2)} &= \frac{1}{V_4} \sum_X  \Big[
   \big\langle p p' \big| U_{(2,2)} \big| X \big\rangle
   \big\langle X \big| U_{(1,1)}^\dagger \big| p p' \big\rangle 
  + \big\langle p p' \big| U_{(1,1)} \big| X \big\rangle
   \big\langle X \big| U_{(2,2)}^\dagger \big| p p' \big\rangle \Big] \,.
\end{align}
At this order in $\alpha_s$ we can either contract both the $O_n^{iA} O_n^{iB}$ and $O_\bn^{jA} O_\bn^{jB}$ in $U_{(2,2)}$ to give a Glauber box diagram as in \fig{glaub_loop} or we could attach the two forward collinear lines in each of $O_n^{iA} O_n^{i'B}$ and $O_\bn^{jA} O_\bn^{j'B}$ to different partons in the incoming $ \langle p_n p'_\bn |$ state. Without additional emissions neither of these contributions has a logarithmic rapidity divergence, and hence it suffices to consider just $T_{(1,1)}$ when deriving the leading-logarithmic renormalization equations. 

\begin{figure}[t!]
	%
	%
%
\begin{center}
	\includegraphics[width=0.3\columnwidth]{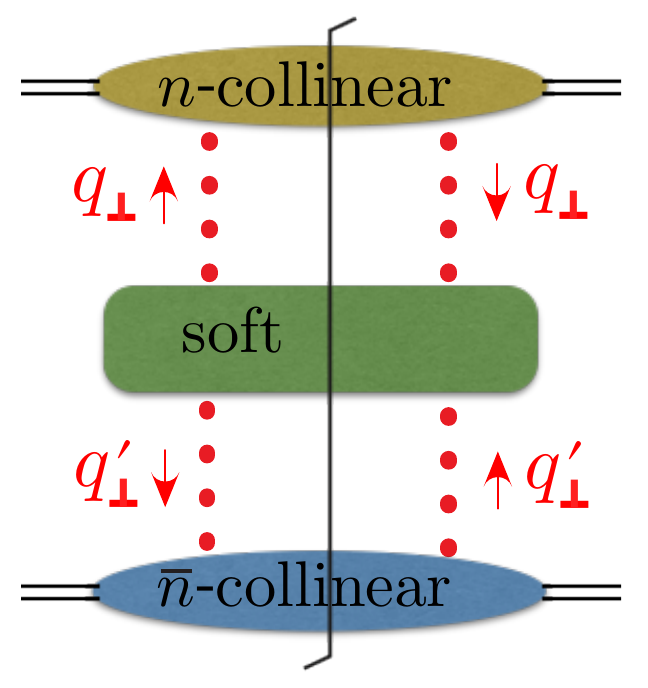} 
\end{center}
	\vspace{-0.6cm}
	\caption{\setcaptionskip
		Pictoral representation of the soft-collinear factorization of the lowest order forward scattering Glauber interaction. This factorization is linear in the Glauber exchange on each side of the cut and leads to soft and collinear functions whose RGEs are given by the BFKL equation.}
	\label{fig:bfkl_factorization}
	\setmainskip
\end{figure}

Introducing the rapidity cutoff $\nu$ and renormalized collinear and soft functions we have
\begin{align}  \label{eq:T11rap}
 T_{(1,1)} &= \int\! d^2q_\perp  d^2q_\perp'\:
 C_n(q_\perp,p^-,\nu)  S_G(q_\perp,q_\perp',\nu) C_\bn(q_\perp',p^{\prime +},\nu)  \,.
\end{align}
The physical picture for this factorization of the forward cross section is given in \fig{bfkl_factorization}.
In the next section we derive the leading-logarithmic evolution equation for the soft function $S_G(q_\perp,q_\perp',\nu)$ and show that it is the BFKL equation. Then in \sec{bfklconsist} we will derive the BFKL equations for $C_n(q_\perp,p^-,\nu)$ and $C_\bn(q_\perp',p^{\prime +}\nu)$ by using renormalization group consistency.


\subsection{BFKL Equation for the Soft Function}
\label{sec:bfkl}

In evaluating matrix elements of the forward scattering operator, large logs arise due to the tension between the collinear modes  whose natural rapidity scale is $\nu_c\sim \sqrt{\hat s}$ and the soft mode for which $\nu_s \sim \sqrt{-t}$. Thus the large logs cannot be minimized with a single choice of the rapidity scale $\nu$ in the SCET matrix elements.  Since the final result is independent of which $\nu$ we choose, we will  take $\nu=\nu_c$ so that all the large logs reside in the soft part of the matrix element. These logs are  summed up by running the soft function in rapidity space
from $\nu_s$ to $\nu_c$.   For the calculations in this section we set the IR mass regulator $m=0$ since infrared divergences will cancel in the sum of real and virtual diagrams. We also set $d=4$ since only the rapidity divergences will be relevant for our RGE analysis.

We will be working in the limit where $(-t)\gg \Lambda^2_{\rm QCD}$ so that we may treat Glauber exchange perturbatively, but do not attempt to factorize these two infrared scales in the EFT explicitly. To sum the logarithms at leading logarithmic order (LL) we only need to consider the $k=k'=1$ term in \eq{Uglabexpn}, and this Glauber operator effectively acts like an external current.  This term yielded the factorization formulae in \eq{T11rap}.

We label the soft piece of the forward scattering operator in terms of the incoming $q_\perp$ and $q_\perp^\prime$ such that the lowest order Feynman rule is given by
\begin{align}
 \big\langle 0  \big|  O_{s(1,1)}^{AB}(q_\perp,q_\perp^\prime)
 \big| 0 \big\rangle
 =
- i\, 8\pi\alpha_s(\mu) \, \delta^{AB}\,  {\vec q_\perp^{\:2}}\:
\delta^{2}( \vec q_\perp + \vec q^{\:\prime}_\perp) \,.
\end{align}
Here $O_{s(1,1)}^{AB}$ was defined in \eq{Os11red}, and this lowest order contribution comes from ${\cal O}_s^{AB}(q_\perp,-q_\perp^\prime)$ which was defined in \eq{Os1q}. Thus at the level of the amplitude squared
\begin{align} \label{eq:G0}
	\raisebox{-1.85cm}{
\includegraphics[width=0.16\columnwidth]{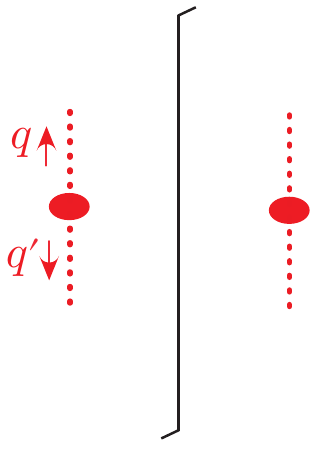}
}  \hspace{0.3cm}\vspace{-20cm} 
	=S_G^{(0)}(q_\perp,q_\perp^\prime)
   & \equiv \frac{(2\pi)^4}{V_2} \frac{1}{(\vec q_\perp^{\:2}\, \vec q_\perp^{\:\prime\,2})}
   \big\langle 0 \big| O_{s(1,1)}^{AB} \big| 0 \big\rangle 
   \big\langle 0\big| O_{s(1,1)}^{AB\dagger} \big| 0 \big\rangle 
  \\[-35pt]
   & = (8\pi\alpha_s)^2 \delta^{AA}
   \: (2\pi)^2 \delta^{2}( \vec q_\perp\! + \vec q^{\:\prime}_\perp)
	. \nn 
\end{align}
Here the solid vertical line denotes the final state cut. The color factor $\delta^{AA}=N_c^2-1$ and the volume factor $V_2=(2\pi)^2\delta^2(0)$.

To renormalize the $S_G(q_\perp,q_\perp')$ matrix element we must consider the ${\cal O}(\alpha_s)$ real and virtual corrections.  The real radiation correction is calculated using the one-soft gluon Feynman rule of ${\cal O}_s^{AB}(q_\perp,q_\perp^\prime)$ (equivalent to the Lipatov vertex) which was given in Fig.~\ref{fig:onegluon}, and implementing the prefactors in the definition in \eq{Gdefn}. We let the outgoing momentum of the soft gluon be $k=-q-q'$, and note that the multipole expansion for collinear particles restricts the ${\cal O}(\lambda)$ momentum flow as discussed in \sec{LGtransverse}. This gives $n\cdot q'=-n\cdot k$ and $\bn\cdot q =- \bn\cdot k$.  Summing over polarizations in Feynman gauge, the square of the one-gluon Feynman rule is
\begin{align} \label{eq:bfkl1gmelt}
 & \frac{(2\pi)^4}{V_2}  
   \frac{1}{(\vec q_\perp^{\:2}\, \vec q_\perp^{\:\prime\,2})}
   \big\langle 0 \big| O_{s(1,1)}^{AB} \big| g(k) \big\rangle 
   \big\langle g(k) \big| O_{s(1,1)}^{AB\dagger} \big| 0 \big\rangle 
  \\
  &= - \frac{(8\pi\alpha_s)^2(4\pi\alpha_s)}{(\vec q_\perp^{\:2}\, \vec q_\perp^{\:\prime\,2})}  f^{ABE} f^{ABE} \bigg(
  q_\perp^\mu \minus q_\perp^{\prime\mu} 
  \plus n\cdot q' \frac{\bn^\mu}{2} \minus \bn\cdot q \frac{n^\mu}{2}
  \plus \frac{n^\mu \vec q_\perp^{\:2}}{n\cdot q'}
  \minus \frac{\bn^\mu \vec q_\perp^{\:\prime\,2}}{\bn\cdot q}\bigg)^2
  \!\! (2\pi)^2 \delta^2(\vec k_\perp \plus \vec q_\perp \plus \vec q_\perp^{\:\prime})
  \nn \\
 &= - \frac{(8\pi\alpha_s)^2(4\pi\alpha_s)}{(\vec q_\perp^{\:2}\, \vec q_\perp^{\:\prime\,2})}  C_A \delta^{AA} \bigg(
  - n\cdot q'\, \bn\cdot q
  + (\vec q_\perp + \vec q_\perp^{\:\prime})^2 
  - \frac{4 \vec q_\perp^{\:\prime\,2}\, \vec q_\perp^{\:2}}
  {n\cdot q'\,\bn\cdot q} \bigg) 
  (2\pi)^2 \delta^2(\vec k_\perp \plus \vec q_\perp \plus \vec q_\perp^{\:\prime})
  \nn  \\
 &= (8\pi\alpha_s)^2 \: \frac{(16 \pi\alpha_s)}{(\vec q_\perp + \vec q_\perp^{\:\prime})^2}
  \: C_A \delta^{AA} (2\pi)^2 \delta^2(\vec k_\perp \plus \vec q_\perp \plus \vec q_\perp^{\:\prime})
 \nn \,,
\end{align}
where in the last equality we used the soft gluon equations of motion $0 = k^2 = (q+q')^2 = n\cdot q'\: \bn\cdot q - (\vec q_\perp + \vec q_\perp^{\:\prime})^2$ to eliminate all but the last term in the large round brackets, and to replace the product $n\cdot q'\, \bn\cdot q$. Note that this squared matrix element is independent of the longitudinal gluon momentum. Since the surviving term in \eq{bfkl1gmelt} was generated by the soft Wilson lines in the operator ${\cal O}_s^{AB}(q_\perp,q_\perp')$ we must also include appropriate factors of the rapidity regulator, giving $w^2 |2k^z|^{-\eta} \nu^\eta$. This factor regulates the soft gluons phase space integral, which is
\begin{align}
  \int \ddslash\!^dk\: C(k)\: w^2 |2k^z|^{-\eta} \nu^\eta
   &= \int \frac{\ddslash\!^{d-1}k}{2E_k}\: w^2 |2k^z|^{-\eta} \nu^\eta
    \bigg|_{E_k^2=\vec k^{\:2}}
   = \int_0^\infty\! \frac{\ddslash\!k^-}{2k^-}\, \ddslash\!^{d-2}k_\perp\: w^2 |2k^z|^{-\eta} \nu^\eta \bigg|_{k^+=\vec k_\perp^{\:2}/k^-} 
  \nn\\
  &= \frac{w^2}{4\pi} \int\!\! \ddslash\!^{d-2}k_\perp\! \int_0^\infty \frac{dk^-}{(k^-)^{1-\eta}} \, \big| (k^-)^2 - \vec k_\perp^{\:2}\big|^{-\eta}
  \nu^\eta 
  \nn\\[5pt]
  &= \frac{w^2}{4\pi} \frac{\Gamma(\frac{\eta}{2})\Gamma(\frac{1-\eta}{2})\nu^\eta}{2^\eta \sqrt{\pi}}
   \int\!\! \ddslash\!^{d-2}k_\perp \:  | \vec k_\perp |^{-\eta}
  \,.
\end{align} 
Here $C(k)=2\pi\delta(k^2)\theta(k^0)$ is the factor from the cut gluon. 
Putting these pieces together, and keeping only the $1/\eta$ divergent contribution, for the real emission contribution to the ${\cal O}(\alpha_s)$ correction to $S_G(q_\perp,q_\perp^\prime)$ we have
\begin{align}  \label{eq:Greal}
\raisebox{-1.5cm}{
\includegraphics[width=0.2\columnwidth]{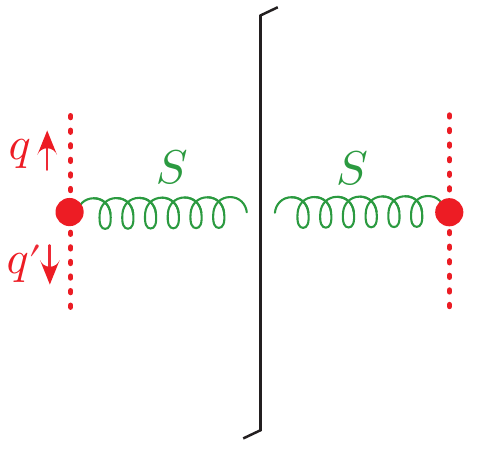}
}  \hspace{0.2cm}\vspace{0cm} 
&= 
 (8\pi\alpha_s)^2 4\alpha_s  
 C_A \delta^{AA} w^2\Gamma\Big(\frac{\eta}{2}\Big)
 \!\int\!\! \frac{ \ddslash^{\!2}k_\perp}
 { \vec k_\perp^{\:2} }
  \: (2\pi)^2 \delta^2( \vec k_\perp +\vec q_\perp + \vec q_\perp^{\:\prime})
  \nn \\[-25pt]
&= 
 (8\pi\alpha_s)^2 4\alpha_s  
 C_A \delta^{AA} w^2\Gamma\Big(\frac{\eta}{2}\Big)
 \!\int\!\! \frac{ \ddslash^{\!2}k_\perp}
 { (\vec k_\perp- \vec q_\perp)^2}
  \: (2\pi)^2 \delta^2( \vec k_\perp  + \vec q_\perp^{\:\prime})
  \nn \\
&=\frac{ C_A \alpha_s }{\pi^2}\, w^2 \Gamma\Big(\frac{\eta}{2}\Big) 
  \int\!\! \frac{ d^{2}k_\perp}
  { ( \vec k_\perp- \vec q_\perp )^2 }
  \: S_G^{(0)}( k_\perp,   q^{\,\prime}_\perp)
\,,
\end{align}
where in the second equality we took $\vec k_\perp \to \vec k_\perp - \vec q_\perp$. In the last equality we used $\ddslash\!^{2}k_\perp = d^2k_\perp/(2\pi)^2$ and the tree level $S_G^{(0)}$ from \eq{G0}. 

For the soft virtual corrections we have contributions from the flower and eye graphs, which we must incorporate at a level where we have not yet performed the $k_\perp$ loop integration. To obtain results for $\big\langle 0  \big|  O_{s(1,1)}^{AB}(q_\perp,q_\perp^\prime) \big| 0 \big\rangle$ we strip off the factor of $(\bar u_n T^A \frac{\bnslash}{2} u_n)(\bar v_\bn \bar T^B\frac{\nslash}{2} v_\bn)/(\vec q_\perp^{\:2})^2$ from the soft loop integrands in \sec{loop2match} in \eqs{soft_eye}{soft_flower} and include a $(2\pi)^2 \delta(\vec q_\perp+\vec q_\perp^{\:\prime})$.  Keeping only the rapidity divergent terms we have 
\begin{align} \label{eq:bfkl_eye}
 \raisebox{-1.7cm}{
\includegraphics[width=0.1\columnwidth]{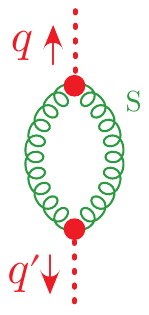}
}
  &=  2 g^4\: C_A \, \delta^{AB}
     \!\! \int\!\! 
   \frac{\ddslash\!^{4} k\  w^2  |2k_z|^{-\eta}\,\nu^\eta }{ [k^2][(k+q)^2]} 
  \: \bigg\{ \frac{4 [k_\perp \cdot (k_\perp+q_\perp)]^2}{\bn\cdot k\: n\cdot k}
    \bigg\} (2\pi)^2 \delta(\vec q_\perp\plus \vec q_\perp^{\:\prime})
 \nn\\[-35pt]
  &= - i\,32\pi^2\alpha_s^2 C_A  \delta^{AB}
  \frac{w^2 \Gamma(\frac{\eta}{2})\Gamma(\frac{1-\eta}{2})}{2^\eta \pi^{3/2}}
  \!\! \int\!\! \ddslash\!^{2} k_\perp
   \frac{[\vec k_\perp \mcdot (\vec k_\perp\plus \vec q_\perp)]^2}{(2\vec k_\perp\mcdot \vec q_\perp \plus \vec q_\perp^{\:2})} \bigg[ \frac{1}{(\vec k_\perp\plus \vec q_\perp)^2} \minus \frac{1}{\vec k_\perp^{\:2}}\bigg]
  (2\pi)^2 \delta(\vec q_\perp\plus \vec q_\perp^{\:\prime})
  \nn\\
 &= -i\,32\pi \alpha_s^2\: C_A \, \delta^{AB}
  w^2 \Gamma\Big(\frac{\eta}{2}\Big)
  \!\! \int\!\! \ddslash\!^{2} k_\perp\,
   \frac{-[\vec k_\perp^{\:2} - \vec q_\perp^{\:2}/4]^2}{(\vec k_\perp + \vec q_\perp/2)^2\, (\vec k_\perp - \vec q_\perp/2)^2}
  \: (2\pi)^2 \delta(\vec q_\perp\plus \vec q_\perp^{\:\prime})
  \nn\\
 &= - i\,16\pi \alpha_s^2 C_A  \delta^{AB}
  w^2 \Gamma\Big(\frac{\eta}{2}\Big)
  \!\! \int\!\! \ddslash\!^{2} k_\perp \bigg[ \frac{\vec q_\perp^{\:2} }{\vec k_\perp^{\:2}} - \frac{ (\vec q_\perp^{\:2})^2 }
   {2 (\vec k_\perp \plus \vec q_\perp/2)^2\, (\vec k_\perp \minus \vec q_\perp/2)^2}
  \bigg]
  (2\pi)^2 \delta(\vec q_\perp\plus \vec q_\perp^{\:\prime})
  .
\end{align}
To obtain the third equality we shifted $\vec k_\perp \to \vec k_\perp - \vec q_\perp/2$ and then simplified the integrand, and to obtain the last line we partial fractioned the numerator and dropped integrands that are odd in $\vec k_\perp$ and which vanish in dimensional regularization because they are power law divergent. Similarly, for the flower graph we have
\begin{align} \label{eq:bfkl_flower}
 \raisebox{-1.2cm}{
\includegraphics[width=0.15\columnwidth]{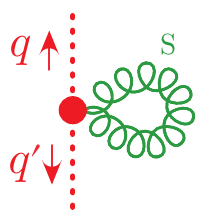}
}
  &= - 4g^4 \vec q_\perp^{\:2}\:  C_A \delta^{AB}
    \!\! \int\!\! \ddslash\!^{4} k 
   \: \frac{ w^2 |2k_z|^{-\eta}\,\nu^\eta }{ [k^2](n\cdot k)(\bn\cdot k)} 
   (2\pi)^2 \delta(\vec q_\perp+\vec q_\perp^{\:\prime})
  \nn\\[-20pt]
  &= \frac{4i (4\pi\alpha_s)^2\vec q_\perp^{\:2}}{(4\pi)}  C_A \delta^{AB}
    \frac{w^2\Gamma(\frac{\eta}{2})\Gamma(\frac{1-\eta}{2})}{2^\eta \sqrt{\pi}}
    \!\! \int\!\! \frac{\ddslash\!^{2} k_\perp}{\vec k_\perp^{\:2}}
   \:  (2\pi)^2 \delta(\vec q_\perp+\vec q_\perp^{\:\prime})
  \nn\\
  &= i\, 16\pi\alpha_s^2\:  C_A \delta^{AB}
    w^2\Gamma\Big(\frac{\eta}{2}\Big)
    \!\! \int\!\! \frac{\ddslash\!^{2} k_\perp\  \vec q_\perp^{\:2}}{\vec k_\perp^{\:2}}
   \:  (2\pi)^2 \delta(\vec q_\perp+\vec q_\perp^{\:\prime})
   \,.
\end{align}
Combining \eqs{bfkl_eye}{bfkl_flower} we see that the self contraction of Wilson lines in the soft flower graph cancels one of the terms in the eye-graph, leaving
\begin{align}
   \raisebox{-1.7cm}{
\includegraphics[width=0.1\columnwidth]{figs/softTOP}
}
 +\!\! \raisebox{-1.2cm}{
\includegraphics[width=0.15\columnwidth]{figs/softW}
}
  &=  i\,8\pi \alpha_s^2 C_A  \delta^{AB}
  w^2 \Gamma\Big(\frac{\eta}{2}\Big)
  \!\! \int\!\! \frac{\ddslash\!^{2} k_\perp\ (\vec q_\perp^{\:2})^2  }
   { (\vec k_\perp \plus \vec q_\perp/2)^2\, (\vec k_\perp \minus \vec q_\perp/2)^2}
  (2\pi)^2 \delta(\vec q_\perp\plus \vec q_\perp^{\:\prime})
  \nn\\[-30pt]
 &=  i\,8\pi \alpha_s^2\: C_A  \delta^{AB}
  w^2 \Gamma\Big(\frac{\eta}{2}\Big)
  \!\! \int\!\! \frac{\ddslash\!^{2} k_\perp\  (\vec q_\perp^{\:2})^2  }
   { \vec k_\perp^{\:2}\, (\vec k_\perp \minus \vec q_\perp)^2}
  (2\pi)^2 \delta(\vec q_\perp\plus \vec q_\perp^{\:\prime})
  .
\end{align}
The contribution coming from the soft Wilson line and the time ordered product 
can be combined to give the full ${\cal O}(\alpha_s)$ virtual correction to $S_G(q_\perp,q_\perp^\prime)$
\begin{align} \label{eq:Gvirt}
& 2\
\raisebox{-1.8cm}{
\includegraphics[width=0.17\columnwidth]{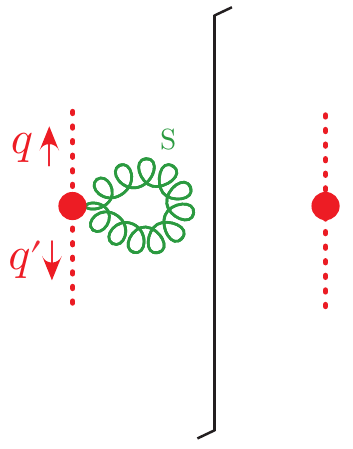}
}  \hspace{0.15cm} + 2\hspace{0.15cm}
\raisebox{-1.8cm}{
\includegraphics[width=0.17\columnwidth]{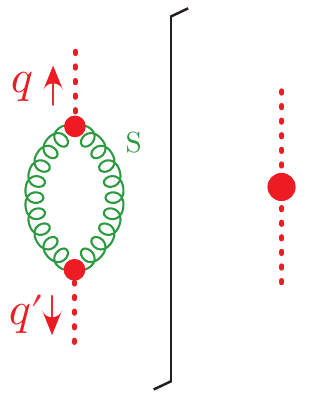}
}  \hspace{0.1cm} 
\\
&\qquad 
 =  -\frac{2(8\pi \alpha_s)^2\alpha_s}{\vec q_\perp^{\:2} }\:
  C_A  \delta^{AB}
  w^2 \Gamma\Big(\frac{\eta}{2}\Big)
  \!\! \int\!\! \frac{\ddslash\!^{2} k_\perp \ (\vec q_\perp^{\:2})^2 }
   { \vec k_\perp^{\:2}\, (\vec k_\perp \minus \vec q_\perp)^2}
  (2\pi)^2 \delta(\vec q_\perp\plus \vec q_\perp^{\:\prime})
  \nn\\
&\qquad 
  =-\frac{ C_A \alpha_s}{2\pi^2}\, w^2 \Gamma\Big(\frac{\eta}{2}\Big)\!
  \int\!\! d^2k_\perp\frac{ \vec q_\perp^{\:2}}{\vec k_\perp^{\:2} (\vec k_\perp-\vec q_\perp)^2}\:
  S_G^{(0)}(q_\perp,q_\perp^\prime) 
  \,,\nn
\end{align}
where in the last line we used $\ddslash\!^{2}k_\perp = d^2k_\perp/(2\pi)^2$ and the tree level $S_G^{(0)}$ from \eq{G0}. 
The factors of $2$ next to the graphs appear because we get the same contribution when the virtual loop appears on either side of the cut.

The results up to ${\cal O}(\alpha_s)$ from Eqs.~(\ref{eq:G0},\ref{eq:Greal},\ref{eq:Gvirt}) can be summarized as yielding the ${\cal O}(\alpha_s)$ rapidity divergent correction to the bare soft function,
\begin{align}
 S_G^{\rm bare}(q_\perp,q^\prime_\perp)
  = S_G^{(0)}(q_\perp,q_\perp^\prime)
  +\frac{\alpha_s C_A}{\pi^2} w^2\Gamma\Big(\frac{\eta}{2}\Big) 
   \!\! \int\!\! \frac{d^2k_\perp}{(\vec k_\perp-\vec q_\perp)^2}
  \bigg[ S_G^{(0)}(k_\perp,q_\perp^\prime)
     -\frac{\vec q_\perp^{\:2}}{2\vec k_\perp^{\:2} }  S_G^{(0)}(q_\perp,q_\perp^\prime)
  \bigg] .
\end{align}
The rapidity divergence in the soft function is renormalized by a standard SCET soft function counterterm $Z_{S_G}(q_\perp,q_\perp')$ through the convolution
\begin{align} \label{eq:rren}
  S_G(\vec q_\perp, \vec q_\perp^{\:\prime},\nu)= \int\!\! d^2k_\perp\: Z_{S_G}(q_\perp,k_\perp)\:  S_G^{\rm bare}(k_\perp,q_\perp^\prime) \,.
\end{align}
To cancel the $1/\eta$ divergence we require
\begin{align} \label{eq:bfklZ1loop}
 Z_{S_G}(q_\perp,k_\perp) &= \delta^2(\vec q_\perp\! - \vec k_\perp) 
  - \frac{2C_A \alpha_s(\mu) w^2(\nu)}{\pi^2\,\eta} \bigg[ \frac{1}{(\vec k_\perp\! -\vec q_\perp)^2} - \delta^2(\vec q_\perp \!-\! \vec k_\perp) \int\!\! 
  \frac{d^2k_\perp'\: \vec q_\perp^{\:2}}{2 \vec k_\perp^{\,\prime 2} (\vec k_\perp^{\:\prime}\! - \vec q_\perp)^2} \bigg]
  .
\end{align}
The rapidity renormalization group (RRG) equation then follows from the $\nu$-independence of the bare soft function,
\begin{align}
 0 = \nu \frac{d}{d\nu}  S_G^{\rm bare}(q_\perp,q_\perp^\prime) 
  =\nu \frac{d}{d\nu}  \int\!\! d^2k_\perp\, Z_{S_G}^{-1}(q_\perp,k_\perp)\,   S_G(k_\perp,q_\perp^\prime,\nu) \,.
\end{align}
Writing out the derivatives of the two terms and inverting, we find that the renormalized soft function obeys the RGE equation
\begin{align}
\nu \frac{d}{d\nu}  S_G(q_\perp,q_\perp^\prime,\nu)
 = \int \!\! d^2k_\perp \:  \gamma_{S_G}(q_\perp, k_\perp) \: 
  S_G(k_\perp,q_\perp^\prime,\nu) \,,
\end{align}
where the anomalous dimension is given by
\begin{align}
\gamma_{S_G} (q_\perp, q_\perp^\prime)
 &= - \int d^2k_\perp Z_{S_G}(q_\perp,k_\perp) \,
  \nu \frac{d}{d\nu}\, Z_{S_G}^{-1}(k_\perp,q_\perp^\prime) 
 \,.
\end{align}
Inserting the one-loop result from \eq{bfklZ1loop} and using $(\nu d/d\nu) w^2(\nu) = -\eta\, w^2(\nu)$ then sending $w^2(\nu)\to 1$ this gives
\begin{align}  \label{eq:gSG}
 \gamma_{S_G}(q_\perp, q_\perp^\prime) 
  &=\frac{2 C_A \alpha_s(\mu)}{\pi^2}\bigg[
 \frac{1}{(\vec q_\perp-\vec q_\perp^{\:\prime})^2} 
  -\delta^2(\vec q_\perp-\vec q_\perp^{\:\prime})\int\!\! d^2k_\perp \frac{\vec q_\perp^{\:2}}{2 \vec k_\perp^{\:2} (\vec k_\perp-\vec q_\perp)^2}
  \bigg]
  \,.
\end{align}
Note that this anomalous dimension is not just a function of the difference $q_\perp-q_\perp'$, but it is easy to see from \eq{gSG} that it is symmetric,
\begin{align}
  \gamma_{S_G}(q_\perp,q_\perp') = \gamma_{S_G}(q_\perp',q_\perp) \,.
\end{align}

The anomalous dimension $\gamma_{S_G}$ yields an RGE for $S_G(q_\perp,q_\perp',\nu)$ which is precisely the leading logarithmic BFKL equation, 
\begin{align} \label{eq:bfkl}
 \nu\frac{d}{d\nu}\,  S_G(q_\perp,q_\perp',\nu)
   &= \frac{2 C_A\alpha_s(\mu)}{\pi^2} \int\!\! d^2k_\perp 
   \bigg[ \frac{ S_G(k_\perp,q_\perp',\nu)}
     {(\vec k_\perp-\vec q_\perp)^2}
   - \frac{\vec q_\perp^{\:2}\:  S_G(q_\perp,q_\perp',\nu)}
   { 2\vec k_\perp^2 (\vec k_\perp-\vec q_\perp)^2}  \bigg]
  \,.
\end{align}
The BFKL equation is often~\cite{ForshawBook:1997,IoffeFadinLipatov:2010,KovchegovBook:2012} written in terms of the derivative of a rapidity, $Y = \ln(\nu^2/\mu^2)\sim \ln s$.  The fact that $\partial/\partial Y = (1/2) \nu d/d\nu$ explains our factor of $2$ in the prefactor on the right-hand side of \eq{bfkl}. Note that in our SCET calculation, the fact that \eq{bfkl} is obtained for the all orders soft function $S_G$ (rather than just the one-loop soft function) follows immediately from the structure of the effective field theory operators and the multiplicative form of the rapidity renormalization in \eq{rren}. In classic derivations of the BFKL equation, this step is often quite involved.

A derivation of the BFKL equation from an SCET based operator construction with Glaubers was considered earlier by Fleming in Ref.~\cite{Fleming:2014rea}. Although the idea of carrying out rapidity renormalization of a squared matrix element of soft fields is common between our two calculations, there are also a few differences, both on the conceptual and calculation sides. The scattering operator considered in~\cite{Fleming:2014rea} is ${\cal O}_G^{n\bn}=(\bar\chi_\bn S_\bn^\dagger T^A \frac{\nslash}{2} S_\bn \chi_\bn) \frac{1}{\cP_\perp^2} (\bar\chi_n S_n^\dagger T^a \frac{\bnslash}{2} S_n \chi_n)$, which differs from our $O_{ns\bn}^{qq}$. In particular, unlike $O_{ns\bn}^{qq}$,  the operator ${\cal O}_G^{n\bn}$ is not soft gauge invariant in \SCETb due to the presence of the $\frac{1}{\cP_\perp^2}$, which does not allow the soft gauge transformation factors from the two sides to cancel.  This distinction also causes differences for the calculations.  In the soft part of our Regge calculation the $t$-dependence is induced by the time ordered product of two collinear-soft scattering operators, through the soft eye diagram in \fig{SCET2_oneloop_matching}c, whereas $O_{ns\bn}^{qq}$ contributes the additional flower diagram. In~\cite{Fleming:2014rea} the soft part of the Regge result calculated in Feynman gauge comes solely from ${\cal O}_G^{n\bn}$ (the collinear calculations, which require both quark and gluon operators, were not considered there). For the BFKL calculation, Ref.~\cite{Fleming:2014rea} uses a rapidity renormalization equation analogous to our \eq{rren}, but with objects depending on the difference of $\perp$-momenta rather than individually on two $\perp$-momenta (the soft operator in our \eq{Gdefn} was not constructed in~\cite{Fleming:2014rea}). Our final result for the soft function's anomalous dimension and the kernel in the BFKL equation, \eqs{gSG}{bfkl}, also differ from~\cite{Fleming:2014rea} by a factor of two.  

It would be interesting to extend the calculation of the soft functions rapidity anomalous dimension beyond the leading logarithmic level to confirm the expectation that it will reproduce at the next order the next-to-leading-logarithmic BFKL equation. At next-to-next-to-leading logarithmic order it is known that the double box diagram~\cite{Caron-Huot:2013fea} breaks the expected form for the Regge factorization of the virtual amplitude~\cite{DelDuca:2001gu,DelDuca:2013ara}. This contribution is precisely the 3 Glauber exchange double box graph in our language.


\subsection{BFKL Equations for the Collinear Functions via Consistency}
\label{sec:bfklconsist}

At leading logarithmic order the $\nu$ dependence in the soft and collinear functions of the transition matrix $T_{(1,1)}$ must cancel, so 
\begin{align}  \label{eq:T11consist}
  \nu \frac{d}{d\nu} \int\! d^2q_\perp  d^2q_\perp'\:
 C_n(q_\perp,p^-,\nu)  S_G(q_\perp,q_\perp',\nu) 
  C_\bn(q_\perp',p^{\prime +},\nu)  = 0 \,.
\end{align}
This result suffices to derive the LL RGE equation for $C_n$ and $C_\bn$, which will also be given by BFKL equations.  Generically, the form of the SCET matrix elements implies that we can have
\begin{align} \label{eq:CnCnbanomdim}
 \nu \frac{d}{d\nu} C_n(q_\perp,p^{-},\nu)
     & = \int\!\! d^2k_\perp \:
    \gamma_C(q_\perp,k_\perp) \: C_n(k_\perp,p^{-},\nu) 
    \,,\\
 \nu \frac{d}{d\nu} C_\bn(q_\perp,p^{\prime +},\nu) 
    & = \int\!\! d^2k_\perp \:
    \gamma_C(q_\perp,k_\perp) \: C_\bn(k_\perp,p^{\prime +},\nu) \,.
  \nn
\end{align}
Note that the same anomalous dimension $\gamma_C(q_\perp,k_\perp)$ appears for both collinear functions. This follows from the fact that $C_n\leftrightarrow C_\bn$ if we take $n\leftrightarrow \bn$, and that the anomalous dimensions cannot involve convolutions in the large conserved collinear momenta, and hence are independent of $n$ and $\bn$.  To exploit \eq{T11consist} it is useful to write the RGE for the soft function in a symmetric form.  As noted in \sec{bfkl}, both $S_G$ and $\gamma_{S_G}$ are symmetric in their two arguments, so the BFKL equation for the soft function can  be written as
\begin{align} \label{eq:bfklsymm}
 \nu\frac{d}{d\nu}\, S_G(q_\perp,q_\perp',\nu)
   &= \frac{1}{2} \nu\frac{d}{d\nu}\, S_G(q_\perp,q_\perp',\nu)
     + \frac{1}{2} \nu\frac{d}{d\nu}\, S_G(q_\perp',q_\perp,\nu) 
   \nn\\
   &= \frac{1}{2} \int\!\! d^2k_\perp \Big[
    \gamma_{S_G}(q_\perp,k_\perp) S_G(k_\perp,q_\perp',\nu)
    +  S_G(q_\perp,k_\perp,\nu) \gamma_{S_G}(k_\perp,q_\perp')
   \Big]
  \,.
\end{align}
Plugging \eqs{CnCnbanomdim}{bfklsymm} into \eq{T11consist} we then have
\begin{align}
  0 &= \int\! d^2q_\perp  d^2q_\perp' d^2k_\perp \bigg[
   C_n(k_\perp,p^{-},\nu) \gamma_C(q_\perp,k_\perp) S_G(q_\perp,q_\perp',\nu) C_\bn(q_\perp',p^{\prime +},\nu) 
  \\
  &\qquad\qquad\qquad\qquad  
   + C_n(q_\perp,p^{-},\nu)S_G(q_\perp,q_\perp',\nu)  \gamma_C(q'_\perp,k_\perp) C_\bn(k_\perp,p^{\prime +},\nu) 
  \nn\\
  & \qquad\qquad\qquad\quad \ 
  + \frac12 C_n(q_\perp,p^{-},\nu) \gamma_{S_G}(q_\perp,k_\perp) S_G(k_\perp,q_\perp',\nu) C_\bn(q_\perp',p^{\prime +},\nu) 
  \nn\\
 & \qquad\qquad\qquad\quad \ 
  + \frac12 C_n(q_\perp,p^{-},\nu)  S_G(q_\perp,k_\perp,\nu) \gamma_{S_G}(k_\perp,q_\perp') C_\bn(q_\perp',p^{\prime +},\nu) \bigg]
  \,. \nn
\end{align}
Swapping the integration variables $k_\perp \leftrightarrow q_\perp$ in the third line, and $k_\perp\leftrightarrow q_\perp^\prime$ in the fourth line, we see that this equation can only be satisfied for arbitrary $C_n$, $S_G$, and $C_\bn$ functions if $\gamma_C(q_\perp,k_\perp) = -\frac{1}{2} \gamma_{S_G}(k_\perp,q_\perp)$ and $\gamma_C(q_\perp',k_\perp)  = -\frac{1}{2} \gamma_{S_G}(q_\perp',k_\perp)$, which implies that $\gamma_C$ is also symmetric in its two arguments and given by 
\begin{align}
   \gamma_C(q_\perp,q_\perp') &= -\frac{1}{2}\: \gamma_{S_G}(q_\perp,q_\perp') \,.
\end{align}
Therefore the RGE equations for $C_n$ and $C_\bn$ are also given by a BFKL equation. Writing this out explicitly we have
\begin{align}
  \nu\frac{d}{d\nu} C_n(q_\perp,p^{-},\nu) &= -\frac{C_A\alpha_s}{\pi^2}
  \int \!\! d^2k_\perp\: 
   \bigg[ \frac{C_n(k_\perp,p^{-},\nu)}
     {(\vec k_\perp-\vec q_\perp)^2}
   - \frac{\vec q_\perp^{\:2}\: C_n(q_\perp,p^{-},\nu)}
   { 2\vec k_\perp^2 (\vec k_\perp-\vec q_\perp)^2}  \bigg] \,,
\end{align}
and we will also have the same BFKL equation for $C_\bn(q_\perp,p^{\prime +},\nu)$.  Note that there is a factor of $(-1/2)$ for these BFKL equations for the collinear functions as compared to the soft function in \eq{bfkl}. The sign comes from the fact that the collinear functions run in the opposite direction in rapidity space, from $\nu\simeq p^- = \sqrt{s}$ down to $\nu\simeq \sqrt{t}$, and the $1/2$ comes from the fact that two collinear functions must balance against a single soft function. Again both virtual and real collinear diagrams contribute if we compute the diagrams needed to directly determine these collinear RGE equations. The direct computation for the virtual contributions was carried out in \sec{regge} and agrees with the factor of $(-1/2)$ that we determined here by the renormalization group consistency argument.

\section{Glauber Exponentiation and (Non-)Eikonalization}
\label{sec:phases}

Below in \sec{exponentiation} we carry out the all order resummation of Glauber boxes in forward scattering, demonstrating that the rapidity regulator yields an eikonal phase. In \sec{forwardgraphs} we derive a spacetime picture along with explicit rules for when graphs with multiple Glauber exchange vanish, and determine general rules for when the eikonal approximation can and cannot be used. A precise connection between the dynamics of Glauber exchange and the semi-classical and shock wave interpretations of this scattering are made in \sec{semiclassical}.

\subsection{Glauber Exponentiation for Boxes with Rapidity Regulator}
\label{sec:exponentiation}

In \sec{GlauberBox} we showed how the rapidity regulator leads to a well defined integral for the one-loop box and cross-box graphs, with the latter vanishing.  In this section we will sum up all the Glauber exchange box diagrams with the  rapidity regulator, and show that the eikonal phase is correctly reproduced. The connection of this sum of diagrams to the classical coherent state generated by each of the collinear partons is explored further in \sec{semiclassical}. In the abelian limit soft contributions vanish and the  phase can be reproduced at the integrand level, as demonstrated explicitly in \app{abelianexp}.

We begin by noting that the argument given in \sec{GlauberBox} for the vanishing of the one-loop cross box holds for all non-ladder type topologies. Rapidity divergences are regulated by factors $|2 k_1^z|^{-\eta}\cdots |2 k_N^z|^{-\eta}$, so we can consider the $k_i^0$ integrals to be done by contours without concern that the remaining integral might be unregulated.  For any diagram with one or more crossed Glauber exchange lines there is one or more $k_i^0$ integrals for which the poles are all on the same side of the real axis (and converge at $\infty$). Thus, all diagrams with crossed Glauber rungs vanish with our rapidity regulator, and we only need to consider the sum of the ladder graphs.  

To show exponentiation we will manipulate an $N$-Glauber exchange diagram into the product of single  exchanges with a factor of $1/N!$. The product form arises when we transform from $q_\perp$ to the impact parameter space $b_\perp$. In impact parameter space we will see that the amplitude from iterated Glauber exchange is simply determined by a phase, given by the Fourier transform of the $1/q_\perp^2$ potential between particles $1$ and $2$:
\begin{align} \label{eq:phi}
 \phi(b_\perp) 
  &=  - {\rm\bf T}^A_1 \otimes {\rm\bf T}^A_2\,g^2(\mu)\!
 \int\! \frac{\ddslash\!^{d-2}q_\perp\, (\iota^\epsilon \mu^{2\epsilon})}{\vec q_\perp^{\:2}} \,
   e^{i \vec q_\perp \cdot \vec b_\perp} 
 \\
 &=  -{\rm\bf T}^A_1 \otimes {\rm\bf T}^A_2\, g^2(\mu)\, \frac{ \Gamma(-\epsilon) }{4\pi}
  \bigg( \frac{\mu|\vec b_\perp|e^{\gamma_E/2}}{2} \bigg)^{2\epsilon}
 \,. \nn
\end{align}
The result is a matrix in the color space with ${\rm\bf T}^A_1$ and ${\rm\bf T}^A_2$ being the color matrix generators that commute with each other, and act on particle $1$ and $2$ respectively. This color matrix notation is by now quite standard, see Appendix A of~\cite{Catani:1996vz} for an introduction to this notation. 
Recall that $d=4-2\epsilon$ and that $\iota^\epsilon = e^{\epsilon\gamma_E}/(4\pi)^\epsilon$ is our notation for the factor that enters with each $\mu^{2\epsilon}$ when the coupling is in the $\overline{\rm MS}$ scheme. The $\Gamma(-\epsilon)$ infrared divergence will be discussed further at the end of this section. 

The exponentiation results derived below hold equally well when iterating Glauber exchange potentials between quark-quark, quark-antiquark, quark-gluon, and gluon-gluon channels, and for cases where the scattering particles are $n$-$\bn$, $n$-$s$, or $\bn$-$s$. 
To be definite we consider quark-antiquark $n$-$\bn$ scattering, where
\begin{align}
  {\rm\bf T}^A_1 \otimes {\rm\bf T}^A_2 =  T^A \otimes \bar T^A 
  \,.
\end{align}
For convenience we define the Fourier transform operation as the application of the integral: \allowdisplaybreaks[1]
\begin{align}
   \underset{{\rm F.T.}_\perp}{\implies} \ 
    = \int\!\! \ddslash\!^{d-2}q_\perp\:  e^{i \vec q_\perp \cdot \vec b_\perp} 
  \,.
\end{align}
The Fourier transform of one Glauber exchange result is then given in terms of $\phi(b_\perp)$ by
\begin{align} \label{eq:1gFT}
  \raisebox{-1.1cm}{
\includegraphics[width=0.14\columnwidth]{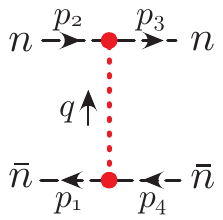}
} = \vspace{-.5in} \frac{-2i g^2}{\vec q_\perp^{\:2}} 
    \Big[ \bar u_n  \frac{\bnslash}{2} T^A u_n \Big]
    \Big[ \bar v_\bn  \frac{\nslash}{2} \bar T^A v_\bn \Big]
  \underset{{\rm F.T.}_\perp}{\implies}
    i\, \phi(b_\perp)\, 2\,{\cal S}^{n\bn} \,
 \,, 
\end{align}
where the spinor factor is 
\begin{align} \label{eq:Snbn}
 {\cal S}^{n\bn} = \Big[ \bar u_n  \frac{\bnslash}{2} u_n \Big]\Big[ \bar v_\bn  \frac{\nslash}{2} v_\bn \Big] \,,
\end{align} 
and the $2$ in \eq{1gFT} comes from $n\cdot \bn=2$, which is the factor needed to make ${\cal S}^{n\bn}$ RPI-III invariant.  In \eq{1gFT} the color matrix inside $\phi(b_\perp)$ operates on this spinor product. In general we will let
\begin{align} \label{eq:SnbnN}
  (T^{A_1}\cdots T^{A_N}) \otimes (\bar T^{A_1}\cdots \bar T^{A_N}) {\cal S}^{n\bn}  
  = \Big[ \bar u_n  \frac{\bnslash}{2} T^{A_1}\cdots T^{A_N} u_n \Big]
    \Big[ \bar v_\bn  \frac{\nslash}{2} \bar T^{A_1}\cdots \bar T^{A_N} v_\bn \Big] 
  \equiv {\cal S}^{n\bn}_{(N)} \,,
\end{align}
which is the color structure that appears from $N$ Glauber rungs. We also define the product rule for the matrix multiplication in $\phi^N(b_\perp)$ via $(T^A\otimes \bar T^A)^N {\cal S}^{n\bn} = {\cal S}^{n\bn}_{(N)}$. These same definitions apply equally well for a general choice of scattering particles in different color representations, using $({\rm\bf T}^A_1 \otimes {\rm\bf T}^A_2)^N$ times a generic ${\cal S}^{n\bn}$. 

The loop integrals are carried out by doing the energy integrals by contours, and then treating the $k^z$ integrals in Fourier space. Therefore we need to transform the $\eta$ regulator to Fourier space, as well as the $k^z$ dependent propagators. To do this we can use the transforms
\begin{align} \label{eq:fteta}
 & \int_{-\infty}^{+\infty}\!\! \ddslash\!k^z\: e^{ i x k^z} |2k^z|^{-\eta}
  =  \kappa_\eta\: \frac{\eta}{2}\: | x |^{-1+\eta}\: 
  \,,
 & \int_{-\infty}^\infty\!\! dx & \: e^{-i x k^z} \kappa_\eta\: \frac{\eta}{2}\: | x |^{-1+\eta} = | 2 k^z |^{-\eta} 
  \,, \\
 &  \int_{-\infty}^{+\infty}\! \frac{\ddslash\!k^z\, e^{ -i \alpha k^z}}{k^z + \Delta+i0} = -i\, \theta(\alpha) e^{i\alpha\Delta} 
  \,,
 & \int_{-\infty}^\infty\!\! d\alpha & \: e^{i \alpha k^z} (-i) \theta(\alpha) e^{i\alpha \Delta} = \frac{1}{k^z+\Delta+i0} 
  \,, \nn 
\end{align}
where
\begin{align}
  \kappa_\eta = 2^{-\eta}\, \Gamma(1-\eta) \frac{\sin(\pi\eta/2)}{(\pi\eta/2)}
      = 1 + {\cal O}(\eta) \,.
\end{align}
Another integral that will be relevant is the Fourier transform of $(N+1)$ Glauber rungs,
\begin{align} \label{eq:FTglaub}
  & \int\!\! \ddslash\!^{d-2}q_\perp\:  e^{i \vec q_\perp \cdot \vec b_\perp}
  \int \frac{ \ddslash\!^{d-2}k_{1\perp}\cdots \ddslash\!^{d-2}k_{N\perp}\,
   \big(\iota^\epsilon \mu^{2\epsilon}\big)^{N+1}} 
   { \big(\vec k_{1\perp}+\vec q_\perp\big)^2 \big(\vec k_{2\perp}-\vec k_{1\perp}\big)^2 \cdots \big(\vec k_{N\perp}-\vec k_{(N-1)\perp}\big)^2\, \vec k_{N\perp}^{\:2} }
  \\
  &= \int\!\! \ddslash\!^{d-2}q_\perp\:  e^{i \vec q_\perp \cdot \vec b_\perp}
  \! \int\!\! \Big[\prod_{i=1}^N \ddslash\!^{d-2}k_{i\perp} \Big] 
  \!\int\!\! \bigg[ \prod_{j=1}^{N+1} \ddslash\!^{d-2}r_{j\perp}  \frac{\Gamma(-\epsilon)}{4\pi} 
  \Big( \frac{\mu |r_{j\perp}|e^{\gamma_E}}{2} \Big)^{2\epsilon} \bigg] 
  \nn \\
  &\qquad \times 
    e^{-i(\vec q_\perp+ \vec k_{1\perp})\cdot \vec r_{1\perp}} 
    e^{-i(\vec k_{2\perp}- k_{1\perp})\cdot \vec r_{2\perp}}  \cdots 
    e^{-i(\vec k_{N\perp}- k_{(N-1)\perp})\cdot \vec r_{N\perp} }
    e^{ i \vec k_{N\perp}\cdot \vec r_{(N+1)\perp} } 
  \nn \\
 &=   \!\int\!\! \bigg[ \prod_{j=1}^{N+1} \ddslash\!^{d-2}r_{j\perp}  \frac{\Gamma(-\epsilon)}{4\pi} 
  \Big( \frac{\mu |r_{j\perp}|e^{\gamma_E}}{2} \Big)^{2\epsilon} \bigg] 
\delta^{d-2}\big(\vec r_{1\perp}-\vec b_\perp\big)
   \delta^{d-2}\big(\vec r_{2\perp}-\vec r_{1\perp}\big)  \cdots
   \delta^{d-2}\big(\vec r_{(N+1)\perp}-\vec r_{N\perp}\big)
  \nn \\
 & = \bigg[  \frac{\Gamma(-\epsilon)}{4\pi} 
  \Big( \frac{\mu |\vec b_{\perp}|e^{\gamma_E}}{2} \Big)^{2\epsilon} \bigg]^{N+1} 
  \,.\nn
\end{align}
 
First consider redoing the box graph considered in \sec{GlauberBox} using this Fourier approach. After performing the energy integral by contours, defining $2 \Delta = p_3^+ + p_4^- - (\vec k_\perp+\vec p_{3\perp})^2/p_3^--(\vec k_\perp-\vec p_{4\perp})^2/p_4^+$, and then using \eq{fteta}, we have
\begin{align}  \label{eq:2g}
 	\raisebox{-1cm}{
      \includegraphics[width=0.2\columnwidth]{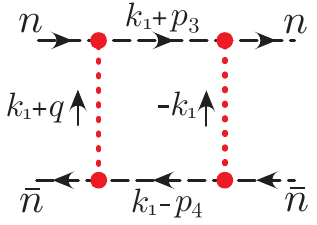}
} 
  \vspace{-.5in}
  &=  -4 i g^4 \big(T^A T^B \otimes \bar T^A \bar T^B \big) {\cal S}^{n\bn}
   \int \frac{\ddslash\!^{d-2}k_{1\perp}\, \ddslash\!k_1^z\, (|2k_1^z|^{-\eta} |2k_1^z|^{-\eta} \nu^{2\eta} \iota^{2\epsilon} \mu^{4\epsilon})}
   {\vec k_{1\perp}^{\:2}(\vec k_{1\perp}+\vec q_\perp)^2\: 2 (- k_1^z + \Delta +i0)}
  \nn\\[-5pt]
  &= -2 g^4 {\cal S}^{n\bn}_{(2)}\:  I^{(1)}_\perp(q_\perp)  \Big(\kappa_\eta \frac{\eta}{2}\Big)^2 \! \int_{-\infty}^{\infty}\!\!\!\! 
   \ddslash\!k_1^z\, dx\, dy\, d\alpha\:
    \theta(\alpha)  	| x y |^{-1+\eta}  
    \, e^{i\alpha(k_1^z+\Delta)+ik_1^z (x-y)}
    \nn \\[5pt]
   & = -2 g^4 {\cal S}^{n\bn}_{(2)}\:  I^{(1)}_\perp(q_\perp)
   \Big(\kappa_\eta \frac{\eta}{2}\Big)^2 \! 
   \int_{-\infty}^{\infty} \!\!\! dx\, dy\, \theta(y-x)\,  | x y|^{-1+\eta}
   \,  e^{i\Delta(y-x)}
    \nn \\[5pt]
   & = 2{\cal S}^{n\bn}_{(2)}\:  i^2 g^4  I^{(1)}_\perp(q_\perp)
   \: \frac{1}{2!}\: \Big[ 1 + {\cal O}(\eta) \Big] 
   \,,
\end{align}
where we defined
\begin{align} \label{eq:I1}
I^{(1)}_\perp(q_\perp)
  = \int\! \frac{\ddslash\!^{d-2}k_{1\perp}\, (\iota^{\epsilon} \mu^{2\epsilon})^2}{\vec k_{1\perp}^2(\vec k_{1\perp} +\vec q_\perp)^2}
  \,.
\end{align}
To get to the third equality in \eq{2g} we performed the $dk_1^z$ to get a $\delta$-function, and then did the $d\alpha$ integral.  For the last equality in \eq{2g} we note that due to the presence of the $\eta^2$ in the prefactor, only the ultraviolet $1/\eta^2$ part of the integrals from $x\to 0$ and $y\to 0$ contributes at leading order in the $\eta$ expansion, and therefore the result is independent of $\Delta$ at this order. The integral can be done directly, or we can note that the limit $x,y\to 0$ allows us to  symmetrize the theta function as, $\theta(y-x) \to [ \theta(y-x)+\theta(x-y)]/(2!) =1/(2!)$.  Performing the $\perp$ Fourier transform of the integral in \eq{I1} using \eq{FTglaub} we find
\begin{align} \label{eq:2gFT}
\raisebox{-1cm}{
\includegraphics[width=0.2\columnwidth]{figs/Glaub_ptnl_1box}
} \vspace{-.5in}
   & \underset{{\rm F.T.}_\perp}{\implies}
   \ \frac{1}{2!} \big[i \phi(b_\perp) \big]^2 \ 2 {\cal S}^{n\bn}
  \,.
\end{align}
As anticipated, comparing \eq{2gFT} to \eq{1gFT} we see that this is the second term in the expansion of an exponential.

Next consider the double box diagram. Again performing the contour integrals over
the energies, and then using \eq{fteta} we find
\begin{align} \label{eq:3g}
 & \raisebox{-1.cm}
    {
\includegraphics[width=0.29\columnwidth]{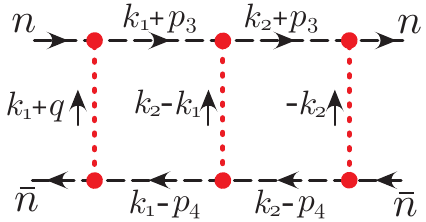}
} 
  =  -8 i g^6 {\cal S}^{n\bn}_{(3)}\: I^{(2)}(q_\perp) \!
   \int\!\! \frac{ \ddslash\!k_1^z\, \ddslash\!k_2^z\
   |2k_1^z|^{-\eta} \,|2k_1^z-2k_2^z|^{-\eta} \, |2k_2^z|^{-\eta} \nu^{3\eta}}
   { 4 (- k_1^z + \Delta_1 +i0)(- k_2^z + \Delta_2 +i0)}
  \nn\\[5pt] 
 &\qquad
   = 2 i g^6 {\cal S}^{n\bn}_{(3)}\: I^{(2)}_\perp(q_\perp)
   \! \int_{-\infty}^{\infty}\!\!\!\! \ddslash\!k_1^z\, \ddslash\!k_2^z\, dx\, dy\, dz\,  d\alpha_1\, d\alpha_2\, \theta(\alpha_1) \theta(\alpha_2)
  \Big(\kappa_\eta \frac{\eta}{2}\Big)^3  | xyz |^{-1+\eta} 
  \nn\\
  &\qquad\qquad\qquad\qquad\qquad\qquad
 \times e^{i k_1^z x} e^{i(k_2^z-k_1^z)y} e^{-i k_2^z z} 
   e^{i\alpha_1(k_1^z +\Delta_1)}  e^{i\alpha_2(k_2^z+\Delta_2)} 
  \nn\\[0pt] 
 &\qquad
   = 2 i g^6 {\cal S}^{n\bn}_{(3)}\: I^{(2)}_\perp(q_\perp)\,
   \Big(\kappa_\eta \frac{\eta}{2}\Big)^3
   \! \int_{-\infty}^{\infty}\!\!\!\!  dx\, dy\, dz\,  \theta(y-x) \theta(z-y)
   \, | xyz |^{-1+\eta} \,
   e^{i(y-x)\Delta_1} e^{i(z-y) \Delta_2} 
  \nn\\[0pt] 
 &\qquad
   = -2 {\cal S}^{n\bn}_{(3)}\: i^3 g^6 \: I^{(2)}_\perp(q_\perp)\, 
  \frac{1}{3!} \, \Big[ 1+ {\cal O}(\eta) \Big]
  \,,
\end{align}
where to obtain the third equality we performed the $k_1^z$ and $k_2^z$ integrals to get $\delta(x-y+\alpha_1)\delta(y-z+\alpha_2)$ and then performed the $\alpha_1$ and $\alpha_2$ integrals. Again due to the $\eta^3$ term in the prefactor only the leading ultraviolet divergent contribution from the $dxdydz$ integral contributes, which comes from the limit $x,y,z\to 0$ where the $\Delta_1=\Delta_1(k_{1\perp})$ and $\Delta_2=\Delta_2(k_{2\perp})$ dependence drops out. In this limit we can either do the integral directly to give the $1/3!$, or note that we can symmetrize as $\theta(z>y>x)\to [\theta(z>y>x)+\theta(y>z>x)+\theta(z>x>y)+\theta(x>z>y)+\theta(x>y>z)+\theta(y>x>z)]/(3!) = 1/(3!)$.  Everywhere in \eq{3g} the $\perp$ integral is contained in
\begin{align} \label{eq:I2}
  I^{(2)}_\perp(q_\perp) 
 &= \int \frac{\ddslash\!^{d-2}k_{1\perp}\ddslash\!^{d-2}k_{2\perp}\, (\iota^\epsilon \mu^{2\epsilon})^3}
   {(\vec k_{1\perp}+\vec q_\perp)^2 (\vec k_{2\perp}-\vec k_{1\perp})^2\, \vec k_{2\perp}^{\:2} } \,.
\end{align}
Performing the $\perp$ Fourier transform of this integral using \eq{FTglaub} gives
\begin{align}
\raisebox{-1.2cm}{
\includegraphics[width=0.29\columnwidth]{figs/Glaub_ptnl_2box}
} \vspace{-.5in} &  \underset{{\rm F.T.}_\perp}{\implies}
\  \frac{1}{3!} \big[ i \phi(b_\perp) \big]^3 \  2 {\cal S}^{n\bn} 
\,,
\end{align}
which is the third term in the expansion of the exponential. 

This naturally generalizes to the case of the $N$-loop box graph with $(N+1)$-rungs. Doing the energy integrals by contours and using \eq{fteta} we have
\begin{align} \label{eq:NgFT}
& \raisebox{-.8cm}
   {
\includegraphics[width=0.35\columnwidth]{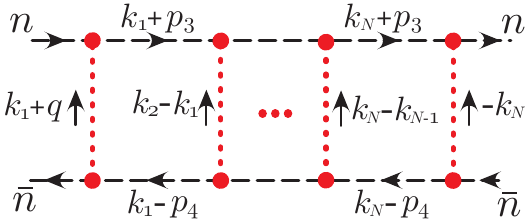}
}  
 \nn \\
 &\qquad
 =   -i (2g^2)^{N+1}  {\cal S}^{n\bn}_{(N+1)}\: I^{(N)}(q_\perp) \!
   \int\!\! \frac{ \ddslash\!k_1^z\cdots \ddslash\!k_N^z\,
   \big|2k_1^z(2k_1^z\minus 2k_2^z) \cdots 
   (2k_{N-1}^z\minus 2k_N^z)2k_N^z \big|^{-\eta} \nu^{N\eta}}
   { 2^N (- k_1^z + \Delta_1 +i0)\cdots (- k_N^z + \Delta_N +i0)}
  \nn\\[5pt]
 &\qquad
 =   -2i (g^2)^{N+1} (-i)^N {\cal S}^{n\bn}_{(N+1)}\, I^{(N)}(q_\perp) 
   \Big(\kappa_\eta \frac{\eta}{2}\Big)^{N+1} \!\!
   \int_{-\infty}^{+\infty}\!
   \bigg[ \prod_{i=1}^N \ddslash\!k_i^z\, d\alpha_i\, \theta(\alpha_i) \bigg] 
   \bigg[ \prod_{j=1}^{N+1} dx_j\: |x_j|^{-1+\eta}\bigg] 
  \nn\\[-5pt]
 & \qquad\qquad\qquad \times 
  e^{i k_1^z x_1 + i (k_2^z-k_1^z)x_2 + \ldots + i(k_N^z-k_{N-1}^z)x_N -ik_N^z x_{N+1}} \exp\bigg[ \sum_{m=1}^N i \alpha_m (k_m^z+\Delta_m)\bigg] 
 \nn\\
 &\qquad
 =   2 (-i g^2)^{N+1} {\cal S}^{n\bn}_{(N+1)}\, I^{(N)}(q_\perp) 
   \Big(\kappa_\eta \frac{\eta}{2}\Big)^{N+1} \!\!
   \int_{-\infty}^{+\infty}\!
   \bigg[ \prod_{j=1}^{N+1} dx_j\: |x_j|^{-1+\eta}\bigg] 
  \nn\\[-5pt]
 & \qquad\qquad\qquad \times 
  \theta(x_2\minus x_1)\theta(x_3\minus x_2)\cdots \theta(x_{N+1}\minus x_N) \exp\bigg[ \sum_{m=1}^N i \Delta_m (x_{m+1}-x_m )\bigg] 
 \nn\\
 &\qquad
 = 2 (-i g^2)^{N+1} {\cal S}^{n\bn}_{(N+1)} I^{(N)}_\perp(q_\perp)\ \frac{1}{(N+1)!}   \Big[1 +{\cal O}(\eta) \Big]
\nn\\
 &\qquad   
 \underset{{\rm F.T.}_\perp}{\implies}
    \	\frac{1}{(N+1)!}\: \big[ i \phi(b_\perp) \big]^{N+1}\ 2 {\cal S}^{n\bn}
  \,,
\end{align}
where to take the final Fourier transform we used \eq{FTglaub} for the integral
\begin{align} \label{eq:IN}
    I^{(N)}_\perp(q_\perp) 
 &= \int \frac{\ddslash\!^{d-2}k_{1\perp}\cdots \ddslash\!^{d-2}k_{N\perp}\, (\iota^\epsilon \mu^{2\epsilon})^{N+1}}
   {(\vec k_{1\perp}+\vec q_\perp)^2 (\vec k_{2\perp}-\vec k_{1\perp})^2\cdots
   (\vec k_{N\perp}-\vec k_{(N-1)\perp})^2\, \vec k_{N\perp}^{\:2} } \,.
\end{align}
The final result in \eq{NgFT} is the $(N+1)$'th term in the expansion of the exponential. Therefore the sum of Glauber box graphs for 2-to-2 $n$-$\bn$ scattering exponentiates to give  
\begin{align} \label{eq:fwdscatterphase}
 \int \ddslash\!^{d-2}q_\perp\: e^{i \vec q_\perp \cdot \vec b_\perp }
  \sum_{N=0}^\infty \text{G.Box\,}_{N}^{2 \rightarrow 2}(q_\perp)
  =  \big( \tilde G(b_\perp) -1 \big) 2  {\cal S}^{n\bn} 
\end{align}
where the position space Glauber function is given by
\begin{align} \label{eq:Gb}
  \tilde G(b_\perp) = e^{i \phi(b_\perp)}  \,,
\end{align}
and where the the color matrix phase $\phi(b_\perp)$ defined in \eq{phi} is a Hermitian matrix.  For convenience we also define the momentum space Glauber function
\begin{align}  \label{eq:Gq}
  G(q_\perp) &= \int d^{2}b_\perp\: e^{-i \vec q_\perp \cdot \vec b_\perp } \ e^{i \phi(b_\perp)} \,.
\end{align}
In SCET the results for the sum of Glauber boxes given by \eqs{Gb}{Gq} are valid for any color channel, simply taking $T^A\otimes \bar T^A\to {\rm\bf T}^A_1 \otimes {\rm\bf T}^A_2$ in $\phi(b_\perp)$. The same $(e^{i\phi(b_\perp)}-1)$ result is also obtained if we consider the sum of box diagrams for the soft-$n$ two-parton scattering since the Glauber light cone momenta will still be parametrically smaller then corresponding soft momentum. 

It is interesting to pause to consider physically what the $|2k_j^z|^{-\eta}$ factors are doing in the $N$-loop box graph in \eq{NgFT}. At finite $\eta$ this regulator implies that the Glauber exchanges are not instantaneous in the corresponding longitudinal position. (They are still instantaneous in time.)  Diagrammatic calculations are easy to interpret in position space, where these regulators were transformed to factors of $|x_j|^{-1+\eta}$. Each of these longitudinal coordinates $x_j$ corresponds to the location of one of the Glauber exchanges. Hence, they spread out with a string of increasing longitudinal coordinates $x_1 < x_2 < \ldots < x_{N+1}$, where the $\theta$-functions inducing these inequalities are provided by the collinear propagators between the Glauber exchanges. However each position space regulator also comes with a factor of $(\kappa_\eta \eta/2)$, and hence only the most divergent part of the $x_j$-integrals contributes to the final result. This divergent contribution comes  from the simultaneous limit where all coordinates $x_j\to 0$, restoring the physical picture of the Glauber exchanges being simultaneously instantaneous in their longitudinal positions. From the calculation in \eq{NgFT} we see that the ordered nature of the instantaneous limit is important for providing the correct $1/(N+1)!$ factor for $(N+1)$ Glauber exchanges. 

While the phase $\phi(b_\perp)$ in \eq{phi} has an infrared divergence, this is simply an overall phase in the scattering amplitude and hence drops out from the physical forward scattering cross section.  To see this explicitly we  switch to using the (slightly simpler) gluon mass IR regulator setting $d=4$, so with unspecified color channels for the forward scattering states
\begin{align}
  \phi(b_\perp) &= -{\rm\bf T}^A_1 \otimes {\rm\bf T}^A_2\,g^2(\mu)\!
 \int\! \frac{\ddslash\!^{2}q_\perp}{\vec q_\perp^{\:2}+m^2} \,
   e^{i \vec q_\perp \cdot \vec b_\perp} 
  \\
  &=  2 {\rm\bf T}^A_1 \otimes {\rm\bf T}^A_2 \, \alpha_s(\mu)\, 
  \ln\bigg(  \frac{|\vec b_\perp| m e^{\gamma_E}}{2} \bigg)
 \,. \nn 
\end{align}
Then taking the inverse Fourier transform of \eq{fwdscatterphase} we get
\begin{align} \label{eq:fwdscatterampl}
  \sum_{N=0}^\infty \text{G.Box\,}_{N}^{2 \rightarrow 2}(q_\perp)
 &= 2 {\cal S}^{n\bn}\, \big[ G(q_\perp) - (2\pi)^2 \delta^2(q_\perp)\big] \,,
\end{align}
The momentum space Glauber function corresponds to the sum of Glauber exchange diagrams, including the diagram with no-exchange,
\begin{align}  \label{eq:Gqperp}
G(q_\perp) &= (2\pi)^2 \delta^2(q_\perp)
  + \int d^{2}b_\perp\: e^{-i \vec q_\perp \cdot \vec b_\perp } \big( e^{i \phi(b_\perp)}-1 \big) 
  \\
 &= (2\pi)^2 \delta^2(q_\perp)
  + \frac{i 4\pi \hat c\, \alpha_s(\mu)}{t} \:
    \frac{\Gamma\big(1+i \hat c\, \alpha_s(\mu)\big)}
      {\Gamma\big(1-i \hat c\, \alpha_s(\mu)\big)}
    \Big(\frac{-t}{m^2 e^{2\gamma_E}}\Big)^{-i\, \hat c\, \alpha_s(\mu)} 
   \nn \\
 &=(2\pi)^2 \delta^2(q_\perp)
  + \frac{i 4\pi \hat c\, \alpha_s(\mu)}{t} \:
 e^{ i \delta(t,\alpha_s) } \,, \nn
\end{align}
where $t=q_\perp^2 = -\vec q_\perp^{\:2}<0$, and we defined the color operator 
\begin{align}
 \hat c = {\rm\bf T}^A_1 \otimes {\rm\bf T}^A_2 \,. 
\end{align}
It is implicit that the $(2\pi)^2 \delta^2(q_\perp)$ term in \eq{Gqperp} has a unit matrix in the color space.  The momentum space phase appearing in \eq{Gqperp} is given by the hermitian expression
\begin{align}
\label{NAphase}
  \delta(t,\alpha_s) &= -\hat c\, \alpha_s(\mu) \ln\Big(\frac{-t}{m^2 }\Big) 
   + 2 \sum_{k=1}^\infty \frac{(-1)^{k+1} \zeta_{2k+1}}{2k+1} \big( \hat c\, \alpha_s(\mu)\big)^{2k+1} \,, 
 \end{align}
and is again an operator in the color space. From \eq{Gqperp} the result for the scattering  is given by the lowest order Glauber exchange potential (tree-level) times a phase. Unlike in position space, this momentum space phase $\delta$ is an infinite series in $\alpha_s$.   Since the infrared divergence only appears in $\delta$, it will drop out of physical predictions for scattering cross sections (just like the IR divergent Coulomb phase for scattering with a Coulomb potential drops out of the cross section). For later convenience we also define a notation for the ${\cal O}(\alpha_s)$ contribution to $G(q_\perp)$ as
\begin{align}  \label{eq:G0defn}
  G^0(q_\perp) &\equiv \frac{i 4\pi\hat c\, \alpha_s(\mu)}{q_\perp^2} 
   = \frac{-i g^2(\mu)\: \hat c}{ \vec q_\perp^{\:2} }
   \,,
\end{align}
where we have the relation
\begin{align}
  \int\!\! \ddslash\!^{d-2}q_\perp\:  e^{i \vec q_\perp \cdot \vec b_\perp}  
   G^0(q_\perp) = i \phi(b_\perp)  
   \,. 
\end{align}

Note that the same results for the summation of box graphs is obtained for situations where the small plus and minus momenta of the collinear lines are not equal, $p_2^+\ne p_3^+$ and $p_1^-\ne p_4^-$ or where the exchanged $\perp$-momentum is not evenly split, $p_2^\perp \ne -p_3^\perp$ and $p_1^\perp \ne -p_4^\perp$.  The only place that $p_{2,3}^\perp$ and $p_{1,4}^\perp$ appeared outside of $q_\perp$ was in the $\Delta_i$ factors in the collinear  propagators, but the result was independent of these factors. When $q^+=p_3^+-p_2^+\ne 0$ and $q^-=p_1^- - p_4^-\ne 0$ we have both a modification to the $\Delta_i$ factors, and nonzero exchanged momenta $q^+$ and $q^-$. The smaller $q^+q^- \ll q_\perp^2$ do not modify the Glauber potentials, and again the change to $\Delta_i$ does not effect the result. So the only possible change induced by the nonzero $q^\pm$ is to the rapidity regulator for (say) the first rung of the ladder graphs. However, as in the case of $\Delta_i$, the dependence on  $q^\pm$ is higher order in $\eta$.   This implies that the same results for this summation are obtained even when the ladder graphs are considered inside of another loop in SCET, as long as that additional loop does not need a rapidity regulator. To leave the $n$ and $\bn$ collinear propagators nearly onshell the extra loop can only have Glauber (or ultrasoft) scaling.  We will exploit this property for some of our calculations in \sec{spectator} below.  

The independence of the $\Delta_i$ in \eq{NgFT} implies that the collinear lines in these box diagrams are effectively behaving as if they were eikonal and hence classical. However, we stress that this is not a general property of collinear propagators in the presence of Glauber exchange. Examples where it is not true include those in the next section, those for spectator interactions with a hard scattering vertex discussed in \sec{spectator}, and for mixed graphs containing a $1/\eta$ from a soft or collinear loop where the ${\cal O}(\eta)$ term from the Glauber loop integral in \eq{2g} or \eq{NgFT} must be considered. The fact that the Glauber box diagrams are classical can be understood by noting that the Glauber potential is classical and that, as long as  we consider only two to two scattering the partons effectively act as classical sources. This will no longer be true when we consider scattering between hadrons where the open ends of the box are closed off by the interpolating field for the hadrons, since in this case the transverse momentum dependence in the collinear lines can no longer be ignored.

It is interesting to ask about higher order corrections to $\phi(b_\perp)$, and in particular about the form of higher order non-abelian corrections to this phase.  Non-abelian corrections at one-loop can be generated by the soft and collinear loop graphs shown in \fig{SCET2_oneloop_matching}. The $\beta_0 \ln(\mu^2/-t)$ logarithm associated to the running of the $\alpha_s(\mu)$ that appears in the lowest order $\phi(b_\perp)$ comes from the soft sector and must exponentiate in the same manner.  For $n$-$\bn$ scattering it is actually clear that the full one-loop soft result in \eq{soft_total_loop} will exponentiate when it is iterated as a kernel for Glauber loops, because the Glauber loop momenta $k_i^\pm\sim {\cal O}(\lambda^2)$ are parametrically smaller than the soft momenta, and hence pass through the soft loops without changing their results. For these graphs the Glauber loop integrals lead to $i^N/N!$ just as they did for the Glauber potential box graphs.  For the collinear loops, it turns out that the parts associated to rapidity divergences will also exponentiate in this same manner, as they must do so to ensure the cancellation of the rapidity divergences. It is not clear whether the full contributions from $n$-collinear loops will exponentiate since the collinear and Glauber $+$-momenta are both ${\cal O}(\lambda^2)$, and hence the fact that Glauber loop momenta pass through the collinear loop integral could change its result.

\subsection{Longitudinal Constraints and Eikonalization}
\label{sec:forwardgraphs}

Let us now consider how collinear and soft corrections, both real and virtual,  affect multiple Glauber exchange contributions.  As we will see below, the possible corrections to Glauber exchanges is restricted by a spacetime constraint, causing many corrections to lead to a vanishing result. We will also determine the general criteria for when a collinear or soft propagator within a Glauber loop may be treated as eikonal.

\begin{figure}[t!]
%
%
\hspace{0.5cm}
\raisebox{0cm}{
	\hspace{-0.2cm} $a$)\hspace{4.2cm} $b$)\hspace{4.7cm} $c$)\hspace{4.5cm}  } 
\\[-10pt]
\includegraphics[width=0.23\columnwidth]{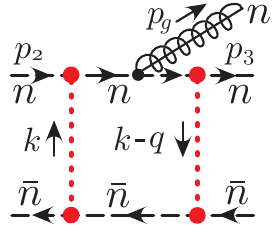} 
\hspace{0.7cm}
\includegraphics[width=0.26\columnwidth]{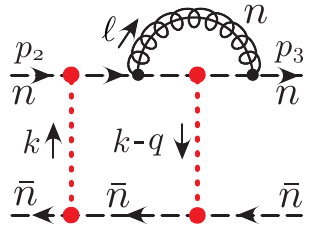} 
\hspace{0.7cm}
\includegraphics[width=0.24\columnwidth]{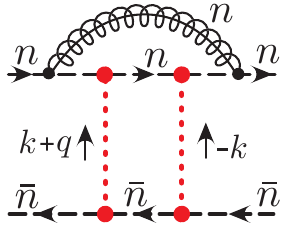} 
\\[0pt] 
\hspace{0.2cm}
\raisebox{-0.1cm}{
	\hspace{-0.2cm} $d$)\hspace{4cm}  $e$) \hspace{11cm}} \\[-10pt]
\hspace{0.1cm}
\raisebox{-0.08cm}{
\includegraphics[width=0.24\columnwidth]{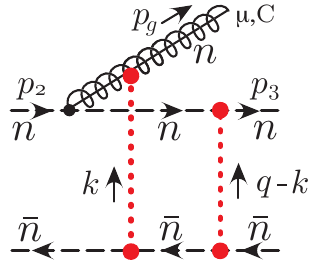}
 }
\hspace{0.3cm}
\includegraphics[width=0.31\columnwidth]{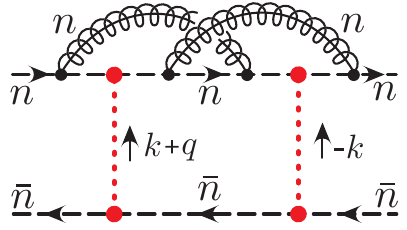}
\hspace{-0.1cm} \raisebox{1cm}{\Large =} \hspace{-0.1cm}
\includegraphics[width=0.35\columnwidth]{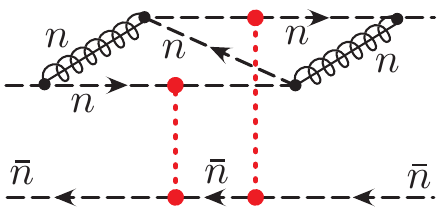}
%
\caption{\setcaptionskip
Graphs with multiple Glauber exchanges that occur at distinct light-cone times vanish, including the real emission graph a) and virtual graph b).  Graphs like c), d), and e) with multiple Glauber exchanges that can be collapsed to the same time and longitudinal position do not vanish. Graph c) contributes to an effective form factor leaving a factorized eikonal form.  Graphs d) and e) are examples where the Glauber exchange attaches to different particles which exist at the same light-cone times. The second figure in e) is the same graph, but is time ordered.} 
\label{fig:formfactor}
\setmainskip
\end{figure}

To build up the physical picture, we start by considering the diagrams in \fig{formfactor} which involve $n$-$\bn$ forward scattering with additional collinear loops or radiation.  In \fig{formfactor}a we have a collinear gluon radiated with Glaubers attached both before and after the radiation. Recall that the Glauber loop momentum scales as $(n\cdot k,\bn\cdot k,k_\perp) \sim (\lambda^2,\lambda^2,\lambda)$, and hence does not change the large momenta of the collinear lines. For this real final state emission we have $\bn\cdot p_2>0$, $\bn\cdot p_g>0$, and $\bn\cdot p_3 = \bn\cdot (p_2-p_g)>0$. Therefore there are two $n$-collinear quark propagators in the Glauber loop, which has the form
\begin{align}  \label{eq:Glaubrad0}
  \text{ \fig{formfactor}a } 
   &= \text{(pre)}
   \!\! \int\!\! \ddslash\!^{d} k \:
   \frac{(|k_z|^{-2\eta}\,\nu^{2\eta})\  \text{Num}(k_\perp) }
  { \vec k_\perp^{\:2} \big(\vec k_\perp\minus \vec q_\perp\big)^2 
   \big[n\cdot k-\Delta_1 +i0\big]
   \big[n\cdot k -\Delta_2+i0\big]
   \big[\bn\cdot k+\bar\Delta_1' - i0\big]} 
   \nn\\
  &=0
  \,,
\end{align}
where the prefactor, (pre), includes the couplings and color structure, and the numerator Num$(k_\perp)$ and $\Delta$ factors only depend on the $k_\perp$ loop momentum. Here the $dk^0 dk^z$ integration gives a vanishing result since there are two $n\cdot k$ propagators with the same $+i0$, as discussed in detail in \app{integrals}. 

Next consider a collinear loop which interrupts two Glauber exchanges, as in  \fig{formfactor}b. If we consider carrying out the collinear $n\cdot \ell$ integral by contours, we find that the integral is nonzero only for $0 < \bn\cdot \ell < \bn\cdot p_2 $, thus ensuring that all collinear quark propagators have positive large momenta, $\bn\cdot p_2>0$ and $\bn\cdot (p_2-\ell)>0$, and that the virtual collinear gluon has positive light cone energy $\bn\cdot\ell >0$ and is traveling forward in light-cone time. Hence once again we have two $n\cdot k$ dependent $n$-collinear quark propagators with the same $+i0$, as in \eq{Glaubrad0}, and the $dk^0 dk^z$ integration vanishes. The collinear gluon vertex in the loop interrupts the Glauber loop in the same manner as for the collinear radiation graph. Note that in either of \fig{formfactor}a,b, if we had instead attached the rightmost Glauber exchange to the $n$-collinear gluon, then the graphs would again vanish for the same reason.

On the other hand the diagram in \fig{formfactor}c  is non-vanishing. Here there is only one pole in $k^+$ and $k^-$ for the Glauber loop, and it gives the same result as for the box diagram in \eq{2g}. Indeed, one is free to add any number of Glauber exchanges between the collinear vertices, which simply builds up the higher order terms in the Glauber function $G(q_\perp)$, so this type of amplitude can be written as
\begin{align}
 \label{eq:facthalf}
 F^n(q_\perp)  \big[ G(q_\perp) -(2\pi)^2 \delta^2(q_\perp) \big],
\end{align}
where $F^{n}(q_\perp)$ is a one-loop abelian form factor for the $n$-collinear line. In this non-vanishing result the eikonal approximation arises in the same manner as in \sec{exponentiation} for the internal collinear propagators participating in the Glauber loops. The collinear propagators outside the Glauber loops are not eikonal.  The same form would also be obtained if we iterated Glauber exchanges solely between the $n$-collinear gluon and the $\bn$-collinear antiquark.

In contrast, non-vanishing diagrams such as \fig{formfactor}d do not have collinear propagators that can all be described by the eikonal approximation. Using the momentum routing shown,
\begin{align} \label{eq:formnonzero}
  \text{ \fig{formfactor}d } 
   &= \text{(pre)}
   \!\! \int\!\! \ddslash\!^{d} k \:
   \frac{(|k_z|^{-2\eta}\,\nu^{2\eta})\  \text{Num}(k_\perp) }
  { \vec k_\perp^{\:2} \big(\vec k_\perp\minus \vec q_\perp\big)^2 
   \big[n\cdot k+\Delta_1' -i0\big]
   \big[n\cdot k -\Delta_1+i0\big]
   \big[\bn\cdot k+\bar\Delta_1' - i0\big]} 
   \nn\\
  &=\frac{\text{(pre)}}{4} 
   \!\! \int\!\! \ddslash\!^{d-2} k_\perp \:
   \frac{ \text{Num}(k_\perp) }
  { \vec k_\perp^{\:2} \big(\vec k_\perp\minus \vec q_\perp\big)^2 
   \big(\Delta_1+\Delta_1' -i0\big) } 
  \,,
\end{align}
where the steps for carrying out the $dk^0 dk^z$ here are described in detail in \app{integrals}. Here  (pre)$=4 g^5 i f^{ABC} T^{D}T^B \otimes \bar T^D \bar T^A$  and   Num$(k_\perp)$ depends only on external momenta and the $k_\perp$ loop momentum. The $\Delta$ factors depend on $k_\perp$ and are given by
\begin{align} \label{eq:Deltari}
 \Delta_1' &\! = 
    \frac{(\vec k_\perp \minus \vec p_{g\perp})^2}{\bn \cdot p_g} 
    - n\cdot p_g \,,
 & \Delta_1 &\! = 
   \frac{(\vec k_\perp\plus \vec p_{3\perp}\minus \vec q_\perp)^2}{\bn \cdot p_3}  - n\cdot p_3 \,, 
 & \bar \Delta_1' &\! = 
    \frac{(\vec k_\perp \minus \vec p_{1\perp})^2}{n\cdot p_1} 
    - \bn\cdot p_1
  \,.
\end{align}
The presence of the $(\Delta_1+\Delta_1')$ propagator in the remaining $k_\perp$ integral in \eq{formnonzero}, indicates that here the non-eikonal nature of the $n$-collinear propagators was important. Since $\bar\Delta_1'$ does not appear, the $\bn$-collinear propagator can still be treated as eikonal. The same conclusion that non-eikonal propagators are necessary is also obtained if we consider the collinear loop graph where the radiated $n$-collinear gluon in \fig{formfactor}d is reabsorbed by the $n$-collinear quark after its Glauber attachment.  Furthermore, this need for non-eikonal collinear propagators is also true even in an abelian theory, where it occurs for the diagram in \fig{formfactor}e. Both of the diagrams in \fig{formfactor}d,e  involve a $k_\perp$ convolution between the Glaubers and the collinear source function.

To determine in a simple manner whether or not a graph with multiple Glauber exchange does or does not vanish, we use time-ordered  perturbation theory (TOPT) to order the vertices in a diagram.  Usually  one would utilize light cone ordered perturbation theory  (LCPT ) when analysing high energy scattering, as it greatly reduces the number of relevant diagrams \cite{Weinberg:1966jm}. However, when we factorize in  rapidity space we necessarily break boost invariance via the rapidity regulator. With our regulator in place we  can perform the energy integrals by contours, but not the light cone momentum, leading to a set of time ordered diagrams.  Notice that the advantage gained using LCPT, via the reduction in the number of diagrams, is maintained in TOPT when working in the EFT because the propagators are linear in energy for these Glauber loops. The regulated Glauber exchanges with $|k_i^z|^{-\eta}$ also remain instantaneous in time. Next we transform the longitudinal integrals $k_i^z$ to position space, and thereby assign a longitudinal position label $x_i$ for each Glauber exchange in a TOPT diagram, as was discussed in the previous section for the example in \eq{NgFT}. Since each transformed Glauber exchange comes with a prefactor of $(\eta/2)$, only the most divergent part of the $x_i$ integrals can contribute. Furthermore, anything that interrupts these longitudinal integrations, causing them to become less divergent, will lead to a result that vanishes as $\eta\to 0$. An interruption of this type occurs if there is a vertex that unavoidably inserts an additional longitudinal position in the midst of the burst of Glauber gluons, and therefore stops them from coming together to yield a leading short distance divergence. The ``collapse rule'' therefore states that:
\begin{quote}
Graphs with more than one Glauber exchange will vanish unless the exchanges can be moved towards each other unimpeded, so that they all occur at the same longitudinal position $x_0$ for both sources.
\end{quote} 
This ordered collapse corresponds to the instantaneous limit $x_i\to x_0$ for every $i$. After taking this limit the Glauber exchanges are now instantaneous in both time and longitudinal position\footnote{For $n$-$\bn$ scattering the longitudinal position is $(n\cdot x-\bn\cdot x)/2$. For the more general case with $n_i$ and $n_j$ collinear particles, the ``longitudinal position'' for this discussion is defined by $(n_i\cdot x-n_j\cdot x)/2$. See also \eqs{decomposep}{measure}.}, or equivalently in $x^+$ and $x^-$. This reproduces our original physical picture regarding the instantaneous nature of Glauber exchange. This  general rule applies for diagrams with any number of loops or with additional radiation.  If we replace one of the collinear sectors by soft particles then the same argument holds, or simultaneously have $\{n,s,\bn\}$ particles, then again the same rule regarding Glauber loops also holds true. 

For the simple diagrams in \fig{formfactor}a,b,c,d there is only one non-trivial time ordered diagram. For the graphs in \fig{formfactor}a,b the collapse to equal longitudinal position of the two Glauber exchanges is impeded by the collinear gluon vertex which sets an intermediate position that stops the Glaubers from coming together, so the graphs vanish. In other words, the integral over the longitudinal positions vanishes unless all the positions collapse to zero, but theta functions from the collinear propagators enforce a definite ordering which forbids this collapse. This is worked out explicitly in \eq{I2001x} of the \app{integrals} yielding for the integral appearing in \eq{Glaubrad0}:
\begin{align}
 & \int\!\!  \ddslash\! k^0 \ddslash\! k^z  \frac{|2k^z|^{-2\eta} \nu^{2\eta}}
   {\big(k^+ -\Delta_1 +i0\big)\big(k^+ -\Delta_2 +i0\big)\big(k^- + \bar\Delta_1' -i0\big)} 
 \nn\\
 &=  -\frac{i}{4} \Big(\kappa_\eta \frac{\eta}{2}\Big)^{\!2}\! 
   \!\int\!\! \ddslash\!^{d-2}\!k\!\!
    \int_{-\infty}^{+\infty}\!\!\!\!\!\!\!\! 
    dx_1 dx_2 d\alpha \, 
   \theta(x_1\minus\alpha)\theta(\alpha\minus x_2) |x_1 x_2|^{-1+\eta} 
   \big[ 1 + {\cal O}(\eta) \big]
  \nn\\
 & = {\cal O}(\eta) \,.
\end{align}
Here $\alpha$ is the intermediate coordinate that interrupts the collapse, leading to a less divergent integral.  In the graphs in \fig{formfactor}c,d,e the collapse to equal longitudinal positions is possible and the results for these diagrams do not vanish as $\eta\to 0$. For \fig{formfactor}e this is made clear with the second way of drawing the same diagram, namely that the non-vanishing contribution occurs when the time ordering is such that the central $n$-collinear propagator corresponds to an antiquark.  

Note that the collapse rule does not imply that a soft exchange between Glaubers will lead to a vanishing result. As an example, if we consider the H-diagram in \fig{hgraph}a,  the light-cone time scale for the soft momenta is short ($\sim \lambda^{-1})$ compared to the Glauber exchange time scale ($\sim \lambda^{-2}$) and thus the longitudinal positions of the Glauber exchanges can coincide. The $\ell^\pm$ loop momenta only appear in the soft gluon  propagator, and thus effectively the soft gluon has a tadpole like integral in these variables. However, this does not  localize the transverse coordinates corresponding to the soft loop momentum $\ell_\perp^\mu\sim \lambda$  and the  H-diagram type topology persists for the $\ell_\perp$ and $k_\perp$ integrals.

It is also straightforward to identify rules for when a collinear or soft propagator inside a Glauber loop can be treated as eikonal.  First let us determine under what conditions the integrals vanish by considering the momentum
space propagator structure. Consider an arbitrary loop graph, with one Glauber loop momentum $k^\mu$, then the general structure of the propagators is
\begin{align} \label{eq:generalGlaubloop}
  \!\! \int\!\! \ddslash\! k^0\: \ddslash\! k^z \:
   \frac{|k^z|^{-2\eta}\,\nu^{2\eta}   }
  { \big[n\cdot k\minus \Delta_{1} \plus i0\big]
    \cdots \big[n\cdot k\minus \Delta_{n_+} \plus i0\big]
    \big[n\cdot k \plus \Delta_1'\minus i0\big]
    \cdots \big[n\cdot k \plus \Delta_{n_-}'\minus i0\big]
    } 
   \nn\\
 \times \frac{1}
  { \big[\bn\cdot k\minus \bar\Delta_{1} \plus i0\big]
    \cdots \big[\bn\cdot k\minus \bar\Delta_{\bn_+} \plus i0\big]
    \big[\bn\cdot k \plus \bar\Delta_1'\minus i0\big]
    \cdots \big[\bn\cdot k \plus \bar\Delta_{\bn_-}'\minus i0\big]
    } \,.
\end{align}
Here the various $\Delta$s depend on the $k_\perp$ loop momentum, but not on $k^0$ or $k^z$.  On the $n$-collinear side we have $n_+$ propagators with a $+i0$ and $n_-$ propagators with a $-i0$, and similarly on the $\bn$-collinear side we have $\bn_\pm$ propagators with a $\pm i0$. 

Let us first enumerate all the situations where the one-loop integral in \eq{generalGlaubloop} vanishes. If any three of the indices $\{n_+, n_-, \bn_+,\bn_-\}$ are zero, so that there are no propagators of that type, then it obviously vanishes. Next consider cases where two of these indices are zero.  If all the poles are on the same side for the $k^0$ contour integral, namely $n_+=\bn_+=0$ or $n_-=\bn_-=0$, then the integral vanishes. If all the poles occur in one of the two collinear sectors, $n_+=n_-=0$ or $\bn_+=\bn_-=0$, then the integral also vanishes. Here performing the $k^0$ integral by contours we either immediately get zero, or we get an integrand that other than the regulator is independent of $k^z$, and hence vanishes since $\int dk^z |k^z|^{-2\eta} = 0$. Finally we could have poles on opposite sides of the $k^0$ contour in the two collinear sectors, $n_+=\bn_-=0$ or $n_-=\bn_+=0$. In this case the only situation where we get a nonzero result is when both of the remaining nonzero indices are $=1$.  If both of the remaining indices are $>1$, such as when $n_+=\bn_-=0$, $n_->1$, $\bn_+>1$, then after closing the $k^0$ contour we are left with an integral in $k^z$ that converges at infinity (so we can drop the regulator), and vanishes by contour integration.
Next consider situations where only one of the indices vanishes. Again in this situation, the other index in that collinear sector must be $=1$ to obtain a non-vanishing result, since after closing the $k^0$ contour in the opposite direction, only this propagator has $k^z$ dependence. Thus the analysis is identical for this case.  So the integral will vanish if ($n_+=0$ and $n_->1$),  ($n_0=0$ and $n_+>1$), ($\bn_+=0$ and $\bn_->1$), or ($\bn_-=0$ and $\bn_+>1$).  To summarize, the non-vanishing cases where either exactly two or one index is zero we have:
\begin{align} \label{eq:vanishingrule}
\bullet\ \text{non-vanishing 1-loop Glauber} & \text{ integral with exactly two indices zero (2 cases):} 
  \hspace{4cm}
\nn\\*
 & n_+=\bn_-=0 \ \text{and}\ (n_-=1 \ \text{and}\ \bn_+=1) 
 \,, \\
 & n_-=\bn_+=0 \ \text{and}\ (n_+=1 \ \text{and}\ \bn_-=1) 
   \nn \\*
\bullet\ \text{non-vanishing 1-loop Glauber} & \text{ integral with exactly one index zero (4 cases):} 
  \hspace{4cm}
\nn\\
 & n_+=0 \ \text{and}\ n_-=1 
 \,, \qquad
  n_-=0  \ \text{and}\ n_+=1 
 \,, \nn \\
 & \bn_+=0 \ \text{and}\ \bn_-=1 
 \,, \qquad
  \bn_-=0  \ \text{and}\ \bn_+=1 
  \,. \nn 
\end{align}
An example of a non-vanishing Glauber loop integral where one of the indices was zero was given in \eq{formnonzero}.  If all four indices are nonzero then the integral will not vanish.

If the Glauber integral in \eq{generalGlaubloop} does not vanish, then we may ask the question when do the propagators that appear in the loop integrand, and depend on $k^0$ and $k^z$, behave as if they are effectively eikonal? 
We will consider eikonalization in each collinear direction separately.
The rules for eikonalization for propagators associated to these nonzero loops are quite simple:
\begin{align} \label{eq:eikonalrule}
 &  n_+ + n_- \ge 2 \,, \qquad \text{non-eikonal} 
  \,,
  &\bn_+ + \bn_- \ge 2 & \,, \qquad \text{non-eikonal} 
  \,,
  \nn\\
 &  n_+ + n_- = 1 \,, \qquad \text{eikonal} 
  \,,
  &  \bn_+ + \bn_- = 1 & \,, \qquad \text{eikonal} 
  \,.
\end{align}
To prove this consider the four cases. 
 If both $n_++n_-\ge 2$ and $\bn_+ + \bn_-\ge 2$ then we write $dk^0 dk^z =(dk^+ dk^-)/2$ and can set $\eta=0$ and perform the $k^+$ and $k^-$ integrals by contours, since both integrals are convergent at infinity. The result will depend on both the $\Delta^{(')}$ and $\bar \Delta^{(')}$ factors, and hence neither side eikonalizes (since we get two factors of $i$ this non-eikonal result is imaginary). If $n_++n_-=1$ and $\bn_+ + \bn_-=1$ then we have precisely the integrals considered in \sec{GlauberBox}, where the $|k^z|^{-2\eta}$ regulator is needed. Here the $k^0$ contour integral gives zero if both poles are on the same size. If the poles are on opposite sides it gives a result that is equivalent to having both the $n$-collinear and $\bn$-collinear propagators be eikonal at ${\cal O}(\eta^0)$. Next consider the case where $n_++n_-=1$ and $\bn_+ + \bn_-\ge 2$. Here the $|k^z|^{-2\eta}$ regulator is required to make the $k^+$ integration well defined, and forces us to consider the $k^0$ contour integral. Considering the $k^0$ contour integral in \eq{generalGlaubloop}, the single $n$-collinear propagator will have a pole either above or below the axis, and we choose to close the contour the other way so that we only select $\bn$-collinear propagator poles. Without loss of generality we take the single $n$-collinear pole to be $[n\cdot k-\Delta_1+i0]$, for which this gives
\begin{align} \label{eq:1Glaubloop}
  i \sum_{i=1}^{\bn_-} \! \int\!\! \ddslash\! k^z \:
   \frac{|k^z|^{-2\eta}\,\nu^{2\eta}  }
  { \big[\minus 2k^z \minus \bar\Delta_i' \minus \Delta_{1} \plus i0\big]
    } \ f_i(\{ \bar\Delta_j, \bar\Delta_k'\} ) 
  =  \frac{i}{4\pi} \Big[ -i\pi +{\cal O}(\eta) \Big]
   \sum_{i=1}^{\bn_-} \ f_i(\{ \bar\Delta_j, \bar\Delta_k'\} )
  \,,
\end{align}
where $f_i(\{ \bar\Delta_j, \bar\Delta_k'\} )$ is a function of the various $\bar \Delta$s. Since the result is independent of $\Delta_1$ the single $n$-collinear propagator is eikonal at ${\cal O}(\eta^0)$, whereas the $\bn$-collinear propagators are non-eikonal. Obviously for the opposite case, where $n_++n_-\ge 2$ and $\bn_+ + \bn_- =1$, we will find by the same logic that the $\bn$-collinear propagator is effectively eikonal and the $n$-collinear propagators are non-eikonal. 

Note that for a $n$--$\bn$, a $n$--$s$, or a $\bn$--$s$ Glauber loop the decomposition in \eq{generalGlaubloop}, the rules for vanishing cases in \eq{vanishingrule} and the rule for eikonalization in \eq{eikonalrule} all apply equally well.

\begin{figure}[t!]
%
%
\begin{center}
\includegraphics[width=0.4\columnwidth]{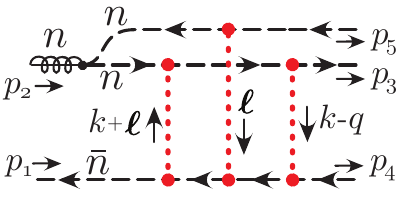} 
\end{center}
\vspace{-0.5cm}
\caption{\setcaptionskip
Two loop example with multiple collinear lines and Glauber exchange.} 
\label{fig:moregraphs}
\setmainskip
\end{figure}

When we consider extending \eqs{vanishingrule}{eikonalrule} for use in  multi-Glauber-loop diagrams, we must address the issue that now collinear or soft propagators can carry more than one $k_i^+$ or $k_i^-$ loop momentum. The number of propagators through which each loop momentum flows will also depend on the loop momentum routing, but whether a graph vanishes or whether a particular propagator can be treated as eikonal will be independent of the routing choice.  We choose to route each Glauber loop momentum through the minimum number of collinear propagators, this  maximizes the number of cases where $n_++n_-=1$ or $\bn_++\bn_-=1$. Essentially this means that we route loop momentum to maximize the number of obviously identifiable eikonal propagators, and then these propagators are removed when considering the eikonal propagator count for the next loop momentum (even if other loop momenta flow through them). Practically this means that if a Glauber loop momentum comes in through one exchange, then we route the momentum out of the collinear (or soft) sector on the next available Glauber exchange vertex. When determining whether a given loop vanishes, terms that are zero due to the energy contour integral $k_i^0$ are exactly analogous to the analysis at one-loop. However, we must be more careful when considering the implications of the $k_i^z$ integrals, since it is not enough to simply consider the convergence when a single $k_i^z$ variable gets large, since we must also ensure that the integral remains regulated when two or more $k_i^z$ variables simultaneously become large. 
For example, if any $k^z$ variable appears in a single propagator, then that propagator should be removed from consideration when considering the convergence for other $k_i^z$ variables. 

As a non-trivial example we consider \fig{moregraphs}.  Using the physical arguments discussed above or the collapse rule, we conclude that this contribution should be non-vanishing since the Glaubers
can accumulate at a single longitudinal point without obstruction.
With the loop momentum routing shown this diagram gives
\begin{align} \label{eq:moregrapha}
  \text{ \fig{moregraphs} } \!
   &=\!\! 
    \int\!\! 
   \frac{\ddslash\!^{d} k\, \ddslash\!^{d}\ell \
     \big(|k^z+\ell^z|^{-\eta}\,|\ell^z|^{-\eta}\,|k^z|^{-\eta}\,
     \nu^{3\eta}\big)\  \text{(pre)} \text{Num}(\ell_\perp) }
  { \big(\vec k_\perp\plus \vec \ell_\perp\big)^2 \vec \ell_\perp^{\:2} \big(\vec k_\perp\minus \vec q_\perp\big)^2 
   \big[\ell^+\minus \Delta_1 \big]
   \bigl[-\ell^+\minus \Delta_1' \big]
   \big[k^+\minus \Delta_2 \big]
   \bigl[-\ell^-\minus k^-\minus \bar\Delta_1'\big]
   \bigl[-k^-\minus \bar\Delta_2' \big]} 
   \nn\\
  &= \!\!
   \int\!\! 
   \frac{(-1) \ddslash\!^{d-1} k\, \ddslash\!^{d-1}\ell \
     \big(|k^z+\ell^z|^{-\eta}\,|\ell^z|^{-\eta}\,|k^z|^{-\eta}\,
     \nu^{3\eta}\big)\  \text{(pre)} \text{Num}(\ell_\perp) }
  { \big(\vec k_\perp\plus \vec \ell_\perp\big)^2 \vec \ell_\perp^{\:2} \big(\vec k_\perp\minus \vec q_\perp\big)^2 
   \big[\minus \Delta_1\minus \Delta_1' \big]
   \bigl[-2\ell^z\minus 2k^z \minus \bar\Delta_1'\minus\Delta_1\minus\Delta_2 \big]
   \bigl[-2k^z\minus \bar\Delta_2'\minus\Delta_2 \big]} 
  \,,
\end{align}
where all propagators inside square brackets $[\cdots ]$ have a $+i0$. In the first line the $\Delta_1$ and $\Delta_1'$ dependent propagators are those next to the collinear gluon pair production vertex, and are not eikonal, as is clear from the $(\Delta_1+\Delta_1')$ denominator in the second line.  Performing the $k^z$ and $\ell^z$ integrals gives a nonzero result, eliminating the corresponding propagators without inducing additional dependence on any $\Delta$s at ${\cal O}(\eta^0)$.  Therefore the three remaining collinear propagators in \fig{moregraphs} are eikonal (the three propagators inside the quark-antiquark $n$-$\bn$ scattering box). Note that for fixed $\ell^z$, the $k^z$ integral would converge at infinity with two poles on the same side of the axis, and hence seem to fall into a category that would vanish by our 1-loop criteria. Nevertheless, we get a finite result due to the additional divergence structure of the $\ell^z$ integral, which does not allow us to drop the $\eta$-regulators, and needs the same $[-2\ell^z\minus 2k^z+\ldots]$ denominator.

\subsection{Semi-classical Eikonal Phase and the Glauber Gluon}
\label{sec:semiclassical}

We have seen in \sec{exponentiation} that the sum of the Glauber boxes for elastic near-forward scattering, at leading power in $t/s\ll 1$, leads to an eikonal amplitude which is dictated by a phase $\delta(t)$. This is the expected behavior for any amplitude which can be approximated as being a semi-classical process. Moreover, in \sec{forwardgraphs} we have seen that the eikonal approximation is valid for collinear propagators that are internal to the interactions in a Glauber burst.  In this section we will put these and other results derived from SCET into the context of known results on forward scattering~\cite{ChengWuBook:1987,ForshawBook:1997,IoffeFadinLipatov:2010,KovchegovBook:2012}. We will also use our EFT to derive the picture of multiple Wilson lines crossing a shockwave~\cite{Mueller:1994gb,Balitsky:1995ub,Kovchegov:1999yj,JalilianMarian:1997jx,Iancu:2000hn}.  The questions that we will address include:
\begin{enumerate}
\item Why do Glauber gluons reproduce the semi-classical expression and what is the range of applicability of this approximation? 
\item In the abelian theory, there are no soft corrections other than light quark bubbles, but in the non-abelian theory soft corrections play an important role. How do these corrections  affect the form of the amplitude? Do the soft corrections simply dress the Glauber kernel in the absence of collinear radiation?
\item  When the semi-classical result breaks down, it is replaced by a picture where one must follow the trajectories of multiple partons crossing a shockwave. The partons are represented as infinite Wilson lines separated in the transverse space, and the number of lines is not fixed and can evolve. What is the precise criteria for the validity of this multi-Wilson line approximation, and how does it emerge from our EFT?
\item  Once a hard interaction is included,
to what extent do we expect the effects of  Glauber exchange to still be governed by a phase? And under what conditions do the Glauber exchanges cancel?
\end{enumerate}
In the remainder of this section we will provide answers to questions 1, 2, 3, referring to results from earlier sections when appropriate. We leave the discussion of point 4 to \secs{properties}{spectator} below.

Let us begin by exploring how the large $s/|t|$ limit appears semi-classically. In the semi-classical approximation, we may write the amplitude as 
\begin{align}
\int [D\phi] e^{iS} \sim e^{iS_0}(1+....)
  \,,
\end{align}
where $S_0$ is the action for the classical field configuration and corrections are down by the appropriate expansion parameter(s).  Quite often the starting point of any semi-classical approximation  involves solving the classical field equations without the presences of sources, such as in the case of instantons or monopoles. However, here we are considering scattering processes, so a necessary condition is that the external lines behave classically at leading order in the relevant expansion parameter(s). Furthermore, if we are to take an approach based upon Feynman diagrams then we must be able to distinguish between classical and quantum contributions. Thus both the sources and the bulk field (gauge fields) must behave in a characteristic fashion in the context of a semi-classical approximation.\footnote{Our definition of a source for the discussion here is the subset of collinear excitations.} Moreover, if this approximation is valid, then we should be able to find a power counting parameter which explicitly shows that quantum fluctuations in both the sources  and bulk fields are suppressed.

Consider the elastic scattering process between quarks in the limit $|t|/s\ll 1$. In what respect do these quarks behave as classical sources? Let us assume that we can follow a single quark through a series of interactions (perhaps by tagging it by a conserved quantum number). If we start by ignoring color, then the criteria for classical behavior is that the quark source currents commute. To see this consider the propagation of the quark source current with a large light-like momentum  $p^-\gg p^+$ in an external background whose Fourier modes $k$ may or may not have a  hierarchy between its components. As seen from our Lagrangians in \secs{SCET}{GlauberEFT} for either soft, ultrasoft, or Glauber gluons (the latter taken in Feynman gauge for the argument being made here), the boosted current will be dominated by the light-cone component  $A^+$, such that
\begin{align}
  L_{int}\approx\int d^4x\: J^-(x_+,x_-,x_\perp) A^+(x_+,x_-,x_\perp)
  \,.
\end{align}
We may write the part of the transition amplitude involving the quark as
\begin{align}
\label{Eres}
iM_{if}= \Big\langle i \Big| T \exp\Big(i \int d^4x  J^-(x_+,x_-,x_\perp)   A^+(x_+,x_-,x_\perp)  \Big) \Big| f \Big\rangle
  \,.
\end{align}
From the point of view of the quark it propagates in the presence of the gauge fields $A^+$ which can be viewed as a background whose internal dynamics, along with that of other sources, we can consider at a separate stage. What are the characteristics of the coupling of these gauge fields which will lead to a convergent semi-classical approximation? Note that if we set one of the light-cone momenta to zero (say $x^+=0$ ) in the current then 
\begin{align} \label{eq:JJcommutator}
  (\vec x_\perp- \vec y_\perp)^2 \neq 0 
  \qquad\text{implies}\qquad [J^-(0,x^-,x_\perp), J^-(0,y^-,y_\perp)]=0
  \,,
\end{align}
as long as the current is constructed from local operators whose algebra obeys micro-causality. Moreover, when  $x_\perp=y_\perp$ and the commutator is on the light cone, current algebra dictates that the commutator vanishes when the currents  are abelian,  highlighting the distinction with the non-abelian case. Thus if we can set $x^+=0$ in the current it will behave classically, in the abelian limit. This does not mean that the non-abelian theory has no classical contribution, only that the non-abelian currents do not have a canonical classical interpretation.  This should not surprise us since non-abelian currents are only covariantly conserved.

Whether or not it is justified to set $x^+=0$ depends upon the relative momenta of the quark source $p$ and gauge field momentum $k$.
In particular, if 
\begin{align}  \label{eq:cond}
p^- \gg  k^- 
\end{align} 
then we can set $x^+=0$ as the conjugate momentum $k^-$ has become irrelevant once we drop power suppressed terms from the propagator.  This does not mean that the propagator is necessarily of eikonal form, since the expansion in \eq{cond} leaves
\begin{align}
\label{eq:expprop}
\frac{1}{(p+k)^2} = \frac{1}{p^-(k^++p^+) -(\vec p_\perp +\vec k_\perp)^2} + \ldots,
\end{align}
as opposed to the eikonal form $1/(p^-k^+)$. Whether or not the transverse momenta  piece matters will depend on the power counting for the remaining terms, and the convergence property of the  $k^+$ integrals. If the low energy $A^+$ gluons with $k^-\ll p^-$ are close to their mass shell with $k^+k^- \sim \vec k_\perp^{\:2}$,  then they are either soft ($k^\mu\sim \lambda$) with $p^-k^+ \gg p^-p^+, (\vec p_\perp +\vec k_\perp)^2$ or ultrasoft ($k^\mu\sim \lambda^2$) with $p^- (k^++p^+) \gg \vec p_\perp\cdot \vec k_\perp, \vec k_\perp^{\:2}$. In either cases we are justified in dropping $x^+$ and can also fix the coordinate $x_\perp$, so the source propagators become purely eikonal $1/(k^+)$, and the corresponding quarks does not recoil. This is identical to the analysis in the standard construction of soft and ultrasoft interactions in SCET, see~\cite{Bauer:2001yt}. For the soft and ultrasoft cases these approximations are justified without referring to the nature of other particles that interact with the $A^+$ gluons. From the point of view of the soft or ultrasoft gluons the quark source becomes a Wilson line along a light-like direction and hence behaves much like a classical source.

In contrast we can consider the case where $p^- (k^++p^+)\sim (\vec p_\perp +\vec k_\perp)^2 \gg (p^++k^+) k^-$ so that transverse momentum  terms in \eq{expprop} are retained. This implies that the $A^+$ gluons are offshell with $k^+k^- \ll \vec k_\perp^{\:2}$, and hence correspond to the exchange of Glauber gluons.  To determine the outcome here we need to know the dynamics of the other particles to which the gluon couples. Assuming that it couples to an energetic particle which is just the mirror image of the quark source, then the abelian forward scattering is described by Glauber exchange, as discussed in \sec{GlauberEFT} and we are led to integrals with log-divergent $k^\pm$ integrations as in \secs{GlauberBox}{exponentiation}. The log divergent nature of these integrals, together with the action of a proper regulator, make the propagators effectively eikonal for forward scattering, despite the presence of the $\perp$-momentum dependent terms. We saw this explicitly in  our calculations in \sec{exponentiation} where the $\Delta_i(k_{i\perp})$ terms all dropped out. Thus once again the quarks (effectively) travel along a straight line and do not fluctuate, and we can treat the source quark as a classical current which yields a Wilson line.  However this eikonalization does not happen more generally for either the abelian or non-abelian cases, for the reasons discussed below. We will also see below that in these more general situations we can still describe the physical situation with what is known as the shock wave solution, including a variable number of eikonal sources. 

We start by reviewing the shock wave solution and its relation to the eikonal sources generated by Glauber exchange in the abelian case, which is obtained in the limit where the boost becomes maximal. Here a gauge is chosen  such that  the gauge field is purely $A_\perp$ and vanishes off the light cone,  given by~\cite{Jackiw:1991ck}
\begin{align}  \label{eq:Ashock}
 A_\pm=0,~~ A^\mu_\perp= -\frac{e}{2\pi} \ln(\mu x_\perp)\delta(x^-).
\end{align}
This configuration is related to the one where $A^+\ne 0$ by a gauge transformation. What is relevant here is that the field is independent of $x^+$ thus we may set $k^-=0$, which is consistent with the expansion in \eq{expprop}.
 Thus there is a very simple picture of the abelian quantum mechanical case with a frozen background and no radiation.  Indeed the well known semi-classical eikonal solution can be obtained by solving for the wave function in the field generated by the shock wave solution in \eq{Ashock}, and then obtaining the scattering amplitude~\cite{tHooft:1987rb, Jackiw:1991ck}. This yields the same result we found above in \eq{Gqperp} for the Glauber function $G(q_\perp)$ (taking its abelian limit with $C_F=1$) by summing the  Glauber gluon ladder graphs.\footnote{This resummation in QED was done fifty years ago by Sucher and Levy \cite{Levy:1969cr} where the issue of rapidity divergences was avoided by using the full photon propagator. By doing so the calculation inherently includes other modes aside from the Glauber, in particular the soft contribution. However, as we have seen in QED the soft contribution cancels and as such the lack of a homogeneous scaling in the calculation \cite{Levy:1969cr} is benign. In the context of gravity a similar resummation was performed in \cite{Kabat:1992tb}.} This picture is also supported by considering the two point function for the potential Glauber exchange
\begin{align} \label{eq:GpropFT}
 \int \frac{d^{d-2}k_\perp}{k_\perp^2} \, e^{i k \cdot x}\propto \delta(x_+)\delta(x_-) \ln(x_\perp^2\mu^2)
\end{align}
thus the two particles interact at a point in light cone time when their respective shock waves cross each other. 

Thus in the abelian limit the iteration of Glauber gluons reproduces the standard semi-classical result, and leads to effectively eikonal propagators for the integrals appearing in forward scattering.  However, even for the abelian theory there is no limit in which the collinear radiation can be treated as subleading. That is, while the propagation in the shock wave background leads to classical source propagation, because the Glauber interactions leave the electrons nearly onshell,  their fluctuations due to  interactions with collinear photons are not suppressed. In the effective field theory this is obvious since the leading power collinear action includes these effects. It is interesting to ask how the collinear quantum corrections change the form of the semi-classical result. In particular, one might think that these corrections could simply dress the eikonal form, decoupling from the Glaubers. If we consider a collinear loop which interrupts two Glauber exchanges, such as in  \fig{formfactor}b, the integral vanishes as explained in \sec{forwardgraphs}.  The analogous graph which has a real collinear gluon emission between two Glauber exchanges shown in \fig{formfactor}a also vanishes.  On the other hand the diagram in \fig{formfactor}c  is non-vanishing,  the Glauber loop needs to have its rapidity divergence regulated and there is only one pole in $k^+$ for the Glauber loop. One is free to add any number of Glauber exchanges between the collinear vertices, which builds up the higher order terms in the Glauber function $G(q_\perp)$. Accounting for such corrections on both the $n$-collinear and $\bn$-collinear side this amplitude can be written as
\begin{align}
 \label{eq:fact}
 M = {\cal S}^{n\bar n}F^n(q_\perp) F^{\bar n}(q_\perp)
 \big[ G(q_\perp) -(2\pi)^2 \delta^2(q_\perp) \big],
\end{align}
where $F^{n,\bn}(q_\perp)$ are the abelian quark (electron) form factors for the top and bottom lines. At one-loop the product $F^{n}(q_\perp)F^{\bn}(q_\perp)$ is given by the abelian part of our \eq{oneloop_scet2_vertwfn}. On the other hand, diagrams such as \fig{formfactor}e modify the result in \eq{fact} by having a $k_\perp$ convolution between the Glaubers and the collinear source.  This fact should not come as a surprise as diagrams  such as  \fig{formfactor}e involve a time ordering where pair creation is manifest as shown in the second way we draw the diagram. For QED this conclusion that only form factors dress the eikonal amplitude if pair creation and annihilation are ignored was reached long ago in Ref.~\cite{Chang:1970kja}.

Thus even in the abelian limit the  energetic quarks do not behave solely like single Wilson lines.\footnote{It is interesting to note that up to the two loop level, single Wilson lines will give the right answer for two-to-two scattering as long as one appends the correct one loop form factor to the result. This includes both semi-classical and quantum   corrections. The notion that loop corrections are necessarily quantum corrections fails here, as it does in other cases where there is a contribution from regions where some fields are behaving like classical sources. In the effective theory each loop may be considered as classical or quantum, but in the full theory, since integrals do not scale homogeneously in the power counting, the result can be mixed quantum and classical. For a discussion on this point see \cite{Neill:2013wsa}.} This becomes even more prevalent when we make the theory non-abelian, for example \fig{formfactor}d also cannot be described by a single Wilson line, and indeed requires non-eikonal propagators for the $n$-collinear quark and gluon. In QED the same is true of a diagram where the radiated photon creates another $e^+e^-$ pair, and then we simultaneously consider Glauber attachments to members of this pair as well as the original $e^-$. In the non-abelian case,  current algebra dictates that the light-cone commutator from \eq{JJcommutator} is nonzero, and thus there is no reason to believe that the semi-classical approximation should hold universally. Indeed, we found in section~\ref{sec:SCops} that in the non-abelian case the interaction with soft gluons is non-vanishing and occurs at leading power, and that there are also one-loop non-abelian collinear graphs that contain rapidity divergences and are not simply form factors.  In general, we will not have a completely eikonal description for sources coupling to Glauber interactions when we include collinear splitting or collinear loop diagrams as discussed in \sec{forwardgraphs}, or once a hard interaction is involved as discussed in \sec{spectator}.  On the other hand, both soft and ultrasoft gluons do continue to have eikonal interactions with collinear particles in these cases.

\begin{figure}[t!]
	%
	%
	\begin{center}
		\raisebox{0cm}{ \hspace{-0.5cm} $a$)\hspace{6.8cm} $b$)\hspace{4.7cm}  } 
		\\[-10pt]
		\raisebox{0.3cm}{
\includegraphics[width=0.4\columnwidth]{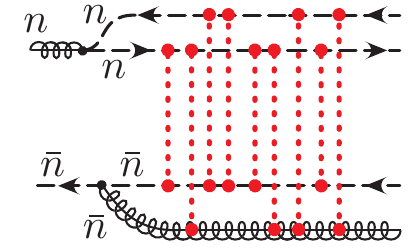}
 } \hspace{1cm}
		\includegraphics[width=0.3\columnwidth]{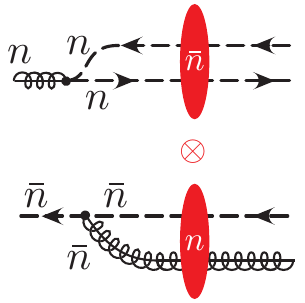} 
	\end{center}
	\vspace{-0.8cm}
	\caption{\setcaptionskip
		Correspondence between multiple Glauber exchange, with an example diagram shown in panel a), and the picture of Wilson lines used to represent partons crossing a shockwave, shown in panel b). The $n$-collinear particles see the other side via Glauber exchange only at an instant in time and longitudinal distance indicated by the location of the shockwave that is drawn as a large shaded red ellipse. In a mirror manner, the $\bn$-collinear particles also see a shockwave representing the Glauber exchanges. } 
	\label{fig:shockwave}
	\setmainskip
\end{figure}

Despite the breakdown of the simplest Wilson line picture for these more general situations with Glauber exchange, the eikonal approximation does still play an important role in the dynamics. For instance, in the non-abelian case  we may still sum the Glauber interactions between pairs of forward scattering partons, as the existence of generators at the vertices  presents no obstruction due to  the simple topology of the ladder  series. This sum was carried out explicitly in \sec{exponentiation}.  Indeed, it is known that in the non-abelian case  many of the contributions to the amplitude are still captured by the shock-wave theory, and can even be calculated in a two-dimensional field theory~\cite{Verlinde:1993te}. In \sec{forwardgraphs} we showed that the eikonal approximation is valid as long as a single source propagator can be associated to each Glauber loop momentum, and breaks down if there are two or more such  propagators of the same source. Furthermore, even in the presence of multiple particles which couple via Glauber exchange, diagrams with multiple Glaubers vanish unless all of these instantaneous exchanges can be collapsed onto a single longitudinal position. Thus, if we consider an arbitrary number of Glauber exchanges in a non-abelian theory, they can interact with any particles that are present at given reference time $t_0$ and longitudinal position $x_0$, which is referred to as the location of the shock-wave seen by these particles. Furthermore, after the first Glauber interaction on each of these propagators, further Glauber attachments yield propagators that are described by the eikonal approximation for that source particle. Each of these sources therefore becomes a Wilson line along the light-cone, located at a different $x_\perp$ coordinate according to the position of the initial particles crossing the shockwave. This analysis based on our framework yields precisely the picture of multiple interacting Wilson lines in~\cite{Mueller:1994gb,Balitsky:1995ub,Kovchegov:1999yj,JalilianMarian:1997jx,Iancu:2000hn}.   An example of the association of the multiple Glauber exchanges with the shockwave is shown in \fig{shockwave}.  Rather than rely on a single Wilson line to describe interactions with the collinear source, one instead considers a picture where multiple Wilson lines are used to describe the color sources that exist at the instant of the shockwave. This is the technique utilized as the starting point for the Wilson line EFT~\cite{Balitsky:1995ub,Balitsky:1998ya,Balitsky:2001gj}, as well as for deriving the BJMWLK equation~\cite{JalilianMarian:1997jx,Iancu:2000hn}. In these frameworks the non-eikonal contributions should only occur in the coefficient functions~\cite{Balitsky:2001gj} for multi-Wilson line matrix elements.

To conclude this section, we give answers to the questions raised at the beginning of this section:
\begin{enumerate}
	\item Glauber iterations reproduce the semi-classical approximation, but in QCD these diagrams are not sufficient by themselves since soft and collinear loops enter at the same order in the coupling and power expansions.  To attempt to make this approximation valid we could remove the non-abelian diagrams by considering an electron scattering observable in QED, and then forbidding pair creation by considering scattering momenta much below the mass of the electron.  Here the semi-classical approximation would dominate the scattering, but in general the form of the amplitude will be more complicated than in \eq{fact} for an infrared safe observable which allows contributions from forward collinear radiation.
	\item Some soft corrections to $n$-$\bn$ scattering do just dress the Glauber kernel, such as those in \fig{SCET2_oneloop_matching}c,d,e which appear as virtual corrections dressing the Glauber exchange from non-abelian and quark-antiquark interactions.  However other soft corrections are not simply dressings of the Glauber kernel, such as the H-graph in \fig{hgraph}.
	\item We have seen above and in section~\ref{sec:forwardgraphs} that the multi-Wilson line shockwave picture emerges because multiple Glauber exchanges collapse so that they occur at a single time and longitudinal position, and because when a graph with a Glauber loop momentum $k$ has only a single propagator that depends on $k^+$ and $k^-$, then these propagators eikonalize.  For other situations with Glauber loops the propagators do not eikonalize, and hence occur before or after the shockwave. 
	\item In the next two sections we take up the issue of Glauber phases produced in the presence of a hard interaction.
\end{enumerate}

\section{Hard Matching: the Cheshire Glauber}
\label{sec:properties}

In this section we consider Glauber gluons in hard scattering processes.  In \sec{hardmatching} we consider Glauber gluon exchange at one-loop in the context of a hard vertex that either annihilates, scatters, or creates two energetic particles.  Then in \sec{softemission} we extend this analysis to include the emission of an additional soft gluon at one-loop. We show that Glauber exchange produces all the low energy imaginary $(i\pi)$ terms, and demonstrate a connection with contributions from soft gluons.  In \sec{higherorder} we extend this analysis to two-loop order with more complicated interactions from the Glauber operators, demonstrating that the same conclusion remains true.

In carrying out hard matching calculations from full QCD onto SCET at one, two, and even three loops, it is known that Glauber exchange graphs are not needed to reproduce the infrared structure of the full theory result and obtain a Wilson coefficient that is independent of the infrared regulator. In this section we demonstrate that the hidden nature of Glauber exchange for these hard scattering calculations is connected to the need to modify soft diagrams by including 0-bin subtractions $S^{(G)}$ from the Glauber region. 
\vskip.1in
\noindent We begin by summarizing our conclusions from two points of view:

1) In \SCETb Glauber exchange contributions $G$ are present as interactions between certain active hard scattering lines. The Glauber subtractions $S^{(G)}$ remove a contribution from the soft diagrams, and in particular are responsible for canceling the contribution arising from direction dependence of soft Wilson lines (whether they extend from $-\infty$ or to $+\infty$), which is related to the sign $(n\cdot k \pm i0)$ used in soft Wilson line induced eikonal propagators. This sign is relevant only in the region where the ${\cal O}(\lambda)$ soft momentum $n\cdot k\to 0$, or more precisely $n\cdot k\sim \lambda^2$. This momentum region is not soft, but is instead correctly accounted for by the Glauber exchange graphs $G$, and hence is removed from the soft diagrams by the subtraction $S^{(G)}$ when forming the complete soft diagram. At one-loop (see \eq{0bins}) the complete soft diagram is $S= \tilde S- S^{(G)}$.  Depending on the choice of the direction for the soft Wilson lines, we may or may not have $G=S^{(G)}$ here. (In contrast, in \SCETa the Glauber exchange contributions $G$ between active lines are scaleless, and are exactly canceled by the ultrasoft 0-bin subtraction on the Glauber graph, $G^{(U)}$.)  

2) Alternatively, for these hard scattering diagrams we can exploit the correspondence between the results for the Glauber exchange graph and the Glauber subtraction for soft graphs in \SCETb. With a specific choice of the direction of the soft Wilson lines we have $G=S^{(G)}$. The correct choice corresponds to directions that agree with the physical direction of the collinear particles probed by the long distance Glauber exchange process. This allows us to consider the alternative but equivalent interpretation, that the Glauber exchange contributions for these hard scattering diagrams can be absorbed into the soft Wilson lines.  This absorption removes both the $G$ and $S^{(G)}$ terms by canceling them, and corresponds to the standard approach that is adopted in typical SCET matching calculations where Glauber exchange is ignored. This absorption does not work in all possible diagrams in SCET (see \sec{ssfactorization}), and hence in general we need to consider the soft and Glauber exchanges as distinct contributions. (In \SCETa the $\perp$-momentum of these Glauber exchanges is at a larger scale than the ultrasoft Wilson lines, and hence is not related to fixing their direction in the same manner.)

We will refer to the above properties of Glaubers in hard scattering diagrams as the Cheshire nature of the Glauber exchange.

\subsection{One Loop Hard Matching with Glaubers}
\label{sec:hardmatching}

In this section we discuss the Cheshire nature of the Glauber at one-loop. 
For \SCETb the Glauber subtractions $S^{(G)}$ are explicitly nonzero for soft diagrams involving pairs of soft Wilson lines that are both outgoing or both incoming, and we will see the precise connection between the subtractions, active-active Glauber graphs, and the direction of soft Wilson lines.  For completeness we also discuss how things change when considering loop graphs in \SCETa.

\begin{figure}[t!]
%
%
\raisebox{-0.2cm}{\hspace{0.3cm} a)\hspace{3.4cm} b)\hspace{3.1cm} c)\hspace{4.2cm} d)\hspace{7cm}} 
  \\[-5pt]
\includegraphics[width=0.18\columnwidth]{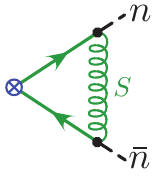}
  \hspace{0.5cm}
\includegraphics[width=0.18\columnwidth]{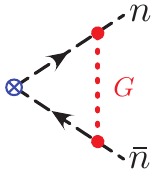}
  \hspace{0.5cm}
\raisebox{0.2cm}{
\includegraphics[width=0.22\columnwidth]{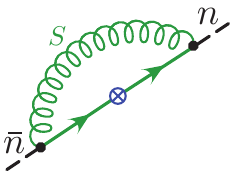}
 }
  \hspace{0.5cm}
\raisebox{0.2cm}{
\includegraphics[width=0.22\columnwidth]{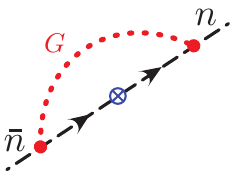}
 }
\caption{\setcaptionskip
One loop soft gluon and Glauber potential exchange with a hard scattering vertex $\otimes$ in \SCETb. The solid green lines denote eikonal propagators from soft Wilson lines. Graphs a) and b) are for 2-particle production, while c) and d) involve hard scattering with one incoming and one outgoing particle. The graph d) is zero.}
\label{fig:hard_Glauber2_oneloop}
\setmainskip
\end{figure}

We begin our discussion in \SCETb, considering the one-loop graphs shown in \fig{hard_Glauber2_oneloop}  with a mass IR regulator $m$.  We take the physical momenta to be  $p$ for the $n$-collinear quark, and $\bar p$ for the $\bn$-collinear (anti)quark. The soft diagrams drawn here arise from the contraction between two gluons taken from the soft Wilson lines that appear in the SCET hard current
\begin{align} \label{eq:current}
J_\Gamma= (\bar \xi_n W_n) S_n^\dagger \Gamma S_{\bn} (W_{\bn}^\dagger \xi_{\bn}) \,.
\end{align}
The usual directions taken for the soft Wilson lines in this current are both $(0,\infty)$ for $n$-$\bn$ production as in \fig{hard_Glauber2_oneloop}a, while we have $S_\bn(-\infty,0)$ and $S_n^\dagger(0,\infty)$ when the $\bn$ quark is in the initial state, as in \fig{hard_Glauber2_oneloop}c. For the $n$-$\bn$ annihilation case (not shown) we would have both lines over $(-\infty,0)$. (See \app{Wdirection} for explicit formulas for the Wilson lines in these cases.) Often in SCET one would draw the soft diagram in \fig{hard_Glauber2_oneloop}a with the eikonal lines contracted to a point. However, for clarity we leave these extended as solid green lines since at higher orders drawing things in this way allows us to make explicit the ordering of the color matrices in our diagrams, and also corresponds with the standard directions for the Wilson lines mentioned above. 

First consider $n$-$\bn$ production in \SCETb, defining the spinor matrix element
\begin{align}
 {\cal S}_\Gamma = \bar u_n \Gamma v_\bn^* \,,
\end{align} 
where the complex conjugation on $v_\bn$ appears due to our convention for the antiquark spinors.
The naive loop integral with a soft gluon exchange is
\begin{align} \label{eq:St1}
   \tilde S({\rm Fig.}~\ref{fig:hard_Glauber2_oneloop}a)
  & = -2 i g^2 C_F   {\cal S}_\Gamma
  \!\! \int\!\! \ddslash\!^{d} k \:
   \frac{(\iota^\epsilon \mu^{2\epsilon}\,|k_z|^{-\eta}\,\nu^\eta) }
  { [k^2-m^2][n\cdot k+i0][\bn\cdot k - i0]} 
   \\
  & = 2 g^2 C_F {\cal S}_\Gamma
  \!\! \int \!\! \ddslash k^z\, \ddslash\!^{d^\prime}\! k_\perp 
  \bigg[ 
   \frac{-(\iota^\epsilon \mu^{2\epsilon}\,|k_z|^{-\eta}\,\nu^\eta)}
    {2(\vec k^{\,2}+m^2)^{1/2} (\vec k_\perp^{\,2}+m^2)}
   + \frac{(\iota^\epsilon \mu^{2\epsilon}\,|k_z|^{-\eta}\,\nu^\eta)}
   {(\vec k_\perp^{\,2}+m^2)(2 k^z -i0)}
   \bigg]
  \nn\\
 &= {\cal S}_\Gamma \frac{C_F \alpha_s}{2\pi}  
  \Bigg\{ \bigg[ \frac{-2 h(\epsilon,\mu^2/m^2)}{\eta}
  + \ln\frac{\mu^2}{\nu^2} \Big( \frac{1}{\epsilon}
 + \ln\frac{\mu^2}{m^2} \Big)  + \frac{1}{\epsilon^2}
  - \frac12 \ln^2\frac{\mu^2}{m^2} - \frac{\pi^2}{12} \bigg]
  \nn\\
  & \qquad\qquad\qquad\qquad
  + \bigg[ (i\pi) \Big(\frac{1}{\epsilon} + \ln\frac{\mu^2}{m^2}\Big) \bigg] 
  \Bigg\}
 \nn\\
 &=  {\cal S}_\Gamma \frac{C_F\alpha_s}{2\pi}
   \bigg[ \frac{-2 h(\epsilon,\mu^2/m^2)}{\eta}
  + \ln\frac{\mu^2}{-\nu^2\minus i0} \Big( \frac{1}{\epsilon}
 + \ln\frac{\mu^2}{m^2} \Big) + \frac{1}{\epsilon^2}
  - \frac12 \ln^2\frac{\mu^2}{m^2} - \frac{\pi^2}{12} \bigg]
 \nn\,,
\end{align}
where $d'=d-2=2-2\epsilon$.  In writing down \eq{St1} we are using the notation where a tilde over a symbol, such as $\tilde S$, denotes a completely unsubtracted integral, which we will refer to as the naive or unsubtracted result. To obtain the second line of \eq{St1} we evaluated the integrand by contours in $k^0$, obtaining the first term from the pole from the relativistic propagator $k^0 = -(\vec k^{\,2}+m^2)^{1/2}+i0$, and the second term  proportional to $(i\pi)$ from the pole in the eikonal propagator $k^0 = -k^z +i0$. The result for these integrals is shown separately in the third equality, and can be combined by introducing a $(-1-i0)$ in the rapidity logarithm, as shown in the final line. If we consider the Glauber zero-bin subtraction integral for this soft loop, then we have 
\begin{align}  \label{eq:SG}
  S^{(G)}({\rm Fig.}~\ref{fig:hard_Glauber2_oneloop}a) 
  &= -2 i g^2 C_F\, {\cal S}_\Gamma
  \!\! \int\!\! \ddslash\!^{d} k \:
   \frac{(\iota^\epsilon \mu^{2\epsilon}\,|k_z|^{-\eta}\,\nu^\eta) }
  { [k_\perp^2-m^2][n\cdot k+i0][\bn\cdot k - i0]} 
  \nn \\
  &=  2 g^2 C_F {\cal S}_\Gamma
  \!\! \int \!\! \ddslash k^z\, \ddslash\!^{d^\prime}\! k_\perp 
  \ \frac{(\iota^\epsilon \mu^{2\epsilon}\,|k_z|^{-\eta}\,\nu^\eta)}
   {(\vec k_\perp^{\,2}+m^2)(2 k^z -i0)}
  \nn\\
 &=  {\cal S}_\Gamma \frac{C_F \alpha_s}{2\pi}  
  \bigg[ (i\pi) \Big(\frac{1}{\epsilon} + \ln\frac{\mu^2}{m^2}\Big) \bigg] 
   \,.
\end{align}
Therefore the full result for the soft graph in a theory with Glauber exchange is given by the result without the $(i\pi)$ contribution
\begin{align}
S({\rm Fig.}~\ref{fig:hard_Glauber2_oneloop}a) 
 &= \tilde S - S^{(G)}
  \\
 &= {\cal S}_\Gamma \frac{C_F\alpha_s}{2\pi}
   \bigg[ \frac{-2 h(\epsilon,\mu^2/m^2)}{\eta}
  + \ln\frac{\mu^2}{\nu^2} \Big( \frac{1}{\epsilon}
 + \ln\frac{\mu^2}{m^2} \Big) + \frac{1}{\epsilon^2}
  - \frac12 \ln^2\frac{\mu^2}{m^2} - \frac{\pi^2}{12} \bigg]
  . \nn
\end{align}

To this we must then also add the result for the Glauber exchange graph in \fig{hard_Glauber2_oneloop}b, which exactly gives the same $(i\pi)$ term
\begin{align} \label{eq:hard1G}
  G({\rm Fig.}~\ref{fig:hard_Glauber2_oneloop}b)
  &=  -2 i g^2 C_F\, {\cal S}_\Gamma
  \!\! \int\!\! \ddslash\!^{d} k \:
   \frac{(\iota^\epsilon \mu^{2\epsilon}\,|k_z|^{-\eta}\,\nu^\eta) }
  { [k_\perp^2-m^2][n\cdot k-\Delta(k_\perp)+i0]
  [\bn\cdot k +\bar \Delta'(k_\perp) - i0]} 
  \nn \\
  &= {\cal S}_\Gamma \frac{C_F \alpha_s}{2\pi}  
  \bigg[ (i\pi) \Big(\frac{1}{\epsilon} + \ln\frac{\mu^2}{m^2}\Big) \bigg] 
  \,,
\end{align}
where $\Delta(k_\perp)= -n\cdot p +(\vec k_\perp+\vec p_\perp)^2/\bn\cdot p$ and $\bar\Delta'(k_\perp)= -\bn\cdot\bar p + (\vec k_\perp-\vec p_\perp)^2/n\cdot \bar p$. Note that this is the same integral evaluated in \eq{Gbox}, so the values of $\Delta(k_\perp)$ and $\bar\Delta'(k_\perp)$ do not affect the result for this integral, and hence it yields precisely the same value as in \eq{SG}. From this analysis we see that 
\begin{align}  \label{eq:SGisG}
  S^{(G)} = G  \,, \qquad\qquad\quad
  S + G =  (\tilde S-S^{(G)}) + G =  \tilde S  \,.
\end{align}
When we include the Glauber gluon in \SCETb the result for the soft graph is $(\tilde S-S^{(G)})$ and is insensitive to the $(i\pi)$ term that was generated from the choice of directions of the soft Wilson lines. Physically,  the $(i\pi)$ term is generated by the Glauber momentum region and hence occurs in $G$. 

Alternatively, when we added up the soft and Glauber graphs in \eq{SGisG} the sum just reproduces the naive soft result $\tilde S$. If we had not considered Glauber gluons as degrees of freedom in \SCETb, then we would arrive at the same result, since the soft graph would simply give $\tilde S$.  Therefore the Glauber gluon is Cheshire, it is not directly visible as a distinct degree of freedom in this loop integrand  at the level of matching. If all loop integrands behaved in this manner, then we could simply absorb the Glauber exchange into our soft degree of freedom. We will see that this pattern persists for hard scattering graphs (active-active graphs), but is not the case once we consider graphs with spectator quarks or gluons, where some Glauber exchange can be absorbed into collinear Wilson lines, while others cannot be absorbed at all. The fact that for partons in hard scattering  the active-active Glauber contributions can be absorbed into soft Wilson lines is consistent with the contour deformation picture of CSS, where the combined soft+Glauber loop integral is deformed away from the Glauber region for active-active diagrams, and then further expanded to leave only contributions from what we call the naive soft region~\cite{Collins:1988ig,Collins:2011zzd}. Similar logic to that of CSS was used to avoid having Glaubers in the amplitude level factorization theorem for final state particle production in~\cite{Feige:2014wja}. Note that with the SCET operators we can still choose to treat the Glauber exchange as specific non-vanishing contributions which describe amplitude level rescattering phases, even for $e^+e^-$ annihilating into just final state strongly interacting particles.

Next consider how the above one-loop \SCETb analysis changes for the case with one incoming and one outgoing collinear quark, hard scattering from $n$ to $\bn$. Repeating the above calculations for the graphs relevant to this case, we have
\begin{align} \label{eq:Sdist1}
   \tilde S({\rm Fig.}~\ref{fig:hard_Glauber2_oneloop}c)
  & = -2 i g^2 C_F  \bar u_n \Gamma u_\bn
  \!\! \int\!\! \ddslash\!^{d} k \:
   \frac{(\iota^\epsilon \mu^{2\epsilon}\,|k_z|^{-\eta}\,\nu^\eta) }
  { [k^2-m^2][n\cdot k+i0][\bn\cdot k + i0]} 
   \\
  & = 2 g^2 C_F \bar u_n \Gamma u_\bn
  \!\! \int \!\! \ddslash k^z\, \ddslash\!^{d^\prime}\! k_\perp 
  \bigg[ 
   \frac{-(\iota^\epsilon \mu^{2\epsilon}\,|k_z|^{-\eta}\,\nu^\eta)}
    {2(\vec k^{\,2}+m^2)^{1/2} (\vec k_\perp^{\,2}+m^2)}
   \bigg]
  \nn\\
 &=  \bar u_n \Gamma u_\bn \frac{\alpha_s C_F}{2\pi} 
  \bigg[ \frac{-2 h(\epsilon,\mu^2/m^2)}{\eta}
  + \ln\frac{\mu^2}{\nu^2} \Big( \frac{1}{\epsilon}
 + \ln\frac{\mu^2}{m^2} \Big)  + \frac{1}{\epsilon^2}
  - \frac12 \ln^2\frac{\mu^2}{m^2} - \frac{\pi^2}{12} \bigg]
 \nn ,\\
  S^{(G)}({\rm Fig.}~\ref{fig:hard_Glauber2_oneloop}c) 
  &= -2 i g^2 C_F\, \bar u_n \Gamma u_\bn
  \!\! \int\!\! \ddslash\!^{d} k \:
   \frac{(\iota^\epsilon \mu^{2\epsilon}\,|k_z|^{-\eta}\,\nu^\eta) }
  { [k_\perp^2-m^2][n\cdot k+i0][\bn\cdot k + i0]} 
  = 0 \,,
  \nn \\
 G({\rm Fig.}~\ref{fig:hard_Glauber2_oneloop}d) 
  &=  -2 i g^2 C_F\, \bar u_n \Gamma u_\bn
  \!\! \int\!\! \ddslash\!^{d} k \:
   \frac{(\iota^\epsilon \mu^{2\epsilon}\,|k_z|^{-\eta}\,\nu^\eta) }
  { [k_\perp^2-m^2][n\cdot k-\Delta(k_\perp)+i0]
  [\bn\cdot k -\bar\Delta(k_\perp) + i0]} 
  = 0 \,.
 \nn
\end{align}
Here both eikonal poles lie on the same side of the $k^0$ contour, and hence do not contribute. This leads to there being no $(i\pi)$ terms in the naive soft graph $\tilde S$, as well as yielding $S^{(G)}=G=0$. Physically, the absence of a Glauber exchange contribution here occurs because there is no time at which freely propagating $n$-collinear and $\bn$-collinear particles exist simultaneously. Therefore, here the full soft result is the same as the naive soft result, $S({\rm Fig.}~\ref{fig:hard_Glauber2_oneloop}c)= \tilde S({\rm Fig.}~\ref{fig:hard_Glauber2_oneloop}c)$. Once again the result is the same whether or not Glauber exchange is included in the theory. For incoming and outgoing collinear antiquarks, or collinear gluons we would also find the same results as in this case, $S=\tilde S$ and $S^{(G)}=G=0$. On the other hand, for an incoming collinear quark and antiquark, or two incoming or outgoing $n$-$\bn$ collinear gluons, the situation is the same as in \eq{SGisG} with $S^{(G)}=G$ both given by a nonzero $(i\pi)$ term. 

Note that the result $S^{(G)}=G$ for these active-active diagrams, and the fact that these Glauber exchanges can be absorbed into the soft region, relies on using the physical directions for the soft Wilson lines $S_n^\dagger$ and $S_\bn$.\footnote{More precisely, the 1-loop calculation distinguishes in-in and out-out from in-out lines. A calculation with an additional soft emission, done below in \sec{softemission}, also distinguishes out-out from in-in.} When considering SCET without Glauber gluons these directions are often determined by those of their parent collinear particles and the structure of the product of operators being considered~\cite{Chay:2004zn,Arnesen:2005nk}.  In the theory with Glauber gluons, the choice for the direction of the soft Wilson lines is not relevant, since this dependence is instead captured by contributions from the Glauber region. In the calculation above we see explicitly that the choice of direction of the soft lines does not change the one-loop result for $\tilde S-S^{(G)}$.  Any change to $\tilde S$, such as picking one incoming and one outgoing line for $n$-$\bn$ production, is compensated by a corresponding change to $S^{(G)}$, since the 0-bin in SCET implies that the original integrand and its subtractions are always defined with Wilson lines in the same directions. Also, there is no choice to be made for the $i0$s in the Glauber propagators, since they simply come from the physically propagating collinear modes.

It is also worth recalling that these $(i\pi)$ terms from SCET play a role in determining the hard matching coefficients $C$ for the current in \eq{current}, which are related to the ultraviolet parts of time-like and space-like form factors. For \SCETb, in addition to the soft and Glauber graphs, the matching calculation for the production current involves the collinear graphs,
\begin{align} \label{eq:Chard}
 & 
\raisebox{-0.97cm}{
\includegraphics[width=0.12\columnwidth]{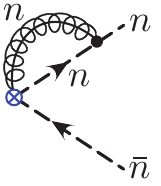}
} \hspace{-0.3cm}+ \hspace{0.2cm}
\raisebox{-1.13cm}{
\includegraphics[width=0.12\columnwidth]{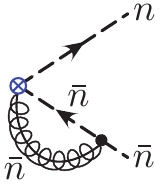}
} \hspace{-0.3cm}+ \hspace{0.2cm} (Z_\xi -1)
\raisebox{-0.95cm}{
\includegraphics[width=0.12\columnwidth]{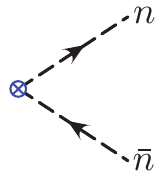}
}  \nn\\[5pt]
&= {\cal S}_\Gamma \frac{\alpha_s C_F}{2\pi} \bigg[ 
  \frac{ 2 h(\epsilon,\mu^2/m^2)}{\eta} 
  + \ln\Big(\frac{\nu^2}{s}\Big)\Big\{ \frac{1}{\epsilon} + \ln\frac{\mu^2}{m^2}\Big\} + \frac{3}{2\epsilon} + \frac{3}{2} \ln\frac{\mu^2}{m^2}  + \frac{9}{4} - \frac{\pi^2}{3} \bigg] ,
\end{align}
where $s=p^-\bar p^+$ and the $p^-$ dependence comes from the $n$-collinear loop, while the $\bar p^+$ dependence comes from the $\bn$-collinear loop. We recall from the calculation in \sec{loop2match} that the subtractions for these collinear graphs all vanish, but are not scaleless, at one-loop with our regulators.  Adding \eq{Chard} together with the result for $S + G = \tilde S$ from \eq{St1}, and simplifying, gives the bare result and $\overline{\rm MS}$ counterterm
\begin{align} \label{eq:hardTotProd}
 \text{SCET}_{\rm prod}
 &= {\cal S}_\Gamma \frac{\alpha_s C_F}{2\pi} \bigg[ 
   \frac{1}{\epsilon^2} + \frac{1}{\epsilon} \ln\frac{\mu^2}{-s} + \frac{3}{2\epsilon} + \frac12 \ln^2\Big(\frac{\mu^2}{-s}\Big)
  - \frac12 \ln^2\Big(\frac{-s}{m^2}\Big) 
  + \frac{3}{2}\ln\frac{\mu^2}{m^2} + \frac{9}{4} - \frac{5\pi^2}{12}
  \bigg] \,,
 \nn\\
Z_C &= 1+ \frac{\alpha_s C_F}{2\pi} \bigg[ -\frac{1}{\epsilon^2} - \frac{1}{\epsilon} \ln\frac{\mu^2}{-s} -\frac{3}{2\epsilon} \bigg] 
 .
\end{align}
Here the $(i\pi)$ SCET terms which arise from the direction of the soft Wilson lines and/or Glauber exchange are necessary to yield the $\ln(-s-i0)$ dependence. This result can be compared to the full theory one-loop vertex and wavefunction graphs with the mass infrared regulator, which combine to give
\begin{align} \label{eq:hardFull}
 \text{Full}_{\rm prod} 
  &= {\cal S}_\Gamma \frac{\alpha_s C_F}{2\pi} \bigg[ 
  - \frac12 \ln^2\Big(\frac{-s}{m^2}\Big) 
  + \frac{3}{2}\ln\Big(\frac{-s}{m^2}\Big) - \frac{7}{4} - \frac{\pi^2}{3}
  \bigg] .
\end{align}
Again the full theory has $\ln(-s-i0)$ dependence, and in particular for $s>0$ we see that there is a $(i\pi)\ln(s/m^2)$ term involving an infrared divergence in both SCET and the full theory, which agree. Subtracting Full$_{\rm prod} -$(SCET$_{\rm prod}$+$(Z_C-1) {\cal S}_\Gamma$) allows us to compute the (standard) result for the SCET Wilson coefficient~\cite{Manohar:2003vb,Bauer:2003di}
\begin{align}  \label{eq:hardC}
  C(s,\mu) &= 1 + \frac{\alpha_s C_F}{2\pi} \bigg[ 
  - \frac12 \ln^2\Big(\frac{\mu^2}{-s}\Big) 
  - \frac{3}{2} \ln\Big(\frac{\mu^2}{-s}\Big) - 4 + \frac{\pi^2}{12} \bigg].
\end{align}
This result for $C(s,\mu)$ is independent of the choice for the IR regulator, and corresponds with the infrared finite parts of the time-like form factor as expected. For the spacelike case with one incoming and one outgoing quark, we use the soft result from \eq{Sdist1}, the collinear results in \eq{Chard} remain unchanged,  while for Eqs.~(\ref{eq:hardTotProd},\ref{eq:hardFull},\ref{eq:hardC}) the corresponding results are obtained by simply taking $(-s)\to s$. The Wilson coefficient $C=C(-s,\mu)$ obtained for this case corresponds with the spacelike form factor. The total SCET result in \eq{hardTotProd} agrees with that in~\cite{Chiu:2007dg} where a different rapidity regulator was used for the soft and collinear components.

Next consider how the analysis of the one-loop $n$-$\bn$ hard production and hard scattering are different for \SCETa. In the case of \SCETa, with ultrasoft and soft modes, and Glauber exchange at the scale of the soft modes, several things change in the above picture. Here, there are no soft Wilson lines in the hard scattering operator, but instead we have ultrasoft Wilson lines $Y_n^\dagger$ and $Y_\bn$ (after the BPS field redefinition)
\begin{align} \label{eq:current1}
J_\Gamma= (\bar \xi_n W_n) Y_n^\dagger \Gamma Y_{\bn} (W_{\bn}^\dagger \xi_{\bn}) \,.
\end{align}
The ultrasoft Wilson lines are generated from the BPS field redefinition~\cite{Bauer:2001yt}. In the calculation of S-matrix elements, the direction of the combined ultrasoft lines is determined by the product of ultrasoft lines generated by $J_\Gamma$ and by the interpolating fields for the incoming and outgoing states (or equivalently those of the external collinear particles)~\cite{Chay:2004zn,Arnesen:2005nk}. The directions for the ultrasoft Wilson lines in the S-matrix elements are then both $(0,\infty)$ for $n$-$\bn$ production, while we have $Y_\bn(-\infty,0)$ and $Y_n^\dagger(0,\infty)$ for the $n$-$\bn$ quark scattering. If we had included both soft and ultrasoft Wilson lines for the current in \eq{current1}, then due to the kinematics present for \SCETa applications, we can (effectively) simply absorb the scaleless graphs involving these soft Wilson lines into analogous graphs involving the ultrasoft lines in the hard production current, replacing $\bar \chi_n Y_n^\dagger S_n^\dagger \Gamma S_\bn Y_\bn \chi_\bn$ by the $J_\Gamma=\bar \chi_n Y_n^\dagger  \Gamma  Y_\bn \chi_\bn $ shown in \eq{current1}. The appropriate IR regulator now sits at the ultrasoft scale, $k^\mu\sim \lambda^2$, so we drop the mass $m$ from the soft loop calculation, and the results for the hard scattering $\tilde S$ in \eqs{St1}{Sdist1} become scaleless.  These scaleless integrals are exactly canceled by the ultrasoft 0-bin subtractions, and hence here the soft modes can be absorbed into the ultrasoft modes, and simply act to pull these ultrasoft modes up so that their ultraviolet divergences occur at the hard scale~\cite{Hoang:2001rr}. So if we did include soft Wilson lines, then the subtractions on the soft graphs always yield zero for these diagrams, $S-S^{(U)}=0$ and $S^{(G)}-S^{(G)(U)}=0$, so that the subtractions cancel the soft lines. 

Next consider the Glauber loop graph in \SCETa. For \SCETa the $(i\pi)$ terms in the EFT are actually entirely reproduced by the ultrasoft region, and not by the Glauber region. Let us reinterpret \fig{hard_Glauber2_oneloop} where now the green gluons and eikonal lines represent the ultrasoft Wilson lines.  Here there are no subtractions on the ultrasoft diagram since the Glaubers have larger $\perp$-momenta, but there are ultrasoft subtractions on the Glauber diagrams, see \eq{scet1subt}. For $n$-$\bn$ production in \SCETa Glauber loop integrals with $k_\perp\sim \lambda$ also become scaleless and hence are also exactly canceled by their ultrasoft 0-bin subtraction, 
\begin{align} \label{eq:Ghardscet1}
  \tilde G({\rm Fig.}~\ref{fig:hard_Glauber2_oneloop}b)
  &=  -2 i g^2 C_F\, {\cal S}_\Gamma
  \!\! \int\!\! \ddslash\!^{d} k \:
   \frac{(\iota^\epsilon \mu^{2\epsilon}\,|k_z|^{-\eta}\,\nu^\eta) }
  { [k_\perp^2][n\cdot k-\Delta(k_\perp)+i0]
  [\bn\cdot k +\bar\Delta'(k_\perp) - i0]} 
  \nn \\
  &= {\cal S}_\Gamma \frac{C_F \alpha_s}{2\pi}  
  \bigg[ (i\pi) \Big(\frac{1}{\epsilon} -\frac{1}{\epsilon_{\rm IR}}\Big) \bigg] 
  = G^{(U)}({\rm Fig.}~\ref{fig:hard_Glauber2_oneloop}b)
  \,,
   \nn\\
  \tilde G - G^{(U)} &=0 
  \,.
\end{align}
Here the $\Delta$s can also depend on the offshellness regulators.  The result in \eq{Ghardscet1} agrees with the \SCETa calculation with Glauber contributions in Ref.~\cite{Bauer:2010cc}. For the $n$-$\bn$ scattering graph or $n$-$\bn$ annihilation we also have $G=G^{(U)}=0$. Therefore for all cases in \SCETa the Glauber graphs $G=\tilde G - G^{(U)}$ do not contribute, and hence the result for the one-loop hard scattering SCET graphs are the same with or without the inclusion of Glauber gluons. In this situation the $(i\pi)$ factors are carried by the ultrasoft diagrams.  Again these factors are necessary to correctly reproduce the hard scattering Wilson coefficients in \eq{hardC}, which for this current are the same in \SCETa as in \SCETb.

\subsection{One Loop Soft Real Emission for Soft-Glauber Correspondence}
\label{sec:softemission}

\begin{figure}[t!]
%
%
\begin{center}
\hspace{0.5cm}
\raisebox{-0.1cm}{
  \hspace{-0.3cm} $S_1^{\rm ee}$)\hspace{2.5cm} $S_2^{\rm ee}$)\hspace{2.5cm} $S_3^{\rm ee}$)\hspace{2.4cm}   $S_4^{\rm ee}$)\hspace{2.6cm} 
  $S_5^{\rm ee}$)\hspace{2.3cm} } 
  \\[-5pt]
\includegraphics[width=0.165\columnwidth]{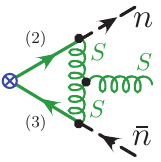}
 \hspace{0.1cm}
\includegraphics[width=0.18\columnwidth]{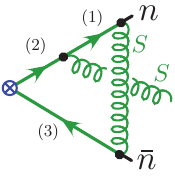}
\hspace{0.1cm}
\includegraphics[width=0.18\columnwidth]{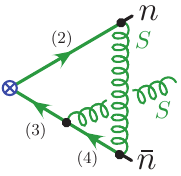}
\hspace{0.1cm}
\includegraphics[width=0.18\columnwidth]{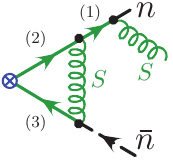}
\hspace{0.1cm}
\includegraphics[width=0.18\columnwidth]{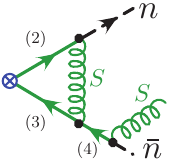}
 \\[0pt]
\raisebox{-0.1cm}{
  \hspace{-0.1cm} $S_6^{\rm ee}$)\hspace{2.7cm} $S_7^{\rm ee}$)\hspace{2.5cm} $G_1^{\rm ee}$)\hspace{2.4cm}   $G_2^{\rm ee}$)\hspace{2.6cm} 
  $G_3^{\rm ee}$)\hspace{2.3cm} } 
  \\[-5pt]
\includegraphics[width=0.18\columnwidth]{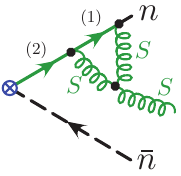}
\hspace{0.1cm}
\includegraphics[width=0.18\columnwidth]{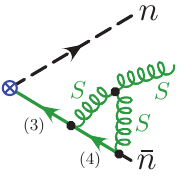}
\hspace{0.1cm}
\includegraphics[width=0.17\columnwidth]{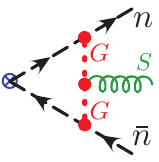}
\hspace{0.15cm}
\includegraphics[width=0.17\columnwidth]{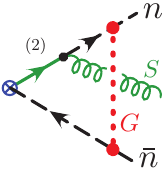}
\hspace{0.15cm}
\includegraphics[width=0.17\columnwidth]{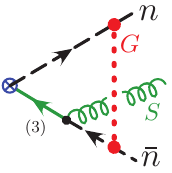}
 \\[0pt]
\raisebox{-0.1cm}{
  \hspace{0.2cm} $G_4^{\rm ee}$)\hspace{2.7cm} $G_5^{\rm ee}$)\hspace{2.5cm} $G_6^{\rm ee}$)\hspace{2.4cm}   $G_7^{\rm ee}$)\hspace{2.6cm} } 
  \\[-5pt]
\includegraphics[width=0.18\columnwidth]{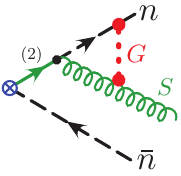}
\hspace{0.1cm}
\includegraphics[width=0.18\columnwidth]{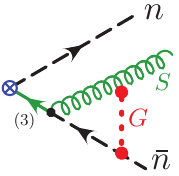}
\hspace{0.1cm}
\includegraphics[width=0.17\columnwidth]{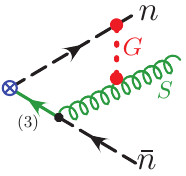}
\hspace{0.1cm}
\includegraphics[width=0.17\columnwidth]{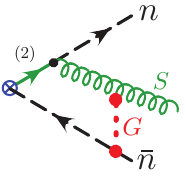}
\end{center}
\vspace{-0.5cm}
\caption{\setcaptionskip
Single soft emission graphs for an $e^+e^-$ annihilation current with a soft or Glauber loop. Solid green lines are eikonal propagators from soft Wilson lines, dashed black lines are collinear propagators, springs are soft gluons, and Glauber exchange is a dotted red line.}
\label{fig:hard_Glauber_oneloop_emission}
\setmainskip
\end{figure}

\begin{figure}[t!]
%
%
\begin{center}
\hspace{0.5cm}
\raisebox{-0.1cm}{
  \hspace{-0.3cm} $S_1^{\rm ep}$)\hspace{2.5cm} $S_2^{\rm ep}$)\hspace{2.5cm} $S_3^{\rm ep}$)\hspace{2.4cm}   $S_4^{\rm ep}$)\hspace{2.6cm} 
  $S_5^{\rm ep}$)\hspace{2.3cm} } 
  \\[-5pt]
\hspace{-0.14cm}
\includegraphics[width=0.185\columnwidth]{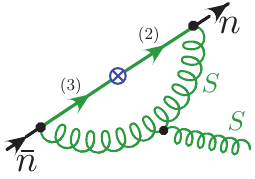}
 \hspace{0.08cm}
\includegraphics[width=0.185\columnwidth]{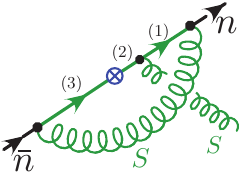}
\hspace{0.08cm}
\includegraphics[width=0.185\columnwidth]{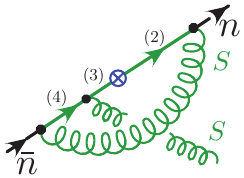}
\hspace{0.08cm}
\includegraphics[width=0.185\columnwidth]{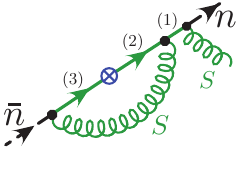}
\hspace{0.08cm}
\includegraphics[width=0.185\columnwidth]{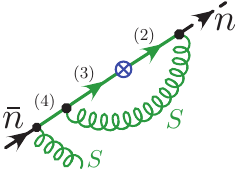}
 \\[0pt]
\raisebox{-0.2cm}{
  \hspace{-0.3cm} $S_6^{\rm ep}$)\hspace{2.8cm} $S_7^{\rm ep}$)\hspace{2.6cm} $G_1^{\rm ep}$)\hspace{2.3cm}   $G_2^{\rm ep}$)\hspace{2.2cm} 
  $G_3^{\rm ep}$)\hspace{2.3cm} } 
  \\[-10pt]
\hspace{-0.2cm}
\raisebox{0.2cm}{
\includegraphics[width=0.18\columnwidth]{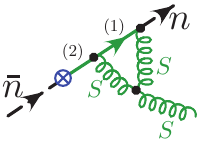}
 }
\hspace{0.08cm}
\raisebox{-0.2cm}{
\includegraphics[width=0.18\columnwidth]{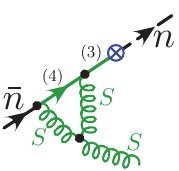}
 }
\hspace{-0.15cm}
\includegraphics[width=0.185\columnwidth]{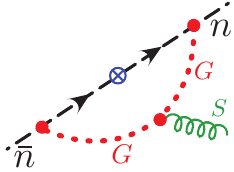}
\hspace{0.04cm}
\includegraphics[width=0.185\columnwidth]{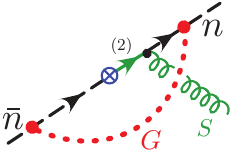}
\hspace{0.04cm}
\includegraphics[width=0.185\columnwidth]{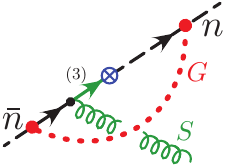}
 \\[5pt]
\raisebox{-0.1cm}{
  \hspace{0.2cm} $G_4^{\rm ep}$)\hspace{2.7cm} $G_5^{\rm ep}$)\hspace{2.5cm} $G_6^{\rm ep}$)\hspace{2.4cm}   $G_7^{\rm ep}$)\hspace{2.6cm} } 
  \\[-8pt]
\includegraphics[width=0.21\columnwidth]{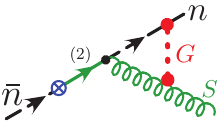}
\hspace{0.1cm}
\includegraphics[width=0.19\columnwidth]{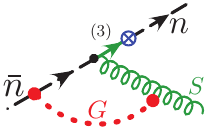}
\hspace{0.1cm}
\includegraphics[width=0.16\columnwidth]{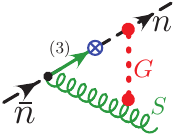}
\hspace{0.1cm}
\includegraphics[width=0.22\columnwidth]{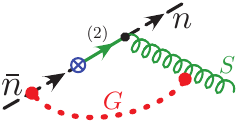}
\end{center}
\vspace{-0.6cm}
\caption{\setcaptionskip
Single soft emission graphs for a $e^-p$ hard scattering current with a soft or Glauber loop. Of the Glauber loop graphs displayed here, only $G_6^{ep}$ is nonzero.}
\label{fig:hard_Glauber_oneloop_emission_ep}
\setmainskip
\end{figure}

\begin{figure}[t!]
%
%
\begin{center}
\hspace{0.5cm}
\raisebox{-0.1cm}{
  \hspace{0.1cm} $S_1^{\rm pp}$)\hspace{2.4cm} $S_2^{\rm pp}$)\hspace{2.3cm} $S_3^{\rm pp}$)\hspace{2.4cm}   $S_4^{\rm pp}$)\hspace{2.cm} 
  $S_5^{\rm pp}$)\hspace{3.3cm} } 
  \\[0pt]
\includegraphics[width=0.17\columnwidth]{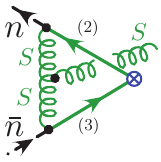} 
\hspace{0.1cm}
\includegraphics[width=0.17\columnwidth]{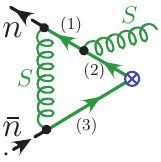}
\hspace{0.1cm}
\includegraphics[width=0.17\columnwidth]{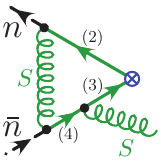}
\hspace{0.1cm}
\raisebox{0.15cm}{
\includegraphics[width=0.14\columnwidth]{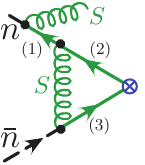}
 }
\hspace{0.1cm}
\includegraphics[width=0.14\columnwidth]{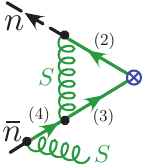}
 \\[5pt]
\raisebox{-0.1cm}{
  \hspace{0.2cm} $S_6^{\rm pp}$)\hspace{2.4cm} $S_7^{\rm pp}$)\hspace{2.2cm} $G_1^{\rm pp}$)\hspace{2.2cm}   $G_2^{\rm pp}$)\hspace{2.1cm} 
  $G_3^{\rm pp}$)\hspace{3.3cm} } 
  \\[-8pt]
\includegraphics[width=0.155\columnwidth]{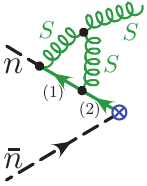}
\hspace{0.1cm}
\raisebox{-0.3cm}{
\includegraphics[width=0.15\columnwidth]{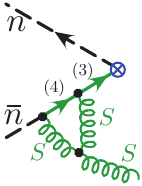}
 }
\hspace{0.1cm}
\includegraphics[width=0.17\columnwidth]{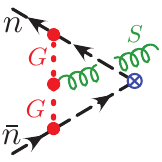}
\hspace{0.1cm}
\includegraphics[width=0.15\columnwidth]{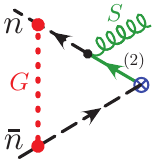}
\hspace{0.1cm}
\includegraphics[width=0.15\columnwidth]{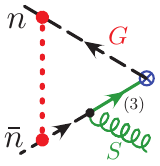}
 \\[5pt]
\raisebox{-0.1cm}{
  \hspace{0.5cm} $G_4^{\rm pp}$)\hspace{2.4cm} $G_5^{\rm pp}$)\hspace{2.1cm} $G_6^{\rm pp}$)\hspace{1.9cm}   $G_7^{\rm pp}$)\hspace{3.4cm} } 
  \\[-5pt]
\includegraphics[width=0.16\columnwidth]{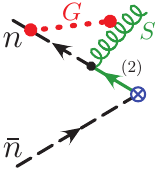}
\hspace{0.15cm}
\includegraphics[width=0.16\columnwidth]{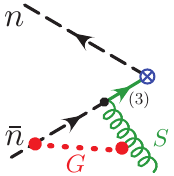}
\hspace{0.15cm}
\includegraphics[width=0.15\columnwidth]{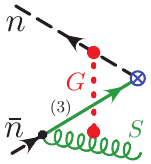}
\hspace{0.15cm}
\includegraphics[width=0.15\columnwidth]{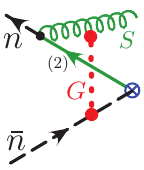}
\end{center}
\vspace{-0.6cm}
\caption{\setcaptionskip
Single soft emission graphs for a $pp$ quark annihilation current with a soft or Glauber loop. Of the Glauber loop graphs shown here, only $G_{1,2,3}^{pp}$ are nonzero.}
\label{fig:hard_Glauber_oneloop_emission_pp}
\setmainskip
\end{figure}

We next show that the correspondence between Glauber contributions and Glauber subtractions for soft graphs discussed in \sec{hardmatching}, also holds for the situation with two active quarks participating in a hard interaction plus one soft gluon emission. In this section we only consider \SCETb.  This soft emission case is interesting because there are three different physical situations, corresponding to an outgoing quark/antiquark pair, an incoming and then outgoing quark, or an incoming quark/antiquark pair. We will refer to these as $ee$, $ep$ and $pp$ respectively, since the underlying hard scattering would be relevant for each of these three hard collision processes. Since our soft gluon is always outgoing, these processes involve either 3 outgoing particles, 2 outgoing and 1 incoming particle, or 1 outgoing and 2 incoming particles. The relevant diagrams with soft or Glauber loops are shown in Figs.~\ref{fig:hard_Glauber_oneloop_emission}, \ref{fig:hard_Glauber_oneloop_emission_ep}, and \ref{fig:hard_Glauber_oneloop_emission_pp}.  As usual, these SCET graphs also contain subtraction contributions as in \eq{0bins}. In the case being considered here these subtractions ensure that the soft propagators in the loop are truly soft, and hence do not give contributions from the region where the propagators momentum becomes Glauber. Based on the physical picture advocated in earlier sections, we could immediately determine that some of the Glauber exchange diagrams are zero. Here we prefer to list all the diagrams and save the discussion of this physical interpretation for determining the nonzero diagrams to the end of this section. 

The contribution of the $i$th diagram from Figs.~\ref{fig:hard_Glauber_oneloop_emission}--\ref{fig:hard_Glauber_oneloop_emission_pp} can be written as
\begin{align} \label{eq:Asoftemission}
i{\cal A}_i^{\rm chan} 
  &=(i\pi) \frac{g^3}{\pi} T^A 
   \Big(\frac{n^\mu}{n\mcdot k} -\frac{\bar n^\nu}{\bn\mcdot k} \Big) 
   \bigg[ -\frac12\, a_i^{\rm chan}\, C_A\, n\mcdot k\, \bn\mcdot k\, I_\perp^{(1)}(k_\perp)
         + b_i^{\rm chan}\, C_F  I_\perp^{(0)} 
         -\frac12\,  c_i^{\rm chan}\, C_A  I_\perp^{(0)} \bigg] 
   .
\end{align}
Here $k$ is the outgoing momentum of the soft gluon which has color $A$ and vector index $\mu$, and the integrals that appear are
\begin{align}
 I_\perp^{(0)} &=\int \frac{\ddslash\!^{d-2} \ell_\perp\, (\iota^{\epsilon} \mu^{2\epsilon})}{\vec \ell_\perp^{\: 2}+m^2}
  \,, 
 & I^{(1)}_\perp(k_\perp)
  &= \int\! \frac{\ddslash\!^{d-2}\ell_\perp\, (\iota^{\epsilon} \mu^{2\epsilon})^2}{(\vec \ell_\perp^{\:2}+m^2)\,\big[(\vec \ell_\perp +\vec k_\perp)^2+m^2\big]}
  \,,
\end{align}
to which we can also freely add a suitable IR regulator. For example, with the displayed gluon mass $m$ the integral $I_\perp^{(0)}$ is not scaleless. The only diagram and channel dependent factors in \eq{Asoftemission} are the constants $a_i^{\rm chan}$, $b_i^{\rm chan}$, $c_i^{\rm chan}$, where $i$ determines which Glauber or soft diagram is being considered, and ${\rm chan}=ee$, $ep$, or $pp$.

\begin{table}[t!]
\centering
\begin{tabular}{|ll|ccccc|cccccccc|}
 \hline
   & & $G_1$ & $\!G_{2+3}\!$ & $\!G_{4+5}\!$ & $G_6$ & $G_7$  
    & $S_1^{(2,3)}\!$
    & $\!S_1^{(23)}\!$ 
    & $\!S_{2+3}^{(13,24)}\!$
    & $\!S_{2+3}^{(23)}\!$ 
    & $\!S_{2+3}^{(1,..,4)}\!$
    & $\!S_{4+5}^{(23)}\!$ 
    & $\!S_{4+5}^{(2,3)}\!$ 
    & $\!S_{6+7}^{(1,4)}\!$
   \\
  \hline
$\!C_A I_\perp^{(1)}\!\!:$ 
 & $a_i^{ee}\!\!$ 
  & $-\frac{1}{4}$ & $0$ & $0$ & $+\frac{1}{4}$ & $+\frac{1}{4}$
    & $+\frac{1}{2}$ & $-\frac{1}{4}$ & $0$ & $0$ & $0$ & $0$ & $0$ & $0$
  \\
 & $a_i^{ep}\!\!$ 
  & $0$ & $0$ & $0$ & $+\frac{1}{4}$ & $0$
    & $+\frac{1}{4}$ & $0$ & $0$ & $0$ & $0$ & $0$ & $0$ & $0$
 \\
 & $a_i^{pp}\!\!$ 
  & $-\frac{1}{4}$ & $0$ & $0$ & $0$ & $0$
    & $0$ & $-\frac{1}{4}$ & $0$ & $0$ & $0$ & $0$ &  $0$ & $0$
\\
 \hline
$\!C_F I_\perp^{(0)}\!\!:$ 
 & $b_i^{ee}\!\!$ 
  & $0$ & $-\frac{1}{2}$ & $0$ & $0$ & $0$
   & $0$ & $0$ & $-\frac{1}{2}$ & $+\frac{1}{2}$ & $-\frac{1}{2}$
   & $-\frac{1}{2}$ & $+\frac{1}{2}$ &$0$
  \\
 & $b_i^{ep}\!\!$ 
  & $0$ & $0$ & $0$ & $0$ & $0$
   & $0$ & $0$ & $0$ & $0$ & $-\frac{1}{2}{}^*$ & $0$ 
   & $+\frac{1}{2}{}^*$ & $0$ 
  \\
 & $b_i^{pp}\!\!$ 
  & $0$ & $-\frac{1}{2}$ & $0$ & $0$ & $0$
   & $0$ & $0$ & $-\frac{1}{2}$ & $+\frac{1}{2}$ & $0$
   & $-\frac{1}{2}$ & $0$ &$0$
  \\
 \hline
$\!C_A I_\perp^{(0)}\!\!:$ 
 & $c_i^{ee}\!\!$ 
  & $+\frac{1}{2}$ & $-\frac{1}{2}$ & $0$ & $0$ & $0$
    & $+\frac{1}{4}$ & $0$ & $-\frac{1}{2}$ & $+\frac{1}{2}$ 
    & $-\frac{1}{2}$ & $0$ & $0$ & $+\frac{1}{4}$ 
  \\
 & $c_i^{ep}\!\!$ 
  & $0$ & $0$ & $0$ & $0$ & $0$
    & $+\frac{1}{4}{}^*$ & $0$ & $0$ & $0$ & $-\frac{1}{2}{}^*$ 
    & $0$ & $0$ & $+\frac{1}{4}{}^*$
 \\
 & $c_i^{pp}\!\!$ 
  & $+\frac{1}{2}$ & $-\frac{1}{2}$ & $0$ & $0$ & $0$
    & $0$ & $0$ & $-\frac{1}{2}$ & $+\frac{1}{2}$ & $0$ & $0$ & $0$ & $0$ 
 \\
 \hline
\end{tabular} 
\caption{\label{table:coeffabc}
Final results for the coefficients appearing in \eq{Asoftemission} for the Glauber graphs and subtractions for Soft graphs in Figs.~\ref{fig:hard_Glauber_oneloop_emission},\ref{fig:hard_Glauber_oneloop_emission_ep}, and \ref{fig:hard_Glauber_oneloop_emission_pp}. Here the ${}^*$ superscript indicates results which should have $\bn^\mu\to 0$ in the prefactor they multiply. These terms are not gauge invariant on their own, but sum to zero.}
\end{table}

In \tab{coeffabc} we show the results for the $a_i^{\rm chan}$, $b_i^{\rm chan}$, and $c_i^{\rm chan}$ coefficients for the Glauber graphs $G_i$ for each of the three processes. We also show results for the terms we wish to compare them to, namely the results for the Glauber subtractions $S_i^{(j)}$ of the soft graphs $S_i$.  As usual, the subtractions $(j)$ are determined by considering all possible $n$-$\bn$, $s$-$n$, and $s$-$\bn$ Glauber limits of the soft gluon propagators (see \tab{modes}).  These subtractions are in one-to-one correspondence with Glauber limits of the soft eikonal propagators, so we enumerate the subtractions by letting the superscript $(j)$ indicate which eikonal propagator(s) are taken to be near mass shell with virtuality of order $\lambda^2$. For example, $S_1^{(2)}$ is the graph $S_1$ with propagator $2$'s momentum taken to have scaling of $\lambda^2$.  In general both the Glauber loop graphs $G_i$ and soft loop graphs $S_i$ will have Glauber subtractions to ensure that soft propagators are truly soft. Since different Glauber limits for the soft graphs may overlap, the soft diagrams may also contain double subtractions that remove the overlapping contributions. We use a double superscript to denote these double subtractions. For example, if we are considering the limit where the $3$ propagator is going on-shell $S^{(3)}$ then we must add back the contribution where $2$ also goes on shell since that contribution is part of $S^{(23)}$. Thus $S^{(3)(2)}$ corresponds to the contribution that must be added back to ensure that we are not over-subtracting. Since our discussion here is focused on the subtractions themselves, it is convenient to include these double subtractions into the soft single subtraction results. With this convention the result for the full soft graph in SCET is obtained by the naive soft graph $\widetilde S_i$ minus just these single subtractions.  The Glauber loop graphs $G_i$ do not have double subtractions and hence are also obtained by removing single subtraction contributions from the naive contribution $\widetilde G_i$,
\begin{align}
  S_i &= \tilde S_i -  \sum_j  S_i^{(j)} \,,
  & G_i & = \tilde G_i - \sum_j G_i^{(j)} \,.
\end{align}
We will detail the double subtraction results contained in each $S_i^{(j)}$ below. In several cases the column labels in \tab{coeffabc} indicate a sum of diagrams:
\begin{align}
 G_{2+3} &= G_2+G_3 
 \,,
 & G_{4+5} &= G_4+G_5 
 \,, \nn\\
 S_1^{(2,3)} &= S_1^{(2)} + S_1^{(3)} 
 \,,
 & S_{2+3}^{(13,24)} &= S_2^{(13)} + S_3^{(24)} \,,
 \nn\\
 S_{2+3}^{(23)} &= S_2^{(23)} + S_3^{(23)} 
 \,,
 & S_{2+3}^{(1,2,3,4)} & = S_2^{(1)} + S_2^{(2)} + S_2^{(3)} 
     + S_3^{(2)} + S_3^{(3)} + S_3^{(4)}   \,,
 \nn\\
 S_{4+5}^{(23)} &= S_4^{(23)} + S_5^{(23)}
 \,,
 & S_{4+5}^{(2,3)} & = S_4^{(2)} + S_4^{(3)} + S_5^{(2)} + S_5^{(3)}  
 \,,
 \nn\\
 S_{6+7}^{(1,4)} & = S_6^{(1)} + S_7^{(4)} 
  \,. 
\end{align}

For the Glauber loop graphs $G_i$, only $G_6$ and $G_7$ have nonzero subtractions. Therefore the results we obtain for the graphs $G_{1,\cdots,5}$ are simply given in \tab{coeffabc}. Note that $G_{2,4}$ and $G_{3,5}$ produce the $n^\mu$ and $\bn^\mu$ structures in \eq{Asoftemission} respectively, while $G_1$ alone produces both of these structures.   For $G_{6,7}$ we have
\begin{align} \label{eq:G67subt}
   &\phantom{xx} &  G_6 & = \tilde G_6 - G_6^{(3)}  \,,
                 &  G_7 & = \tilde G_7 - G_7^{(4)} 
    \nn \\
   &a_i^{ee}:    &  \frac{1}{4} & = \ 0\  +\ \frac{1}{4}\  \,,
                 &  \frac{1}{4} & = \ 0\  +\ \frac{1}{4}\  \,,
    \nn\\
   &a_i^{ep}:    &  \frac{1}{4} & = \ \frac{1}{4}\  +\ 0\  \,,
                 &  0 & = \ 0\  +\ 0\  \,,
    \\
   &a_i^{pp}:    &  0 & = -\frac{1}{4} + \frac{1}{4}\  \,,
                 &  0 & = -\frac{1}{4} + \frac{1}{4}\  \,.
    \nn
\end{align}
The $b_i^{\rm chan}$ and $c_i^{\rm chan}$ coefficients are all zero for $\tilde G_6$, $\tilde G_7$, $G_6^{(3)}$, and $G_7^{(4)}$.

For the soft subtractions the nonzero double subtractions for $S_1$ are given by
\begin{align}
 &\phantom{xx} &  S_1^{(2)} & = \tilde S_1^{(2)} - \tilde S_1^{(2)(3)}  \,,
               &  S_1^{(3)} & = \tilde S_1^{(3)} - \tilde S_1^{(3)(2)} 
    \nn \\
   &a_i^{ee}:    &  \frac{1}{4}\ & = \ 0\ \ +\ \frac{1}{4}\  \,,
                 &  \frac{1}{4}\ & = \ 0\ \ +\ \frac{1}{4}\  \,,
    \nn\\
   &a_i^{ep}:    &  \frac{1}{4}\ & = \ \frac{1}{4}\ \: +\ 0\  \,,
                 &  0\ & = \ 0\ \:\,  +\ 0\  \,,
    \nn\\
   &a_i^{pp}:    &  0\ & = -\frac{1}{4}\, +\ \frac{1}{4}\  \,,
                 &  0\ & = -\frac{1}{4}\, +\ \frac{1}{4}\  \,,
 \\[5pt]
   &c_i^{ee}:    &  \frac{1^n}{4} \ & = \ \frac{1^n}{4}\ \ +\ 0\  \,,
                 &  \frac{1^\bn}{4}\ & = \ \frac{1^\bn}{4}\ \ +\ 0\  \,,
    \nn\\
   &c_i^{ep}:    &  \frac{1^n}{4} & = \ \frac{1^n}{4}{}\:  +\ 0\  \,,
                 &  0\ & = \ 0\ \:\,  +\ 0\  \,,
    \nn\\
   &c_i^{pp}:    &  0\ & = \ 0\ \ \, +\ \, 0\  \,,
                 &  0\ & = \ 0\ \  +\ \, 0\  \,.
 \nn
\end{align}
Here the coefficients with a $n$ or $\bn$ superscript only contribute to the $n^\mu$ or $\bn^\mu$ structures in \eq{Asoftemission} respectively, while all others give the full $(n^\mu/n\cdot k - \bn^\mu/\bn\cdot k)$ combination.  The $S_{2}$ and $S_{3}$ graphs and their subtractions only contribute to $n^\mu$ and $\bn^\mu$ respectively, and do not have any $a_i^{\rm chan}$ contributions. For $S_2$, the terms with nonzero double subtractions are 
\begin{align}
 &\phantom{xx} &  S_2^{(3)} & = \tilde S_2^{(3)} - \tilde S_2^{(3)(1)}
                     - \tilde S_2^{(3)(2)}   \,,
               &  S_2^{(2)} & = \tilde S_2^{(2)} - \tilde S_2^{(2)(3)} \,,
               &  S_2^{(1)} & = \tilde S_2^{(1)} - \tilde S_2^{(1)(3)} 
   \nn  \\
   &b_i^{ee}=c_i^{ee}: 
       &  0\ & = \ 0\ \ \: +\ \ \frac{1}{2}\ \ \ -\ \ \frac{1}{2}\  \,,
       &  -\frac{1}{2}\ & = \ 0\ \ \: -\ \ \frac{1}{2}\  \,,
       &  0\ & = -\frac{1}{2} \  +\ \ \frac{1}{2}\  \,,
    \nn\\
   &b_i^{ep}=c_i^{ep}:    
       &  0\ & = \ 0\ \ \: +\ \ 0\ \ \ +\ \ 0\  \,,
       &  -\frac{1}{2}\ & = -\frac{1}{2} \  +\ \ 0\  \,,
       &  0\ & = \ \ 0 \ \  +\ \ 0\  \,,
    \\
   &b_i^{pp}=c_i^{ep}:    
       &  0\ & = \ 0\ \ \: +\ \ \frac{1}{2}\ \ \ -\ \ \frac{1}{2}\  \,,
       &  0\ & = \ \frac{1}{2} \ \ \: -\ \ \frac{1}{2}\  \,,
       &  0\ & = -\frac{1}{2} \  +\ \ \frac{1}{2}\  
  \,, \nn
\end{align}
and except for the $ep$ terms we have the same results for $S_3$,
\begin{align}
 &\phantom{xx} &  S_3^{(2)} & = \tilde S_3^{(2)} - \tilde S_3^{(2)(4)}
                     - \tilde S_3^{(2)(3)}   \,,
               &  S_3^{(3)} & = \tilde S_3^{(3)} - \tilde S_3^{(3)(2)} \,,
               &  S_3^{(4)} & = \tilde S_3^{(4)} - \tilde S_3^{(4)(2)} 
   \nn  \\
   &b_i^{ee}=c_i^{ee}: 
       &  0\ & = \ 0\ \ \: +\ \ \frac{1}{2}\ \ \ -\ \ \frac{1}{2}\  \,,
       &  -\frac{1}{2}\ & = \ 0\ \ \: -\ \ \frac{1}{2}\  \,,
       &  0\ & = -\frac{1}{2} \  +\ \ \frac{1}{2}\  \,,
    \nn\\
   &b_i^{ep}=c_i^{ep}:    
       &  0\ & = \ 0\ \ \: +\ \ 0\ \ \ +\ \ 0\  \,,
       &  0\ & =\ 0 \ \ \: +\ \ \: 0\  \,,
       &  0\ & = \ \ 0 \ \  +\ \ 0\  \,,
    \\
   &b_i^{pp}=c_i^{ep}:    
       &  0\ & = \ 0\ \ \: +\ \ \frac{1}{2}\ \ \ -\ \ \frac{1}{2}\  \,,
       &  0\ & = \ \frac{1}{2} \ \ \: -\ \ \frac{1}{2}\  \,,
       &  0\ & = -\frac{1}{2} \  +\ \ \frac{1}{2}\  
 \,. \nn
\end{align}
Together these $S_{2,3}$ results give the anticipated gauge invariant combination, $(n^\mu/n\cdot k - \bn^\mu/\bn\cdot k)$, except in the $ep$ channel (where the full results sum to zero). 
The $S_{4}$ and $S_{5}$ graphs and their subtractions only give contributions to $n^\mu$ and $\bn^\mu$ respectively, and due to their color structure their coefficients $a_i^{\rm chan}$ and $c_i^{\rm chan}$ are all zero. The nonzero double subtractions for $S_4$ are
\begin{align}
 &\phantom{xx} &  S_4^{(2)} & = \tilde S_4^{(2)} - \tilde S_4^{(2)(3)} \,,
               &  S_4^{(3)} & = \tilde S_4^{(3)} - \tilde S_4^{(3)(2)} 
   \nn  \\
   &b_i^{ee}:    &  \frac{1}{2}\ & = \ 0\ \ +\ \frac{1}{2}\  \,,
                 &  0\ & = -\frac{1}{2}\ \ +\ \frac{1}{2}\  \,,
    \nn\\
   &b_i^{ep}:    &  \frac{1}{2}\ & = \ \frac{1}{2}\ \: +\ 0\  \,,
                 &  0\ & = \ 0\ \:\,  +\ 0\  \,,
    \\
   &b_i^{pp}:    &  0\ & = -\frac{1}{2}\, +\ \frac{1}{2}\  \,,
                 &  0\ & = -\frac{1}{2}\, +\ \frac{1}{2}\  \,,
    \nn
\end{align}
and again except for the $ep$ terms we have the same nonzero double subtractions for $S_5$,
\begin{align} \label{eq:S5subt}
 &\phantom{xx} &  S_5^{(2)} & = \tilde S_5^{(2)} - \tilde S_5^{(2)(3)} \,,
               &  S_5^{(3)} & = \tilde S_5^{(3)} - \tilde S_5^{(3)(2)} 
   \nn  \\
   &b_i^{ee}:    &  \frac{1}{2}\ & = \ 0\ \ +\ \frac{1}{2}\  \,,
                 &  0\ & = -\frac{1}{2}\ \ +\ \frac{1}{2}\  \,,
    \nn\\
   &b_i^{ep}:    &  0\ & = \ 0\ \: +\ 0\  \,,
                 &  0\ & = \ 0\ \:\,  +\ 0\  \,,
    \\
   &b_i^{pp}:    &  0\ & = -\frac{1}{2}\, +\ \frac{1}{2}\  \,,
                 &  0\ & = -\frac{1}{2}\, +\ \frac{1}{2}\  
   \,.\nn
\end{align}
Again, together the $S_{4,5}$ results give the anticipated gauge invariant combination, $(n^\mu/n\cdot k - \bn^\mu/\bn\cdot k)$, except in the $ep$ channel (where the full results sum to zero). 

Note that when there is a nonzero subtraction term, there is always an SCET diagram for that same region which the subtraction is ensuring we do not double count. All together the results detailed in Eqs.~(\ref{eq:G67subt}-\ref{eq:S5subt}) contribute to the final results given in \tab{coeffabc}.

Recall that the physical picture for the Glauber exchange was that of an instantaneous interaction in both time and longitudinal position, or equivalently in the light-like time for each of the forward scattering particles. This picture allows us to immediately predict which of the entries in \tab{coeffabc} could be nonzero, since the scattering particles must be allowed to interact through an instantaneous exchange of this type, and hence must both be present on trajectories that can interact in spacetime. To carry out this analysis we shrink the green eikonal lines down to the same point as the hard scattering operator.  For the $ee$ diagrams in \fig{hard_Glauber_oneloop_emission} we have exchanges between $n$-$\bn$, $n$-$s$, or $\bn$-$s$ particles, all of which exist on a final state trajectory, and hence all the Glauber exchanges are physically allowed. In this case all Glauber exchange diagrams have nonzero contributions in \tab{coeffabc} except for $G_4^{ee}$ and $G_5^{ee}$. These two vanish because we are using Feynman gauge for the soft gluon propagator and the $n$ (or $\bn$) polarization from the Wilson line has a vanishing contraction with the polarizations for the forward scattering soft gluon in the Glauber exchange operator.  For the $ep$ diagrams the graphs $G_1^{ep}$, $G_2^{ep}$, $G_3^{ep}$, $G_5^{ep}$, and $G_7^{ep}$ all involve Glaubers between lines that are in the initial and final states, and hence vanish. Here $G_6^{ep}$ is allowed and nonzero, and $G_4^{ep}$ is allowed but vanishes for the same reason as in the $ee$ case. Finally, for $pp$ we have initial-final state interactions for graphs $G_4^{pp}$, $G_5^{pp}$, $G_6^{pp}$, and $G_7^{pp}$ which all give vanishing contributions. Here it is particularly obvious that we need to shrink the soft eikonal line to a point before coming to this conclusion. On the other hand the graphs $G_1^{pp}$, $G_2^{pp}$, and $G_3^{pp}$ are all allowed and are nonzero.

Separately adding the results for the $G_i$ and $S_i^{(j)}$ in the rows of \tab{coeffabc}, we see that the net contribution obtained from the sum of graphs with Glauber operators is the same as the sum of the subtraction limits of the soft graphs, as anticipated. Thus once again $G = S^{(G)}$ and the same result will be obtained for these amplitudes in the theory with or without Glauber operators. To summarize these results we can define
\begin{align}
  a^{\rm chan} &\equiv \sum_i a_{G_i}^{\rm chan}
               = \sum_{i,j} a_{S_{i}^{(j)}}^{\rm chan} 
   ,
  & b^{\rm chan} & \equiv \sum_i b_{G_i}^{\rm chan}
               = \sum_{i,j} b_{S_{i}^{(j)}}^{\rm chan}
   , \nn\\
   c^{\rm chan} & \equiv \sum_i c_{G_i}^{\rm chan}
               = \sum_{i,j} c_{S_{i}^{(j)}}^{\rm chan}
   ,
\end{align}
where chan$=ee, ep, pp$ and the values in \tab{coeffabc} give
\begin{align}  \label{eq:abcresult}
 a^{ee}&= +\frac{1}{4}\,, & b^{ee}&= -\frac{1}{2}\,, & c^{ee}&= 0 \,,
   \\
 a^{ep}&= +\frac{1}{4}\,, & b^{ep}&= 0\,, & c^{ep}&= 0 \,,
   \nn\\
 a^{pp}&= -\frac{1}{4}\,, & b^{pp}&= -\frac{1}{2}\,, & c^{pp}&= 0 \,.
   \nn
\end{align}

In Ref.~\cite{Catani:2000pi}, the results for the soft graphs without subtractions, $\widetilde S_i$ were calculated for the processes denoted as $ee$, $ep$, and $pp$ in the context of computing the one-loop soft current. For the non-abelian channel they were found to contain both real contributions and $(i\pi)$ terms, while the abelian channel obeys the expected soft-theorem.\footnote{In particular, the nonzero results for $b_{\tilde S}^{ee}$ and $b_{\tilde S}^{pp}$ in \eq{Sabcresult} correspond to the values needed to yield the product of the soft one-loop amplitude with no emissions from \eqs{St1}{Sdist1}, times the ${\cal O}(g)$ tree level soft emission.} In our notation their results for the $(i\pi)$ terms can be summarized by defining
\begin{align}
a^{\rm chan}_{\tilde S} &= \sum_i a_{\tilde S_i}^{\rm chan} \,, 
 & b^{\rm chan}_{\tilde S} &= \sum_i b_{\tilde S_i}^{\rm chan} \,, 
 & c^{\rm chan}_{\tilde S} & = \sum_i c_{\tilde S_i}^{\rm chan} \,,
\end{align}
for which the results are
\begin{align} \label{eq:Sabcresult}
  a^{ee}_{\tilde S} &= +\frac{1}{4} \,,
 & b^{ee}_{\tilde S} &= -\frac{1}{2} \,,
 & a^{ep}_{\tilde S} &= +\frac{1}{4} \,,
 & b^{ep}_{\tilde S} &= 0  \,,
 & a^{pp}_{\tilde S} &= -\frac{1}{4} \,,
 & b^{pp}_{\tilde S} &= -\frac{1}{2}  \,,
\end{align}
while $c^{\rm chan}_{\tilde S}=0$ in all cases.
These same $(i\pi)$ terms were recently also obtained by Ref.~\cite{Kang:2015moa} in the context of computing the channel dependence of the two-loop ultrasoft function involving lines in two collinear directions.\footnote{The terms with $b^{\rm chan}$ and $c^{\rm chan}$ were not discussed in Ref.~\cite{Kang:2015moa} because the $I_\perp^{(0)}$ integral is scaleless for the \SCETa regulators used there. } Comparing \eqs{abcresult}{Sabcresult} we see that 
\begin{align}
 a^{\rm chan} &= a^{\rm chan}_{\tilde S} \,, 
 & b^{\rm chan} &= b^{\rm chan}_{\tilde S} \,, 
 & c^{\rm chan} & = c^{\rm chan}_{\tilde S} \,, 
\end{align}
so the $(i\pi)$ terms from the naive soft diagrams agrees exactly with the soft subtractions.  Therefore, in SCET with Glauber operators, the full soft diagrams, $S_i = \tilde S_i - S^{(G)}$, are completely free of $(i\pi)$ contributions as anticipated.  All $(i\pi)$ terms are correctly reproduced by the Glauber exchange diagrams in the \SCETb calculation. 

Let us take as a given that the two correspondences discussed here, that $S^{(G)}=G$ and that the graphs with Glauber exchange give the $(i\pi)$ terms in $\tilde S$, remain true for active partons to all orders.  It is then interesting to note that the simplest method of computing these $(i\pi)$ terms is by making use of the the $G_i$ Glauber diagrams. These diagrams have loop integrals that are very simple to evaluate since the $(i\pi)$ contribution is always directly obtained by the $k^\pm$ integrations. 

The physical picture that the $(i\pi)$ terms that are in the EFT are generated from the region of momenta described by Glauber loops in \SCETb makes it very plausible that for hard scattering diagrams this correspondence remains true to all orders.

We could also reconsider this single emission calculation in \SCETa.  Since there are phase space restrictions that do not allow soft emissions in this theory, we take the emission to be ultrasoft. Analogs of the purely soft diagrams $S_i^{\rm chan}$  in \SCETb now exist as purely ultrasoft diagrams in \SCETa. If we include soft Wilson lines in the current, and consider a soft loop with an ultrasoft emission, then the emission only occurs outside the loop at leading power, and the loop is canceled by subtractions, consistent with just using \eq{current1}.  Finally, we can consider graphs with a single ultrasoft gluon emission in the presence of a Glauber exchange. These graphs are either zero or are fully canceled by their ultrasoft subtraction, reproducing a similar pattern to what  we saw for \SCETa in \sec{hardmatching}, namely that the contributions are all contained in the purely ultrasoft diagrams.

\subsection{Hard Matching at Two-Loops and Higher Orders}
\label{sec:higherorder}

Let us consider how the observations of the previous two sections generalize to higher orders and to hard vertices with additional real emissions. In particular we wish to show that for higher order active-active graphs in \SCETb, diagrams with Glauber operator insertions continue to go hand-in-hand with the Glauber 0-bin subtractions on soft diagrams. Again, the same results will be obtained for the sum of these SCET loop graphs whether or not Glauber operators and the subtractions are included or neglected. Our focus here will be on two-loop graphs with soft or Glauber loop momenta, which brings in new types of diagrams with Glauber operators, such as those with the Lipatov vertex or soft-collinear forward scattering.\footnote{While there is no correspondence between naive Collinear graphs and Glauber subtractions at one-loop for hard scattering graphs, a correspondence can appear at two-loops. We will discuss this correspondence for active-spectator graphs in \sec{asfactorization}, since in this situation it appears already at one-loop.} 

For this analysis we are interested in the equality of Glauber loop graphs with the Glauber subtractions on soft graphs. Since this can be established at the   integrand level we will not bother to write out explicitly the infrared regulator $m^2$ in this section.\footnote{It is also well known that care must be taken with a gluon mass infrared regulator at 2-loops, since a simple gluon mass spoils gauge invariance.} It is important to note that the \SCETb Glauber and soft graphs considered below should be considered to have an infrared scale set by $m\sim \lambda$, and hence are not scaleless.

\subsubsection{Two Loop abelian Soft-Glauber Correspondence} 
\label{sec:twoloopabelian}

\begin{figure}[t!]
%
%
\begin{center}
\hspace{0.5cm}
\raisebox{-0.2cm}{
  \hspace{0.0cm} $S_1$)\hspace{2.9cm} $S_2$)\hspace{3.1cm} $SG$)\hspace{3.1cm} $G$)\hspace{3.3cm} } 
  \\[-5pt]
\includegraphics[width=0.18\columnwidth]{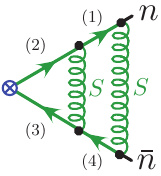}
 \hspace{0.5cm}
\includegraphics[width=0.18\columnwidth]{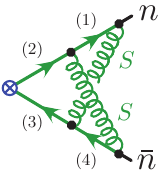}
 \hspace{0.5cm}
\includegraphics[width=0.18\columnwidth]{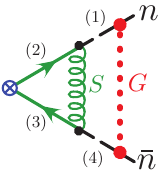}
\hspace{0.5cm}
\includegraphics[width=0.18\columnwidth]{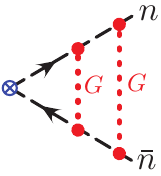}
\end{center}
\vspace{-0.5cm}
\caption{\setcaptionskip
Two loop graphs that have abelian contributions and either soft gluons or Glauber exchange with a hard scattering vertex $\otimes$ in \SCETb. We refer to these graphs as $S_1$, $S_2$, $SG$, and $G$, and we number the collinear/eikonal fermion propagators (1), (2), (3), (4) as shown. }
\label{fig:hard_Glauber2_twoloop}
\setmainskip
\end{figure}

First we consider the \SCETb diagrams for $n$-$\bn$ production  that have abelian contributions at two-loops,  which are shown in \fig{hard_Glauber2_twoloop}.   Both the soft box and soft cross-box graphs have abelian contributions with color factor $C_F^2$, while the cross-box also contributes to the non-abelian $C_FC_A$ terms to be considered below. A crossed graph with two Glauber exchanges does exist as discussed in \sec{GlauberBox}, but is not shown since it still evaluates to zero for the reasons discussed there, even within higher order loop graphs.  Mixed soft Glauber graphs can only arise from time ordered products of a Glauber operator and the hard matching current in \eq{current} so the Glauber vertex can only enter outside the soft interactions. Therefore there are no graphs where a Glauber gluon is nested inside of the soft loop at leading power, and\footnote{This can be seen easily by noting that the soft emission could also be written as coming from a Wilson line at the vertex, often pictured by contracting the eikonal lines to a point. Glauber interactions are not part of this soft Wilson line.}
\begin{align}
  \includegraphics[width=0.14\columnwidth]{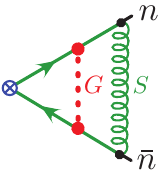}
\hspace{0.5cm}  
  \raisebox{1.3cm}{\text{does not exist.}}  \nn
\end{align}
For the same reason there are no crossed boxes involving mixed Glauber and soft gluon rungs. 

The two-loop soft diagrams have subtractions when the various soft gluon propagators scale into a Glauber region. As in the one-loop soft emission analysis of the previous section,  an equivalent enumeration of the relevant zero-bin limits can be given with the eikonal propagators appearing in these soft diagrams. We therefore enumerate the subtractions by considering cases where eikonal propagators with momentum $k^\pm \sim \lambda$ scale into the Glauber region with $k^\pm\sim \lambda^2$. For eikonal factors in the $S_n$ line one takes $k^+\sim \lambda^2$ in order to scale into the Glauber region, whereas for eikonal factors in the $S_\bn$ line one takes $k^-\sim \lambda^2$. Referring to the propagator numbering in \fig{hard_Glauber2_twoloop}a, the relevant Glauber limits for the two-loop diagrams can be enumerated as (14), (23), (13), (24), (1234), where the numbers refer to which eikonal propagators have a modified scaling for their momentum.  Once again the relevant Glauber subtraction limits occur for cases where there is a corresponding SCET Glauber diagram, and hence ensure there is no double counting.  The results for the graphs with soft momenta after subtractions are then
\begin{align} \label{eq:subt2loop}
  S_i &= \tilde S_i 
     - \big[ S_i^{(G_{23})} - S_i^{(G_{23})(G_{14})}\big] 
     - \big[ S_i^{(G_{14})} - S_i^{(G_{14})(G_{23})} \big]
     - \big[ S_i^{(G_{13})} - S_i^{(G_{13})(G_{24})} \big]
  \nn\\
  &\qquad
     - \big[ S_i^{(G_{24})} - S_i^{(G_{24})(G_{13})} \big]
     - S_i^{(G_{1234})} 
  \,, \nn\\
  SG &= \widetilde {SG} - SG^{(G_{23})} \,.
\end{align}
Here, for example,  $S_i^{(G_{23})}$ denotes the soft graphs integrand with propagators (23) scaled into the Glauber region. This subtraction itself has its own  subtraction $S_i^{(G_{23})(G_{14})}$, which takes the integrand $S_i^{(G_{23})}$ and then subtracts the result where (14) have Glauber scaling. This ensures that in the difference, $S_i^{(G_{23})} - S_i^{(G_{23})(G_{14})}$, the (14) propagators are truly soft. (See~\cite{Manohar:2006nz} for further discussion.) The subtraction $S_i^{(G_{1234})}$ simultaneously considers both loop momenta to have Glauber scaling. The SCET graph $SG$ shown in \fig{hard_Glauber2_twoloop}c contains a soft loop, and hence also has a Glauber subtraction given by $SG^{(G_{23})}$.

Since the abelian soft graphs have trivial numerators, it suffices to study these overlaps by listing the denominator propagators for the  integrands for the graphs shown in \fig{hard_Glauber2_twoloop}, and for their 0-bin subtractions. For the original graphs these are
\begin{align} \label{eq:S1S2props}
  \tilde S_1: \quad &  
    \big[n\cdot k_1\big]\big[n\cdot (k_1\plus k_2)\big]
    \big[-\bn\cdot (k_1\plus k_2)\big]\big[-\bn\cdot k_1\big] 
    \big[k_{1}^2\big] \big[k_2^2 \big] 
  \\
 =& \big[n\cdot k_1\big] \big[n\cdot k_2^\prime \big]
    \big[-\bn\cdot k_2^\prime\big]\big[-\bn\cdot k_1\big] 
    \big[k_{1}^2\big] \big[(k_2^{\prime}\minus k_1)^2\big] 
   \,, \nn \\
  \tilde S_2: \quad &  
   \big[n\cdot k_1\big]\big[n\cdot (k_1\plus k_2)\big]
    \big[-\bn\cdot (k_1\plus k_2)\big]\big[-\bn\cdot k_2\big] 
    \big[k_{1}^2\big] \big[k_2^2\big] 
  \nn\\
 =& \big[n\cdot k_1\big] \big[n\cdot k_2^\prime \big]
    \big[-\bn\cdot k_2^\prime\big]\big[\bn\cdot (k_1\minus k_2^\prime) \big] 
    \big[k_{1}^2 \big] \big[(k_2^{\prime}\minus k_1)^2\big] 
  \nn\\
 =& \big[n\cdot (k_2^\prime\minus k_1^\prime)\big] \big[n\cdot k_2^\prime \big]
    \big[-\bn\cdot k_2^\prime\big]\big[-\bn\cdot k_1^\prime) \big] 
    \big[k_{1}^{\prime 2}\big] \big[(k_2^{\prime}\minus k_1^\prime)^2 \big] 
   \,, \nn \\
 \widetilde {SG}: \quad &
    \big[n\cdot k_1\minus \Delta_1 \big]\big[n\cdot k_2 \big]
    \big[-\bn\cdot k_2 \big]\big[-\bn\cdot k_1\minus \Delta'_1 \big] 
    \big[k_{1\perp}^2\big] \big[k_2^2\big] 
   \,,\nn\\
 G: \quad & 
    \big[n\mcdot k_1\minus \Delta_1 \big]
    \big[n\mcdot (k_1\plus k_2)\minus \Delta_2 \big]
    \big[-\bn\mcdot (k_1\plus k_2)\minus \Delta'_2\big]
    \big[-\bn\mcdot k_1\minus \Delta'_1 \big] 
    \big[k_{1\perp}^2\big] \big[k_{2\perp}^2\big] 
   .\nn
\end{align}
where we show the eikonal propagators listed from (1) to (4), and multiple momentum routings are shown for the purely soft graphs for later convenience.  Here and below, all propagators in square brackets include a $+i0$. The results are all regulated with $|k_1^z|^{-\eta}|k_2^z|^{-\eta}$ (using the notation of the first momentum routings) and these regulator factors are not modified when taking the 0-bin limits, and hence need not be written out explicitly in the analysis below.  It should be evident from \fig{hard_Glauber2_twoloop} that the $\widetilde{SG}$ diagram has the same scaling structure for propagators  as $S_1^{(G_{14})}$, while the $G$ diagram has the same structure as $S_1^{(G_{1234})}$. 

First consider the abelian terms in the (23) limit. Since there are no Glauber graphs that correspond to this limit we anticipate that the soft box and cross-box diagrams will cancel. Using \eq{S1S2props} we find
\begin{align}
 S_1^{(G_{23})}:\quad
  & \big[n\cdot k_1\big] \big[n\cdot k_2^\prime \big]
    \big[-\bn\cdot k_2^\prime\big]\big[-\bn\cdot k_1\big] 
    \big[k_{1}^2 \big] 
    \big[k_1^+k_1^- \minus (\vec k_{2\perp}^{\prime}\minus \vec k_{1\perp})^2 \big]
   \,,\\
 S_2^{(G_{23})}:\quad 
  & \big[n\cdot k_1\big] \big[n\cdot k_2^\prime \big]
    \big[-\bn\cdot k_2^\prime\big]\big[+ \bn\cdot k_1 \big] 
    \big[k_{1}^2 \big] 
    \big[k_1^+k_1^- \minus (\vec k_{2\perp}^{\prime}\minus \vec k_{1\perp})^2 \big]
   \,. \nn
\end{align}
Therefore in the sum relevant to the abelian contribution, $ S_1^{(G_{23})}+ S_2^{(G_{23})}$, we get a $\delta(\bn\cdot k_1)$, which causes the two gluon propagators to only depend on transverse momenta. This sum is therefore identical to the sum of Glauber subtraction terms $S_1^{(G_{23})(G_{14})}+S_2^{(G_{23})(G_{14})}$ and there is no contribution from the (23) limit,
\begin{align}
  S_1^{(G_{23})}+S_2^{(G_{23})} - S_1^{(G_{23})(G_{14})}-S_2^{(G_{23})(G_{14})}
   = 0 \,.
\end{align}  
The situation is similar for the (13) and (24) limits, where
\begin{align}
 S_1^{(G_{13})}:\quad & \big[n\cdot k_1\big] \big[n\cdot k_2^\prime \big]
    \big[-\bn\cdot k_2^\prime\big]\big[-\bn\cdot k_1\big] 
    \big[k_{1\perp}^2 \big] 
    \big[-k_2^{\prime+}k_1^- \minus (\vec k_{2\perp}^{\prime}\minus \vec k_{1\perp})^2 \big]
   \,,\\
 S_2^{(G_{13})}:\quad & 
    \big[n\cdot k_1\big] \big[n\cdot k_2^\prime \big]
    \big[-\bn\cdot k_2^\prime\big]\big[+ \bn\cdot k_1 \big] 
    \big[k_{1\perp}^2 \big] 
    \big[-k_2^{\prime+}k_1^- \minus (\vec k_{2\perp}^{\prime}\minus \vec k_{1\perp})^2 \big]
   \,, \nn \\
 S_1^{(G_{24})}:\quad & 
    \big[+n\cdot k_1^\prime \big] \big[n\cdot k_2^\prime \big]
    \big[-\bn\cdot k_2^\prime\big]\big[-\bn\cdot k_1^\prime\big] 
    \big[k_{1\perp}^{\prime 2} \big] 
    \big[-k_1^{\prime+}k_2^{\prime -} \minus (\vec k_{2\perp}^{\prime}\minus \vec k_{1\perp}^\prime)^2 \big]
   \,,\nn \\
 S_2^{(G_{24})}:\quad & 
    \big[-n\cdot k_1^\prime\big] \big[n\cdot k_2^\prime \big]
    \big[-\bn\cdot k_2^\prime\big]\big[- \bn\cdot k_1^\prime \big] 
    \big[k_{1\perp}^2 \big] 
    \big[-k_1^{\prime+}k_2^{\prime -} \minus (\vec k_{2\perp}^{\prime}\minus \vec k_{1\perp}^\prime)^2 \big]
   \,. \nn 
\end{align}
So $S_1^{(G_{13})}+ S_2^{(G_{13})}$ gives a $\delta(\bn\cdot k_1)$, and $S_1^{(G_{24})}+S_2^{(G_{24})}$ gives a $\delta(n\cdot k_1^\prime)$, making these combinations equal to the sum of their subtractions
\begin{align}
S_1^{(G_{13})}+S_2^{(G_{13})} - S_1^{(G_{13})(G_{24})}-S_2^{(G_{13})(G_{24})}
   &= 0 \,,
   \\
S_1^{(G_{24})}+S_2^{(G_{24})} - S_1^{(G_{24})(G_{13})}-S_2^{(G_{24})(G_{13})}
   &= 0 \,. 
  \nn
\end{align}

For the abelian graphs this leaves only (14) and (1234). The full Glauber limit of the soft cross box,  $S_2^{(G_{1234})}=0$, because the $k_1^0$ contour integral vanishes (considering the $k_1$-$k_2'$ routing). In the (14) limit the cross box gluon propagators depend on only $\perp$-momenta,
\begin{align}
  S_2^{(G_{14})}:\quad &  \big[n\cdot k_1\big]\big[n\cdot k_2\big]
   \big[-\bn\cdot k_1\big]\big[-\bn\cdot k_2\big] 
   \big[k_{1\perp}^2 \big]\big[k_{2\perp}^2 \big] \,,
\end{align}
which is identical to its subtraction. So there is no contribution from the cross box in this limit
\begin{align}
 S_2^{(G_{14})}  - S_2^{(G_{14})(G_{23})}  = 0 \,.
\end{align}
Thus for abelian contributions, the only important subtractions come from the box topology. 
For the full Glauber limit of the soft box graph in these two limits we have
\begin{align}
  S_1^{(G_{1234})}: \quad & 
    \big[n\cdot k_1\big]\big[n\cdot (k_1+k_2)\big]
    \big[-\bn\cdot (k_1+k_2)\big]\big[-\bn\cdot k_1\big] 
    \big[k_{1\perp}^2\big] \big[k_{2\perp}^2 \big] 
   \,, \\
  S_1^{(G_{14})}: \quad &  \big[n\cdot k_1\big]\big[n\cdot k_2\big]
    \big[-\bn\cdot k_2\big] \big[-\bn\cdot k_1\big] 
    \big[k_{1\perp}^2\big] \big[k_{2}^2 \big] 
   \,. \nn
\end{align}
Since the results for Glauber graphs $G$ and $\widetilde {SG}$ with the integrands in \eq{S1S2props} do not depend on the $\Delta_i$ or $\Delta_i'$ factors appearing in the propagators, they are equal to the results for these two subtractions, respectively.  The subtractions have the same integrands just setting all the $\Delta$s to zero. Thus
\begin{align} \label{eq:SisG2loopabel}
  S_1^{(G_{1234})} &= G \,,
  & S_1^{(G_{14})}& = \widetilde {SG} \,,
  & S_1^{(G_{14})(G_{23})}& =  {SG}^{(G_{23})} 
  \,.
\end{align}
The third result follows from the second. Just like at one-loop, the choice of soft Wilson line directions in the hard operator are important for the correspondence in \eq{SisG2loopabel} to be true. Furthermore, in this two-loop analysis, the precise relative powers of $\eta$ used in our soft Wilson line and Glauber potential rapidity regulators are also important in order to obtain these correspondences. Combining all the above results, we find that the sum of all abelian diagrams with their subtractions are equal to the abelian part of the naive soft graphs
\begin{align} \label{eq:abelcorr}
  S_1 + S_2 + SG + G =  \tilde S_1 + \tilde S_2  \,,
\end{align}
as anticipated.

We can also consider this abelian two-loop analysis in \SCETa. Here if we included soft Wilson lines in the current, then there would be contributions from boxes and cross-boxes with either (soft-soft, soft-ultrasoft, soft-Glauber, ultrasoft-Glauber, or ultrasoft-ultrasoft) loops. All cases except the ultrasoft-ultrasoft loops either give integrals that are zero, or that are exactly canceled by their subtractions. Thus there is no impediment to simply using the current which absorbs the soft Wilson lines, given in \eq{current1}, for this calculation.

\subsubsection{Two Loop Non-abelian Soft-Glauber Correspondence}
\label{sec:twoloopnonabelian}

\begin{figure}[t!]
	%
	%
	\begin{center}
		\hspace{0.cm}\raisebox{-0.2cm}{
			\hspace{-0.6cm} $S_3$)\hspace{2.6cm} $S_4$)\hspace{2.7cm} $S_5$)\hspace{2.7cm} $S_6$)\hspace{2.9cm} 
			\hspace{2.7cm}} 
		\\[-5pt]
		\hspace{-0.2cm}
		\includegraphics[width=0.17\columnwidth]{figs/Hard_n-nb_2loop_soft1}
 \hspace{0.25cm}
		\includegraphics[width=0.17\columnwidth]{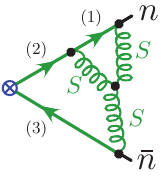}
		\hspace{0.3cm}
		\includegraphics[width=0.17\columnwidth]{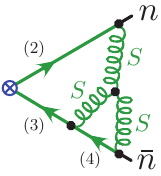}
		\hspace{0.25cm}
		\includegraphics[width=0.17\columnwidth]{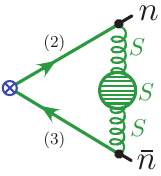}
		\hspace{0.2cm}
		\raisebox{1.1cm}{
		\includegraphics[width=0.19\columnwidth]{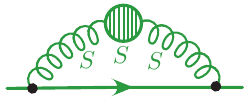}
        }
		\\
		\hspace{0.cm} \raisebox{-0.2cm}{
			\hspace{-0.6cm} $GS_1$)\hspace{2.3cm} $GS_2$)\hspace{2.3cm} $GS_3$)\hspace{2.8cm} $LS_1$)\hspace{2.4cm} $LS_2$)\hspace{2.cm}  } 
		\\[0pt]
		\hspace{-0.2cm}
		\includegraphics[width=0.17\columnwidth]{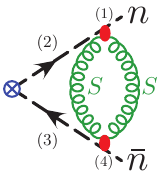}
 \hspace{0.3cm}
		\includegraphics[width=0.17\columnwidth]{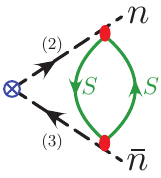}
 \hspace{0.3cm}
		\includegraphics[width=0.19\columnwidth]{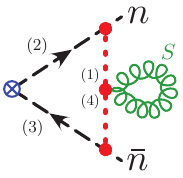}
 \hspace{0.3cm}
		\includegraphics[width=0.17\columnwidth]{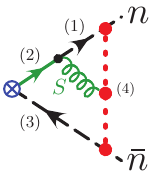}
		\hspace{0.3cm}
		\includegraphics[width=0.17\columnwidth]{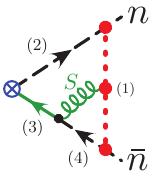}
	\end{center} 
	\vspace{-0.5cm}
	\caption{\setcaptionskip
		Non-abelian two loop graphs with soft gluons and Glauber exchange with a hard scattering vertex $\otimes$ in \SCETb. Only graphs that are non-vanishing in Feynman gauge are shown. We will refer to them as $S_3$, $S_4$, $S_5$, $S_6$, $GS_1$, $GS_2$, $GS_3$, $LS_1$, $LS_2$, and we number the collinear/eikonal fermion propagators (1), (2), (3), (4) as shown.  }
	\label{fig:hard_Glauber2_twoloop_na}
	\setmainskip
\end{figure}

Next we consider generalizing the analysis of the previous section to the \SCETb graphs with the non-abelian $C_F C_A$ or $C_F n_f$ color factors. Using Feynman gauge for the soft gluons, the nonzero graphs  are shown in \fig{hard_Glauber2_twoloop_na}.  Here $S_3$ denotes the same cross box graph called $S_2$ above, just now with the non-abelian part of its color factor.  It has four eikonal propagators. Therefore we have $S_3^{(G_{1234})}=0$ and $S_3^{(G_{14})}-S_3^{(G_{14})(G_{23})}=0$, and must only consider the (23), (13), and (24) limits for $S_3$. For $S_4$ the 3-gluon vertex yields two terms in the numerator that cancel one or the other of the two $n$-eikonal propagators, and likewise for $S_5$ with the two $\bn$-eikonal propagators. Therefore these graphs each have two terms, both with two eikonal propagators.  It is convenient to consider these pieces separately so we write
\begin{align}
  S_4 &= S_{4h} + S_{4r} \,,
  & S_5 &= S_{5h} + S_{5r} \,, 
\end{align}
where the ``h'' subscript indicate terms with two eikonal propagators next to the hard vertex which can have a (23) subtraction limit, whereas the two remaining eikonal propagators in $S_{4r}$ and $S_{5r}$ are such that they only have (13) and (24) limits respectively.
Finally, $S_6$ includes both the vertex graph and Wilson line self energy graphs, and only has a nontrivial (23) limit. The role of the self energy contribution here is to cancel the $k_{2}^\mu k_{2}^\nu$ vacuum polarization numerator term in the vertex graph.  

The graphs in the second row of \fig{hard_Glauber2_twoloop_na} involve Glauber operators.  Here $GS_1$ always has two collinear propagators, and has an internal eye-graph involving terms with both zero and two eikonal propagators. For the numerator from the eye-graph vertices in $GS_{1}$ (see \eq{soft_eye}), we write
\begin{align} \label{eq:eyenumerator}
  & (d-2) n\cdot k_1\: \bn\cdot k_1 + 2 (k_{1\perp}+k_{2\perp})^2 + 2 k_{1\perp}^2  + \frac{4 [k_{1\perp} \cdot (k_{1\perp}+k_{2\perp})]^2}{\bn\cdot k_1\: n\cdot k_1}
  \nn \\
  &=  \Big\{ (d-2) n\cdot k_1\: \bn\cdot k_1  + 4 k_{2\perp}^2 - 2 (k_1+k_{2\perp})^2 -2 k_1^2 \Big\} 
  + \frac{4 [k_1 \cdot (k_1+k_{2\perp})]^2}{\bn\cdot k_1\: n\cdot k_1}
  \,,
\end{align}
and then split the two-loop $GS_1$ graph into two parts by defining 
\begin{align}
  GS_1 &= GS_{1h} + GS_{1f} \,.
\end{align} 
Here $GS_{1h}$ is the result involving the terms in curly brackets in \eq{eyenumerator}, while $GS_{1f}$ refers to the term on the second line with the $(\bn\cdot k_1\, n\cdot k_1)$ eikonal propagators.  The graphs with a Lipatov vertex, $LS_1$ and $LS_2$, have two collinear propagators and terms with both two and zero eikonal propagators (depending on whether the Lipatov vertex cancels the soft eikonal propagator or adds an additional one).  Since these terms also need to be considered separately we divide the graphs up as
\begin{align}
  LS_1 & = LS_{1r} + LS_{1f} \,,
  & LS_2 & = LS_{2r} + LS_{2f} \,,
\end{align} 
where the ``f'' subscript refers to terms with two eikonal plus two collinear propagators, and the ``r'' subscript refers to terms with just the two (black-dashed) collinear propagators that are explicit in \fig{hard_Glauber2_twoloop_na}.  To summarize the nontrivial Glauber subtractions for these contributions we write
\begin{align} \label{eq:subt2loopna}
S_3 &= \tilde S_3
- \big[ S_3^{(G_{23})} - S_3^{(G_{23})(G_{14})}\big] 
- \big[ S_3^{(G_{13})} - S_3^{(G_{13})(G_{24})} \big]
- \big[ S_3^{(G_{24})} - S_3^{(G_{24})(G_{13})} \big]
\,, \nn\\
S_{4h} &= \tilde S_{4h} -  S_{4h}^{(G_{23})} \,, \qquad
S_{4r} =  \tilde S_{4r} -  S_{4r}^{(G_{13})} \,, \qquad
S_{5h} = \tilde S_{5h} -  S_{5h}^{(G_{23})} \,, \qquad
S_{5r} =  \tilde S_{5r} -  S_{5r}^{(G_{24})} \,,
 \nn\\
 S_6 &= \tilde S_6 - S_6^{(G_{23})} \,, \qquad
GS_{1f} = \widetilde {GS}_{1f} - GS_{1f}^{(G_{14})} \,, \qquad
GS_{3} = \widetilde {GS}_{3} - GS_{3}^{(G_{14})} \,,
\nn\\
LS_{1f}  &= \widetilde {LS}_{1f} - LS_{1f}^{(G_{24})} \,, \qquad
LS_{2f}  = \widetilde {LS}_{2f} - LS_{2f}^{(G_{13})} 
 \,,
\end{align}
whereas there are no nontrivial subtractions for $GS_{1h}$, $GS_2$, $LS_{1r}$, or $LS_{2r}$.

The simplest soft two loop contributions are those that only have eikonal propagators next to the hard vertex, for (2) and (3).  This includes the entire $S_6$, as well as $S_{4h}$ and $S_{5h}$ where the momentum factor from the 3-gluon vertex cancels propagators (1) and (4) respectively. For these terms, $(G_{23})$ is the only nontrivial Glauber subtraction on these soft graphs, and the equivalence of these subtractions with the SCET Glauber operator diagrams is directly analogous to what we observed in \sec{hardmatching} at one-loop. Carrying out the calculations we find that
\begin{align} \label{eq:StoGcorrNA1}
S_{4h}^{(G_{23})} + S_{5h}^{(G_{23})} + S_{6}^{(G_{23})}  
&= GS_{1h} + GS_2 \,.
\end{align}
Note that $GS_{1h}$ has collinear propagators for (2) and (3), while the remaining propagators for the other loop are relativistic, and hence do not themselves have a nontrivial Glauber subtraction.  The equality in \eq{StoGcorrNA1} once again relies on the independence of the graphs $GS_{1h}$ and $GS_2$ on the $\Delta$'s that appear in the (2) and (3) collinear propagators.

The remaining part of the Y-graphs in \fig{hard_Glauber2_twoloop_na} include the term $S_{4r}$ where the 3-gluon vertex cancels the eikonal (2) and there is only a $(G_{13})$ subtraction, and a term in $S_{5r}$ where its 3-gluon vertex cancels the eikonal (3) and there is only a $(G_{24})$ subtraction.  These subtraction terms are exactly equal to $LS_{1r}$ and $LS_{2r}$ where the numerator momentum factor from the Lipatov vertex cancels the soft eikonal propagator from the hard vertex Wilson line. Thus,
\begin{align}
  S_{4r}^{(G_{13})} &=  LS_{1r} \,,
  & S_{5r}^{(G_{24})} &=  LS_{2r} \,.
\end{align} 
Once again these are the only relevant subtractions for these terms.

The above considerations account for two loop soft graphs with two eikonal propagators, so the remaining graphs to analyze are those with a total of four eikonal and/or collinear propagators that come from $S_3$, $GS_1$, $GS_3$, $LS_1$, and $LS_2$. Considering first the (23) subtraction we find the propagators
\begin{align} \label{eq:S3G23}
 S_3^{(G_{23})}: \quad &     
   \big[n\cdot k_1\big]\big[ n\cdot k_2'\big]\big[-\bn\cdot k_2'\big]\big[\bn\cdot k_1\big]
   \big[ k_1^2 \big] \big[ k_1^+k_1^- \minus (\vec k_{1\perp}\minus \vec k_{2\perp}')^2 \big] \,.
\end{align}
To identify the corresponding result in the graphs with Glauber exchange, we partial fraction the $GS_{1f}$ integrand by writing,
\begin{align} \label{eq:eyenum}
  \frac{4 [k_1 \cdot (k_1+k_{2\perp})]^2}{\bn\cdot k_1\: n\cdot k_1}
 &=  \frac{ [ 2m^2 \minus k_{2\perp}^2]^2}{\bn\cdot k_1\: n\cdot k_1} 
 + \frac{2 [ k_1^2\minus m^2][(k_1\plus k_{2\perp})^2\minus m^2]}
   {\bn\cdot k_1\: n\cdot k_1}  + \frac{ [ (k_1\plus k_{2\perp})^2\minus m^2]^2}{\bn\cdot k_1\: n\cdot k_1}
  \\
 & + \frac{ [ k_1^2\minus m^2]^2}{\bn\cdot k_1\: n\cdot k_1} 
  + \frac{2 [2m^2\minus k_{2\perp}^2][k_1^2\minus m^2]}
   {\bn\cdot k_1\: n\cdot k_1} 
  + \frac{2 [2m^2\minus k_{2\perp}^2][(k_1+k_{2\perp})^2\minus m^2]}
   {\bn\cdot k_1\: n\cdot k_1} \,.
  \nn
\end{align}
and then dividing by the remaining propagators:
\begin{align}
   [k_1^2-m^2][(k_1+k_{2\perp})^2-m^2][k_{2\perp}^2-m^2]^2[\bn\cdot k_2 -\Delta(k_{2\perp})][-n\cdot k_2-\bar\Delta'(k_{2\perp})] \,.
\end{align}
The second term in \eq{eyenum} gives a vanishing integral, the 3rd, 4th, and 5th terms cancel against each other, and the 6th term in \eq{eyenum} cancels exactly against the flower graph $GS_3$. This leaves only the 1st term. 
After dropping the $\Delta$ and $\Delta'$ factors which again drop out for the integrals over $n\cdot k_2$ and $\bn\cdot k_2$, the result for this remaining term in $GS_{1f}+GS_3$ has the same form as \eq{S3G23} for its propagators, and also the same prefactor. The only possible difference are the directions associated with the soft eikonal factors $(n\cdot k_1)$ and $(\bn\cdot k_1)$, which we did not assign for $GS_{1f}+GS_3$. However, since these eikonal factors have soft scaling, which is ensured by Glauber 0-bin subtractions, the results are identical irrespective of the signs $\pm i0$ used for these eikonal propagators in $GS_{1f}+GS_3$. If both are $+i0$ this gives
\begin{align}
 GS_{1f}+GS_3 - S_3^{(G_{23})} 
&
\propto 
   \frac{1}{[n\cdot k_1][\bn\cdot k_1]} 
  -  \frac{1}{[n\cdot k_1][\bn\cdot k_1]} =0 \,.
\end{align}
In this case we also have $GS_{1f}^{(G_{14})}+GS_3^{(G_{14})} - S_3^{(G_{23})(G_{14})}=0$. The same is true if both are $-i0$ since the rest of the integrand is symmetric under $k_1\to -k_1$. If one eikonal is $+i0$ and the other is $-i0$ (or visa versa) then this gives
\begin{align}
& GS_{1f}+GS_3 - S_3^{(G_{23})}  \nn\\
&\hspace{1cm} \propto 
  \frac12 \bigg[  \frac{-1}{[n\cdot k_1][-\bn\cdot k_1]} -  \frac{1}{[-n\cdot k_1][\bn\cdot k_1]}  \bigg]
  -  \frac12 \bigg[ \frac{1}{[n\cdot k_1][\bn\cdot k_1]} + 
   \frac{1}{[-n\cdot k_1][-\bn\cdot k_1]} \bigg]  
  \nn\\
 & \hspace{1cm}
  = -\frac12 (-2\pi i)^2\, \delta(n\cdot k_1) \delta(\bn\cdot k_1) \,,
\end{align}
which forces the soft $k_1$ momentum into the Glauber region. This contribution is therefore exactly canceled when the $(G_{14})$ subtraction from the $n\cdot k_1\sim \bn\cdot k_1\sim\lambda^2$ region is applied to these terms. This result can be rearranged to yield a relation between the Glauber subtractions of the soft graph and the original Glauber graph that applies for any choice for the eikonal propagators in the Glauber operator vertex,
\begin{align}
  S_3^{(G_{23})} - S_3^{(G_{23})(G_{14})}  &= GS_{1f} +GS_3 - GS_{1f}^{(G_{14})} -GS_3^{(G_{14})} \,.
\end{align}

A similar result will be obtained for the (13) and (24) limits, except now the correspondence is with the $LS_1$ and $LS_2$ graphs. Here we have
\begin{align}
 S_3^{(G_{13})}: \qquad &
  \big[n\cdot k_1\big]\big[ n\cdot k_2'\big]\big[-\bn\cdot k_2'\big]\big[\bn\cdot k_1\big]
  \big[ k_{1\perp}^2 \big] \big[ -k_2^{\prime+}k_1^- \minus (\vec k_{1\perp}\minus \vec k_{2\perp}')^2 \big] 
  \\
   &= \big[n\cdot k\big]\big[ -n\cdot \ell\big]\big[-\bn\cdot k\big]\big[\bn\cdot \ell\big]
   \big[ (k_{\perp}+\ell_\perp)^2 \big] \big[\ell^2 \big] 
   \,,
   \nn\\
  S_3^{(G_{24})}: \qquad &
  \big[ -n\cdot k_1'\big]\big[ n\cdot k_2'\big]\big[-\bn\cdot k_2'\big]\big[-\bn\cdot k_1'\big]
  \big[ k_{1\perp}^{\prime 2} \big] \big[ - k_1^{\prime +}k_2^{\prime -} \minus (\vec k_{1\perp}\minus \vec k_{2\perp}')^2 \big] 
   \nn \\
    &= \big[n\cdot \ell\big]\big[ n\cdot k\big]\big[-\bn\cdot \ell\big]\big[-\bn\cdot k \big]
    \big[ k_{\perp}^2 \big] \big[\ell^2 \big] 
  \,.  \nn
\end{align}
In the first equality we took $k_1 = k +(n/2)\bn\cdot\ell+\ell_\perp$ and $k_2'=k-(\bn/2)n\cdot\ell$, while in the second equality we took $k_1' = k - (\bn/2)n\cdot\ell$ and $k_2=k+(n/2)\bn\cdot\ell+\ell_\perp$. With this change of variables the (13) limit is simply $k\sim (\lambda^2,\lambda^2,\lambda)$ while the momentum $\ell$ remains soft. This change of variables also makes it easier to see that 
the result for $LS_{1f}$ is the same as for $S_3^{(G_{13})}$. For the $(-\bn\cdot\ell)$ propagator in $LS_{1f}$ that arises from the Glauber operator vertex, we have not yet specified whether it is $\pm i0$. Once again the same result is obtained with either choice, either $LS_{1f} - S_3^{(G_{13})}=0$ or 
$LS_{1f} - S_3^{(G_{13})}$ is proportional to $\delta(\bn\cdot\ell)$ which is killed by the terms with a further Glauber 0-bin subtraction on this momentum.  Similarly the result for $LS_{2f}$ is the same as $S_3^{(G_{24})}$. Here when we subtract, $LS_{2f}-S_3^{(G_{24})}$ is zero or proportional to $\delta(n\cdot \ell)$. In both cases these $\delta$-functions force the $\ell$-momentum in these differences into a Glauber region, making the results equal to their (24) and (13) subtractions respectively.  Rearranging, these results we have
\begin{align}
   S_3^{(G_{13})} - S_3^{(G_{13})(G_{24})}  &= LS_{1f} - LS_{1f}^{(G_{24})} \,,
   & S_3^{(G_{24})} - S_3^{(G_{24})(G_{13})}  &= LS_{2f} - LS_{2f}^{(G_{13})} \,.
\end{align}

Putting all these results together  we find that
\begin{align} \label{eq:nonabelcorr}
  S_3 + S_4 + S_5 + S_6 + GS_1 +GS_2 +GS_3 + LS_1 + LS_2 
   &=  \tilde S_3 + \tilde S_4 + \tilde S_5 + \tilde S_6 \,.
\end{align}
So the non-abelian two-loop result is again simply given by the sum of the naive soft graph results. 
From \eqs{abelcorr}{nonabelcorr} we see that, just as in the one loop case, the same result is obtained for hard production graphs at two-loops in theories with or without the inclusion of Glauber gluon exchange, as 
long as the proper subtractions are performed on the soft graphs.  

It is clear that the pattern established above continues to all orders in the abelian diagrams which involve soft and Glauber rungs that go between an active $n$-collinear and active $\bn$-collinear line. The nontrivial Glauber regions of the soft diagrams occur when the momenta of one or more pairs of propagators (one from the $n$ line and one from the $\bn$ line) scale into the Glauber region. For the purely abelian graphs, the box and cross-box subtraction terms continue to cancel unless the soft loops all occur on the internal side next to the hard vertex, with Glauber loops on the outside. When we consider Glauber 0-bin subtractions on any soft graph, we must do so by considering soft gluons from the outside-in, otherwise we again have vanishing contributions. These 0-bin subtractions are then in one-to-one correspondence with a graph where that rung is replaced by a Glauber gluon from the start. 0-bin contributions from simultaneous Glauber limits of two rungs again are only nonzero when considered from the outside-in, and correspond precisely with the replacement of  those two rungs by Glauber gluons.  The same is true if we consider the 0-bin subtractions for the simultaneous limit of $N$-rungs. Given this correspondence for the once subtracted loop integrals, we can also immediately conclude that there is a correspondence on additional iterated subtractions that are considered for these integrands (as in our two-loop example of $S^{(G_2)(G_1)}=SG^{(G_1)}$ above). While the topologies and subtractions are more complicated for the nonabelian graphs, we can again see from our two loop analysis that there is a tight connection between the subtraction terms and the direct Glauber integrands, and thus also anticipate that the correspondence will remain true at higher orders. The key feature of being able to ignore the dependence on the $\perp$-momentum dependent terms $\Delta_i$ remains true for these active-active diagrams. Thus the result at any order for the $n$-$\bn$ production graphs should be the same in the theories with or without the inclusion of Glauber gluon exchange. This implies that from the perspective of these diagrams, the Glauber could be absorbed into the soft gluon degree of freedom.  This result does not however hold for all graphs in \SCETb, as previously mentioned. For example, the results in \sec{ssfactorization} will depend on the $\Delta_i$ terms in the denominators.

One may also consider the extension of the above non-abelian 2-loop analysis to the case of \SCETa. There are now exact analogs of the graphs $S_i$ with two soft loops, that have two ultrasoft loops.  Here if we allowed soft Wilson lines in the production current, then we would still have the diagrams LS$_1$ and LS$_2$ from \fig{hard_Glauber2_twoloop_na}, which involve a contraction between the current and the Lipatov vertex. However, these graphs become scaleless when the IR regulator is only at the smaller scale in \SCETa, and we have checked that they are canceled by the additional ultrasoft subtractions, hence validating the   use of  \eq{current1}.  There are graphs involving one Glauber exchange dressed by an ultrasoft gluon, which are again exactly canceled by their ultrasoft 0-bin subtractions.  Finally, in \SCETa there are also soft loop graphs that do not involve soft Wilson lines from the production current, GS$_1$, GS$_2$, and GS$_3$ from \fig{hard_Glauber2_twoloop_na}, and we expect that they will also be canceled by their 0-bin subtractions.

\subsubsection{Two Loop Soft-Glauber Correspondence For More Than Two Active Lines}

\begin{figure}[t!]
%
%
\begin{center}
\hspace{0.5cm}
\raisebox{-0.2cm}{
  \hspace{0cm} S)\hspace{3.5cm} SG)\hspace{3.6cm} G)\hspace{3.3cm} } 
  \\[-5pt]
\includegraphics[width=0.18\columnwidth]{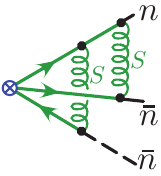}
 \hspace{0.8cm}
\includegraphics[width=0.18\columnwidth]{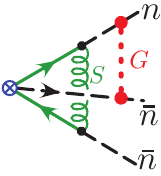}
\hspace{0.8cm}
\includegraphics[width=0.18\columnwidth]{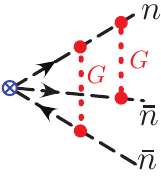}
\end{center}
\vspace{-0.5cm}
\caption{\setcaptionskip
Two loop graphs with two active $\bn$-collinear lines from the hard scattering vertex $\otimes$ in \SCETb, dressed by soft gluons and Glauber exchange. Again the inner rung carries $k_1$ and the outer rung carries $k_2$ for each case, the green lines are eikonal propagators from the soft Wilson lines, and the dashed black lines are collinear propagators. The three diagrams where
the inner gluon contracts with the outgoing $\bn$ quark line give analogous results.}
\label{fig:hard_Glauber2_twoloop_3lines}
\setmainskip
\end{figure}

Next we continue our analysis of hard scattering vertices, by considering possible difference that might occur when we have more collinear lines. To be definite we consider the case where there are two active $\bn$-collinear lines and one active $n$-collinear line in \SCETb.  The relevant diagrams with Glauber and soft rungs which involve all three collinear lines and have a relation to the Glauber exchange are shown in \fig{hard_Glauber2_twoloop_3lines} (the diagrams with the rungs in the other order work in the same manner).  The analysis for these diagrams proceeds in a very similar manner to that of the graphs in \fig{hard_Glauber2_twoloop}. The corresponding propagators and their subtractions are
\begin{align}
  \tilde S: \quad &  
    \big[n\cdot (k_1\plus k_2)\big]\big[-\bn\cdot k_1 \big]
    \big[n\cdot k_2\big]\big[-\bn\cdot k_2\big] 
    \big[k_{1}^2\big] \big[k_2^2\big] 
   \,,\\
 S^{(G_2)}: \quad & 
     \big[n\cdot k_1\big]\big[-\bn\cdot k_1\big]
    \big[n\cdot k_2\big]\big[-\bn\cdot k_2\big] 
    \big[k_{1}^2\big] \big[k_{2\perp}^2\big] 
   \,,\nn\\
 S^{(G_2G_1)}: \quad & 
     \big[n\cdot (k_1+k_2)\big]\big[-\bn\cdot k_1\big]
    \big[n\cdot k_2\big]\big[-\bn\cdot k_2\big] 
    \big[k_{1\perp}^2\big] \big[k_{2\perp}^2\big] 
   \,,\nn\\
 S^{(G_2)(G_1)}: \quad & 
     \big[n\cdot k_1\big]\big[-\bn\cdot k_1\big]
    \big[n\cdot k_2\big]\big[-\bn\cdot k_2\big] 
    \big[k_{1\perp}^2\big] \big[k_{2\perp}^2\big] 
   \,,\nn\\
 \widetilde {SG}: \quad &
    \big[n\cdot k_1 \big]\big[-\bn\cdot k_1 \big]
    \big[n\cdot k_2\plus \Delta_2 \big]\big[-\bn\cdot k_2\plus \Delta'_2 \big] 
    \big[k_{1}^2\big] \big[k_{2\perp}^2\big] 
   \,,\nn\\
 SG^{(G_1)}: \quad &
    \big[n\cdot k_1 \big]\big[-\bn\cdot k_1 \big]
    \big[n\cdot k_2\plus \Delta_2 \big]\big[-\bn\cdot k_2\plus \Delta'_2 \big] 
    \big[k_{1\perp}^2\big] \big[k_{2\perp}^2\big] 
   \,,\nn\\
 G: \quad & 
    \big[n\mcdot (k_1\plus k_2)\plus \Delta_1 \big]\big[-\bn\mcdot k_1\plus \Delta'_1\big]
    \big[n\mcdot k_2\plus \Delta_2 \big]\big[-\bn\mcdot k_2\plus \Delta'_2 \big] 
    \big[k_{1\perp}^2\big] \big[k_{2\perp}^2\big] 
   .\nn
\end{align}
Once again, due to the independence of these loop integrals to $\Delta_i$ and $\Delta'_i$, we have the same relations as before $S^{(G_2)}=\widetilde {SG}$, $S^{(G_2)(G_1)}=SG^{(G_1)}$, and $S^{(G_2 G_1)}=G$. For these correspondences to be valid, the choice of outgoing soft Wilson lines for the hard scattering operator is important. Putting these results together, we once again find
\begin{align}
  \Big( \tilde S - S^{(G_2)} - S^{(G_2G_1)} + S^{(G_2)(G_1)} \Big) 
  + \Big( \widetilde {SG} - SG^{(G_1)}  \Big) + G 
  & = \tilde S \,.
\end{align}
Thus we obtain the same result for this hard scattering calculation in the theory with or without Glauber gluons.

The key feature of being able to ignore the dependence on the $\Delta_i$ remains true when we have additional active lines in the hard scattering diagram, and hence do not change the correspondence between subtractions and Glauber exchange contributions.

If we consider the analogous computation with more than two active lines in \SCETa, then the pattern we have seen in previously subsections repeats once again. We have checked explicitly that graphs with Glauber or soft exchanges are either zero or canceled by the their subtractions. Therefore the dynamics here are once again described by a hard current with only ultrasoft Wilson lines.

\section{Glauber Effects with Spectators in Hard Scattering}
\label{sec:spectator}

In our calculation for near forward scattering  in \sec{exponentiation} there was no hard interaction. The collinear lines effectively acted as classical sources with only a small recoil from an exchanged $q_\perp$ and the source propagators were effectively eikonal because of the structure of the rapidity regulated integrals. In addition, in our analysis of Glauber exchange in purely active hard scattering diagrams in \sec{properties} eikonalization was manifest as well. However, in general we know that Glauber exchange does not lead to purely eikonal propagators for the scattering particles. Examples of non-eikonal situations were discussed above in \sec{forwardgraphs}, and occur when there are two or more $n$-collinear propagators that are sensitive to the Glauber loop momentum, with at least one pole on either side of the axis in the appropriate complex momentum plane (or simply the energy plane).  In the presence of hard interactions we will also see that certain collinear or soft propagators attached to a Glauber interaction are also not exclusively eikonal. Despite the presence of non-eikonal propagators, indicating that the associated propagators are not captured in Wilson lines (i.e. classical sources), we will still see that  Glauber iterations can exponentiate.  

In this section we will consider effects of Glauber exchanges in the hard scattering of color singlet bound states that we treat with interpolating fields.  We consider a hard momentum $q$ with $| q^2 | \gg \Lambda_{\rm QCD}^2$  to flow into the \SCETb electro-magnetic current $J_\Gamma$ in \eq{current}, which involves both collinear and soft Wilson lines. A matrix element is then taken with one or two hadrons which are interpolated for by collinear quark bilinears that have (for simplicity) the quantum numbers of a longitudinal vector meson
\begin{align}
  \Phi_n(x)   &= \Big(\bar\xi_n \frac{\bnslash}{2} \xi_n\Big)(x) \,,
 &\Phi_\bn(y) &= \Big( \bar\xi_\bn \frac{\nslash}{2} \xi_\bn\Big)(y) \,.
\end{align}
In order to have a proxy for the incoming color singlet hadrons that we can handle simply in perturbation theory, we can couple these interpolating fields to light neutral ``hadron'' fields $\rho_n(x)$ and $\rho_\bn(y)$ that are longitudinal vectors and which annihilate ``hadron'' states $| h_n\rangle$ and $|h_\bn\rangle$ via
\begin{align}
 {\cal L}_{\phi_n\Phi_n} &=  \rho_n \Phi_n+  \rho_\bn \Phi_\bn ,
\end{align}
with
\begin{align}
 \rho_n(x') \big|h_n(P) \big\rangle & 
    = n\cdot\varepsilon \: e^{-iP\cdot x'} \,,
 & \rho_\bn(x') \big|h_\bn(\bar P) \big\rangle& 
    = \bn\cdot\varepsilon\:  e^{-i\bar P\cdot x'} \,.
\end{align}
Since the $\rho_{n,\bn}$ fields have no dynamics they have $Z=1$ in the LSZ formula. Since they are light we can take $P^2=\bar P^2=0$ at lowest order. We then study the matrix elements ${\cal M}_\Gamma^{\rm DY}$ and ${\cal M}_\Gamma^{\rm DIS}$  defined by
\begin{align}  \label{eq:ends}
& \delta^4(P+\bar P-q-p_X) \: {\cal M}_\Gamma^{\rm DY}
\equiv 
   \int\!\! d^4x\, d^4y\, 
   \, d^4z\, e^{i q\cdot z}\,  \Big\langle X \Big| 
  T {\cal L}_{\rho_n\Phi_n}(x)\, {\cal L}_{\rho_\bn\Phi_\bn}(y)\, J_\Gamma(z) \Big| h_n(P) h_\bn(\bar P) \Big\rangle  
  \nn\\
 & =\!\!\!
   \lim_{ \parbox{1.2cm}{\footnotesize \hspace{0.05cm} $P^2\!\!\to\! 0$ 
    \\[-5pt] \phantom{x}\hspace{-0.12cm} $\bar P^2\!\!\to\! 0$}}
   \!\!\!  \frac{ P^2 \bar P^2}{i^2} \!\!\!
   \int\!\! d^4x' d^4y' e^{-i P\cdot x'}e^{-i \bar P\cdot y'} \!\!
   \int\!\! d^4x d^4y 
    d^4z e^{i q\cdot z}  \Big\langle X \Big| 
  T {\cal L}_{\rho_n\Phi_n}(x) {\cal L}_{\rho_\bn\Phi_\bn}(y) J_\Gamma(z) \rho_n(x') \rho_\bn(y') \Big| 0 \Big\rangle  
  \nn\\[-5pt]
  &\qquad\qquad\qquad\qquad\qquad\quad\
  = 
   \int\!\! d^4x\, d^4y\, d^4z\,  e^{-i P\cdot x}e^{-i \bar P\cdot y} 
   e^{i q\cdot z}\,  \Big\langle X \Big| 
  T \Phi_n(x)\, \Phi_\bn(y)\, J_\Gamma(z) \Big| 0 \Big\rangle  
   \,,
 \nn\\[10pt]
& \delta^4(\bar P-q-p_X) \: {\cal M}_\Gamma^{\rm DIS} 
   \equiv \int\!\! d^4y\, 
   \, d^4z\, e^{i q\cdot z}\,  \Big\langle X \Big| 
  T {\cal L}_{\rho_\bn\Phi_\bn}(y)\, J_\Gamma(z) \Big| h_\bn(\bar P) \Big\rangle  
  \nn\\
 &\qquad\qquad\qquad\qquad\quad\
  =\!\!\!
   \lim_{ \parbox{1.2cm}{\footnotesize \hspace{0.05cm} $P^2\!\!\to\! 0$ 
    }}
   \!\!\!  \frac{ \bar P^2}{i} \!\!\!
   \int\!\! d^4y' e^{-i \bar P\cdot y'} \!\!
   \int\!\! d^4y \,
    d^4z \, e^{i q\cdot z}  \Big\langle X \Big| 
  T {\cal L}_{\rho_\bn\Phi_\bn}(y) J_\Gamma(z) \rho_\bn(y') 
   \Big| 0 \Big\rangle  
  \nn\\
  &\qquad\qquad\qquad\qquad\quad\
   = 
   \int\!\! d^4y\, d^4z\,  e^{-i \bar P\cdot y}
   e^{i q\cdot z}\,  \Big\langle X \Big| 
  T \Phi_\bn(y)\, J_\Gamma(z) \Big| 0 \Big\rangle  
   \,.
\end{align}
Thus after accounting for momentum conservation, for our perturbative calculations the vertex involving the $\Phi_n$ and $\Phi_\bn$ fields will simply give factors of $(\bnslash/2)$ and $(\nslash/2)$ respectively. Adopting $n$-collinear scaling for the momentum $P$ and $\bn$-collinear scaling for $\bar P$ we can work out the scaling of ${\cal M}_\Gamma^{\rm DY}$ and ${\cal M}_\Gamma^{\rm DIS}$. The outgoing state has the same scaling as a $q\bar q$ pair, $\langle X |\sim \lambda^{-2}$, the fields $\Phi_n\sim \Phi_\bn\sim J_\Gamma\sim \lambda^2$, and the measures $d^4x\sim d^4y \sim \lambda^{-4}$. After shifting coordinates in the matrix elements to $\Phi_n(x-z)$, $\Phi_\bn(y-z)$ and $J_\Gamma(0)$, and shifting $x\to x+z$ and $y\to y+z$, we get momentum conserving $\delta$-function from the $z$-integration. Therefore it is not surprising that the scaling of the momentum conserving $\delta$-functions on the left-hand side is the same as the $\int d^4z$ on the right-hand side. All together we therefore have the power counting results
\begin{align}
  &  {\cal M}_\Gamma^{\rm DY}\sim \lambda^{-4} \,,
  &  {\cal M}_\Gamma^{\rm DIS}\sim \lambda^{-2} & 
  \,.
\end{align}

We will see below that in hard scattering diagrams the source propagators in Glauber loop graphs do not all eikonalize. However, despite this fact,  an overall phase will still be generated if we sum over Glauber exchange rungs (ignoring here soft and collinear radiation). We will also show under what circumstances the phase cancels. Of course this  cancellation is a necessary but not sufficient condition for the proof of a factorization theorem that does not include Glauber exchange. For the full non-abelian case with radiation and quantum corrections there may be contributions that could break factorization and which are not simple pure phases. A complete proof of factorization in SCET entails proving that the Glauber Lagrangian does not contribute to a hard scattering process, and a demonstration of how complete proofs  of factorization can be carried out using our theory for Glauber exchange will be given elsewhere.  

In our perturbative analysis of the diagrams with spectators we will treat only a single scale $t$. It could be taken to be at the hadronic scale $t\sim \Lambda_{\rm QCD}^2$, or rather say $\mu= 2\,{\rm GeV}$ so that perturbation theory still makes sense. Or it could be taken to be at a perturbative scale $Q^*$ of a final state hadronic measurement, in which case $t\sim Q^{*2}\gg \Lambda_{\rm QCD}^2$. (If we wanted to consider our calculations below for the latter case we would replace our $\rho_{n,\bn}$ couplings by perturbative gluon splitting with invariant mass $\sim Q^{*2}$, and would also have to keep $P^2,\bar P^2\ne 0$. The results in \eq{ends} are still valid for this case, but appear without the limits on the RHS.) A complete proof of factorization with our formalism must treat both of these cases.  For the sake of the discussion here we consider ourselves to be in one of the two cases, but we do not treat the mixed situation.

To organize our discussion we divide up the collinear lines in the matrix elements in \eq{ends} into spectator and active lines. At lowest order this division is simple. Considering the base graph in \fig{end_SS}b, the collinear lines contracted with the hard scattering operator are active, and those that are contracted with the hadron interpolating fields which do not directly participate in the hard scattering are spectators. Below in \sec{ssfactorization} we consider Glauber exchange between spectator lines. Then in \sec{asfactorization} we consider Glauber exchange between a spectator line and an active line.  In \sec{aafactorization} we reconsider Glauber exchange between active lines in the presence of hadronic interpolating fields. The generalization of these results to \SCETa is discussed in \sec{scet1spectators}. Finally, we also propose a definition of spectators and active exchanges valid at any order in perturbation theory in \sec{defnSA}.

\subsection{Spectator-Spectator}  
\label{sec:ssfactorization}

\begin{figure}[t!]
	%
	%
	\includegraphics[width=0.21\columnwidth]{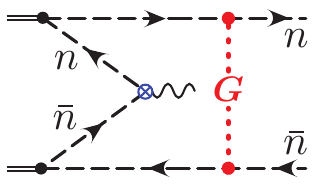} 
       	\hspace{0.1cm} \raisebox{0.9cm}{\large =} \hspace{0.1cm}
    \raisebox{-0.1cm}{
	\includegraphics[width=0.19\columnwidth]{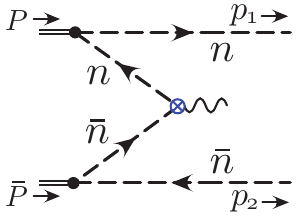}
 }
       	\hspace{-0.3cm} \raisebox{0.9cm}{\large +} \hspace{-0.2cm}
    \raisebox{0.05cm}{
	\includegraphics[width=0.19\columnwidth]{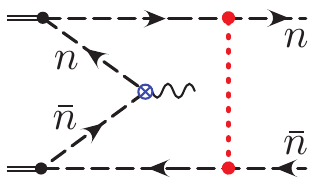}
  } 
       	\hspace{-0.2cm} \raisebox{0.9cm}{\large +}  \hspace{-0.2cm}
    \raisebox{0.1cm}{
	\includegraphics[width=0.2\columnwidth]{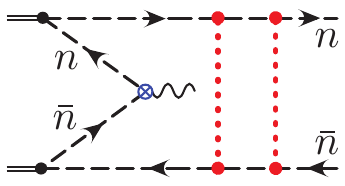}
 }
    	\hspace{-0.3cm} \raisebox{0.9cm}{\large + $\ldots$}
    \\[-5pt]
	\hspace{1.3cm} a)\hspace{4cm} b) 
    \hspace{3.2cm} c) \hspace{3.3cm} d)  \hspace{2.cm}
    \vspace{-0.2cm}
	\caption{\setcaptionskip
		Spectator-specator interactions for the hard scattering correlator 
        in \eq{ends}. The Glauber interaction labeled $G$ indicates the sum of all ladder diagrams including the graph with $0$ Glaubers as indicated.
		} 
	\label{fig:end_SS}
	\setmainskip
	\label{end}
\end{figure}

We  begin by considering the spectator-spectator (SS) interaction diagrams in \fig{end_SS}. Since the hard scattering case with ${\cal M}_\Gamma^{\rm DIS}$ has only a single hadron, these SS contributions only exist for the hard annihilation case with ${\cal M}_\Gamma^{\rm DY}$, where the two participating spectators are created by $\Phi_n$ and $\Phi_\bn$ respectively. In these graphs the hard interaction is indicated by the $\otimes$, and our routing for incoming and outgoing external momentum is shown in \fig{end_SS}b.  For simplicity we take the limit where the mass of the incoming hadrons  is ignored, so that $P^2=\bar P^2=0$.~\footnote{The generalization to the case with $P^2, \bar P^2\ne 0$ is discussed in \eq{end_tree2} below.} This is accomplished by taking $P^\mu = \bn \cdot P\, n^\mu/2$ and
$\bar P= n \cdot \bar P\, \bn^\mu/2$ respectively.
The  tree level result for \fig{end_SS}b is then given by
\begin{align}  \label{eq:end_tree}
{\rm Fig.}~\ref{fig:end_SS}b
  &= -{\cal S}^\gamma \: 
     \frac{ i\, \bn\cdot (p_1\minus P)}{(P-p_1)^2} \:
     \frac{ i\, n\cdot (\bar P-p_2)}{(\bar P-p_2)^2}
  \\
  &= -{\cal S}^\gamma 
  \left[\frac{1}{\vec p_{1\perp}^{\:2}} \frac{1}{\vec p_{2\perp}^{\:2}}\right]
   \left[\frac{\bn \cdot p_1\, \bn \cdot(P\minus p_1)}{\bn \cdot P}
    \frac{n \cdot p_2\, n \cdot(\bar P \minus p_2)}{n \cdot \bar P}\right] 
   \nn\\
  & \equiv  S^\gamma \, E(p_{1\perp},p_{2\perp})
  , \nn
\end{align}
where this defines the function $E$, and we have defined the spinor factor for the outgoing quark-antiquark as 
\begin{align} \label{eq:Sgamma}
 {\cal S}^\gamma = \bar u_n \gamma_\perp^\mu v^*_\bn
  \,.
\end{align}
The $v^*_\bn$ appears here because of our convention for the antiquark spinors, see the discussion near \eq{treennbresult}.
Note that $\bn\cdot p_1>0$, $\bn\cdot (P-p_1)>0$, $n\cdot p_2>0$, and $n\cdot (\bar P-p_2)>0$. To obtain the second line of \eq{end_tree} we used momentum conservation, and the equation of motion to remove the small momentum components, $n\cdot p_1 = \vec p_{1\perp}^{\:2}/\bn\cdot p_1$ and $\bn\cdot p_2 = \vec p_{2\perp}^{\:2}/n\cdot p_2$. The final momentum dependence of the result in \eq{end_tree} is defined as the ``end-function'' $E(p_{1\perp},p_{2\perp})$. We suppress the dependence on the light cone momenta in its arguments since it is the $\perp$-momenta that will play the prominent role for our discussion here. The factor involving light-cone momenta that appears in $E$ will often occur at intermediate steps, so we define
\begin{align} \label{eq:kappa}
\kappa\equiv  \left[\frac{\bn \cdot p_1\, \bn \cdot(P-p_1)}{\bn \cdot P}
\frac{n \cdot p_2\, n \cdot(\bar P-p_2)}{n \cdot \bar P}\right] \,.
\end{align}
In terms of power counting we note that the tree level amplitude scales as $E(p_{1\perp},p_{2\perp})\sim \lambda^{-4}$ just as expected for the scaling of ${\cal M}_\Gamma^{\rm DY}$. 

Next we dress the end $E$ with SS Glauber exchanges as in \fig{end_SS}c,d.  To do this we may utilize the results from \sec{exponentiation} for Glauber exchange in forward scattering diagrams. Here the hard scattering end produces a pair of quarks that are then fed into the forward scattering. In particular, the one-loop hard scattering graph in \fig{end_SS}c is just the tree-level forward scattering graph tied off with an extra loop on the end and{\tiny } the two-loop hard scattering graph in \fig{end_SS}d is  the one-loop box-graph for forward scattering tied off with an extra loop on the end, etc. Due to the extra loop present in hard scattering, the incoming quarks are offshell, with ${\cal O}(\lambda^2)$ nonzero  $\pm$ loop momenta flowing through the forward scattering part of the graph, and unrelated $\perp$-momenta for the two incoming lines. However, as discussed in \sec{exponentiation}, the presence of these modifications from the additional loop do not change the result for the sum of forward scattering ladder graphs.  Thus we can first perform all the forward scattering loop integrals to give $2 G(k_\perp)$, where $G(k_\perp)$ is taken from \eq{Gqperp} setting ${\rm\bf T}^A_1 \otimes {\rm\bf T}^A_2 =  T^A \otimes \bar T^A $. This leaves only the loop-integral with momentum that flows through the end, and corresponds to evaluating \fig{end_SS}a.  The result is
\begin{align} \label{eq:SSendloop}
{\rm Fig.}~\ref{fig:end_SS}a 
  &= {\cal S}^\gamma i^4 \int\!\! \ddslash\!^d k\: \frac{2 G(k_\perp)\ 
     (-1)^2 }
  {[ k^+ \minus \Delta_1\plus i0]
   [ - k^+ \minus \Delta_1'\plus i0]
   [ k^- \minus \bar\Delta_1 \plus i0]
   [ -k^- \minus \bar\Delta_1'\plus i0]   }
  \nn\\
  &= \frac{2(-i)^2}{2} {\cal S}^\gamma \!\! \int\!\! \ddslash\!^{d-2}k_\perp\: 
  \frac{ G(k_\perp) }{ (\Delta_1+\Delta_1')(\bar\Delta_1+\bar\Delta_1') }
  \nn \\
&= -{\cal S}^\gamma \kappa \int\!\! \ddslash\!^{d-2}k_\perp \: 
  \frac{ G(k_\perp) }{ (\vec k_\perp + \vec p_{1\perp})^2\,
   (\vec k_\perp - \vec p_{2\perp})^2}
  \nn\\
&=  {\cal S}^\gamma \int\!\! \ddslash\!^{d-2}k_\perp\: G(k_\perp)\:    
    E(p_{1\perp}+k_\perp,p_{2\perp}-k_\perp) \,.
\end{align}
To obtain the first line, note that the small $k^\pm$ loop momenta do not appear in the numerator of the collinear propagators, so we can group these factors into the denominators, for example
\begin{align}
  \frac{ \bn\cdot p_1}{ \bn\cdot p_1 (k^+ \plus n\cdot p_1) - (\vec k_\perp \plus \vec p_{1\perp})^2 + i0 }   &=  \frac{1}{k^+ - \Delta_1 +i0} \,.
\end{align}
Using momentum conservation and $n\cdot P=\bn\cdot\bar P=0$, and the fact that the incoming hadrons have
vanishing $\perp$-momenta so $(P-p_1)_\perp= -p_{1\perp}$ and $(P-p_2)_\perp= -p_{2\perp}$, the various $k_\perp$ dependent factors in \eq{SSendloop} include
\begin{align} \label{eq:Deltai}
 \Delta_1 &= 
    \frac{(\vec k_\perp+ \vec p_{1\perp})^2}{\bn \cdot p_1} 
    - n\cdot p_1 \,,
 & \Delta_1' &= 
   \frac{(\vec k_\perp+ \vec p_{1\perp})^2}{\bn \cdot (P\minus p_1)} 
    + n\cdot p_1 \,, 
   \\
  \bar\Delta_1' &= 
    \frac{(\vec k_\perp - \vec p_{2\perp})^2}{n\cdot p_2} 
    - \bn\cdot p_2 \,,
 & \bar\Delta_1 &= 
   \frac{(\vec k_\perp - \vec p_{2\perp})^2}{n\cdot (\bar P\minus p_2)} 
    + \bn\cdot p_2 
  \,. \nn
\end{align}
To obtain the second line of \eq{SSendloop} we note that there are no rapidity divergences and hence we simply perform the $k^+$ and $k^-$ integrals by contours. The final lines simply follow from the definitions in \eq{Deltai} and \eq{kappa}. Note that unlike in the forward scattering loop integrals that the final result here depends on the non-vanishing $\Delta_1+\Delta_1'$ and $\Delta_2+\Delta_2'$, so the collinear fermions that appear outside of $G$ here are not eikonal. 

To exhibit the rescattering phase it is convenient to express \eq{SSendloop} in Fourier space. If we hold the photons $q_\perp = -p_{1\perp}-p_{2\perp}$ fixed, then we can consider Fourier transforming in $\Delta p_\perp = (p_{2\perp}-p_{1\perp})/2$, to give
\begin{align}
  {\cal A}_{SS}(\Delta p_\perp,q_\perp) &= {\rm Fig.}~\ref{fig:end_SS}a
   \nn\\
  &= {\cal S}^\gamma \int\!\! \ddslash\!^{d-2}k_\perp\: G(k_\perp)\:    
    E\Big(k_\perp-\Delta p_\perp-\frac{q_\perp}{2}, \Delta p_{\perp}-k_\perp-\frac{q_\perp}{2}\Big) 
   \nn \\
 & \equiv {\cal S}^\gamma \int\!\! \ddslash\!^{d-2}k_\perp\: G(k_\perp)\:    
    E'(\Delta p_\perp-k_\perp, q_\perp ) 
  \nn \\
 &= {\cal S}^\gamma \int\!\! \ddslash\!^{d-2}k_\perp\:
 \int d^{d-2}b_\perp\: e^{-i \vec k_\perp \cdot \vec b_\perp } \ \tilde G(b_\perp)
 \int d^{d-2}b_\perp'\: e^{-i (\Delta \vec p_\perp- \vec k_\perp) \cdot \vec b_\perp' }
    \tilde E'(b_\perp', q_\perp ) 
  \nn \\
& = {\cal S}^\gamma \int d^{d-2}b_\perp\: e^{-i \Delta \vec p_\perp \cdot \vec b_\perp }   \ \tilde E'(b_\perp, q_\perp) \ e^{i\phi(b_\perp)} .
\end{align}
In the third line we have defined a related two argument end function $E'$ which allows us to keep the expressions more compact. From the final result we see that the iterations of the spectator-spectator Glauber potentials produce a final state rescattering phase $\phi(b_\perp)$, where the distance $b_\perp$ is conjugate to the difference of the $\perp$-momenta of the two spectators undergoing the scattering.

It is interesting to ask: under what conditions does this Glauber induced phase cancel?
Considering the modulus squared of the amplitude, the phase cancels as long as we carry out the phase space integral over  $\Delta p_\perp$,
\begin{align} \label{eq:phasecancels}
  & \int\!\! \ddslash\!^{d-2}\Delta p_\perp\: 
   \big|  {\cal A}_{SS}(\Delta p_\perp,q_\perp) \big|^2 
  \nn \\
   &\qquad
  =  | {\cal S}^\gamma |^2 \int\!\! \ddslash\!^{d-2}\Delta p_\perp 
   \int d^{d-2}b_\perp\: d^{d-2}b_\perp'\:  e^{i \Delta \vec p_\perp \cdot (\vec b_\perp'-\vec b_\perp) }   \ \tilde E'(b_\perp, q_\perp) 
   \tilde E^{\prime \dagger}(b_\perp', q_\perp) \ e^{i\phi(b_\perp)-i\phi(b_\perp')}
  \nn \\
  &\qquad
  =  | {\cal S}^\gamma |^2 
   \int d^{d-2}b_\perp\:   \big| \tilde E'(b_\perp, q_\perp) \big|^2
  \nn \\
  &\qquad
  =  | {\cal S}^\gamma |^2 
   \int \ddslash\!^{d-2}\Delta p_\perp\:   
   \big| E'(\Delta p_\perp, q_\perp) \big|^2 \,,
\end{align}
where the final result is just the integral over the squared tree level result in \eq{end_tree}.
Thus the Glauber exchange for these SS graphs cancel as long as the limits of integration for $\Delta p_\perp$ are taken to infinity in the effective theory. As long as the measurement made on the final state particles takes place at a perturbative scale $Q^*$ with $\sqrt{t}\sim \Delta p_\perp \ll Q^*$, then the measurement does not see the spectator particles at leading power, and the $\Delta p_\perp$ integration is unrestricted  for the leading power analysis. This need to integrate over $\Delta p_\perp$ also appears in the CSS Drell-Yan factorization proof~\cite{Collins:1988ig,Aybat:2008ct}. Although the result in \eq{phasecancels} does exhibit the cancellation of final state interactions, taken alone it is far from a proof of factorization, even in the abelian case. What this resummation does do however is to highlight the importance of the $\Delta p_\perp$ integration and illuminate the semi-classical nature of the physics.

If on the other hand, we would like to address the case with a single $t\sim Q^{*2}\gg \Lambda_{\rm QCD}^2$, then as mentioned above, we should replace the $\rho_{n,\bn}$ by a perturbative gluon. In this case $p_{i\perp}, \Delta p_\perp \sim Q^*$ and we have a Glauber loop momentum $k_\perp\sim Q^*$, and the calculations above need to be redone with $P^2,\bar P^2\ne 0$ and $P_\perp, \bar P_\perp\ne 0$. For this case one still obtains \eq{SSendloop}, with the same $k_\perp$ convolution with a $G(k_\perp)$, but where the $E$ function is now given by
\begin{align} \label{eq:end_tree2}
  E(p_{1\perp},p_{2\perp}) &= \dfrac{-1}{
   \Bigg[ \dfrac{\vec p_{1\perp}^{\:2}}{p_1^-} 
   + \dfrac{\big(\vec P_\perp- \vec p_{1\perp}\big)^2}{\big(P^--p_1^-\big)} 
   - P^+  \Bigg]
   \Bigg[ \dfrac{\vec p_{2\perp}^{\:2}}{p_2^+} 
   + \dfrac{\big(\vec {\bar P}_\perp- \vec p_{2\perp}\big)^2}
     {\big(\bar P^+-p_2^+\big)}  - \bar P^-  \Bigg]
   }
  \,,
\end{align}
and we have a different prefactor $S^\gamma$. 
Assuming that the measurement restricts the $\Delta p_\perp\sim Q^*$, then these spectator-spectator scattering Glauber exchange diagrams will not cancel out.  In this case the measurement spoils the factorization of the perturbative $n$- and $\bn$-collinear initial state beam radiation, but only starting at ${\cal O}(\alpha_s^4)$. To get a nonzero cross section level result we need one perturbative Glauber exchange from $G(k_\perp)$ on each side of the cut, giving an $\alpha_s^2$, and two splittings (one for each collinear direction) on each side of the cut, giving another $\alpha_s^2$. (For the case with a Glauber exchange on only one side of the cut, the amplitude level contribution is nonzero, but this contribution vanishes upon  squaring to obtain the cross section because it is purely imaginary.)  For event shape observable such as beam thrust~\cite{Stewart:2009yx,Stewart:2010qs,Stewart:2010pd}, transverse thrust~\cite{Banfi:2004nk,Banfi:2010xy}, and $E_T$~\cite{Papaefstathiou:2010bw,Tackmann:2012bt,Grazzini:2014uha} the importance of this type of diagram for the violation of factorization was first discussed in~\cite{Gaunt:2014ska}.  This diagram has also been computed numerically for a double spin asymmetry beam thrust observable, demonstrating explicitly that it is nonzero~\cite{Zeng:2015iba}. Using the formalism developed here it is straightforward to obtain an essentially analytic result for this type of calculation, and it is clear that it applies to any observable where $\Delta p_\perp$ is constrained, including for example beam thrust without the spin asymmetries.\footnote{The cancellation of Glauber gluons was discussed in the original beam thrust paper~\cite{Stewart:2009yx}, and divided into two categories, those with $t\sim \Lambda_{\rm QCD}^2$ and those with $t\sim Q^{*2}$ as we do here. The argument there for the cancellation of leading power $t\sim \Lambda_{\rm QCD}^2$ Glaubers is correct and agrees with the results here and from Ref.~\cite{Aybat:2008ct}, but the discussion of the lack of leading power $t\sim Q^{*2}$ Glaubers was too naive. Although beam thrust is a \SCETa observable, and in this section we are considering calculations in \SCETb, the results here are still applicable. In particular, with Glauber exchange the \SCETa theory consists of \SCETb plus ultrasoft modes, and somewhat different subtractions which, although they change the result for the Glauber phase (see \eqs{glauber_loop1}{glauber_loop1subt}), do not change the conclusions drawn here.  }   Note that this ${\cal O}(\alpha_s^4)$ perturbative effect is beyond the order considered in beam thrust resummation~\cite{Berger:2010xi} or other Higgs jet veto resummed calculations~\cite{Banfi:2012jm,Becher:2013xia,Stewart:2013faa}, or in transverse thrust resummation~\cite{Becher:2015lmy}. This effect alone does not spoil the use of perturbative factorization with beam functions to carry out double logarithmic resummation, until one considers resummation where this non-logarithmic $\alpha_s^4$ correction enters, which is at the next-to-next-to-next-to-next-to-leading logarithmic order (N$^4$LL). This graph alone also does not explain the sensitivity to underlying event observed for beam thrust or transverse thrust, in agreement with~\cite{Gaunt:2014ska}. We leave for the future the exploration of other Glauber induced factorization violating effects using our formalism.

Notice that for these spectator-spectator interactions, as opposed to the active-active case previously discussed in \sec{hardmatching}, that there are no analogous diagrams in \SCETb where the Glauber gluons are replaced by soft gluons. If one of the Glauber gluons here became soft then it would knock multiple fermion lines in the end loop integral offshell (not yielding just leading power Wilson lines), and hence such interactions are power suppressed. There are also no diagrams where a spectator-spectator Glauber exchange is replaced by and $n$-collinear or $\bn$-collinear gluon, again these are power suppressed.  Thus once we consider matrix elements involving spectators lines the Glauber mode is necessary to reproduce the full theory \SCETb result.

\subsection{Active-Spectator and the Collinear Overlap}  
\label{sec:asfactorization}

Next we consider Glauber exchange for the lowest order active-spectator type diagrams.  We will show that the Glaubers here can be absorbed into the direction of collinear Wilson lines, since there is an exact overlap between these Glauber diagrams and the Glauber 0-bin subtractions of graphs involving collinear Wilson lines from the hard scattering vertex. This Glauber--collinear Wilson line correspondence is analogous to the Glauber correspondence with soft Wilson lines in the hard scattering diagrams considered in \sec{properties} (and reconsidered below in \sec{aafactorization} as active-active diagrams).

\begin{figure}[t!]
	%
	%
\begin{center}
	\hspace{-3.3cm} a) \hspace{3.4cm} b) \hspace{3.5cm} c) \hspace{3.5cm} d) \hspace{2.9cm}\\
	\includegraphics[width=0.2\columnwidth]{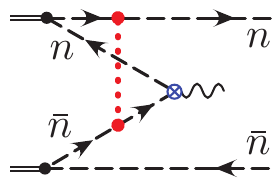} 
	\hspace{0.5cm}
	\includegraphics[width=0.21\columnwidth]{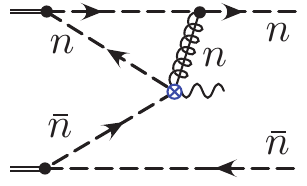}
	\hspace{0.5cm}
	\includegraphics[width=0.2\columnwidth]{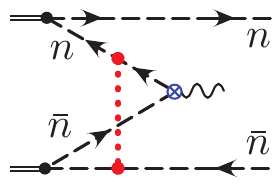} 
	\hspace{0.5cm}
	\includegraphics[width=0.21\columnwidth]{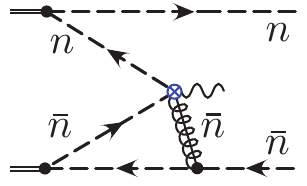}
	\\
\end{center}
\vspace{-0.5cm}
	\caption{\setcaptionskip
		One-loop graphs with Active-Spectator interactions related to the Glauber-Collinear overlap for the hard annihilation Drell-Yan correlator in \eq{ends}.  a) and c) involve Glauber exchange, while b) and d) are the corresponding graphs with Wilson line interactions involving a collinear gluon.  }	
	\label{fig:end_AS}
	\setmainskip
\end{figure}

We start by considering hard production with ${\cal M}_\Gamma^{\rm DY}$, that is, two incoming hadrons. The single Glauber graphs are shown by the diagrams in \fig{end_AS}a,c. Unlike the single Glauber exchange graph with a spectator-spectator interaction, the results here need the rapidity regulator to be well defined. The active-spectator Glauber exchange graph in \fig{end_AS}a is given by
\begin{align} \label{eq:GAS}
 {\rm Fig.}\ref{fig:end_AS}a 
   &= 2 S^\gamma\, \frac{n\mcdot(\bar P\minus p_2)}{(\bar P\minus p_2)^2}
  \!\int\!\! \ddslash\!^{d}k  \frac{G^0(k_\perp) |2 k^z|^{-\eta} \nu^\eta}
  {[k^- \minus \bar\Delta_1\plus i0][-k^+\minus\Delta_1'\plus i0] [k^+ \minus \Delta_1\plus i0]} \,,
\end{align}
where $S^\gamma$ is given in \eq{Sgamma} and a single Glauber exchange yields $2 G^0(k_\perp)$, where $G^0$ is given by \eq{G0defn}, and for the $qq$ channel relevant here is equal to
\begin{align}  \label{eq:G0qq}
 G^0(k_\perp) &=  \frac{-ig^2}{\vec k_\perp^{\:2}+m^2 }\:  T^A\otimes T^A \,.
\end{align}
The other $k_\perp$ dependent factors $\Delta_1$, $\Delta_1'$, $\bar\Delta_1$ are given above in \eq{Deltai}.  Performing the $k^0$ integration by contours gives
\begin{align}  \label{eq:GAS_result}
  {\rm Fig.}\ref{fig:end_AS}a
 &= 2i\,S^\gamma\frac{n\mcdot p_2\, n\mcdot(\bar P\minus p_2)}
   {n\mcdot \bar P \, \vec p_{2\perp}^{\:2}}
  \int\!\! \ddslash\! k^z \ddslash\!^{d'}\!k_\perp 
  \frac{G^0(k_\perp) |2 k^z|^{-\eta} \nu^\eta}
  {[2k^z \minus\Delta_1^\prime\minus \bar\Delta_1\plus i0][-\Delta_1\minus \Delta_1'\plus i0]}
\nn\\
 &= - \frac{1}{2}\,S^\gamma \frac{n\mcdot p_2\, n\mcdot(\bar P\minus p_2)}
  {n\mcdot \bar P \, \vec p_{2\perp}^{\:2}} 
  \int\!\! \ddslash\!^{d'}\!k_\perp 
  \frac{G^0(k_\perp) }
  {\Delta_1+\Delta_1'}
\nn\\
 &= - \frac{1}{2}\,S^\gamma \frac{n\mcdot p_2\, n\mcdot(\bar P\minus p_2)}
   {n\mcdot \bar P \, \vec p_{2\perp}^{\:2}} 
  \frac{\bn\mcdot p_1\,\bn\mcdot (P\minus p_1)}{\bn\mcdot P}
  \int\!\! \ddslash\!^{d'}\!k_\perp 
  \frac{G^0(k_\perp) }
  {(\vec k_\perp+\vec p_{1\perp})^2}
\nn\\
 &= \frac{1}{2}\,S^\gamma 
  \int\!\! \ddslash\!^{d'}\!k_\perp 
  \: G^0(k_\perp) E(p_{1\perp}+k_\perp,p_{2\perp})
  \,,
\end{align}
where $d'=d-2$. To obtain the second line, the $k^z$ integral was performed using \eq{kzint}. The final result here is written in terms of the end function defined in \eq{end_tree}. 

Now consider the collinear loop graph in \fig{end_AS}b.  Here the gluon entering the hard vertex has momentum $k$ and is generated by the Wilson line $W_n[\bn\cdot A_n]$ from the current in \eq{current}. We take it to be $W_n(-\infty,0)$ since in this case it is generated in the QCD to \SCETb matching calculation from integrating out offshell propagators along the incoming quark line plus non-abelian graphs. We have
\begin{align} \label{eq:CnAS}
 \tilde C_n({\rm Fig.}\ref{fig:end_AS}b)
   &= S^\gamma \, \frac{n\mcdot(\bar P\minus p_2)}{(\bar P\minus p_2)^2}
  \!\int\!\! \ddslash\!^{d}k \:
   \frac{(-2ig^2C_F)}{(k^2\minus m^2\plus i0)}\frac{ \bn\cdot(k\minus P\plus p_1)\, \bn\cdot(k\plus p_1)\,|\bn\cdot k|^{-\eta} \nu^\eta}
  {[k^- +i0][(k\minus P\plus p_1)^2+i0] [(k\plus p_1)^2+i0]}.
\end{align}
From \eq{0bins} this collinear loop graph potentially has both soft and Glauber subtractions.  For the soft subtraction we find that the soft limit $k^\mu \sim \lambda$ of \eq{CnAS} gives
\begin{align}
 C_n^{(S)}({\rm Fig.}\ref{fig:end_AS}b)
   &= S^\gamma \, \frac{n\mcdot(\bar P\minus p_2)}{(\bar P\minus p_2)^2}
  \!\int\!\! \ddslash\!^{d}k \:
  \frac{(-2ig^2C_F)}{(k^2\minus m^2\plus i0)}\frac{ (-1) |\bn\cdot k|^{-\eta} \nu^\eta}
  {[k^- +i0][-k^+ +i0] [k^+ +i0]}
 \,,
\end{align}
which scales as $\sim \lambda^4/\lambda^7=\lambda^{-3}$ and hence is dropped since it is power suppressed relative to the leading amplitude $E\sim {\cal O}(\lambda^{-4})$ (the overlap subtraction $C_n^{(S)(G)}$ vanishes for the same reason). The reason for the vanishing of this soft subtraction is clear once we recall that the soft gluons cannot couple to collinear lines without knocking them offshell, and hence are only leading power for the active attachments which generate soft Wilson lines. Thus there is no leading power soft diagram that is analogous to  the active-spectator interaction in \fig{end_AS}b. 

On the other hand, there is a leading power Glauber subtraction, given by taking the $k^\pm \ll \vec k_\perp$ limit of \eq{CnAS},
\begin{align}  \label{eq:CnGAS}
  C_n^{(G)}({\rm Fig.}\ref{fig:end_AS}b)
   &= 2 S^\gamma\, \frac{n\mcdot(\bar P\minus p_2)}{(\bar P\minus p_2)^2}
  \!\int\!\! \ddslash\!^{d}k   \frac{ G^0(k_\perp)\, |\bn\cdot k|^{-\eta} \nu^\eta}
  {[k^- +i0][-k^+ -\Delta_1' +i0] [k^+-\Delta_1+i0]}
 \,.
\end{align}
Comparing this integral with the active-spectator Glauber result in \eq{GAS} we see that the two are the same up to the presence of different rapidity regulators and the absence of $\bar\Delta_1(k_\perp)$ in \eq{CnGAS}. Decomposing $d^dk = (1/2) dk^+ dk^- d^{d'}k_\perp$, performing the $k^+$ contour integral, and then using $\int dk^- |k^-|^{-\eta}/(k^-+i0)= -i/2 +{\cal O}(\eta)$ gives
\begin{align}  \label{eq:CnGAS_result}
  C_n^{(G)}({\rm Fig.}\ref{fig:end_AS}b)
   &= -\frac{1}{2} S^\gamma\, \frac{n\mcdot p_2\, n\mcdot(\bar P\minus p_2)}
   {n\mcdot \bar P \, \vec p_{2\perp}^{\:2}} 
  \!\int\!\! \ddslash\!^{d'}\!k_\perp   \frac{ G^0(k_\perp)}
  {\Delta_1 +\Delta_1^\prime } 
  \nn \\
 &= \frac{1}{2}\,S^\gamma 
  \int\!\! \ddslash\!^{d'}\!k_\perp 
  \: G^0(k_\perp) E(p_{1\perp}+k_\perp,p_{2\perp})
 \,.
\end{align}
This result for the subtraction on the collinear graph is the same as the Glauber graph result in \eq{GAS_result}, despite the lack of $\Delta_2$ and difference in rapidity regulators,
\begin{align}  \label{eq:CGcorr}
  C_n^{(G)}({\rm Fig.}\ref{fig:end_AS}b) = G({\rm Fig.}\ref{fig:end_AS}a) \,.
\end{align}
This equality is similar to the result obtained in our analysis of soft and Glauber exchange for active-active lines in \sec{hardmatching}. In particular, this type of Glauber exchange can be absorbed into the collinear Wilson lines, in an analogous manner to the way we discussed absorbing certain Glauber exchanges into soft Wilson lines in \sec{hardmatching}. The fact that these active-spectator Glauber exchanges can be absorbed is consistent with the contour deformation picture in CSS, where the combined collinear+Glauber loop integral can be deformed away from the Glauber region for these types of diagrams~\cite{Collins:1988ig,Collins:2011zzd}.

In SCET the collinear subtraction result is sensitive to the direction of the Wilson line $W_n$ which is encoded by the sign in the propagator $[k^-+i0]$, and the Glauber subtraction $C_n^{(G)}$ precisely removes this dependence. In order for the correspondence in \eq{CGcorr} to be true it is important for $n$-$\bn$ annihilation that the $W_n=W_n(-\infty,0)$ Wilson line in the $J_{\Gamma}$ current is taken to extend from $(-\infty,0)$ in the Wilson line integration variable $n\cdot x$. If instead we had taken this Wilson line to extend from $(0,\infty)$ then we would replace $[k^-+i0]\to [k^- -i0]$ in \eqs{CnAS}{CnGAS}. Since $\int dk^- |k^-|^{-\eta}/(k^--i0)= +i/2 +{\cal O}(\eta)$, this flips the overall sign of the final result for $C_n^{(G)}$ in \eq{CnGAS_result}. In this case the Glauber subtraction on the collinear graph would not be equal to the Glauber graph itself, and we could not simply absorb the Glauber graph into the collinear Wilson line. (The direction dependence is still canceled in $C_n-C_n^{(G)}$, and only encoded by $G$ in this case.) 

For the graphs in \fig{end_AS}c,d the results can be obtained by swapping $n\leftrightarrow \bn$, $p_1\leftrightarrow p_2$, $n\cdot \bar P \to \bn\cdot P$, and $T^A\otimes T^A \to \bar T^A\otimes \bar T^A$ in the analysis above. Therefore we find
\begin{align}  \label{eq:CGcorr2}
  C_\bn^{(G)}({\rm Fig.}\ref{fig:end_AS}d) = G({\rm Fig.}\ref{fig:end_AS}c) \,.
\end{align}
Here the $W_\bn^\dagger=W_\bn^\dagger(-\infty,0)$ Wilson line in the $J_{\Gamma}$ current has to extend from $(-\infty,0)$ in order for the correspondence in \eq{CGcorr2} to be true. For easy reference we record the Feynman rules for collinear Wilson lines in various directions in \app{Wdirection}.  We see that the correspondence between Glauber subtractions on the collinear graphs, and the Glauber graphs themselves is sensitive to the direction of each of the $W_n$ and $W_\bn^\dagger$ Wilson lines in the hard current $J_\Gamma$. Again, if the Wilson line in the hard scattering current were taken to extend out to $+\infty$, then the two amplitudes in \eq{CGcorr2} would differ by a sign.

\begin{figure}[t!]
	%
	%
\begin{center}
	\hspace{-3cm} a) \hspace{4.cm} b) \hspace{4.cm} c) \hspace{6.5cm} 
    \\[-5pt]
	\includegraphics[width=0.23\columnwidth]{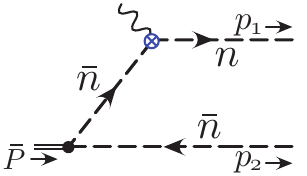} 
      \hspace{0.5cm}
	\includegraphics[width=0.23\columnwidth]{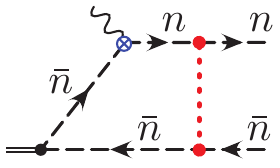} 
      \hspace{0.5cm}
	\includegraphics[width=0.23\columnwidth]{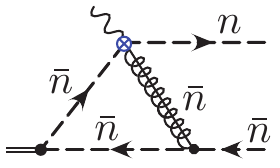}
    \\[-5pt]
\end{center}
\vspace{-0.5cm}
	\caption{\setcaptionskip
	One-loop graphs with Active-Spectator interactions related to the Glauber-Collinear overlap for the DIS hard scattering correlator in \eq{ends}. a) is the lowest order end, b) involves Glauber exchange, c) is the corresponding graphs with a Wilson line interaction involving a collinear gluon.  }	
    \label{fig:end_ASdis}
	\setmainskip
\end{figure}

Next we consider active-spectator scattering for the ${\cal M}_\Gamma^{\rm DIS}$ amplitude of \eq{ends}, which has active quarks in the initial and final states, and only $\bn$-collinear spectators from the one incoming hadron. The relevant diagrams are shown in \fig{end_ASdis}. We let the incoming momentum of the hadron be $\bar P= n \cdot \bar P\, \bn^\mu/2$ and label the outgoing quark momenta as $p_1$ and $p_2$ as shown.  At tree level the correlator is
\begin{align}  \label{eq:Edis}
 {\rm Fig.}\ref{fig:end_ASdis}a 
 &= {\cal S}^\gamma \: 
     \frac{ i\, n\cdot (\bar P-p_2)}{(\bar P-p_2)^2}
 = -i {\cal S}^\gamma 
   \frac{1}{\vec p_{2\perp}^{\:2}}\:
    \frac{n \cdot p_2\, n \cdot(\bar P \minus p_2)}{n \cdot \bar P} 
   \nn\\
  & \equiv  S^\gamma \, E(p_{2\perp})
  , 
\end{align}
which defines the end factor $E(p_{2\perp})$, and again we suppress the dependence on $n\cdot p_2$ in its arguments. We distinguish this function from that in \eq{end_tree} by its dependence on only a single $\perp$-variable. Note that $E(p_{2\perp})\sim \lambda^{-2}$ just as expected for the scaling of ${\cal M}_\Gamma^{\rm DIS}$.

In the hard scattering case, for the single Glauber exchange diagram we have
\begin{align} \label{eq:GASdis}
 {\rm Fig.}\ref{fig:end_ASdis}b 
   &= -2i\, S^\gamma
  \!\int\!\! \ddslash\!^{d}k  \frac{G^0(k_\perp) |2 k^z|^{-\eta} \nu^\eta}
  {[k^+ \minus \Delta_1\plus i0][k^- \minus \bar\Delta_1\plus i0] [-k^- \minus \bar\Delta_1^\prime \plus i0]} 
\nn\\
 &= -2\,S^\gamma
  \int\!\! \ddslash\! k^z \ddslash\!^{d'}\!k_\perp 
  \frac{G^0(k_\perp) |2 k^z|^{-\eta} \nu^\eta}
  {[-2k^z \minus \Delta_1\minus \bar\Delta_1'\plus i0][-\bar\Delta_1\minus \bar\Delta_1']}
  \nn\\
 &= -\frac{i}{2}\,S^\gamma \frac{n\mcdot p_2\, n\mcdot(\bar P\minus p_2)}
  {n\mcdot \bar P \, \vec p_{2\perp}^{\:2}} 
  \int\!\! \ddslash\!^{d'}\!k_\perp 
  \frac{G^0(k_\perp) }
  {(\vec k_\perp -\vec p_{2\perp})^2}
\nn\\
 &= \frac{1}{2}\,S^\gamma 
  \int\!\! \ddslash\!^{d'}\!k_\perp 
  \: G^0(k_\perp) E(p_{2\perp}-k_\perp)
  \,,
\end{align}
where $d'=d-2$. Here using \eq{G0defn} for $q\bar q$ scattering we have
\begin{align}  \label{eq:G0qqb}
 G^0(k_\perp) &=  \frac{-ig^2}{\vec k_\perp^{\:2}+m^2 }\:   T^A\otimes \bar T^A \,.
\end{align}
The result in \eq{GASdis} is similar to the result for the hard annihilation case, just with a different color factor and the opposite overall sign.

Now consider the collinear loop graph in  \fig{end_ASdis}c.  Here the gluon exiting the hard vertex has momentum $k$ and is generated by the Wilson line $W_\bn^\dagger[n\cdot A_\bn]$ from the current in \eq{current}. We take it to be $W_\bn^\dagger(0,\infty)$ since here it is generated in the full theory to \SCETb matching calculation from integrating out offshell fluctuations for the outgoing quark line plus non-abelian graphs. For this $\bn$-collinear loop we then have
\begin{align} \label{eq:CnbASdis}
 \tilde C_\bn({\rm Fig.}\ref{fig:end_ASdis}c)
   &= i S^\gamma
  \!\int\!\! \ddslash\!^{d}k  \frac{(2ig^2 C_F)}{(k^2\minus m^2\plus i0)}\frac{ n\cdot(k\minus  p_2)\, n\cdot(k\plus \bar P\minus p_2)\,|n\cdot k|^{-\eta} \nu^\eta}
  {[k^+ +i0][(k\minus p_2)^2+i0] [(k\plus \bar P-p_2)^2+i0]}.
\end{align}
Once again, for this active-spectator loop graph the soft subtraction is zero, since it is power suppressed. There is a nonzero Glauber subtraction, which can be determined by taking the $k^\pm \ll \vec k_\perp$ limit of \eq{CnbASdis},
\begin{align}  \label{eq:CnbGASdis}
  C_\bn^{(G)}({\rm Fig.}\ref{fig:end_ASdis}c)
   &=  -2i S^\gamma\, 
  \!\int\!\! \ddslash\!^{d}k   \frac{ G^0(k_\perp)\, |n\cdot k|^{-\eta} \nu^\eta}
  {[k^+ +i0][k^- -\bar\Delta_1 +i0] [-k^--\bar\Delta_1'+i0]}
   \nn\\
   &=  S^\gamma\,
  \!\int\!\! \ddslash\!k^+ \ddslash\!^{d'}\!k_\perp  
  \frac{ G^0(k_\perp)\, |n\cdot k|^{-\eta} \nu^\eta}
  {(k^+\plus i0)(\bar\Delta_1\plus \bar\Delta_1'-i0) } 
  \nn \\
 &= -\frac{i}{2}\,S^\gamma \frac{n\mcdot p_2\, n\mcdot(\bar P\minus p_2)}
  {n\mcdot \bar P \, \vec p_{2\perp}^{\:2}}
   \!\int\!\!  \ddslash\!^{d'}\!k_\perp  
  \frac{ G^0(k_\perp)}
  {(\vec k_\perp - \vec p_{2\perp})^2 } 
 \nn\\
 &=
 \frac{1}{2} S_\gamma  \int\!\! \ddslash\!^{d'}\!k_\perp 
  \: G^0(k_\perp) E(p_{2\perp}-k_\perp)
 \,.
\end{align}
This result for the subtraction on the collinear graph is the same as the Glauber graph result in \eq{GASdis}, despite the lack of $\Delta_1$ in the $k^+$ propagator, and the difference in rapidity regulators, so
\begin{align}  \label{eq:CGcorr2dis}
  C_\bn^{(G)}({\rm Fig.}\ref{fig:end_ASdis}c) = G({\rm Fig.}\ref{fig:end_ASdis}b) \,.
\end{align}
Once again this equality only works out with the proper direction for the Wilson line in the $J_\Gamma$ current, which here is $W_\bn^\dagger(0,\infty)$.

If we consider the iteration of active-spectator Glauber exchanges we can also show that the result yields a phase, similar to our spectator-spectator analysis. A key difference is that for the active-spectator graphs we required the $\eta$-regulator already for the single-exchange graph. Therefore the ladder sum cannot be carried out independent from the loop involving the hard scattering vertex. We carry out this calculation in detail in \app{ASexp}, finding for the annihilation that the sum of graphs gives
\begin{align} \label{eq:ASsum}
& \raisebox{-1cm}{ 	
\includegraphics[width=0.18\columnwidth]{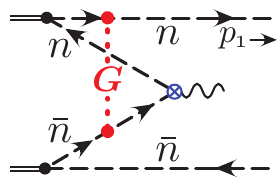}
  }
 = S^\gamma
   \int\!\! d^{d'}b_\perp  e^{i\vec p_{1\perp}\cdot \vec b_\perp}\:
   e^{i\phi(b_\perp)/2} \:\tilde E(-b_\perp,p_{2\perp}) 
\,.
\end{align}
Thus the phase for this sum of active-spectator exchanges is $\phi(b_\perp)/2$, where $b_\perp$ is the transverse distance between the upper spectator quark and the hard annihilation vertex (which is at the position $0_\perp$). Similarly for active-spectator exchanges between the antiquark entering the hard vertex, and antiquark spectator, we obtain an analogous phase $\phi(b_\perp)/2$, where $b_\perp$ is now the transverse distance between the lower spectator antiquark and the hard annihilation vertex.

We can also consider the iteration of active-spectator Glauber exchanges for the DIS-like hard scattering case. From the calculations done in \app{ASexp} we find that the sum of graphs gives
\begin{align} \label{eq:ASdissum}
& \raisebox{-0.9cm}{ 
	\includegraphics[width=0.18\columnwidth]{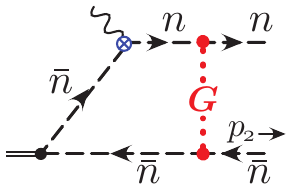}
  }
 = S^\gamma
   \int\!\! d^{d'}b_\perp  e^{i\vec p_{2\perp}\cdot \vec b_\perp}\:
   e^{i\phi(b_\perp)/2} \:\tilde E(-b_\perp) 
\,.
\end{align}
Thus the phase for the active-spectator exchanges here is $\phi(b_\perp)/2$, where $b_\perp$ is the distance between the hard vertex (at $0_\perp$) and the spectator quark.  Due to the factors of $1/2$ these active-spectator phases differs from what we found for spectator-spectator scattering.

Although the active-spectator Glauber exchanges may be absorbed into specifying directions for collinear Wilson lines in the hard scattering operators, we may instead wish to leave them separate and then consider their cancellation. If these Glauber exchanges do cancel, then it enables the factorization to be insensitive to the directions of collinear Wilson lines.  Just as we did for spectator-spectator scattering, we can consider the modulus squared of the active-spectator amplitude alone. For the result in \eq{ASsum}, this cancellation requires integration over $p_{1\perp}$,
\begin{align}
  & \int\!\! \ddslash\!^{d-2} p_{1\perp}\: 
   \big|  {\cal A}_{AS}(p_{1\perp},p_{2\perp}) \big|^2 
  \nn  \\
   &\qquad
  =  | {\cal S}^\gamma |^2 \int\!\! \ddslash\!^{d-2} p_{1\perp} 
   \int d^{d-2}b_\perp\: d^{d-2}b_\perp'\:  e^{i \vec p_{1\perp} \cdot (\vec b_\perp-\vec b_\perp') }   \ \tilde E(b_\perp, p_{2\perp}) 
   \tilde E^{\prime \dagger}(b_\perp', p_{2\perp}) \ e^{\frac{i}{2}[\phi(b_\perp)-\phi(b_\perp')]}
  \nn \\
  &\qquad
  =  | {\cal S}^\gamma |^2 
   \int d^{d-2}b_\perp\:   \big| \tilde E(b_\perp, p_{2\perp}) \big|^2
  \nn \\
  &\qquad
  =  | {\cal S}^\gamma |^2 
   \int \ddslash\!^{d-2} p_{1\perp}\:   
   \big| E( p_{1\perp}, p_{2\perp}) \big|^2 \,,
\end{align}
where the final result is just the integral over the squared tree level result in \eq{end_tree}. If we consider the square of the other type of active-spectator graph (again alone), then we would need to integrate over $p_{2\perp}$. Again, as long as the measurement made on the final state particles takes place at a perturbative scale $Q^*$ with $p_{1\perp}\ll Q^*$ and $p_{2\perp}\ll Q^*$, then the $p_{i\perp}$ phase space integrations will not be constrained at leading power, and one will freely integrate over these variable.  Note that these cancellations require integration over more $\perp$-momenta variables than the spectator-spectator scattering.  This shows that the directions of collinear Wilson lines are important when considering $p_T$-factorization, in agreement with~\cite{Collins:2011zzd}. 
The directions of our collinear Wilson lines for the Drell-Yan-like and DIS-like cases, $W_\bn^\dagger(-\infty,0)$ and $W_\bn^\dagger(0,\infty)$ respectively, also agree with those of Collins~\cite{Collins:2011zzd}.  It would be interesting to compare the SCET subtraction formalism with the subtractions utilized in the CSS approach, such as those in transverse momentum dependent PDFs~\cite{Collins:2004nx,Collins:2011zzd}.  For processes where collinear factorization is valid (i.e.~only integrated PDFs appear), the Glauber contributions cancel and the infinite collinear Wilson lines combine to lines of a finite length which are insensitive to the $\pm \infty$ appearing at intermediate steps. 

Again our analysis in this section is merely indicative of the necessary elements for a proof of factorization, but additional contributions must be considered for a full proof of factorization using our framework.

\subsection{Active-Active and the Soft Overlap}
\label{sec:aafactorization}

Finally we will consider Glauber interactions between two partons that participate in the hard scattering, namely active-active terms. In  Secs.~\ref{sec:hardmatching},~\ref{sec:softemission}, and~\ref{sec:higherorder} we showed that in hard scattering graphs without spectators, such Glauber interactions give the same contributions as the Glauber  zero-bin subtractions of soft Wilson line graphs. The Glauber exchange could therefore be absorbed into these soft graphs as long as the correct directions for the soft Wilson lines are employed. In this section we will demonstrate that all the results and conclusions about active-active Glauber interactions from those sections carry over to the case when we include the interpolating fields for the incoming hadrons. 

\begin{figure}[t!]
	%
	%
\begin{center}
	\includegraphics[width=0.23\columnwidth]{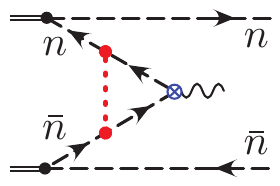} 
           \hspace{0.7cm}
	\includegraphics[width=0.23\columnwidth]{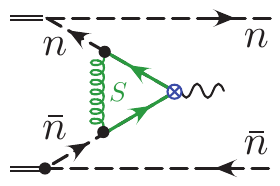} 
    \\[-5pt]
	\hspace{2.2cm} a) \hspace{4.cm} b) \hspace{2.3cm}
    \vspace{-0.2cm}
\end{center}
\vspace{-0.5cm}
	\caption{\setcaptionskip
	a) Active-Active interaction for the hard scattering correlator in \eq{ends}. b) Corresponding graph with two Wilson line interactions involving a soft gluon.  }
	\setmainskip
	\label{fig:endAA}
\end{figure}

The general reason for this can be discussed by looking at the example given in \fig{endAA}. In any purely active-active loop graph with spectators present, the hadron interpolating fields are always external to the loops. From the $n$- and $\bn$-collinear propagators that are outside of the loop, we immediately get the same tree-level end factor $E(p_{1\perp},p_{2\perp})$ as in \eq{end_tree}. The only possible changes to the calculations done in Secs.~ \ref{sec:hardmatching}, \ref{sec:softemission}, and \ref{sec:higherorder} are due to the fact that the active collinear propagators entering the loops are now offshell. This does not affect any soft propagator from a Wilson line (solid green in \fig{endAA}), since here only the soft gluon loop momentum appears. This is immediate from the SCET Feynman rules, and is also clear from expanding a full-theory propagator, since $(p_n+p_s)^2 = \bn\cdot p_n\, n\cdot p_s + \ldots$, where the displayed leading ${\cal O}(\lambda)$ term gives precisely the eikonal propagator of the soft Wilson line, and the offshellness of the external collinear propagator only enters at ${\cal O}(\lambda^2)$. Thus, the only possible effect on the active-active loop graphs could be to modify the collinear propagators appearing in loop integrals with Glauber or collinear momentum scaling. For Glauber loops like \fig{endAA}a the nonzero offshellness of external lines will change the formula for the $\Delta_i(k_\perp)$ terms that appear from collinear propagators with Glauber loop momenta running through them. However, for active-active Glauber loops these $\Delta_i(k_\perp)$s all drop out when we expand to ${\cal O}(\eta^0)$. This fact was a key ingredient in making the correspondence between Glauber contributions and the Glauber 0-bin subtractions from soft Wilson line graphs. Here it suffices to ensure that this correspondence remains true even when the external collinear lines are offshell. 

As an explicit example, for \fig{endAA}a we have
\begin{align} \label{eq:AAG}
{\rm Fig.}~\ref{fig:endAA}a 
&= -2 S^\gamma E(p_{1\perp},p_{2\perp}) \int\!\! \ddslash\!^{d}k\: 
 \frac{  G^0(k_\perp)\:|2 k^z|^{-\eta} \nu^\eta }{[ -k^+ \minus \Delta_1'\plus i0][k^- \minus \bar\Delta_1 \plus i0]}    
\nn\\
&= 2i\, S^\gamma  E(p_{1\perp},p_{2\perp}) 
  \int\!\! dk^z\,\ddslash\!^{d-2}k_\perp 
  \frac{G^0(k_\perp)\:|2 k^z|^{-\eta} \nu^\eta }{(2k^z\minus \Delta_1' \minus \bar\Delta_1 \plus i0)}    
  \nn \\
&=  \frac{1}{2}\, S^\gamma  E(p_{1\perp},p_{2\perp})
   \int \!\! \ddslash\!^{d-2}k_\perp\: G^0(k_\perp) 
    + {\cal O}(\eta) 
  \nn\\
&=  - E(p_{1\perp},p_{2\perp})  \frac{i}2 {\bar T^A\otimes T^A}  \alpha_s \Big( \frac{1}{\epsilon} + \ln\frac{\mu^2}{m^2} \Big) \ S^\gamma
  \,,
\end{align}
where after using momentum conservation $\Delta_1'$ and $\bar\Delta_1$ are given in \eq{Deltai}, and the $k^z$ integral was performed using \eq{kzint}. We also used the fact that up to the spinor factors a single Glauber exchange yields $2 G^0(k_\perp)$, where for this incoming $\bar q q$ pair we have
\begin{align}  \label{eq:G0qbq}
 G^0(k_\perp) &=  \frac{-ig^2}{\vec k_\perp^{\:2}+m^2 }\:  \bar T^A\otimes T^A \,,
\end{align}
and we have included the mass IR regulator. Since there is no dependence on the $\Delta_i$, the result in \eq{AAG} is identical to that in \eq{hard1G} multiplied by $E(p_{1\perp},p_{2\perp})$, and so as anticipated, the correspondence $G= S^{(G)}$ goes through in the same manner here. The various correspondences also remain true for active-active graphs where the hard vertex involves scattering or production, rather than annihilation, and for higher loop orders.

From the second to last line in \eq{AAG} we also see that the contribution of the active-active Glauber graph corresponds to $ E(p_{1\perp},p_{2\perp})\: \bigl( -i\phi(0)/2 \bigr) S^\gamma$ in the notation of \eq{phi}, where $\phi(0)=\phi(b_\perp=0)$. If we consider the iteration of active-active Glauber exchanges, the result again yields a phase.  Similar to the active-spectator graphs, the $\eta$-regulator was already required for the single-exchange graph, so the ladder sum cannot be carried out independent of considering the loop involving the hard scattering vertex. In \app{AAexp} we carry out this calculation, finding
\begin{align}
& \raisebox{-0.9cm}{ 
	\includegraphics[width=0.18\columnwidth]{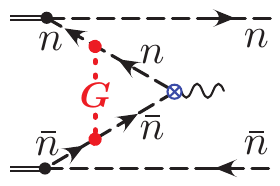}
  }
 = S^\gamma\, E(p_{1\perp},p_{2\perp}) \:
   e^{i\phi(0)/2} 
\,.
\end{align}
Thus we see that the phase for this sum of active-active exchanges is $\phi(0)/2$.  For hard $n$-$\bn$ scattering with ${\cal M}_\Gamma^{\rm DIS}$ the Glauber graphs give zero, so there is no Glauber phase in this case.

\subsection{Spectators in \SCETa}
\label{sec:scet1spectators}

In \SCETa we consider Glaubers at a $\perp$-momentum scale of ${\cal O}(\lambda)$ that is much larger than the  corresponding  $\perp$-momentum  (${\cal O}(\lambda^2)$ ) of the ultrasoft mode. In this setup the infrared regulator is also introduced at the ${\cal O}(\lambda^2)$ scale.  The addition of the ultrasoft mode, and its corresponding subtractions, modifies the \SCETb calculations from Secs.~\ref{sec:ssfactorization}, \ref{sec:asfactorization}, and \ref{sec:aafactorization}. The corresponding one-loop \SCETa calculation  with an offshellness regulator and scalar propagators was carried out in Ref.~\cite{Bauer:2010cc}, and the relevant formula enumerating subtractions at one-loop was given in \eq{scet1subt}. We summarize the changes from our \SCETb discussion below.

For active-active diagrams the situation is the same as was already described at the end of \sec{hardmatching}. The Glauber loop integral becomes scaleless as in \eq{Ghardscet1} and is canceled by its ultrasoft zero-bin subtraction. For the active-active one-loop topology our soft mode is scaleless and is absorbed into the ultrasoft mode (serving to pull it up to the hard scale). Here the ultrasoft gluon Wilson lines directions can matter, and there is no direct correspondence with Glauber gluon subtractions (unless we were to add an additional Glauber gluon at the ${\cal O}(\lambda^2)$ scale, which then would make the theory at the IR scale \SCETb like). 

For the active-spectator topology there are now Glauber, ultrasoft, and collinear Wilson line diagrams. Just as in \sec{asfactorization} the Glauber graph exactly agrees with the Glauber 0-bin of the collinear Wilson line graph. This was noted in the calculation of Ref.~\cite{Bauer:2010cc}, though the important role that the Wilson lines directions play in enabling this correspondence was not observed there. The ultrasoft-Glauber subtraction $(G)(U)$ of the collinear diagram gives a scaleless integral which is equal to the ultrasoft subtraction of the Glauber graph, and hence cancel. The ultrasoft diagram itself is nonzero and plays an important role when considering how the SCET graphs reproduce the full theory result in a matching calculation~\cite{Bauer:2010cc}.

For the spectator-spectator topology there are both Glauber and ultrasoft diagrams. Just as in \SCETb neither soft or collinear graphs contribute. The ultrasoft subtraction on the Glauber diagram is scaleless, proportional to the difference between UV and IR $1/\epsilon$ poles. Again both the Glauber and the ultrasoft contributions are needed to reproduce the full theory spectator-spectator result in a matching calculation~\cite{Bauer:2010cc}.

\subsection{Definition of Spectator and Active at Higher Orders}
\label{sec:defnSA}

Having considered the differences between the cases of spectator-spectator, active-spectator, and active-active Glauber exchange at lowest order, we may also envision higher order extensions. The first thing to decide is whether the active and spectator language remains useful. The key attribute that differentiates a spectator from an active line for the calculations in Secs.~\ref{sec:ssfactorization}, \ref{sec:asfactorization}, and \ref{sec:aafactorization} is that $n$-collinear active particles were effectively eikonal inside the  Glauber loop integral involving the hard vertex, and the same was separately true for $\bn$-collinear active particles. On the other hand, spectator collinear particles always involved a pole on each side of the axis in the  $k^+$ or $k^-$ variable showing up in their propagators, and yielded results where the eikonal approximation is invalid.

Therefore to extend the active and spectator language to all orders in perturbation theory for propagators inside Glauber loops, we adopt the definition that cases where the propagator outside the Glauber burst may be treated as effectively eikonal are called active, while those that are not are called spectators. This definition implies that spectator gluons or quarks may be created by collinear radiation from active lines.

\section{Conclusion}
\label{sec:conclusion}

In this paper we have constructed  an effective field theory of high-energy forward  scattering within the framework of the Soft Collinear Effective Theory (SCET). This provides a common framework for calculating near forward scattering observables and addressing the question of factorization violation in hard scattering processes, which occurs from near forward sub-processes. This framework incorporates the exchange of Glauber gluons in non-local potential operators that connect collinear and soft fields, that is, connecting modes which have the same offshellness but live in different rapidity sectors.  These operators mediate forward scattering at leading order in a $\lambda\sim \sqrt{|t|/s}$ power expansion. They also can violate factorization since they couple together collinear and soft modes in a leading power Lagrangian, unlike in canonical SCET without Glauber exchange potentials, where the leading Lagrangian can be written as a sum of terms, each with fields from a single sector.

The power counting in our EFT is such that each operator scales homogeneously in the power counting parameter $\lambda$. Time ordered products often simply scale as the sum of the scalings of their component operators, but not always. In the presence of loop momenta with Glauber, collinear, soft, or ultrasoft scalings, we derived a general power counting formula, given in \eq{delta}, that determines the overall $\lambda$ scaling of an amplitude at any order in $\alpha_s$ and any order in the power expansion (and in both \SCETa or \SCETb).  The only input needed for this formula are the type and scaling for the inserted operators, and general connectedness information about the resulting diagram.  This formula was used to show that all time ordered products scale at least as the sum of contributions from its constituent operators. 

To construct the Glauber Lagrangian  we matched various QCD amplitudes onto forward scattering operators in SCET with final state particles in various sectors.  These operators were constructed using power counting, gauge symmetry, and calculations at tree level (see \sec{GlauberEFT} for an overview and \sec{match} for the details). Due to the presences of Wilson lines they encode contributions to all orders in the gauge coupling. There are two types of operators which are generated in this way: those with two rapidity sectors, soft and $n$-collinear, and those with three rapidity sectors $n$-collinear, $n'$-collinear and soft (where $n\cdot n'\gg\lambda^2$).   Both two and three rapidity sector operators are composed of gauge invariant building blocks, and can be written as
\begin{align}  \label{eq:LGconclusion}
{\cal L}_G^{{\rm II}(0)}  
& = 
e^{-i x\cdot \cP} \sum_{n,n'} \sum_{i,j=q,g}  \:  O_{nsn'}^{i j}
+  e^{-i x\cdot \cP} \sum_n \sum_{i,j=q,g} \:  O_{ns}^{i j} 
\nn\\
& \equiv
e^{-i x\cdot \cP} \sum_{n,n'} \sum_{i,j=q,g}  \:  
{\cal O}_n^{i B} \frac{1}{\cP_\perp^2} {\cal O}_s^{BC}  \frac{1}{\cP_\perp^2} {\cal O}_{n'}^{j C} 
+ e^{-i x\cdot \cP} \sum_n \sum_{i,j=q,g} \:  {\cal O}_n^{i B} \frac{1}{\cP_\perp^2} {\cal O}_s^{j_n B}   \,.
\end{align}
Although for our explicit calculations we focused almost entirely on the case where $n'=\bn$, with $\bn\cdot n=2$, \eq{LGconclusion} and our results apply for the general situation with any two collinear directions $n_1$ and $n_2$, by adding appropriate factors of $n_1\cdot n_2$. The ${\cal O}_n^{i B},~{\cal O}_s^{BC}$ and ${\cal O}_s^{j_n B}$ appearing in \eq{LGconclusion} are defined in \tab{opsummary}. The operator $O_{S}^{BC}$ encodes the well known Lipatov vertex for 1-soft gluon emission, but also contains additional structures that give vertices with $\ge 2$ soft gluons. In \sec{basis} the soft operator ${\cal O}_s^{BC}$ was decomposed into a general basis based on symmetry, dimensional analysis, and power counting constraints, to give  a polynomial in $\perp$-derivatives and gluon building block operators dressed by Wilson lines. The coefficients of the operators in this basis were then fixed by carrying out a tree level matching calculation with zero, one, and two external soft gluons in \sec{2sgluon}.  \eq{LGconclusion} is the complete Glauber exchange Lagrangian for \SCETb, and also applies for \SCETa where we call it ${\cal L}_G^{{\rm I}(0)}$. The corrections to this result  are  power suppressed. For any given process, traditional hard scattering factorization will be violated unless the contributions from ${\cal L}_G^{{\rm II}(0)}$ (or ${\cal L}_G^{{\rm I}(0)}$) can be shown to vanish.

Several technical ingredients play an important role in our construction of the Glauber exchange EFT:  The multipole expansion enables subleading momentum components to flow through a  diagram, even though these momenta may not show up in intermediate propagators.  The use of a rapidity regulator for both the Glauber exchange potential, and soft and collinear Wilson lines is crucial to obtain well defined and physically meaningful results. Our implementation of the rapidity renormalization is analogous to the $\overline{\rm MS}$ renormalization in dimensional regularization, with $1/\eta$ poles analogous to $1/\epsilon$ poles, and a rapidity cutoff $\nu$ analogous to the invariant mass cutoff $\mu$.  Finally, SCET is formulated with a subtraction formalism that ensures there is no double counting of infrared regions from loop graphs in the EFT. These technical ingredients are discussed in \sec{multiglauber}.
 
A remarkable property of the forward matching procedure used to derive ${\cal L}_G^{\rm II(0)}$ is that there are no hard matching corrections, making \eq{LG} the {\it exact} Lagrangian for Glauber exchange at leading power. The reason for this  simplification is that when there is no hard scattering involved in an S-matrix element, then for any leading power contribution there are no closed set of hard lines forming loops, and that the offshell Glauber lines are localized in tree level sub-diagrams. Working to one loop, in both \SCETa and \SCETb, we have explicitly shown in \secs{loop2match}{loop1match} that our effective theory exactly reproduces the full theory leading power forward scattering result for all color structures, logarithms, and constants. The fact that the tree level matching coefficients are exact imposes strong constraints on the renormalization of the theory.  It implies that the operators in \eq{LG} have no overall renormalization group anomalous dimensions. Therefore, beyond the strong coupling, there is no running in $\mu$ in the leading power SCET Lagrangian unless there is a hard scattering.  

This does not mean that there are no large logs in the forward scattering amplitude, since the soft and collinear modes still yield loop level amplitudes with logarithms that are minimized at different rapidity cutoffs, $\nu=\sqrt{-t}$ for soft modes and $\nu=\sqrt{s}$ for collinear modes.\footnote{In \SCETa there are also different scales $\mu$ for the modes, since ultrasoft modes live at a parametrically smaller invariant mass scale than the collinear and soft modes. This distinction becomes apparent below the collinear scale.} The logarithms that are generated due to this hierarchy   are summed with the rapidity renormalization group.  At the amplitude level the structure of the anomalous dimensions and their explicit one-loop computation can be found in \sec{regge}. The lack of matching beyond tree level implies that the sum of quark and gluon operators mixes back into itself. For example, for the $n$-collinear operators we have
\begin{align} \label{eq:rel1conclusion}
& \nu \frac{d}{d\nu} ({\cal O}_n^{qA} + {\cal O}_n^{gA}) = \gamma_{n\nu}
({\cal O}_n^{qA} + {\cal O}_n^{gA}) 
\,.
\end{align}
Solving this equation at one-loop yields gluon Reggeization from the rapidity renormalization group flow of octet operators,
\begin{align}
& 
({\cal O}_n^{qA} + {\cal O}_n^{gA})(\nu_1) = \Big(\frac{\nu_0}{\nu_1}\Big)^{-\gamma_{n\nu}}
({\cal O}_n^{qA} + {\cal O}_n^{gA})(\nu_0) \,,
\end{align}
when we take $\nu_1=\sqrt{s}$ and $\nu_2=\sqrt{-t}$, see \sec{RRGEsoln}.
At one loop the anomalous dimension is the same as the Regge exponent,  $\gamma_{n\nu}=\frac{\alpha_s(\mu) C_A}{2\pi} \ln(-t/m^2)$. Here $m$ is an IR regulator which appears because the virtual amplitude alone is not a physical observable. 

To perform a more physically relevant resummation we can use the rapidity evolution to sum logarithms in the inclusive forward scattering cross section.  In \sec{BFKL} we factorized the squared single Glauber exchange amplitude into $\perp$-convolutions of all orders soft and collinear functions, $C_n(q_\perp,p^-,\nu)\otimes S_G(q_\perp,q_\perp',\nu) \otimes C_\bn(q_\perp',p^{\prime +},\nu)$, where each of these functions is defined by field theory matrix elements of soft and collinear operators. We then explicitly calculated the rapidity anomalous dimension for the soft function at 1-loop, obtaining after a very simple calculation the standard BFKL equation,
\begin{align} \label{eq:bfklconclusion}
 \nu\frac{d}{d\nu}\,  S_G(q_\perp,q_\perp',\nu)
   &= \frac{2 C_A\alpha_s(\mu)}{\pi^2} \int\!\! d^2k_\perp 
   \bigg[ \frac{ S_G(k_\perp,q_\perp',\nu)}
     {(\vec k_\perp-\vec q_\perp)^2}
   - \frac{\vec q_\perp^{\:2}\:  S_G(q_\perp,q_\perp',\nu)}
   { 2\vec k_\perp^2 (\vec k_\perp-\vec q_\perp)^2}  \bigg]
  \,.
\end{align}
The collinear functions $C_n$ and $C_\bn$ also obey BFKL-like equations which ensure that the physical forward scattering amplitude is $\nu$ independent. One powerful property of the formulation of anomalous dimensions in SCET is the ability to systematically derive the structure of anomalous dimensions at higher orders in $\alpha_s$ (often to all orders), which should facilitate in the future exploring the BFKL-type resummation beyond the next-to-leading logarithms. 

We have utilized the effective theory to explore the eikonalization (or lack thereof) of propagators in high energy forward scattering.  The canonical eikonal scattering phase or semi-classical phase arises from an infinite ladder sum of Glauber exchanges. This sum was computed in \sec{exponentiation} for an arbitrary color channel, and is obtained from an ordered collapse of the iterated Glaubers onto exchanges that occurs at equal time and equal longitudinal distance. This collapse is handled properly by the rapidity regulator. In \sec{forwardgraphs} we considered the general properties of Glauber loops in the presence of virtual and real collinear fluctuations, including general rules for when propagators inside Glauber loops do or do not eikonalize.  Glauber loops will in general vanish unless all of the Glauber exchanges can be slid without impediment to an equal longitudinal position when considering diagrams in  time ordered perturbation theory.\footnote{Often high-energy scattering is studied in light cone ordered perturbation theory, but our need for rapidity regularization between particles with two different light-cone times, enforced the use of time ordering instead. Nevetheless, many of the simplifying features of light-cone ordering are retained in our EFT.} In \sec{semiclassical} we consider the correspondence of our EFT derived results with the semi-classical picture of eikonal scattering, as well as with the shockwave picture and the multi-Wilson line EFT framework of~\cite{Balitsky:1995ub}.  While the exclusive two to two scattering amplitude can be thought of as a semi-classical process, virtual collinear corrections, even in the purely abelian case, violate this picture for this exclusive rate.  Once splitting and pair production is included the source terms for Glauber interactions do not in general eikonalize. However in multi-Glauber exchange diagrams in the non-abelian theory, after the first Glauber attachment to a source, it does eikonalize for subsequent attachments. The fact that these exchanges occur at the same time and longitudinal position directly yields the setup where multiple Wilson lines cross a shockwave. The source propagators outside the shockwave are non-eikonal.

Another significant component of this paper was dedicated to studying the role of Glauber gluons in hard scattering cross sections, where they will violate factorization unless their effects cancel out. For a proton-proton collision, non-perturbative Glauber exchanges (with $|t|\sim\Lambda_{\rm QCD}^2$) couple together the fields in matrix elements that we would like to factorize into distinct parton distribution functions. Perturbative Glauber exchange (with $|t|\gg \Lambda_{\rm QCD}^2$) can couple together modes in the matrix elements defining collinear and soft functions, which are then no longer universal perturbative ingredients that can be used for resummation (beyond some order).   We have also seen in \secs{GlauberBox}{exponentiation} that our Glauber loop results are imaginary, proportional to exactly the $(i\pi)$ factors that are commonly associated to amplitude level factorization violation.  For a given observable, it is therefore important to understand whether the Glauber Lagrangian will or will not contribute.  In particular, one would like to go beyond the small number of proton-proton collision observables where insensitivity to the Glauber region has been demonstrated so far, in order to perturbatively incorporate Glauber exchange or demonstrate cancellations.

In \secs{properties}{spectator} we have demonstrated that Glauber interactions between two active partons or between active and spectator partons can be absorbed into soft and collinear Wilson lines that appear in hard scattering operators.  This is done by picking directions for these soft and collinear Wilson lines that agree with the physical allowed scattering of nearly onshell particles from the Glauber exchange. This absorption was derived by showing that the zero bin subtraction of soft or collinear loops involving these Wilson lines are exactly equal to the contribution from a corresponding set of Glauber graphs. In particular, in \sec{properties} we showed this for soft loops which dress $n$-$\bn$ hard operators at 1-loop and 2-loop order, and for $n$-$\bn$ hard production/annihilation/scattering with an additional soft emission at 1-loop (see also \sec{aafactorization}). In \sec{asfactorization} we showed that active-spectator Glauber exchanges are equal to the zero-bin subtraction of a collinear 1-loop graph involving a collinear Wilson line. Establishing the fact  that these Glauber exchanges can be absorbed into Wilson lines  gains a measure of control, as it implies that they appear in objects that can still be factorized into independent matrix elements, and thus do not necessarily invalidate the factorization program. However, spectator-spectator Glauber exchanges cannot be absorbed into Wilson lines or other properties of soft or collinear modes, for the reasons discussed in \sec{ssfactorization}.

While in this paper we have not provided full factorization proofs, we have used our results to determine sufficient criteria for cross section level {\it factorization  violation}. In particular, in section \ref{sec:ssfactorization} we showed how iterated spectator-spectator  Glauber exchanges  exponentiate into the eikonal phase within the confines of a hard scattering amplitude. In order for these Glauber phases to cancel in the cross section requires an unconstrained integration over the relative $\perp$ momentum of the spectators. This cancellation will occur if the observable of choice operates at a scale $Q^*$ that is much larger than that of the considered Glauber's transverse momentum, $Q^{*2}\gg q_\perp^2$, or if the measurement is inclusive enough that these transverse momenta are not restricted. 

In this paper we have accomplished our main goal of setting up an effective theory for describing forward scattering and factorization violation with a universal Glauber exchange Lagrangian. We have also enumerated many of its subtleties, features, and properties.  There remain many avenues to explore further in the future.


\begin{acknowledgments}

We thank Duff Neill for discussions and Ian Moult, Frank Tackmann, and Chris Lee for helpful comments on the text.
We also thank Aditya Pathak for carrying out a gauge invariance cross check on the SCET diagrams in \fig{match_Lipatov_2gluon}b. This enabled us to fix a typo in the ordering of our color matrices in the adjoint Wilson line, which influences the 2-soft gluon Feynman rule of ${\cal O}_s^{AB}$. 
I.S is supported by the U.S. Department of Energy (DOE) and the Office of Nuclear Physics under DE-SC0011090.  I.S. was also supported by the Simons Foundation through the Investigator grant 327942. I.Z.R. is supported by  DOE grants DOE DE-FG02-04ER41338 and FG02- 06ER41449. I.Z.R would also like to thank the Caltech theory group for its hospitality and the Moore Foundation for support.

\end{acknowledgments}

\appendix

\section{SCET Power Counting Formula including Glaubers}  
\label{app:powercount}

In this appendix we derive a general power counting formula for SCET including Glauber loops. For simplicity we consider a case with two collinear sectors since the generalization to $N$ distinct collinear sectors will be obvious.  We start with \SCETb. Consider a graph built from the insertion of an arbitrary number of different operators, which may or may not involve Glauber potentials, and which may be leading order or power suppressed at some order. We may consider an infinite number of insertions from operators in the leading order Lagrangians that yield $\lambda^0$ contributions, but we start with a finite number for simplicity and only take the infinite limit at the end (when it becomes trivial).  We start by enumerating the number of vertices that appear that involve operators of order $\lambda^k$ as follows, there are:
\begin{itemize}
  \item[]  $V_k^n$ vertices with operators involving only $n$-collinear fields,
  \item[]  $V_k^\bn$ vertices with only $\bn$-collinear fields,
  \item[]  $V_k^S$ vertices with only soft fields,
  \item[]  $V_k^{nS}$ vertices that have both $n$-collinear and soft fields but do not have $\bn$ fields,
  \item[]  $V_k^{\bn S}$ vertices with both $\bn$-collinear and soft fields but not $n$ fields,
  \item[]  $V_k^{n\bn}$ vertices with both $n$ and $\bn$-collinear fields (with or without soft fields).
\end{itemize}  
For example, the leading power 3-rapidity sector Glauber Lagrangian contributes to $V_{k=2}^{n\bn}$ and the leading power 2-rapidity sector Glauber Lagrangians contribute to $V_{k=3}^{nS}$ or $V_{k=3}^{\bn S}$. We also introduce variables that count the number of loop integrals and the number of internal lines of various types:
\begin{align}
   L^n  \  & : \ \ \ n\text{--collinear loops with }  
      k^\mu \sim Q(\lambda^2,1,\lambda)  \text{ loop momenta, } 
    \nn\\
   L^\bn \ & : \ \ \ \bn\text{--collinear loops with }  
      k^\mu \sim Q(1,\lambda^2,\lambda)  \text{ loop momenta, } 
   \nn\\
   L^S \ & : \ \ \ \text{soft loops with }  
      k^\mu \sim Q(\lambda,\lambda,\lambda)   \text{ loop momenta,} 
  \nn\\
   L^{nS} \ & : \ \ \ \text{$s$--$n$ Glauber loops with }  
      k^\mu \sim Q(\lambda^2,\lambda,\lambda)  \text{ loop momenta}
       \,, \nn\\
   L^{\bn S} \ &: \ \ \ \text{$s$--$\bn$ Glauber loops with }  
       k^\mu \sim Q(\lambda,\lambda^2,\lambda) \text{ loop momenta}
        \,,\nn\\
   L^{n\bn} \ & : \ \ \ \text{$n$--$\bn$ Glauber loops with }  
       k^\mu \sim Q(\lambda^2,\lambda^2,\lambda)  \text{ loop momenta}
       \,,\nn\\
    I^n \ & : \ \ \ \text{ internal } n\text{--collinear propagators},
    \nn\\  
    I^\bn \ & : \ \ \ \text{ internal } \bn\text{--collinear propagators,}
     \nn\\
     I^S \ & : \ \ \ \text{ internal } \text{ soft propagators.} 
\end{align}
Note that the $I$'s only include the propagating (nearly on-shell) particles\footnote{Wilson lines scale as $\sim \lambda^0$, and do not modify the power counting of the operators that contain them. The eikonal propagators from these Wilson lines are not counted by the $I^x$s.} and that the loop momentum scaling is determined by the maximum allowed value which leaves all propagating modes near  their mass shell.

\begin{figure}[t!]
	%
	%
\begin{center}
 \includegraphics[width=0.55\columnwidth]{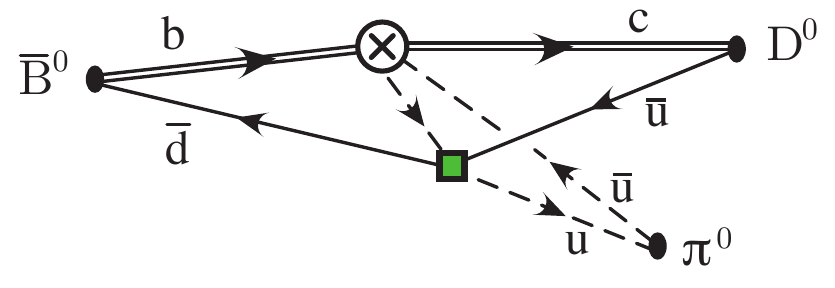} 
\end{center}	
\vspace{-0.5cm}
	\caption{\setcaptionskip
	Example of a contribution to the process $\bar B^0\to D^0\pi^0$ from Ref.~\cite{Mantry:2003uz} that involves a loop with Glauber scaling without having a Glauber exchange potential operator. The $\otimes$ is the $W$-mediated hard interaction, solid black dots are interpolating field insertions for the mesons, double lines are soft heavy quarks, single solid lines are light soft quarks, and dashed lines are light collinear quarks, all in \SCETb. This long distance  contribution involves a (non-Glauber exchange) power suppressed operator (the green square) that causes an interaction between soft and collinear particles. The left most loop in this diagram involves both soft and collinear quark propagators and hence has a loop momentum with Glauber scaling, $k^\mu\sim Q(\lambda^2,\lambda,\lambda)$.  }
	\setmainskip
	\label{fig:bdpi}
\end{figure}

Then to determine the overall $\lambda$ scaling of the diagram we count up the powers of $\lambda$ for these operators, add the powers of $\lambda$ from momentum space loop integrals, and subtract the powers of $\lambda$ generated by turning some of the fields in the operators into propagators.  This gives that a general graph in \SCETb with Glauber operators scales as $\lambda^\delta$ where the power
\begin{align}  \label{eq:delta2original}
   \delta 
 & = \sum_k k \Big( V_k^n  \plus V_k^\bn \plus V_k^S \plus V_k^{nS}
                 \plus V_k^{\bn S} \plus V_k^{n\bn} \Big)
  \\
 &\qquad
    \plus 4 L^n \plus 4 L^\bn \plus 4 L^S \plus 5 L^{n S} \plus 5 L^{\bn S} \plus 6 L^{n\bn}
     \minus 4 I^n \minus 4 I^\bn \minus  4 I^S  .
 \nn
\end{align}
Note that $\delta$ is at the operator level. It includes the scaling of fields for all non-contracted external lines, but it does not account for the scaling associated to external states. The scaling for the external states in a matrix element can be trivially added as well, as discussed for example in Ref.~\cite{Bauer:2002uv}.
While this formula can be used to determine the power counting, it is more useful to have a formula that only depends on the vertex indices and on topological properties of the graph. To this end, note that  
the topological Euler identity between vertices, loops, and propagators for the overall diagram implies
\begin{align}  \label{eq:eulerfull2}
1 &=\sum_k\big(V_k^n + V_k^\bn + V_k^S + V_k^{nS} + V_k^{\bn S} + V_k^{n\bn} 
  \big) 
  \\
 &\qquad
 + L^n + L^\bn +L^S + L^{n S}+ L^{\bn S} + L^{n\bn} - I^n - I^\bn - I^S 
  \,,\nn
\end{align}
which allows us to write \eq{delta2original} as
\begin{align} \label{eq:pc2step1}
   \delta 
 & = 4+ \sum_k (k-4) \Big( V_k^n + V_k^\bn + V_k^S 
     + V_k^{nS} + V_k^{\bn S} + V_k^{n\bn} \Big)
     +  L^{n S} +  L^{\bn S}  + 2  L^{n\bn} \,.
\end{align}
This result is still inconvenient for power counting since we have to determine the number of Glauber loops, so a few further manipulations are useful. Following Ref.~\cite{Luke:1999kz} 
we can exploit individual topological identities for the various sectors to generate a simpler power counting formula. If we erase all $n$-collinear propagators then we define
 $N^{\bn S}$ as the number of disconnected subgraphs that appear, if we erase
all $\bn$-collinear propagators then $N^{n S}$ is the number of disconnected subgraphs that appear, and if we erase
all soft propagators then $N^{n\bn}$ is the number of disconnected subgraphs that appear. In this counting procedure a vertex is not erased unless all types of fields that appear in it are erased (so two point vertices can appear). For these we have the topological identities
\begin{align} \label{eq:N2index}
  N^{n S} &=\sum_k \big( V_k^n + V_k^S + V_k^{nS} + V_k^{\bn S} + V_k^{n\bn} 
          \big) + L^n + L^S + L^{nS} - I^n - I^S 
        \,, \nn\\
 N^{\bn S} &=\sum_k \big( V_k^\bn + V_k^S + V_k^{n S} + V_k^{\bn S} + V_k^{n\bn} 
         \big) + L^\bn + L^S + L^{\bn S} - I^\bn - I^S 
       \,, \nn\\
 N^{n\bn} &=\sum_k \big( V_k^n + V_k^\bn + V_k^{n S} + V_k^{\bn S} + V_k^{n\bn} 
         \big) + L^n + L^\bn + L^{n\bn} - I^n - I^\bn 
       \,.
\end{align} 
In addition, if we erase all sectors but one then we count the number of connected components of that type alone. For the Glauber Lagrangian we do not erase the 3-rapidity sector vertex even if both $n$ and $\bn$ collinear lines are erased. So the number of connected soft, $n$-collinear, and $\bn$-collinear components is 
\begin{align} \label{eq:N1index}
  N^S &= \sum_k \big( V_k^S + V_k^{n S}+V_k^{\bn S} +V_k^{n\bn} \big) + L^S - I^S\,,
   \nn\\
  N^n &= \sum_k \big( V_k^n + V_k^{n S}+V_k^{n\bn} \big) + L^n - I^n\,,
   \nn\\
   N^\bn &= \sum_k \big( V_k^\bn + V_k^{\bn S}+V_k^{n\bn} \big) + L^\bn - I^\bn \,,
\end{align}
respectively.  Note that these indices obey the inequalities
\begin{align} \label{eq:Nbounds}
 &  N^n +N^S \ge N^{nS} \,, \qquad 
  N^n +N^\bn \ge N^{n\bn} \,, \qquad
  N^\bn +N^S \ge N^{\bn S} 
    \,.
\end{align}
The Euler identities in \eq{N1index} can be combined to give results for $L^{nS}$, $L^{\bn S}$ and $L^{n\bn}$, and a further result that follows from combining them with the original Euler identity in \eq{eulerfull2},
\begin{align} \label{eq:Loopindex}
   L^{n S} &= -N^{[nS]} + \sum_k \big( V_k^{n S} + V_k^{n\bn}\big) \,,  
     \qquad \text{where }  N^{[nS]} \equiv N^n + N^S - N^{n S} 
     \,, \nn\\
   L^{\bn S} &= -N^{[\bn S]} + \sum_k \big( V_k^{\bn S} + V_k^{n\bn}\big) \,,
       \qquad  \text{where }  N^{[\bn S]} \equiv N^\bn + N^S - N^{\bn S} 
     \,, \nn\\
   L^{n\bn} &= -N^{[n\bn]} + \sum_k  V_k^{n\bn} \,,
       \qquad  \text{where }  N^{[n\bn]} \equiv N^n + N^\bn - N^{n\bn} 
     \,, \nn\\
   1 & + N^{[nS]} + N^{[\bn S]} + N^{[n\bn]} - N^n - N^\bn -N^S 
      =  \sum_k V_k^{n\bn} \,.
\end{align}
Using these results in \eq{pc2step1} gives the final power counting formula for \SCETb:
 \begin{align}  \label{eq:delta2}
   \delta 
 & = 6\minus N^n\minus N^\bn\minus N^{nS} \minus N^{\bn S}
  \plus \sum_k \Big[ (k\minus 4) \big( V_k^n + V_k^\bn + V_k^S \big) 
       + (k\minus 3)\big(  V_k^{nS} + V_k^{\bn S} \big) 
       + (k\minus 2) V_k^{n\bn} \Big] 
     \,.
\end{align}
In this formula the power counting is obtained entirely from the power counting of the inserted operators (through the various $V$'s) plus topological information about how connected the graph is in different sectors (encoded in the $N$'s).  One does not have to consider the power counting for loops or propagators, which easily allows the result to be applied when an infinite number of operators are considered.   In the special case that there are no $\bn$ fields this result reduces to the power counting formula  derived earlier in Ref.~\cite{Stewart:2003gt} which included the $n$-$s$ Glauber loops because they appear at subleading power (this earlier result is obtained by setting $N^\bn=0$, $N^{\bn S}=N^S$, $N^{nS}=1$, $V_k^\bn=V_k^{\bn S}=V_k^{n\bn}=0$). Indeed, even when the Glauber Lagrangian is not included these Glauber loops play an important role in power suppressed time ordered products, such as in the factorization formula for long distance corrections to color-suppressed $B\to D \pi$ decays derived in Ref.~\cite{Mantry:2003uz}. As shown in \fig{bdpi}, this occurs through a subleading Lagrangian interaction in \SCETb, which mediates forward scattering of soft and $n$-collinear particles without involving a long distance Glauber potential. Replacing the Glauber potential by an interaction that is localized at a harder scale leads to the power suppression in this case.

To extend the above power counting to \SCETa we must add ultrasoft fields. We otherwise keep the same starting point as above, so in addition to ultrasoft fields we will have $n$-collinear, $\bn$-collinear, and soft fields (again the generalization to more collinear fields will be obvious). Since the ultrasoft fields have scaling $\psi_{us}\sim \lambda^3$, $A_{us}^\mu \sim \lambda^2$, which is greater or equal to all other field components, we only need an additional index $V_k^{us}$ which counts the number of operators with only ultrasoft fields that have scaling $\lambda^k$. All other indices retain their definitions and are now allowed to contain power counting contributions from ultrasoft fields that appear in their operators. Thus we start by obtaining a formula for $\delta$ for a general \SCETa diagram by simply adding $\sum_k k V_k^{us} + 8 L^{us} - 8 I^{us}$ to the \SCETb power counting formula  in \eq{delta2original}.  We have to distinguish the special case where a graph contains only ultrasoft fields, vertices, and loops. After using the overall Euler identity,  $1=\sum_k V_k^{us} + L^{us} - I^{us}$, the final result for this special case is $\delta = 8 + \sum_k (k-8) V_k^{us}$.  In contrast, in the presence of one or more collinear or soft fields we retain the Euler identity in \eq{eulerfull2} because the graph must remain connected if we erase all ultrasoft fields. Furthermore the identies in Eqs.~(\ref{eq:N2index},\ref{eq:N1index},\ref{eq:Loopindex}) are not modified in the presence of ultrasoft fields (except that we must add $+1$ to the right-hand side of the last identity in \eq{Loopindex} for the special case with graphs that have only ultrasoft fields). There is an addition an overall Euler identity in the presence of ultrasoft fields which can be obtained by adding $\sum_k V_k^{us} + L^{us} - I^{us}$ to the right-hand side of \eq{eulerfull2}.  Together these two Euler identities imply $\sum_k V_k^{us} + L^{us} - I^{us}=0$ for diagrams that have at least some non-ultrasoft fields.  These identities allows us to remove the loop and propagator counting factors for the ultrasoft loops without changing any other part of the derivation for \SCETb done above.  

Thus the final power counting formula for either \SCETa or \SCETb is
\begin{align} \label{eq:delta1}
   \delta 
 & = 6 - N^n - N^\bn - N^{nS} - N^{\bn S} + 2 u 
   \,,\\
  & + 
 \sum_k \Big[ 
    (k-8) V_k^{us} +  (k\minus 4) \big( V_k^n + V_k^\bn + V_k^S \big) 
       + (k\minus 3)\big(  V_k^{nS} + V_k^{\bn S} \big) 
       + (k\minus 2) V_k^{n\bn}  \Big]
     \,. \nn
\end{align}
Here $u=1$ for purely ultrasoft graphs where $0=N^n=N^\bn=N^{nS}=N^{\bn S}$ and $0=V_k^n=V_k^\bn=V_k^S=V_k^{n S}=V_k^{\bn S}=V_k^{n\bn}$, and otherwise $u=0$.  If no ultrasoft fields are present then $V_k^{us}=0$ and $u=0$, so \eq{delta1} reduces to \eq{delta2}, demonstrating explicitly that the formula in \eq{delta1} is valid for both \SCETa and for \SCETb.  In the special case for \SCETa where there are no Glauber loops present and no $\bn$ or soft fields this result reduces to the power counting formula derived in Ref.~\cite{Bauer:2002uv}, $\delta = 4 + 4u + \sum_k (k-8) V_k^{us} + (k-4) V_k^n$. To obtain this result we note that if $u=0$ then $N^n=N^{nS}=1$ and $N^\bn=N^{\bn S}=0$, while if $u=1$ then $N^n=N^{nS}=N^\bn=N^{\bn S}=0$.

To see how \eq{delta1} works in practice, lets consider several simple examples in \SCETb with $u=0$.
\begin{enumerate}
\item \raisebox{-0.3cm}{
\includegraphics[width=0.13\columnwidth]{figs/GlaubOp_tree_qqqq}
}
\quad
\parbox{0.77\columnwidth}{\setcaptionskip
Here $N^n=N^\bn=N^{nS}=N^{\bn S}=1$ and $V_2^{n\bn}=1$, so $\delta =2$. This agrees with the scaling of four external $\xi_n'$ fields times a $1/q_\perp^2$ potential, $\lambda^\delta =\lambda^4 \lambda^{-2}=\lambda^2$. }
\item \raisebox{-0.3cm}{
\includegraphics[width=0.13\columnwidth]{figs/GlaubOp_treeS_qqqq}
}
\quad
\parbox{0.77\columnwidth}{\setcaptionskip
Here $N^n=N^{nS}=N^{\bn S}=1$, $N^{\bn}=0$ and $V_3^{nS}=1$, so $\delta =3$. This agrees with the scaling of two $\xi_n$  and two $q_s$ external fields times a $1/q_\perp^2$ potential, $\lambda^\delta = \lambda^2\lambda^3 \lambda^{-2}=\lambda^3$. }
\item \raisebox{-0.6cm}{
\includegraphics[width=0.15\columnwidth]{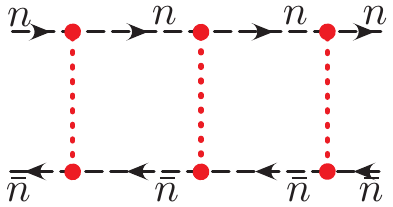}
}
\quad
\parbox{0.73\columnwidth}{\setcaptionskip
Here $N^n=N^{nS}=N^{\bn S}=N^{\bn}=1$ and $V_2^{n\bn}=3$, so $\delta =2$.  }
\item \raisebox{-0.6cm}{
\includegraphics[width=0.1\columnwidth]{figs/EFT_loop3_qqqq_ext}
}
\quad
\parbox{0.73\columnwidth}{\setcaptionskip
Here $N^n=N^{nS}=N^{\bn S}=N^{\bn}=1$ and $V_3^{nS}=V_3^{\bn S}=1$, so $\delta =2$.  }
\item \raisebox{-0.7cm}{
\includegraphics[width=0.12\columnwidth]{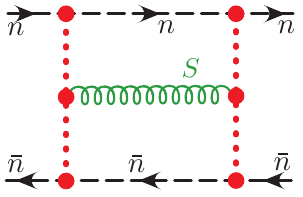}
}
\quad
\parbox{0.73\columnwidth}{\setcaptionskip
Here $N^n=N^{nS}=N^{\bn S}=N^{\bn}=1$ and $V_2^{n\bn}=2$, so $\delta =2$.}
\item \raisebox{-0.7cm}{
\includegraphics[width=0.15\columnwidth]{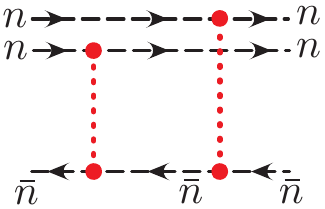}
}
\quad
\parbox{0.73\columnwidth}{\setcaptionskip
Here $N^n=N^{nS}=2$, $N^{\bn S}=N^{\bn}=1$ and $V_2^{n\bn}=2$, so $\delta =0$. This agrees with the scaling for six $\xi_{n'}$ external fields, two $1/q_\perp^2$s, and one collinear propagator.  }
\item \raisebox{-0.5cm}{
\includegraphics[width=0.16\columnwidth]{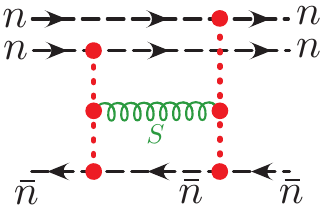}
}
\quad
\parbox{0.73\columnwidth}{\setcaptionskip
Here $N^n=2$, $N^{nS}=N^{\bn S}=N^{\bn}=1$ and $V_2^{n\bn}=2$, so $\delta =1$. Note that this is suppressed relative to the graph in example 6 by ${\cal O}(\lambda)$, even though it involves only the leading power Lagrangian. This ${\cal O}(\lambda)$ agrees with the direct counting, where we add a $n$-$s$ Glauber loop measure $\sim \lambda^5$, two Glauber potentials giving $\sim \lambda^{-4}$, a soft propagator $\sim\lambda^{-2}$, and two Lipatov vertices each giving a momentum factor that is $\sim \lambda$.  }
\end{enumerate}
To fully compare results with a different number (or type) of external particles, we start with $\delta$ and then add contributions from the scaling of external states, phase space integrals, and the overal momentum conserving delta-function for each case.  Although all of our examples here are for forward scattering graphs, the formula for $\delta$ works in an identical manner for cases including a hard interaction operator.

In general \eq{delta1} gives a simple way to determine the scaling of a given graph, by simply adding up the powers $k$ associated to the various operators that the graph is built from.  Noting that the sum on $k$ runs over operators with a different power counting, and the $V_k$s give the total count of all operators of a particular type with power counting $\lambda^k$, we can rewrite the result for $\delta$ as a sum over the power counting contribution $\Delta_i$ of each individual operator $i$,
\begin{align} \label{eq:deltaDelta}
  \delta =  \delta^{\rm conn} + \sum_i \Delta_i \,,
\end{align}
where 
\begin{align}
\delta^{\rm conn}= 6 - N^n - N^\bn - N^{nS} - N^{\bn S} + 2 u 
\end{align}
is a factor that depends on the sector-based connectedness of the diagram, and for an operator $i$ which is of order $\lambda^{k_i}$ we have
\begin{align}
  \Delta_i = \left\{ \begin{array}{l} 
     (k_i-8)\quad \text{pure ultrasoft operator} \\
     (k_i-4)\quad \text{pure soft, $n$-collinear, or $\bn$-collinear operator}\\
     (k_i-3)\quad \text{mixed soft-$n$ or mixed soft-$\bn$ operator}\\
     (k_i-2)\quad \text{mixed $n$-$\bn$ operator with/without soft fields} .
  \end{array}  \right.
\end{align}
Intuitively, the subtractions in $(k_i-8)$, $(k_i-4)$, $(k_i-3)$, and $(k_i-2)$ are just associated to the largest momentum for each operator $O_i(x)$, which determine the scaling of the corresponding measure $d^4x$. The measure is either purely ultrasoft ($\lambda^{-8}$), purely collinear or soft ($\lambda^{-4}$), mixed collinear-soft ($\lambda^{-3})$, or mixed $n$--$\bn$ collinear ($\lambda^{-2}$).  In general
\begin{align}
\label{canon}
  T\: O_1(0) \prod_{i=2}^{N} \int\!\! d^4x_i\: O_i(x_i)
 \ \sim\  \big( \lambda^{\sum_i \Delta_i} \big) \ \big( \lambda^{\delta^{\rm conn}} \big)
   \,,
\end{align}
where $T$ denotes the time-ordered product. The spacetime position of one operator has been fixed since we are not including the overall momentum conserving delta function in the power counting determined by $\delta$ in \eq{delta1} or \eq{deltaDelta}. 

\eq{deltaDelta} is particularly simple to use if we are interested in comparing a set of diagrams where the connectedness term $\delta^{\rm conn}$ is fixed, since then we simply add up the factors $\Delta_i$ for each operator. In this situation the leading power contributions are just determined by the leading power Lagrangian, which consists of operators with $\Delta_i=0$, perhaps supplemented by a hard scattering operator that provides a base for the process and only appears once.  However, sometimes the connectedness term can play an important role. This is the case in NRQCD, where for example, a graph similar to the soft eye graph in figure \fig{SCET1_oneloop_matching}c but with the collinear fermions replaced by heavy non-relativistic fermions, is built from $\Delta_i=0$ operators, but is enhanced by a single power of $1/v$ from the connectedness term. In that theory the enhancement occurs because the higher energy soft sector mediates a potential between the lower energy heavy non-relativistic fermions.\footnote{The reason for this enhancement in NRQCD beyond the powers obtained from the $v^0$ interaction Lagrangian, is that the large ${\cal O}(mv)$ soft energy carried by the soft gluons or quarks can only run in the loop. No soft energy can be carried away by the fermions on the external legs. In perturbative NRQCD this enhancement does not endanger the power counting because the $v$ expansion and coupling expansion are tied together by the virial relation $\alpha_s \sim v$, which implies that  the graph in \fig{SCET1_oneloop_matching}c  is suppressed by a single $\alpha_s$ relative to the leading order Coulomb potential.} In the NRQCD action the heavy fermion kinetic term is $\sim v^0$, but the tree level Coulomb potential operator is $\sim 1/v$. The enhancement of the soft eye loop graph in NRQCD makes it the same order in $v$ as this tree level Coulomb exchange, so that the only difference is an extra power of $\alpha_s$.  

In SCET, the connectedness term does not yield an analogous enhancement for loop graphs. To discuss this we can safely set $u=0$. To prove that there are no connectedness enhancements in loop graphs, first consider a graph with $i_n$ initial state $n$-collinear particles, and $i_\bn$ initial state $\bn$-collinear particles. The simplest contribution is to connect all these fields with tree level Glauber potentials, which gives $N^n = N^{nS}=i_n$ and $N^\bn=N^{\bn S}=i_\bn$, and the power counting result is $\delta = \delta^{\rm conn} = 6-2 i_n -2 i_\bn$.  This provides a baseline for the scattering operator with these external states, and we can then ask whether any graphs with the same external fields can have a smaller $\delta^{\rm conn}$.   Since for the baseline graph $N^n=N^{nS}$ and $N^\bn=N^{\bn S}$ with $N^S=0$, the bounds in \eq{Nbounds} imply that the only way we can decrease $\delta^{\rm conn}$ is by increasing $N^n$ or $N^{\bn}$, or by increasing $N^S$ with  a simultaneous increase to $N^{nS}$ or $N^{\bn S}$. Holding the external fields fixed, it is not possible to increase these indices at tree level, so we must consider adding loops to do so.  

The prototype for a loop correction which could decrease $\delta^{\rm conn}$ is to start with a graph having a single $n$-collinear sector, and split it into two sectors by joining lines together with either a soft loop or $\bn$-collinear loop, such as in
\begin{align}
  \raisebox{0.4cm}{
\includegraphics[width=0.14\columnwidth]{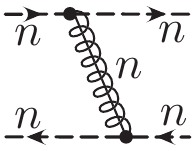}
 }
  \qquad \raisebox{1.2cm}{\Large $\Longrightarrow$}\qquad
  \includegraphics[width=0.15\columnwidth]{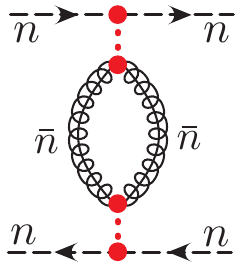}
  \quad\raisebox{1.2cm}{$,$}\quad
  \includegraphics[width=0.15\columnwidth]{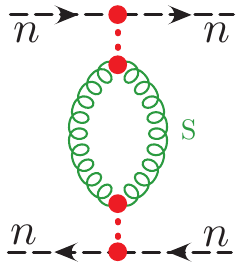}  
   \quad\raisebox{1.2cm}{$.$}
\end{align}
Here all three graphs are built from leading power Lagrangian interactions. The graph on the left has $\delta^{\rm conn}=4$ from $N^n=N^{nS}=1$ and $N^{\bn}=N^{\bn S}=0$, while the graph in the middle has $\delta^{\rm conn}=0$ (from $N^n=N^{nS}=2$, $N^{\bn}=N^{\bn S}=1$) and the graph on the far right has $\delta^{\rm conn}=2$ (from $N^n=2$, $N^{nS}=1$, $N^{\bn}=0$, and $N^{\bn S}=1$). (Although we have drawn these graphs with external collinear quarks, the same results are obtained if they are collinear gluons.) Thus the two graphs on the right seem to be enhanced by their value of $\delta^{\rm conn}$ relative to what we expect for leading power interactions in the $n$-collinear sector. However, all such enhanced graphs, which attempt to use Glauber interactions to join a disconnected set of $n$-collinear fields, vanish by RPI-III invariance symmetry~\cite{Manohar:2002fd}. Both collinear operators $O_n^i$ that contain the external fields scale as a single $\bn$, and the soft or $\bn$-collinear loops cannot yield dependence on any external $n$-collinear momentum to compensate this. Since scattering results must be invariant under an RPI-III transformation, enhanced contributions of this sort always vanish. This remains true even if power suppressed Lagrangian interactions are considered in these loops in order to make them have $\delta=4$ in the presence of the connectedness enhancement. The reason is that in the end we would have to obtain a result involving a $O_n^i$ and $O_n^j$ at leading power (modulated by logarithms), and we have already argued that no such leading power operator exists. Effectively since the Glauber operators only involve $n$-collinear fields in operators with a specific RPI-III scaling, they cannot be used to generate any loop diagram that leads to an enhanced contribution. This argument is also consistent with the fact that all scattering between collinear lines in the same direction is via collinear exchanges and collinear loop graphs, since this is just a boosted version of QCD. The same RPI-III argument applies if we consider external soft fields with $O_s^{i_n}$ and $O_s^{j_n}$, or with $O_\bn^i$ with $O_\bn^j$, since these products again have a non-trivial RPI-III scaling.  Note that it is possible for the connectedness term to cause a suppression, as it does for the graph in example 7 discussed above.

\section{Useful formulae}
\label{app:useful}

\subsection{Expansion of Adjoint Wilson Lines and Gluon Building Blocks} 

In \sec{basis} we made use of various coupling expansions of Wilson lines and composite operators in order to obtain the one and two soft gluon Feynman rules for the operators, which were used for the calculations in \sec{2sgluon}. This includes the Feynman rules for the final operator $O_S^{AB}$ given in \fig{twosoftgluon_feynrule}. Here we give results for some of those expansions for easy reference. 

We start with the expansion of the adjoint Wilson line in momentum space, where for brevity here we use $A_{sk}^C$ to denote the soft field $A_{s}^C$ with incoming momentum $k^\mu$. This gives
\begin{align}
{\cal S}_{\bn}^{AB} 
 & = \delta^{AB} + i g f^{ABC} \frac{\bn\cdot A_{sk}^C}{\bn\cdot k} 
 + g^2 \biggl[
 \frac{f^{C_2 AE}f^{C_1 BE}}{\bn\mcdot k_1\, \bn\mcdot (k_1\plus k_2)}
 +\frac{f^{C_1 AE}f^{C_2 BE}}{\bn\mcdot k_2\, \bn\mcdot (k_1\plus k_2)} 
  \biggr] \frac{\bn\cdot A_{sk_1}^{C_1}\bn\cdot A_{sk_2}^{C_2}}{2!}
 + \ldots 
  \,.
\end{align}
The analogous expansion for ${\cal S}_n^{T\,AB}$ is easily obtained by taking $A\leftrightarrow B$ and $\bn\to n$. In constructing the two-gluon Feynman rule from this operator either of the fields $A_{sk_1}^{C_1}$ or $A_{sk_2}^{C_2}$ can be contracted with either gluon, which cancels the $2!$.   For the adjoint building block gluon field we have the expansion
\begin{align}
 \big(g \tilde {\cal B}_{s\perp}^{n\mu} \big)^{AB} 
  &= -i g f^{ABC} \Bigl(A_{s\perp k}^{C\mu}
    - \frac{k_\perp^\mu}{n\cdot k}  \, n\cdot A^C_{sk} \Bigr) 
  + g^2 \Big( f^{C_1BE} f^{C_2AE} - f^{C_2BE} f^{C_1AE} \Big)
  \frac{A_{s\perp}^{C_1\mu} \, n\cdot A^{C_2}_{sk}}{n\cdot k} 
  \nn \\
  &\quad
   + g^2 \big(k_{1\perp}^\mu \plus k_{2\perp}^\mu \big)  \biggl[
   \frac{f^{C_2AE} f^{C_1BE}}{n\mcdot k_1\, n\mcdot(k_1\plus k_2)} +
   \frac{f^{C_1AE} f^{C_2BE}}{n\mcdot k_2\, n\mcdot(k_1\plus k_2)}\biggr] 
    \frac{n\cdot A^{C_1}_{sk_1}\: n\cdot A^{C_2}_{sk_2}}{2!} 
   \nn\\
  &\quad
- g^2 f^{C_1 AE} f^{C_2 B E}\: \frac{k_{2\perp}^\mu \, n\cdot A_{sk_1}^{C_1} 
    n\cdot A_{sk_2}^{C_2} }{n\cdot k_1\, n\cdot k_2}+ \ldots 
  \,,
\end{align}
with the analogous result for $g \tilde {\cal B}_{s\perp}^{\bn\mu}$ obtained by taking $n\to \bn$. Finally we note that slightly simpler results are obtained for certain combinations due to cancellations between the explicit Wilson line and those internal to the adjoint gluon building blocks, such as
\begin{align}
  \big(g \tilde {\cal B}_{s\perp}^{n\mu} {\cal S}_n^T\big)^{AB} 
  &= -i g f^{ABC} \Big( A_{s\perp}^{C\mu} - \frac{k_\perp^\mu}{n\cdot k} n\cdot A^C_{sk} \Big) 
  + g^2 f^{C_1BE} f^{C_2 AE}\, \frac{A_{s\perp}^{C_1\mu} \, n\cdot A^{C_2}_{sk_2}}{n\cdot k_2} 
  \\
  &\quad
  - g^2 \bigl(k_{1\perp}^\mu \plus k_{2\perp}^{\mu}\bigr)  \biggl[
   \frac{f^{C_1AE} f^{C_2BE}}{n\cdot k_1\, n\cdot(k_1\plus k_2)} +
   \frac{f^{C_2AE} f^{C_1BE}}{n\cdot k_2\, n\cdot(k_1\plus k_2)}\biggr]
    \frac{n\cdot A^{C_1}_{sk_1}\: n\cdot A^{C_2}_{sk_2}}{2!} + \ldots 
  \,,\nn\\[5pt]
  \big( {\cal S}_\bn g \tilde {\cal B}_{s\perp}^{\bn\mu} \big)^{AB} 
  &= -i g f^{ABC} \Big( A_{s\perp}^{C\mu} - \frac{k_\perp^\mu}{\bn\cdot k} \bn\cdot A^C_{sk} \Big) 
  - g^2 f^{C_1AE}f^{C_2BE}\, \frac{A_{s\perp}^{C_1\mu}\, \bn\cdot A^{C_2}_{sk_2}}{\bn\cdot k_2} 
   \nn \\
  &\quad
  + g^2 \bigl(k_{1\perp}^\mu \plus k_{2\perp}^{\mu}\bigr)  \biggl[
   \frac{f^{C_2AE} f^{C_1BE}}{\bn\cdot k_1\, \bn\cdot(k_1\plus k_2)} +
   \frac{f^{C_1AE} f^{C_2BE}}{\bn\cdot k_2\, \bn\cdot(k_1\plus k_2)}\biggr]
    \frac{\bn\cdot A^{C_1}_{sk_1}\: \bn\cdot A^{C_2}_{sk_2}}{2!} + \ldots 
  \,.\nn
\end{align}

\subsection{Useful Rapidity Regulated Integrals} \label{app:integrals}

In this section we tabulate results for a few integrals that required the rapidity regulator. We consider both the Glauber type loop integrals, which must be regulated, but do not introduce logarithmic divergences, as well as collinear and soft loop integrals that do induce logarithmic divergences.

The rapidity divergent $k^z$ integral that shows up in Glauber loops is
\begin{align}  \label{eq:kzint}
  \int_{-\infty}^{+\infty} \!\!\!  \ddslash\! k^z \:
   \frac{   |2k^z|^{-2\eta}\: \nu^{2\eta} }{( \minus 2k^z\plus 2A \plus i0)  }  
  &=  \int_0^\infty \!\! \frac{ \ddslash\! k^z \ (k^z)^{-2\eta}\: 
   (\nu/2)^{2\eta} }{2}
   \bigg[  \frac{1}{ \minus k^z\plus A \plus i0 }  + \frac{1}{  k^z\plus A \plus i0 }  \bigg]
\nn\\
  &= \frac{1}{4\pi} (\nu/2)^{2\eta}\, \pi \csc(2\pi \eta) 
   \Big[ (A+i0)^{-2\eta} - (-A-i0)^{-2\eta} \Big]
\nn\\
  &= \frac{1}{4\pi} \Big[ (\nu/2)^{2\eta}\, (-2i\pi)  \csc(2\pi \eta) 
    \sin(\pi\eta) \: (-i A)^{-2\eta} \Big]
\nn\\
  &= \Big(\frac{1}{4\pi}\Big) 
    \: \Big[ -i\pi +{\cal O}(\eta) \Big]   \,.
\end{align}
Note that the result at leading order in the limit $\eta\to 0$ is independent of $A$. This term is therefore also independent of the exact power in $\eta$, giving the same result whether the regulator is $|2k^z|^{-2\eta}$ or $|2k^z|^{-\eta}$ for this integral. We also obtain the same result if we swap $k^z\to -k^z$ in the original integral. We obtain the opposite sign for the ${\cal O}(\eta^0)$ term if the original integral appears instead with a $-i0$, which is the complex conjugate of the result in \eq{kzint}.  One common case where this integral appear is in evaluating
\begin{align}  \label{eq:k0kzint}
  \int\!\!  \ddslash\! k^0 \ddslash\! k^z  \frac{|2k^z|^{-2\eta} \nu^{2\eta}}
   {\big(k^+ -\Delta +i0\big)\big(k^- + \Delta' -i0\big)}
 = +\frac14 +{\cal O}(\eta) \,.
\end{align}

If we have two $k^+$ dependent propagators with poles on the same side, opposite to the $k^-$ propagator, then the integral vanishes at ${\cal O}(\eta^0)$
\begin{align}  \label{eq:k0kzint2}
 I_{2001} &=  \int\!\!  \ddslash\! k^0 \ddslash\! k^z  \frac{|2k^z|^{-2\eta} \nu^{2\eta}}
   {\big(k^+ -\Delta_1 +i0\big)\big(k^+ -\Delta_2 +i0\big)\big(k^- + \bar\Delta_1' -i0\big)}
 \\
 &= i \int\!\! \ddslash\! k^z  \frac{|2k^z|^{-2\eta} \nu^{2\eta}}
   {\big(-2k^z-\Delta_1-\bar\Delta_1' +i0\big)
    \big(-2k^z-\Delta_2-\bar\Delta_1' +i0\big)}
 = {\cal O}(\eta) 
  \,. \nn
\end{align}
To see this note that after doing the $k^0$ contour integral we have two poles on the same side of a $k^z$ integral that is convergent without the regulator. We could also keep the regulator, and transform the $k^z$ dependent integrand in \eq{k0kzint2} to longitudinal position space with \eq{fteta}. To obtain the spacetime picture we separately transform the regulator factor for each of the two Glauber exchanges contributing to \eq{k0kzint2}, obtaining 
\begin{align} \label{eq:I2001x}
 I_{2001}  
  & = -\frac{i}{4} \Big(\kappa_\eta \frac{\eta}{2}\Big)^{\!2}\! 
    \int_{-\infty}^{+\infty}\!\!\!\!\!\!\!\! 
    dx_1 dx_2 d\alpha_1 d\alpha_2\, 
   \theta(\alpha_1)\theta(\alpha_2) |x_1 x_2|^{-1+\eta} 
    e^{ik^z (x_1-x_2) -i\alpha_1(k^z\minus \Delta_{1\bar 1'})
     -i\alpha_2(k^z\minus \Delta_{2\bar 1'})}
\nn\\
  &=  -\frac{i}{4} \Big(\kappa_\eta \frac{\eta}{2}\Big)^{\!2}\! 
    \int_{-\infty}^{+\infty}\!\!\!\!\!\!\!\! 
    dx_1 dx_2 d\alpha_1 \, 
   \theta(\alpha_1)\theta(x_1\minus x_2\minus\alpha_1) |x_1 x_2|^{-1+\eta} 
    e^{i\alpha_1(\Delta_{1\bar 1'}+ i(x_1-x_2-\alpha_1) \Delta_{2\bar 1'}}
 \nn\\
 &=  -\frac{i}{4} \Big(\kappa_\eta \frac{\eta}{2}\Big)^{\!2}\! 
    \int_{-\infty}^{+\infty}\!\!\!\!\!\!\!\! 
    dx_1 dx_2 \, 
   \theta(x_1\minus x_2) |x_1 x_2|^{-1+\eta} 
    e^{ i(x_1-x_2) \Delta_{2\bar 1'}} 
   \Big[ (x_1\minus x_2) + {\cal O}(\eta)\Big]
 \nn\\ 
  &= {\cal O}(\eta) \,.
\end{align}
The presence of a vertex between the two Glauber exchanges at $x_1$ and $x_2$ leads to the extra restricted integral over $\alpha_1$ that ranges over the interval between these two Glauber exchanges. At leading order in the $x_j\to 0$ limit it simply gives a factor of $(x_1-x_2)$. This factor reduces the  divergences from $x_j\to 0$, such that they can no longer overcome the $\eta^2$ prefactor.
 
On the other hand if the $k^+$ dependent propagators are on opposite sides, then the result is nonzero at ${\cal O}(\eta^0)$,
\begin{align}  \label{eq:k0kzint3}
 I_{1101} &= \int\!\!  \ddslash\! k^0 \ddslash\! k^z  \frac{|2k^z|^{-2\eta} \nu^{2\eta}}
   {\big(k^+ -\Delta +i0\big)\big(k^+ +\Delta'' -i0\big)\big(k^- + \Delta' -i0\big)} 
  \nn\\
 &
 = \frac{-i}{\big(\Delta +\Delta'' -i0\big)}
   \int\!\!   \ddslash\! k^z  \frac{|2k^z|^{-2\eta} \nu^{2\eta}}
   {\big(2k^z + \Delta+ \Delta' -i0\big)} 
 = \frac{1}{4\big(\Delta +\Delta'' -i0\big)}  +{\cal O}(\eta) 
\,.
\end{align}

We can also consider rapidity regulated integrals that lead to $1/\eta$ divergences with corresponding logarithms. This occurs in both soft and collinear loops.   The basic rapidity divergent loop integral that appears in the soft eye graph after reducing the numerator (and including the Glauber 0-bin subtraction which removes the dependence on whether we use $\pm i0$ in the eikonal propagators), is
\begin{align} \label{eq:soft_eye_scalar_integral}
  & 
    \!\! \int\!\! \ddslash\!^{d} k\:
   \frac{(\iota^\epsilon \mu^{2\epsilon} |2k_z|^{-\eta}\,\nu^\eta) }
   { [k^2-m^2][(k+q)^2-m^2](\bn\cdot k)(n\cdot k)} 
 \nn\\
  &= - \frac{i}{8\pi^2 t}\: \bigg\{
    \frac{2\, g(\epsilon,\mu^2/t)}{\eta}\,  + \frac{1}{\epsilon^2} +\frac{1}{\epsilon} \ln\Big(\frac{\mu^2}{\nu^2}\Big) 
    + \ln\Big(\frac{\mu^2}{\nu^2}\Big) \ln\Big(\frac{\mu^2}{-t}\Big)
    - \frac12 \ln^2\Big(\frac{\mu^2}{-t}\Big) + \frac{\pi^2}{12}
    \bigg\}
  \,,
\end{align}
where $g(\epsilon,\mu^2/t)$ is defined in \eq{g}. The full loop integral for the soft eye graph was given in \eq{soft_eye}. Similarly, the rapidity divergent loop integral appearing in the soft flower graph is
\begin{align} \label{eq:soft_flower_integral}
   & \!\! \int\!\! \ddslash\!^{d} k
   \: \frac{ (\iota^\epsilon \mu^{2\epsilon} |2k_z|^{-\eta}\,\nu^\eta) }{ [k^2-m^2](n\cdot k)(\bn\cdot k)} 
   \\
   &= -\frac{i}{16\pi^2}\, 
    \bigg\{ \frac{2\, h(\epsilon,\mu^2/m^2)}{\eta} 
   - \frac{1}{\epsilon^2}- \frac{1}{\epsilon} \ln\Big(\frac{\mu^2}{\nu^2}\Big) 
   -  \ln\Big(\frac{\mu^2}{\nu^2}\Big) \ln\Big(\frac{\mu^2}{m^2}\Big)
   \!+\! \frac12 \ln^2\Big(\frac{\mu^2}{m^2}\Big) \!+\! \frac{\pi^2}{12} \bigg\}
   \,, \nn 
\end{align}
where $h(\epsilon,\mu^2/m^2)$ is defined in \eq{h}.
The rapidity divergent loop integral that appears for the collinear V-graph is
\begin{align} \label{eq:ncollinearV_integral}
&   \int\!\! \ddslash\!^{d} k\:
  \frac{(\iota^\epsilon \mu^{2\epsilon} |\bn\cdot k|^{-\eta} \nu^\eta) \: k_\perp\!\cdot\! (k_\perp\plus q_\perp)\,\bn\!\cdot\!(k\plus p_3)}{[k^2-m^2][(k+q)^2-m^2](k+p_3)^2\, ( \bn\cdot k)}
  \\
&=  - \frac{i}{16\pi^2 }\: \bigg\{
    \frac{g(\epsilon,\mu^2/t)}{\eta}\,  -\frac{1}{\epsilon} \ln\Bigl(\frac{\nu}{\bn\cdot p_3}\Bigr)
    - \ln\Big(\frac{\nu}{\bn\cdot p_3}\Big) \ln\Big(\frac{\mu^2}{-t}\Big)
    -\frac{3}{2\epsilon} - \frac{3}{2} \ln\Big(\frac{\mu^2}{-t}\Big)  
 \nn\\
 &\qquad\qquad \quad
   + \frac{3}{4} \ln\Big(\frac{m^2}{-t}\Big)  - \frac{15}{8}  + \frac{\pi^2}{3}
    \bigg\}
 \,. \nn
\end{align}
In the full result for the loop integral for the V-graph, given in \eq{oneloop_collinear_V}, the dependence on $\ln(m^2)$ cancels. 
Similarly, the rapidity divergent loop integral for the collinear Wilson line graph is
\begin{align}  \label{eq:oneloop_collinear_W_loop}
  & 
   \int\!\! \ddslash\!^{d} k\:
  \frac{(\iota^\epsilon \mu^{2\epsilon} |\bn\cdot k|^{-\eta} \nu^\eta) \ \bn\cdot (k+p_3)}{[k^2-m^2](k+p_3)^2\, (\bn\cdot k)}
   \\
  &
  =  \frac{i}{16\pi^2}\,
  \bigg\{ \frac{h(\epsilon,\mu^2/m^2)}{\eta} 
   +\frac{1}{\epsilon} \ln\Big(\frac{\nu}{\bn\cdot p_3}\Big) 
   + \ln\Big(\frac{\nu}{\bn\cdot p_3}\Big) \ln\Big(\frac{\mu^2}{m^2}\Big)
   + \frac{1}{\epsilon} + \ln\Big(\frac{\mu^2}{m^2}\Big) 
   + 1 - \frac{\pi^2}{6} \bigg\}
  \,. \nn
\end{align}

\subsection{Three Gluon Feynman Rule for ${\cal O}_n^{gA}$}

The three collinear gluon Feynman rule for the ${\cal O}_n^{gA}$ operator, which was used in \sec{oneloopregge}, is given in terms of a vertex function $V_1^{\mu\nu\lambda}$ by
\begin{align}  \label{eq:feynrule_ggg}
\includegraphics[width=0.25\columnwidth]{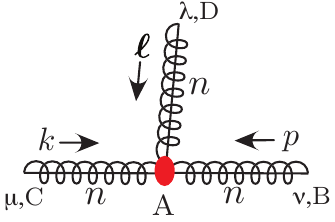} 
 \hspace{0.cm} 
\raisebox{1.2cm}{
\begin{minipage}{4.65in}
  $=- g\,\bn\cdot k\, f^{ACE}f^{BDE}\, V_1^{\mu\nu\lambda}(k,\ell,p)
   - g\,\bn\cdot p\, f^{ABE}f^{DCE}\, V_1^{\nu\lambda\mu}(p,k,\ell)$ 
   \\[4pt]
   \phantom{x}\hspace{0.1cm}
   $\ \ - g\,\bn\cdot \ell\, f^{ADE}f^{CBE}\, V_1^{\lambda\mu\nu}(\ell,p,k)
   \hspace{+4.3cm}$
\end{minipage}
\hspace{0.cm} }
  \nn \\
 \hspace{-0.3cm} 
\raisebox{-.2cm}{
\begin{minipage}{6.4in}
 $V_1^{\mu\nu\lambda}(k,\ell,p)
  =
   \mbox{\Large $\frac{g_\perp^{\mu\nu}\bn^\lambda}{\bn\cdot\ell}$}  
  -\mbox{\Large $\frac{g_\perp^{\mu\lambda}\bn^\nu}{\bn\cdot p}$}  
  -\mbox{\Large $\frac{\ell_\perp^\mu \bn^\nu\bn^\lambda}
     {\bn\cdot k\,\bn\cdot p}$} 
  +\mbox{\Large $\frac{p_\perp^\mu \bn^\nu\bn^\lambda}{\bn\cdot k\,\bn\cdot\ell}$}  
  +\mbox{\Large $\frac{k_\perp^\lambda \bn^\mu\bn^\nu}{\bn\cdot k\,\bn\cdot p}$} 
   -\mbox{\Large $\frac{k_\perp^\nu \bn^\mu\bn^\lambda}{\bn\cdot k\,\bn\cdot\ell}$}
   +\mbox{\Large $\frac{\bn^\mu\bn^\nu \bn^\lambda}{(\bn\cdot k)^2}
    \big(\frac{k_\perp\!\cdot\ell_\perp}{\bn\cdot p} $}
   -\mbox{\Large $\frac{k_\perp\!\cdot p_\perp}{\bn\cdot \ell} \big).$} 
   $
\end{minipage}
\hspace{0.cm} }
\end{align}
Here the operator ${\cal O}_n^{gA}$ has an incoming Glauber exchange momentum $q_\perp$ and  $p=-k-\ell-q_\perp$.

\subsection{Wilson Line Feynman Rules for Various Directions}
\label{app:Wdirection}

For easy reference we record here some notation and Feynman rules for various Wilson lines. For $n$-collinear Wilson lines $W_n=W_n[\bn\cdot A_n]$ the lines with various specified directions are:
\begin{align}
  W_n(-\infty,0)  &= {\textrm{P}} 
    \exp \Big( ig\! \int^0_{-\infty }
 \!\!\!\!\!
  ds \ \bn \mcdot
 A_{n}(x_s^\mu ) \Big) 
 \,,
  & W_n(0,\infty) &= \overline {\textrm{P}} 
    \exp \Big( \!\!-\!\! ig\! \int_0^{\infty }
 \!\!\!\!\!
  ds \: \bn \mcdot
 A_{n}(x_s^\mu) \Big)
   \,, \\
  W_n^\dagger(-\infty,0)  &=  \overline {\textrm{P}} 
    \exp \Big(\!\!-\!\! ig\! \int^0_{-\infty }
 \!\!\!\!\!\!
  ds \: \bn \mcdot
 A_{n}(x_s^\mu  ) \Big)  
  \,, 
 & W^\dagger_n(0,\infty)
 &= {\textrm{P}} 
    \exp \Big(  ig\! \int_0^{\infty }
 \!\!\!\!\!
  ds \ \bn \mcdot
 A_{n}(x_s^\mu  ) \Big)
   \,, \nn
\end{align}
where $x_s^\mu = x^\mu + s (\bn^\mu/2)$. Thus the lines have one end at the space-time point $x^\mu=(n\cdot x,\bn\cdot x,x_\perp)$, extend along the $\bn^\mu$ light-cone, and have the other end at $(\pm\infty,\bn\cdot x,x_\perp)$. 
Here, as the notation implies, we have $[W_n(-\infty,0)]^\dagger = W_n^\dagger(-\infty,0)$ and $[W_n(0,\infty)]^\dagger = W_n^\dagger(0,\infty)$.

For an incoming gluon with momentum $k$ the 1-gluon Feynman rules for $n$-collinear Wilson lines in various directions are
\begin{align}
 & \text{$n$-line:}
 & W_n(-\infty,0) &
 & W^\dagger_n(0,\infty) &
 & W_n^\dagger(-\infty,0) &
 & W_n(0,\infty) & 
\nn \\
 & \text{1-gluon:}
 &\frac{-g\,\bn^\mu T^A}{\bn\cdot k + i0} &
 & \frac{-g\,\bn^\mu T^A}{-\bn\cdot k + i0} &
 & \frac{-g\,\bn^\mu T^A}{-\bn\cdot k - i0} &
 & \frac{-g\,\bn^\mu T^A}{\bn\cdot k - i0} & 
\\
 & \text{$s$-line:}
 & S_\bn(-\infty,0) &
 & S^\dagger_\bn(0,\infty) &
 & S_\bn^\dagger(-\infty,0) &
 & S_\bn(0,\infty) & 
\nn 
\,.
\end{align}
As indicated, the results for the soft Wilson lines $S_\bn[\bn\cdot A_s]$ are the same as for $W_n[\bn\cdot A_n]$ but just involve the soft gluon field $A_s^\mu$. The Wilson lines and Feynman rules for the $\bn$-collinear Wilson lines $W_\bn[n\cdot A_\bn]$ are obtained by taking $n\leftrightarrow \bn$ in these results, and those for $S_n[n\cdot A_s]$ are the same as those for $W_\bn[n\cdot A_\bn]$.

\subsection{Conventions for Antiquark Feynman Rules}
\label{app:Antiquarks}

Throughout this paper we have adopted the convention for antiquarks which yields Feynman rules that directly parallel those for quarks, removing the need for the inclusion of extra minus signs. This differs from the convention often used in SCET (following eg.~Peskin~\cite{Peskin}) where the large label momentum $\bn\cdot p$ follows the fermion number flow, and quarks ($\bn\cdot p>0$) and antiquarks ($\bn\cdot p<0$) are combined in a single collinear propagator as
\begin{align}
  i \frac{\nslash}{2} \frac{\bn\cdot p}{p^2+i0}
  =  i \frac{\nslash}{2} \frac{\theta(\bn\cdot p)}{n\cdot p + \frac{p_\perp^2}{\bn\cdot p}+i0}
   + i \frac{\nslash}{2} \frac{\theta(-\bn\cdot p)}{n\cdot p + \frac{p_\perp^2}{\bn\cdot p}-i0}  \,.
\end{align}
Here we instead follow the convention commonly used in the spinor-helicity community (eg.~\cite{Dixon:1996wi}) where charge conjugate fields are used for the antiquarks. This leads to the following differences in Feynman rules
\begin{align}
& \text{\underline{Antiquark rules:}} \nn\\ 
 & 
 & \text{Peskin convention}\qquad &
 & \text{Spinor-helicity convention (here)} &
\nn \\
 & \text{color matrix:}
 & T^A\qquad\qquad\quad &
 & \bar T^A  \qquad\qquad\qquad\qquad &
\nn\\
 & \text{$n$-propagator (physical $p^\mu$):}
 & \frac{i\nslash/2}{-n\cdot p - \frac{p_\perp^2}{\bn\cdot p}-i0 }\qquad &
 & \frac{i\nslash/2}{n\cdot p + \frac{p_\perp^2}{\bn\cdot p}+i0 }\qquad\qquad\quad &
\nn\\
 & \text{external antiquark factor:} 
 & (-1)\qquad\qquad\quad &
 & (+1)\qquad\qquad\qquad\qquad &  
\end{align}
With the spinor-helicity convention there is a direct parallel between quark and antiquark results since the only difference is the switch of the color representation from $3$ to $\bar 3$. 

\section{Glauber Exponentiation Calculations}  
\label{app:expcalcs}

\subsection{Abelian Exponentiation of Glaubers at the Integrand level}  \label{app:abelianexp}

In this section we briefly discuss how the summation of forward scattering Glauber exchange graphs can be done in an abelian theory, without the need to introduce the $\eta$ rapidity regulator. For the abelian theory the box and cross-box Glauber exchange graphs can be combined  at the integrand level to obtain obtain well defined results as long as we impose that the integrand is regulated in a manner that retains an invariance under using different possible routings for the loop momenta. The final result for the sum of graphs obtained here is the same as the abelian limit of \eq{fwdscatterphase}, taking $T^A\otimes \bar T^A\to -1$ in $\phi(b_\perp)$. For simplicity we work with the kinematics specified in \eqs{fwdkinematics}{perpkinematics}. 
 
First consider the sum of the one-loop box and crossed box graphs symmetrized over two momentum routings,
\begin{align} \label{eq:abbox1}
 &   \raisebox{0.9cm}{$\dfrac1{2!} \Bigg\{$ }
 \includegraphics[width=0.17\columnwidth]{figs/Glaub_ptnl_1box}
  \raisebox{0.9cm}{+}
 \includegraphics[width=0.17\columnwidth]{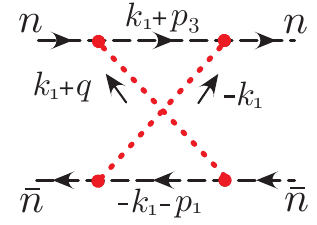}
   \raisebox{0.9cm}{$+(k_1\to -k_1-q)\ \Bigg\}$ }
   \nn \\
   &= 2  {\cal S}^{n\bn}\:   \frac{g^4}{2!} I^{(1)}_\perp(q_\perp)
   \!\! \int\!\! \ddslash\!k_1^+\ddslash\!k_1^-\, 
   \bigg[\frac{1}{ (k_1^+ \minus \Delta_1^n)(-k_1^- \minus \Delta_1^\bn)}
      + \frac{1}{ (k_1^+ \minus \Delta_1^n)(k_1^- \minus \Delta_1^\bn)} 
  + (k_1^\pm \to -k_1^\pm)
 \bigg]
    \nn\\[0pt]
  &= 2  {\cal S}^{n\bn}\:   \frac{g^4}{2!} I^{(1)}_\perp(q_\perp)
   \!\! \int\!\! \ddslash\!k_1^+\ddslash\!k_1^-\,
  \bigg[\frac{-2 \Delta_1^n}
        { (k_1^+ \minus \Delta_1^n)(-k_1^+ \minus \Delta_1^n)}
       \frac{-2 \Delta_1^\bn}
        { (k_1^- \minus \Delta_1^\bn)(-k_1^- \minus \Delta_1^\bn)} 
 \bigg]
    \nn\\[0pt]
   & =  2  {\cal S}^{n\bn}\: \frac{(ig^2)^2}{2!}\: I^{(1)}_\perp(q_\perp)
   \,,
\end{align}
where ${\cal S}^{n\bn}$ is given in \eq{Snbn}, $I_\perp^{(1)}$ is given in \eq{I1}, and here $\Delta_1^n = (\vec k_\perp + \vec p_{3\perp})^2/p_2^- - p_3^+ -i0$ and $\Delta_1^\bn = (\vec k_\perp -\vec p_{4\perp})^2/p_1^+ - p_4^- -i0$. The two displayed diagrams symmetrize over attachments to the bottom line while holding the momentum routing through the top collinear propagator fixed, whereas the $k_1\to -k_1-q$ analogs symmetrize over attachments to the top line holding the momentum routing in the bottom propagator fixed. The sum of propagators falls off quadratically for both $k_1^+\to \pm \infty$ and $k_1^- \to \pm \infty$, and hence both of these can be done by contours, simply giving a $(-i)^2$. The result in \eq{abbox1} is the same as that found with the $\eta$ regulator  in the abelian limit for the box graph alone, see \eq{2g}.

For three Glauber exchanges we must symmetrize over $3!$ choices of momentum routings through the Glauber lines in order to simultaneously symmetrize over attachments to the top and bottom collinear lines
\begin{align}
 \label{eq:abbox2}
 &   \raisebox{0.6cm}{$\dfrac1{3!} \Bigg\{$ } \hspace{-0.1cm}
 \includegraphics[width=0.15\columnwidth]{figs/Glaub_ptnl_2box}
  \hspace{-0.1cm}\raisebox{0.6cm}{+}
 \includegraphics[width=0.15\columnwidth]{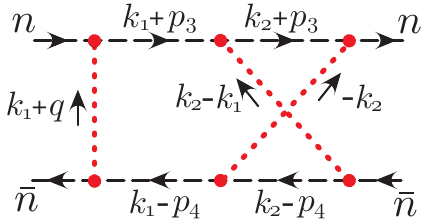}
  \hspace{-0.05cm}\raisebox{0.6cm}{+}\hspace{-0.05cm}
 \includegraphics[width=0.15\columnwidth]{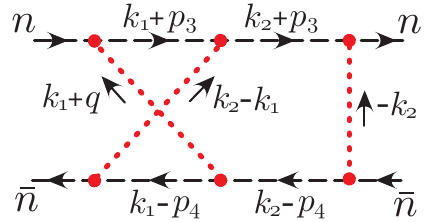}
  \raisebox{0.6cm}{+} \hspace{-0.1cm}
 \includegraphics[width=0.15\columnwidth]{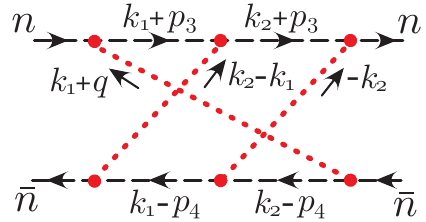}
  \hspace{-0.05cm}\raisebox{0.6cm}{+}\hspace{-0.05cm}
 \includegraphics[width=0.15\columnwidth]{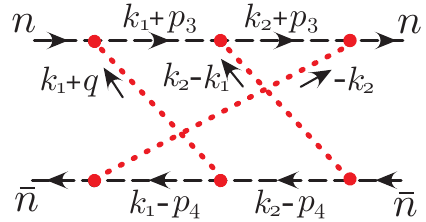}
  \hspace{-0.05cm}\raisebox{0.6cm}{+}\hspace{-0.15cm}
 \includegraphics[width=0.15\columnwidth]{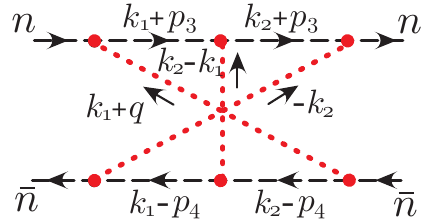}
 \nn\\
 &\qquad   \raisebox{0.6cm}{$+ \text{5 momentum routing permutations swapping\ } \{ k_1+q, k_2-k_1, -k_2 \} \Bigg\}$ }
   \nn \\
   &= -2 i  {\cal S}^{n\bn}\,  g^6 I^{(2)}_\perp(q_\perp) \,
   \frac{1}{3!} \! \int\!\! \ddslash\!k_1^+\ddslash\!k_1^-\ddslash\!k_2^+\ddslash\!k_2^-\, 
\bigg[
   \frac{1}{(k_1^+ \minus \Delta_1^n)(k_2^+ \minus \Delta_2^n)}
 + \frac{1}{(k_1^+ \minus \Delta_1^n)(k_1^+\minus k_2^+ \minus \Delta_{12}^n)} 
\nn\\
 &\qquad
 + \frac{1}{(-k_1^+\plus k_2^+ \minus \Delta_{12}^n)(k_2^+ \minus \Delta_2^n)} 
 + \frac{1}{(-k_1^+\plus k_2^+ \minus \Delta_{12}^n)(-k_1^+\minus \Delta_1^n)}  
 + \frac{1}{(k_2^+\minus \Delta_{2}^n)(k_1^+\minus k_2^+ \minus\Delta_{12}^n)} 
\nn\\
&\qquad
 + \frac{1}{(-k_2^+\minus \Delta_{2}^n)(-k_1^+\minus \Delta_{1}^n)} 
  \bigg] \bigg[ \text{same 6-terms with } k_{1,2}^+ \to k_{1,2}^-,\ \Delta_i^n\to \Delta_i^\bn \bigg]
    \nn\\[4pt]
&= 2  {\cal S}^{n\bn}\:   \frac{(ig^2)^3}{3!} I^{(2)}_\perp(q_\perp)
  \,,
\end{align}
where $I_N^{(2)}$ is given in \eq{I2}.
Again the four $k_i^\pm$ integrals each give a factor of $(-i)$. To see this note that the sum of six terms depending on $k_{1,2}^+$ gives a result that falls off with the 4'th power as $k_{1,2}^+\to \pm\infty$, and with two powers if either $k_1^+\to\infty$ or $k_2^+\to \infty$ individually. Therefore with these sums of terms the integrals are all well defined and can be done by contours. We can do the contour integral without combining denominators as long as we close the contour the same way for all 6-terms. Closing both contours below, only the first term contributes. (The same result is of course obtained if we first combine denominators.) The result in \eq{abbox2} matches \eq{3g} in the abelian limit.

The pattern is clear, so the graphs with $N+1$ Glauber rungs are no more difficult,
\begin{align} \label{eq:abboxN}
 &   \raisebox{0.55cm}{$\dfrac1{(N\plus 1)!} \Bigg\{ \Bigg[$ } \hspace{-0.1cm}
 \includegraphics[width=0.19\columnwidth]{figs/Glaub_ptnl_nbox}
  \hspace{0.15cm}\raisebox{0.55cm}{+ $[(N\plus 1)!\minus 1]$ \!\!\!\!
  \text{ crossed graphs} \!\!\!\! $\Bigg]$ }
 \raisebox{0.6cm}{$\!\!\! + [(N\plus 1)!\minus 1]$
  \parbox{2.8cm}{\text{\small momentum routing}\\[-5pt] \text{\small permutations}}  
   $\Bigg\}$ }
   \nn \\
   &= -2 {\cal S}^{n\bn}\, (-ig^2)^{N+1} I^{(N)}_\perp(q_\perp) \,
   \frac{1}{(N\plus 1)!} \! \int\!\! \ddslash\!k_1^+\ddslash\!k_1^-\cdots \ddslash\!k_N^+\ddslash\!k_N^-\, 
\bigg[
   \frac{1}{(k_1^+ \minus \Delta_1^n)\cdots (k_N^+ \minus \Delta_N^n)}
\nn\\
 &\qquad
  \raisebox{0.cm}{$+ [(N\plus 1)!\minus 1]$
  \parbox{2.8cm}{\text{\small crossed graph}\\[-5pt] \text{\small propagator terms}}  $\bigg]$ } 
  \!\! \bigg[ \text{same terms with } k_{i}^+ \to k_{i}^-,\ \Delta_i^n\to \Delta_i^\bn \bigg]
    \nn\\[4pt]
&= 2  {\cal S}^{n\bn}\:   \frac{(ig^2)^{N+1}}{(N\plus 1)!}   
  I^{(N)}_\perp(q_\perp)
  \,,
\end{align}
where $I^{(N)}_\perp$ is given in \eq{IN}. There are $(N+1)!$ terms in the sum of $k_i^+$ dependent propagators, and each one has poles only either above or below for each $k_i^+$ or $k_i^-$.\footnote{To see this for $k_i^+$ note that there are only two Glauber exchanges that depend on $k_i^+$, with incoming $k_i^+-k_{i-1}^+$ or incoming $k_{i+1}^+-k_i^+$. Which ever one of these vertices occurs first from the left induces a dependence on $k_i^+$ in the collinear propagators which follow this vertex, until the attachment of the other one, after which the collinear propagators no longer have any dependence on $k_i^+$. So the sign for the $k_i^+$ poles that appear in a given propagator term are determined by which of the two vertices attach first, and there are never simultaneously poles on both sides of the contour for a single term. The proof for $k_i^-$ is the same.} Again the integrals converge as $k_i^\pm\to \pm\infty$ once we consider the sum of all terms.  Finally, there is a single term with the poles in all $k_i^\pm$ below so closing the global contour in this manner simply gives $(-i)^{2N}$, and thus the result in \eq{abboxN}.  The result in \eq{abboxN} matches \eq{NgFT} in the abelian limit.  Thus we also obtain the exponentiated results of \sec{exponentiation} for the abelian case by adding integrands in this manner. 

This analysis demonstrates that in the more general non-abelian context, the $\eta$ regulator is properly regulating the Glauber sector of the theory, separating it from the soft and collinear sectors, as opposed to enforcing a particular result on the sum of Glauber exchange graphs.

\subsection{Exponentiation for Active-Spectator Exchanges}  
\label{app:ASexp}

In this appendix we derive the result for iterated Glauber exchange on active-spectator lines. Unlike for spectator-spectator scattering, this result requires the $\eta$-regulator both for the single exchange integral, and for the ladder iterations, therefore it cannot simply be inferred from the exponentiation result in \sec{exponentiation}.
 
Starting with the Drell-Yan case, we first repeat the single Glauber exchange calculation in \eq{GAS_result}, but carry out the intermediate regulator dependent integral in Fourier space following the notation introduced in \sec{exponentiation}, in order to setup the procedure for the $N$-exchange diagram. Writing the final result in the transverse Fourier space we have
\begin{align}
& \raisebox{-1cm}{ 	
\includegraphics[width=0.18\columnwidth]{figs/end_AS}
  }
 = 2i\,S^\gamma\frac{n\mcdot p_2\, n\mcdot(\bar P\minus p_2)}
   {n\mcdot \bar P \, \vec p_{2\perp}^{\:2}}
  \int_{-\infty}^{+\infty}\!\! \ddslash\! k^z \ddslash\!^{d'}\!k_\perp 
  \frac{G^0(k_\perp) |2 k^z|^{-\eta} \nu^{\eta} }
  {[2k^z \minus\Delta_1'\minus \bar\Delta_1\plus i0]
   [-\Delta_1\minus \Delta_1'\plus i0]}
\nn\\
& = -S^\gamma\frac{n\mcdot p_2\, n\mcdot(\bar P\minus p_2)}
   {n\mcdot \bar P \, \vec p_{2\perp}^{\:2}} 
   \!\int\! \frac{\ddslash\!^{d'}\!k_\perp G^0(k_\perp\!)}
   {\Delta_1\plus\Delta_1'}\,
   \Big(\kappa_\eta \frac{\eta}{2}\Big)\! 
    \int_{-\infty}^{+\infty}\!\! \ddslash\! k^z\, dx\, d\alpha\, 
   \theta(\alpha) |x|^{-1+\eta} e^{i\alpha(k^z\plus \Delta)-ik^z x}
\nn\\
& = -S^\gamma\frac{n\mcdot p_2\, n\mcdot(\bar P\minus p_2)}
   {n\mcdot \bar P \, \vec p_{2\perp}^{\:2}} 
   \frac{\bn\mcdot p_1\, \bn\mcdot(P\minus p_1)}{\bn\mcdot P} 
  \!\int\! \frac{\ddslash\!^{d'}\!k_\perp G^0(k_\perp\!)}
   {(\vec k_\perp \plus \vec p_{1\perp})^2}\,
   \Big(\kappa_\eta \frac{\eta}{2}\Big)\! 
    \int_{-\infty}^{+\infty}\!\!  dx\,  
   \theta(x) |x|^{-1+\eta} e^{i x \Delta}
\nn\\
& = S^\gamma
  \!\int\! \ddslash\!^{d'}\!k_\perp G^0(k_\perp\!)
   E(p_{1\perp}+k_\perp,p_{2\perp})
   \Big(\kappa_\eta \frac{\eta}{2}\Big)\! 
   \bigg[ \frac{1}{\eta} + {\cal O}(\eta^0) \bigg]
\nn\\
& = \frac{S^\gamma}{2}
  \!\int\! \ddslash\!^{d'}\!k_\perp 
   \int\!\! d^{d'}b_\perp  e^{-i\vec k_\perp\cdot \vec b_\perp}
   \tilde G^0(b_\perp\!)
   \int\!\! d^{d'}b_\perp' e^{-i(\vec k_\perp\plus\vec p_{1\perp})\cdot \vec b_\perp'}  \tilde E(b_\perp',p_{2\perp})
\nn\\
& = S^\gamma
   \int\!\! d^{d'}b_\perp  e^{i\vec p_{1\perp}\cdot \vec b_\perp}\,
   \Big[ \frac12 \tilde G^0(b_\perp\!)\Big] \,\tilde E(-b_\perp,p_{2\perp})
 \,.
\end{align}
Recall that we use the notation $d'=d-2$, and that the Fourier transform $\tilde G^0(b_\perp)=i\phi(b_\perp)$ with $\phi(b_\perp)$ given by \eq{phi}. The end factor $E(p_{1\perp},p_{2\perp})$ is defined in \eq{end_tree}.  The dependence on $\Delta=-(\Delta_1'+\bar\Delta_1)/2$ drops out at leading order in $\eta$.  Next we carry out the analogous calculation but with $N$ Glauber exchanges between these two lines. Closing all the $k_i^0$ contours above we have
\begin{align} \label{eq:AS_Nrungs}
& \raisebox{-1cm}{ 	
\includegraphics[width=0.18\columnwidth]{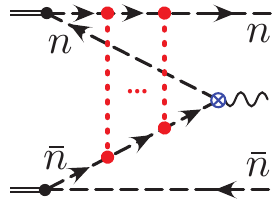}
  }
 =-(i^2 2)^N S^\gamma\frac{n\mcdot p_2 n\mcdot(\bar P\minus p_2)}
   {n\mcdot \bar P \,(- \vec p_{2\perp}^{\:2})}
  \!\! \int\!\! \ddslash\!^{d}\!k_1\cdots  \ddslash\!^{d}\!k_N\,
  \frac{G^0(k_{1}^\perp)G^0(k_{2}^\perp\minus k_{1}^\perp)\cdots 
    G^0(k_{N}^\perp\minus k_{N-1}^\perp)}
  { [k_1^-\minus\bar\Delta_1\plus i0]\cdots [k_N^-\minus\bar\Delta_N\plus i0] }
  \nn\\[-15pt]
 &\hspace{4cm} \times
  \frac{|2 k_1^z|^{-\eta}|2 k_2^z\minus 2k_1^z|^{-\eta} \cdots
    |2 k_N^z\minus 2k_{N\minus 1}^z|^{-\eta}\: \nu^{N\eta} }
  { [-k_N^+\minus\Delta_1'\plus i0][k_N^+\minus\Delta_1\plus i0]
   [k_N^+\minus k_1^+ \minus \Delta_2\plus i0]\cdots 
   [k_N^+\minus k_{N-1}^+ \minus \Delta_N\plus i0]}
\nn\\[4pt]
& = (2i)^N  S^\gamma \frac{n\mcdot p_2 n\mcdot(\bar P\minus p_2)}
   {n\mcdot \bar P \,\vec p_{2\perp}^{\:2}}
 \!\! \int\!\! \ddslash\!^{d\minus 1}\!k_1\cdots  \ddslash\!^{d\minus 1}\!k_N\,
 \frac{G^0(k_{1}^\perp)G^0(k_{2}^\perp\minus k_{1}^\perp) \cdots  
 G^0(k_{N}^\perp\minus k_{N-1}^\perp) } { [-\Delta_1\minus\Delta_1'] }
 \nn\\
 &\hspace{4cm} \times
 \frac{ |2 k_1^z|^{-\eta}|2 k_2^z\minus 2k_1^z|^{-\eta} \cdots 
  |2 k_N^z\minus 2k_{N\minus 1}^z|^{-\eta}\: \nu^{N\eta}} 
  {[2k_1^z\plus 2\Delta_{\bar 1 1' 2}\plus i0] \cdots [2k_N^z \plus 2 \Delta_{1'\bar N} \plus i0]}
  \nn\\
& =   S^\gamma 
 \!\! \int\!\! \ddslash\!^{d'}\!k_{1\perp}\cdots  \ddslash\!^{d'}\!k_{N\perp}\,
 E(p_{1\perp}\plus k_{1\perp},p_{2\perp}) 
 G^0(k_{1}^\perp)G^0(k_{2}^\perp\minus k_{1}^\perp) \cdots  
 G^0(k_{N}^\perp\minus k_{N-1}^\perp)  
\Big(\kappa_\eta \frac{\eta}{2} \nu^\eta\Big)^N 
 \nn\\*
 &\hspace{0.1cm} \times 
 \!\!\int_{-\infty}^{+\infty}\!\!\!\!\!\!
  \ddslash\! k_1^z\cdots \ddslash\! k_N^z
  dx_1\cdots dx_N \, d\alpha_1 \cdots d\alpha_N
  \frac{\theta(\alpha_1)\cdots\theta(\alpha_N)}
  {| x_1\cdots x_N |^{1-\eta} } \,
 e^{i(\alpha_1+ x_1-x_2)k_1^z + \ldots+ i(\alpha_N+x_N)k_N^z }\:
 e^{i\alpha_1 \Delta_{\bar 1 1'2}+\ldots}
  \nn\\
& =  S^\gamma 
 \!\! \int\!\! \ddslash\!^{d'}\!k_{1\perp}\cdots  \ddslash\!^{d'}\!k_{N\perp}\,
 E(p_{1\perp}\plus k_{1\perp},p_{2\perp}) 
 G^0(k_{1}^\perp)G^0(k_{2}^\perp\minus k_{1}^\perp) \cdots  
 G^0(k_{N}^\perp\minus k_{N-1}^\perp)  
 \nn\\
 &\hspace{0.1cm} \times \!
 \Big(\kappa_\eta \frac{\eta}{2} \nu^\eta\Big)^N \!\!
  \int_{-\infty}^{+\infty}\!\!\!\! dx_1\cdots dx_N \, 
  \frac{\theta(x_2\minus x_1)\theta(x_3\minus x_2)\cdots\theta(x_{N}\minus x_{N-1})\theta(-x_N)}  {| x_1\cdots x_N |^{1-\eta} } \,
 e^{i(x_2-x_1) \Delta_{\bar 1 1'2}+\ldots  -i x_N\Delta_{1'\bar N} }
 \nn\\
& =  S^\gamma 
 \!\! \int\!\! \ddslash\!^{d'}\!k_{1\perp}\cdots  \ddslash\!^{d'}\!k_{N\perp}\,
 E(p_{1\perp}\plus k_{1\perp},p_{2\perp}) 
 G^0(k_{1}^\perp)G^0(k_{2}^\perp\minus k_{1}^\perp) \cdots  
 G^0(k_{N}^\perp\minus k_{N-1}^\perp)  
 \bigg[ \frac{1}{N!\, 2^N} + {\cal O}(\eta) \bigg]
 \nn\\
& = S^\gamma
   \int\!\! d^{d'}b_\perp  e^{i\vec p_{1\perp}\cdot \vec b_\perp}\,
   \frac{1}{N!} \bigg[ \frac12 \tilde G^0(b_\perp\!) \bigg]^N \,\tilde E(-b_\perp,p_{2\perp})
 \nn\\
& = S^\gamma
   \int\!\! d^{d'}b_\perp  e^{i\vec p_{1\perp}\cdot \vec b_\perp}\,
   \frac{1}{N!} \bigg[ \frac{i\phi(b_\perp)}{2} \bigg]^N \,\tilde E(-b_\perp,p_{2\perp})
 \,.
\end{align}
At intermediate steps we defined $\Delta_{\bar 1 1'2}=-(\bar\Delta_1+\Delta_1'+\Delta_2)/2$, etc., but to leading order in $\eta$ the answer is independent of these factors. The only such factors which contribute are $\Delta_1+\Delta_1'$, which contributes to give the $E(p_{1\perp}+k_{1\perp},p_{2\perp})$ (these $\Delta$s were defined in \eq{Deltai}). At an intermediate step we see that the longitudinal coordinates are ordered as $x_1 < x_2 <\cdots < x_N < 0$, that is, they occur before the hard scattering vertex at $x=0$. As $\eta\to 0$ the ${\cal O}(\eta^0)$ result comes from the limit where all $x_j\to 0$. Summing over the number of Glauber exchanges in \eq{AS_Nrungs} from $N=0$ to $N=\infty$ we get
\begin{align}
& \raisebox{-1cm}{ 	
\includegraphics[width=0.18\columnwidth]{figs/end_AS_sum}
  }
 = S^\gamma
   \int\!\! d^{d'}b_\perp  e^{i\vec p_{1\perp}\cdot \vec b_\perp}\:
   e^{i\phi(b_\perp)/2} \:\tilde E(-b_\perp,p_{2\perp}) 
\,.
\end{align}
Thus we see that the phase for this sum of active-spectator exchanges is $\phi(b_\perp)/2$.

For the hard scattering case, we can sum up the ladder graphs for active-spectator scattering in a similar manner. For one exchange we again repeat the calculation of \eq{GASdis} using the regulator in position space, and writing the result in the transverse Fourier space:
\begin{align} \label{eq:GASdis2}
 & \raisebox{-0.7cm}{ 
\includegraphics[width=0.18\columnwidth]{figs/enddis_AS}
  }
  = -2\,S^\gamma
  \int\!\! \ddslash\! k^z \ddslash\!^{d'}\!k_\perp 
  \frac{G^0(k_\perp) |2 k^z|^{-\eta} \nu^\eta}
  {[-2k^z \minus \Delta_1\minus \bar\Delta_1'\plus i0][-\bar\Delta_1\minus \bar\Delta_1']}
  \nn\\
 &\hspace{0.2cm}
  = -i S^\gamma\, \frac{n\mcdot p_2\, n\mcdot(\bar P\minus p_2)}
  {n\mcdot \bar P \, \vec p_{2\perp}^{\:2}} 
  \int\!\! \ddslash\!^{d'}\!k_\perp 
  \frac{G^0(k_\perp) }{(\vec k_\perp -\vec p_{2\perp})^2}
  \Big( \kappa_\eta \frac{\eta}{2} \Big)\!\!
  \int_{-\infty}^{+\infty}\!\!\!\! \ddslash\! k^z dx\, d\alpha\,
  \theta(\alpha)\, |x|^{-1+\eta}\, e^{i\alpha(-k^z+\Delta)+ik^z x}
\nn\\
 &=  S^\gamma 
  \int\!\! \ddslash\!^{d'}\!k_\perp  G^0(k_\perp) E(p_{2\perp}-k_\perp) 
\Big( \kappa_\eta \frac{\eta}{2} \Big)\!\!
 \int_{-\infty}^{+\infty}\!\!\!\!  dx\,
  \theta(x)\, |x|^{-1+\eta}\, e^{i\alpha\Delta}
\nn\\
 &= \frac{S^\gamma }{2}
  \int\!\! \ddslash\!^{d'}\!k_\perp 
 \int\!\! d^{d'}b_\perp  e^{-i\vec k_\perp\cdot \vec b_\perp}
   \tilde G^0(b_\perp\!)
   \int\!\! d^{d'}b_\perp' e^{-i(\vec p_{2\perp}-\vec k_\perp)\cdot \vec b_\perp'}  \tilde E(b_\perp')
\nn\\
 &= S^\gamma 
 \int\!\! d^{d'}b_\perp \: e^{i\vec p_{2\perp}\cdot \vec b_\perp}
  \,\Big[ \frac{1}2 \tilde G^0(b_\perp\!)\Big]\, \tilde E(-b_\perp)
  \,.
\end{align}
The dependence on $\Delta = -(\Delta_1+\bar\Delta_1')/2$ drops out at leading order in $\eta$. The function $E(p_{2\perp})$ appearing here is defined in \eq{Edis}. Continuing with the corresponding calculation with $N$ Glauber exchanges between these collinear lines we have
\begin{align} \label{eq:ASdis_Nrungs}
& \raisebox{-.7cm}{
\includegraphics[width=0.18\columnwidth]{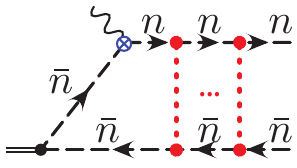}
}
 =i^{2N+1} (2)^N S^\gamma
  \!\! \int\!\! \ddslash\!^{d}\!k_1\cdots  \ddslash\!^{d}\!k_N\,
  \frac{G^0(k_{1}^\perp)G^0(k_{2}^\perp\minus k_{1}^\perp)\cdots 
    G^0(k_{N}^\perp\minus k_{N-1}^\perp)}
  { [k_1^+\minus\Delta_1\plus i0]\cdots [k_N^+\minus\Delta_N\plus i0] }
  \nn\\[-7pt]
 &\hspace{4cm} \times
  \frac{|2 k_1^z|^{-\eta}|2 k_2^z\minus 2k_1^z|^{-\eta} \cdots
    |2 k_N^z\minus 2k_{N\minus 1}^z|^{-\eta}\: \nu^{N\eta} }
 { [-k_1^- \minus \bar\Delta_1'\plus i0]\cdots 
   [- k_N^- \minus \bar\Delta_N'\plus i0][k_N^-\minus\bar\Delta_N\plus i0]}
\nn\\[6pt]
& = i\, 2^N (i)^N S^\gamma 
 \!\! \int\!\! \ddslash\!^{d\minus 1}\!k_1\cdots  \ddslash\!^{d\minus 1}\!k_N\,
 \frac{G^0(k_{1}^\perp) \cdots  
  G^0(k_{N}^\perp\minus k_{N-1}^\perp) \: 
  |2 k_1^z|^{-\eta} \cdots |2 k_N^z\minus 2k_{N\minus 1}^z|^{-\eta}\: \nu^{N\eta} }
 { [-\bar\Delta_N\minus\bar\Delta_N']
 [-2k_1^z\plus 2\Delta_{1 1'}\plus i0] \cdots [-2k_N^z \plus 2 \Delta_{NN'} \plus i0] }
 \nn\\
& =  S^\gamma 
 \!\! \int\!\! \ddslash\!^{d'}\!k_{1\perp}\cdots  \ddslash\!^{d'}\!k_{N\perp}\,
 G^0(k_{1}^\perp)G^0(k_{2}^\perp\minus k_{1}^\perp) \cdots  
 G^0(k_{N}^\perp\minus k_{N-1}^\perp)   E(p_{2\perp}\minus k_{N\perp}) 
\Big(\kappa_\eta \frac{\eta}{2} \nu^\eta\Big)^N 
 \nn\\*
 &\hspace{0.1cm} \times 
 \!\!\int_{-\infty}^{+\infty}\!\!\!\!\!\!
  \ddslash\! k_1^z\cdots \ddslash\! k_N^z
  dx_1\cdots dx_N \, d\alpha_1 \cdots d\alpha_N
  \frac{\theta(\alpha_1)\cdots\theta(\alpha_N)}
  {| x_1\cdots x_N |^{1-\eta} } \,
 e^{i(x_1-x_2-\alpha_1)k_1^z+ \ldots+ i(-\alpha_N+x_N)k_N^z }\:
 e^{i\alpha_1 \Delta_{1 1'}+\ldots}
  \nn\\
& =   S^\gamma 
 \!\! \int\!\! \ddslash\!^{d'}\!k_{1\perp}\cdots  \ddslash\!^{d'}\!k_{N\perp}\,
 G^0(k_{1}^\perp)G^0(k_{2}^\perp\minus k_{1}^\perp) \cdots  
 G^0(k_{N}^\perp\minus k_{N-1}^\perp)   E(p_{2\perp}\minus k_{N\perp}) 
 \nn\\
 &\hspace{0.1cm} \times \!
 \Big(\kappa_\eta \frac{\eta}{2} \nu^\eta\Big)^N \!\!
  \int_{-\infty}^{+\infty}\!\!\!\!\!\! dx_1\cdots dx_N \, 
  \frac{\theta(x_1\minus x_2)\theta(x_2\minus x_3)\cdots\theta(x_{N-1}\minus x_{N})\theta(x_N)}  {| x_1\cdots x_N |^{1-\eta} } \,
 e^{i(x_1-x_2) \Delta_{1 1'}+\ldots  +i x_N\Delta_{N N'} }
 \nn\\
& =  S^\gamma 
 \!\! \int\!\! \ddslash\!^{d'}\!k_{1\perp}\cdots  \ddslash\!^{d'}\!k_{N\perp}\,
 G^0(k_{1}^\perp)G^0(k_{2}^\perp\minus k_{1}^\perp) \cdots  
 G^0(k_{N}^\perp\minus k_{N-1}^\perp)  E(p_{2\perp}\minus k_{N\perp})  
 \bigg[ \frac{1}{N!\, 2^N} + {\cal O}(\eta) \bigg]
 \nn\\
& = S^\gamma
   \int\!\! d^{d'}b_\perp  e^{i\vec p_{2\perp}\cdot \vec b_\perp}\,
   \frac{1}{N!} \bigg[ \frac{1}{2} \tilde G^0(b_\perp\!) \bigg]^N \,\tilde E(-b_\perp)
 \nn\\
& = S^\gamma
   \int\!\! d^{d'}b_\perp  e^{i\vec p_{2\perp}\cdot \vec b_\perp}\,
   \frac{1}{N!} \bigg[ \frac{i\phi(b_\perp)}{2} \bigg]^N \,\tilde E(-b_\perp)
 \,.
\end{align}
Here $\bar\Delta_N = \bn\cdot p_2 + (\vec k_{N}^\perp - \vec p_{2\perp})^2/n\cdot(\bar P-p_2)$ and $\bar\Delta_N' = -\bn\cdot p_2 + (\vec k_{N}^\perp - \vec p_{2\perp})^2/n\cdot p_2$, and the sum $\bar\Delta_N+\bar\Delta_N'$ contributed to $E(p_{2\perp}-k_{N\perp})$. At an intermediate step we see that the longitudinal coordinates are ordered as $0 < x_N < x_{N-1} <\cdots < x_1$, that is after the hard scattering vertex at $x=0$. As $\eta\to 0$ the ${\cal O}(\eta^0)$ result once again comes from the limit where all $x_j\to 0$.  Summing over the number of Glauber exchanges in \eq{ASdis_Nrungs} from $N=0$ to $N=\infty$ we get
\begin{align}
& \raisebox{-0.9cm}{ 
	\includegraphics[width=0.18\columnwidth]{figs/enddis_AS_sum}
  }
 = S^\gamma
   \int\!\! d^{d'}b_\perp  e^{i\vec p_{2\perp}\cdot \vec b_\perp}\:
   e^{i\phi(b_\perp)/2} \:\tilde E(-b_\perp) 
\,.
\end{align}
Thus the phase for the active-spectator exchanges in this case is $\phi(b_\perp)/2$.

\subsection{Exponentiation for Active-Active Exchanges}  
\label{app:AAexp}

In this appendix we derive the result for iterated Glauber exchange on active-active lines. Unlike for spectator-spectator scattering, this result requires the $\eta$-regulator both for the single exchange integral, and for the ladder iterations, therefore it cannot simply be inferred from the exponentiation result in \sec{exponentiation}.

Recall that the Glauber loop vanishes for active lines in the hard-scattering case (DIS), so we only have to consider the annihilation case (Drell-Yan).  We repeat the single Glauber exchange calculation in \eq{AAG}, but carry out the intermediate regulator dependent integral in Fourier space following the notation introduced in \sec{exponentiation}, in order to setup the procedure for the $N$-exchange diagram. Writing the final result in the transverse Fourier space we have
\begin{align}
& \raisebox{-1cm}{ 	
\includegraphics[width=0.18\columnwidth]{figs/end_AA}
  }
 = 2i\,S^\gamma \, E(p_{1\perp},p_{2\perp})
  \int_{-\infty}^{+\infty}\!\! \ddslash\! k^z \ddslash\!^{d'}\!k_\perp 
  \frac{G^0(k_\perp) |2 k^z|^{-\eta} \nu^{\eta} }
  {[2k^z \minus\Delta_1'\minus \bar\Delta_1\plus i0]}
\nn\\
&\hspace{0.4cm}
 = S^\gamma\, E(p_{1\perp},p_{2\perp})
   \!\int\! \ddslash\!^{d'}\!k_\perp G^0(k_\perp\!)
   \Big(\kappa_\eta \frac{\eta}{2}\Big)\! 
    \int_{-\infty}^{+\infty}\!\! \ddslash\! k^z\, dx\, d\alpha\, 
   \theta(\alpha) |x|^{-1+\eta} e^{i\alpha(k^z\plus \Delta)-ik^z x}
\nn\\
&\hspace{0.4cm}
  = S^\gamma\, E(p_{1\perp},p_{2\perp})
  \!\int\! \ddslash\!^{d'}\!k_\perp G^0(k_\perp\!)
   \Big(\kappa_\eta \frac{\eta}{2}\Big)\! 
    \int_{-\infty}^{+\infty}\!\!  dx\,  
   \theta(x) |x|^{-1+\eta} e^{i x \Delta}
\nn\\
&\hspace{0.4cm}
  = S^\gamma\, E(p_{1\perp},p_{2\perp})
  \!\int\! \ddslash\!^{d'}\!k_\perp G^0(k_\perp\!)
   \Big(\kappa_\eta \frac{\eta}{2}\Big)\! 
   \bigg[ \frac{1}{\eta} + {\cal O}(\eta^0) \bigg]
\nn\\
&\hspace{0.4cm}
  = S^\gamma\, E(p_{1\perp},p_{2\perp})\
   \Big[ \frac{1}{2}\, \tilde G^0(b_\perp=0) \Big]
  \,.
\end{align}
Recall that we use the notation $d'=d-2$, and that the Fourier transform $\tilde G^0(0)=i\phi(0)$ with $\phi(0)$ given by \eq{phi}. The end factor $E(p_{1\perp},p_{2\perp})$ is defined in \eq{end_tree}.  The dependence on $\Delta=-(\Delta_1'+\bar\Delta_1)/2$ drops out at leading order in $\eta$.  

Next we carry out the analogous calculation but with $N$ Glauber exchanges between these two lines. Closing all the $k_i^0$ contours above we have
\begin{align} \label{eq:AA_Nrungs}
& \raisebox{-1cm}{ 	
\includegraphics[width=0.18\columnwidth]{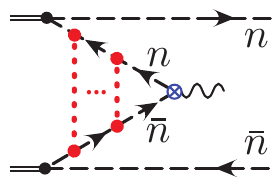}
  }
 = (-2)^N S^\gamma\, E(p_{1\perp},p_{2\perp})
  \!\! \int\!\! \ddslash\!^{d}\!k_1\cdots  \ddslash\!^{d}\!k_N\:
    G^0(k_{1}^\perp)G^0(k_{2}^\perp\minus k_{1}^\perp)\cdots 
    G^0(k_{N}^\perp\minus k_{N-1}^\perp)
  \nn\\[-15pt]
 &\hspace{4cm} \times
  \frac{|2 k_1^z|^{-\eta}|2 k_2^z\minus 2k_1^z|^{-\eta} \cdots
    |2 k_N^z\minus 2k_{N\minus 1}^z|^{-\eta}\: \nu^{N\eta} }
  { [k_1^-\minus\bar\Delta_1\plus i0]\cdots [k_N^-\minus\bar\Delta_N\plus i0]
  [- k_1^+ \minus \Delta_1\plus i0]\cdots [- k_N^+ \minus \Delta_N\plus i0]}
\nn\\[6pt]
& = (2i)^N  S^\gamma E(p_{1\perp},p_{2\perp})
 \!\! \int\!\! \ddslash\!^{d\minus 1}\!k_1\cdots  \ddslash\!^{d\minus 1}\!k_N\,
 \frac{G^0(k_{1}^\perp)\cdots   G^0(k_{N}^\perp\minus k_{N-1}^\perp) \,
  |2 k_1^z|^{-\eta} \cdots |2 k_N^z\minus 2k_{N\minus 1}^z|^{-\eta}\:  } 
  { [2k_1^z\plus 2\Delta_{1 \bar 1}\plus i0] \cdots [2k_N^z \plus 2 \Delta_{N \bar N} \plus i0]\, \nu^{-N\eta} }
 \nn\\
%
& =   S^\gamma \, E(p_{1\perp},p_{2\perp})
 \!\! \int\!\! \ddslash\!^{d'}\!k_{1\perp}\cdots  \ddslash\!^{d'}\!k_{N\perp}\,
 G^0(k_{1}^\perp)G^0(k_{2}^\perp\minus k_{1}^\perp) \cdots  
 G^0(k_{N}^\perp\minus k_{N-1}^\perp)  
\Big(\kappa_\eta \frac{\eta}{2} \nu^\eta\Big)^N 
 \nn\\*
 &\hspace{0.1cm} \times 
 \!\!\int_{-\infty}^{+\infty}\!\!\!\!\!\!
  \ddslash\! k_1^z\cdots \ddslash\! k_N^z
  dx_1\cdots dx_N \, d\alpha_1 \cdots d\alpha_N
  \frac{\theta(\alpha_1)\cdots\theta(\alpha_N)}
  {| x_1\cdots x_N |^{1-\eta} } \,
 e^{i(x_1-x_2+\alpha_1)k_1^z+ \ldots+ i(\alpha_N+x_N)k_N^z }\:
 e^{i\alpha_1 \Delta_{1 \bar 1}+\ldots}
  \nn\\
& =  S^\gamma \, E(p_{1\perp},p_{2\perp})
 \!\! \int\!\! \ddslash\!^{d'}\!k_{1\perp}\cdots  \ddslash\!^{d'}\!k_{N\perp}\,
 G^0(k_{1}^\perp)G^0(k_{2}^\perp\minus k_{1}^\perp) \cdots  
 G^0(k_{N}^\perp\minus k_{N-1}^\perp)  
 \nn\\
 &\hspace{0.1cm} \times \!
 \Big(\kappa_\eta \frac{\eta}{2} \nu^\eta\Big)^N \!\!
  \int_{-\infty}^{+\infty}\!\!\!\! dx_1\cdots dx_N \, 
  \frac{\theta(x_2\minus x_1)\theta(x_3\minus x_2)\cdots\theta(x_{N}\minus x_{N-1})\theta(-x_N)}  {| x_1\cdots x_N |^{1-\eta} } \,
 e^{i(x_2-x_1) \Delta_{1 \bar 1}+\ldots  -i x_N\Delta_{N\bar N} }
 \nn\\
& =  S^\gamma \, E(p_{1\perp},p_{2\perp})
  \bigg[ \int\!\! \ddslash\!^{d'}\!k_{\perp}\, G^0(k_\perp) \bigg]^N  
 \bigg[ \frac{1}{N!\, 2^N} + {\cal O}(\eta) \bigg]
 \nn\\
& = S^\gamma\, E(p_{1\perp},p_{2\perp})
   \frac{1}{N!} \bigg[ \frac{1}2 \tilde G^0(b_\perp=0) \bigg]^N 
 \nn\\
& = S^\gamma\, E(p_{1\perp},p_{2\perp})
   \frac{1}{N!} \bigg[ \frac{i\phi(0)}{2} \bigg]^N 
 \,.
\end{align}
At intermediate steps we defined $\Delta_{1 \bar 1}=-(\Delta_1+\bar\Delta_1)/2$, etc., but to leading order in $\eta$ the answer is independent of these factors. Summing \eq{AA_Nrungs} over the number of Glauber exchanges from $N=0$ to $N=\infty$ we get
\begin{align}
& \raisebox{-0.9cm}{ 
	\includegraphics[width=0.18\columnwidth]{figs/end_AA_sum}
  }
 = S^\gamma\, E(p_{1\perp},p_{2\perp}) \:
   e^{i\phi(0)/2} 
\,.
\end{align}
Thus we see that the phase for this sum of active-active exchanges is $\phi(0)/2$.


\bibliographystyle{JHEP}
\bibliography{../glauberbib}

\end{document}